\documentclass{thesis}
\usepackage{thesis}

\makeatletter
\@ifpackageloaded{hyperref}{%
\hypersetup{%
pdftitle = {Wrapping effects in supersymmetric gauge theories},
pdfsubject = {PhD thesis},
pdfkeywords = {Wrapping, Finite-size, Integrability, Supersymmetry},
pdfauthor = {Frank}
}
}{}
\makeatother

\begin{document}
\title{Wrapping effects in supersymmetric gauge theories}
\author{Francesco Fiamberti}

\begin{fmffile}{graphs}
\fmfcmd{
wiggly_len := 2mm;
vardef wiggly expr p_arg =
 save wpp,len;
 numeric wpp,alen;
 wpp = ceiling (pixlen (p_arg, 10) / wiggly_len);
 len=length p_arg;
 for k=0 upto wpp - 1:
  point arctime k/(wpp-1)*arclength(p_arg) of p_arg of p_arg
    {direction arctime k/(wpp-1)*arclength(p_arg) of p_arg of p_arg rotated wiggly_slope} ..
  point  arctime (k+.5)/(wpp-1)*arclength(p_arg) of p_arg of p_arg
 {direction arctime (k+.5)/(wpp-1)*arclength(p_arg) of p_arg of p_arg rotated - wiggly_slope} ..
 endfor
 if cycle p_arg: cycle else: point infinity of p_arg fi
enddef;
}
\fmfcmd{%
marksize=2mm;
def draw_mark(expr p,a) =
  begingroup
    save t,tip,dma,dmb; pair tip,dma,dmb;
    t=arctime a of p;
    tip =marksize*unitvector direction t of p;
    dma =marksize*unitvector direction t of p rotated -45;
    dmb =marksize*unitvector direction t of p rotated 45;
    linejoin:=beveled;
    draw (-.5dma.. .5tip-- -.5dmb) shifted point t of p;
  endgroup
enddef;
style_def derplain expr p =
    save amid;
    amid=.5*arclength p;
    draw_mark(p, amid);
    draw p;
enddef;
def draw_point(expr p,a) =
  begingroup
    save t,tip,dma,dmb,dmo; pair tip,dma,dmb,dmo;
    t=arctime a of p;
    tip =marksize*unitvector direction t of p;
    dma =marksize*unitvector direction t of p rotated -45;
    dmb =marksize*unitvector direction t of p rotated 45;
    dmo =marksize*unitvector direction t of p rotated 90;
    linejoin:=beveled;
    draw (-.5dma.. .5tip-- -.5dmb) shifted point t of p withcolor 0white;
    draw (-.5dmo.. .5dmo) shifted point t of p;
  endgroup
enddef;
style_def derplainpt expr p =
    save amid;
    amid=.5*arclength p;
    draw_point(p, amid);
    draw p;
enddef;
style_def dblderplain expr p =
    save amidm;
    save amidp;
    amidm=.5*arclength p-0.75mm;
    amidp=.5*arclength p+0.75mm;
    draw_mark(p, amidm);
    draw_point(p, amidp);
    draw p;
enddef;
}

\begin{frontmatter}
\begin{flushright}
\end{flushright}
\mbox{ }
\vspace{7ex}
\Large
\begin {center}     
\renewcommand{\thefootnote}{\fnsymbol{footnote}}
{\bf
Wrapping effects in supersymmetric gauge theories\footnote[1]{Based on the author's Ph.D. thesis at Universit\`a degli Studi di Milano}
}
\end {center}

\renewcommand{\thefootnote}{\fnsymbol{footnote}}

\large
\vspace{1cm}
\centerline{Francesco Fiamberti\footnote[7]{\noindent \tt
francesco.fiamberti@mi.infn.it }}
\vspace{4ex}
\normalsize
\begin{center}
\emph{Dipartimento di Fisica, Universit\`a degli Studi di Milano, \\
Via Celoria 16, 20133 Milano, Italy}\\
\vspace{0.2cm}
\emph{INFN--Sezione di Milano,\\
Via Celoria 16, 20133 Milano, Italy}\\
\end{center}
\vspace{0.5cm}
\rm
\begin{center}
\textbf{Abstract}
\end{center}
Several perturbative computations of finite-size effects, performed on the gauge side of the AdS/CFT correspondence by means of superspace techniques, are presented. First, wrapping effects are analyzed in the standard $\N=4$ theory, by means of the calculation of the four-loop anomalous dimension of the Konishi operator. Then, a similar computation at five loops is described. Afterwards, finite-size effects are studied in the $\beta$-deformed case, where thanks to the reduced number of supersymmetries the simpler class of single-impurity operators can be considered, so that the leading corrections to the anomalous dimensions at generic order can be reduced to the computation of a class of integrals. Explicit results are given up to eleven loops. A further chapter is dedicated to the computation of the leading finite-size effects on operators dual to open strings. In the end, some comments are made and proposals for future developments are discussed.
\normalsize 
\noindent 
\vspace{1cm}
\vfill
\thispagestyle{empty}


\begin{acknowledgements} 
First of all, I would like to thank my advisor, Dr. Alberto Santambrogio, for his constant helpfulness and careful supervision of my work, for his useful advice and for all our interesting and fruitful discussions. 

I am also very grateful to my former advisor, Prof. Daniela Zanon, for all her advice and for directing me to challenging research topics.

Then, I would like to tank Christoph Sieg, for his very useful explanations, for all our interesting discussions and for all the work done together. 
\end{acknowledgements}

\tableofcontents

\end{frontmatter}

\begin{mainmatter}
\chapter{Introduction}
\pagenumbering{arabic}
\label{chapter:introduction}
One of the main open problems in modern theoretical physics concerns the search for a quantum theory of gravity. With the current knowledge, gravitational interactions can be successfully described by the theory of General Relativity on large scales, where they are dominating with respect to the other three fundamental forces, namely the strong, electromagnetic and weak ones, but they still lack a quantum description, which would be needed when studying systems with large masses at subatomic scales, where gravitational effects cannot be neglected. On the other hand, the other three fundamental forces can be consistently described in terms of quantum field theories, and have been unified into the Standard Model of elementary particles. However, the standard quantization procedure that can be successfully applied to these interactions fails with gravity because the latter is not renormalizable.

The typical example of a system that cannot be fully studied without a complete quantum theory of gravity is a black hole. Near the singularity, the gravitational force becomes relevant even at small scales, and the standard General Relativity description breaks down. Moreover, an exact analysis of the process of Hawking radiation emission and black hole evaporation requires that quantum effects on the gravitational background too are taken into consideration.
Thus, a quantum theory of gravity is indeed needed. Since the standard quantization procedure cannot be used, new approaches have been considered and are under research, the most important ones being those based on String Theory on one side and on Loop Quantum Gravity on the other one. The former has the additional valuable feature that, besides leading to a quantum description of gravitational interactions, it would offer a unified description of all the four fundamental interactions. In this thesis, only the String Theory approach will be considered.

In the framework of String Theory, the standard concept of point particle is extended by introducing new fields to describe extended objects, namely strings and branes. Then, the techniques of quantum field theory are used to quantize the model on a fixed gravitational background. During the years, several improvements have been obtained, in particular with the introduction of supersymmetry. Different formulations of the theory are possible, depending on the matter content and the constraints on the fields, but a feature that is common to all the possibilities is the need for additional spacetime dimensions, whose number is fixed to be $26$ in the older bosonic theory and $10$ in the most recent superstring formulations. The traditional four-dimensional description is then recovered by compactifying the theory on some appropriate manifold. In any case, the exact quantization of the theory on a generic curved background is a very difficult task.

A big progress in the development of String Theory was obtained with the proposal of the AdS/CFT conjecture~\cite{Maldacena:1997re}, claiming the equivalence of type-IIB superstrings on the $AdS_5\times S^5$ background and $\N=4$ Super Yang-Mills (SYM) in four spacetime dimensions, which is a non-Abelian gauge field theory with $\N=4$ supersymmetries, invariant under conformal transformations. This duality is very interesting because it relates opposite regimes of the two related theories, so that the strong-coupling behaviour of one of the two theories can be studied by performing a simpler analysis on the opposite side of the correspondence. In particular, the strong-coupling regime of $\N=4$ SYM corresponds to the supergravity limit of the string theory. The drawback of this strong/weak nature of the duality is that it makes the conjecture very difficult to prove, because in general only perturbative computations are simple enough to deal with. Up to now, several strong tests have been performed, all confirming the validity of the correspondence, but a rigorous proof is still missing. Anyway, the conjecture has been later extended to relate other pairs of theories, modifying the string background on one side and the field theory action and matter content on the opposite side.

In the last few years, new findings appeared, concerning the possible integrability of both sides of the duality. Such results revealed to be very useful to perform further tests of the conjecture and may even lead to a final demonstration in the future. Their power derives from the fact that in an integrable theory an infinite number of conserved charges exists, from which in principle the exact spectrum can be determined. So, if the two theories are really integrable, the calculation of the full spectra may be within reach, and the comparison of the results would represent a very strong test for AdS/CFT. The first integrability properties~\cite{Minahan:2002ve} were discovered in the simplest approximations of the two dual theories, namely one-loop $\N=4$ SYM on one side and classic strings on the other one. In particular, at first only the reduction of $\N=4$ SYM to the $\sutwo$ sector, which contains operators built using only two out of the three available complex scalar fields, was considered. These simple cases suggested the very useful description according to which composite operators of the field theory are associated to states of closed spin chains. In this picture, operators whose dimensions are not protected by supersymmetry correspond to excitations, called \emph{magnons}, over the ferromagnetic ground state of the chain, which is protected. 

The integrability results have been then extended to all the other sectors~\cite{Beisert:2003jj,Beisert:2003yb,Beisert:2004ag,SchaferNameki:2004ik,Beisert:2005di} and to higher orders~\cite{Beisert:2003tq,Serban:2004jf,Beisert:2004hm,Eden:2004ua,Kotikov:2004er}, and are now believed to hold at any order~\cite{Beisert:2003xu,Beisert:2003jb,Beisert:2003ys,Staudacher:2004tk,Beisert:2005fw,Beisert:2005tm}. The conjectured all-order integrability indeed offers powerful tools for the computation of part of the spectra of anomalous dimensions in $\N=4$ SYM and of energies on the string side. However, for a fixed perturbative order $L$, they cannot be applied to composite operators with classical dimension less than $(L+1)$, because of the appearance of the so-called \emph{wrapping} interactions~\cite{Beisert:2004hm}, which are finite-size effects not encoded in the asymptotic quantities derived from integrability. In terms of Feynman diagrams, such effects correspond to interactions whose range is greater than the length of the operators, and that can thus wrap around it so that the asymptotic states required by the integrability techniques no longer exist.

A systematic study of finite-size contributions is therefore required in order to find the complete spectra. 
The first exact study of such effects on the field theory side consisted in the calculation of the four-loop anomalous dimension of a length-four descendant of the Konishi operator, performed by means of field-theoretical superspace techniques~\cite{us,uslong}. This result was soon after confirmed by an independent computation~\cite{Bajnok:2008bm} from the string side based on the L\"uscher approach~\cite{Luscher:1985dn}, thus realizing another non-trivial check of the correspondence. 
The perturbative approach has been applied also to the study of finite-size effects in the $\beta$-deformation of $\N=4$ SYM, which retains only $\N=1$ supersymmetry, and where the reduced symmetry allows a class of previously-protected operators to acquire a non-trivial anomalous dimension, providing a simpler framework for the analysis of wrapping contributions. As a consequence, perturbative calculations could be performed up to higher loop orders~\cite{betadef,Fiamberti:2008sn}. 

During the last years, finite-size effects have been studied on the string side too in a number of papers
~\cite{Ambjorn:2005wa,SchaferNameki:2006gk,Arutyunov:2006gs,SchaferNameki:2006ey,Astolfi:2007uz,Ramadanovic:2008qd,Janik:2007wt,Gromov:2008ie,Minahan:2008re,Heller:2008at,Hatsuda:2008gd,Hatsuda:2008na,Sax:2008in}, based mainly on direct string methods or on the L\"uscher approach. The computations focused in particular on the string duals of magnons of the $\sutwo$ sector, the so-called \emph{giant magnons}, which are spinning strings moving in a particular subspace of $AdS_5\times S^5$.  However, before the appearance of~\cite{Bajnok:2008bm}, the corrections were generally obtained only in the large $\lambda$ limit, so that a direct comparison with perturbative results from the gauge side was not possible. 

The L\"uscher approach was successfully applied to the study of wrapping corrections in several other situations
~\cite{Ambjorn:2005wa,Beccaria:2009eq,Bajnok:2008qj,Penedones:2008rv}, and to the computation of the five-loop anomalous dimension of the Konishi operator~\cite{Bajnok:2009vm}, but it is difficult to generalize it to deal with generic states.
More recently, some proposals appeared for a general description of finite-size effects, based on the thermodynamic Bethe ansatz
~\cite{Yang:1968rm,Zamolodchikov:1989cf,Arutyunov:2007tc,deLeeuw:2008ye,Arutyunov:2009zu,deLeeuw:2009hn,Arutyunov:2009ga,Arutyunov:2009mi,Arutyunov:2009ce,Arutyunov:2009ux,Arutyunov:2009ax,Arutyunov:2010gb,Balog:2010xa}, from which the so-called Y-system~\cite{Gromov:2009tv,Gromov:2009bc,Bombardelli:2009ns,Arutyunov:2009ur,Hegedus:2009ky} has been derived. The new techniques successfully reproduce the known wrapping contribution to the Konishi anomalous dimension, and have also been applied to analyze explicitly wrapping effects at strong coupling in the CFT~\cite{Gromov:2009zb,Roiban:2009aa,Gromov:2009tq,Gromov:2010vb}.  
Therefore, it will be important to test the new proposals against further direct results. Apart from the Konishi anomalous dimension, the L\"uscher approach has been tested on twist-two operators at four loops~\cite{Bajnok:2008qj} finding agreement with a direct perturbative result~\cite{Velizhanin:2009zz}. As for the Y-system, its prediction for the five-loop anomalous dimensions of twist-three operators in the $\sltwo$ sector matches the value conjectured in~\cite{Beccaria:2009eq}, which was confirmed by means of a perturbative calculation in~\cite{usfive}.

The thermodynamic Bethe ansatz and Y-system have not been extended yet to the $\beta$-deformed theory, which thus lacks a proposal for the general description of finite-size effects. Some checks of the correctness of the known perturbative results are however possible in particular cases, starting from the results for the undeformed case on the string side~\cite{Gunnesson:2009nn,Beccaria:2009hg}.

This thesis contains the presentations of the perturbative computations of finite-size effects that have been performed on the field-theory side by means of $\N=1$ superspace techniques, both in $\N=4$ SYM~\cite{us,uslong,usfive} and in its deformed version~\cite{betadef,Fiamberti:2008sn}. In Chapter~\ref{chapter:N4SYM}, the main features of $\N=4$ SYM and of the deformed theory are presented, and a general result on the divergence of relevant Feynman supergraphs is demonstrated. Chapter~\ref{chapter:integrability} contains the summary of the main integrability results and a more detailed introduction to wrapping effects. In Chapter~\ref{chapter:fourloop}, the computation of the four-loop anomalous dimension of the Konishi operator, comprehensive of finite-size effects, is performed. Chapter~\ref{chapter:fiveloop} deals with the similar analysis of five-loop wrapping effects on length-five operators, and the final result is found to agree with the prediction of the Y-system. Chapter~\ref{chapter:wrapbetadef} is dedicated to the analysis of wrapping contributions in the $\beta$-deformed theory, where the power of superspace techniques and the possibility to study a simpler class of operators allow to find explicitly anomalous dimensions up to any desired order. In Chapter~\ref{chapter:open}, leading finite-size effects in the undeformed theory are studied on a different class of operators, dual to open strings, and a perturbative calculation is presented confirming the results obtained in~\cite{Correa:2009mz} using the thermodynamic Bethe ansatz. Finally, in Chapter~\ref{chapter:conclusions}, some conclusions and comments are presented. 

The most technical details of each calculation have been collected in several appendices. More precisely, Appendix~\ref{app:non-maximal} contains the demonstration of a quite general result on the cancellation of particular classes of Feynman supergraphs. In Appendices~\ref{app:fourR5} and~\ref{app:fourloopwrap}, all the explicit results needed for the four-loop computation of Chapter~\ref{chapter:fourloop} are presented. Appendices~\ref{app:fiveR6} and~\ref{app:fivewrap} contain the results for all the diagrams that are relevant for the five-loop analysis of Chapter~\ref{chapter:fiveloop}. In Appendix~\ref{app:gegenbauer} the Gegenbauer polynomial $x$-space technique for momentum integrals is shortly reviewed. Finally, in Appendix~\ref{app:triangles} the generalized triangle rules needed in Chapter~\ref{chapter:wrapbetadef} are derived.

\chapter{\texorpdfstring{$\mathcal{N}=4$}{N=4} Super Yang-Mills}
\label{chapter:N4SYM}
In this chapter, the main features of the four-dimensional $\mathcal{N}=4$ Super Yang-Mills (SYM) theory with $SU(N)$ gauge group are presented. First, the action and Feynman rules are given using the superspace formalism, both in the standard and in the $\beta$-deformed case.
Then, the proof of a useful result for the perturbative computation of anomalous dimensions is given. Finally, the AdS/CFT correspondence is outlined.

\section{The model}
\label{sec:model}
The four-dimensional $\mathcal{N}=4$ SYM theory with gauge group $SU(N)$ is a maximally supersymmetric gauge theory that is believed to be ultraviolet finite at all the perturbative orders, and therefore conformally invariant both at the classical and at the quantum level~\cite{Ferrara:1974pu,Poggio:1977ma,Caswell:1980ru,Caswell:1980yi,Grisaru:1980jc,Grisaru:1980nk,Gates:1983nr}. 

The matter content is completely fixed by the high number of supersymmetries: the theory includes, as physical particles, one spin-1 Yang-Mills vector, four spin-1/2 Majorana spinors, three scalar and three pseudo-scalar fields. The six spin-0 fields are organized into the \textbf{6} representation of the R-symmetry group $SU(4)$. All the particles are massless and transform under the adjoint representation of the gauge group $SU(N)$. 

The full symmetry group for $\N=4$ SYM is the supergroup $SU(2,2\vert4)$. It is the result of the combination of the conformal group $SO(2,4)\sim SU(2,2)$ (generated by translations, Lorentz transformations, dilatations and special conformal transformations), the $\N=4$ Poincar\'e supersymmetries, the $SO(6)_R\sim SU(4)_R$ R-symmetry group for the Poincar\'e supersymmetries and the additional conformal supersymmetries that are required to close the superalgebra.
The theory is conjectured to be dual to type II B superstrings on the $AdS_5\times S^5$ background, according to the first and most studied version of the AdS/CFT correspondence~\cite{Maldacena:1997re}, which will be briefly presented in Section~\ref{sec:AdSCFT}.  

As it generally happens in gauge theories, perturbative computations based on the standard component-field approach are very difficult to deal with, especially at high loop orders. Therefore, as usual when supersymmetry is present, it is very useful to resort to the superspace formalism~\cite{Gates:1983nr}, where all the ordinary fields of a supermultiplet are combined into a single superfield. Perturbative calculations can then be dramatically simplified by the use of Feynman supergraphs, each encoding the information on a large number of standard diagrams involving ordinary fields. Moreover, the superspace approach often helps to discover simplifications and cancellations directly related to supersymmetry.
\subsection{The action}
\label{subsec:action}
In the case of $\mathcal{N}=4$ SYM, it is convenient to take advantage of the $\mathcal{N}=1$ superspace description, where the field content is given in terms of one real vector superfield $V$ and three chiral superfields $\Phi^i$. The three scalar superfields, organized into the $\mathbf{3}\times\mathbf{\bar{3}}$ representation of $SU(3)\subset SU(4)$, will also be denoted explicitly by $\phi$, $Z$ and $\psi$.  
The action for the theory is given, in the notation of~\cite{Gates:1983nr}, by
\begin{equation}
\label{action}
\begin{aligned}
S &= S_{GF}+\int\de^4 x\de^4 \theta \, \tr \left(e^{-gV} \bar \Phi_i e^{gV}
\Phi^i\right) + \frac 1{2g^2} \int \de^4 x \de^2 \theta
\,\tr \left(W^\alpha W_\alpha\right)\\
&\phantom{{}={}}
+i \frac{g}{3!}  \int\de^4 x\de^2 \theta \,\epsilon^{ijk}\,\tr \left(\Phi_i
\left[\Phi_j , \Phi_k\right]\right) + 
i \frac{g}{3!}  \int\de^4 x\de^2 \bar\theta \,\epsilon^{ijk}\,\tr \left(\bar\Phi_i
\left[\bar\Phi_j , \bar\Phi_k\right]\right)
\col
\end{aligned}
\end{equation}
where $W_\alpha = i\bar \covder^2 \left(e^{-gV} \covder_\alpha\,e^{gV}\right)$.
The gauge-fixing part $S_{GF}$ is 
\begin{equation}
\label{actionGF}
\begin{aligned}
S_{GF} &= -\frac{1}{16\alpha}\int\de^4 x\de^4 \theta\,\tr\left[\left(\covder^2V\right)\left(\bar\covder^2V\right)\right] \\
&\phantom{{}={}}+\frac{1}{\alpha}\int\de^4 x\de^4 \theta\,\tr\left[(\bar{c}'-c')L_{gV/2}\left[(c+\bar{c})+\mathrm{coth}(L_{gV/2})(c-\bar{c})\right]\right]
\col
\end{aligned}
\end{equation}
where $c$ and $c'$ are chiral ghost fields, $\alpha$ is the gauge parameter and $L_Y X=[Y,X]$ is the Lie derivative. 

The superfields can be written as
$V=V^aT_a$, $\Phi^i=\Phi_i^aT_a$, $c=c^a T_a$, {\small $i=1,2,3$}, 
where the $T_a$ are matrices that obey the rules of the $SU(N)$ algebra
\begin{equation}
\label{Tcomm}
[T_a, T_b] = i f_{abc} T_c \col
\end{equation}
and are normalized as
\begin{equation}
\tr(T_a T_b) =\delta_{ab}
\pnt
\end{equation}
The structure constants $f_{abc}$ can be written in terms of the $T_a$ according to the inverse of~\eqref{Tcomm}
\begin{equation}
f_{abc}=-i\,\mathrm{tr}([T_a,T_b]T_c) \col
\end{equation}
which, together with the fundamental relation
\begin{equation}
T_a^{ij}T_a^{kl}=\left(\delta_{il}\delta_{jk}-\frac{1}{N}\delta_{ij}\delta_{kl}\right) \col
\end{equation}
allows to compute the colour structure of Feynman diagrams.

In the action~\eqref{action}, $g$ is the gauge coupling of the theory. It is useful to introduce also the rescaled 't~Hooft coupling $\lambda$, defined as 
\begin{equation}
\lambda=\frac{g^2N}{(4\pi)^2} \pnt
\end{equation}
The Feynman rules for supergraphs can be derived from the action~\eqref{action}: in momentum space, the superfield propagators are
\begin{equation}
\label{propagators}
\langle V^a V^b\rangle=-\frac{\delta^{ab}}{p^2} ,\qquad\langle\Phi^a_i\bar{\Phi}^b_j\rangle=\delta_{ij}\frac{\delta^{ab}}{p^2} \col
\end{equation}
whereas the vertex factors can be read directly from the interaction terms in~\eqref{action}, with an additional $\bcovder^2$ or $\covder^2$ for each chiral or antichiral line respectively. In the three-scalar chiral vertices, derivatives are present only on two out of the three lines, since the third double derivative must be used to build the $\de^4\theta$ integration out of the $\de^2\theta$ that is present in the third term of~\eqref{action}. The same happens with the $\de^2\bar\theta$ at antichiral vertices. For the calculations presented in this thesis, the only vertex terms that will be needed explicitly are the chiral one ($V_C$), the antichiral one ($V_A$), and the ones with a chiral and an antichiral line plus one ($V_V^{(1)}$) or two ($V_V^{(2)}$) vectors
\begin{equation}
\begin{aligned}
\label{vertices}
&V_C=-\frac{g}{3!}\epsilon^{ijk}f_{abc}\Phi^a_i\Phi^b_j\Phi^c_k\ ,& &V_A=-\frac{g}{3!}\epsilon^{ijk}f_{abc}\bar{\Phi}^a_i\bar{\Phi}^b_j\bar{\Phi}^c_k\ , \\
&V_V^{(1)}=i g f_{abc}\delta^{ij}\bar{\Phi}^a_i V^b \Phi^c_j\ ,& &V_V^{(2)}=\frac{g^2}{2}\delta^{ij}f_{adm}f_{bcm}V^a V^b \bar{\Phi}^c_i\Phi^d_j \pnt
\end{aligned}
\end{equation}
Note that, in the superspace formalism, interactions involving fermionic matter do not appear explicitly, thus greatly simplifying the calculations. Moreover, in this work ghosts, which couple only to the vector multiplet, will not be relevant.

\subsection{Anomalous dimensions}
\label{subsec:anomalous}
In $\mathcal{N}=4$ SYM, the quantum dimensions of composite operators, being related to energies of string states by the AdS/CFT correspondence, constitute a class of particularly interesting quantities for the analysis of the duality. 
Let $\{\mathcal{O}_1,\ldots,\mathcal{O}_n\}$ be a set of bare composite operators mixing under renormalization, so that their renormalized versions read
\begin{equation}
\mathcal{O}_i^{\mathrm{ren}}=\mathcal{Z}_i^j\mathcal{O}_j^{\mathrm{bare}} \col
\end{equation}
where $\mathcal{Z}_i^j$ is the one-point function matrix, which gets contributions from all the Feynman diagrams with a single insertion of one of the composite operators $\mathcal{O}_i$. A set of operators with well-defined anomalous dimensions can be found by diagonalizing this matrix.
In a dimensional regularization approach, with $d=4-2\varepsilon$ spacetime dimensions, the anomalous dimensions will be equal to the eigenvalues of the mixing matrix $\mathcal{M}$ defined by
\begin{equation}
\label{anomalous}
\gamma_k=\mathrm{eig}(\mathcal{M})_k\ ,\qquad\mathcal{M}_i^j=\lim_{\varepsilon\rightarrow0}\left[\varepsilon g\frac{\de}{\de g}\mathrm{log}\mathcal{Z}_i^j(g,\varepsilon)\right] \pnt
\end{equation}
Note that for the computation of anomalous dimensions, only the divergent part in the $1/\varepsilon$ expansion of the $\mathcal{Z}_i^j$ function is needed. This results in a great simplification thanks to the general argument described in Section~\ref{sec:proof}. \\

\subsection{Simple sectors}
The study of anomalous dimensions is simpler if one restricts to operators belonging to a sector of the theory that is closed under renormalization, because of the reduced number of different operators that can mix with each other. The most interesting sectors, also for their importance related to the integrability results described in Chapter~\ref{chapter:integrability}, are the $\sutwo$, $\sltwo$ and $\suoneone$ ones. The $\sutwo$ sector contains operators built using only two out of the three scalar superfields, for example $\phi$ and $Z$. The operators of the $\sltwo$ sector are constructed using only one kind of scalar superfield ($Z$) and covariant derivatives $\covder$. Finally, in the $\suoneone$ sector the operators contain one flavour of scalar field ($Z$) and fermions.

\subsection{The dilatation operator}
\label{subsec:dilatation}
In a conformal theory, the scaling dimensions of local operators are strictly related to the \emph{dilatation operator} $\mathcal{D}$, which represents the generator of dilatations on the operator algebra~\cite{Beisert:2004ry}. More precisely, the eigenvalues of the dilatation operator are all the possible dimensions of operators in the theory, and the corresponding eigenvectors are the composite operators that are multiplicatively renormalized. So 
\begin{equation}
\mathcal{D}\,\mathcal{O}=\Delta(\lambda)\mathcal{O}\ ,\qquad\Delta(\lambda)=\Delta_0+\gamma(\lambda) \col 
\end{equation}
where $\gamma$ is the anomalous dimension, and the classical dimension $\Delta_0$ is the eigenvalue of the tree-level component $\mathcal{D}_0$ of $\mathcal{D}$ in a perturbative expansion
\begin{equation}
\mathcal{D}(\lambda)=\sum_{k=0}^\infty\lambda^k \mathcal{D}_k \pnt
\end{equation}
For a generic theory, this point of view adds nothing to the standard approach of~\eqref{anomalous}. However, if the theory is integrable, the dilatation operator formalism turns out to be very useful, as will be explained in Chapter~\ref{chapter:integrability}. \\

\newpage
\section{The \texorpdfstring{$\beta$}{beta}-deformation of  \texorpdfstring{$\mathcal{N}=4$}{N=4} SYM}
\label{sec:betadef}
The $\beta$-deformed $\mathcal{N}=4$ SYM theory is a deformation of the standard one, preserving $\mathcal{N}=1$ supersymmetry. It can be obtained by a modification of the original $\mathcal{N}=4$ superpotential for the chiral superfields 
\begin{equation}
ig\,\tr\left(\phi\,\psi\,Z -  \phi\,Z\,\psi\right)~\longrightarrow ~ih\,\tr\left(e^{i\pi\beta} \phi\,\psi\,Z - e^{-i\pi\beta} \phi\,Z\,\psi\right)\col\qquad q\equiv e^{i\pi\beta}\col
\end{equation}
where $h$ and $\beta$ are in general complex constants. As discussed in~\cite{Leigh:1995ep}, the deformation becomes exactly marginal, so that the deformed theory is still conformally invariant, if one condition is satisfied by the constants $h$ and $\beta$. More precisely, in~\cite{Mauri:2005pa} it has been shown that for \emph{real} values of $\beta$ the deformed theory is superconformal, in the planar limit, at all perturbative orders if
\begin{equation}
h\bar{h}=g^2 \col
\end{equation}
where $g$ is the Yang-Mills coupling constant. In this thesis, only the case of real $\beta$, where $\bar{q}=1/q$, will be considered. 

The $\beta$-deformation breaks the $SU(4)$ R-symmetry of the $\mathcal{N}=4$ theory to $U(1)_R$. The other symmetries that are still present after the deformation are the $Z_3$ associated to cyclic permutations of the scalar superfields, and two non-R-symmetry $U(1)$ factors~\cite{Mauri:2006uw}. \\
The Feynman rules for the propagators and the vertices with vectors remain the same as for the undeformed theory, whereas the contributions for chiral and antichiral vertices get an additional factor of $q$ or $\bar{q}$, depending on the order of the fields
\begin{equation}
\label{betadef-Feynman}
\begin{aligned}
V_C&= - h f_{abc} (e^{i\pi\beta} \Phi_1^a\,\Phi_2^b\,\Phi_3^c - e^{-i\pi\beta} \Phi_1^a\,\Phi_3^b\,\Phi_2^c) \col \\
V_A&= - \bar{h} f_{abc} (e^{-i\pi\beta}\bar{\Phi}_1^a\, \bar{\Phi}_2^b\, \bar{\Phi}_3^c-e^{i\pi\beta} \bar{\Phi}_1^a\, \bar{\Phi}_3^b\, \bar{\Phi}_2^c) \pnt
\end{aligned}
\end{equation}

The interest in the $\beta$-deformed theory derives mainly from the fact that, even though it is still superconformal, it may present some new features because of the reduced number of supersymmetries. In particular, classes of operators exist that are protected in the undeformed theory but not in the deformed one, and that may allow to extract useful information. Moreover, according to the AdS/CFT correspondence, also the deformed theory is conjectured to have a dual, represented by superstring theory on the Lunin-Maldacena background~\cite{Lunin:2005jy}.

\section{A theorem about supergraphs}
\label{sec:proof}
The first step in the computation of a Feynman supergraph is represented by the \mbox{\emph{D-algebra}} procedure~\cite{Gates:1983nr}, in which the covariant derivatives $\covder$, $\bcovder$ acting on propagators are rearranged by means of integrations by parts at the vertices, in order to bring the diagram into a form that is suitable for the integrations on the superspace Grassmann variables. Once such integrations have been performed, one is left with a standard momentum-space integral, which can be analyzed through dimensional regularization in $d=4-2\varepsilon$ dimensions. For each integration by parts, in general several terms will be produced, leading to a rather large number of contributions to the final result. 
However, in the calculation of anomalous dimensions, only the divergent part of the expansion of a diagram in powers of $1/\varepsilon$ is needed. Hence, it would be very useful to be able to discard all the finite contributions, which turn out to represent the majority of the terms generated by integrations by parts, as soon as they appear. This is indeed possible in most of the situations thanks to the argument explained in the following.

Consider a planar Feynman supergraph with a single insertion of a composite operator made only of chiral superfields, and where the final operator is made of the same fields as the initial one. Then, a divergent contribution can be found only if, during the D-algebra, none of the covariant derivatives of type $D$ is moved out of the diagram, onto the external fields (an exception is represented by derivatives acting on propagators that do not belong to any loop, as will be explained later).
To demonstrate this assertion, consider the quantities defined in Table~\ref{tab:proof}. 
Note that from the gauge-fixing part of the action~\eqref{actionGF}, it follows that a vertex with two ghosts and any number $n$ of vectors has the same derivative structure as the one with two scalar superfields and $n$ vectors. So, in the following demonstration there is no need to consider ghosts explicitly: the numbers of possible vertices involving ghosts must be simply added to the corresponding $V_V^{(n)}$.

Several relations can be found between the numbers of the different kinds of vertices and the numbers of propagators. First of all, from each chiral or antichiral vertex, three scalar propagators start. Moreover, two scalar and $n$ vector lines start from each of the $V_V^{(n)}$ vertices. Finally, from every $\tilde{V}_V^{(n)}$ vertex, $n$ vector lines start. Since in this way all the propagators are counted twice, and by hypothesis the number of outgoing fields equals the number of fields in the composite operator, the following relations can be written
\begin{table}[t]
\capstart
\begin{tabular}{ll}
\toprule
$V_C$ & number of chiral vertices \\
$V_A$ & number of antichiral vertices \\
$V_V^{(n)}$ & number of vertices with a chiral, an antichiral and $n$ vector lines \\
$\tilde{V}_V^{(n)}$ & number of vertices with $n$ vector lines \\
$p_S$ & number of scalar propagators belonging to at least one loop \\
$p_V$ & number of vector propagators \\
$p$ & total number of propagators belonging to at least one loop \\
$p_E$ & number of scalar propagators not belonging to any loop \\
$V_{EC}$ & number of chiral vertices not belonging to any loop \\
$N_\ell$ & number of loops \\
$N_\covder$, $N_{\bcovder}$ & numbers of $\covder$ and $\bcovder$ derivatives \\
\bottomrule
\end{tabular}
\caption{Definitions needed to prove the result of Section~\ref{sec:proof}}
\label{tab:proof}
\end{table}

\begin{equation}
\label{proof:propagators}
\begin{aligned}
p_S&=\frac{1}{2}\left[3(V_C+V_A)+2\sum_{n\geq1} V_V^{(n)}\right]-p_E \col \\
p_V&=\frac{1}{2}\left[\sum_{n\geq1} nV_V^{(n)}+\sum_{m\geq3}m\tilde{V}_V^{(m)}\right] \pnt
\end{aligned}
\end{equation}
Moreover, the number of scalar propagators not belonging to any loop is equal to the one of chiral vertices with the same feature, that is $p_E=V_{EC}
$, because all the outgoing fields must be chiral.
Note in addition that for the final operator to contain the same fields as the original one, the numbers of chiral and antichiral vertices must be equal
\begin{equation}
V_C=V_A \pnt
\end{equation}

At every chiral vertex, two out of the three lines have a $\bcovder^2$. The same is true for $\covder^2$ terms at antichiral vertices. A pair $\covder^2$, $\bcovder^2$ is present at every $V_V^{(n)}$ vertex, whereas at $\tilde{V}_V^{(n)}$ vertices complex derivative structures appear, always involving two $\covder$s and two $\bcovder$s. Thus, the numbers of covariant derivatives $N_\covder$ and $N_{\covder}$ can be related to the numbers of vertices
\begin{equation}
\label{proof:ND}
\begin{aligned}
N_\covder&=4V_C+2\sum_{n\geq1}V_V^{(n)}+2\sum_{m\geq3}\tilde{V}_V^{(m)} \col \\ 
N_{\bcovder}&=4V_A+2\sum_{n\geq1}V_V^{(n)}+2\sum_{m\geq3}\tilde{V}_V^{(m)} \col
\end{aligned}
\end{equation}
and since $V_C=V_A$, one finds $N_\covder=N_{\bcovder}$. So, 
thanks to~\eqref{proof:propagators} and~\eqref{proof:ND}, the total number of propagators $p=p_S+p_V$ can be written as
\begin{equation}
\label{proof:p}
p=\frac{1}{2}\left[N_\covder+V_C+V_A+\sum_{n\geq1}n V_V^{(n)}+\sum_{m\geq3}(m-2)\tilde{V}_V^{(m)}\right]-p_E \pnt
\end{equation}
According to a simple power counting, in order to result in a momentum integral that is superficially divergent, the D-algebra must produce a number of 
momenta in the numerator of the integral equal to at least
\begin{equation}
\label{proof:bound}
2p-4N_\ell \pnt
\end{equation}
To build a momentum in the numerator, a $\covder$ (together with a $\bcovder$) is required. Moreover, for every loop a $\covder^2$ and a $\bcovder^2$ are needed to complete the D-algebra with a non-vanishing $\de^4\theta$ integration. Hence, the minimum number of $\covder$s that must be kept inside the diagram to have a divergent result is
\begin{equation}
\label{proof:boundD}
2p-2N_\ell \pnt
\end{equation}
Consider now what happens on a scalar propagator that does not belong to any loop: either a factor of $\covder^2$ is already there from the beginning, or it can be brought there by integrating by parts at the vertex at which the propagator is attached to the rest of the diagram. After performing this operation for all the $p_E$ propagators of this kind, the total number of $D$ derivatives left inside the diagram, which can thus be effectively used for the D-algebra, is
\begin{equation}
\label{proof:effectiveND}
N_\covder-2p_E \pnt
\end{equation}
It will be possible to move one more derivative of type $D$ out of the diagram only if this number exceeds the minimum number required by~\eqref{proof:boundD}, that is
\begin{equation}
\label{proof:ineq}
N_\covder-2p_E>2p-2N_\ell \col
\end{equation}
which with the use of~\eqref{proof:p} can be rewritten as
\begin{equation}
\label{proof:ineq2}
N_\ell>\frac{1}{2}\left[V_C+V_A+\sum_{n\geq1}n V_V^{(n)}+\sum_{m\geq3}(m-2)\tilde{V}_V^{(m)}\right] \pnt
\end{equation}
This condition can never be satisfied, as can be seen by using Euler's formula for planar, connected graphs
\begin{equation}
\label{proof:Euler}
\mathcal{V}-\mathcal{E}+\mathcal{F}=2 \col
\end{equation}
where $\mathcal{V}$ is the total number of vertices, $\mathcal{E}$ is the number of edges and $\mathcal{F}$ is the number of faces, including the external infinite region. For a Feynman diagram, as the operator insertion behaves as an additional vertex,
\begin{equation}
\mathcal{V}=V_C+V_A+\sum_{n\geq1}V_V^{(n)}+\sum_{m\geq3}\tilde{V}_V^{(m)}+1\ ,\qquad \mathcal{E}=p+p_E\ ,\qquad \mathcal{F}=N_\ell+1\pnt
\end{equation}
As a consequence,
\begin{equation}
N_\ell=\frac{1}{2}\left[V_C+V_A+\sum_{n\geq1}n V_V^{(n)}+\sum_{m\geq3}(m-2)\tilde{V}_V^{(m)}\right] \col
\end{equation}
which is not compatible with~\eqref{proof:ineq}.
So for the considered class of diagrams all the covariant derivatives of type $D$ are needed to produce a divergence. In the course of the D-algebra, this result allows to discard all the contributions where at least one of the derivatives would act on an external field, greatly reducing the number of relevant terms. Note that this is true also in the $\beta$-deformed theory, since the only difference in the Feynman rules with respect to the undeformed case is represented by~\eqref{betadef-Feynman}.

\section{The AdS/CFT correspondence}
\label{sec:AdSCFT}
The AdS/CFT correspondence~\cite{Maldacena:1997re} conjectures the equivalence between $\mathcal{N}=4$ SYM and type IIB superstring theory on the $AdS_5\times S^5$ background. In particular, anomalous dimensions of composite operators on the field theory side are related to energies of string states. It is a so-called strong/weak duality, since in the parameter range where one of the two theories is weakly-coupled, and can be studied perturbatively, the other one is strongly coupled, and vice-versa. On one hand, this fact makes the correspondence very interesting, because it would allow to investigate the non-perturbative regime of a theory by means of perturbative computations performed on the opposite side of the duality. On the other hand, however, it also makes the correspondence very difficult to prove. In fact, no rigorous proof of the conjecture exists at the moment, even if it has passed several non-trivial checks.

In its strongest form, the correspondence claims the exact equivalence of the two theories for any values of the parameters. Weaker formulations exist in addition, that are more tractable as they concern particular simplified limits.
The main example of such weaker versions is represented by the 't~Hooft limit, in which $N\to\infty$ while the 't Hooft coupling $\lambda$ is kept fixed. On the field theory side, non-planar Feynman diagrams are suppressed in this regime (if the classical dimensions of the operators under consideration do not grow too fast with $N$), and perturbative computations can thus be performed in the \emph{planar} limit, where all the non-planar contributions are neglected. Since only non-planar diagrams can modify the trace structure of an operator, in this regime it is enough to restrict to single-trace operators.  
In the 't~Hooft limit, the field-theory coupling $\lambda$ is fixed but arbitrary, with the case $\lambda\ll1$ corresponding to the perturbative regime of the field theory side. When $\lambda\gg1$ the field theory is strongly coupled, whereas the string side can be approximated by classical type IIB supergravity on the $AdS_5\times S^5$ background.

The first necessary condition for the possible validity of the conjecture is the matching of the two symmetry groups, which is fulfilled because both theories have a $SU(2,2\vert4)$ symmetry. The origin of this supergroup on the field theory side has been outlined briefly in Section~\ref{sec:model}. On the string side, the isometry groups $SO(2,4)$ and $SO(6)$ of the $AdS_5$ and $S^5$ parts of the background can be combined into the group $SO(2,4)\times SO(6)\sim SU(2,2)\times SU(4)$, which is the maximal bosonic subgroup of $SU(2,2\vert4)$. Moreover, it is possible to show that the whole $SU(2,2\vert4)$ is actually a symmetry for superstring theory on this background~\cite{D'Hoker:2002aw}.
Besides this matching of the symmetry groups, several further tests, all supporting the validity of the conjecture, have been performed. They rely on properties that do not depend on the coupling, such as a class of correlation functions protected from quantum corrections and the spectrum of chiral primary operators~\cite{Aharony:1999ti}.

The original formulation of the AdS/CFT correspondence has been later extended to relate new pairs of field and string theories. As an example, the $\beta$-deformation of $\N=4$ SYM is supposed to be equivalent to superstring theory on the Lunin-Maldacena background~\cite{Lunin:2005jy}, with geometry $AdS_5\times\tilde{S}^5$, where $\tilde{S}^5$ is a deformation of the five-dimensional sphere. Moreover, the duality has been applied to theories in a different number of dimensions, as in the case of three-dimensional $\N=6$ superconformal Chern-Simons theory~\cite{Aharony:2008ug}, which is believed to be equivalent to a string theory on $AdS_4\times CP^3$~\cite{Arutyunov:2008if,Stefanski:2008ik,Chen:2008qq}.

A renewed interest in the original version of the AdS/CFT correspondence arose recently, with the discovery of several results suggesting that both the involved theories may be integrable in the planar limit, as described in Chapter~\ref{chapter:integrability}. The powerful tools available for the study of integrable systems revealed to be very useful in the determination of large parts of the spectra of the two theories, and there is hope that they can be further extended to permit their full calculation, which would represent a very strong test of the validity of the correspondence, at least in the 't~Hooft limit.

\section{Conclusions}
In this chapter the main properties of the $\mathcal{N}=4$ SYM theory and of its $\beta$-deformed version have been presented. Both the theories are superconformal, and can be studied using a superspace approach, which greatly simplifies the perturbative analysis. In particular, if one is interested in the calculation of anomalous dimensions, the result demonstrated in Section~\ref{sec:proof} turns out to be a very useful tool, since it allows to discard immediately most of the irrelevant finite contributions. 
In the next chapter the integrability properties of these two theories will be described, together with their application to the computation of the spectra.

\chapter{Integrability}
\label{chapter:integrability}
This chapter is dedicated to the presentation of the main results related to the integrability of both standard and $\beta$-deformed $\mathcal{N}=4$ SYM.
First, the historical development of the integrability techniques for $\N=4$ SYM will be briefly outlined. Then, after describing the Bethe ansatz for the theory, its connection with the dilatation operator will be shown. Afterwards, the breakdown of the asymptotic Bethe ansatz due to wrapping interactions will be introduced, followed by a short introduction to the thermodynamic Bethe ansatz technique. In the end, some information on the integrability of the $\beta$-deformed theory will be given.

\section{The discovery of integrability in \texorpdfstring{$\mathcal{N}=4$}{N=4} SYM}
\label{sec:integrability}
\subsection{Integrable systems}
A quantum system is said to be \emph{integrable} if it is characterized by an infinite number of mutually commuting charges $\mathcal{Q}_k$, the Hamiltonian $\mathcal{H}\equiv\mathcal{Q}_2$ being one of them~\cite{Beisert:2004ry}. This property is equivalent to the fact that any multi-particle scattering process can be factorized into a product of two-particle interactions, as described by the \emph{Yang-Baxter} functional equation. The unknown in such an equation is the \emph{$R$-matrix}, which depends on a spectral parameter and governs the scattering of a pair of particles. From the $R$-matrix, monodromy and transfer matrices can be obtained, which can be used to construct the conserved charges, in a way that resembles the standard solution for the Heisenberg spin chain model~\cite{Bethe:1931hc,Nepomechie:1998jf,Beisert:2004ry,Faddeev:1996iy}.

For the analysis of integrable systems, very powerful techniques exist, which allow in principle to determine the full energy spectrum. Such techniques may prove to be useful to achieve a better understanding of the AdS/CFT correspondence.

\subsection{Spin chains and integrability}
\label{subsec:chains}
As explained in Chapter~\ref{chapter:N4SYM}, in the planar limit of $\mathcal{N}=4$ SYM, only single-trace operators need to be studied explicitly. So, consider a single-trace operator in the $\sutwo$ sector, that is the trace over the $N$-dimensional colour space (to make the operator gauge-invariant) of the product of a given number $L$ of $\phi$ and $Z$ superfields. Such an operator can be related to a state of an $\sutwo$ closed spin chain of length $L$, simply by stating a correspondence between the flavour of the fields and the spin projection along a fixed axis. This analogy allows to refer to operators of the field theory simply as \emph{states} of the associated spin chain, making the identification
\begin{equation}
\mathrm{tr}(\cdots\phi ZZ\phi Z\cdots)\leftrightarrow\vert\cdots\phi ZZ\phi Z\cdots\rangle\pnt
\end{equation}

This spin-chain point of view turned out to be very fruitful since it led to the first result on integrability in $\mathcal{N}=4$ SYM, with the discovery, by Minahan and Zarembo, that the planar, one-loop dilatation operator in the $\sutwo$ sector of the theory, once it is translated into the spin-chain formalism, is equal to the Hamiltonian of the Heisenberg XXX$_{1/2}$ spin chain ~\cite{Minahan:2002ve,Minahan:2006sk}. This is an $\sutwo$ spin chain with nearest-neighbour interactions that was already known to be integrable, and its full analytic spectrum can be found explicitly by means of the so called \emph{Bethe ansatz}~\cite{Bethe:1931hc,Nepomechie:1998jf,Faddeev:1996iy,Beisert:2004ry}.
The integrability property was then recognized to hold good, at one loop, for the whole theory, without sector restrictions~\cite{Beisert:2003jj,Beisert:2003yb,Beisert:2004ag,SchaferNameki:2004ik,Beisert:2005di}. 
Afterwards, strong evidence for integrability, and even exact proofs for particular sectors of the theory, were found at two and three loops~\cite{Beisert:2003tq,Serban:2004jf,Beisert:2004hm,Eden:2004ua,Kotikov:2004er}. These results suggested that integrability
may hold good at \emph{all} orders, and encouraged an intense work on the subject, which led to the formulation of a proposal for an all-loop Bethe ansatz for the whole $\N=4$ SYM~\cite{Beisert:2003xu,Beisert:2003jb,Beisert:2003ys,Staudacher:2004tk,Beisert:2005fw,Beisert:2005tm}, made possible by the discovery that the symmetry properties of the theory completely determine the structure of the S-matrix up to a phase factor~\cite{Beisert:2005tm}. 
Despite all this progress, however, a rigorous proof of the all-order integrability of $\mathcal{N}=4$ SYM has not been found yet. 

Several results were found concerning the possible integrability also of superstring theory on the $AdS_5\times S^5$ background, which is related to $\N=4$ SYM by the AdS/CFT correspondence.
First of all, it was shown that classical strings on that background are integrable~\cite{Mandal:2002fs,Bena:2003wd,Kazakov:2004qf,Arutyunov:2004yx,Alday:2005gi}. Then, a Bethe ansatz was proposed to describe strings on the $\mathds{R}\times S^3$ subspace of $AdS_5\times S^5$~\cite{Arutyunov:2003za,Arutyunov:2004vx,Gromov:2006cq}, and it was realized that, for the AdS/CFT correspondence to be valid, the S-matrices of the gauge and string sides must be related by a global \emph{dressing factor}~\cite{Serban:2004jf,Arutyunov:2004vx,Callan:2003xr,Callan:2004uv,Hernandez:2006tk,Beisert:2006ib,Beisert:2006ez,Beisert:2007hz,Eden:2006rx}
\begin{equation}
\label{Sdressing}
S_{\mathrm{string}}=\hat{S}_{\mathrm{dressing}}S_{\mathrm{gauge}}\col
\end{equation}
that is constrained to reduce to a single phase by symmetry. Afterwards, the proposals for the Bethe ansatz and the S-matrix have been extended to the full theory~\cite{Beisert:2003ea,Arutyunov:2003rg,Arutyunov:2004xy,Kazakov:2004nh,Swanson:2004qa,Beisert:2005bm,Beisert:2005cw,SchaferNameki:2005is,Staudacher:2004tk,Freyhult:2006vr}.

The importance of the possible integrability of $\N=4$ SYM and superstrings on $AdS_5\times S^5$ comes from the fact that, if confirmed, it would provide powerful techniques for the study of the spectra, which in the future may allow to perform a much stronger check of the AdS/CFT correspondence. 
That is why even though the complete integrability of neither theory has been rigorously demonstrated yet, it is nowadays customary to assume its validity and exploit all the consequent tools to perform non-trivial computations on both sides of the correspondence. 
The main such tool is the Bethe ansatz, which will be briefly reviewed in Section~\ref{sec:bethe}.

\subsection{Parity degeneracy}
\label{subsec:parity}
A general property of integrable spin chains is the appearance of degeneracy in the spectrum of the Hamiltonian. The behaviour of the commuting charges $\mathcal{Q}_k$ under the parity transformation $\mathfrak{p}$, which reverses the order of the spins in a closed chain, is~\cite{Beisert:2004ry}
\begin{equation}
\mathfrak{p}\mathcal{Q}_k\mathfrak{p}^{-1}=(-1)^k\mathcal{Q}_k \col
\end{equation}
so that the Hamiltonian $\mathcal{H}$ is parity-even, and the first higher charge $\mathcal{Q}_3$ is parity-odd. 
Since the charge $\mathcal{Q}_3$ commutes with $\mathcal{H}$, for any eigenstate $\vert E\rangle$ of $\mathcal{H}$ with definite parity $\mathcal{Q}_3\vert E\rangle$ must be an eigenstate with the same energy, unless it vanishes. Therefore, as $\mathcal{Q}_3$ reverses the parity of states, pairs of states with the same energy but opposite parity exist in general. The same reasoning applied to the next parity-odd charge $\mathcal{Q}_5$ typically does not add information, reproducing the same degeneracy as $\mathcal{Q}_3$.

\section{The Bethe ansatz for \texorpdfstring{$\mathcal{N}=4$}{N=4} SYM}
\label{sec:bethe}
In the correspondence between the field theory and the spin-chain system, the dilatation operator of the conformal theory is mapped onto the Hamiltonian of the chain. Thus the anomalous dimensions of composite operators become the energies of states of the chain. 

The maximum number of neighbouring interacting fields in a \emph{planar} Feynman diagram, which is usually denoted as the \emph{range} of the diagram, increases with the number of loops. More precisely, an $\ell$-loop graph has a range that is less than or equal to $\ell+1$. So the $\ell$-loop dilatation operator will have range $\ell+1$, and one must deal with long-range spin chains. Indeed, working at \emph{finite} values of the 't~Hooft coupling would require the knowledge of all the perturbative orders and consequently of an infinite-range dilatation operator~\cite{Bargheer:2009xy}. 

From all the results on integrability, an \emph{Asymptotic Bethe Ansatz} (ABA) has been developed~\cite{Beisert:2005fw,Staudacher:2004tk}. It is an adaptation of the original ansatz by Bethe~\cite{Bethe:1931hc} to the more complicated case of long-range chains, and constitutes the main tool for the perturbative analysis of the theory. This ABA is believed to be valid at any loop order, and is available for the whole theory, in its most general formulation which can be found in~\cite{Beisert:2005fw}. For the present thesis, it is enough to consider its restriction to the $\sutwo$ sector, which has a much simpler form, and whose main features are outlined here for convenience, taken from~\cite{Beisert:2005fw}.

\subsection{Derivation of the Bethe equations}
Consider the operator $\mathrm{tr}(Z^L)$, given by the trace of the product of $L$ superfields of the same flavour $Z$. From the field-theoretical point of view, it is protected by supersymmetry, and its anomalous dimension vanishes at all orders. From the spin-chain point of view, this is the ferromagnetic ground state of the chain 
\begin{equation}
\vert Z^L\rangle\equiv\vert0\rangle\col
\end{equation}
and its energy is zero
\begin{equation}
\mathcal{H}\vert0\rangle=0\pnt
\end{equation}
It is now possible to study the excited states obtained by substitution of some of the $Z$ fields with impurities, which are $\phi$ fields in the $\sutwo$ sector or covariant derivatives $\covder$ in $\sltwo$. In view of the relationship with ferromagnetic spin chains, a state with $n$ impurities is commonly referred to as a $n$-\emph{magnon} state. 
Consider the case of a single impurity at the chain site $j$ in the $\sutwo$ sector, 
\begin{equation}
\vert\cdots Z\phi Z\cdots\rangle=a_j^{+}\vert0\rangle\col
\end{equation}
where $a_j^{+}$ is a creation operator turning a $Z$ field into a $\phi$. From the knowledge of the one-loop case, it follows that the natural Bethe ansatz for an exact eigenvector of the Hamiltonian is a plane-wave superposition of such states with momentum $p$
\begin{equation}
\vert p\rangle=\sum_j e^{i p j}a_j^{+}\vert0\rangle\pnt
\end{equation}
This must be an eigenstate corresponding to the all-loop prediction for the eigenvalue~\cite{Beisert:2005fw,Serban:2004jf}
\begin{equation}
\mathcal{E}(p)=\frac{1}{2\lambda}\left[\sqrt{1+16\lambda\sin^2\left(\frac{p}{2}\right)}-1\right] \col
\end{equation}
which was found by guessing a general formula whose perturbative expansion matched the known explicit results for the lowest orders.

After solving the problem for single-impurity states, two-impurity ones must be considered. The ansatz in this case is
\begin{equation}
\vert p_1,p_2\rangle=\sum_{j_1,j_2} \Psi_{j_1,j_2}(p_1,p_2)a_{j_1}^{+}a_{j_2}^{+}\vert0\rangle\pnt
\end{equation}
At fixed loop order, the range of interaction is finite. If the length of the chain is greater than this range, then when the two impurities are far enough from each other they cannot interact directly and therefore the wave function $\Psi$ must factorize into the product of two single-impurity functions
\begin{equation}
\Psi_{j_1,j_2}(p_1,p_2)=e^{i p_1 j_1 + i p_2 j_2}A\qquad\mathrm{if}\qquad j_1\ll j_2\col
\end{equation}
and the same form must be valid with a different coefficient $A'$ for $j_1\gg j_2$. The ratio	$A'/A$ is by definition the S-matrix for a two-particle scattering
\begin{equation}
\frac{A'}{A}=S(p_1,p_2) \pnt
\end{equation}
Thanks to the integrability hypothesis, the information on two-body interactions is enough to reconstruct the general $n$-particle scattering: given a state with $K$ impurities, the eigenstates have the form
\begin{equation}
\vert\{p_1,\ldots,p_K\}\rangle=\sum_{j_1,\ldots,j_K}A_{\{j_1,\ldots,j_K\}}\prod_{k=1}^{K}e^{i p_k j_k}a_{j_k}^{+}\vert0\rangle\col
\end{equation}
and the total energy is equal to the sum of the single-magnon energies
\begin{equation}
E=\sum_{k=1}^K \mathcal{E}(p_k) \pnt
\end{equation}
Since the chain has a large but finite length $L$, periodicity conditions must be imposed on the wave functions, leading to the \emph{momentum constraint}
\begin{equation}
\label{momentum-constraint}
\prod_{k=1}^K e^{i p_k}=1\col
\end{equation}
and to a set of $K$ \emph{asymptotic Bethe equations}
\begin{equation}
\label{bethe-equations}
e^{i L p_k}=\prod_{\substack{j=1\\j\neq k}}^K S(p_k,p_j) \pnt
\end{equation}
As already anticipated, the explicit form of the S-matrix for $\mathcal{N}=4$ SYM is completely determined up to a phase by the symmetry properties of the theory. It is hence possible to obtain a form of the Bethe equations that is suitable for explicit computations. 
It is best written in terms of the \emph{rapidities} $u_k$, defined as 
\begin{equation}
u_k=u(p_k)\col\qquad u(p)=\frac{1}{2}\mathrm{cot}\left(\frac{p}{2}\right)\sqrt{1+16\lambda\sin^2\left(\frac{p}{2}\right)}\col
\end{equation}
and of the corresponding \emph{spectral parameters} $x_k$, related to the rapidities by~\cite{Beisert:2005wv}
\begin{equation}
\label{xu}
x_k=x(u_k)\col\qquad x(u)=\frac{u}{2}\left[1+\sqrt{1-4\frac{\lambda}{u^2}}\ \right]\col\qquad u(x)=x+\frac{\lambda}{x}\pnt
\end{equation}
With these definitions, the Bethe equations for the $\sutwo$ sector now read~\cite{Beisert:2005fw,Beisert:2005wv}
\begin{equation}
\label{bethe-equations2}
\left[\frac{x(u_k+i/2)}{x(u_k-i/2)}\right]^L=\prod_{\substack{j=1\\j\neq k}}^K\frac{u_k-u_j+i}{u_k-u_j-i}e^{2i\theta(u_k,u_j)}\col
\end{equation}
where $e^{2i\theta}$ is the undetermined dressing phase factor, and the momentum constraint is
\begin{equation}
\label{momentum-constraint2}
\prod_{j=1}^K\frac{x(u_k+i/2)}{x(u_k-i/2)}=1\pnt
\end{equation}
An explicit expression for the infinite set of conserved charges is also available
\begin{equation}
Q_s=\sum_{k=1}^K q_s(u_k)\col\qquad q_s(u)=\frac{i}{s-1}\left[\frac{1}{x(u+i/2)^{s-1}}-\frac{1}{x(u-i/2)^{s-1}}\right]\pnt
\end{equation}
The charge $Q_2$ is the Hamiltonian of the chain.

It is fundamental to remember that the derivation of the ABA requires that the length of the spin chain is greater than the maximum range of interactions for a fixed perturbative order. Therefore, for a state of given length $L$, the asymptotic Bethe equations~\eqref{bethe-equations2} give the correct result for the anomalous dimension only up to order $L-1$. 

\subsection{The dressing phase}
The dressing factor $e^{2i\theta(u_k,u_j)}$ of~\eqref{bethe-equations2}
cannot be determined by symmetry considerations. The most general form for $\theta$ has been given in~\cite{Arutyunov:2004vx,Beisert:2005wv}
\begin{equation}
\label{dressingphase}
\theta(u_k,u_j)=\sum_{r=2}^\infty\sum_{s=r+1}^\infty \beta_{r,s}(\lambda)[q_r(u_k)q_s(u_j)-q_s(u_k)q_r(u_j)]\col
\end{equation}
where the functions $\beta_{r,s}(\lambda)$ can be expanded as
\begin{equation}
\beta_{r,s}(\lambda)=\sum_{k=s-1}^\infty\lambda^k\beta_{r,s}^{(k)}\pnt
\end{equation}
The dressing phase does not appear up to three loops, so $\beta_{2,3}^{(2)}=0$. The first non-vanishing component was found to be  $\beta_{2,3}^{(3)}=4\zeta(3)$ in~\cite{Beisert:2007hz}. A conjecture on the values of the $\beta_{r,s}(\lambda)$ up to seven loops and on their transcendentality behaviour at all orders has been presented in~\cite{Eden:2006rx,Beisert:2006ez}.

\subsection{Bethe equations for two-impurity states}
\label{subsec:bethetwo}
For a single-impurity state the momentum constraint~\eqref{momentum-constraint2} forces the momentum of the magnon to be zero, and consequently the energy of the state vanishes. This is consistent with the fact that the corresponding operator in the field theory is protected by supersymmetry. As a consequence, any non-trivial computation in $\mathcal{N}=4$ SYM must involve states with at least two impurities. 

Two-impurity states of one of the three rank-one sectors $\sutwo$, $\sltwo$ and $\suoneone$ have the remarkable property that they belong to multiplets with representatives in all these three sectors~\cite{Beisert:2002tn,Beisert:2005fw}. Thus, the spectra of the restrictions of the full theory to such sectors must be the same. 
Moreover, the Bethe equations for two-impurity states are simple enough to admit an explicit, analytic perturbative solution. In fact, if $u_1$ and $u_2$ are the rapidities associated to the two impurities, the momentum constraint~\eqref{momentum-constraint2} requires
\begin{equation}
u_2=-u_1\col
\end{equation}
and so the system~\eqref{bethe-equations2} reduces to a single equation
\begin{equation}
\label{bethe-equations2imp}
\left[\frac{x(u+i/2)}{x(u-i/2)}\right]^{L-1}=e^{2i\theta(u,-u)}\col
\end{equation}
where $u\equiv u_1$. This equation can be easily solved order by order in $\lambda$ for a generic value of $L$.

Since two-impurity states of the $\sutwo$ and $\sltwo$ sectors are invariant under parity, they cannot be used to observe the parity degeneracy described in Section~\ref{sec:integrability}.

\section{The dilatation operator up to three loops}
For any fixed perturbative order, the integrability tools allow to determine the planar asymptotic dilatation operator, whose eigenvalues are the anomalous dimensions of long operators up to that order. 
As only planar interactions are considered, it is useful to write the dilatation operator in a basis of operators built from permutations of spin chain sites
\begin{equation}\label{permstrucdef}
\pthree{a_1}{\dots}{a_n}=\sum_{r=0}^{L-1}\perm_{a_1+r,a_1+r+1}\cdots
\perm_{a_n+r,a_n+r+1}
\col
\end{equation}
where $\perm_{a,a+1}$ swaps the fields at sites $a$ and $a+1$. For a chain of length $L$, the cyclic identification $\perm_{a,a+1}\simeq\perm_{a+L,a+L+1}$ must be considered. 
The range of such an operator can be found from the list of integers $a_1,\ldots a_n$ according to
\begin{equation}\label{nneighbourint}
\kappa=2+\max\{a_1,\dots, a_n\}-\min\{a_1,\dots, a_n\}\pnt
\end{equation}
Under a parity transformation, the permutation operators~\eqref{permstrucdef} transform as
\begin{equation}
\mathfrak{p}\{a_1,\ldots,a_n\}\mathfrak{p}^{-1}=\{-a_1,\ldots,-a_n\}\col
\end{equation}
and their behaviour under Hermitian conjugation is
\begin{equation}
\{a_1,\ldots,a_n\}^\dagger=\{a_n,\ldots,a_1\}\pnt
\end{equation}
Moreover, they fulfill several simplification identities, which follow from the properties of permutations
\begin{equation}
\label{property-subtract}
\{a_1+k,\ldots,a_n+k\}=\{a_1,\ldots,a_n\} \col
\end{equation}
\begin{equation}
\{\ldots,k,k,\ldots\}=\{\ldots,\ldots\} \col
\end{equation}
\begin{equation}
\label{permutations:swap}
\{\ldots,p,q,\ldots\}=\{\ldots,q,p,\ldots\}\qquad\mathrm{if}\ \ \vert p-q\vert\geq2 \col
\end{equation}
\begin{equation}
\label{property-simplify}
\begin{aligned}
\{\ldots,n,n\pm1,n,\ldots\}&=\{\ldots,\ldots\}-\{\ldots,n,\ldots\}-\{\ldots,n\pm1,\ldots\}\\
&\qquad+\{\ldots,n,n\pm1,\ldots\}+\{\ldots,n\pm1,n,\ldots\} \pnt
\end{aligned}
\end{equation}
From the property $\perm_{i,i+1}\perm_{i+2,i+3}=\perm_{i+2,i+3}\perm_{i,i+1}$ of permutations, a formula for the commutator between two permutation operators can be obtained as
\begin{equation}
\label{permcomm}
\begin{aligned}
\big[\{a_1,\ldots,a_n\},\{b_1,\ldots,b_m\}\big]=\sum_{k=s-1}^{S+1}(&\{k+a_1,\ldots,k+a_n,b_1,\ldots,b_m\}\\
&-\{b_1,\ldots,b_m,k+a_1,\ldots,k+a_n\}) \col
\end{aligned}
\end{equation}
where $s=\min\{b_i\}-\max\{a_i\}$ and $S=\max\{b_i\}-\min\{a_i\}$.

The explicit procedure to find the dilatation operator exploiting the hypothesis of integrability can be found in~\cite{Beisert:2004ry} and will be outlined for the five-loop case in Chapter~\ref{chapter:fiveloop}. Here, the explicit expressions for the components of the dilatation operator up to three loops are presented~\cite{Beisert:2004ry,Beisert:2007hz,Beisert:2003tq} (see Chapters~\ref{chapter:fourloop} and~\ref{chapter:fiveloop} for the meaning of the unknown coefficient $\epsilon_{2a}$)
\begin{equation}
\label{Duptothree}
\begin{aligned}
\mathcal{D}_0&={}+{}\{\} \col \\
\mathcal{D}_1&=2\left(\{\}-\{1\}\right) \col \\
\mathcal{D}_2&=2\left(-4\{\}+6\{1\}-\left(\{1,2\}+\{2,1\}\right)\right) \col \\
\mathcal{D}_3&=60\{\}-104\{1\}+4\{1,3\}+24\left(\{1,2\}+\{2,1\}\right) \\
&-4i\epsilon_{2a}\{1,3,2\}+4i\epsilon_{2a}\{2,1,3\}-4\left(\{1,2,3\}+\{3,2,1\}\right) \pnt
\end{aligned}
\end{equation}

Since single-impurity states are protected in $\mathcal{N}=4$ SYM, and because of the $\sosix$ symmetry relating the scalar fields, the shortest, non-protected operator of the $\sutwo$ sector has length four and wrapping effects cannot show up before the fourth loop order. So up to three loops the asymptotic dilatation operator is actually the full, correct one. 

\section{Finite-size effects and wrapping}
For a spin chain with $L$ sites, when the loop order becomes greater than or equal to $L$, the range of the Hamiltonian exceeds the length of the chain. In this situation, the asymptotic regime where the impurities cannot interact directly is not available and the asymptotic Bethe ansatz breaks down. Thus, the Bethe equations~\eqref{bethe-equations2} give the correct anomalous dimensions only up to $L-1$ loops. Alternatively, for a fixed perturbative order $\ell$, they can be applied only to states of length greater than $\ell$, which are referred to as \emph{asymptotic} states, or simply \emph{long} states.

From the field-theoretical point of view, the breakdown of the asymptotic description is caused by the appearance of the so called \emph{wrapping interactions}~\cite{Beisert:2004hm}. The name comes from the fact that the corresponding Feynman diagrams, having a range greater than the length of the state, can wrap around it. Wrapping interactions are therefore finite-size effects that modify the anomalous dimensions of short operators, that is operators whose length is less than or equal to the perturbative order of interest. For a given operator length $L$, the loop order $L$, which is the first one where wrapping interactions become relevant, is called the \emph{critical} order.

Obviously, it is interesting to study finite-size effects both in $\mathcal{N}=4$ SYM and in its string dual, to investigate if they are still compatible with the AdS/CFT correspondence. 
On the field theory side, the properties of wrapping interactions have been first analyzed in terms of Feynman diagrams in~\cite{Sieg:2005kd}, whereas in~\cite{Ambjorn:2005wa,Janik:2007wt} some proposal were considered for their general analysis, and in particular it was argued for the first time that they may be understood using the Thermodynamic Bethe Ansatz (TBA). In~\cite{Fischbacher:2004iu} finite-size effects have been discussed in a different theory, the BMN matrix model. Some proposals were formulated for a description of wrapping contributions by means of the Hubbard model~\cite{Rej:2005qt} or of the BFKL equation~\cite{Lipatov:1976zz,Kuraev:1977fs,Balitsky:1978ic,Kotikov:2007cy}, but they have been later ruled out by the appearance of the first exact computation of finite-size effects in $\mathcal{N}=4$ SYM, concerning the four-loop anomalous dimension of the Konishi operator~\cite{us,uslong}, which represents the main subject of this thesis, together with similar results found in the $\beta$-deformed theory. This result was later confirmed by a computer-made, perturbative calculation in the component-field formalism, presented in~\cite{Velizhanin:2009zz}, and, most importantly, it agrees with an independent result obtained through the L\"uscher approach~\cite{Luscher:1985dn} applied to the string side~\cite{Bajnok:2008bm}, thus representing a further non-trivial check of the AdS/CFT correspondence. 

In the meanwhile, the problem of finite-size corrections had been addressed on the string side too, even before the appearance of~\cite{Bajnok:2008bm}, 
in particular in the case of giant magnons, which are spinning strings moving in a $\mathds{R}\times S^3$ subspace of the $AdS_5\times S^5$ background, the $\mathds{R}$ factor corresponding to the time direction of $AdS_5$, and which are dual to states of the $\sutwo$ sector in $\N=4$ SYM. 
The computations were performed mainly either directly by means of stringy techniques~\cite{Arutyunov:2006gs,SchaferNameki:2006ey,SchaferNameki:2006gk,Astolfi:2007uz,Ramadanovic:2008qd} or by the application of the L\"uscher approach~\cite{Ambjorn:2005wa,Gromov:2008ie,Minahan:2008re,Heller:2008at,Hatsuda:2008gd,Hatsuda:2008na,Janik:2007wt,Sax:2008in}.
In most of the cases, however, the $\lambda\gg1$ limit was considered from the beginning, and therefore the final results cannot be checked against field-theoretical predictions.

The L\"uscher method has been used to analyze wrapping contributions also on different operators~\cite{Beccaria:2009eq,Bajnok:2008qj,Penedones:2008rv} on the gauge side, even beyond the critical order~\cite{Bajnok:2009vm},
but its generalization to arbitrary states is very difficult, and thus alternative approaches have been attempted. In particular,
a great progress has been made towards a general description of finite-size effects in terms of integrable structures, both on the gauge and on the string side, by means of the thermodynamic Bethe ansatz
~\cite{Yang:1968rm,Zamolodchikov:1989cf,deLeeuw:2008ye,Arutyunov:2007tc,Arutyunov:2009zu,Arutyunov:2009ce,Arutyunov:2009mi,Arutyunov:2009ga,deLeeuw:2009hn,Arutyunov:2009ux,Arutyunov:2009ax,Arutyunov:2010gb,Balog:2010xa}, which led to the formulation of the Y-system approach~\cite{Gromov:2009tv,Gromov:2009bc,Bombardelli:2009ns,Arutyunov:2009ur,Hegedus:2009ky}. These techniques have been then applied to compute wrapping effects at strong coupling in the CFT~\cite{Gromov:2009zb,Roiban:2009aa}. Because of its relevance for the derivation of the Y-system, the thermodynamic Bethe ansatz will be briefly reviewed in Section~\ref{sec:thermo}.

The new techniques for the analysis of finite-size effects appear to be very powerful, and so it is important to test them as much as possible by comparing their predictions with direct perturbative computations.
This is the main motivation to the analysis of wrapping at critical order for length-five states in $\mathcal{N}=4$ SYM, presented in Chapter~\ref{chapter:fiveloop}.

Integrability properties and finite size effects can be studied also in different versions of the AdS/CFT correspondence. For example, several papers recently appeared concerning the duality between the planar $\N=6$ superconformal Chern-Simons theory in three dimensions~\cite{Aharony:2008ug} and type II A string theory on $AdS_4\times CP^3$~\cite{Arutyunov:2008if,Stefanski:2008ik,Chen:2008qq}. As in the previous case, hints suggesting the integrability of both theories were found first~\cite{Benna:2008zy,Minahan:2008hf,Gaiotto:2008cg,Gromov:2008qe,Ahn:2008aa,Bak:2008cp,Ahn:2008hj,McLoughlin:2008ms,Alday:2008ut,Krishnan:2008zs,Gromov:2008fy,McLoughlin:2008he,Ahn:2008tv,Grignani:2008is,Zwiebel:2009vb,Minahan:2009te,Kalousios:2009mp,Suzuki:2009sc,Beccaria:2009wb,Kalousios:2009ey}, and then also finite-size effects have been analyzed~\cite{Grignani:2008te,Astolfi:2008ji,Shenderovich:2008bs,Bombardelli:2008qd,Lukowski:2008eq,Ahn:2008wd,Abbott:2008qd,Abbott:2009um,Minahan:2009aq,Minahan:2009wg}.
However, a discrepancy between the string theory and Bethe ansatz results still exists, which will have to be investigated in the future~\cite{McLoughlin:2008ms,Alday:2008ut,Krishnan:2008zs,McLoughlin:2008he}. In this thesis, this version of the AdS/CFT conjecture will not be considered.

Another theory where integrability and finite-size effects can be analyzed is 
the $\beta$-deformed version of $\mathcal{N}=4$ SYM, as outlined in Section~\ref{sec:betadef-int}. A detailed discussion of wrapping effects in this theory will be presented in Chapter~\ref{chapter:wrapbetadef}.

\section{The thermodynamic Bethe ansatz}
\label{sec:thermo}
The thermodynamic Bethe ansatz~\cite{Yang:1968rm} is an extension of the standard Bethe ansatz method that allows to study systems at finite temperature in the thermodynamic limit. Its relevance in the context of finite-size interactions relies on the general idea originally developed in~\cite{Zamolodchikov:1989cf}, and then recently applied to the AdS/CFT correspondence~\cite{Ambjorn:2005wa,Arutyunov:2007tc,deLeeuw:2008ye,Arutyunov:2009zu,deLeeuw:2009hn,Arutyunov:2009ga,Arutyunov:2009mi,Arutyunov:2009ce,Arutyunov:2009ux,Arutyunov:2009ax,Arutyunov:2010gb,Balog:2010xa}.

Given a field theory defined on a torus generated by two orthogonal circles $\mathcal{C}$ and $\mathcal{B}$, with circumferences $R$ and $L\gg R$, 
two interpretations are possible, depending on the meaning assigned to the $x$ and $y$ coordinates, parameterizing $\mathcal{C}$ and $\mathcal{B}$ respectively:
\begin{itemize}
\item if $x$ is the space coordinate and $y$ represents Euclidean time, in the limit $L\rightarrow\infty$ one has a standard theory at zero temperature, defined on a compact one-dimensional space of length $R$, where finite-size effects appear. The partition function will be dominated by the term corresponding to the ground state
\begin{equation}
\label{partition1}
Z(R,L)\sim e^{-E(R)L}\Rightarrow\mathrm{log}Z(R,L)\sim-E(R)L\pnt
\end{equation}

\item If, on the contrary, $y$ is space and $x$ is time in the Euclidean formalism, the latter can be interpreted as the inverse of a temperature, and a theory at finite temperature, named the \emph{mirror} theory of the original one, is found. In the thermodynamic limit $L\rightarrow\infty$ the logarithm of the partition function becomes
\begin{equation}
\label{partition2}
\mathrm{log}Z^*(R,L)\sim-LRf(R) \col
\end{equation}
where $f(R)$ is the minimum of the free energy of the system at temperature $T=1/R$.
\end{itemize}
The Hamiltonian $\mathcal{H}^*$ of the mirror theory will in general be different from the original one, but it is easy to show~\cite{Arutyunov:2007tc} that the partition functions must be the same. So the ground state energy, comprehensive of finite-size effects, is given by
\begin{equation}
\label{partition3}
E(R)=Rf(R) \col
\end{equation}
and the problem is reduced to the analysis of a system at finite temperature in the thermodynamic limit, which can be realized by means of the thermodynamic Bethe ansatz.

First of all, the standard Bethe ansatz must be formulated for the mirror theory, which is obtained through a double Wick rotation of the actual one. In the case of the AdS/CFT correspondence, this task has been performed in a series of recent papers~\cite{deLeeuw:2008ye,Arutyunov:2007tc,Arutyunov:2009zu,Arutyunov:2009ce,Arutyunov:2009mi,Arutyunov:2009ga,deLeeuw:2009hn,Arutyunov:2009ux,Bombardelli:2009ns,Arutyunov:2009ur}. Then, the thermodynamic limit is considered, and the Hamiltonian $\mathcal{H}^*$ and entropy $\mathcal{S}^*$ of the model are written in terms of the density of the Bethe roots, so that the free energy for the mirror theory can be calculated
\begin{equation}
f(R)=\mathcal{H}^*-\frac{1}{R}\mathcal{S}^* \pnt
\end{equation}
By minimising the free energy, it is possible to derive the thermodynamic Bethe equations, from which
the ground-state energy $E(R)$ for the original theory can be found thanks to~\eqref{partition3}~\cite{Arutyunov:2009ur}
\begin{equation}
E(R)=-\int_{-\infty}^{\infty}\di u\sum_{k=1}^\infty\frac{1}{2\pi}\frac{\di p^*_k(R)}{\di u}\mathrm{log}\left[1+e^{-\epsilon^*_k(R)}\right] \col
\end{equation}
where the $p^*_k$ and $\epsilon^*_k$ are the momenta and energies of the solutions of the thermodynamic Bethe equations in the mirror theory.

The described technique can be restated in a more compact form in terms of integrable structures, leading to the formulation of the Y-system~\cite{Gromov:2009tv,Gromov:2009bc,Hegedus:2009ky,Bombardelli:2009ns,Arutyunov:2009ur}, which will be briefly reviewed in Chapter~\ref{chapter:fiveloop}.

\section{Integrability of the \texorpdfstring{$\beta$}{beta}-deformed \texorpdfstring{$\mathcal{N}=4$}{N=4} SYM}
\label{sec:betadef-int}
The $\beta$-deformed $\mathcal{N}=4$ SYM theory is believed to be integrable as well, at least for real values of the phase $\beta$, according to the results of~\cite{Roiban:2003dw,Berenstein:2004ys,Beisert:2005if}, and in addition integrable structures have been singled out in its deformed string dual, too
~\cite{Frolov:2005dj,Frolov:2005ty}. In particular, an all-loop asymptotic Bethe ansatz has been proposed~\cite{Beisert:2005if} in analogy with the undeformed theory.

As in the standard case, the asymptotic Bethe ansatz breaks down for short operators because of wrapping corrections. In the deformed string theory, such finite-size effects have been studied in~\cite{Bykov:2008bj}, whereas on the field-theory side the first exact results appeared in~\cite{Fiamberti:2008sn,betadef}, then followed by~\cite{Gunnesson:2009nn,Beccaria:2009hg,Bajnok:2009vm}. At the moment, a general solution to the wrapping problem in the deformed theory, possibly in the form of a modification of the undeformed Y-system, is still missing.

An interesting feature of the deformed Bethe ansatz derived in~\cite{Beisert:2005if} is the appearance of phase factors in the momentum constraint. As a consequence, the momentum of a single-impurity state is no longer forced to vanish, consistently with the fact that single-impurity operators are no longer protected by supersymmetry. Such operators are much easier to study than multiple-impurity ones, and so they allow in general to perform computations at higher loop orders. In fact, in the asymptotic case, the anomalous dimension of a single-impurity operator $\mathcal{O}_{\mathrm{as}}$ in the deformed theory can be found at any perturbative order, even without the need for the Bethe ansatz, from the all-order result~\cite{Mauri:2005pa}
\begin{equation}
\label{single-all-orders}
\gamma(\mathcal{O}_\text{as})=-1+\sqrt{1+4\lambda\Big\vert q-\frac{1}{q}\Big\vert^2}=-1+\sqrt{1+16\lambda\sin^2(\pi\beta)} \pnt
\end{equation}
Even in presence of wrapping corrections, the perturbative study of single-impurity states is considerably simplified, because of the reduced number and complexity of the relevant Feynman diagrams, as will be discussed in detail in Chapter~\ref{chapter:wrapbetadef}.

\section{Conclusions}
Several recent results suggest that both $\mathcal{N}=4$ SYM and type-IIB superstring theory on the $AdS_5\times S^5$ background, which are supposed to be equivalent according to the AdS/CFT correspondence, are integrable. For an integrable theory, powerful tools for the analysis of the spectrum are available, and may be used to test the correspondence. However, the standard integrability techniques break down on short operators, because of wrapping interactions. The determination of the full spectra thus requires the analysis of finite-size effects, for which new tools have been developed very recently. In order to check their validity, some explicit perturbative computations in the field theory have been performed, and will be presented in the following chapters. Finally, the integrability results and the wrapping problem can be extended to the $\beta$-deformed theory, where new unprotected operators can be more easily studied, thus making calculations at higher orders possible.

\chapter{The four-loop anomalous dimension of the Konishi operator}
\label{chapter:fourloop}
In this chapter, the field-theoretical computation of the four-loop anomalous dimension of a descendant of the Konishi operator is presented. This is the easiest case where finite-size effect appear in $\mathcal{N}=4$ SYM. First of all, the general approach, which will be applied also in the following chapters, is explained.

\section{The general approach}
\label{section:approach}
Since single-impurity states in the $\sutwo$ sector of $\mathcal{N}=4$ SYM are protected, to study finite-size effects one is forced to consider operators with at least two impurities. Thus, the shortest states with non-vanishing anomalous dimension will have length $L=4$. Two independent states in this sector are
\begin{equation}
\label{Opbasis}
\basisop{4}{1}=\tr(\phi Z\phi Z)\col\qquad\basisop{4}{2}=\tr(\phi\phi ZZ)\col
\end{equation}
that mix under renormalization. As will become clear in the following, one of the two linear combinations of the operators~\eqref{Opbasis} that are multiplicatively renormalized is protected, whereas the other one is a descendant of the Konishi operator. Note that three-impurity operators of length four are equivalent to single-impurity ones, and thus are protected, as can be seen immediately by exchanging the roles of $Z$ and $\phi$. For the same reason, a four-impurity length-four operator is actually the same thing as the zero-impurity one, and so it is protected as well. Hence, at four loops in the $\sutwo$ sector of $\N=4$ SYM, two-impurity states are the only case where finite-size effects show up.

Because of the relationship between the perturbative order and the maximum range of the interactions, the asymptotic Bethe equations will give the correct components of the Konishi anomalous dimension only up to three loops. In particular, it is possible to apply the three-loop asymptotic dilatation operator to the basis~\eqref{Opbasis}, to find the linear combinations with well-defined behaviour under renormalization by diagonalizing the corresponding $2\times 2$ matrix.
Starting from four loops, wrapping effects appear and the asymptotic dilatation operator obtained from the Bethe ansatz cannot be used.

On the other hand, a complete, four-loop computation in terms of Feynman diagrams, even with the help of superspace techniques, would be very complicated, because of the very large number and complex structures of the involved graphs. However, a procedure exists that allows to avoid the worst part of the calculation. In fact, even though the asymptotic, four-loop dilatation operator cannot be applied directly, it is still possible to exploit its knowledge in order to extract the information on the most difficult and large classes of diagrams that should be considered.
Since the validity of this strategy extends to any order, it is useful to present it here in its general form, which would allow in principle to determine the $\ell$-loop anomalous dimension of length-$L$ states:

\begin{itemize}
\item find the $\ell$-loop asymptotic dilatation operator $\mathcal{D}_{\ell}$ using the integrability hypothesis and the results from the Bethe equations, according to the procedure described in~\cite{Beisert:2004ry} and explicitly applied in the five-loop case in Chapter~\ref{chapter:fiveloop}.
The asymptotic operator gets contributions from Feynman supergraphs belonging to two different classes: the first one is made of diagrams with range less than or equal to $L$, which are relevant also in the length-$L$ case, possibly with modified combinatorial factors. The second class contains all the higher-range diagrams, which do not appear in the finite-length computation.
\item Subtract from $\mathcal{D}_{\ell}$ all the contributions from the diagrams with range greater than $L$. What remains of the dilatation operator contains all the information on the graphs of the first class. The key point in this approach is that after the subtraction, the dilatation operator can be applied to the length-$L$ subsector to find the correct contribution of the diagrams with range not greater than $L$, because the difference in the combinatorial factors of the graphs with respect to the asymptotic case is automatically accounted for by the functional form of $\mathcal{D}_{\ell}$ as an operator on the states of the $\sutwo$ sector, with the action described by the basis~\eqref{permstrucdef}. 
\item Add the contributions of wrapping diagrams, which must be computed explicitly.
\end{itemize}

At last, the correct expression for the dilatation operator on the length-$L$ sector is found. 
This procedure is particularly useful because for a fixed loop order, the number and complexity of Feynman diagrams typically increase when the interaction range decreases, so that the lower-range graphs, whose computation is completely avoided, would be actually the most difficult to analyze explicitly.
As will be shown, this approach becomes even simpler when it is applied to the critical perturbative order $L+1$ for states of length $L$.
The general procedure will be now explicitly applied to the four-loop case.

\section{Subtraction of range-five diagrams}
\subsection{The asymptotic dilatation operator}
The four-loop asymptotic dilatation operator $\mathcal{D}_4$ is already known in the literature~\cite{Beisert:2004ry,Beisert:2007hz} in the basis of permutation operators
\begin{equation}\label{D4}
\begin{aligned}
\mathcal{D}_4&={}-{}(560+4\beta)\pid\\
&\phantom{{}={}}+(1072+12\beta+8\epsilon_{3a})\pone1\\
&\phantom{{}={}}-(84+6\beta+4\epsilon_{3a})\ptwo13
-4\ptwo14-(302+4\beta+8\epsilon_{3a})(\ptwo12+\ptwo21)\\
&\phantom{{}={}}
+(4\beta+4\epsilon_{3a}+2i\epsilon_{3c}-4i\epsilon_{3d})\pthree132
+(4\beta+4\epsilon_{3a}-2i\epsilon_{3c}+4i\epsilon_{3d})\pthree213\\
&\phantom{{}={}}+(4-2i\epsilon_{3c})(\pthree124+\pthree143)
+(4+2i\epsilon_{3c})(\pthree134+\pthree214)\\
&\phantom{{}={}}+(96+4\epsilon_{3a})(\pthree123+\pthree321)\\
&\phantom{{}={}}-(12+2\beta+4\epsilon_{3a})\pfour2132
+(18+4\epsilon_{3a})(\pfour1324+\pfour2143)\\
&\phantom{{}={}}-(8+2\epsilon_{3a}+2i\epsilon_{3b})(\pfour1243+\pfour1432)\\
&\phantom{{}={}}
-(8+2\epsilon_{3a}-2i\epsilon_{3b})(\pfour2134+\pfour3214)\\
&\phantom{{}={}}-10(\pfour1234+\pfour4321)
\col
\end{aligned}
\end{equation}
where $\epsilon_{3a}$, $\epsilon_{3b}$, $\epsilon_{3c}$ and $\epsilon_{3d}$
parameterize the behaviour of the Hamiltonian under similarity transformations (see Chapter~\ref{chapter:fiveloop}  and ~\cite{Beisert:2007hz} for details). They do not enter the spectrum of $\mathcal{D}_4$, and in a perturbative approach their values depend on the choice of the renormalization scheme.
The parameter $\beta=\beta_{2,3}^{(3)}$ is the first non-vanishing component of the dressing phase~\cite{Arutyunov:2004vx,Hernandez:2006tk,Beisert:2006ib,Beisert:2006ez,Beisert:2007hz}, whose value $\beta=4\zeta(3)$ has been computed in~\cite{Beisert:2007hz}.

\subsection{The basis of chiral structures}
The first step in the derivation of the correct dilatation operator on the length-four subsector is the subtraction of range-five diagrams. In order to perform this task, it is useful to change the basis, from the permutation operators~\eqref{permstrucdef} to a new set of functions that are directly related to the chiral structures of Feynman supergraphs. At four loops, one can define the following chiral functions
\begin{equation}
\label{chistruc}
\begin{aligned}
\chi(a,b,c,d)&=\pid-4\pone1
+\ptwo ab+\ptwo ac+\ptwo ad+\ptwo bc+\ptwo bd+\ptwo cd\\
&\phantom{{}={}}
-\pthree abc-\pthree abd-\pthree acd-\pthree bcd
+\pfour abcd\col\\
\chi(a,b,c)&=-\pid+3\pone1
-\ptwo ab-\ptwo ac-\ptwo bc+\pthree abc\col\\
\chi(a,b)&=\pid-2\pone1+\ptwo ab\col\\
\chi(1)&=-\pid+\pone1\col\\
\chi() &=\pid
\pnt
\end{aligned}
\normalsize
\end{equation}
The new functions are more suitable for diagrammatic calculations, since any Feynman supergraph acts on the flavours of the scalar superfields as a sequence of permutations exactly described by one of them. They are called \emph{chiral} because their structure is determined only by the interactions among chiral and antichiral fields in the diagram, being totally insensitive to vector interactions. 

\begin{figure}[t]
\vspace{1cm}
\capstart
\unitlength=0.75mm
\settoheight{\eqoff}{$\times$}%
\setlength{\eqoff}{0.5\eqoff}%
\addtolength{\eqoff}{-12.5\unitlength}%
\settoheight{\eqofftwo}{$\times$}%
\setlength{\eqofftwo}{0.5\eqofftwo}%
\addtolength{\eqofftwo}{-7.5\unitlength}%
\begin{equation*}
\begin{aligned}
\chi(1):\quad-&
\raisebox{1.5\eqoff}{%
\fmfframe(1,1)(3,4){%
\begin{fmfchar*}(30,30)
\WoneplainB
\fmfipair{w[]}
\svertex{w1}{p4}
\fmfiv{l=\footnotesize{$\phi$},l.a=-90,l.d=5}{vloc(__v6)}
\fmfiv{l=\footnotesize{$Z$},l.a=-90,l.d=5}{vloc(__v7)}
\fmfiv{l=\footnotesize{$\phi$},l.a=90,l.d=5}{vloc(__v2)}
\fmfiv{l=\footnotesize{$Z$},l.a=90,l.d=5}{vloc(__v3)}
\fmfiv{l=\footnotesize{$\psi$},l.a=0,l.d=5}{w1}
\end{fmfchar*}}}
&+&
\raisebox{1.5\eqoff}{%
\fmfframe(3,1)(1,4){%
\begin{fmfchar*}(30,30)
\WoneplainB
\fmfipair{w[]}
\svertex{w1}{p4}
\fmfiv{l=\footnotesize{$\phi$},l.a=-90,l.d=5}{vloc(__v6)}
\fmfiv{l=\footnotesize{$Z$},l.a=-90,l.d=5}{vloc(__v7)}
\fmfiv{l=\footnotesize{$Z$},l.a=90,l.d=5}{vloc(__v2)}
\fmfiv{l=\footnotesize{$\phi$},l.a=90,l.d=5}{vloc(__v3)}
\fmfiv{l=\footnotesize{$\psi$},l.a=0,l.d=5}{w1}
\end{fmfchar*}}}
\end{aligned}
\end{equation*}
\caption{Building block for chiral structures}
\label{buildingblock}
\end{figure}

The explicit expressions~\eqref{chistruc} for the chiral functions in terms of the old permutation operators can be easily found. First, $\chi()$ is the identity, which is the chiral structure associated to Feynman diagrams with only vector interactions. The simplest non-trivial function is $\chi(1)$, corresponding to the building block of Figure~\ref{buildingblock}.
This diagram contains one chiral and one antichiral vertices connected by a $\langle\psi\bar{\psi}\rangle$ propagator, which is cancelled by the D-algebra. It behaves as an effective four-vertex, similar to the one which describes scalar interactions in the component formalism. The $\chi(1)$ function can be seen as a building block for all the more complicated structures, as the chiral part of any four-loop supergraph contributing to the one-point function of a composite operator in the $\sutwo$ sector can be built by assembling together up to four copies of it. The number $n$ of arguments of a chiral function $\chi$ equals the number of blocks required to construct the corresponding structure. In addition, every diagram will also contain $(4-n)$ vector interactions.
Note that in~\eqref{chistruc} the coefficients of the permutation operators in the expression for each chiral structure sum up to zero. This is consistent with the fact that the operators of the form $\mathrm{tr}(Z^L)$ and $\mathrm{tr}(\phi Z^L)$ are protected for any value of $L$.

In order to rewrite the four-loop dilatation operator in terms of the new basis, the inverse relations of~\eqref{chistruc}, giving the old permutation basis in terms of the new one, are required
\begin{equation}
\label{invchistruc}
\begin{aligned}
\pfour abcd &= \chi(a,b,c,d) + \chi(a,b,c) + \chi(a,b,d) + \chi(a,c,d)
+ \chi(b,c,d)\\
&\phantom{{}={}}
+ \chi(a,b) + \chi(a,c) + \chi(a,d) + \chi(b,c) + \chi(b,d) + \chi(c,d)\\
&\phantom{{}={}}
+4 \chi(1) + \chi()\col\\
\pthree abc &= \chi(a,b,c) + \chi(a,b) + \chi(a,c) + \chi(b,c) + 3\chi(1)
+\chi()\col\\
\ptwo ab &= \chi(a,b) + 2\chi(1) + \chi()\col\\
\pone1 &= \chi(1) + \chi()\col\\
\pid &= \chi()\pnt
\end{aligned}
\end{equation}
The dilatation operator~\eqref{D4} can now be written in the new basis as
\begin{equation}
\label{D4chi}
\begin{aligned}
\mathcal{D}_4&={}+{}200\chi(1)
-150[\chi(1,2)+\chi(2,1)]
+8(10+\epsilon_{3a})\chi(1,3)
-4\chi(1,4)\\
&\phantom{{}={}}
+60[\chi(1,2,3)+\chi(3,2,1)]\\
&\phantom{{}={}}
+(8+2\beta+4\epsilon_{3a}-4i\epsilon_{3b}+2i\epsilon_{3c}-4i\epsilon_{3d})
\chi(1,3,2)\\
&\phantom{{}={}}
+(8+2\beta+4\epsilon_{3a}+4i\epsilon_{3b}-2i\epsilon_{3c}+4i\epsilon_{3d})
\chi(2,1,3)\\
&\phantom{{}={}}
-(4+4i\epsilon_{3b}+2i\epsilon_{3c})[\chi(1,2,4)+\chi(1,4,3)]\\
&\phantom{{}={}}
-(4-4i\epsilon_{3b}-2i\epsilon_{3c})[\chi(1,3,4)+\chi(2,1,4)]\\
&\phantom{{}={}}-(12+2\beta+4\epsilon_{3a})\chi(2,1,3,2)\\
&\phantom{{}={}}
+(18+4\epsilon_{3a})[\chi(1,3,2,4)+\chi(2,1,4,3)]\\
&\phantom{{}={}}
-(8+2\epsilon_{3a}+2i\epsilon_{3b})[\chi(1,2,4,3)+\chi(1,4,3,2)]\\
&\phantom{{}={}}
-(8+2\epsilon_{3a}-2i\epsilon_{3b})[\chi(2,1,3,4)+\chi(3,2,1,4)]\\
&\phantom{{}={}}
-10[\chi(1,2,3,4)+\chi(4,3,2,1)]
\pnt
\end{aligned}
\end{equation}

In~\eqref{D4chi} the coefficient of each chiral function gets contributions from all the diagrams with that particular chiral structure. 
As the chiral structures are completely insensitive to vector interactions, the range of a diagram may be different from the one of the associated chiral function. When this happens, a diagrams is said to be \emph{non-maximal}.

\subsection{Subtraction of range-five diagrams}
The range-five Feynman diagrams, whose contribution must be subtracted from the asymptotic four-loop dilatation operator, can be divided into two classes:
\begin{enumerate}
\item\label{class-five} diagrams whose chiral structure has range equal to five, and which therefore either contain only scalar interactions, or have a disconnected scalar structure completed to a connected graph by vector interactions,
\item\label{class-ltfive} diagrams built on a chiral structure of shorter range, that become of range five with the addition of a vector interaction.
\end{enumerate}
The contribution from the graphs of the first class can be read directly from the expression for the asymptotic dilatation operator written in the new basis~\eqref{D4chi}, since it is exactly given by the sum of the terms with range-five chiral structures. 
For the second class this is no longer true, because the corresponding diagrams contribute to the coefficients of the structures with lower range, together with all the diagrams with the same chiral structure but range less than five, which must be conserved. So in principle an explicit computation of all the diagrams of the second class would be required. However, thanks to the fact that here the perturbative order is exactly the critical one, the situation is greatly simplified, because the non-maximal diagrams fulfill the hypothesis on which the cancellation result of Appendix~\ref{app:non-maximal} is based, and thus their divergent parts sum up to zero. 

All the range-five contributions can hence be subtracted from the four-loop dilatation operator by simply deleting the terms with range-five chiral structures
\begin{equation}
\begin{aligned}
\label{r5D}
\delta \mathcal{D}_4
&={}-{}10[\chi(1,2,3,4)+\chi(4,3,2,1)]+(18+4\epsilon_{3a})[\chi(1,3,2,4)+\chi(2,1,4,3)]\\
&\phantom{{}={}}
-(8+2\epsilon_{3a}+2i\epsilon_{3b})[\chi(1,2,4,3)+\chi(1,4,3,2)]\\
&\phantom{{}={}}
-(8+2\epsilon_{3a}-2i\epsilon_{3b})[\chi(2,1,3,4)+\chi(3,2,1,4)]\\
&\phantom{{}={}}
-(4+4i\epsilon_{3b}+2i\epsilon_{3c})[\chi(1,2,4)+\chi(1,4,3)]\\
&\phantom{{}={}}
-(4-4i\epsilon_{3b}-2i\epsilon_{3c})[\chi(1,3,4)+\chi(2,1,4)]\\
&\phantom{{}={}}
-4\chi(1,4)
\pnt
\end{aligned}
\end{equation}
It is important to stress that, owing to the non-trivial mixing described by~\eqref{chistruc}, this subtraction is not equivalent to the simple cancellation of the range-five permutation operators, as attempted in~\cite{Fischbacher:2004iu} for the BMN matrix model. This is the first example of the simplifications following from the choice of the basis of chiral functions.

The described simple subtraction procedure can be applied to any critical loop order $\ell$, thanks to the general validity of the argument of Appendix~\ref{app:non-maximal} on the vanishing of the total contribution of non-maximal, range-$(\ell+1)$ diagrams at $\ell$ loops. This allows to compute the $\ell$-loop subtracted dilatation operator on the length-$\ell$ subsector directly from the asymptotic operator $\mathcal{D}_{\ell}$ without the need for explicit diagrammatic calculations.

From the previous considerations, it follows that the required subtracted four-loop dilatation operator on the length-four subsector is
\begin{equation}
\label{D4sub}
\begin{aligned}
\mathcal{D}_4^\text{sub}\equiv \mathcal{D}_4 - \delta \mathcal{D}_4 &=
{}+{}200\chi(1)
-150[\chi(1,2)+\chi(2,1)]
+8(10+\epsilon_{3a})\chi(1,3)\\
&\phantom{{}={}}
+60[\chi(1,2,3)+\chi(3,2,1)]\\
&\phantom{{}={}}
+(8+2\beta+4\epsilon_{3a}-4i\epsilon_{3b}+2i\epsilon_{3c}-4i\epsilon_{3d})
\chi(1,3,2)\\
&\phantom{{}={}}
+(8+2\beta+4\epsilon_{3a}+4i\epsilon_{3b}-2i\epsilon_{3c}+4i\epsilon_{3d})
\chi(2,1,3)\\
&\phantom{{}={}}-(12+2\beta+4\epsilon_{3a})\chi(2,1,3,2)
\pnt
\end{aligned}
\end{equation}
In the two-dimensional vector space generated by the length-four operators $\basisop{4}{1}$ and $\basisop{4}{2}$ given in~\eqref{Opbasis}, all the chiral structures with range up to four can be written in terms of a single mixing matrix
\begin{equation}
M_4=\left(
\begin{array}{cc}
4 & -2 \\
-4 & 2
\end{array}
\right)\col
\end{equation}
as
\begin{align}
\label{chiM}
&\chi(1)\to -M_4\col &&\chi(1,2)\to M_4\col &&\chi(1,3)\to 2M_4\col \notag \\
&\chi(2,1)\to M_4\col &&\chi(1,2,3)\to -M_4\col &&\chi(3,2,1)\to -M_4\col \\
&\chi(2,1,3)\to-2M_4\col &&\chi(1,3,2)\to-2M_4\col &&\chi(2,1,3,2)\to 2M_4\pnt \notag
\end{align}
In this basis, the subtracted dilatation operator reads
\begin{equation}
\label{D4submatrix1}
\mathcal{D}_4^{\,\text{sub}}\to
-4(129+2\epsilon_{3a}+12\zeta(3))M_4
\pnt
\end{equation}
Even though the similarity coefficients $\epsilon_{3x}$ do not appear in the spectrum of the asymptotic dilatation operator, they may apparently affect the final spectrum after the subtraction of range-five interactions and the addition of wrapping diagrams. However, this is not in contrast with the fact that these coefficients are non-physical. In fact, the wrapping correction must depend on them as well, in such a way that the total result does not, but, as will be clear later, such dependence is hidden because the explicit computation of wrapping diagrams requires the choice of a renormalization scheme, where the similarity coefficients have well-defined values. As a consequence, the values of the relevant similarity coefficients in the chosen renormalization scheme, which in this thesis is minimal subtraction (MS) with G-scheme~\cite{Chetyrkin:1980pr}, are required
\begin{equation}
\label{epsilons}
\epsilon_{3a}=-4\col\qquad
\epsilon_{3b}=-i\frac{4}{3}\col\qquad
\epsilon_{3c}=i\frac{4}{3}\pnt
\end{equation}
Their calculation is presented in Appendix~\ref{app:fourR5}. After the substitution of the explicit value for $\epsilon_{3a}$, the matrix form of $\mathcal{D}_4^{\,\text{sub}}$ becomes
\begin{equation}
\label{D4submatrix}
\mathcal{D}_4^{\,\text{sub}}\to
-4(121+12\zeta(3))M_4
\pnt
\end{equation}
The subtracted dilatation operator $\mathcal{D}_4^{\,\text{sub}}$ is now free of any range-five interaction. In order to obtain the full operator for the length-four subsector, wrapping interactions will now be analyzed.

\section{Wrapping contributions}
Unlike the subtraction of range-five diagrams, the analysis of wrapping contributions requires the explicit calculation of the relevant Feynman supergraphs, and thus constitutes the most difficult part of the determination of the four-loop anomalous dimension of the Konishi operator.

As a preliminary step, all the possible relevant supergraphs must be drawn, according to the results of~\cite{Sieg:2005kd}. Then, for every diagram the D-algebra procedure is performed, producing a standard four-loop integral in momentum space. A powerful method for the computation of such integrals, using dimensional regularization in $d=4-2\varepsilon$ dimensions, is the Gegenbauer Polynomial $x$-space Technique (GPXT)~\cite{Chetyrkin:1980pr,Kotikov:1995cw,Celmaster:1980ji}, which will be briefly outlined later in this chapter and with more details in Appendix~\ref{app:gegenbauer}.

\subsection{Wrapping diagrams}
\label{subsec:wrapping}
\begin{figure}[t]
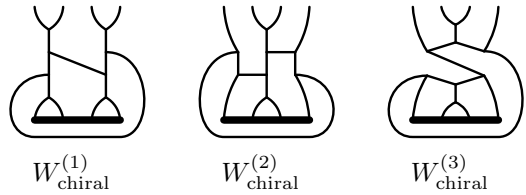

\capstart
\renewcommand*{\thesubfigure}{}
\unitlength=0.75mm
\settoheight{\eqoff}{$\times$}%
\setlength{\eqoff}{0.5\eqoff}%
\addtolength{\eqoff}{-12.5\unitlength}%
\settoheight{\eqofftwo}{$\times$}%
\setlength{\eqofftwo}{0.5\eqofftwo}%
\addtolength{\eqofftwo}{-7.5\unitlength}%
\raisebox{\eqoff}{%
\subfigure[$W_{\mathrm{chiral}}^{(1)}$]{
\fmfframe(3,1)(1,4){%
\begin{fmfchar*}(20,20)
\fmftop{v1}
\fmfbottom{v5}
\fmfforce{(0.125w,h)}{v1}
\fmfforce{(0.125w,0)}{v5}
\fmffixed{(0.25w,0)}{v1,v2}
\fmffixed{(0.25w,0)}{v2,v3}
\fmffixed{(0.25w,0)}{v3,v4}
\fmffixed{(0.25w,0)}{v5,v6}
\fmffixed{(0.25w,0)}{v6,v7}
\fmffixed{(0.25w,0)}{v7,v8}
\fmffixed{(0,0.9w)}{v5,vh1}
\fmf{plain,tension=0.5,right=0.25}{v1,vc1}
\fmf{plain,tension=0.5,left=0.25}{v2,vc1}
\fmf{plain,tension=0.5,right=0.25}{v3,vc2}
\fmf{plain,tension=0.5,left=0.25}{v4,vc2}
\fmf{plain}{vc1,vc3}
\fmf{plain}{vc3,vc7}
\fmf{plain}{vc7,vc5}
\fmf{plain}{vc2,vc8}
\fmf{plain}{vc8,vc4}
\fmf{plain}{vc4,vc6}
\fmf{plain,tension=0}{vc3,vc4}
\fmf{plain,tension=0.5,left=0.25}{v5,vc5}
\fmf{plain,tension=0.5,right=0.25}{v6,vc5}
\fmf{plain,tension=0.5,left=0.25}{v7,vc6}
\fmf{plain,tension=0.5,right=0.25}{v8,vc6}
\fmf{plain,tension=0.5,right=0,width=1mm}{v5,v8}
\fmffreeze
\fmfposition
\plainwrap{vc7}{v5}{v8}{vc8}
\end{fmfchar*}}}
\subfigspace
\subfigure[$W_{\mathrm{chiral}}^{(2)}$]{
\fmfframe(3,1)(1,4){%
\begin{fmfchar*}(20,20)
\fmftop{v1}
\fmfforce{(0.125w,h)}{v1}
\fmfforce{(0.125w,0)}{v5}
\fmffixed{(0.25w,0)}{v1,v2}
\fmffixed{(0.25w,0)}{v2,v3}
\fmffixed{(0.25w,0)}{v3,v4}
\fmffixed{(0.25w,0)}{v5,v6}
\fmffixed{(0.25w,0)}{v6,v7}
\fmffixed{(0.25w,0)}{v7,v8}
\fmffixed{(whatever,0)}{vc1,vc3}
\fmffixed{(whatever,0)}{vc5,vc7}
\fmffixed{(whatever,0)}{vc3,vc4}
\fmffixed{(whatever,0)}{vc7,vc8}
\fmf{plain,tension=1,right=0.125}{v1,vc1}
\fmf{plain,tension=0.5,right=0.25}{v2,vc2}
\fmf{plain,tension=0.5,left=0.25}{v3,vc2}
\fmf{plain,tension=1,left=0.125}{v4,vc4}
\fmf{plain,tension=1,left=0.125}{v5,vc5}
\fmf{plain,tension=0.5,left=0.25}{v6,vc6}
\fmf{plain,tension=0.5,right=0.25}{v7,vc6}
\fmf{plain,tension=1,right=0.125}{v8,vc8}
\fmf{plain}{vc1,vc5}
\fmf{plain}{vc4,vc8}
\fmf{plain}{vc2,vc3}
\fmf{plain}{vc6,vc7}
\fmf{plain,tension=3}{vc3,vc7}
\fmf{plain,tension=0.5}{vc3,vc4}
\fmf{plain,tension=0.5}{vc5,vc7}
\fmf{phantom,tension=0.5}{vc7,vc8}
\fmf{phantom,tension=0.5}{vc1,vc3}
\fmf{plain,tension=0.5,right=0,width=1mm}{v5,v8}
\fmffreeze
\fmfposition
\plainwrap{vc1}{v5}{v8}{vc8}
\fmffreeze
\end{fmfchar*}}}
\subfigspace
\subfigure[$W_{\mathrm{chiral}}^{(3)}$]{
\fmfframe(3,1)(1,4){%
\begin{fmfchar*}(20,20)
\fmftop{v1}
\fmfbottom{v5}
\fmfforce{(0.125w,h)}{v1}
\fmfforce{(0.125w,0)}{v5}
\fmffixed{(0.25w,0)}{v1,v2}
\fmffixed{(0.25w,0)}{v2,v3}
\fmffixed{(0.25w,0)}{v3,v4}
\fmffixed{(0.25w,0)}{v5,v6}
\fmffixed{(0.25w,0)}{v6,v7}
\fmffixed{(0.25w,0)}{v7,v8}
\fmffixed{(0,whatever)}{vc1,vc5}
\fmffixed{(0,whatever)}{vc2,vc3}
\fmffixed{(0,whatever)}{vc3,vc6}
\fmffixed{(0,whatever)}{vc6,vc7}
\fmffixed{(0,whatever)}{vc4,vc8}
\fmffixed{(0.5w,0)}{vc1,vc4}
\fmffixed{(0.5w,0)}{vc5,vc8}
\fmf{plain,tension=1,right=0.125}{v1,vc1}
\fmf{plain,tension=0.25,right=0.25}{v2,vc2}
\fmf{plain,tension=0.25,left=0.25}{v3,vc2}
\fmf{plain,tension=1,left=0.125}{v4,vc4}
\fmf{plain,tension=1,left=0.125}{v5,vc5}
\fmf{plain,tension=0.25,left=0.25}{v6,vc6}
\fmf{plain,tension=0.25,right=0.25}{v7,vc6}
\fmf{plain,tension=1,right=0.125}{v8,vc8}
\fmf{plain,tension=0.5}{vc1,vc3}
\fmf{plain,tension=0.5}{vc2,vc3}
\fmf{plain,tension=0.5}{vc3,vc4}
\fmf{plain,tension=0.5}{vc5,vc7}
\fmf{plain,tension=0.5}{vc6,vc7}
\fmf{plain,tension=0.5}{vc7,vc8}
\fmf{plain,tension=2}{vc1,vc8}
\fmf{phantom,tension=2}{vc5,vc4}
\fmffreeze
\fmfposition
\plainwrap{vc5}{v5}{v8}{vc4}
\fmf{plain,tension=1,left=0,width=1mm}{v5,v8}
\fmffreeze
\end{fmfchar*}}}
}
\caption{Wrapping diagrams with only chiral interactions}
\label{diagrams-chi}
\end{figure}
First of all, there is a set of wrapping graphs made entirely of scalar interactions, presented in Figure~\ref{diagrams-chi}, where the thick line represents the composite operator, and straight lines correspond to scalar propagators. In the following, graphs with vector interactions will appear, where the vector propagator is denoted by a wiggly line.

With the identification of the first and fifth lines, the chiral functions~\eqref{chistruc}, based on the permutation operators~\eqref{permstrucdef}, can still be used to describe the chiral structure of diagrams, even in the wrapping case, and for the graphs of Figure~\ref{diagrams-chi} one finds
\begin{equation}
\label{chiMr4}
\begin{aligned}
W_{\mathrm{chiral}}^{(1)}\quad&\sim\quad\chi(2,4,1,3)\quad\rightarrow\quad 4M_4\col\\
W_{\mathrm{chiral}}^{(2)}\quad&\sim\quad\chi(4,1,2,3)\quad\rightarrow\quad M_4\col\\
W_{\mathrm{chiral}}^{(3)}\quad&\sim\quad\chi(4,3,1,2)\quad\rightarrow\quad 2M_4\pnt
\end{aligned}
\end{equation}
A planar wrapping diagram would be best drawn on the surface of a cylinder, with the circumference of one of the bases representing the composite operator. Such a three-dimensional representation is suggested by the cyclicity of the trace, and is useful for the determination of the symmetries of the diagram, which can be somehow hidden when the graph is projected onto a plane, as in Figure~\ref{diagrams-chi}. 
In particular, the cylindrical representation helps to understand the behaviour under a parity transformation, which reverses the order of fields inside the trace, thus leaving two-impurity operators of the $\sutwo$ sector unchanged. It turns out that the diagrams $W_{\mathrm{chiral}}^{(1)}$ and $W_{\mathrm{chiral}}^{(3)}$ are symmetric under parity, but $W_{\mathrm{chiral}}^{(2)}$ is not. However, its reflection, with structure $\chi(1,4,3,2)\rightarrow M_4$, will obviously produce the same contribution, which can be taken into account by doubling the result from $W_{\mathrm{chiral}}^{(2)}$.

Note that, differently from the asymptotic case, here the exchange $1\leftrightarrow4$ according to~\eqref{permutations:swap} is not allowed, because the identification of the first and fifth lines makes lines 1 and 4 to be neighbours. Moreover, the choice~\eqref{chiMr4} for the structures of the diagrams of Figure~\ref{diagrams-chi} is not unique:
shifts of all the arguments by an integer generate four different chiral functions for $W_{\mathrm{chiral}}^{(1)}$ and $W_{\mathrm{chiral}}^{(3)}$ and two each for $W_{\mathrm{chiral}}^{(2)}$ and for its reflection. A cyclic rotation of the external legs of $W_{\mathrm{chiral}}^{(3)}$ generates an additional chiral function. However, when restricted to the length-four subsector, all the structures associated to the same diagram reduce consistently to the same mixing matrix, as expected, so it is possible to choose a single representative from each class as in~\eqref{chiMr4}. 

After completion of the D-algebra, the results for the diagrams of Figure~\ref{diagrams-chi} are
\begin{eqnarray}
W_{\mathrm{chiral}}^{(1)}&\rightarrow&\cf{4} (\jint{4}{11}/2)\,\chi(2,4,1,3)\rightarrow 2\cf{4} \jint{4}{11} M_4\col  \\
W_{\mathrm{chiral}}^{(2)}&\rightarrow&\cf{4} \jint{4}{9}\,\chi(1,4,3,2)\rightarrow \cf{4} \jint{4}{9} M_4\col  \\
W_{\mathrm{chiral}}^{(3)}&\rightarrow&\cf{4} \jint{4}{10}\,\chi(4,3,1,2)\rightarrow 2\cf{4} \jint{4}{10} M_4
\col
\end{eqnarray}
where the $\jint{4}{i}$ are the momentum-space integrals given in Section~\ref{app:fourloopwrapintegrals}, and only the relevant, divergent terms have been kept. So the total contribution of completely chiral wrapping diagrams, taking the reflection of $W_{\mathrm{chiral}}^{(2)}$ into account, is
\begin{equation}
\label{wrapchiral}
\begin{aligned}
\sum_i W_{\mathrm{chiral}}^{(i)} &\to \cf{4} [(\jint{4}{11}/2)\,\chi(2,4,1,3)+\jint{4}{9}\,(\chi(1,4,3,2)+\chi(4,1,2,3))+\jint{4}{10}\,\chi(4,3,1,2)]\\
&\rightarrow 2\cf{4} (\jint{4}{11}+\jint{4}{9}+\jint{4}{10})M_4
\pnt
\end{aligned}
\end{equation}
The D-algebra for each diagram always produces a $(g^2 N)^4$ colour factor, which combines with the $1/(4\pi)^8$ from the four-loop momentum integral into the $\lambda^4$ power of the 't~Hooft coupling. Thus, there is no need to show such factors explicitly.

Once completely chiral graphs have been analyzed, diagrams with vector interactions, which are conveniently classified according to their chiral structure, must be considered. Their number is huge in principle, growing with the number of vectors. However, only a small number of them is relevant, thanks to the result demonstrated in Appendix~\ref{app:non-maximal}. In fact, for a diagram to be possibly relevant, all its scalar lines with no chiral or antichiral vertices can contain only a single, double-vector vertex. Moreover, diagrams with a vector propagator starting from an outgoing scalar line connected to a three-scalar vertex do not contribute either. The number of graphs that must be computed explicitly is thus dramatically reduced. All the contributing diagrams are listed in Appendix~\ref{app:fourloopwrap}.

Note that, when projected onto a plane, a wrapping diagram at the critical order containing at least one vector propagator can always be drawn in such a way that the wrapping line is a vector one. So one simply needs to consider all the chiral structures with range less than or equal to four (which have at most three loops), and then complete each one to a four-loop wrapping diagram by adding the right number of vectors.

The structures requiring only one vector are $\chi(1,2,3)$, $\chi(1,3,2)$ and $\chi(2,1,3)$. All the relevant wrapping diagrams for these structures, according to the results on cancellations, are shown in Figures~\ref{diagrams-123}, \ref{diagrams-132} and \ref{diagrams-213} in Appendix~\ref{app:fourloopwrap} respectively, together with the results of D-algebra. Here, only the total contribution from each structure will be given
\begin{equation}
\label{wrapthree}
\begin{aligned}
&\sum_i \wgraph{4}{i}{1,2,3}\rightarrow-2\cf{4}(\jint{4}{14}-\jint{4}{16})\chi(1,2,3) \rightarrow 2\cf{4}(\jint{4}{14}-\jint{4}{16})\,M_4 \col \\
&\sum_i \scgraph{4}{i}{1,3,2}\wgraph{4}{i}{1,3,2}\rightarrow -2\cf{4} \jint{4}{13} \chi(1,3,2) \rightarrow 4\cf{4} \jint{4}{13} \,M_4 \col \\
&\sum_i \scgraph{4}{i}{2,1,3}\wgraph{4}{i}{2,1,3}\rightarrow -2\cf{4} \jint{4}{15} \chi(2,1,3) \rightarrow 4\cf{4} \jint{4}{15} \,M_4 \pnt
\end{aligned}
\end{equation}

Unlike $\chi(1,3,2)$ and $\chi(2,1,3)$, $\chi(1,2,3)$ is not symmetric under parity.
Since this transformation changes $\chi(1,2,3)$ into $\chi(3,2,1)$, the diagrams associated to $\chi(3,2,1)$ are simply the reflections of the ones related to $\chi(1,2,3)$, and therefore they do not have to be considered explicitly.
The same observation applies to the other non-symmetric structure $\chi(1,2)$, which is the reflection of $\chi(2,1)$. On the other hand, for symmetric structures, a generic diagram will be symmetric or not depending on the configuration of vector lines. If it is not, by doubling its contribution one automatically includes the one from its reflection, which is itself a valid diagram for the same chiral structure, without the need to consider it explicitly. So the non-trivial symmetry factors $\scgraph{4}{i}{\cdots}$ will be given for non-symmetric diagrams built starting from a symmetric chiral structure, as in Figures~\ref{diagrams-132}, \ref{diagrams-213}, \ref{diagrams-1} and \ref{diagrams-13}.

There are two structures requiring two vectors, $\chi(1,3)$ and $\chi(2,1)$. The corresponding diagrams are listed in Figures~\ref{diagrams-21} and \ref{diagrams-13}. Note that some of the diagrams for the structure $\chi(1,3)$ have a symmetry factor of 4, caused by the additional symmetry of this chiral structure under rotations. The results of D-algebra are
\begin{equation}
\label{wraptwo}
\begin{aligned}
&\sum_i \scgraph{4}{i}{1,3}\wgraph{4}{i}{1,3}\rightarrow 2\cf{4} (\jint{4}{17}+\jint{4}{18}-2\jint{4}{19}) \chi(1,3) \rightarrow 4\cf{4} (\jint{4}{17}+\jint{4}{18}-2\jint{4}{19}) \,M_4 \col \\
&\sum_i \wgraph{4}{i}{2,1}\rightarrow 2\cf{4} (\jint{4}{14}-\jint{4}{16}) \chi(2,1) \rightarrow 2\cf{4} (\jint{4}{14}-\jint{4}{16}) \,M_4 \pnt
\end{aligned}
\end{equation}

Finally, the only one-loop chiral structure, which must be completed with three vectors, is $\chi(1)$, and the associated diagrams are listed in Figure~\ref{diagrams-1}. Note that the number of graphs for this structure would be much larger if the cancellation result of Appendix~\ref{app:non-maximal} had not been used. The total contribution of $\chi(1)$ is
\begin{equation}
\label{wrapone}
\sum_i \scgraph{4}{i}{1}\wgraph{4}{i}{1}\rightarrow 2\cf{4} (\jint{4}{1}-\jint{4}{9}) \chi(1) \rightarrow -2\cf{4} (\jint{4}{1}-\jint{4}{9}) \,M_4 \pnt
\end{equation}

After D-algebra, the outcome of each diagram is given in terms of standard momentum-space integrals, which can be computed using dimensional regularization in $d=4-2\varepsilon$ dimensions. The actual method used for all the integrals is the Gegenbauer Polynomial $x$-space Technique. The main features of this technique that make it ideal for this kind of calculations, will be described in the next section, whereas a short technical review of the method can be found in Appendix~\ref{app:gegenbauer}. Some integrals have also been checked by means of the method of uniqueness~\cite{Kazakov:1983ns} or of the {\tt MINCER} computer program~\cite{Larin:1998}.

\subsection{The Gegenbauer Polynomial \texorpdfstring{$x$}{x}-space Technique}
\label{subsec:GPXT}
The Gegenbauer Polynomial $x$-space Technique (GPXT)~\cite{Chetyrkin:1980pr,Kotikov:1995cw,Celmaster:1980ji}, for the computation of integrals in a space with $d=4-2\varepsilon$ dimensions, is based on the observation that in coordinate space all the propagators depend only on the difference of two coordinates. Thus, all the propagators can be expanded in the basis of Gegenbauer polynomials. The integration with respect to each $d$-dimensional variable can be split into a radial integration and an angular one, and the orthogonality property of the Gegenbauer polynomials can be exploited to perform the angular integrations. In the end, one is left with a given number of series involving the radial integrals. 

The method is effective when the infinite summations can be actually performed exactly, which happens for example when at most one summation survives after the angular integrations. This makes the technique particularly useful for integrals coming from diagrams where a large number of propagators end at the same vertex, which can be chosen as the root vertex (see Appendix~\ref{app:gegenbauer} and~\cite{Chetyrkin:1980pr}). A further simplification is possible if the derivatives in the numerators, when present, can be moved through integration by parts so that at the end of the procedure only the propagators coming out from the root vertex have non-trivial numerators. Another great simplification can be obtained if only the divergent parts of the integrals  are of interest.

Therefore, this technique is in general suitable for the integrals required in the computation of anomalous dimensions, because only the pole part in $\varepsilon$ is needed and the insertion of the composite operator offers a good candidate for the root vertex. In fact, all the integrals that will be found in the present thesis can be calculated in this way. However, it is not possible to show in general that the GPXT can be successfully applied to all the integrals required for the study of anomalous dimensions of composite operators in the planar limit at a generic order, since non-trivial numerators at high orders may make it impossible to perform all the summations exactly.

\subsection{The four-loop anomalous dimension}
Once the values of the integrals $J_i^{(4)}$ have been computed, it is possible to write down the explicit expression for the wrapping correction to the four-loop dilatation operator, simply by summing the partial results given by the equations from~\eqref{wrapchiral} to~\eqref{wrapone}, remembering 
to add the reflections of the non-symmetric structures.
According to the definition of anomalous dimension~\eqref{anomalous}, the contribution of an $\ell$-loop diagram to the coefficient of the corresponding chiral structure is equal to the coefficient of the $1/\varepsilon$ pole of the value of the diagram, multiplied by $(-2\ell)$. At four loops the result is
\vspace{-0.1cm}
\begin{equation}
\begin{aligned}
\label{D4w}
\mathcal{D}_4^{\,\text{w}}
&=-{}8\Big(2\zeta(3)\chi(1)-(3\zeta(3)-5\zeta(5))[\chi(1,2)+\chi(2,1)]
-(1+3\zeta(3)-5\zeta(5))\chi(1,3)\\
&\phantom{{}={}-8}
+(3\zeta(3)-5\zeta(5))[\chi(1,2,3)+\chi(3,2,1)]
+\frac{7}{6}\chi(1,3,2)
+\frac{11}{6}\chi(2,1,3)\\
&\phantom{{}={}-8}
-\frac{1}{2}(1-\zeta(3))\chi(2,4,1,3)
+\Big(\frac{5}{4}-\zeta(3)\Big)[\chi(1,4,3,2)+\chi(4,1,2,3)]\\
&\phantom{{}={}-8}-\Big(\frac{1}{2}-\zeta(3)\Big)\chi(4,1,3,2)
\Big)
\col
\end{aligned}
\end{equation}
\vspace{-0.25cm}
which in the basis~\eqref{Opbasis} reads
\begin{equation}
\label{D4wrapping}
\mathcal{D}_4^{\,\text{w}}\to 8\Big(\frac{17}{2}+18\zeta(3)-30\zeta(5)\Big)\,M_4
\pnt
\end{equation}
This is the correction that must be added to the subtracted operator~\eqref{D4sub} in order to find the exact expression for the dilatation operator on the length-four subsector
\begin{equation}
\label{D4exact}
\begin{aligned}
\mathcal{D}_4^{\,\text{sub}}+\mathcal{D}_4^{\,\text{w}}
&=(200-16\zeta(3))\chi(1)-(150-24\zeta(3)+40\zeta(5))[\chi(1,2)+\chi(2,1)]\\
&\phantom{{}={}}
+(88+8\epsilon_{3a}+24\zeta(3)-40\zeta(5))\chi(1,3)\\
&\phantom{{}={}}+(60-24\zeta(3)+40\zeta(5))[\chi(1,2,3)+\chi(3,2,1)]\\
&\phantom{{}={}}-\Big(\frac{4}{3}-2\beta-4\epsilon_{3a}+4i\epsilon_{3b}-2i\epsilon_{3c}+4i\epsilon_{3d}\Big)\chi(1,3,2)\\
&\phantom{{}={}}-\Big(\frac{20}{3}-2\beta-4\epsilon_{3a}-4i\epsilon_{3b}+2i\epsilon_{3c}-4i\epsilon_{3d}\Big)\chi(2,1,3)\\
&\phantom{{}={}}+4(1-\zeta(3))\chi(2,4,1,3)
-(10-8\zeta(3))[\chi(1,4,3,2)+\chi(4,1,2,3)]\\
&\phantom{{}={}}-(12+2\beta+4\epsilon_{3a})\chi(2,1,3,2)
+(4-8\zeta(3))\chi(4,1,3,2)
\pnt
\end{aligned}
\end{equation}
In the basis \eqref{Opbasis} it gives
\begin{equation}\label{D4matrix}
\mathcal{D}_4^\text{sub}+\mathcal{D}_4^{\,\text{w}}
\to-\big(416-96\zeta(3)+240\zeta(5)\big)\,M_4 \pnt
\end{equation}
To compute the four-loop eigenvalue of the dilatation operator, in principle all the lower-order components should be considered, because of a possible mixing between different orders. However, according to~\eqref{chiM} all the chiral structures up to four loops are proportional to the mixing matrix $M_4$ in the basis~\eqref{Opbasis}, and so also all the lower-order components $\mathcal{D}_1$, $\mathcal{D}_2$ and $\mathcal{D}_3$ of the dilatation operator will share this feature. As a consequence, the eigenstates of the dilatation operator, which are the operators with well-defined anomalous dimension, do not change passing from one to four loops. They can be written in terms of the basis~\eqref{Opbasis} as
\begin{equation}
\begin{aligned}
&\mathcal{O}_{\mathrm{protected}}=\mathcal{O}_{4,1}+2\mathcal{O}_{4,2}=\frac{1}{2}[3\,\mathrm{tr}(\phi\{Z,\phi\}Z)-\mathrm{tr}(\phi[Z,\phi]Z)] \col \\
&\mathcal{O}_K=\basisop{4}{1}-\basisop{4}{2}=\mathrm{tr}(\phi[Z,\phi]Z) \pnt
\end{aligned}
\end{equation}
The first one is protected, whereas $\mathcal{O}_K$ is a descendant of the Konishi operator. The four-loop component of the corresponding eigenvalue of $\mathcal{D}$ is simply the non-vanishing eigenvalue of $(\mathcal{D}_4^{\,\text{sub}}+\mathcal{D}_4^{\,\text{w}})$
\begin{equation}
\gamma_4=-2496+576\zeta(3)-1440\zeta(5) \pnt
\end{equation}
The final result for the anomalous dimension of the Konishi operator up to four loops is thus
\begin{equation}
\label{finalgamma}
\gamma=12\lambda-48\lambda^2+336\lambda^3+\lambda^4(-2496+576\zeta(3)-1440\zeta(5))
\col
\end{equation}
where the dependence on the coupling constant $\lambda$ has been restored, and the lower-order contributions~\cite{Beisert:2006ez} have been included.

Several comments are in order at this point. First of all, the $\zeta(5)$ term, produced entirely by wrapping interactions, increases the degree of transcendentality of the final result with respect to what would be expected from the asymptotic regime, where the only transcendental quantity that can appear at four loops is $\zeta(3)$. In the exact, finite-size case~\eqref{finalgamma}, the $\zeta(3)$ term gets contributions both from wrapping interactions and from the asymptotic dressing phase. Note that the wrapping correction~\eqref{D4wrapping} changes the rational term too.

Before the publication~\cite{us,uslong} of the computation of the four-loop result~\eqref{finalgamma}, several conjectures had been formulated on its possible value, based on the Hubbard model~\cite{Rej:2005qt} and on an analogy with the BFKL equation~\cite{Kotikov:2007cy}. The exact result~\eqref{finalgamma} rules out these early proposals.

The anomalous dimension~\eqref{finalgamma}, obtained through a perturbative, field-theoretical approach, agrees with a later, independent result from the string theory side~\cite{Bajnok:2008bm}, which constitutes a further, non-trivial check of the AdS/CFT correspondence.  More recently, this result has been successfully reproduced by the Y-system approach~\cite{Gromov:2009tv}. 
Moreover, it has also been confirmed by a computer-made perturbative calculation performed in the component-field formalism~\cite{Velizhanin:2009zz}, which has been later extended to find the non-planar contribution~\cite{Velizhanin:2009gv}.

\section{Conclusions}
In this chapter, the perturbative computation of the four-loop anomalous dimension of the length-four Konishi descendant in the $\sutwo$ sector has been presented. At this order, wrapping interactions appear and the asymptotic Bethe equations no longer give the correct spectrum. The knowledge of the asymptotic dilatation operator can still be exploited in order to extract the contribution of a large class of relevant diagrams, after the subtraction of range-five interactions. Then, all the possible wrapping diagrams can be considered. All the computations are greatly simplified by the use of $\N=1$ superspace techniques, which allow to find powerful cancellation rules and reduce the number of Feynman graphs that must be analyzed. 
The approach based on the asymptotic dilatation operator, combined with superspace methods, revealed to be so powerful that the whole calculation could be performed by hand, with the explicit analysis of few tens of diagrams. This should be compared with the full calculation in the component-field formalism, which involves more than 130\!~000 graphs~\cite{Velizhanin:2009zz}.

\chapter{Five-loop leading wrapping corrections in \texorpdfstring{$\mathcal{N}=4$}{N=4} SYM}
\label{chapter:fiveloop}
This chapter contains the analysis of five-loop finite-size corrections on length-five operators in the $\sutwo$ sector of $\N=4$ SYM. The interest in such effects derives from the need for explicit tests of the recently proposed Y-system for the description of the full spectrum of $\N=4$ SYM, inclusive of short operators. Therefore, the perturbative superspace approach will be followed first, as in Chapter~\ref{chapter:fourloop}, and then the result will be compared with the prediction from the Y-system.

\section{Length-five operators in the \texorpdfstring{$\sutwo$}{SU(2)} sector}
The general procedure described in Section~\ref{section:approach}, for the computation of wrapping corrections to the asymptotic results produced by the Bethe ansatz, can be applied to determine the exact five-loop anomalous dimension of the length-five operators of the $\sutwo$ sector. 
Since the loop number equals the length of the states, the perturbative order is again critical, and all the cancellation results of Appendix~\ref{app:non-maximal} still apply.

Two length-five operators with two impurities exist in the $\sutwo$ sector, mixing under renormalization
\begin{equation}
\label{Opbasis5}
\basisop{5}{1}=\mathrm{tr}(\phi Z\phi ZZ)\col\qquad\basisop{5}{2}=\mathrm{tr}(\phi\phi ZZZ) \pnt
\end{equation}
Again, all the chiral structures with range less than or equal to five are proportional to the same $2\times2$ mixing matrix $M_5$ 
\begin{equation}
M_5=\left(
\begin{array}{cc}
1 & -1 \\
-1 & 1
\end{array}
\right)\col
\end{equation}
when they are written in the basis~\eqref{Opbasis5}
\begin{align}
\label{chiM5}
&\chi(1)\rightarrow -2M_5\col &&\chi(2,1)\rightarrow M_5\col &&\chi(1,4)\rightarrow M_5\col \notag \\
&\chi(2,1,4)\rightarrow M_5\col &&\chi(1,2,3)\rightarrow M_5\col &&\chi(2,1,3)\rightarrow -2M_5\col \\
&\chi(1,3,2)\rightarrow -2M_5\col &&\chi(1,4,3,2)\rightarrow -2M_5\col &&\chi(1,2,3,4)\rightarrow -2M_5\col \notag \\
&\chi(3,2,1,4)\rightarrow -2M_5\col &&\chi(2,4,1,3)\rightarrow M_5\col &&  \notag
\end{align}
where all the structures that are not shown explicitly are either the parity reflection of or equivalent to one of these.

As in the length-four case, two-impurity states with length five are all one needs to consider in order to complete the study of \emph{leading} wrapping effects at five loops. To determine the full five-loop spectrum on the $\sutwo$ sector, the only additional quantity that would be required is the five-loop anomalous dimension of the Konishi operator, which gets contributions from finite-size effects beyond the critical order. Its value has been recently computed in~\cite{Bajnok:2009vm,Lukowski:2009ce} by means of the L\"uscher approach and in~\cite{Arutyunov:2010gb,Balog:2010xa} using the Thermodynamic Bethe Ansatz.
So, it would be very interesting to compute this anomalous dimension through a field-theoretical calculation. Such a task, however, would be much more difficult than in the critical case, because of the very large number of relevant diagrams, and could be undertaken only if new supergraph cancellation identities were found. Hence, in this thesis, this quantity will not be studied.

The first step for the five-loop computation of wrapping effects on the length-five subsector is the subtraction of range-six interactions from the asymptotic dilatation operator.

\section{The asymptotic dilatation operator}
First of all, the asymptotic five-loop dilatation operator on the $\sutwo$ sector is needed. This operator has already been obtained in the past~\cite{Beisert:2004ry}, but the calculation was carried out before the discovery of the existence of the dressing phase.
According to~\cite{Beisert:2006ez}, this phase should have two relevant components at five loops: the four-loop contribution $\bdp{3}=\beta_{2,3}^{(3)}$ and a new one, $\bdp{4}=\beta_{2,3}^{(4)}$. Thus, the derivation of~\cite{Beisert:2004ry} must be remade, in order to find the dependence of the coefficients of the dilatation operator on the components of the dressing phase. As a further extension of the known result for the dilatation operator, the general behaviour under similarity transformations must be studied, so that computations in any renormalization scheme are possible. 

\subsection{The general procedure}
The general procedure to find the explicit expression for the asymptotic dilatation operator of $\N=4$ SYM in the $\sutwo$ sector has been described in~\cite{Beisert:2004ry}. Here, the main steps will be reviewed.

The starting point is the assumption that the theory is integrable. This implies the existence of an infinite set of conserved charges, commuting with the Hamiltonian, which is identified with the dilatation operator as explained in Chapter~\ref{chapter:integrability}. In particular, the dilatation operator is the second charge $\mathcal{Q}_2$, the first one being related to the translation operator generated by momentum. The fundamental property for the whole procedure is the existence of the first higher charge $\mathcal{Q}_3$. On the length-$L$ subsector, the perturbative expansions of $\mathcal{D}$ and $\mathcal{Q}_3$ read
\begin{equation}
\mathcal{D}(\lambda)\equiv \mathcal{Q}_2=L+\sum_{k=1}^\infty \lambda^k \mathcal{D}_k\col\qquad\mathcal{Q}_3=\sum_{k=1}^\infty \lambda^k \mathcal{Q}_3^{(k)} \pnt
\end{equation}

The $k$-loop dilatation operator $\mathcal{D}_k$ must have range $(k+1)$. So, it can be written as a linear combination of the permutation operators with range less than or equal to $(k+1)$, which form a finite set. In the same way, $\mathcal{Q}_3^{(k)}$, whose range must be equal to $(k+2)$, can be written in the basis of permutation operators as well. The unknown coefficients can be determined by imposing several constraints:
\begin{enumerate}
\item $\mathcal{D}_k$ and $\mathcal{Q}_3^{(k)}$ have even and odd parity respectively
\label{point-parity}
\begin{equation}
\mathfrak{p}\,\mathcal{D}_k\,\mathfrak{p}=\mathcal{D}_k\col\qquad\mathfrak{p}\,\mathcal{Q}_3^{(k)}\,\mathfrak{p}=-\mathcal{Q}_3^{(k)} \pnt
\end{equation}
\item $\mathcal{D}_k$ is symmetric under transposition, $\mathcal{Q}_3^{(k)}$ is antisymmetric
\begin{equation}
\mathcal{D}_k^{\,T}=\mathcal{D}_k\col\qquad\left(\mathcal{Q}_3^{(k)}\right)^T=-\mathcal{Q}_3^{(k)} \pnt
\end{equation}
\item Both $\mathcal{D}_k$ and $\mathcal{Q}_3^{(k)}$ satisfy the BMN scaling~\cite{Berenstein:2002jq,Gross:2002su} on single-impurity states in the thermodynamic limit of large length $L$ of the spin chain~\cite{Beisert:2004ry}
\begin{equation}
\mathcal{D}_k\sim L^{-(k+1)}\col\qquad\mathcal{Q}_3^{(k)}\sim L^{-(k+2)} \qquad\mathrm{for}\ L\to\infty\pnt
\end{equation}
\item The charge $\mathcal{Q}_3$ is perturbatively conserved up to order $(k+1)$. \\
Since
\label{point-comm}
\begin{equation}
[\mathcal{D},\mathcal{Q}_3]=\sum_{n=1}^\infty\lambda^n\sum_{m=1}^{n-1} [\mathcal{D}_m,\mathcal{Q}_3^{(n-m)}]\col
\end{equation}
this requires 
\begin{equation}
\sum_{m=1}^{k} [\mathcal{D}_m,\mathcal{Q}_3^{(k+1-m)}]=0 \pnt
\end{equation}
\item The spectrum of $\mathcal{D}$ agrees with the asymptotic Bethe equations up to order $k$.
\label{point-bethe}
\end{enumerate}

After these constraints have been fully exploited, some coefficients may remain undetermined. Because of point~\ref{point-bethe}, such unknown coefficients cannot enter the spectrum of the dilatation operator. They are in fact related to similarity transformations, and their actual values depend on the renormalization scheme. Therefore, it is necessary to generalize the result by applying the most general similarity transformation, so that the final expression for the dilatation operator can be used in explicit perturbative computations with any scheme. 
As a consequence, new unknown coefficients typically appear.
Note that the sequence of steps~\ref{point-parity}-\ref{point-bethe} would force the new coefficients to vanish, thus somehow selecting a class of renormalization schemes. When the components of $\mathcal{D}$ and $\mathcal{Q}_3$ are reused at point~\ref{point-comm} to find the higher-order ones, the reduced forms with the new coefficients equal to zero (that is, the ones obtained directly from steps~\ref{point-parity}-\ref{point-bethe}), must be used, instead of the most general ones. Only after the required higher-loop components have been found, the latter can be generalized by taking similarity transformations into account.

By applying this procedure up to four loops, one finds the components of the dilatation operator shown in~\eqref{Duptothree} and~\eqref{D4} and the corresponding ones for $\mathcal{Q}_3$~\cite{Beisert:2004ry,Beisert:2007hz}. In particular, the four-loop component $\mathcal{Q}_3^{(4)}$ of $\mathcal{Q}_3$, which is needed for the five-loop computation but had not been presented in the literature before~\cite{usfive}, can be determined. It is shown in Table~\ref{Q3_4}. 

\begin{table}
\capstart
\begin{equation*}
\footnotesize
\begin{aligned}
\mathcal{Q}_3^{\,(4)} &= {}-{}2i \,(373+2\bdp{3}+\epsilon _{3 a})\,(\{1,2\}-\{2,1\}) \\
&\phantom{{}={}} +2i\,(180+\bdp{3}+2\epsilon _{3 a})\,(\{1,2,3\}-\{3,2,1\}) \\
&\phantom{{}={}} +i\,(40+3\bdp{3})\,(\{1,2,4\}-\{1,4,3\}+\{1,3,4\}-\{2,1,4\}) \\
&\phantom{{}={}} +2i\,(\{1,2,5\}-\{1,5,4\}+\{1,4,5\}-\{2,1,5\}) \\
&\phantom{{}={}} -2i\,(57+\epsilon _{3 a})\,(\{1,2,3,4\}-\{4,3,2,1\}) \\
&\phantom{{}={}} +i\,(23-2\bdp{3}-\epsilon _{3 a})\,(\{1,2,4,3\}-\{1,4,3,2\}+\{2,1,3,4\}-\{3,2,1,4\}) \\
&\phantom{{}={}} +4i\,(8-\bdp{3})\,(\{1,3,2,4\}-\{2,1,4,3\}) \\
&\phantom{{}={}} -4i\,(\{1,2,3,5\}-\{1,5,4,3\}+\{1,2,4,5\}-\{2,1,5,4\}+\{1,3,4,5\}-\{3,2,1,5\}) \\
&\phantom{{}={}} +i\,(1+\bdp{3}+\epsilon _{3 a})\,(\{1,3,2,4,3\}-\{2,1,4,3,2\}+\{2,1,3,2,4\}-\{3,2,1,4,3\}) \\
&\phantom{{}={}} +i \,(1+\epsilon _{3 a})\,(\{1,2,3,5,4\}-\{1,5,4,3,2\}+\{2,1,3,4,5\}-\{4,3,2,1,5\}) \\
&\phantom{{}={}} -i\,(7+\epsilon _{3 a})\,(\{1,2,4,3,5\}-\{2,1,5,4,3\}+\{1,3,2,4,5\}-\{3,2,1,5,4\}) \\
&\phantom{{}={}} +2i\,(4+\epsilon _{3 a})\,(\{1,4,3,2,5\}-\{2,1,3,5,4\}) \\
&\phantom{{}={}} +14i\,(\{1,2,3,4,5\}-\{5,4,3,2,1\}) \\
\end{aligned}
\normalsize
\end{equation*}
\caption{Four-loop component of the first higher conserved charge $\mathcal{Q}_3$}
\label{Q3_4}
\end{table}

\subsection{The five-loop case}

The five-loop operator $\mathcal{D}_5$ must have range equal to six. There are 63 permutation operators with range less than or equal to this value. As for $\mathcal{Q}_3^{(5)}$, in its expansion all the 180 operators with range not greater than seven can appear. The constraints discussed in the previous section can now be applied to fix most of the 63 unknown coefficients $a_i$ for $\mathcal{D}_5$ and of the 180 coefficients $b_j$ for $\mathcal{Q}_3^{(5)}$.
\begin{enumerate}
\item The parity constraints give 24 conditions on the $a_i$ and 100 on the $b_j$.
\item The symmetry requirements produce 12 conditions on the $a_i$ and 34 on the $b_j$.
\item Remarkably, as already noted in~\cite{Beisert:2004ry}, some of the conditions from the vanishing of the commutator between $\mathcal{D}$ and $\mathcal{Q}_3$ involve only $a_i$ coefficients, thus imposing 17 new constraints on them, together with 43 relationships between $b_j$ and $a_i$.
\item The BMN scaling constraint adds 5 new independent conditions on the $a_i$ and 4 relating $a_i$ and $b_j$.
\item Up to this point, only 5 out of the 63 $a_i$ coefficients are still unknown. Two more conditions can be found by requiring agreement of the spectrum with the Bethe equations in the simplest case of two-impurity, length-six states. This is the only step where the components of the dressing phase can enter the coefficients of the dilatation operator. Afterwards, the Bethe equations do not give any more conditions.
\end{enumerate}
Three coefficients of $\mathcal{D}_5$ remain unknown after all the steps. They are related to similarity transformations and can be written in terms of the undetermined coefficients of~\cite{Beisert:2004ry}. In order to obtain an expression for $\mathcal{D}_5$ that can be used in a generic renormalization scheme, it is now necessary to consider the most general similarity transformation, to restore the dependence on the possible additional similarity coefficients that were forced to vanish by the previous procedure.

\subsection{Similarity transformations}
A similarity transformation does not alter the spectrum of the dilatation operator. It acts as
\begin{equation}
\mathcal{D}\to\mathcal{D}\,'=e^{-i\chi}\,\mathcal{D}\,e^{i\chi} \col
\end{equation}
where the generating function $\chi$ can be expanded perturbatively as
\begin{equation}
\chi=\sum_{k=0}^\infty\lambda^k\chi_k \pnt
\end{equation}
The components of $\mathcal{D}'$ up to five loops are given in terms of those of $\mathcal{D}$ by
\begin{equation}
\label{sim}
\begin{aligned}
&\mathcal{D}_0\,'= \mathcal{D}_0 \col \\
&\mathcal{D}_1\,'= \mathcal{D}_1 \col \\
&\mathcal{D}_2\,'= \mathcal{D}_2 \col \\
&\mathcal{D}_3\,'= \mathcal{D}_3 + i\,[\mathcal{D}_1,\chi_2] +i\,[\mathcal{D}_2,\chi_1] \col \\
&\mathcal{D}_4\,'= \mathcal{D}_4 + i\,[\mathcal{D}_1,\chi_3] + i\,[\mathcal{D}_2,\chi_2] +i\,[\mathcal{D}_3,\chi_1]+\frac{1}{2}\big[\chi_1,[\mathcal{D}_1,\chi_2]+[\mathcal{D}_2,\chi_1]\big] \col \\
&\mathcal{D}_5\,'= \mathcal{D}_5 + i\,[\mathcal{D}_1,\chi_4] + i\,[\mathcal{D}_2,\chi_3] + i\,[\mathcal{D}_3,\chi_2] + i\,[\mathcal{D}_4,\chi_1] \\ 
&\qquad\qquad\qquad+ \frac{1}{2}\big[\chi_1,[\mathcal{D}_3,\chi_1]+[\mathcal{D}_2,\chi_2]+[\mathcal{D}_1,\chi_3]\big]+\frac{1}{2}\big[\chi_2,[\mathcal{D}_1,\chi_2]+[\mathcal{D}_2,\chi_1]\big] \\
&\qquad\qquad\qquad -\frac{i}{6}\Big[\chi_1,\big[\chi_1,[\mathcal{D}_1,\chi_2]+[\mathcal{D}_2,\chi_1]\big]\Big]
\pnt
\end{aligned}
\end{equation}
In order not to modify the range of the dilatation operator after a similarity transformation, the expression for the $k$-loop component $\chi_k$ in the basis of permutation operators can contain only contributions with range less than or equal to $(k+1)$. Hence, $\chi_0$ is forced to be proportional to the identity $\{\}$ and is not relevant, whereas
\begin{equation}
\chi_1=\tilde{\epsilon}_1\{1\} \pnt
\end{equation}
To explicitly preserve the invariance of $\mathcal{D}$ under parity, it is possible to look for a parity-even $\chi$. This forces $\chi_2$ to be of the form
\begin{equation}
\label{chisim2}
\chi_2=\tilde{\epsilon}_{2a}(\{1,2\}+\{2,1\})+\tilde{\epsilon}_{2b}\{1\} \pnt
\end{equation}
Similarly, one can take $\chi$ to be Hermitian for real values of the similarity coefficients, thus preserving the Hermiticity of the dilatation operator. However, such Hermiticity is broken if some of the coefficients actually turn out to be complex, which is common when explicit computations are performed in a particular renormalization scheme.
This happens already at four loops, where the most general Hermitian form for $\chi_2$ and $\chi_3$ is~\cite{Beisert:2007hz}
\begin{equation}
\label{chisim3}
\begin{aligned}
\chi_3 &= i\,\tilde{\epsilon}_{3a}(\{2,1,3\}-\{1,3,2\})+\tilde{\epsilon}_{3b}(\{1,2,3\}+\{3,2,1\})\\
&\qquad\qquad\qquad+\tilde{\epsilon}_{3c}\{1,3\}+\tilde{\epsilon}_{3d}(\{1,2\}+\{2,1\}) + \tilde{\epsilon}_{3e}\{1\}\pnt
\end{aligned}
\end{equation}
With this choice for $\chi_3$, the transformed dilatation operator is still Hermitian only for real values of the $\tilde{\epsilon}_{3x}$ coefficients, but a Feynman diagram computation in a typical scheme will in general produce complex values for them~\cite{uslong,Beisert:2007hz}. 
The same happens at five loops.
Anyway, the apparent non-Hermiticity of the dilatation operator depends on the renormalization scheme and can be removed by an appropriate choice of the scalar product, as explained in~\cite{Beisert:2007hz}.

Note that the coefficient $\tilde{\epsilon}_{2a}$ in~\eqref{chisim2} is not the same as the $\epsilon_{2a}$ that enters the three-loop dilatation operator given in~\eqref{Duptothree}. The latter is a useful redefinition of the former that removes the dependence on $\tilde{\epsilon}_1$ in order to show explicitly that $\mathcal{D}_3$ depends on a single similarity coefficient. In the same way, the coefficients $\epsilon_{3x}$ of~\eqref{D4} are redefinitions of the $\tilde{\epsilon}_{3x}$ appearing in~\eqref{chisim3}.

The most general parity-invariant and Hermitian expression for $\chi_4$ is
\begin{equation}
\chi_4=\sum_{\alpha\in\{a,b,\ldots,n\}}\tilde{\epsilon}_{4\alpha}\chi_{4\alpha} \col
\end{equation}
where the components $\chi_{4k}$ are shown in Table~\ref{chi4}. The general form of the five-loop dilatation can now be found by application of~\eqref{sim}, and the result in the basis of permutation operators is given in Table~\ref{D5-perm}. 
This expression can be easily transformed to the basis of chiral structures through the extension of the rules~\eqref{chistruc} and~\eqref{invchistruc} to five loops 
\begin{equation}
\footnotesize
\begin{aligned}
\chi(a,b,c,d,e)&=
-\pid+5\pone{1}-\ptwo{a}{b}-
\ptwo{a}{c}-\ptwo{a}{d}-\ptwo{a}{e}-\ptwo{b}{c}-\ptwo{b}{d}-\ptwo{b}{e} \\
&\phantom{{}={}}
-\ptwo{c}{d}-\ptwo{c}{e}-\ptwo{d}{e}+\pthree{a}{b}{c}+
\pthree{a}{b}{d}+\pthree{a}{b}{e}+\pthree{a}{c}{d} \\
&\phantom{{}={}}
+\pthree{a}{c}{e}+
\pthree{a}{d}{e}+\pthree{b}{c}{d}+\pthree{b}{c}{e}+\pthree{b}{d}{e}+
\pthree{c}{d}{e} \\
&\phantom{{}={}}
-\pfour{a}{b}{c}{d}-\pfour{a}{b}{c}{e}-\pfour{a}{b}{d}{e}-
\pfour{a}{c}{d}{e}-\pfour{b}{c}{d}{e} \\
&\phantom{{}={}}
+\{a,b,c,d,e\}\col \\
\{a,b,c,d,e\} &=
\chi (a,b,c,d,e)+\chi (b,c,d,e)+\chi (a,c,d,e)+
\chi (a,b,d,e)+\chi (a,b,c,e) \\
&\phantom{{}={}}
+\chi (a,b,c,d)+\chi (c,d,e)+
\chi (b,d,e)+\chi (b,c,e)+\chi (b,c,d)+\chi (a,d,e) \\
&\phantom{{}={}}
+\chi (a,c,e)+\chi (a,c,d)+\chi (a,b,e)+\chi (a,b,d)+
\chi (a,b,c)+\chi (d,e) \\
&\phantom{{}={}}
+\chi (c,e)+\chi (c,d)+\chi (b,e)+
\chi (b,d)+\chi (b,c)+\chi (a,e)+\chi (a,d) \\
&\phantom{{}={}}
+\chi (a,c)+
\chi (a,b)+5\chi (1)+\chi ()\col 
\end{aligned}
\normalsize
\end{equation}
and the result is presented in Table~\ref{table:D5}.

\begin{table}[h!]
\capstart
\newcommand{\fwchi}[1]{\makebox[0.72cm][l]{$\chi_{#1}$}}
\begin{flalign*}
&\fwchi{4a}\!=\{1,2,3,4\}+\{4,3,2,1\} &&\fwchi{4h}\!=i(\{1,3,2\}-\{2,1,3\}) \\
&\fwchi{4b}\!=i(\{1,2,4,3\}+\{1,4,3,2\}-\{2,1,3,4\}-\{3,2,1,4\}) &&\fwchi{4i}\!=\{1,3,2\}+\{2,1,3\} \\
&\fwchi{4c}\!=\{1,2,4,3\}+\{1,4,3,2\}+\{2,1,3,4\}+\{3,2,1,4\} && \fwchi{4j}\!=\{1,2,3\}+\{3,2,1\} \\
&\fwchi{4d}\!=\{1,3,2,4\}+\{2,1,4,3\} && \fwchi{4k}\!=\{1,2\}+\{2,1\} \\
&\fwchi{4e}\!=\{2,1,3,2\} && \fwchi{4l}\!=\{1,4\} \\
&\fwchi{4f}\!=i(\{1,2,4\}+\{1,4,3\}-\{1,3,4\}-\{2,1,4\}) && \fwchi{4m}\!=\{1,3\} \\
&\fwchi{4g}\!=\{1,2,4\}+\{1,4,3\}+\{1,3,4\}+\{2,1,4\} && \fwchi{4n}\!=\{1\} 
\end{flalign*}
\caption{Components of the generating function}
\label{chi4}
\end{table}

\begin{table}[p]
\capstart
\vspace{-1.5cm}
\footnotesize
\newcommand{\eqspace}{\phantom{D_5={}}}
\begin{equation*}
\begin{aligned}
\mathcal{D}_5&= 
+4(1479+14\bdp{3}-\bdp{4})\redperm{}
-4(2902+42\bdp{3}-3\bdp{4}+32\epsilon_{4h})\redperm{1}\\
&\phantom{{}={}}
+4(816+13\bdp{3}-\bdp{4}-16\epsilon_{4b}+32\epsilon_{4h})(\redperm{1,2}+\redperm{2,1})\\
&\phantom{{}={}}
+2(512+48\bdp{3}-3\bdp{4}+4\epsilon_{2a}^2+4\epsilon_{3a}+64\epsilon_{4f}+32\epsilon_{4h})\redperm{1,3}\\
&\phantom{{}={}}
+8(20+\bdp{3}+2\epsilon_{2a}^2+2\epsilon_{3a}-16\epsilon _{4b}-16\epsilon_{4f})\redperm{1,4}
+4\redperm{1,5}\\
&\phantom{{}={}}
-4(326-\bdp{3}+2\epsilon_{2a}^2+2\epsilon_{3a}-32\epsilon_{4b}+16\epsilon_{4h})(\redperm{1,2,3}+\redperm{3,2,1})\\
&\phantom{{}={}}
+4(94-15\bdp{3}+\bdp{4}+16\epsilon_{4b}-16\epsilon_{4h}+16i\epsilon_{4k})\redperm{1,3,2}\\
&\phantom{{}={}}
+4(94-15\bdp{3}+\bdp{4}+16\epsilon_{4b}-16\epsilon_{4h}-16i\epsilon_{4k})\redperm{2,1,3}\\
&\phantom{{}={}}
-8(12+\bdp{3}+\epsilon _{2a}^2+\epsilon_{3a}-8\epsilon_{4b}+4i\epsilon_{4m})(\redperm{1,3,4}+\redperm{2,1,4})\\
&\phantom{{}={}}
-8(12+\bdp{3}+\epsilon_{2a}^2+\epsilon _{3a}-8\epsilon_{4b}-4i\epsilon_{4m})(\redperm{1,2,4}+\redperm{1,4,3})\\
&\phantom{{}={}}
-4(1-8i\epsilon_{4l})(\redperm{1,2,5}+\redperm{1,5,4})\\
&\phantom{{}={}}
-4(1+8i\epsilon_{4l})(\redperm{1,4,5}+\redperm{2,1,5})
-8\redperm{1,3,5}\\
&\phantom{{}={}}
-2(40-12\bdp{3}+\bdp{4}+4\epsilon_{2a}^2+4\epsilon_{3a}-32\epsilon_{4h})\redperm{2,1,3,2}\\
&\phantom{{}={}}
+8(35+\epsilon_{2a}^2+\epsilon_{3a}-8\epsilon_{4b})(\redperm{1,2,3,4}+\redperm{4,3,2,1})\\
&\phantom{{}={}}
-8(21-2\bdp{3}-8\epsilon_{4f}+8\epsilon_{4h})(\redperm{1,3,2,4}+\redperm{2,1,4,3})\\
&\phantom{{}={}}
+4(\epsilon_{2a}^2+\epsilon_{3a}-8(3\epsilon_{4b}+\epsilon_{4f}-\epsilon_{4h}+i\epsilon_{4j}))(\redperm{2,1,3,4}+\redperm{3,2,1,4})\\
&\phantom{{}={}}
+4(\epsilon_{2a}^2+\epsilon_{3a}-8(3\epsilon_{4b}+\epsilon_{4f}-\epsilon_{4h}-i\epsilon_{4j}))(\redperm{1,2,4,3}+\redperm{1,4,3,2})\\
&\phantom{{}={}}
+32(1-2 \epsilon_{4f})(\redperm{1,2,4,5}+\redperm{2,1,5,4})\\
&\phantom{{}={}}
-8 (3-8\epsilon_{4f}-8i\epsilon_{4g})\redperm{1,2,5,4}
-8(3-8\epsilon_{4f}+8i\epsilon_{4g})\redperm{2,1,4,5}\\
&\phantom{{}={}}
-2(2i\epsilon_{2a}+i\epsilon_{3c}-16\epsilon_{4f}-16i\epsilon_{4g})(\redperm{1,2,3,5}+\redperm{1,5,4,3})\\
&\phantom{{}={}}
+2(2i\epsilon_{2a}+i\epsilon_{3c}+16\epsilon_{4f}-16i\epsilon _{4g})(\redperm{1,3,4,5}+\redperm{3,2,1,5})\\
&\phantom{{}={}}
+2 (4-2i\epsilon_{2a}-i\epsilon_{3c}-16\epsilon_{4f}-16i\epsilon_{4g})(\redperm{1,4,3,5}+\redperm{2,1,3,5})\\
&\phantom{{}={}}
+2(4+2i\epsilon_{2a}+i\epsilon_{3c}-16\epsilon_{4f}+16i\epsilon_{4g})(\redperm{1,3,2,5}+\redperm{1,3,5,4})\\
&\phantom{{}={}}
+2(10-\bdp{3}+16\epsilon_{4b}+16i\epsilon_{4e})(\redperm{1,3,2,4,3}+\redperm{2,1,4,3,2})\\
&\phantom{{}={}}
+2(10-\bdp{3}+16\epsilon_{4b}-16i\epsilon_{4e})(\redperm{2,1,3,2,4}+\redperm{3,2,1,4,3})\\
&\phantom{{}={}}
+4(4+\epsilon _{2a}^2+\epsilon_{3a}-16i\epsilon_{4d})\redperm{2,1,4,3,5}\\
&\phantom{{}={}}
+4(4+\epsilon_{2a}^2+\epsilon_{3a}+16i\epsilon_{4d})\redperm{1,3,2,5,4}\\
&\phantom{{}={}}
+4(2+\epsilon_{2a}^2+2i\epsilon_{2a}+\epsilon_{3a}-i\epsilon_{3b}-8\epsilon_{4b}-8i\epsilon_{4c}+8i\epsilon_{4d})(\redperm{1,2,4,3,5}+\redperm{2,1,5,4,3})\\
&\phantom{{}={}}
+4(2+\epsilon_{2a}^2-2i\epsilon_{2a}+\epsilon_{3a}+i\epsilon_{3b}-8\epsilon_{4b}+8i \epsilon_{4c}-8i\epsilon_{4d})(\redperm{1,3,2,4,5}+\redperm{3,2,1,5,4})\\
&\phantom{{}={}}
+4(2-\epsilon_{2a}^2-\epsilon_{3a}+16\epsilon_{4b}+16i\epsilon_{4c})\redperm{1,2,5,4,3}\\
&\phantom{{}={}}
+4(2-\epsilon_{2a}^2-\epsilon_{3a}+16\epsilon_{4b}-16i\epsilon_{4c})\redperm{3,2,1,4,5}\\
&\phantom{{}={}}
-4(7+16\epsilon _{4b})(\redperm{1,4,3,2,5}+\redperm{2,1,3,5,4})\\
&\phantom{{}={}}
-4(\epsilon_{2a}^2+\epsilon_{3a}-8i\epsilon_{4a}-8\epsilon_{4b})(\redperm{1,2,3,5,4}+\redperm{1,5,4,3,2})\\
&\phantom{{}={}}
-4(\epsilon_{2a}^2+\epsilon_{3a}+8i\epsilon_{4a}-8\epsilon_{4b})(\redperm{2,1,3,4,5}+\redperm{4,3,2,1,5})\\
&\phantom{{}={}}
-28(\redperm{1,2,3,4,5}+\redperm{5,4,3,2,1})
\end{aligned}
\end{equation*}
\normalsize
\caption{Asymptotic five-loop dilatation operator in the permutation basis}
\label{D5-perm}
\end{table}
\begin{table}
\capstart
\vspace{-1.5cm}
\footnotesize
\newcommand{\eqspace}{\phantom{D_5={}}}
\begin{equation*}
\begin{aligned}
\mathcal{D}_5&=
-1960\redchi{1}
+1568(\redchi{1,2}+\redchi{2,1})
-16(40-\bdp{3}-\epsilon_{2a}^2-\epsilon_{3a}-8\epsilon_{4f}+8\epsilon_{4h})\redchi{1,3}\\
&\phantom{{}={}}
+16(4+\epsilon_{2a}^2+\epsilon_{3a}-16\epsilon_{4b}-8\epsilon_{4f})\redchi{1,4}
-4\redchi{1,5}
-784(\redchi{1,2,3}+\redchi{3,2,1})\\
&\phantom{{}={}}
+2(64-8\bdp{3}+\bdp{4}
+8\epsilon_{2a}^2+4i\epsilon_{2a}+8\epsilon_{3a}+2i\epsilon_{3c}\\
&\phantom{{}={}+2(}
-32(\epsilon_{4b}+\epsilon_{4h})+32i(\epsilon_{4a}+\epsilon_{4c}+\epsilon_{4d}+\epsilon_{4e}+\epsilon_{4g}+\epsilon_{4j}+\epsilon_{4k}))\redchi{1,3,2}\\
&\phantom{{}={}}
+2(64-8\bdp{3}+\bdp{4}
+8\epsilon_{2a}^2-4i\epsilon_{2a}+8\epsilon_{3a}-2i\epsilon_{3c}\\
&\phantom{{}={}+2(}
-32(\epsilon_{4b}+\epsilon_{4h})-32i(\epsilon_{4a}+\epsilon_{4c}+\epsilon_{4d}+\epsilon_{4e}+\epsilon_{4g}+\epsilon_{4j}+\epsilon_{4k}))\redchi{2,1,3}\\
&\phantom{{}={}}
+2(30+\bdp{3}
+4\epsilon_{2a}^2-4i\epsilon_{2a}+4\epsilon_{3a}+4i\epsilon_{3b}+2i\epsilon_{3c}\\
&\phantom{{}={}+2(}
-48\epsilon_{4b}-16i(2\epsilon_{4a}+2\epsilon_{4c}+2\epsilon_{4d}+\epsilon_{4e}+2\epsilon_{4g}+2\epsilon_{4j}+\epsilon_{4m}))(\redchi{1,3,4}+\redchi{2,1,4})\\
&\phantom{{}={}}
+2(30+\bdp{3}
+4\epsilon_{2a}^2+4i\epsilon_{2a}+4\epsilon_{3a}-4i\epsilon_{3b}-2i\epsilon_{3c}\\
&\phantom{{}={}+2(}
-48\epsilon_{4b}+16i(2\epsilon_{4a}+2\epsilon_{4c}+2\epsilon_{4d}+\epsilon_{4e}+2\epsilon_{4g}+2\epsilon_{4j}+\epsilon_{4m}))(\redchi{1,2,4}+\redchi{1,4,3})\\
&\phantom{{}={}}
-4(1-8i(2\epsilon_{4a}+2\epsilon_{4c}+2\epsilon_{4d}+4\epsilon_{4g}+\epsilon_{4l}))(\redchi{1,2,5}+\redchi{1,5,4})\\
&\phantom{{}={}}
-4(1+8i(2\epsilon_{4a}+2\epsilon_{4c}+2\epsilon_{4d}+4\epsilon_{4g}+\epsilon_{4l}))(\redchi{1,4,5}+\redchi{2,1,5})
-8\redchi{1,3,5}\\
&\phantom{{}={}}
+2(8\bdp{3}-\bdp{4}-4\epsilon_{2a}^2-4\epsilon_{3a}+64\epsilon_{4b}+32\epsilon_{4h})\redchi{2,1,3,2}
+224(\redchi{1,2,3,4}+\redchi{4,3,2,1})\\
&\phantom{{}={}}
-4(20-3\bdp{3}-4\epsilon_{2a}^2-4\epsilon_{3a}-16\epsilon_{4f}+16\epsilon_{4h})(\redchi{1,3,2,4}+\redchi{2,1,4,3})\\
&\phantom{{}={}}
+2(4-\bdp{3}-4i\epsilon_{2a}+2i\epsilon_{3b}\\
&\phantom{{}={}+2(}
-32\epsilon_{4b}-16\epsilon_{4f}+16\epsilon_{4h}-16i(\epsilon_{4a}+\epsilon_{4c}+\epsilon_{4d}+\epsilon_{4e}+\epsilon_{4j}))(\redchi{2,1,3,4}+\redchi{3,2,1,4})\\
&\phantom{{}={}}
+2(4-\bdp{3}+4i\epsilon_{2a}-2i\epsilon_{3b}\\
&\phantom{{}={}+2(}
-32\epsilon_{4b}-16\epsilon_{4f}+16\epsilon_{4h}+16i(\epsilon_{4a}+\epsilon_{4c}+\epsilon_{4d}+\epsilon_{4e}+\epsilon_{4j}))(\redchi{1,2,4,3}+\redchi{1,4,3,2})\\
&\phantom{{}={}}
-8(1-\epsilon_{2a}^2-\epsilon_{3a}+16\epsilon_{4b}+8\epsilon_{4f})(\redchi{1,2,4,5}+\redchi{2,1,5,4})\\
&\phantom{{}={}}
-8(\epsilon_{2a}^2+\epsilon_{3a}
-16\epsilon_{4b}-8\epsilon_{4f}-8i(\epsilon_{4a}+\epsilon_{4c}+\epsilon_{4d}+\epsilon_{4g}))\redchi{1,2,5,4}\\
&\phantom{{}={}}
-8(\epsilon_{2a}^2+\epsilon_{3a}
-16\epsilon_{4b}-8\epsilon_{4f}+8i(\epsilon_{4a}+\epsilon_{4c}+\epsilon_{4d}+\epsilon_{4g}))\redchi{2,1,4,5}\\
&\phantom{{}={}}
-2(6
+2\epsilon_{2a}^2-2i\epsilon_{2a}+2\epsilon_{3a}+2i\epsilon_{3b}+i\epsilon_{3c}\\
&\phantom{{}={}-2(}
-32\epsilon_{4b}-16\epsilon_{4f}
-16i(\epsilon_{4a}+\epsilon_{4c}+\epsilon_{4d}+\epsilon_{4g}))(\redchi{1,2,3,5}+\redchi{1,5,4,3})\\
&\phantom{{}={}}
-2(6
+2\epsilon_{2a}^2+2i\epsilon_{2a}+2\epsilon_{3a}-2i\epsilon_{3b}-i\epsilon_{3c}\\
&\phantom{{}={}-2(}
-32\epsilon_{4b}-16\epsilon_{4f}
+16i(\epsilon_{4a}+\epsilon_{4c}+\epsilon_{4d}+\epsilon_{4g}))(\redchi{1,3,4,5}+\redchi{3,2,1,5})\\
&\phantom{{}={}}
+2(2
+2\epsilon_{2a}^2+2i\epsilon_{2a}+2\epsilon_{3a}-2i\epsilon_{3b}-i\epsilon_{3c}\\
&\phantom{{}={}+2(}
-32\epsilon_{4b}-16\epsilon_{4f}
-16i(\epsilon_{4a}+\epsilon_{4c}+\epsilon_{4d}+\epsilon_{4g}))(\redchi{1,4,3,5}+\redchi{2,1,3,5})\\
&\phantom{{}={}}
+2(2
+2\epsilon_{2a}^2-2i\epsilon_{2a}+2\epsilon_{3a}+2i\epsilon_{3b}+i\epsilon_{3c}\\
&\phantom{{}={}+2(}
-32\epsilon_{4b}-16\epsilon_{4f}
+16i(\epsilon_{4a}+\epsilon_{4c}+\epsilon_{4d}+\epsilon_{4g}))(\redchi{1,3,2,5}+\redchi{1,3,5,4})\\
&\phantom{{}={}}
+2(10-\bdp{3}+16\epsilon_{4b}+16i\epsilon_{4e})(\redchi{1,3,2,4,3}+\redchi{2,1,4,3,2})\\
&\phantom{{}={}}
+2(10-\bdp{3}+16\epsilon_{4b}-16i\epsilon_{4e})(\redchi{2,1,3,2,4}+\redchi{3,2,1,4,3})\\
&\phantom{{}={}}
+4(4+\epsilon_{2a}^2+\epsilon_{3a}-16i\epsilon_{4d})\redchi{2,1,4,3,5}
+4(4+\epsilon_{2a}^2+\epsilon_{3a}+16i\epsilon_{4d})\redchi{1,3,2,5,4}\\
&\phantom{{}={}}
+4(2+\epsilon_{2a}^2+2i\epsilon_{2a}+\epsilon_{3a}-i\epsilon_{3b}-8\epsilon_{4b}-8i(\epsilon_{4c}-\epsilon_{4d}))(\redchi{1,2,4,3,5}+\redchi{2,1,5,4,3})\\
&\phantom{{}={}}
+4(2+\epsilon_{2a}^2-2i\epsilon_{2a}+\epsilon_{3a}+i\epsilon_{3b}-8\epsilon_{4b}+8i(\epsilon_{4c}-\epsilon_{4d}))(\redchi{1,3,2,4,5}+\redchi{3,2,1,5,4})\\
&\phantom{{}={}}
+4(2-\epsilon_{2a}^2-\epsilon_{3a}+16\epsilon_{4b}+16i\epsilon_{4c})\redchi{1,2,5,4,3}\\
&\phantom{{}={}}
+4(2-\epsilon_{2a}^2-\epsilon_{3a}+16\epsilon_{4b}-16i\epsilon_{4c})\redchi{3,2,1,4,5}\\
&\phantom{{}={}}
-4(7+16\epsilon _{4b})(\redchi{1,4,3,2,5}+\redchi{2,1,3,5,4})\\
&\phantom{{}={}}
-4(\epsilon _{2a}^2+\epsilon_{3a}-8i\epsilon_{4a}-8\epsilon_{4b})(\redchi{1,2,3,5,4}+\redchi{1,5,4,3,2})\\
&\phantom{{}={}}
-4(\epsilon_{2a}^2+\epsilon_{3a}+8i\epsilon_{4a}-8\epsilon_{4b})(\redchi{2,1,3,4,5}+\redchi{4,3,2,1,5})\\
&\phantom{{}={}}
-28(\redchi{1,2,3,4,5}+\redchi{5,4,3,2,1})
\end{aligned}
\end{equation*}
\normalsize
\vspace{-0.5cm}
\caption{Asymptotic five-loop dilatation operator in the chiral basis}
\label{table:D5}
\end{table}

\subsection{Subtraction of range-six contributions}
As in the four-loop case, thanks to the cancellation result of Appendix~\ref{app:non-maximal}, the task of subtraction of range-six interactions can be performed by simply deleting from $\mathcal{D}_5$ all the terms whose chiral structures have range six, and thus contain both 1 and 5 among the arguments. In fact, the divergent parts of all the non-maximal range-six graphs, where a shorter-range chiral structure is extended to range six by vector propagators, sum up to zero. In this way one obtains the subtracted operator of Table~\ref{table:D5sub}, which in the length-five basis~\eqref{Opbasis5} reads
\begin{equation}
\label{D5subM}
\mathcal{D}_5^{\,\mathrm{sub}}\to 16(210+13\bdp{3}-\bdp{4}+24\epsilon_{4b}+24\epsilon_{4f})\,M_5 \pnt
\end{equation}
The values of $\epsilon_{4b}$ and $\epsilon_{4f}$ are determined in Appendix~\ref{app:fiveR6} from range-six diagrams.
The $\bdp{3}$ component is equal to $4\zeta(3)$, as already seen at four loops, and the value of $\bdp{4}$ has been conjectured to be equal to $-40\zeta(5)$ in~\cite{Beisert:2006ez}. Substituting into~\eqref{D5subM}, one gets
\begin{equation}
\label{D5subM2}
\mathcal{D}_5^{\,\mathrm{sub}}\to 2(1665+416\zeta(3)+320\zeta(5))\,M_5 \pnt
\end{equation}
Now, wrapping contributions must be considered through a direct computation of all the possible Feynman supergraphs.

\begin{table}[h!]
\capstart
\newcommand{\eqsubspace}{\phantom{D_5^{\mathrm{sub}}={}}}
\footnotesize
\begin{equation*}
\begin{aligned}
\mathcal{D}_5^{\,\mathrm{sub}}&=
-1960\redchi{1}
+1568(\redchi{1,2}+\redchi{2,1})
-16(40-\bdp{3}-\epsilon_{2a}^2-\epsilon_{3a}-8\epsilon_{4f}+8\epsilon_{4h})\redchi{1,3}\\
&\phantom{{}={}}
+16(4+\epsilon_{2a}^2+\epsilon_{3a}-16\epsilon_{4b}-8\epsilon_{4f})\redchi{1,4}
-784(\redchi{1,2,3}+\redchi{3,2,1})\\
&\phantom{{}={}}
+2(64-8\bdp{3}+\bdp{4}
+8\epsilon_{2a}^2+4i\epsilon_{2a}+8\epsilon_{3a}+2i\epsilon_{3c}\\
&\phantom{{}={}+2(}
-32(\epsilon_{4b}+\epsilon_{4h})+32i(\epsilon_{4a}+\epsilon_{4c}+\epsilon_{4d}+\epsilon_{4e}+\epsilon_{4g}+\epsilon_{4j}+\epsilon_{4k}))\redchi{1,3,2}\\
&\phantom{{}={}}
+2(64-8\bdp{3}+\bdp{4}
+8\epsilon_{2a}^2-4i\epsilon_{2a}+8\epsilon_{3a}-2i\epsilon_{3c}\\
&\phantom{{}={}+2(}
-32(\epsilon_{4b}+\epsilon_{4h})-32i(\epsilon_{4a}+\epsilon_{4c}+\epsilon_{4d}+\epsilon_{4e}+\epsilon_{4g}+\epsilon_{4j}+\epsilon_{4k}))\redchi{2,1,3}\\
&\phantom{{}={}}
+2(30+\bdp{3}
+4\epsilon_{2a}^2-4i\epsilon_{2a}+4\epsilon_{3a}+4i\epsilon_{3b}+2i\epsilon_{3c}\\
&\phantom{{}={}+2(}
-48\epsilon_{4b}-16i(2\epsilon_{4a}+2\epsilon_{4c}+2\epsilon_{4d}+\epsilon_{4e}+2\epsilon_{4g}+2\epsilon_{4j}+\epsilon_{4m}))(\redchi{1,3,4}+\redchi{2,1,4})\\
&\phantom{{}={}}
+2(30+\bdp{3}
+4\epsilon_{2a}^2+4i\epsilon_{2a}+4\epsilon_{3a}-4i\epsilon_{3b}-2i\epsilon_{3c}\\
&\phantom{{}={}+2(}
-48\epsilon_{4b}+16i(2\epsilon_{4a}+2\epsilon_{4c}+2\epsilon_{4d}+\epsilon_{4e}+2\epsilon_{4g}+2\epsilon_{4j}+\epsilon_{4m}))(\redchi{1,2,4}+\redchi{1,4,3})\\
&\phantom{{}={}}
+2(8\bdp{3}-\bdp{4}-4\epsilon_{2a}^2-4\epsilon_{3a}+64\epsilon_{4b}+32\epsilon_{4h})\redchi{2,1,3,2}\\
&\phantom{{}={}}
+224(\redchi{1,2,3,4}+\redchi{4,3,2,1})\\
&\phantom{{}={}}
-4(20-3\bdp{3}-4\epsilon_{2a}^2-4\epsilon_{3a}-16\epsilon_{4f}+16\epsilon_{4h})(\redchi{1,3,2,4}+\redchi{2,1,4,3})\\
&\phantom{{}={}}
+2(4-\bdp{3}-4i\epsilon_{2a}+2i\epsilon_{3b}\\
&\phantom{{}={}+2(}
-32\epsilon_{4b}-16\epsilon_{4f}+16\epsilon_{4h}-16i(\epsilon_{4a}+\epsilon_{4c}+\epsilon_{4d}+\epsilon_{4e}+\epsilon_{4j}))(\redchi{2,1,3,4}+\redchi{3,2,1,4})\\
&\phantom{{}={}}
+2(4-\bdp{3}+4i\epsilon_{2a}-2i\epsilon_{3b}\\
&\phantom{{}={}+2(}
-32\epsilon_{4b}-16\epsilon_{4f}+16\epsilon_{4h}+16i(\epsilon_{4a}+\epsilon_{4c}+\epsilon_{4d}+\epsilon_{4e}+\epsilon_{4j}))(\redchi{1,2,4,3}+\redchi{1,4,3,2})\\
&\phantom{{}={}}
+2(10-\bdp{3}+16\epsilon_{4b}+16i\epsilon_{4e})(\redchi{1,3,2,4,3}+\redchi{2,1,4,3,2})\\
&\phantom{{}={}}
+2(10-\bdp{3}+16\epsilon_{4b}-16i\epsilon_{4e})(\redchi{2,1,3,2,4}+\redchi{3,2,1,4,3})\\
\end{aligned}
\end{equation*}
\normalsize
\caption{Subtracted five-loop dilatation operator}
\label{table:D5sub}
\end{table}

\section{Wrapping diagrams}
The procedure to deal with wrapping supergraphs resembles the one followed in Chapter~\ref{chapter:fourloop} for the four-loop case. First, 
the cancellation result of Appendix~\ref{app:non-maximal}
is exploited to recognize all the possibly relevant diagrams, listed in Appendix~\ref{app:fivewrap}.
Then, the D-algebra procedure is applied to each graph, resulting in a standard momentum-space integral that can be computed using the GPXT. 

There are three wrapping diagrams containing only scalar interactions, shown in Figure~\ref{wrap5-chiral}. Through the identification of the first and sixth lines, their chiral structures can still be written in terms of chiral functions, as $\chi(2,1,3,4,5)$, $\chi(1,2,3,4,5)$ and $\chi(1,3,2,5,4)$ respectively. Their parity reflections will produce the same numerical contributions, associated to the reflected structures $\chi(4,5,3,2,1)$, $\chi(5,4,3,2,1)$ and $\chi(5,3,4,1,2)$. Note that, unlike in the four-loop case, here none of the completely-chiral wrapping diagrams is parity-symmetric. 

All the other possible wrapping supergraphs contain at least one vector propagator, which can always be taken as the wrapping line when the diagram is drawn, because the perturbative order is critical. So, once again the wrapping diagrams with vector interactions can be built starting from the chiral structures with range less than or equal to five. A minimal set of \emph{independent} structures is made of 
\begin{gather}
\chi(2,4,1,3)\col\  \chi(3,2,1,4)\col\  \chi(1,2,3,4)\col\  \chi(1,4,3,2)\col\  \chi(1,3,2) \col\\
\chi(2,1,3)\col\  \chi(1,2,3)\col\  \chi(2,1,4)\col\  \chi(2,1)\col\  \chi(1,4)\col\  \chi(1) \pnt\quad 
\end{gather}
In fact, all the other structures are either the reflection under parity or the same thing as one of the structures of the minimal set. In particular, on length-five states the structure $\chi(1,2,4)$ is equivalent to $\chi(2,1,4)$. The same happens for the structures $\chi(1,3)$ and $\chi(1,4)$. This is a new feature of the five-loop case, absent at four loops where all the structures with the correct range were independent, apart from parity reflections, and is similar to the freedom in the choice of the chiral structures associated to completely-chiral wrapping graphs. The contributions of the non-symmetric structures will be doubled to account for the corresponding parity reflections, whereas the structures that are equivalent to another one must be simply discarded. The wrapping diagrams for the structures of the minimal set are shown in Figures~\ref{wrap5-2413}-\ref{wrap5-14}.

Note that the chiral structures $\chi(2,1,4)$ (with its reflection $\chi(1,3,4)$) and $\chi(1,4)$ produce, among the others, two terms involving the antisymmetric tensor $\epsilon_{\mu\nu\rho\sigma}$, and the integral $\jintind{5}{35}{\mu\nu\rho\sigma}$. Their separate calculation would require the knowledge of $\jintind{5}{35}{\mu\nu\rho\sigma}$ for generic values of the indices, which would be very difficult to determine. When combined, however, their total contribution vanishes
\begin{equation}
4i\epsilon_{\mu\nu\rho\sigma}(\jintind{5}{35}{\mu\sigma\rho\nu}+\jintind{5}{35}{\mu\nu\rho\sigma})M_5=0 \col
\end{equation}
thanks to the antisymmetry of $\epsilon_{\mu\nu\rho\sigma}$. 

When all the wrapping contributions are collected, the correction to the dilatation operator on the length-five subsector is found to be
\begin{equation}
\begin{aligned}
\label{D5w}
\mathcal{D}_5^{\,\text{w}}
=-10\Big[&{}
-{}\Big(\frac{1}{6}-\frac{4}{5}\zeta(3)\Big) (\redchi{2,1,3,4,5}+\redchi{4,5,3,2,1}) \\
&
+\Big(\frac{14}{5}-4 \zeta(5)\Big) (\redchi{1,2,3,4,5}+\redchi{5,4,3,2,1}) \\
&
-\Big(\frac{19}{30}-\frac{4}{5} \zeta(3)\Big) (\redchi{1,3,2,5,4}+\redchi{5,3,4,1,2}) \\
&
-\Big(\frac{1}{3}+\frac{12}{5} \zeta(3)-4 \zeta(5)\Big) (\redchi{2,4,1,3}+\redchi{1,3,2,4}) \\
&
+\Big(\frac{1}{3}-\frac{12}{5} \zeta(3)+4 \zeta(5)\Big) (\redchi{3,2,1,4}+\redchi{2,1,3,4}) \\
&
+(8 \zeta(5)-14 \zeta(7)) (\redchi{1,2,3,4}+\redchi{4,3,2,1}) \\
&
-\Big(\frac{2}{5}+\frac{12}{5} \zeta(3)-4 \zeta(5)\Big) (\redchi{1,4,3,2}+\redchi{1,2,4,3}) \\
&
+\Big(\frac{13}{10}+\frac{8}{5} \zeta(3)\Big) \redchi{1,3,2}
+\Big(\frac{19}{10}+\frac{8}{5} \zeta(3)\Big) \redchi{2,1,3} \\ 
&
+\Big(\frac{18}{5}+\frac{44}{5} \zeta(3)-12 \zeta(5)\Big) (\redchi{2,1,4}+\redchi{1,3,4}) \\
&
-(8 \zeta(5)-14 \zeta(7)) (\redchi{2,1}+\redchi{1,2}) \\
&
-\Big(\frac{18}{5}+8 \zeta(3)+8 \zeta(5)-28 \zeta(7)\Big) \redchi{1,4}
+8 \zeta(5) \redchi{1}
\Big]
\pnt
\end{aligned}
\end{equation}
In the basis~\eqref{Opbasis5}, this expression reads
\begin{equation}
\mathcal{D}_5^{\,\mathrm{w}}\rightarrow 2(1-128\zeta(3)+640\zeta(5)-560\zeta(7))M_5 \pnt
\end{equation}
By adding this correction to the subtracted operator in matrix form~\eqref{D5subM2}, one finds
\begin{equation}
\mathcal{D}_5^{\,\text{sub}}+\mathcal{D}_5^{\,\text{w}}\rightarrow 4(833+144\zeta(3)+480\zeta(5)-280\zeta(7))\,M_5 \pnt
\end{equation}
From~\eqref{chiM5} it follows that all the components of the dilatation operator up to four loops are proportional to $M_5$ on this subsector, and the five-loop components of the anomalous dimensions are simply given by the eigenvalues of $(\mathcal{D}_5^{\,\text{sub}}+\mathcal{D}_5^{\,\text{w}})$. One of them vanishes, as expected, whereas the other one is
\begin{equation}
\label{gamma5}
\gamma_5=6664+1152\zeta(3)+3840\zeta(5)-2240\zeta(7) \col
\end{equation}
so that the correct anomalous dimension up to five loops is
\begin{multline}
\label{fullgamma5}
\gamma=8\lambda-24\lambda^2+136\lambda^3-8[115+16\zeta(3)]\lambda^4 \\
+[6664+1152\zeta(3)+3840\zeta(5)-2240\zeta(7)]\lambda^5 \pnt
\end{multline}
The linear combination of the basis~\eqref{Opbasis5} with this anomalous dimension is
\begin{equation}
\label{op5mult}
\basisop{5}{1}'=\mathrm{tr}(\phi Z\phi ZZ)-\mathrm{tr}(\phi\phi ZZZ)\col
\end{equation}
and the protected combination is
\begin{equation}
\basisop{5}{2}'=\mathrm{tr}(\phi Z\phi ZZ)+\mathrm{tr}(\phi\phi ZZZ)\pnt
\end{equation}

The result~\eqref{fullgamma5} shares the main features with the four-loop one~\eqref{finalgamma}. In particular, all the coefficients of the $\zeta$ functions are integer numbers. Moreover, the transcendentality degree is increased by the appearance of the $\zeta(7)$ term, which is generated by finite-size effects and cannot show up in the asymptotic spectrum. Here, the coefficients of $\zeta(3)$ and $\zeta(5)$ get contributions both from wrapping and from the asymptotic dressing phase.

This anomalous dimension was first computed in~\cite{Beccaria:2009eq} from a conjectured form of the five-loop scaling function in the $\sltwo$ sector. The explicit perturbative analysis confirms that result, which can serve as a useful test for the recently-proposed Y-system approach~\cite{Gromov:2009tv,Gromov:2009bc,Bombardelli:2009ns,Arutyunov:2009ur}, as outlined in the next section.

\section{Computation using the Y-system}
\label{section:ysystem5}
The so-called Y-system~\cite{Gromov:2009tv,Gromov:2009bc,Bombardelli:2009ns,Arutyunov:2009ur} is a set of functional equations proposed to describe the full spectrum of $\N=4$ SYM, comprehensive of wrapping effects on short operators. It can be derived from the thermodynamic Bethe ansatz approach~\cite{deLeeuw:2008ye,Arutyunov:2007tc,Arutyunov:2009zu,Arutyunov:2009ce,Arutyunov:2009mi,Arutyunov:2009ga,deLeeuw:2009hn,Arutyunov:2009ux,Arutyunov:2009ax}.
In~\cite{Gromov:2009tv} it was applied to find the four-loop anomalous dimension of the Konishi operator, and the result was found to be in agreement with the previous computations~\cite{us,uslong,Bajnok:2008bm,Velizhanin:2009zz}. Here, this technique will be used to determine the five-loop anomalous dimension of length-five states~\cite{Beccaria:2009eq,usfive}, which will be explicitly checked against the result found in the previous section. Notice of the agreement between the two results, without an explicit proof, was given in~\cite{Beccaria:2009eq}.

Following~\cite{Gromov:2009tv}, it is simpler to work in the $\sltwo$ sector instead of the $\sutwo$ one. This is possible because, as already explained in Chapter~\ref{chapter:integrability}, the two-impurity states of the $\sutwo$ sector belong to multiplets with representatives also in $\sltwo$. In particular, the two-impurity, length-$L$ subsector of $\sutwo$ is mapped onto the two-impurity, length-$(L-2)$ subsector of $\sltwo$. Thus, the operator~\eqref{op5mult} will have the same anomalous dimension as the non-protected linear combination of the two length-three $\sltwo$ states with two covariant derivatives.

The fundamental quantities in the Y-system formalism are the $Y_{a,s}(u)$ functions of the rapidity $u$, solutions of the functional equations
\begin{equation}
\frac{Y_{a,s}^+Y_{a,s}^-}{Y_{a+1,s}Y_{a-1,s}}=\frac{(1+Y_{a,s+1})(1+Y_{a,s-1})}{(1+Y_{a+1,s})(1+Y_{a-1,s})} \col
\end{equation}
where $f^{\pm}(u)=f(u\pm i/2)$.
The finite-size corrections to the asymptotic anomalous dimensions can be computed in terms of the $Y_{a,s}(u)$ functions according to
\begin{equation}
\label{ywrap}
\gamma^{\mathrm{w}}_{L+2}=\sum_{a=1}^\infty\int_{-\infty}^\infty\frac{\mathrm{d}u}{2\pi i}\frac{\partial\epsilon_a^*}{\partial u}\mathrm{log}[1+Y_{a,0}^*(u)] \pnt
\end{equation}
Here $\epsilon_a(u)$ is the energy dispersion relation
\begin{equation}
\epsilon_{a}(u)=a+2i \lambda\left[\frac{1}{x^{[+a]}(u)}-\frac{1}{x^{[-a]}(u)}\right] \col
\end{equation}
where $x$ is a function of $u$ defined as
\begin{equation}
\frac{u}{\sqrt{\lambda}}=x+\frac{1}{x} \pnt
\end{equation}
The notation $f^{[\pm a]}(u)=f(u\pm a i/2)$ has been introduced, and the $*$ on $\epsilon_a$ and $Y_{a,0}$ indicates that these quantities must be evaluated in the mirror kinematics, where $\vert x^{[s]}\vert>1$ for $-a+1\leq s\leq a$ and $\vert x^{[-a]}\vert<1$, rather than in the physical one where $\vert x^{[s]}\vert>1$ for every value of $s$.

The most general states in the theory can be described in terms of seven different kinds of excitations, as explained in~\cite{Beisert:2005fw}. Let $K_m$ be the excitation numbers and $x_{m,j}$ the Bethe roots describing a particular state. 
The corresponding general solution for $Y_{a,0}(u)$ is~\cite{Gromov:2009tv}
\begin{equation}
\label{Ya0-general}
Y_{a,0}(u)=\Big(\frac{x^{[-a]}}{x^{[+a]}}\Big)^L\frac{\phi^{[-a]}}{\phi^{[+a]}}T_{a,-1}^L T_{a,1}^R \col
\end{equation}
where $T_{a,-1}^L$ and $T_{a,1}^R$ are the transfer matrix eigenvalues of the antisymmetric irreducible representations of the $SU(2\vert2)_L$ and $SU(2\vert2)_R$ subgroups of the full $SU(2,2\vert4)$ symmetry, and $\phi$ is a function of $u$ defined by
\begin{equation}
\label{yphi}
\frac{\phi^{[-a]}}{\phi^{[+a]}}=S^2\frac{B^{[+a](+)}R^{[-a](-)}}{B^{[-a](-)}R^{[+a](+)}}\frac{B^{[+a]}_{1L}B^{[-a]}_{3L}}{B^{[-a]}_{1L}B^{[+a]}_{3L}}\frac{B^{[+a]}_{1R}B^{[-a]}_{3R}}{B^{[-a]}_{1R}B^{[+a]}_{3R}} \pnt
\end{equation}
Here, $S$ is the dressing factor, and the $B_x$ and $R_x$ functions are defined as
\begin{equation}
R_m^{(\pm)}(u)=\prod_{j=1}^{K_m}\frac{x(u)-x_{m,j}^\mp}{\sqrt{x_{m,j}^\mp}}\col\qquad B_m^{(\pm)}(u)=\prod_{j=1}^{K_m}\frac{1/x(u)-x_{m,j}^\mp}{\sqrt{x_{m,j}^\mp}} \col
\end{equation}
with the additional convention that $B^{(\pm)}\equiv B_4^{(\pm)}$, $R^{(\pm)}\equiv R_4^{(\pm)}$, whereas the corresponding expressions without the $(\pm)$ superscript are obtained through the substitution $x_{m,j}^\mp\to x_{m,j}$. It is useful to define also the functions
\begin{equation}
Q_m(u)=-R_m(u)B_m(u) \col
\end{equation}
subject to the same conventions.

The expression~\eqref{yphi} reduces to a much simpler form for states in the $\sutwo$ or $\sltwo$ sectors, where the only non-vanishing roots $u_{m,j}$ have $m=4$. Moreover, as will be clear in the following, the dressing factor $S$ can be replaced by unity, which is its zeroth-order approximation, because the 
remaining factor in \eqref{yphi} already produces the required power of the 't~Hooft coupling. So \eqref{yphi} reduces to
\begin{equation}
\frac{\phi^{[-a]}}{\phi^{[+a]}}=\frac{B^{[+a](+)}R^{[-a](-)}}{B^{[-a](-)}R^{[+a](+)}} \pnt
\end{equation}
Since for a two-impurity state the momentum constraint requires $u_{4,2}=-u_{4,1}$, the contribution from the first two factors of~\eqref{Ya0-general} can be written as
\begin{equation}
\Big(\frac{x^{[-a]}}{x^{[+a]}}\Big)^L\frac{\phi^{[-a]}}{\phi^{[+a]}}\rightarrow9\frac{2^{2L}\lambda^{L}(4u_{4,1}^2+1)^2}{(a^2+4u^2)^L\tilde{y}_{-a}(u)} \col
\end{equation}
where
\begin{equation}
\tilde{y}_a(u)=9[((1-a)^2+4u^2)^2+8u_{4,1}^2((1-a)^2-4u^2+2u_{4,1}^2)] \pnt
\end{equation}
The next step is the calculation of the $T_{a,-1}^L T_{a,1}^R$ factor. Again, the restriction to the $\sltwo$ or $\sutwo$ sector considerably simplifies the task, since on these sectors the action of the two left and right $SU(2\vert2)$ subgroups is the same, and the transfer matrices can be extracted from the simplified generating functional~\cite{Gromov:2009tv}
\begin{equation}
\mathcal{W}=\Big[1-\frac{B^{+(+)}R^{-(+)}}{B^{+(-)}R^{-(-)}}\mathfrak{D}\Big]\Big[1-\frac{R^{-(+)}}{R^{-(-)}}\mathfrak{D}\Big]^{-2}[1-\mathfrak{D}]\col\qquad \mathfrak{D}=e^{-i\partial_u} \col
\end{equation}
using
\begin{equation}
\mathcal{W}^{-1}=\sum_{a=0}^\infty(-1)^a T_{a,1}^{[1-a]}\mathfrak{D}^a \pnt
\end{equation}
In this way the contribution from the third factor of~\eqref{Ya0-general} is found to be
\begin{equation}
T_{a,-1}^LT_{a,1}^R\rightarrow2^{10}\lambda^2\frac{[12a(u^2-u_{4,1}^2)+3a(a^2-1)]^2}{(4u_{4,1}^2+1)^2(a^2+4u^2)^2\tilde{y}_a(u)} \col
\end{equation}
so that the required expression for $Y_{a,0}^*(u)$ is
\begin{equation}
\label{ysol}
Y_{a,0}^*(u)=9\lambda^{L+2}2^{10+2L}\frac{[12a(u^2-u_{4,1}^2)+3a(a^2-1)]^2}{(a^2+4u^2)^{L+2}\tilde{y}_a(u)\tilde{y}_{-a}(u)} \pnt
\end{equation}
The Bethe root $u_{4,1}$ is the solution of the two-impurity Bethe equation in the $\sltwo$ sector. As $Y_{a,0}^*(u)$ is proportional to $\lambda^{L+2}$, to determine the wrapping correction at the critical order, which is the leading finite-size contribution, it is enough to take $u_{4,1}$ as the zeroth-order solution of the $\sltwo$ Bethe equations
\begin{equation}
\Big(\frac{u_{4,1}+i/2}{u_{4,1}-i/2}\Big)^{L+1}=1 \pnt
\end{equation}
For $L=2$, computing $u_{4,1}$ and substituting into~\eqref{ysol}, one finds the result of~\cite{Gromov:2009tv} for the four-loop case. 
In the five-loop case of interest here, $L=3$ and consequently $u_{4,1}=1/2$. Now the explicit expression~\eqref{ysol} can be substituted into the general formula~\eqref{ywrap}, where only the leading $\lambda^5$ contribution must be kept. In particular, 
\begin{equation}
\frac{\partial\epsilon_a^*}{\partial u}=-2i+o(1) \col\qquad\log[1+Y_{a,0}^*(u)]= Y_{a,0}^*(u)+o(\lambda^{L+2}) \col
\end{equation}
and the expression~\eqref{ywrap} simplifies to
\begin{equation}
\gamma^{\mathrm{w}}_{L+2}=-\frac{1}{\pi}\sum_{a=1}^\infty\int_{-\infty}^\infty Y_{a,0}^*(u) \pnt
\end{equation}
The value of the integral can be found by means of the residue theorem in the complex plane. If the integration path is closed with a half-circle in the upper half plane and the limit for the corresponding radius going to infinity is taken, the wrapping contribution to the anomalous dimension reads
\begin{equation}
\gamma^{\mathrm{w}}_{5}=-2i\sum_{a=1}^\infty\sum_j\mathrm{Res}\left[Y_{a,0}^*(u),\tilde{u}_j\right] \col
\end{equation}
where the $\tilde{u}_j$ are the locations of the poles of $Y_{a,0}^*(u)$. For $L=3$ and $u_{4,1}=1/2$ they are
\begin{equation}
\begin{aligned}
\pm\frac{1}{2}+i\col\quad\frac{i}{2} \qquad & \mathrm{for}\ a=1 \col \\
\pm\frac{1}{2}+\frac{i}{2}(a\pm 1)\col\quad a\frac{i}{2} \qquad & \mathrm{for}\ a>1 \pnt
\end{aligned}
\end{equation}
After the summations, the result is
\begin{equation}
\gamma_5^{\mathrm{w}}=-\lambda^5[128+512 \zeta(3)-2560\zeta(5)+2240\zeta(7)] \pnt
\end{equation}
This is the correction that must be added to the asymptotic five-loop anomalous dimension given by the Bethe equations
\begin{equation}
\gamma_5^{\mathrm{as}}=\lambda^5[6792+1664\zeta(3)+1280\zeta(5)] \pnt
\end{equation}
The exact five-loop contribution to the anomalous dimension of the two-impurity, length-five operator is therefore
\begin{equation}
\gamma_5=\gamma_5^{\mathrm{as}}+\gamma_5^{\mathrm{w}}=\lambda^5[6664+1152\zeta(3)+3840\zeta(5)-2240\zeta(7)] \col
\end{equation}
which agrees with the result~\eqref{gamma5} of the direct field-theoretical analysis presented in the first part of this chapter. This is another explicit test of the validity of the Y-system for the computation of finite-size effects on short operators in $\N=5$ SYM, in addition to the four-loop anomalous dimension of the Konishi operator. Note that this approach can be applied to find the leading wrapping correction to the asymptotic spectrum of two-impurity operators at a generic order in the $\sutwo$ and $\sltwo$ sectors, since the general solution~\eqref{ysol} is given as a function of both the operator length $L$ and the Bethe root $u_{4,1}$.

\section{Conclusions}
The computation of the five-loop anomalous dimensions of length-five operators in the $\sutwo$ sector of $\N=4$ SYM can be performed by means of a simple extension of the four-loop procedure applied to the Konishi operator. As the order is still critical, all the cancellation results can be exploited. The final result confirms the transcendentality pattern already seen at four loops, with the new $\zeta(7)$ contribution coming entirely from wrapping diagrams and the lower-transcendentality terms $\zeta(3)$ and $\zeta(5)$ getting contributions also from the dressing phase components of the asymptotic spectrum.

Like the four-loop result, also the five-loop one can be used to test the validity of the recently-proposed Y-system, whose prediction agrees with the result of the direct calculation in terms of Feynman diagrams.

\chapter{Wrapping in the \texorpdfstring{$\beta$}{beta}-deformed \texorpdfstring{$\mathcal{N}=4$}{N=4} SYM}
\label{chapter:wrapbetadef}

In this chapter, finite-size effects in the $\beta$-deformed version of $\N=4$ SYM are studied. The reduced number of supersymmetries allows single-impurity operators, which are protected in the undeformed theory, to acquire an anomalous dimension. These operators are easier to study than multiple-impurity ones, and therefore it is possible to perform perturbative computations up to higher orders.

\section{The dilatation operator}
\label{sec:defdilop}
The general procedure presented in Chapter~\ref{chapter:fourloop}, which allows to find the wrapping corrections to the anomalous dimensions of short operators, can be applied also to the deformed theory described in Section~\ref{sec:betadef}. This approach requires the knowledge of the asymptotic dilatation operator for the deformed theory, and it is possible to show that such an operator does not have to be computed from scratch, but can be derived directly from the undeformed case.

The permutation that swaps two fields at sites $i$ and $j$ of the spin chain can be written in terms of the $2\times 2$ Pauli matrices as
\begin{equation}
\perm_{i,j}=\frac{1}{2}\left[\unitmatrix_{i,j}+\vec{\sigma}_i\cdot\vec{\sigma}_j\right]=\frac{1}{2}\left[\unitmatrix_{i,j}+\sigma_i^3\sigma_j^3+\sigma_i^+\sigma_j^-+\sigma_i^-\sigma_j^+\right]
\col
\end{equation}
where $\sigma_j^\pm=\sigma_j^1\pm i\sigma_j^2$.
Consider now
the deformed permutations~\cite{Berenstein:2004ys}
\begin{equation}
\dperm_{i,j}=\frac{1}{2}\left[\unitmatrix_{i,j}+\sigma_i^3\sigma_j^3+q^2\,\sigma_i^+\sigma_j^-+\bar{q}^2\,\sigma_i^-\sigma_j^+\right] 
\col\qquad\qquad q\equiv e^{i\pi\beta} \col
\label{defperm}
\end{equation}
and define the corresponding deformations of the standard basis operators~\eqref{permstrucdef}
\begin{equation}
\label{defbasisops}
\dpthree{a_1}{\dots}{a_n}=\sum_{r=0}^{L-1}\dperm_{a_1+r,\;a_1+r+1}\cdots\dperm_{a_n+r,\;a_n+r+1}
\pnt
\end{equation}
These definitions are very useful, as they capture all the phase dependence of the scalar vertices. In particular, the deformed chiral structure $\dchi{1}$ can be written as
\begin{equation}
\dchi{1}=\dpone{1}-\dpid \pnt
\end{equation}
In the deformed theory too, $\dchi{1}$ is still the building block for all the possible non-trivial chiral structures, since the deformation only affects the coefficients of the scalar vertices. Therefore, with the definition~\eqref{defbasisops} the chiral structures can be written as linear combinations of the operators of the permutation basis~\eqref{defbasisops} with the same coefficients as in the undeformed case: the relations~\eqref{chistruc} and~\eqref{invchistruc} are still valid provided that all the objects are replaced by their deformed counterparts. 

All the dependence on $q$ and $\bar{q}$ is encoded in the deformed basis operators~\eqref{defbasisops}, and hence the dilatation operator must be writable in terms of the deformed basis with coefficients that are independent of $q$ and $\bar{q}$. Note at this point that
\begin{equation}
\lim_{q,\bar{q}\rightarrow1}\dpthree ab\dots=\{a,b,\dots\} \col
\end{equation}
that is, the deformed operators reduce to the undeformed ones for $\beta\to0$. In this limit, the deformed dilatation operator must reduce to the one for standard $\N=4$ SYM, but since its coefficients do not depend on $\beta$, they must be the same as in the undeformed case. The conclusion is that the dilatation operator for the deformed theory can be obtained from the one for undeformed $\N=4$ SYM simply by replacing the standard permutation operators with their deformed versions. It is possible to check that its eigenvalues as functions of $\beta$ agree with the outcomes of the deformed Bethe equations given in~\cite{Beisert:2005if}.

The most interesting feature of the deformed theory is the fact that single-impurity states of the $\sutwo$ sector are no longer protected. For such states, one does not even need to know the asymptotic dilatation operator, thanks to the all-loop result~\eqref{single-all-orders} for the asymptotic anomalous dimensions. As a consequence, complete computations at much higher orders become feasible. Indeed, even though the asymptotic dilatation operator is known only up to five loops at the moment, wrapping effects on single-impurity states will be studied here explicitly up to eleven loops.
In the next section, the easiest single-impurity state will be considered.

\section{Three loops}
\label{sec:three}
The shortest unprotected operator in the $\sutwo$ sector of the $\beta$-deformed $\N=4$ SYM theory is the single-impurity, length-three one
\begin{equation}
\mathcal{O}_3=\mathrm{tr}(\phi ZZ) \pnt
\end{equation}
All the length-two operators, in fact, are known to be protected~\cite{Penati:2005hp,Freedman:2005cg}. Thus, wrapping effects may appear already at three loops, unlike in the undeformed case where all the length-three operators were protected.
The three-loop component of the anomalous dimension for an asymptotic single-impurity state, that is a state with length greater than or equal to four, can be obtained by expanding the all-order solution~\eqref{single-all-orders} up to order $\lambda^3$
\begin{equation}
\label{3loop-asymptotic}
\gamma_3(\mathcal{O}_{L>3})=
256\,\lambda^3\sin^6(\pi\beta)
\pnt
\end{equation}
To determine the exact anomalous dimension of $\mathcal{O}_3$ starting from this result, one must subtract the contributions of range-four diagrams and add the ones from wrapping graphs. Thanks to the fact that the cancellation results demonstrated in Appendix~\ref{app:non-maximal} hold good in the deformed theory too, only three-loop diagrams with a range-four chiral structure need to be considered for subtraction. The only such structures that can be applied to a single-impurity operator are $\dchi{1,2,3}$ and its reflection $\dchi{3,2,1}$. Note that the structures $\dchi{2,1,3}$ and $\dchi{1,3,2}$, both requiring at least two impurities, are not relevant in this case. This is another example of the simplifications concerning single-impurity operators.

In the deformed theory, the contribution of a diagram that is not symmetric under parity is in general a complex number, because of the $q$ and $\bar{q}$ factors in the scalar vertices. The value of the corresponding reflected diagram will be the complex conjugate of that number.
In fact, in a planar diagram every scalar vertex has a factor of $q$ or $\bar{q}$ depending on the clockwise or counter-clockwise order in which the three superfields $\phi$, $Z$ and $\psi$ appear, and a parity reflection reverses this order turning each $q$ into $\bar{q}$ and vice-versa. Therefore, for non-symmetric diagrams, the contribution of the reflected graph can be taken into account by simply adding the complex conjugate of the result of D-algebra. 

\begin{figure}[h]
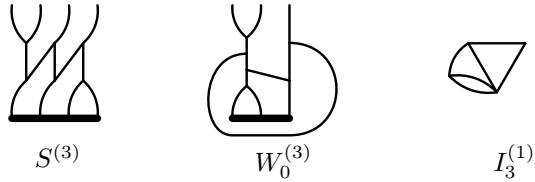

\capstart
\renewcommand*{\thesubfigure}{\ \ }
\unitlength=0.75mm
\settoheight{\eqoff}{$\times$}%
\setlength{\eqoff}{0.5\eqoff}%
\addtolength{\eqoff}{-12.5\unitlength}%
\settoheight{\eqofftwo}{$\times$}%
\setlength{\eqofftwo}{0.5\eqofftwo}%
\addtolength{\eqofftwo}{-7.5\unitlength}%
\subfigure[$\subthree{3}$]{
\raisebox{\eqoff}{%
\fmfframe(3,1)(1,4){%
\begin{fmfchar*}(20,20)
\Wfourplain
\end{fmfchar*}}}
}
\subfigspace
\subfigspace
\subfigure[$\swrapthree{3}{0}$]{
\raisebox{\eqoff}{%
\fmfframe(3,1)(1,4){%
\begin{fmfchar*}(20,20)
\fmftop{v1}
\fmfbottom{v4}
\fmfforce{(0.125w,h)}{v1}
\fmfforce{(0.125w,0)}{v4}
\fmffixed{(0.25w,0)}{v1,v2}
\fmffixed{(0.25w,0)}{v2,v3}
\fmffixed{(0.25w,0)}{v4,v5}
\fmffixed{(0.25w,0)}{v5,v6}
\fmffixed{(0,whatever)}{vc1,vc2}
\fmffixed{(0,whatever)}{vc2,vc3}
\fmffixed{(0,whatever)}{vc4,vc5}
\fmf{plain,tension=0.5,right=0.25}{v1,vc1}
\fmf{plain,tension=0.5,left=0.25}{v2,vc1}
\fmf{plain,tension=2}{vc1,vc7}
\fmf{plain,tension=2}{vc7,vc2}
\fmf{plain,tension=2}{vc2,vc3}
\fmf{plain,tension=0.5,left=0.25}{v4,vc3}
\fmf{plain,tension=0.5,right=0.25}{v5,vc3}
\fmf{plain,tension=0.5,left=0}{v3,vc4}
\fmf{plain,tension=0.5,right=0}{v6,vc5}
\fmf{plain,tension=0}{vc2,vc5}
\fmf{plain,tension=0.5}{vc4,vc5}
\fmfposition
\plainwrap{vc7}{v4}{v6}{vc4}
\fmf{plain,tension=0.5,right=0,width=1mm}{v4,v6}
\fmffreeze
\end{fmfchar*}}}
}
\subfigspace
\subfigspace
\subfigure[$\sint{3}{1}$]{
\raisebox{\eqoff}{%
\fmfframe(3,1)(1,4){%
\begin{fmfchar*}(20,15)
\fmfleft{in}
\fmfright{out}
\fmf{plain}{in,v1}
\fmf{plain,left=0.25}{v1,v2}
\fmf{plain,left=0}{v2,v4}
\fmf{plain,tension=0.5,right=0.25}{v1,v0,v1}
\fmf{phantom,right=0.25}{v0,v3}
\fmf{plain}{v0,v2}
\fmf{plain}{v0,v4}
\fmffixed{(w,0)}{v1,v3}
\fmffixed{(0.5w,0)}{v2,v4}
\fmfpoly{phantom}{v4,v2,v0}
\fmffreeze
\end{fmfchar*}}}
}
\caption{Completely chiral diagrams (wrapping and subtraction)}
\label{def3-scalar}
\end{figure}

According to the previous considerations, only the diagram $\subthree{3}$ of Figure~\ref{def3-scalar}, with structure $\dchi{1,2,3}$, has to be computed to get rid of range-four interactions.
After D-algebra, and adding the reflected graph $\bsubthree{3}$, one finds
\begin{equation}
\subthree{3}+\bsubthree{3}\rightarrow(g^2 N)^3\left(q-\bar{q}\right)^2\left(q^4+\bar{q}^4\right)\sint{3}{1}
\col
\end{equation}
where $\sint{3}{1}$ is the value of the momentum integral shown in Figure~\ref{def3-scalar}.
\\
Now, wrapping diagrams must be analyzed. One of them, the $\swrapthree{3}{0}$ of Figure~\ref{def3-scalar}, is made only of scalar interactions. Its contribution, once the reflected graph is considered too, is
\begin{equation}
\swrapthree{3}{0}+\bswrapthree{3}{0}\rightarrow(g^2 N)^3\left(q-\bar{q}\right)^2\left(q^4+\bar{q}^4\right)\sint{3}{1}
\pnt
\end{equation}
\begin{figure}[t]
\capstart
\renewcommand*{\thesubfigure}{\ \ }
\addtolength{\subfigcapskip}{5pt}
\unitlength=0.75mm
\settoheight{\eqoff}{$\times$}%
\setlength{\eqoff}{0.5\eqoff}%
\addtolength{\eqoff}{-12.5\unitlength}%
\settoheight{\eqofftwo}{$\times$}%
\setlength{\eqofftwo}{0.5\eqofftwo}%
\addtolength{\eqofftwo}{-7.5\unitlength}%
\subfigure[$\swgraph{3}{1}{2,1}$]{
\raisebox{\eqoff}{%
\fmfframe(3,1)(1,4){%
\begin{fmfchar*}(20,20)
\WthreeplainB
\fmfipair{w[]}
\fmfipair{wd[]}
\svertex{w3}{p3}
\svertex{w6}{p6}
\wigglywrap{w3}{v4}{v5}{w6}
\end{fmfchar*}}}
}
\subfigspace
\subfigure[$\swgraph{3}{2}{2,1}$]{
\raisebox{\eqoff}{%
\fmfframe(3,1)(1,4){%
\begin{fmfchar*}(20,20)
\WthreeplainB
\fmfipair{w[]}
\fmfipair{wd[]}
\svertex{w5}{p5}
\svertex{w3}{p3}
\wigglywrap{w3}{v4}{v5}{w5}
\end{fmfchar*}}}
}
\subfigspace
\subfigure[$\swgraph{3}{3}{2,1}$]{
\raisebox{\eqoff}{%
\fmfframe(3,1)(1,4){%
\begin{fmfchar*}(20,20)
\WthreeplainB
\fmfipair{w[]}
\fmfipair{wd[]}
\svertex{w6}{p6}
\svertex{w2}{p2}
\wigglywrap{w2}{v4}{v5}{w6}
\end{fmfchar*}}}
}
\subfigspace
\subfigure[$\swgraph{3}{4}{2,1}$]{
\raisebox{\eqoff}{%
\fmfframe(3,1)(1,4){%
\begin{fmfchar*}(20,20)
\WthreeplainB
\fmfipair{w[]}
\fmfipair{wd[]}
\svertex{w2}{p2}
\svertex{w5}{p5}
\wigglywrap{w2}{v4}{v5}{w5}
\end{fmfchar*}}}
}
\caption{Wrapping diagrams with chiral structure $\dchi{2,1}$}
\label{def3-vect}
\end{figure}
\begin{figure}[h]
\capstart
\renewcommand*{\thesubfigure}{\ \ }
\addtolength{\subfigcapskip}{5pt}
\unitlength=0.75mm
\settoheight{\eqoff}{$\times$}%
\setlength{\eqoff}{0.5\eqoff}%
\addtolength{\eqoff}{-12.5\unitlength}%
\settoheight{\eqofftwo}{$\times$}%
\setlength{\eqofftwo}{0.5\eqofftwo}%
\addtolength{\eqofftwo}{-7.5\unitlength}%
\subfigure[$\swgraph{3}{1}{1}$]{
\raisebox{\eqoff}{%
\fmfframe(3,1)(1,4){%
\begin{fmfchar*}(20,20)
\WoneplainC
\fmfipair{w[]}
\fmfipair{wd[]}
\svertex{w5}{p5}
\svertex{w6}{p6}
\svertex{w3}{p3}
\vvertex{w7}{w6}{p3}
\fmfi{wiggly}{w6..w7}
\wigglywrap{w5}{v4}{v5}{w7}
\end{fmfchar*}}}
}
\subfigspace
\subfigure[$\swgraph{3}{2}{1}(\times2)$]{
\raisebox{\eqoff}{%
\fmfframe(3,1)(1,4){%
\begin{fmfchar*}(20,20)
\WoneplainC
\fmfipair{w[]}
\fmfipair{wd[]}
\svertex{w4}{p4}
\svertex{w6}{p6}
\svertex{w3}{p3}
\vvertex{w7}{w6}{p3}
\fmfi{wiggly}{w6..w7}
\wigglywrap{w4}{v4}{v5}{w7}
\end{fmfchar*}}}
}
\subfigspace
\subfigure[$\swgraph{3}{3}{1}(\times2)$]{
\raisebox{\eqoff}{%
\fmfframe(3,1)(1,4){%
\begin{fmfchar*}(20,20)
\WoneplainC
\fmfipair{w[]}
\fmfipair{wd[]}
\fmfipair{wu[]}
\dvertex{wu4}{wd4}{p4}
\vvertex{w7}{wu4}{p3}
\fmfi{wiggly}{wu4..w7}
\wigglywrap{wd4}{v4}{v5}{w7}
\end{fmfchar*}}}
}
\subfigspace
\subfigure[$\swgraph{3}{4}{1}$]{
\raisebox{\eqoff}{%
\fmfframe(3,1)(1,4){%
\begin{fmfchar*}(20,20)
\WoneplainC
\fmfipair{w[]}
\fmfipair{wd[]}
\fmfipair{wu[]}
\svertex{w4}{p4}
\svertex{w3}{p3}
\fmfi{wiggly}{w4..w3}
\wigglywrap{w4}{v4}{v5}{w3}
\end{fmfchar*}}}
}
\caption{Wrapping diagrams with chiral structure $\dchi{1}$}
\label{def3-vect2}
\end{figure}
Wrapping diagrams with vector interactions can be built starting from the chiral structures $\dchi{2,1}$ or $\dchi{1}$, and the corresponding possibilities are shown in Figures~\ref{def3-vect} and~\ref{def3-vect2}. 
For the first group, after D-algebra one finds that, as far as the divergent parts are concerned,
\begin{equation}
\swgraph{3}{1}{2,1}+\swgraph{3}{2}{2,1}=0\col\qquad \swgraph{3}{3}{2,1}+\swgraph{3}{4}{2,1}=0 \col
\end{equation}
and thus the total contribution of the structure $\dchi{2,1}$ vanishes. The same happens for $\dchi{1}$, because
\begin{equation}
\swgraph{3}{2}{1}+\swgraph{3}{3}{1}=0\col
\end{equation}
whereas $\swgraph{3}{1}{1}$ and $\swgraph{3}{4}{1}$ are finite. So the divergent parts of wrapping diagrams with vectors sum up to zero and the only relevant terms come from $\subthree{3}$ and $\swrapthree{3}{0}$. Now, the values of $\subthree{3}$ and $\swrapthree{3}{0}$ after D-algebra are equal. As $\subthree{3}$ must be subtracted, but $\swrapthree{3}{0}$ has to be added, the total correction vanishes
\begin{equation}
\delta\gamma_3=-6\lim_{\varepsilon\to0}[-(\subthree{3}+\bsubthree{3})+(\swrapthree{3}{0}+\bswrapthree{3}{0})]=0 \col
\end{equation}
and the asymptotic result~\eqref{3loop-asymptotic} happens to be equal to the exact one
\begin{equation}
\gamma_3(\mathcal{O}_{3})=256\,\lambda^3\sin^6(\pi\beta)
\pnt
\end{equation}
Hence, finite-size effects do not show up at three loops in the deformed theory either. Note that there is no reason why the length-three operator should be protected against wrapping corrections. The cancellation between range-four and wrapping contributions is very likely to be a consequence of the simple structure of three-loop diagrams, and wrapping effects are expected to appear beyond the critical order, once the four-loop anomalous dimension of $\mathcal{O}_3$ is considered. Their actual computation, however, would be much more complicated than a typical four-loop analysis at the critical order, such as the one presented in Chapter~\ref{chapter:fourloop}, because of the large number and complicated structures of the involved diagrams. Thus, it would become feasible only if new and more general cancellation results were found.

\section{Four loops}
Wrapping effects at four loops can be easily studied on length-four states, in order to preserve the criticality of the perturbative order. In the deformed theory, three non-trivial length-four operators exist, the single-impurity one and the pair of two-impurity states~\eqref{Opbasis} already analyzed in the undeformed case. Thanks to the considerations of Section~\ref{sec:defdilop}, no additional diagrammatic calculations are required to compute the exact four-loop anomalous dimensions of these operators. In fact, it is enough to deform the final result~\eqref{D4exact} for the four-loop dilatation operator comprehensive of wrapping effects. By applying this deformed version to the single-impurity operator, one finds the four-loop component of the exact anomalous dimension
\begin{equation}
\begin{aligned}
\label{dim14}
\gamma_4(\mathcal{O}_{4})&=
-16
\lambda^4
\big[160\sin^8(\pi\beta)-\zeta(3)\cos(8\pi\beta)+5\big(\zeta(3)-\zeta(5)\big)\cos(6\pi\beta)\\
&\phantom{{}={}-16\lambda^4\big[{}}-\big(10\zeta(3)-15\zeta(5)\big)\cos(4\pi\beta)+\big(11\zeta(3)-15\zeta(5)\big)\cos(2\pi\beta)\\
&\phantom{{}={}-16\lambda^4\big[{}}-5\big(\zeta(3)-\zeta(5)\big)\big]
\pnt
\end{aligned}
\end{equation}
Note that the only rational term in this result comes entirely from the expansion of the asymptotic anomalous dimension, and all the transcendental terms originate from wrapping interactions, since it follows from~\eqref{single-all-orders} that the dressing phase never enters the spectrum of single-impurity operators. 

As for the two-impurity states, it turns out that, unlike in the undeformed case, here the linear combinations of the length-four basis~\eqref{Opbasis} that are multiplicatively renormalized change with the perturbative order. This is due to the lack of a protected combination of the two basis operators. Consequently, to determine the four-loop anomalous dimensions, besides the exact deformed four-loop dilatation operator, it is necessary to consider also the lower-order components. Expanding the eigenvalues in powers of $\lambda$ according to
\begin{equation}
\gamma^{(\pm)}=\lambda\gamma_1^{(\pm)}+\lambda^2\gamma_2^{(\pm)}+\lambda^3\gamma_3^{(\pm)}+\lambda^4\gamma_4^{(\pm)}
\col
\end{equation}
and with the definition of the function
\begin{equation}
\Delta(\beta)=\frac{\sqrt{5+4\cos(4\pi\beta)}}{3}
\col
\end{equation}
the components of the two anomalous dimensions read
\begin{equation}
\label{deffive}
\begin{aligned}
\gamma_1^{(\pm)}&=6(1\mp\Delta(\beta))\col\\
\gamma_2^{(\pm)}&={}-{}3\big(5+3\Delta(\beta)^2\big)\pm\frac{3}{\Delta(\beta)}\big(1+7\Delta(\beta)^2\big)\col\\
\gamma_3^{(\pm)}&=6\big(19+9\Delta(\beta)^2\big)\pm\frac{3}{4\Delta(\beta)^3}\big(1-51\Delta(\beta)^2-165\Delta(\beta)^4-9\Delta(\beta)^6\big)\col\\
\gamma_4^{(\pm)}&={}-{}8\big(151+45\zeta(5)\big)+297\zeta(3)
-6\Delta(\beta)^2\big(41-39\zeta(3)+60\zeta(5)\big)\\
&\phantom{{}={}}
+81\Delta(\beta)^4\big(2-3\zeta(3)\big)\\
&\phantom{{}={}}
\pm\frac{1}{8\Delta(\beta)^5}\big[3-132\Delta(\beta)^2+2\Delta(\beta)^4\big(1723+36\zeta(3)\big)\\
&\hphantom{{}={}\pm\frac{1}{8\Delta(\beta)^5}\big[{}}
+4\Delta(\beta)^6\big(2405-1404\zeta(3)+1440\zeta(5)\big)
-9\Delta(\beta)^8\big(289-360\zeta(3)\big)\big]\pnt
\end{aligned}
\end{equation}
The eigenvalue $\gamma^{(+)}$ vanishes in the limit $\beta\to0$, whereas $\gamma^{(-)}$ reduces to the Konishi anomalous dimension~\eqref{finalgamma} as expected. Note the remarkable feature that all the dependence on $\beta$ is encoded in the $\Delta(\beta)$ function. As far as transcendentality is concerned, here, as in the undeformed case, the $\zeta(5)$ term comes entirely from wrapping interactions, but the $\zeta(3)$ one gets contributions also from the dressing phase.

From~\eqref{deffive} it evident that working with two-impurity states in the deformed theory is even more difficult than in the standard case. 
Hence, it is more useful to focus on single-impurity operators, which will be discussed in general in the next section.

\section{Single-impurity states at generic order}
\subsection{The general result}
The simplifications that appear when single-impurity operators are considered allow to perform explicit perturbative computations up to higher orders. Indeed, in the following the general expression for the $L$-loop wrapping correction to the anomalous dimension of the length-$L$ single-impurity operator $\mathcal{O}_L=\mathrm{tr}(\phi Z^{L-1})$ will be given in terms of a particular class of momentum integrals. A set of recursive rules exists for the calculation of such integrals, so that the finite-size correction can be fully determined for any fixed order. Explicit results will be presented up to eleven loops.

\begin{figure}[t]
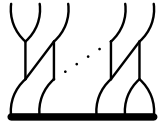

\capstart
\unitlength=0.75mm
\addtolength{\subfigcapskip}{5pt}
\settoheight{\eqoff}{$\times$}%
\setlength{\eqoff}{0.5\eqoff}%
\addtolength{\eqoff}{-12.5\unitlength}%
\settoheight{\eqofftwo}{$\times$}%
\setlength{\eqofftwo}{0.5\eqofftwo}%
\addtolength{\eqofftwo}{-7.5\unitlength}%
\centering
\raisebox{\eqoff}{%
\fmfframe(3,1)(1,4){%
\begin{fmfchar*}(25,20)
\fmftop{v1}
\fmfbottom{v5}
\fmfforce{(0w,h)}{v1}
\fmfforce{(0w,0)}{v5}
\fmffixed{(0.2w,0)}{v1,v2}
\fmffixed{(0.2w,0)}{v2,v3}
\fmffixed{(0.2w,0)}{v3,w1}
\fmffixed{(0.2w,0)}{w1,v4}
\fmffixed{(0.2w,0)}{v4,v9}
\fmffixed{(0.2w,0)}{v5,v6}
\fmffixed{(0.2w,0)}{v6,v7}
\fmffixed{(0.2w,0)}{v7,w2}
\fmffixed{(0.2w,0)}{w2,v8}
\fmffixed{(0.2w,0)}{v8,v10}
\fmffixed{(0,whatever)}{vc1,vc2}
\fmffixed{(0,whatever)}{vc3,vc4}
\fmffixed{(0,whatever)}{vc5,vc6}
\fmffixed{(0,whatever)}{vc7,vc8}
\fmf{plain,tension=0.25,right=0.25}{v1,vc1}
\fmf{plain,tension=0.25,left=0.25}{v2,vc1}
\fmf{plain,left=0.25}{v5,vc2}
\fmf{plain,tension=1,left=0.25}{v3,vc3}
\fmf{phantom,tension=1,left=0.25}{w1,wc1}
\fmf{plain,tension=1,left=0.25}{v4,vc5}
\fmf{plain,tension=1,left=0.25}{v9,vc7}
\fmf{plain,left=0.25}{w2,vc6}
\fmf{plain,tension=0.25,left=0.25}{v8,vc8}
\fmf{plain,tension=0.25,right=0.25}{v10,vc8}
\fmf{plain,left=0.25}{v6,vc4}
\fmf{phantom,left=0.25}{v7,wc2}
\fmf{plain,tension=0.5}{vc1,vc2}
\fmf{phantom,tension=0.5}{wc1,wc2}
\fmf{plain,tension=0.5}{vc2,vc3}
\fmf{phantom,tension=0.5}{wc2,vc5}
\fmf{plain,tension=0.5}{vc3,vc4}
\fmf{phantom,tension=0.5}{vc4,wc1}
\fmf{plain,tension=0.5}{vc5,vc6}
\fmf{plain,tension=0.5}{vc6,vc7}
\fmf{plain,tension=0.5}{vc7,vc8}
\fmf{plain,tension=0.5,right=0,width=1mm}{v5,v10}
\fmffreeze
\fmf{dots,tension=0.5}{vc4,vc5}
\fmfposition
\end{fmfchar*}}}
\caption{$L$-loop, range-$(L+1)$ diagram $\sub{L}$}
\label{diagram-SL}
\end{figure}

For the sake of clarity, it is useful to summarize the main reasons why single-impurity states are easier to analyze:
\begin{enumerate}
\item only one single-impurity length-$L$ state exists for each value of $L$. Thus, no mixing among operators occurs and Feynman diagrams produce simple numbers instead of matrices.
\item The asymptotic dilatation operator is not needed. The $L$-loop component of the asymptotic anomalous dimension can be extracted from the all-loop result~\eqref{single-all-orders} as
\begin{equation}
\gamma^{as}_L= \alpha_L\,\lambda^L \sin^{2L}(\pi\beta)\col\qquad\alpha_L=-(-8)^L \frac{(2L-3)!!}{L!} \pnt
\end{equation}
\item The relevant diagrams are a subset of those required for computations with more than one impurity, and the most difficult classes are excluded from this set. 
In fact, apart from completely chiral wrapping graphs, the only chiral structures that do not require more than one impurity are those of the form
\begin{equation}
\label{chi1imp}
\dchi{1,2,\ldots,k} \col
\end{equation}
and their parity reflections. Note that the only parity-symmetric structure in this subset is $\dchi{1}$.
\end{enumerate}
As a consequence of the last point, for a fixed operator length $L$ the only completely chiral range-$(L+1)$ diagrams to subtract are the one with structure $\dchi{1,2,\ldots,L}$, named $\sub{L}$ and shown in Figure~\ref{diagram-SL}, and its reflection $\dchi{L,\ldots,2,1}$. Thanks to their simple structure, the result of D-algebra for any perturbative order can easily be found to be
\begin{equation}
\begin{aligned}
\sub{L}+\bsub{L}&\rightarrow
(g^2 N)^L \kint{L}{1}\,[\dchi{1,2,\dots,L}+
\dchi{L,\dots,2,1}]\\
&\phantom{{}\rightarrow{}}
=(g^2 N)^L \kint{L}{1}\, 
(q-\bar{q})^2\left[q^{2(L-1)}+\bar{q}^{2(L-1)}
\right]\col
\end{aligned}
\end{equation}
where $\kint{L}{1}$ is the $L$-loop momentum integral shown in Figure~\ref{JL}.
\begin{figure}[h]
\capstart
\addtolength{\subfigcapskip}{5pt}
\unitlength=0.75mm
\settoheight{\eqoff}{$\times$}%
\setlength{\eqoff}{0.5\eqoff}%
\addtolength{\eqoff}{-12.5\unitlength}%
\settoheight{\eqofftwo}{$\times$}%
\setlength{\eqofftwo}{0.5\eqofftwo}%
\addtolength{\eqofftwo}{-7.5\unitlength}%
\centering
\subfigure[$\kint{L}{1}$]{
\label{JL}
\raisebox{\eqoff}{%
\fmfframe(0,0)(0,0){%
\begin{fmfchar*}(30,30)
\fmfleft{in}
\fmfright{out}
\fmf{plain,tension=1}{in,vi}
\fmf{phantom,tension=1}{out,v7}
\fmfpoly{phantom}{v12,v11,v10,v9,v8,v7,v6,v5,v4,v3,v2,v1}
\fmffixed{(0.9w,0)}{v1,v7}
\fmffixed{(0.3w,0)}{vi,v0}
\fmffixed{(0,whatever)}{v0,v3}
\fmf{plain}{v3,v4}
\fmf{plain}{v4,v5}
\fmf{plain}{v5,v6}
\fmf{dashes}{v6,v7}
\fmf{dashes}{v7,v8}
\fmf{dashes}{v8,v9}
\fmf{plain}{v9,v10}
\fmf{plain}{v10,v11}
\fmf{phantom}{v3,vi}
\fmf{plain}{vi,v11}
\fmf{phantom}{v0,v3}
\fmf{plain}{v0,v4}
\fmf{plain}{v0,v5}
\fmf{plain}{v0,v6}
\fmf{dashes}{v0,v7}
\fmf{dashes}{v0,v8}
\fmf{plain}{v0,v9}
\fmf{plain}{v0,v10}
\fmf{plain}{v0,v11}
\fmffreeze
\fmfposition
\fmf{plain}{vi,v0}
\fmf{plain,left=0.25}{v0,v3}
\fmf{plain,left=0.25}{v3,v0}
\fmfipair{w[]}
\fmfiequ{w1}{(xpart(vloc(__v3)),ypart(vloc(__v3)))}
\fmfiequ{w2}{(xpart(vloc(__v4)),ypart(vloc(__v4)))}
\fmfiequ{w3}{(xpart(vloc(__v5)),ypart(vloc(__v5)))}
\fmfiequ{w4}{(xpart(vloc(__v6)),ypart(vloc(__v6)))}
\fmfiequ{w5}{(xpart(vloc(__v7)),ypart(vloc(__v7)))}
\fmfiequ{w6}{(xpart(vloc(__v8)),ypart(vloc(__v8)))}
\fmfiequ{w7}{(xpart(vloc(__v9)),ypart(vloc(__v9)))}
\end{fmfchar*}}
}
}
\qquad\qquad
\subfigure[$\kint{L}{2}$]{
\label{KL}
\raisebox{\eqoff}{%
\fmfframe(0,0)(0,0){%
\begin{fmfchar*}(30,30)
\fmfleft{in}
\fmfright{out}
\fmf{plain,tension=1}{in,vi}
\fmf{phantom,tension=1}{out,v7}
\fmfpoly{phantom}{v12,v11,v10,v9,v8,v7,v6,v5,v4,v3,v2,v1}
\fmffixed{(0.9w,0)}{v1,v7}
\fmffixed{(0.3w,0)}{vi,v0}
\fmffixed{(0,whatever)}{v0,v3}
\fmf{plain}{v3,v4}
\fmf{plain}{v4,v5}
\fmf{plain}{v5,v6}
\fmf{dashes}{v6,v7}
\fmf{dashes}{v7,v8}
\fmf{dashes}{v8,v9}
\fmf{plain}{v9,v10}
\fmf{plain}{v10,v11}
\fmf{plain}{v3,vi}
\fmf{plain}{vi,v11}
\fmf{phantom}{v0,v3}
\fmf{plain}{v0,v4}
\fmf{plain}{v0,v5}
\fmf{plain}{v0,v6}
\fmf{dashes}{v0,v7}
\fmf{dashes}{v0,v8}
\fmf{plain}{v0,v9}
\fmf{plain}{v0,v10}
\fmf{plain}{v0,v11}
\fmffreeze
\fmfposition
\fmf{plain,left=0.25}{v0,v3}
\fmf{plain,left=0.25}{v3,v0}
\fmfipair{w[]}
\fmfiequ{w1}{(xpart(vloc(__v3)),ypart(vloc(__v3)))}
\fmfiequ{w2}{(xpart(vloc(__v4)),ypart(vloc(__v4)))}
\fmfiequ{w3}{(xpart(vloc(__v5)),ypart(vloc(__v5)))}
\fmfiequ{w4}{(xpart(vloc(__v6)),ypart(vloc(__v6)))}
\fmfiequ{w5}{(xpart(vloc(__v7)),ypart(vloc(__v7)))}
\fmfiequ{w6}{(xpart(vloc(__v8)),ypart(vloc(__v8)))}
\fmfiequ{w7}{(xpart(vloc(__v9)),ypart(vloc(__v9)))}
\end{fmfchar*}}
}
}
\caption{$L$-loop integrals from diagrams $\sub{L}$ and $\swrap{L}{0}$}
\label{integrals-SWL}
\end{figure}
\\
The usual argument of Appendix~\ref{app:non-maximal}, on the cancellation of non-maximal diagrams of maximum range, guarantees that this is the only range-$(L+1)$ contribution to the asymptotic anomalous dimension.

After the subtraction of range-$(L+1)$ interactions, wrapping diagrams must be considered. First of all, a single graph $\swrap{L}{0}$ with only scalar interactions, shown in Figure~\ref{diagram-WL0}, exists (together with its reflection) for every $L$. After the identification of the first and $(L+1)$-th lines, its chiral structure can be written as $\dchi{L,1,2,\ldots,L-1}$. The general result of D-algebra for this diagram is
\begin{equation}
\begin{aligned}
\swrap{L}{0}+\bswrap{L}{0}&\rightarrow(g^2 N)^L \kint{L}{2}\,[\dchi{L,1,2,\dots,L-1}+\dchi{1,L,L-1,\dots,2}]\\
&\phantom{{}\rightarrow{}}=(g^2 N)^L \kint{L}{2}\,
(q-\bar{q})^2\left[q^{2(L-1)}+\bar{q}^{2(L-1)}
\right]\col
\end{aligned}
\end{equation}
where the $L$-loop integral $\kint{L}{2}$ is shown in Figure~\ref{KL}.

\begin{figure}[t]
\capstart
\addtolength{\subfigcapskip}{5pt}
\unitlength=0.75mm
\settoheight{\eqoff}{$\times$}%
\setlength{\eqoff}{0.5\eqoff}%
\addtolength{\eqoff}{-12.5\unitlength}%
\settoheight{\eqofftwo}{$\times$}%
\setlength{\eqofftwo}{0.5\eqofftwo}%
\addtolength{\eqofftwo}{-7.5\unitlength}%
\centering
\raisebox{\eqoff}{%
\fmfframe(3,1)(1,4){%
\begin{fmfchar*}(30,20)
\fmftop{v1}
\fmfbottom{v5}
\fmfforce{(0w,h)}{v1}
\fmfforce{(0w,0)}{v5}
\fmffixed{(0.17w,0)}{v1,v2}
\fmffixed{(0.17w,0)}{v2,v3}
\fmffixed{(0.17w,0)}{v3,w1}
\fmffixed{(0.17w,0)}{w1,w3}
\fmffixed{(0.17w,0)}{w3,v4}
\fmffixed{(0.17w,0)}{v4,v9}
\fmffixed{(0.17w,0)}{v5,v6}
\fmffixed{(0.17w,0)}{v6,v7}
\fmffixed{(0.17w,0)}{v7,w2}
\fmffixed{(0.17w,0)}{w2,w4}
\fmffixed{(0.17w,0)}{w4,v8}
\fmffixed{(0.17w,0)}{v8,v10}
\fmffixed{(0,whatever)}{vc1,vc2}
\fmffixed{(0,whatever)}{vc3,vc4}
\fmffixed{(0,whatever)}{vc5,vc6}
\fmffixed{(0,whatever)}{vc7,vc8}
\fmf{phantom,tension=0.25,right=0.25}{v1,vc1}
\fmf{plain,tension=0.25,left=0.25}{v2,vc1}
\fmf{plain,left=0.25}{v5,vc2}
\fmf{plain,tension=1,left=0.25}{v3,vc3}
\fmf{phantom,tension=1,left=0.25}{v3,wc3}
\fmf{phantom,tension=1,left=0.25}{w1,wc3}
\fmf{plain,tension=0.5,right=0.25}{v4,vc7}
\fmf{plain,tension=0.5,left=0.25}{v9,vc7}
\fmf{plain,tension=0.5,left=0.25}{v8,vc8}
\fmf{plain,tension=0.5,right=0.25}{v10,vc8}
\fmf{plain,left=0.25}{v6,vc4}
\fmf{phantom,right=0.25}{v7,vc4}
\fmf{phantom}{vc4,wc3}
\fmf{plain,tension=0.5}{vc1,vc2}
\fmf{plain,tension=0.5}{vc2,vc3}
\fmf{plain,tension=0.5}{vc3,vc4}
\fmf{plain,tension=0.5}{vc7,vc8}
\fmf{plain,tension=0.5,right=0,width=1mm}{v5,v10}
\fmf{plain,tension=0.5}{vc5,vc6}
\fmffreeze
\fmf{plain,tension=0.25,left=0.25}{w2,vc6}
\fmf{phantom,tension=0.25,right=0.25}{w4,vc6}
\fmf{phantom,tension=0.25,right=0.25}{w1,wc1}
\fmf{plain,tension=0.25,left=0.25}{w3,wc1}
\fmf{plain,tension=0.5}{vc6,wc1}
\fmffreeze
\fmf{dots}{vc4,wc1}
\fmfposition
\fmfipath{p[]}
\fmfiset{p1}{vpath(__vc8,__vc7)}
\fmfipair{wz[]}
\fmfiequ{wz1}{point length(p1)/3 of p1}
\fmfiequ{wz2}{point 2*length(p1)/3 of p1}
\fmfiequ{wz3}{(xpart(vloc(__vc6)),ypart(vloc(__vc6)))}
\fmfiequ{wz4}{(xpart(vloc(__vc1)),ypart(vloc(__vc1)))}
\fmf{plain}{vc6,v12}
\fmfforce{(xpart(wz1),ypart(wz1))}{v11}
\fmfforce{(xpart(wz2),ypart(wz2))}{v12}
\plainwrap{vc1}{v5}{v10}{v11}
\end{fmfchar*}}}
\caption{$L$-loop wrapping diagram $\swrap{L}{0}$}
\label{diagram-WL0}
\end{figure}

\begin{figure}[t]
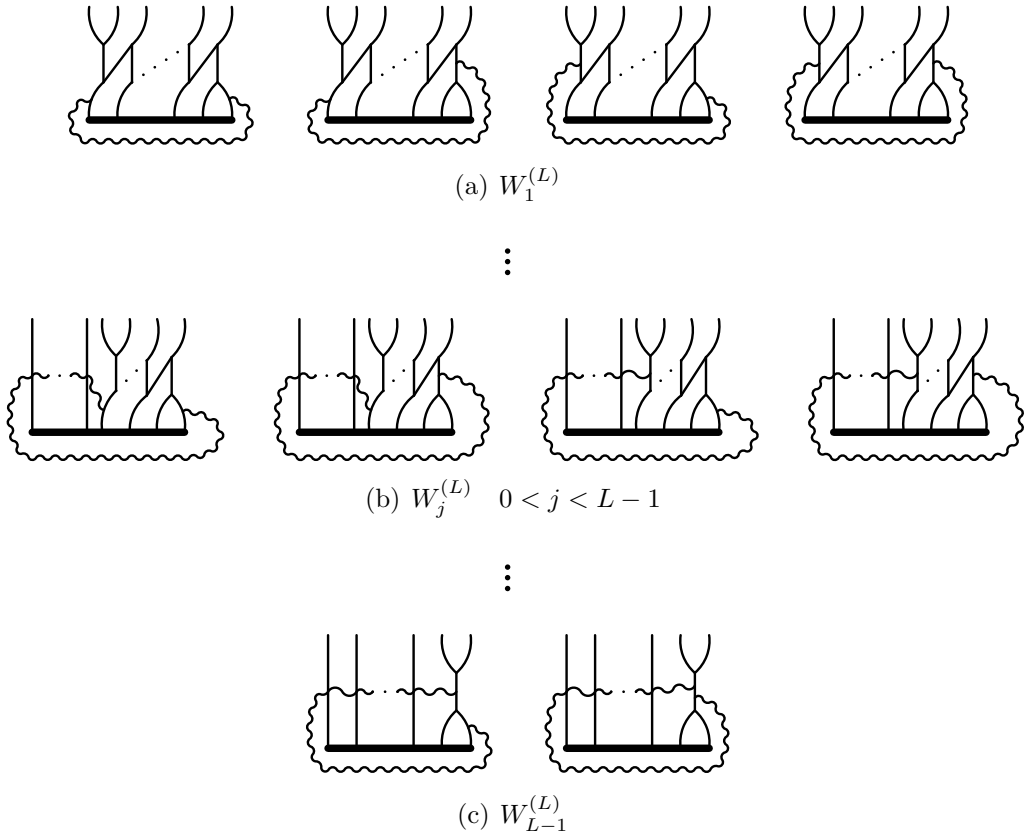

\capstart
\addtolength{\subfigcapskip}{5pt}
\unitlength=0.75mm
\settoheight{\eqoff}{$\times$}%
\setlength{\eqoff}{0.5\eqoff}%
\addtolength{\eqoff}{-12.5\unitlength}%
\settoheight{\eqofftwo}{$\times$}%
\setlength{\eqofftwo}{0.5\eqofftwo}%
\addtolength{\eqofftwo}{-7.5\unitlength}%
\centering
\raisebox{\eqoff}{%
\subfigure[$\swrap{L}{1}$]
{
\fmfframe(3,1)(1,4){%
\begin{fmfchar*}(25,20)
\fmftop{v1}
\fmfbottom{v5}
\fmfforce{(0w,h)}{v1}
\fmfforce{(0w,0)}{v5}
\fmffixed{(0.2w,0)}{v1,v2}
\fmffixed{(0.2w,0)}{v2,v3}
\fmffixed{(0.2w,0)}{v3,w1}
\fmffixed{(0.2w,0)}{w1,v4}
\fmffixed{(0.2w,0)}{v4,v9}
\fmffixed{(0.2w,0)}{v5,v6}
\fmffixed{(0.2w,0)}{v6,v7}
\fmffixed{(0.2w,0)}{v7,w2}
\fmffixed{(0.2w,0)}{w2,v8}
\fmffixed{(0.2w,0)}{v8,v10}
\fmffixed{(0,whatever)}{vc1,vc2}
\fmffixed{(0,whatever)}{vc3,vc4}
\fmffixed{(0,whatever)}{vc5,vc6}
\fmffixed{(0,whatever)}{vc7,vc8}
\fmf{plain,tension=0.25,right=0.25}{v1,vc1}
\fmf{plain,tension=0.25,left=0.25}{v2,vc1}
\fmf{plain,left=0.25}{v5,vc2}
\fmf{plain,tension=1,left=0.25}{v3,vc3}
\fmf{phantom,tension=1,left=0.25}{w1,wc1}
\fmf{plain,tension=1,left=0.25}{v4,vc5}
\fmf{plain,tension=1,left=0.25}{v9,vc7}
\fmf{plain,left=0.25}{w2,vc6}
\fmf{plain,tension=0.25,left=0.25}{v8,vc8}
\fmf{plain,tension=0.25,right=0.25}{v10,vc8}
\fmf{plain,left=0.25}{v6,vc4}
\fmf{phantom,left=0.25}{v7,wc2}
\fmf{plain,tension=0.5}{vc1,vc2}
\fmf{phantom,tension=0.5}{wc1,wc2}
\fmf{plain,tension=0.5}{vc2,vc3}
\fmf{phantom,tension=0.5}{wc2,vc5}
\fmf{plain,tension=0.5}{vc3,vc4}
\fmf{phantom,tension=0.5}{vc4,wc1}
\fmf{plain,tension=0.5}{vc5,vc6}
\fmf{plain,tension=0.5}{vc6,vc7}
\fmf{plain,tension=0.5}{vc7,vc8}
\fmf{plain,tension=0.5,right=0,width=1mm}{v5,v10}
\fmffreeze
\fmf{dots,tension=0.5}{vc4,vc5}
\fmfposition
\fmfipath{p[]}
\fmfiset{p1}{vpath(__v5,__vc2)}
\fmfiset{p2}{vpath(__v10,__vc8)}
\fmfipair{wz[]}
\fmfiequ{wz2}{point length(p2)/2 of p2}
\vvertex{wz1}{wz2}{p1}
\svertex{wz2}{p2}
\wigglywrap{wz1}{v5}{v10}{wz2}
\fmfposition
\end{fmfchar*}}
\qquad
\fmfframe(3,1)(1,4){%
\begin{fmfchar*}(25,20)
\fmftop{v1}
\fmfbottom{v5}
\fmfforce{(0w,h)}{v1}
\fmfforce{(0w,0)}{v5}
\fmffixed{(0.2w,0)}{v1,v2}
\fmffixed{(0.2w,0)}{v2,v3}
\fmffixed{(0.2w,0)}{v3,w1}
\fmffixed{(0.2w,0)}{w1,v4}
\fmffixed{(0.2w,0)}{v4,v9}
\fmffixed{(0.2w,0)}{v5,v6}
\fmffixed{(0.2w,0)}{v6,v7}
\fmffixed{(0.2w,0)}{v7,w2}
\fmffixed{(0.2w,0)}{w2,v8}
\fmffixed{(0.2w,0)}{v8,v10}
\fmffixed{(0,whatever)}{vc1,vc2}
\fmffixed{(0,whatever)}{vc3,vc4}
\fmffixed{(0,whatever)}{vc5,vc6}
\fmffixed{(0,whatever)}{vc7,vc8}
\fmf{plain,tension=0.25,right=0.25}{v1,vc1}
\fmf{plain,tension=0.25,left=0.25}{v2,vc1}
\fmf{plain,left=0.25}{v5,vc2}
\fmf{plain,tension=1,left=0.25}{v3,vc3}
\fmf{phantom,tension=1,left=0.25}{w1,wc1}
\fmf{plain,tension=1,left=0.25}{v4,vc5}
\fmf{plain,tension=1,left=0.25}{v9,vc7}
\fmf{plain,left=0.25}{w2,vc6}
\fmf{plain,tension=0.25,left=0.25}{v8,vc8}
\fmf{plain,tension=0.25,right=0.25}{v10,vc8}
\fmf{plain,left=0.25}{v6,vc4}
\fmf{phantom,left=0.25}{v7,wc2}
\fmf{plain,tension=0.5}{vc1,vc2}
\fmf{phantom,tension=0.5}{wc1,wc2}
\fmf{plain,tension=0.5}{vc2,vc3}
\fmf{phantom,tension=0.5}{wc2,vc5}
\fmf{plain,tension=0.5}{vc3,vc4}
\fmf{phantom,tension=0.5}{vc4,wc1}
\fmf{plain,tension=0.5}{vc5,vc6}
\fmf{plain,tension=0.5}{vc6,vc7}
\fmf{plain,tension=0.5}{vc7,vc8}
\fmf{plain,tension=0.5,right=0,width=1mm}{v5,v10}
\fmffreeze
\fmf{dots,tension=0.5}{vc4,vc5}
\fmfposition
\fmfipath{p[]}
\fmfiset{p1}{vpath(__v5,__vc2)}
\fmfiset{p2}{vpath(__vc7,__vc8)}
\fmfipair{wz[]}
\svertex{wz1}{p1}
\svertex{wz2}{p2}
\wigglywrap{wz1}{v5}{v10}{wz2}
\fmfposition
\end{fmfchar*}}
\qquad
\fmfframe(3,1)(1,4){%
\begin{fmfchar*}(25,20)
\fmftop{v1}
\fmfbottom{v5}
\fmfforce{(0w,h)}{v1}
\fmfforce{(0w,0)}{v5}
\fmffixed{(0.2w,0)}{v1,v2}
\fmffixed{(0.2w,0)}{v2,v3}
\fmffixed{(0.2w,0)}{v3,w1}
\fmffixed{(0.2w,0)}{w1,v4}
\fmffixed{(0.2w,0)}{v4,v9}
\fmffixed{(0.2w,0)}{v5,v6}
\fmffixed{(0.2w,0)}{v6,v7}
\fmffixed{(0.2w,0)}{v7,w2}
\fmffixed{(0.2w,0)}{w2,v8}
\fmffixed{(0.2w,0)}{v8,v10}
\fmffixed{(0,whatever)}{vc1,vc2}
\fmffixed{(0,whatever)}{vc3,vc4}
\fmffixed{(0,whatever)}{vc5,vc6}
\fmffixed{(0,whatever)}{vc7,vc8}
\fmf{plain,tension=0.25,right=0.25}{v1,vc1}
\fmf{plain,tension=0.25,left=0.25}{v2,vc1}
\fmf{plain,left=0.25}{v5,vc2}
\fmf{plain,tension=1,left=0.25}{v3,vc3}
\fmf{phantom,tension=1,left=0.25}{w1,wc1}
\fmf{plain,tension=1,left=0.25}{v4,vc5}
\fmf{plain,tension=1,left=0.25}{v9,vc7}
\fmf{plain,left=0.25}{w2,vc6}
\fmf{plain,tension=0.25,left=0.25}{v8,vc8}
\fmf{plain,tension=0.25,right=0.25}{v10,vc8}
\fmf{plain,left=0.25}{v6,vc4}
\fmf{phantom,left=0.25}{v7,wc2}
\fmf{plain,tension=0.5}{vc1,vc2}
\fmf{phantom,tension=0.5}{wc1,wc2}
\fmf{plain,tension=0.5}{vc2,vc3}
\fmf{phantom,tension=0.5}{wc2,vc5}
\fmf{plain,tension=0.5}{vc3,vc4}
\fmf{phantom,tension=0.5}{vc4,wc1}
\fmf{plain,tension=0.5}{vc5,vc6}
\fmf{plain,tension=0.5}{vc6,vc7}
\fmf{plain,tension=0.5}{vc7,vc8}
\fmf{plain,tension=0.5,right=0,width=1mm}{v5,v10}
\fmffreeze
\fmf{dots,tension=0.5}{vc4,vc5}
\fmfposition
\fmfipath{p[]}
\fmfiset{p1}{vpath(__vc1,__vc2)}
\fmfiset{p2}{vpath(__v10,__vc8)}
\fmfipair{wz[]}
\svertex{wz1}{p1}
\svertex{wz2}{p2}
\wigglywrap{wz1}{v5}{v10}{wz2}
\fmfposition
\end{fmfchar*}}
\qquad
\fmfframe(3,1)(1,4){%
\begin{fmfchar*}(25,20)
\fmftop{v1}
\fmfbottom{v5}
\fmfforce{(0w,h)}{v1}
\fmfforce{(0w,0)}{v5}
\fmffixed{(0.2w,0)}{v1,v2}
\fmffixed{(0.2w,0)}{v2,v3}
\fmffixed{(0.2w,0)}{v3,w1}
\fmffixed{(0.2w,0)}{w1,v4}
\fmffixed{(0.2w,0)}{v4,v9}
\fmffixed{(0.2w,0)}{v5,v6}
\fmffixed{(0.2w,0)}{v6,v7}
\fmffixed{(0.2w,0)}{v7,w2}
\fmffixed{(0.2w,0)}{w2,v8}
\fmffixed{(0.2w,0)}{v8,v10}
\fmffixed{(0,whatever)}{vc1,vc2}
\fmffixed{(0,whatever)}{vc3,vc4}
\fmffixed{(0,whatever)}{vc5,vc6}
\fmffixed{(0,whatever)}{vc7,vc8}
\fmf{plain,tension=0.25,right=0.25}{v1,vc1}
\fmf{plain,tension=0.25,left=0.25}{v2,vc1}
\fmf{plain,left=0.25}{v5,vc2}
\fmf{plain,tension=1,left=0.25}{v3,vc3}
\fmf{phantom,tension=1,left=0.25}{w1,wc1}
\fmf{plain,tension=1,left=0.25}{v4,vc5}
\fmf{plain,tension=1,left=0.25}{v9,vc7}
\fmf{plain,left=0.25}{w2,vc6}
\fmf{plain,tension=0.25,left=0.25}{v8,vc8}
\fmf{plain,tension=0.25,right=0.25}{v10,vc8}
\fmf{plain,left=0.25}{v6,vc4}
\fmf{phantom,left=0.25}{v7,wc2}
\fmf{plain,tension=0.5}{vc1,vc2}
\fmf{phantom,tension=0.5}{wc1,wc2}
\fmf{plain,tension=0.5}{vc2,vc3}
\fmf{phantom,tension=0.5}{wc2,vc5}
\fmf{plain,tension=0.5}{vc3,vc4}
\fmf{phantom,tension=0.5}{vc4,wc1}
\fmf{plain,tension=0.5}{vc5,vc6}
\fmf{plain,tension=0.5}{vc6,vc7}
\fmf{plain,tension=0.5}{vc7,vc8}
\fmf{plain,tension=0.5,right=0,width=1mm}{v5,v10}
\fmffreeze
\fmf{dots,tension=0.5}{vc4,vc5}
\fmfposition
\fmfipath{p[]}
\fmfiset{p1}{vpath(__vc1,__vc2)}
\fmfiset{p2}{vpath(__vc7,__vc8)}
\fmfipair{wz[]}
\svertex{wz1}{p1}
\svertex{wz2}{p2}
\wigglywrap{wz1}{v5}{v10}{wz2}
\fmfposition
\end{fmfchar*}}}
}
\\
\vspace{0.2cm}
$\Huge{\vdots}$
\vspace{0.3cm}
\\
\subfigure[$\swrap{L}{j}\quad 0<j<L-1$]{
\fmfframe(3,1)(1,4){%
\begin{fmfchar*}(30,20)
\fmftop{v1}
\fmfbottom{v5}
\fmfforce{(0w,h)}{v1}
\fmfforce{(0w,0)}{v5}
\fmffixed{(0.25w,0)}{v1,v2}
\fmffixed{(0.16w,0)}{v2,v3}
\fmffixed{(0.16w,0)}{v3,w1}
\fmffixed{(0.16w,0)}{w1,v4}
\fmffixed{(0.16w,0)}{v4,v9}
\fmffixed{(0.25w,0)}{v5,v6}
\fmffixed{(0.16w,0)}{v6,v7}
\fmffixed{(0.16w,0)}{v7,w2}
\fmffixed{(0.16w,0)}{w2,v8}
\fmffixed{(0.16w,0)}{v8,v10}
\fmffixed{(0,whatever)}{vc1,vc2}
\fmffixed{(0,whatever)}{vc3,vc4}
\fmffixed{(0,whatever)}{vc5,vc6}
\fmffixed{(0,whatever)}{vc7,vc8}
\fmf{phantom,tension=0.25,right=0.25}{v2,vc1}
\fmf{phantom,tension=0.25,left=0.25}{v3,vc1}
\fmf{phantom,left=0.25}{v6,vc2}
\fmf{plain,tension=1,left=0.25}{w1,vc3}
\fmf{phantom,tension=1,left=0.25}{v4,wc1}
\fmf{plain,tension=1,left=0.25}{v4,vc5}
\fmf{plain,tension=1,left=0.25}{v9,vc7}
\fmf{plain,left=0.25}{w2,vc6}
\fmf{plain,tension=0.25,left=0.25}{v8,vc8}
\fmf{plain,tension=0.25,right=0.25}{v10,vc8}
\fmf{plain,left=0.25}{v7,vc4}
\fmf{phantom,left=0.25}{w2,wc2}
\fmf{phantom,tension=0.5}{vc1,vc2}
\fmf{phantom,tension=0.5}{vc2,vc3}
\fmf{phantom,tension=0.5}{wc2,vc5}
\fmf{plain,tension=0.5}{vc3,vc4}
\fmf{phantom,tension=0.5}{vc4,wc1}
\fmf{plain,tension=0.5}{vc5,vc6}
\fmf{plain,tension=0.5}{vc6,vc7}
\fmf{plain,tension=0.5}{vc7,vc8}
\fmf{plain}{v1,v5}
\fmf{plain,tension=0.5,right=0,width=1mm}{v5,v10}
\fmffreeze
\fmffixed{(0.32w,0)}{v1,v11}
\fmffixed{(0.32w,0)}{v5,v12}
\fmf{plain,tension=0.25,right=0.25}{v3,vc3}
\fmf{plain}{v11,v12}
\fmf{dots,tension=0.5}{vc4,vc5}
\fmfposition
\fmfipath{p[]}
\fmfipair{wz[]}
\fmfiset{p1}{vpath(__v7,__vc4)}
\fmfiset{p2}{vpath(__v1,__v5)}
\fmfiset{p3}{vpath(__v10,__vc8)}
\fmfiset{p4}{vpath(__v11,__v12)}
\svertex{wz1}{p1}
\vvertex{wz2}{wz1}{p2}
\svertex{wz3}{p2}
\svertex{wz4}{p3}
\svertex{wz5}{p4}
\fmfiequ{wz6}{(xpart(wz3)+6,ypart(wz3))}
\fmfiequ{wz7}{(xpart(wz5)-6,ypart(wz5))}
\fmfi{wiggly}{wz5..wz1}
\fmfi{wiggly}{wz6..wz3}
\fmfi{wiggly}{wz7..wz5}
\fmfi{dots}{wz6..wz7}
\wigglywrap{wz3}{v5}{v10}{wz4}
\end{fmfchar*}}
\qquad
\fmfframe(3,1)(1,4){%
\begin{fmfchar*}(30,20)
\fmftop{v1}
\fmfbottom{v5}
\fmfforce{(0w,h)}{v1}
\fmfforce{(0w,0)}{v5}
\fmffixed{(0.25w,0)}{v1,v2}
\fmffixed{(0.16w,0)}{v2,v3}
\fmffixed{(0.16w,0)}{v3,w1}
\fmffixed{(0.16w,0)}{w1,v4}
\fmffixed{(0.16w,0)}{v4,v9}
\fmffixed{(0.25w,0)}{v5,v6}
\fmffixed{(0.16w,0)}{v6,v7}
\fmffixed{(0.16w,0)}{v7,w2}
\fmffixed{(0.16w,0)}{w2,v8}
\fmffixed{(0.16w,0)}{v8,v10}
\fmffixed{(0,whatever)}{vc1,vc2}
\fmffixed{(0,whatever)}{vc3,vc4}
\fmffixed{(0,whatever)}{vc5,vc6}
\fmffixed{(0,whatever)}{vc7,vc8}
\fmf{phantom,tension=0.25,right=0.25}{v2,vc1}
\fmf{phantom,tension=0.25,left=0.25}{v3,vc1}
\fmf{phantom,left=0.25}{v6,vc2}
\fmf{plain,tension=1,left=0.25}{w1,vc3}
\fmf{phantom,tension=1,left=0.25}{v4,wc1}
\fmf{plain,tension=1,left=0.25}{v4,vc5}
\fmf{plain,tension=1,left=0.25}{v9,vc7}
\fmf{plain,left=0.25}{w2,vc6}
\fmf{plain,tension=0.25,left=0.25}{v8,vc8}
\fmf{plain,tension=0.25,right=0.25}{v10,vc8}
\fmf{plain,left=0.25}{v7,vc4}
\fmf{phantom,left=0.25}{w2,wc2}
\fmf{phantom,tension=0.5}{vc1,vc2}
\fmf{phantom,tension=0.5}{vc2,vc3}
\fmf{phantom,tension=0.5}{wc2,vc5}
\fmf{plain,tension=0.5}{vc3,vc4}
\fmf{phantom,tension=0.5}{vc4,wc1}
\fmf{plain,tension=0.5}{vc5,vc6}
\fmf{plain,tension=0.5}{vc6,vc7}
\fmf{plain,tension=0.5}{vc7,vc8}
\fmf{plain}{v1,v5}
\fmf{plain,tension=0.5,right=0,width=1mm}{v5,v10}
\fmffreeze
\fmffixed{(0.32w,0)}{v1,v11}
\fmffixed{(0.32w,0)}{v5,v12}
\fmf{plain,tension=0.25,right=0.25}{v3,vc3}
\fmf{plain}{v11,v12}
\fmf{dots,tension=0.5}{vc4,vc5}
\fmfposition
\fmfipath{p[]}
\fmfipair{wz[]}
\fmfiset{p1}{vpath(__v7,__vc4)}
\fmfiset{p2}{vpath(__v1,__v5)}
\fmfiset{p3}{vpath(__vc7,__vc8)}
\fmfiset{p4}{vpath(__v11,__v12)}
\svertex{wz1}{p1}
\vvertex{wz2}{wz1}{p2}
\svertex{wz3}{p2}
\svertex{wz4}{p3}
\svertex{wz5}{p4}
\fmfiequ{wz6}{(xpart(wz3)+6,ypart(wz3))}
\fmfiequ{wz7}{(xpart(wz5)-6,ypart(wz5))}
\fmfi{wiggly}{wz5..wz1}
\fmfi{wiggly}{wz6..wz3}
\fmfi{wiggly}{wz7..wz5}
\fmfi{dots}{wz6..wz7}
\wigglywrap{wz3}{v5}{v10}{wz4}
\end{fmfchar*}}
\qquad
\fmfframe(3,1)(1,4){%
\begin{fmfchar*}(30,20)
\fmftop{v1}
\fmfbottom{v5}
\fmfforce{(0w,h)}{v1}
\fmfforce{(0w,0)}{v5}
\fmffixed{(0.25w,0)}{v1,v2}
\fmffixed{(0.16w,0)}{v2,v3}
\fmffixed{(0.16w,0)}{v3,w1}
\fmffixed{(0.16w,0)}{w1,v4}
\fmffixed{(0.16w,0)}{v4,v9}
\fmffixed{(0.25w,0)}{v5,v6}
\fmffixed{(0.16w,0)}{v6,v7}
\fmffixed{(0.16w,0)}{v7,w2}
\fmffixed{(0.16w,0)}{w2,v8}
\fmffixed{(0.16w,0)}{v8,v10}
\fmffixed{(0,whatever)}{vc1,vc2}
\fmffixed{(0,whatever)}{vc3,vc4}
\fmffixed{(0,whatever)}{vc5,vc6}
\fmffixed{(0,whatever)}{vc7,vc8}
\fmf{phantom,tension=0.25,right=0.25}{v2,vc1}
\fmf{phantom,tension=0.25,left=0.25}{v3,vc1}
\fmf{phantom,left=0.25}{v6,vc2}
\fmf{plain,tension=1,left=0.25}{w1,vc3}
\fmf{phantom,tension=1,left=0.25}{v4,wc1}
\fmf{plain,tension=1,left=0.25}{v4,vc5}
\fmf{plain,tension=1,left=0.25}{v9,vc7}
\fmf{plain,left=0.25}{w2,vc6}
\fmf{plain,tension=0.25,left=0.25}{v8,vc8}
\fmf{plain,tension=0.25,right=0.25}{v10,vc8}
\fmf{plain,left=0.25}{v7,vc4}
\fmf{phantom,left=0.25}{w2,wc2}
\fmf{phantom,tension=0.5}{vc1,vc2}
\fmf{phantom,tension=0.5}{vc2,vc3}
\fmf{phantom,tension=0.5}{wc2,vc5}
\fmf{plain,tension=0.5}{vc3,vc4}
\fmf{phantom,tension=0.5}{vc4,wc1}
\fmf{plain,tension=0.5}{vc5,vc6}
\fmf{plain,tension=0.5}{vc6,vc7}
\fmf{plain,tension=0.5}{vc7,vc8}
\fmf{plain}{v1,v5}
\fmf{plain,tension=0.5,right=0,width=1mm}{v5,v10}
\fmffreeze
\fmffixed{(0.32w,0)}{v1,v11}
\fmffixed{(0.32w,0)}{v5,v12}
\fmf{plain,tension=0.25,right=0.25}{v3,vc3}
\fmf{plain}{v11,v12}
\fmf{dots,tension=0.5}{vc4,vc5}
\fmfposition
\fmfipath{p[]}
\fmfipair{wz[]}
\fmfiset{p1}{vpath(__vc3,__vc4)}
\fmfiset{p2}{vpath(__v1,__v5)}
\fmfiset{p3}{vpath(__v10,__vc8)}
\fmfiset{p4}{vpath(__v11,__v12)}
\svertex{wz1}{p1}
\vvertex{wz2}{wz1}{p2}
\svertex{wz3}{p2}
\svertex{wz4}{p3}
\svertex{wz5}{p4}
\fmfiequ{wz6}{(xpart(wz3)+6,ypart(wz3))}
\fmfiequ{wz7}{(xpart(wz5)-6,ypart(wz5))}
\fmfi{wiggly}{wz5..wz1}
\fmfi{wiggly}{wz6..wz3}
\fmfi{wiggly}{wz7..wz5}
\fmfi{dots}{wz6..wz7}
\wigglywrap{wz3}{v5}{v10}{wz4}
\end{fmfchar*}}
\qquad
\fmfframe(3,1)(1,4){%
\begin{fmfchar*}(30,20)
\fmftop{v1}
\fmfbottom{v5}
\fmfforce{(0w,h)}{v1}
\fmfforce{(0w,0)}{v5}
\fmffixed{(0.25w,0)}{v1,v2}
\fmffixed{(0.16w,0)}{v2,v3}
\fmffixed{(0.16w,0)}{v3,w1}
\fmffixed{(0.16w,0)}{w1,v4}
\fmffixed{(0.16w,0)}{v4,v9}
\fmffixed{(0.25w,0)}{v5,v6}
\fmffixed{(0.16w,0)}{v6,v7}
\fmffixed{(0.16w,0)}{v7,w2}
\fmffixed{(0.16w,0)}{w2,v8}
\fmffixed{(0.16w,0)}{v8,v10}
\fmffixed{(0,whatever)}{vc1,vc2}
\fmffixed{(0,whatever)}{vc3,vc4}
\fmffixed{(0,whatever)}{vc5,vc6}
\fmffixed{(0,whatever)}{vc7,vc8}
\fmf{phantom,tension=0.25,right=0.25}{v2,vc1}
\fmf{phantom,tension=0.25,left=0.25}{v3,vc1}
\fmf{phantom,left=0.25}{v6,vc2}
\fmf{plain,tension=1,left=0.25}{w1,vc3}
\fmf{phantom,tension=1,left=0.25}{v4,wc1}
\fmf{plain,tension=1,left=0.25}{v4,vc5}
\fmf{plain,tension=1,left=0.25}{v9,vc7}
\fmf{plain,left=0.25}{w2,vc6}
\fmf{plain,tension=0.25,left=0.25}{v8,vc8}
\fmf{plain,tension=0.25,right=0.25}{v10,vc8}
\fmf{plain,left=0.25}{v7,vc4}
\fmf{phantom,left=0.25}{w2,wc2}
\fmf{phantom,tension=0.5}{vc1,vc2}
\fmf{phantom,tension=0.5}{vc2,vc3}
\fmf{phantom,tension=0.5}{wc2,vc5}
\fmf{plain,tension=0.5}{vc3,vc4}
\fmf{phantom,tension=0.5}{vc4,wc1}
\fmf{plain,tension=0.5}{vc5,vc6}
\fmf{plain,tension=0.5}{vc6,vc7}
\fmf{plain,tension=0.5}{vc7,vc8}
\fmf{plain}{v1,v5}
\fmf{plain,tension=0.5,right=0,width=1mm}{v5,v10}
\fmffreeze
\fmffixed{(0.32w,0)}{v1,v11}
\fmffixed{(0.32w,0)}{v5,v12}
\fmf{plain,tension=0.25,right=0.25}{v3,vc3}
\fmf{plain}{v11,v12}
\fmf{dots,tension=0.5}{vc4,vc5}
\fmfposition
\fmfipath{p[]}
\fmfipair{wz[]}
\fmfiset{p1}{vpath(__vc3,__vc4)}
\fmfiset{p2}{vpath(__v1,__v5)}
\fmfiset{p3}{vpath(__vc7,__vc8)}
\fmfiset{p4}{vpath(__v11,__v12)}
\svertex{wz1}{p1}
\vvertex{wz2}{wz1}{p2}
\svertex{wz3}{p2}
\svertex{wz4}{p3}
\svertex{wz5}{p4}
\fmfiequ{wz6}{(xpart(wz3)+6,ypart(wz3))}
\fmfiequ{wz7}{(xpart(wz5)-6,ypart(wz5))}
\fmfi{wiggly}{wz5..wz1}
\fmfi{wiggly}{wz6..wz3}
\fmfi{wiggly}{wz7..wz5}
\fmfi{dots}{wz6..wz7}
\wigglywrap{wz3}{v5}{v10}{wz4}
\end{fmfchar*}}
}
\\
\vspace{0.2cm}
$\Huge{\vdots}$
\vspace{0.3cm}
\\
\subfigure[$\swrap{L}{L-1}$]
{
\fmfframe(3,1)(1,4){%
\begin{fmfchar*}(25,20)
\fmftop{v1}
\fmfbottom{v5}
\fmfforce{(0w,h)}{v1}
\fmfforce{(0w,0)}{v5}
\fmffixed{(0.2w,0)}{v1,v2}
\fmffixed{(0.2w,0)}{v2,v3}
\fmffixed{(0.2w,0)}{v3,w1}
\fmffixed{(0.2w,0)}{w1,v4}
\fmffixed{(0.2w,0)}{v4,v9}
\fmffixed{(0.2w,0)}{v5,v6}
\fmffixed{(0.2w,0)}{v6,v7}
\fmffixed{(0.2w,0)}{v7,w2}
\fmffixed{(0.2w,0)}{w2,v8}
\fmffixed{(0.2w,0)}{v8,v10}
\fmffixed{(0,whatever)}{vc7,vc8}
\fmf{plain,tension=0.25,right=0.25}{v4,vc7}
\fmf{plain,tension=0.25,left=0.25}{v9,vc7}
\fmf{plain,tension=0.25,left=0.25}{v8,vc8}
\fmf{plain,tension=0.25,right=0.25}{v10,vc8}
\fmf{plain}{v2,v6}
\fmf{plain}{w2,w1}
\fmf{plain,tension=0.5}{vc7,vc8}
\fmf{plain}{v1,v5}
\fmf{plain,tension=0.5,right=0,width=1mm}{v5,v10}
\fmffreeze
\fmfposition
\fmfipath{p[]}
\fmfipair{wz[]}
\fmfiset{p1}{vpath(__v2,__v6)}
\fmfiset{p2}{vpath(__w1,__w2)}
\fmfiset{p3}{vpath(__v10,__vc8)}
\fmfiset{p4}{vpath(__v1,__v5)}
\fmfiset{p5}{vpath(__vc8,__vc7)}
\svertex{wz1}{p1}
\svertex{wz2}{p2}
\svertex{wz3}{p3}
\svertex{wz4}{p4}
\svertex{wz5}{p5}
\fmfiequ{wz6}{(xpart(wz1)+6,ypart(wz1))}
\fmfiequ{wz7}{(xpart(wz2)-6,ypart(wz2))}
\fmfi{wiggly}{wz2..wz5}
\fmfi{dots}{wz6..wz7}
\fmfi{wiggly}{wz4..wz1}
\fmfi{wiggly}{wz1..wz6}
\fmfi{wiggly}{wz7..wz2}
\wigglywrap{wz4}{v5}{v10}{wz3}
\end{fmfchar*}}
\qquad
\fmfframe(3,1)(1,4){%
\begin{fmfchar*}(25,20)
\fmftop{v1}
\fmfbottom{v5}
\fmfforce{(0w,h)}{v1}
\fmfforce{(0w,0)}{v5}
\fmffixed{(0.2w,0)}{v1,v2}
\fmffixed{(0.2w,0)}{v2,v3}
\fmffixed{(0.2w,0)}{v3,w1}
\fmffixed{(0.2w,0)}{w1,v4}
\fmffixed{(0.2w,0)}{v4,v9}
\fmffixed{(0.2w,0)}{v5,v6}
\fmffixed{(0.2w,0)}{v6,v7}
\fmffixed{(0.2w,0)}{v7,w2}
\fmffixed{(0.2w,0)}{w2,v8}
\fmffixed{(0.2w,0)}{v8,v10}
\fmffixed{(0,whatever)}{vc7,vc8}
\fmf{plain,tension=0.25,right=0.25}{v4,vc7}
\fmf{plain,tension=0.25,left=0.25}{v9,vc7}
\fmf{plain,tension=0.25,left=0.25}{v8,vc8}
\fmf{plain,tension=0.25,right=0.25}{v10,vc8}
\fmf{plain}{v2,v6}
\fmf{plain}{w2,w1}
\fmf{plain,tension=0.5}{vc7,vc8}
\fmf{plain}{v1,v5}
\fmf{plain,tension=0.5,right=0,width=1mm}{v5,v10}
\fmffreeze
\fmfposition
\fmfipath{p[]}
\fmfipair{wz[]}
\fmfiset{p1}{vpath(__v2,__v6)}
\fmfiset{p2}{vpath(__w1,__w2)}
\fmfiset{p3}{vpath(__v10,__vc8)}
\fmfiset{p4}{vpath(__v1,__v5)}
\fmfiset{p5}{vpath(__vc8,__vc7)}
\svertex{wz1}{p1}
\svertex{wz2}{p2}
\fmfiequ{wz3}{point 2*length(p5)/3 of p5}
\svertex{wz4}{p4}
\fmfiequ{wz5}{point length(p5)/3 of p5}
\fmfiequ{wz6}{(xpart(wz1)+6,ypart(wz1))}
\fmfiequ{wz7}{(xpart(wz2)-6,ypart(wz2))}
\fmfi{wiggly}{wz2..wz5}
\fmfi{dots}{wz6..wz7}
\fmfi{wiggly}{wz4..wz1}
\fmfi{wiggly}{wz1..wz6}
\fmfi{wiggly}{wz7..wz2}
\wigglywrap{wz4}{v5}{v10}{wz3}
\end{fmfchar*}}
}
\caption{Relevant diagrams after cancellations}
\label{defwrapgraphs}
\end{figure}

Wrapping diagrams with vector interactions can always be drawn starting from a structure with the form~\eqref{chi1imp} and adding the required number of vector lines to complete the $L$ loops, with one of the vector propagators being the wrapping line. In general, the structure $\dchi{1,2,\ldots,k}$ will require $(L-k)$ vector lines and will contain the same number of $Z$ lines directly interacting only with vectors, the single $\phi$ impurity being required to build the non-trivial part of the chiral structure. Note that for a given number $k$ such that $0\leq k\leq L-1$, only one structure of the form~\eqref{chi1imp} exists requiring exactly that number of vectors, and the sum of the corresponding diagrams will be denoted $\swrap{L}{k}$. 
From the results of Appendix~\ref{app:non-maximal} it follows that no single-vector vertices can appear on scalar lines interacting only through vector propagators or leaving the graph. It is easy to verify that, as a consequence, for every chiral structure the number of relevant diagrams is reduced to at most four, as shown in Figure~\ref{defwrapgraphs}.

\begin{figure}[t]
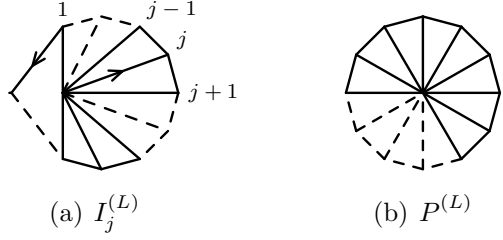

\capstart
\addtolength{\subfigcapskip}{5pt}
\unitlength=0.75mm
\settoheight{\eqoff}{$\times$}%
\setlength{\eqoff}{0.5\eqoff}%
\addtolength{\eqoff}{-14.5\unitlength}%
\settoheight{\eqofftwo}{$\times$}%
\setlength{\eqofftwo}{0.5\eqofftwo}%
\addtolength{\eqofftwo}{-7.5\unitlength}%
\centering
\subfigure[$\sint{j}{L}$]{
\label{ILk}
\raisebox{\eqoff}{%
\fmfframe(0,0)(0,0){%
\begin{fmfchar*}(30,30)
\fmfleft{in}
\fmfright{out}
\fmf{plain,tension=1}{in,vi}
\fmf{phantom,tension=1}{out,v7}
\fmfpoly{phantom}{v12,v11,v10,v9,v8,v7,v6,v5,v4,v3,v2,v1}
\fmffixed{(0.9w,0)}{v1,v7}
\fmffixed{(0.3w,0)}{vi,v0}
\fmffixed{(0,whatever)}{v0,v3}
\fmf{dashes}{v3,v4}
\fmf{dashes}{v4,v5}
\fmf{plain}{v5,v6}
\fmf{plain}{v6,v7}
\fmf{dashes}{v7,v8}
\fmf{dashes}{v8,v9}
\fmf{plain}{v9,v10}
\fmf{plain}{v10,v11}
\fmf{derplain}{v3,vi}
\fmf{dashes}{vi,v11}
\fmf{plain}{v0,v3}
\fmf{dashes}{v0,v4}
\fmf{plain}{v0,v5}
\fmf{derplain}{v0,v6}
\fmf{plain}{v0,v7}
\fmf{dashes}{v0,v8}
\fmf{plain}{v0,v9}
\fmf{plain}{v0,v10}
\fmf{plain}{v0,v11}
\fmfposition
\fmfipair{w[]}
\fmfiequ{w1}{(xpart(vloc(__v3)),ypart(vloc(__v3)))}
\fmfiequ{w2}{(xpart(vloc(__v4)),ypart(vloc(__v4)))}
\fmfiequ{w3}{(xpart(vloc(__v5)),ypart(vloc(__v5)))}
\fmfiequ{w4}{(xpart(vloc(__v6)),ypart(vloc(__v6)))}
\fmfiequ{w5}{(xpart(vloc(__v7)),ypart(vloc(__v7)))}
\fmfiequ{w6}{(xpart(vloc(__v8)),ypart(vloc(__v8)))}
\fmfiequ{w7}{(xpart(vloc(__v9)),ypart(vloc(__v9)))}
\fmfiv{l=\scriptsize{$1$},l.a=90,l.d=4}{w1}
\fmfiv{l=\scriptsize{$j-1$},l.a=45,l.d=4}{w3}
\fmfiv{l=\scriptsize{$j$},l.a=30,l.d=4}{w4}
\fmfiv{l=\scriptsize{$j+1$},l.a=0,l.d=4}{w5}
\end{fmfchar*}}}
}
\qquad\qquad
\subfigure[$\pint{L}_{\protect\phantom{j}}$]{
\label{PL}
\raisebox{\eqoff}{%
\fmfframe(0,0)(0,0){%
\begin{fmfchar*}(30,30)
\fmfleft{in}
\fmfright{out}
\fmf{phantom,tension=1}{in,v1}
\fmf{phantom,tension=1}{out,v7}
\fmfpoly{phantom}{v12,v11,v10,v9,v8,v7,v6,v5,v4,v3,v2,v1}
\fmffixed{(0.9w,0)}{v1,v7}
\fmfforce{(0.5w,0.5h)}{v0}
\fmfposition
\fmffreeze
\fmf{plain}{v1,v2}
\fmf{plain}{v2,v3}
\fmf{plain}{v3,v4}
\fmf{plain}{v4,v5}
\fmf{plain}{v5,v6}
\fmf{plain}{v6,v7}
\fmf{plain}{v7,v8}
\fmf{plain}{v8,v9}
\fmf{dashes}{v9,v10}
\fmf{dashes}{v10,v11}
\fmf{dashes}{v11,v12}
\fmf{dashes}{v12,v1}
\fmf{plain}{v1,v0}
\fmf{plain}{v2,v0}
\fmf{plain}{v3,v0}
\fmf{plain}{v4,v0}
\fmf{plain}{v5,v0}
\fmf{plain}{v6,v0}
\fmf{plain}{v7,v0}
\fmf{plain}{v8,v0}
\fmf{plain}{v9,v0}
\fmf{dashes}{v10,v0}
\fmf{dashes}{v11,v0}
\fmf{dashes}{v12,v0}
\end{fmfchar*}}}
}
\caption{$L$-loop momentum integrals}
\label{integrals}
\end{figure}

Again, these diagrams are simple enough to allow to perform the D-algebra for any $L$. In terms of the deformation factors
\begin{equation}
\label{colfact}
\cfact{L}{j}=(q-\bar{q})^2\left[q^{2(L-j-1)}+\bar{q}^{2(L-j-1)}\right]=-8\sin^2(\pi\beta)\cos[2\pi\beta(L-j-1)] \col
\end{equation}
for $j\in\{0,\ldots,L-1\}$, the results read
\begin{equation}
\label{defres}
\begin{aligned}
(\swrap{L}{0}+\bswrap{L}{0})-(\sub{L}+\bsub{L})&\to(g^2 N)^L\ \cfact{L}{0}(\kint{L}{2}-\kint{L}{1})\col\\
\vdots\\
\swrap{L}{j}+\bswrap{L}{j}&\to2(g^2 N)^L\ \cfact{L}{j}\sint{j+1}{L}\col\\
\vdots\\
\swrap{L}{L-1}+\bswrap{L}{L-1}&\to-(g^2 N)^L\ \cfact{L}{L-1}(\kint{L}{2}-\kint{L}{1})\col
\end{aligned}
\end{equation}
where the subtraction of $\sub{L}$ has been combined with the completely chiral diagram $\swrap{L}{0}$. The momentum integrals $\sint{j}{L}$ are shown in Figure~\ref{ILk}, where the pair of arrows indicates that the numerator of the integral contains the scalar product of the two corresponding momenta. Since the integrals $I_j^{(L)}$ fulfill the relation
\begin{equation}
\label{ILrel}
\sint{j}{L}=-\sint{L-j+1}{L}
\col
\end{equation}
the total contribution of the diagrams with $j$ vectors is proportional to the same integral as the class requiring $(L-j-1)$ vectors, and the total number of integrals to consider is reduced. Note the non-trivial fact that, apart from the cases with zero or $(L-1)$ vectors, all the other classes produce a contribution that can be written in terms of a single integral.

The combination $(\kint{L}{2}-\kint{L}{1})$, which appears in the results for the first and last classes, can be written in terms of $\sint{1}{L}$ and of the integral $\pint{L}$, shown in Figure~\ref{PL}, as
\begin{equation}
\kint{L}{2}-\kint{L}{1}=\pint{L}-2\sint{1}{L}
\pnt
\end{equation}
All the D-algebra results can now be collected to find the exact $L$-loop anomalous dimension of $\mathcal{O}_L$
\begin{equation}
\gamma_L(\mathcal{O}_{L})=\gamma_L^{as}+\delta\gamma_L(\mathcal{O}_{L})
\pnt
\end{equation}
As $\pint{L}$ and all the $\sint{j}{L}$ are free of subdivergences, their Laurent expansion in $\varepsilon$ will present only poles of the first order. 
Thus the wrapping correction can be written as 
\begin{equation}
\delta\gamma_L(\mathcal{O}_{L})=-2L(g^2 N)^L\lim_{\varepsilon\rightarrow0}\varepsilon\Bigg[(\cfact{L}{0}-\cfact{L}{L-1})\pint{L}(\varepsilon)-2\!\sum_{j=0}^{[\frac{L}{2}]-1}\!(\cfact{L}{j}-\cfact{L}{L-j-1})\sint{j+1}{L}(\varepsilon)\Bigg]
\pnt
\end{equation}
So, the computation of the $L$-loop anomalous dimension has been reduced to the calculation of $[L/2]$ momentum integrals with a simple general structure.

\newpage
\subsection{Computation of the integrals}
The divergent parts of the integrals $\pint{L}$ and $\sint{j}{L}$, with $1\leq j\leq [L/2]$, are required to find the actual numeric value of the wrapping correction to the $L$-loop anomalous dimension of the operator $\mathcal{O}_L$.\\
For $\pint{L}$, the result is known as a function of $L$~\cite{Broadhurst:1985vq,Usyukina:1991cp}
\begin{equation}
\pint{L}\sim\frac{1}{\varepsilon}\frac{1}{(4\pi)^{2L}}\frac{2}{L}\binom{2L-3}{L-1}\zeta(2L-3)
\col
\end{equation}
where the symbol $\sim$ means that only the divergent part is of interest here.
The analogous solution for $\sint{j}{L}$ as a function of $L$ and $j$ has not been found yet. However, $\sint{j}{L}$ can be computed exactly for \emph{any} fixed values of $L$ and $j$ thanks to a set of recurrence relations, which can be obtained by means of the technique of integration by parts as in~\cite{Broadhurst:1985vq}, where integrals with the same topology as $\sint{j}{L}$, but without the scalar products of momenta in the numerator, were studied. This approach requires the generalization of the triangle identity of~\cite{Broadhurst:1985vq} to the case of lines with non-trivial numerators. The generalized rules are shown in Appendix~\ref{app:triangles}, together with their derivation.

The standard triangle rule can be used to write any $\sint{j}{L}$ in terms of a finite set of integrals with at most seven loops and with generic propagator weights for the lines leaving the operator insertion. An example of this procedure is given in Appendix~\ref{app:triangles}. The integrals of this reduced set can then be fully calculated by repeatedly reducing them to lower-loop ones, using the generalized triangle rules. Therefore, the whole task of integral computation can be performed as soon as $L$ and $j$ have been fixed. Indeed, this recursive procedure is suitable for automation on a computer.

Another independent set of recurrence relations can be found by means of the GPXT in momentum space. In this approach, a set of master $L$-loop integrals must be determined first, from which all the required integrals can be extracted. The detailed description of this method, whose actual computer implementation becomes more efficient than the previous one for large values of $L$, has been presented in~\cite{Fiamberti:2008sn}.

In addition to the general recursive procedure, for $\sint{1}{L}$ also the technique described in~\cite{Usyukina:1991cp}, based on the method of uniqueness~\cite{Kazakov:1983ns}, can be attempted, leading to a different recurrence relation. Even though it cannot be resolved in closed form, its global structure suggests the following ansatz for the closed form of $\sint{1}{L}$ as a function of $L$
\begin{equation}
\label{defIL1}
\begin{aligned}
\sint{1}{L}&=\frac{1}{2}\pint{L}+\frac{1}{L}\sum_{k=3}^{L-1}\binom{a(L,k)}{L-k}\zeta(a(L,k)+1)\\
&=\frac{1}{2}\pint{L}+\frac{1}{L}\sum_{k=L-1-[\frac{L-1}{2}]}^{L-3}\binom{2k+1}{2k+3-L}\zeta(2k+1)+\frac{1}{2L}[1+(-1)^L](L-2)\zeta(L-1)
\col
\end{aligned}
\end{equation}
where $a(L,k)=2(L-k-1+[k/2])$. This proposal has been successfully verified up to eleven loops.

The approach of~\cite{Usyukina:1991cp} cannot be applied to the other classes of integrals, because of the required computation of complicated higher-loop integrals with generic propagator weights. However, for $\sint{2}{L}$ it is possible to formulate an ansatz by looking for simple modifications of~\eqref{defIL1} fitting the known explicit results
\begin{equation}
\begin{aligned}
\sint{2}{L}\vert_{L=2m}&=\frac{1}{2}\pint{L}-\frac{1}{L}\sum_{k=L-1-[\frac{L-1}{2}]}^{L-3}{\left[\frac{2L}{L-1}(L-2-k)-1\right]}\binom{2k+1}{2k+3-L}\zeta(2k+1)\\
&\quad-\frac{1}{L}(L-2)(L-1)\zeta(L-1)\col\\
& \\
\sint{2}{L}\vert_{L=2m+1}&=\frac{1}{2}\pint{L}-\frac{1}{L}\sum_{k=L-[\frac{L-1}{2}]}^{L-3}{\left[\frac{2L}{L-1}(L-2-k)-1\right]}\binom{2k+1}{2k+3-L}\zeta(2k+1)\\
&\quad-\frac{1}{2}(L-3)(L-1)\zeta(L)
\pnt
\end{aligned}
\end{equation}
This formula too has been checked up to $L=11$.

In the special case where $\beta=1/2$ and $L$ is even, the final result can also be found through an analysis on the string theory side, by using the L\"uscher technique for the undeformed theory~\cite{Gunnesson:2009nn,Beccaria:2009hg}. This method exploits the correspondence with the undeformed case with unphysical momentum $p=\pi$. By relaxing the momentum constraint, and using this non-vanishing value of the momentum in the computation of finite-size effects~\cite{Bajnok:2008bm}, it is possible to find a closed formula for the anomalous dimension as a function of $L$, which agrees with the results presented here. However, the knowledge of such exact result is not enough to extract the full information on the exact values of all the integrals.

\subsection{Results up to \texorpdfstring{$L=11$}{L=11}}
The explicit values of the integrals $\sint{j}{L}$ have been found up to $L=11$, and the full results are listed in Table~\ref{defintegrals}.
In all the considered cases, the subtraction of the range-$(L+1)$ contribution and the addition of the wrapping interactions produce only transcendental terms in the final exact $L$-loop anomalous dimension, with the single rational contribution coming entirely from the asymptotic result. Moreover, the transcendental terms appear according to a precise pattern, as shown in Table~\ref{table:zetas}. In fact, for every value of $L$, only $[L/2]$ different values of the $\zeta$ function enter the final result, all of the form
\begin{equation}
\zeta(2L-2k-1)\col\qquad k\in\{1,2,\ldots,[L/2]\} \pnt
\end{equation}
These are the $\zeta$ functions of the $[L/2]$ consecutive odd arguments from $(2L-3)$ downwards. It is a non-trivial fact that the lowest odd arguments disappear when $L$ increases. For example, the five-loop result for two-impurity states in the undeformed theory contains $\zeta(3)$, $\zeta(5)$ and $\zeta(7)$, but only $\zeta(5)$ and $\zeta(7)$ appear in the five-loop anomalous dimension of $\mathcal{O}_5$. 

It would be interesting to check what happens of this transcendentality pattern when wrapping interactions beyond the critical order are taken into account, for example in the $(L+1)$-loop anomalous dimension of the length-$L$ operator $\mathcal{O}_L$. However, the direct field-theoretical analysis in terms of Feynman diagrams becomes very complicated, because of the non-critical nature of the graphs. In particular, the possible presence of three-vector vertices makes the total number of relevant diagrams huge. Recently, the five-loop anomalous dimension of the length-four, single-impurity operator has been computed using the L\"uscher approach~\cite{Bajnok:2009vm}. The result only partially preserves the transcendentality pattern: the wrapping correction is entirely transcendental, and contains $\zeta(3)$, $\zeta(5)$ and $\zeta(7)$, in addition to the new term $\zeta(3)^2$. This may be interpreted as some kind of mixing between the $\zeta(3)$ and $\zeta(5)$ expected for $L=4$ and the $\zeta(5)$ and $\zeta(7)$ that should appear at five loops, plus the new $\zeta(3)^2$ term.

\begin{table}
\capstart
\begin{center}
\begin{tabular}{cc}
\toprule
$L$ & Transcendental terms \\
\midrule
$4$ & $\zeta(3)$, $\zeta(5)$\\
$5$ & $\zeta(5)$, $\zeta(7)$\\
$6$ & $\zeta(5)$, $\zeta(7)$, $\zeta(9)$\\
$7$ & $\zeta(7)$, $\zeta(9)$, $\zeta(11)$\\
$8$ & $\zeta(7)$, $\zeta(9)$, $\zeta(11)$, $\zeta(13)$\\
$9$ & $\zeta(9)$, $\zeta(11)$, $\zeta(13)$, $\zeta(15)$\\
$10$ & $\zeta(9)$, $\zeta(11)$, $\zeta(13)$, $\zeta(15)$, $\zeta(17)$\\
$11$ & $\zeta(11)$, $\zeta(13)$, $\zeta(15)$, $\zeta(17)$, $\zeta(19)$\\
\bottomrule
\end{tabular}
\end{center}
\caption{Transcendental terms produced for different values of $L$}
\label{table:zetas}
\end{table}

\begin{table}[t]
\capstart
\begin{equation*}
\footnotesize
\begin{aligned}
\sint{1}{4}&=\frac{1}{(4\pi)^8}\frac{1}{\varepsilon}\Big[\ \frac{1}{2}\,\zeta(3)+\frac{5}{2}\,\zeta(5)\ \Big]\col\\
\sint{1}{5}&=\frac{1}{(4\pi)^{10}}\frac{1}{\varepsilon}\Big[\ 2\,\zeta(5)+7\,\zeta(7)\ \Big]\col\\
\sint{1}{6}&=\frac{1}{(4\pi)^{12}}\frac{1}{\varepsilon}\Big[\ \frac{2}{3}\,\zeta(5)+\frac{35}{6}\,\zeta(7)+21\,\zeta(9)\ \Big]\col\\
\sint{1}{7}&=\frac{1}{(4\pi)^{14}}\frac{1}{\varepsilon}\Big[\ 3\,\zeta(7)+18\,\zeta(9)+66\,\zeta(11)\ \Big]\col\\
\sint{1}{8}&=\frac{1}{(4\pi)^{16}}\frac{1}{\varepsilon}\Big[\ \frac{3}{4}\,\zeta(7)+\frac{21}{2}\,\zeta(9)+\frac{231}{4}\,\zeta(11)+\frac{429}{2}\,\zeta(13)\ \Big]\col\\
\sint{1}{9}&=\frac{1}{(4\pi)^{18}}\frac{1}{\varepsilon}\Big[\ 4\,\zeta(9)+\frac{110}{3}\,\zeta(11)+\frac{572}{3}\,\zeta(13)+715\,\zeta(15)\ \Big]\col\\
\sint{1}{10}&=\frac{1}{(4\pi)^{20}}\frac{1}{\varepsilon}\Big[\ \frac{4}{5}\,\zeta(9)+\frac{33}{2}\,\zeta(11)+\frac{1287}{10}\,\zeta(13)+\frac{1287}{2}\,\zeta(15)+2431\,\zeta(17)\ \Big]\col\\
\sint{1}{11}&=\frac{1}{(4\pi)^{22}}\frac{1}{\varepsilon}\Big[\ 5\,\zeta(11)+65\,\zeta(13)+455\,\zeta(15)+2210\,\zeta(17)\ +8398\,\zeta(19)\ \Big]\col\\
& \\
\sint{2}{4}&=\frac{1}{(4\pi)^8}\frac{1}{\varepsilon}\Big[\ -\frac{3}{2}\,\zeta(3)+\frac{5}{2}\,\zeta(5)\ \Big]\col\\
\sint{2}{5}&=\frac{1}{(4\pi)^{10}}\frac{1}{\varepsilon}\Big[\ -4\,\zeta(5)+7\,\zeta(7)\ \Big]\col\\
\sint{2}{6}&=\frac{1}{(4\pi)^{12}}\frac{1}{\varepsilon}\Big[\ -\frac{10}{3}\,\zeta(5)-\frac{49}{6}\,\zeta(7)+21\,\zeta(9)\ \Big]\col\\
\sint{2}{7}&=\frac{1}{(4\pi)^{14}}\frac{1}{\varepsilon}\Big[\ -12\,\zeta(7)-24\,\zeta(9)+66\,\zeta(11)\ \Big]\col\\
\sint{2}{8}&=\frac{1}{(4\pi)^{16}}\frac{1}{\varepsilon}\Big[\ -\frac{21}{4}\,\zeta(7)-\frac{75}{2}\,\zeta(9)-\frac{297}{4}\,\zeta(11)+\frac{429}{2}\,\zeta(13)\ \Big]\col\\
\sint{2}{9}&=\frac{1}{(4\pi)^{18}}\frac{1}{\varepsilon}\Big[\ -24\,\zeta(9)-\frac{385}{3}\,\zeta(11)-\frac{715}{3}\,\zeta(13)+715\,\zeta(15)\ \Big]\col\\
\sint{2}{10}&=\frac{1}{(4\pi)^{20}}\frac{1}{\varepsilon}\Big[\ -\frac{36}{5}\,\zeta(9)-\frac{187}{2}\,\zeta(11)-\frac{4433}{10}\,\zeta(13)-\frac{1573}{2}\,\zeta(15)+2431\,\zeta(17)\ \Big]\col\\
\sint{2}{11}&=\frac{1}{(4\pi)^{22}}\frac{1}{\varepsilon}\Big[\ -40\,\zeta(11)-364\,\zeta(13)-1547\,\zeta(15)-2652\,\zeta(17)\ +8398\,\zeta(19)\ \Big]\col\\
\end{aligned}
\end{equation*}
\caption{Momentum integrals for single-impurity states}
\label{defintegrals}
\end{table}

\addtocounter{table}{-1}
\begin{table}[t]
\begin{equation*}
\footnotesize
\begin{aligned}
\sint{3}{6}&=\frac{1}{(4\pi)^{12}}\frac{1}{\varepsilon}\Big[\ \frac{20}{3}\,\zeta(5)-\frac{14}{3}\,\zeta(7)\ \Big]\col\\
\sint{3}{7}&=\frac{1}{(4\pi)^{14}}\frac{1}{\varepsilon}\Big[\ 15\,\zeta(7)-6\,\zeta(9)\ \Big]\col\\
\sint{3}{8}&=\frac{1}{(4\pi)^{16}}\frac{1}{\varepsilon}\Big[\ \frac{63}{4}\,\zeta(7)+\frac{81}{2}\,\zeta(9)-\frac{99}{4}\,\zeta(11)\ \Big]\col\\
\sint{3}{9}&=\frac{1}{(4\pi)^{18}}\frac{1}{\varepsilon}\Big[\ 56\,\zeta(9)+\frac{440}{3}\,\zeta(11)-\frac{286}{3}\,\zeta(13)\ \Big]\col\\
\sint{3}{10}&=\frac{1}{(4\pi)^{20}}\frac{1}{\varepsilon}\Big[\
\frac{144}{5}\,\zeta(9)+209\,\zeta(11)+\frac{2431}{5}\,\zeta(13)
-\frac{715}{2}\,\zeta(15)\ \Big]\col\\
\sint{3}{11}&=\frac{1}{(4\pi)^{22}}\frac{1}{\varepsilon}\Big[\ 135\,\zeta(11)+819\,\zeta(13)+1638\,\zeta(15)-1326\,\zeta(17)\ \Big]\col\\
& \\
\sint{4}{8}&=\frac{1}{(4\pi)^{16}}\frac{1}{\varepsilon}\Big[\ -\frac{105}{4}\,\zeta(7)-\frac{15}{2}\,\zeta(9)+\frac{165}{4}\,\zeta(11)\ \Big]\col\\
\sint{4}{9}&=\frac{1}{(4\pi)^{18}}\frac{1}{\varepsilon}\Big[\ -56\,\zeta(9)-55\,\zeta(11)+143\,\zeta(13)\ \Big]\col\\
\sint{4}{10}&=\frac{1}{(4\pi)^{20}}\frac{1}{\varepsilon}\Big[\
-\frac{336}{5}\,\zeta(9)-231\,\zeta(11)-\frac{429}{5}\,\zeta(13)
+\frac{1001}{2}\,\zeta(15)\ \Big]\col\\
\sint{4}{11}&=\frac{1}{(4\pi)^{22}}\frac{1}{\varepsilon}\Big[\ -240\,\zeta(11)-936\,\zeta(13)-182\,\zeta(15)+1768\,\zeta(17)\ \Big]\col\\
& \\
\sint{5}{10}&=\frac{1}{(4\pi)^{20}}\frac{1}{\varepsilon}\Big[\
\frac{504}{5}\,\zeta(9)+154\,\zeta(11)-\frac{1144}{5}\,\zeta(13)\ \Big]\col\\
\sint{5}{11}&=\frac{1}{(4\pi)^{22}}\frac{1}{\varepsilon}\Big[\ 210\,\zeta(11)+546\,\zeta(13)-637\,\zeta(15)\ \Big]
\pnt
\end{aligned}
\end{equation*}
\normalfont
\normalsize
\caption{Momentum integrals for single-impurity states \textit{(continued)}}
\end{table}

\section{Conclusions}
Finite-size effects can be studied also in the $\beta$-deformed version of $\N=4$ SYM, by following the same approach applied in the undeformed case. In particular, a clever choice of a set of deformed basis operators allows to find the asymptotic dilatation operator directly from the undeformed one, simply by replacing the old basis operators with the new ones, without any change in the coefficients.

In the deformed case, the new class of single-impurity operators can be considered. Since dealing with such states is much easier than working on multiple-impurity ones, perturbative computations can be performed up to higher orders. In particular, the anomalous dimension of any length-$L$ operator at the critical loop order $L$ can be written explicitly in terms of a single class of momentum integrals, whose exact values can always be obtained by means of recurrence relations. The found results show a precise transcendentality pattern.

As single-impurity states are easier to analyze, the $\beta$-deformed theory may be in the future the preferred environment for a possible perturbative study of wrapping effect beyond the critical order. In particular, the simplest calculation should concern the four-loop anomalous dimension of the length-three operator $\mathcal{O}_3$ analyzed in Section~\ref{sec:three}, which unlike the three-loop one is expected to receive a non-trivial correction. After that, the next simplest and most interesting quantity would be the five-loop anomalous dimension of $\mathcal{O}_4$, for which a prediction, derived by means of the L\"uscher approach, already exists~\cite{Bajnok:2009vm}. However, the actual computation of these two results will necessarily require the discovery of more general and powerful cancellation identities in order to be able to deal with the huge number and highly complicated structures of the relevant Feynman supergraphs.

\chapter{Finite-size effects on open strings}
\label{chapter:open}
This chapter contains the presentation of the perturbative analysis that reproduces the results found in~\cite{Correa:2009mz} for a new class of operators on the string theory side. In the first section, the general framework will be described. Then, the explicit computation for the leading, finite-size corrections at any perturbative order will be presented.

\section{Open strings}
In the previous chapters, only single-trace operators, corresponding to states of closed spin chains, have been considered in $\N=4$ SYM or in its deformed version. However, taking traces on the colour space is not the only possibility to build gauge-invariant quantities. In particular, 
a new interesting class of operators of the $\sutwo$ sector has the form
\begin{equation}
\label{op-open}
\tilde{\mathcal{O}}_{L}=\epsilon^{i_1,\ldots,i_N}\epsilon_{j_1,\ldots,j_N}Z_{i_1}^{j_1}\cdots Z_{i_{N-1}}^{j_{N-1}}(\phi Z^{(L-2)}\phi)_{i_N}^{j_N} \pnt
\end{equation}
Such operators can be obtained from $\mathrm{det}(Z)$ by replacing one of the $Z$ fields with an open chain with two impurities at the boundaries, and so they are interpreted as excitations over the ground state $\mathrm{det}(Z)$, which is the field-theory dual of a giant graviton in the string theory~\cite{Balasubramanian:2001nh,Corley:2001zk,Berenstein:2003ah}. For this reason, the operators~\eqref{op-open} are believed to be dual to states of open strings attached to a giant graviton~\cite{Balasubramanian:2001nh,Berenstein:2002ke,Balasubramanian:2002sa,Balasubramanian:2004nb,Berenstein:2005vf}, for which several integrability results have been found~\cite{Berenstein:2005vf,Mann:2006rh,Hofman:2007xp,Galleas:2009ye}.

It is therefore interesting to study finite-size effects on the open chains~\eqref{op-open}. The first results in this field appeared in~\cite{Correa:2009mz,Correa:2009dm}, where the boundary version of the thermodynamic Bethe ansatz was applied on the string theory side in order to find the leading and next-to-leading finite-size corrections at general order. The results for the leading case were checked perturbatively on the field theory side up to five loops. 
In the following, a perturbative calculation based on $\N=1$ superspace techniques will be presented to check the results for the leading correction at all orders.

One must be careful when dealing with operators whose dimension scales as $N$, as in the present case, because in general non-planar contributions are no longer suppressed with respect to planar ones in the large $N$ limit~\cite{Berenstein:2005vf,Berenstein:2006qk}. However, as discussed in~\cite{Correa:2009mz}, it is still possible to work in the planar approximation as long as one is only interested in the \emph{leading corrections} generated by finite-size effects, rather than in the full anomalous dimensions or in finite-size effects beyond the leading order. 

In the approach followed in~\cite{Correa:2009mz}, the planar asymptotic dilatation operator at the required order was applied to the two cases, and the difference between the two results was computed. To extend the calculation to generic order, the need for the full dilatation operator, which is known only up to five loops, must be removed, working at a lower level in terms of planar Feynman supergraphs.

Under renormalization, a generic operator $\tilde{\mathcal{O}}_L$ of the form~\eqref{op-open} will in general undergo a complicate mixing with chains where at least one of the impurities has been moved away from the corresponding boundary, where it has been substituted by a $Z$ field. However, 
a generic operator of that type can always be written as the sum of a term with the same structure~\eqref{op-open} but larger classical dimension, and the product of a closed chain and a determinant~\cite{Balasubramanian:2004nb,Berenstein:2006qk}. The mixing of such a combination with the original state can be neglected in the large $N$ limit~\cite{Correa:2009mz,Berenstein:2005fa,Berenstein:2006qk}, and thus in that limit the operators~\eqref{op-open} actually renormalize multiplicatively.

\newpage
\section{The superspace computation}
Superspace techniques will be now used to find the leading finite-size correction to the anomalous dimension of one of the operators~\eqref{op-open}. Note that since the chain is open, these corrections are not due to wrapping interactions. Instead, finite-size effects on open strings appear as soon as the loop order is high enough to let the two boundary impurities feel each other. Because of the relationship between the loop order of a Feynman diagram and its interaction range, the critical order for the length-$L$ open chain is $\ell=L-1$. So for fixed order $\ell$ one must study what happens when the length of the chain 
is reduced from $(\ell+2)$, which is the asymptotic case, to $(\ell+1)$.

When the length is $(\ell+2)$, the interaction range is too short for the two impurities to interact with each other, and thus the possible Feynman diagrams are a subset of the ones that can be applied to single-impurity states. Since only diagrams that do not move the impurities from the boundaries can be relevant, the underlying chiral structures must have this property. It is easy to see that the only structures fulfilling these constraints are $\chi()$ and $\chi(1)$. Now, at any order the coefficient of $\chi()$ must vanish, because the dilatation operator is the same on the whole $\sutwo$ sector and standard single-trace, zero-impurity operators must be protected. Hence, the only relevant structure in the asymptotic case is $\chi(1)$, which acts onto one of the two impurities and must be completed with the addition of $(\ell-1)$ vector lines.

\begin{figure}[t]
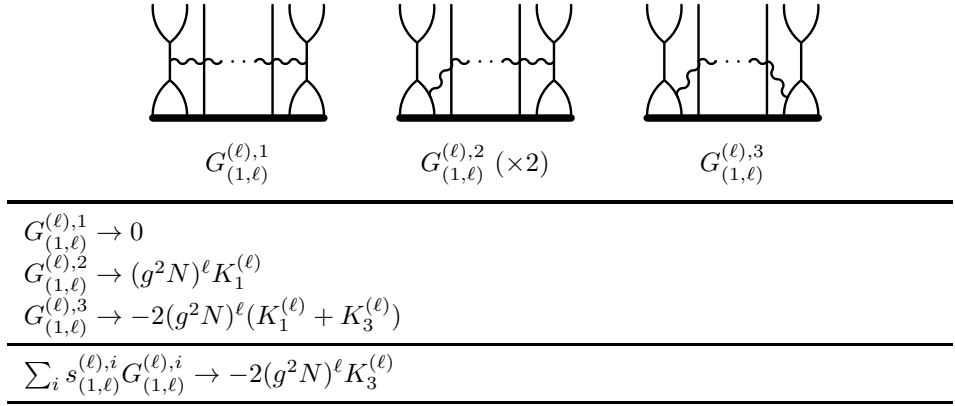

\capstart
\renewcommand*{\thesubfigure}{}
\footnotesize
\centering
\unitlength=0.75mm
\settoheight{\eqoff}{$\times$}%
\setlength{\eqoff}{0.5\eqoff}%
\addtolength{\eqoff}{-12.5\unitlength}%
\settoheight{\eqofftwo}{$\times$}%
\setlength{\eqofftwo}{0.5\eqofftwo}%
\addtolength{\eqofftwo}{-7.5\unitlength}%
\subfigure[$\nwgraph{\ell}{1}{1,\ell}$]{
\raisebox{\eqoff}{%
\fmfframe(3,1)(1,4){%
\begin{fmfchar*}(30,20)
\fmftop{v1}
\fmfbottom{v7}
\fmfforce{(0,h)}{v1}
\fmfforce{(0,0)}{v7}
\fmffixed{(0.2w,0)}{v1,v2}
\fmffixed{(0.1w,0)}{v2,v3}
\fmffixed{(0.4w,0)}{v3,v4}
\fmffixed{(0.1w,0)}{v4,v5}
\fmffixed{(0.2w,0)}{v5,v6}
\fmffixed{(0.2w,0)}{v7,v8}
\fmffixed{(0.1w,0)}{v8,v9}
\fmffixed{(0.4w,0)}{v9,v10}
\fmffixed{(0.1w,0)}{v10,v11}
\fmffixed{(0.2w,0)}{v11,v12}
\fmffixed{(0.1w,0)}{v3,va1}
\fmffixed{(0.1w,0)}{va3,v4}
\fmffixed{(0.1w,0)}{v9,va2}
\fmffixed{(0.1w,0)}{va4,v10}
\fmf{plain,tension=0.25,right=0.25}{v1,vc1}
\fmf{plain,tension=0.25,left=0.25}{v2,vc1}
\fmf{plain,tension=0.25,left=0.25}{v7,vc2}
\fmf{plain,tension=0.25,right=0.25}{v8,vc2}
\fmf{plain}{v3,v9}
\fmf{plain}{v4,v10}
\fmf{plain,tension=0.25,right=0.25}{v5,vc3}
\fmf{plain,tension=0.25,left=0.25}{v6,vc3}
\fmf{plain,tension=0.25,left=0.25}{v11,vc4}
\fmf{plain,tension=0.25,right=0.25}{v12,vc4}
\fmf{plain,tension=0.5}{vc1,vc2}
\fmf{plain,tension=0.5}{vc3,vc4}
\fmf{plain,tension=0.5,right=0,width=1mm}{v7,v12}
\fmf{phantom}{va1,va2}
\fmf{phantom}{va3,va4}
\fmffreeze
\fmfposition
\fmfipath{p[]}
\fmfipair{w[]}
\fmfiset{p1}{vpath(__vc1,__vc2)}
\fmfiset{p2}{vpath(__v3,__v9)}
\fmfiset{p3}{vpath(__va1,__va2)}
\fmfiset{p4}{vpath(__va3,__va4)}
\fmfiset{p5}{vpath(__v4,__v10)}
\fmfiset{p6}{vpath(__vc3,__vc4)}
\svertex{w1}{p1}
\svertex{w2}{p2}
\svertex{w3}{p3}
\svertex{w4}{p4}
\svertex{w5}{p5}
\svertex{w6}{p6}
\fmfi{wiggly}{w1..w2}
\fmfi{wiggly}{w2..w3}
\fmfi{wiggly}{w4..w5}
\fmfi{wiggly}{w5..w6}
\fmfi{dots}{w3..w4}
\end{fmfchar*}}}
}
\subfigspace
\subfigure[$\nwgraph{\ell}{2}{1,\ell}$ $(\times 2)$]{
\raisebox{\eqoff}{%
\fmfframe(3,1)(1,4){%
\begin{fmfchar*}(30,20)
\fmftop{v1}
\fmfbottom{v7}
\fmfforce{(0,h)}{v1}
\fmfforce{(0,0)}{v7}
\fmffixed{(0.2w,0)}{v1,v2}
\fmffixed{(0.1w,0)}{v2,v3}
\fmffixed{(0.4w,0)}{v3,v4}
\fmffixed{(0.1w,0)}{v4,v5}
\fmffixed{(0.2w,0)}{v5,v6}
\fmffixed{(0.2w,0)}{v7,v8}
\fmffixed{(0.1w,0)}{v8,v9}
\fmffixed{(0.4w,0)}{v9,v10}
\fmffixed{(0.1w,0)}{v10,v11}
\fmffixed{(0.2w,0)}{v11,v12}
\fmffixed{(0.1w,0)}{v3,va1}
\fmffixed{(0.1w,0)}{va3,v4}
\fmffixed{(0.1w,0)}{v9,va2}
\fmffixed{(0.1w,0)}{va4,v10}
\fmf{plain,tension=0.25,right=0.25}{v1,vc1}
\fmf{plain,tension=0.25,left=0.25}{v2,vc1}
\fmf{plain,tension=0.25,left=0.25}{v7,vc2}
\fmf{plain,tension=0.25,right=0.25}{v8,vc2}
\fmf{plain}{v3,v9}
\fmf{plain}{v4,v10}
\fmf{plain,tension=0.25,right=0.25}{v5,vc3}
\fmf{plain,tension=0.25,left=0.25}{v6,vc3}
\fmf{plain,tension=0.25,left=0.25}{v11,vc4}
\fmf{plain,tension=0.25,right=0.25}{v12,vc4}
\fmf{plain,tension=0.5}{vc1,vc2}
\fmf{plain,tension=0.5}{vc3,vc4}
\fmf{plain,tension=0.5,right=0,width=1mm}{v7,v12}
\fmf{phantom}{va1,va2}
\fmf{phantom}{va3,va4}
\fmffreeze
\fmfposition
\fmfipath{p[]}
\fmfipair{w[]}
\fmfiset{p1}{vpath(__v8,__vc2)}
\fmfiset{p2}{vpath(__v3,__v9)}
\fmfiset{p3}{vpath(__va1,__va2)}
\fmfiset{p4}{vpath(__va3,__va4)}
\fmfiset{p5}{vpath(__v4,__v10)}
\fmfiset{p6}{vpath(__vc3,__vc4)}
\svertex{w1}{p1}
\svertex{w2}{p2}
\svertex{w3}{p3}
\svertex{w4}{p4}
\svertex{w5}{p5}
\svertex{w6}{p6}
\fmfi{wiggly}{w1..w2}
\fmfi{wiggly}{w2..w3}
\fmfi{wiggly}{w4..w5}
\fmfi{wiggly}{w5..w6}
\fmfi{dots}{w3..w4}
\end{fmfchar*}}}
}
\subfigspace
\subfigure[$\nwgraph{\ell}{3}{1,\ell}$]{
\raisebox{\eqoff}{%
\fmfframe(3,1)(1,4){%
\begin{fmfchar*}(30,20)
\fmftop{v1}
\fmfbottom{v7}
\fmfforce{(0,h)}{v1}
\fmfforce{(0,0)}{v7}
\fmffixed{(0.2w,0)}{v1,v2}
\fmffixed{(0.1w,0)}{v2,v3}
\fmffixed{(0.4w,0)}{v3,v4}
\fmffixed{(0.1w,0)}{v4,v5}
\fmffixed{(0.2w,0)}{v5,v6}
\fmffixed{(0.2w,0)}{v7,v8}
\fmffixed{(0.1w,0)}{v8,v9}
\fmffixed{(0.4w,0)}{v9,v10}
\fmffixed{(0.1w,0)}{v10,v11}
\fmffixed{(0.2w,0)}{v11,v12}
\fmffixed{(0.1w,0)}{v3,va1}
\fmffixed{(0.1w,0)}{va3,v4}
\fmffixed{(0.1w,0)}{v9,va2}
\fmffixed{(0.1w,0)}{va4,v10}
\fmf{plain,tension=0.25,right=0.25}{v1,vc1}
\fmf{plain,tension=0.25,left=0.25}{v2,vc1}
\fmf{plain,tension=0.25,left=0.25}{v7,vc2}
\fmf{plain,tension=0.25,right=0.25}{v8,vc2}
\fmf{plain}{v3,v9}
\fmf{plain}{v4,v10}
\fmf{plain,tension=0.25,right=0.25}{v5,vc3}
\fmf{plain,tension=0.25,left=0.25}{v6,vc3}
\fmf{plain,tension=0.25,left=0.25}{v11,vc4}
\fmf{plain,tension=0.25,right=0.25}{v12,vc4}
\fmf{plain,tension=0.5}{vc1,vc2}
\fmf{plain,tension=0.5}{vc3,vc4}
\fmf{plain,tension=0.5,right=0,width=1mm}{v7,v12}
\fmf{phantom}{va1,va2}
\fmf{phantom}{va3,va4}
\fmffreeze
\fmfposition
\fmfipath{p[]}
\fmfipair{w[]}
\fmfiset{p1}{vpath(__v8,__vc2)}
\fmfiset{p2}{vpath(__v3,__v9)}
\fmfiset{p3}{vpath(__va1,__va2)}
\fmfiset{p4}{vpath(__va3,__va4)}
\fmfiset{p5}{vpath(__v4,__v10)}
\fmfiset{p6}{vpath(__v11,__vc4)}
\svertex{w1}{p1}
\svertex{w2}{p2}
\svertex{w3}{p3}
\svertex{w4}{p4}
\svertex{w5}{p5}
\svertex{w6}{p6}
\fmfi{wiggly}{w1..w2}
\fmfi{wiggly}{w2..w3}
\fmfi{wiggly}{w4..w5}
\fmfi{wiggly}{w5..w6}
\fmfi{dots}{w3..w4}
\end{fmfchar*}}}
}
\begin{tabular}{m{12cm}}
\toprule
$\nwgraph{\ell}{1}{1,\ell}\rightarrow 0$ \\
$\nwgraph{\ell}{2}{1,\ell}\rightarrow (g^2N)^\ell K_{1}^{(\ell)}$ \\
$\nwgraph{\ell}{3}{1,\ell}\rightarrow -2(g^2 N)^\ell (K_{1}^{(\ell)}+K_{3}^{(\ell)})$ \\
\midrule
$\sum_{i}\scgraph{\ell}{i}{1,\ell}\nwgraph{\ell}{i}{1,\ell}\rightarrow -2 (g^2 N)^\ell K_{3}^{(\ell)}$ \\
\bottomrule
\end{tabular}
\normalsize
\caption{Diagrams with structure $\chi(1,\ell)$ at generic loop order $\ell$}
\label{chi-open}
\end{figure}

\begin{figure}
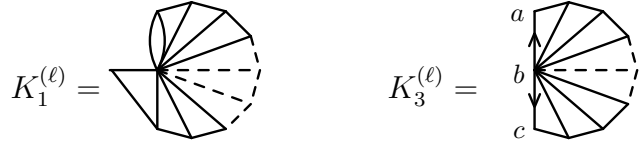

\capstart
\settoheight{\eqoff}{$\times$}%
\setlength{\eqoff}{0.5\eqoff}%
\addtolength{\eqoff}{-7.5\unitlength}
\begin{align*}
K_{1}^{(\ell)}=
\raisebox{\eqoff}{%
\begin{fmfchar*}(20,20)
\fmfleft{in}
\fmfright{out}
\fmf{plain,tension=1}{in,vi}
\fmf{phantom,tension=1}{out,v7}
\fmfpoly{phantom}{v12,v11,v10,v9,v8,v7,v6,v5,v4,v3,v2,v1}
\fmffixed{(0.9w,0)}{v1,v7}
\fmffixed{(0.3w,0)}{vi,v0}
\fmffixed{(0,whatever)}{v0,v3}
\fmf{plain}{v3,v4}
\fmf{plain}{v4,v5}
\fmf{plain}{v5,v6}
\fmf{dashes}{v6,v7}
\fmf{dashes}{v7,v8}
\fmf{dashes}{v8,v9}
\fmf{plain}{v9,v10}
\fmf{plain}{v10,v11}
\fmf{phantom}{v3,vi}
\fmf{plain}{vi,v11}
\fmf{phantom}{v0,v3}
\fmf{plain}{v0,v4}
\fmf{plain}{v0,v5}
\fmf{plain}{v0,v6}
\fmf{dashes}{v0,v7}
\fmf{dashes}{v0,v8}
\fmf{plain}{v0,v9}
\fmf{plain}{v0,v10}
\fmf{plain}{v0,v11}
\fmffreeze
\fmfposition
\fmf{plain}{vi,v0}
\fmf{plain,left=0.25}{v0,v3}
\fmf{plain,left=0.25}{v3,v0}
\fmfipair{w[]}
\fmfiequ{w1}{(xpart(vloc(__v3)),ypart(vloc(__v3)))}
\fmfiequ{w2}{(xpart(vloc(__v4)),ypart(vloc(__v4)))}
\fmfiequ{w3}{(xpart(vloc(__v5)),ypart(vloc(__v5)))}
\fmfiequ{w4}{(xpart(vloc(__v6)),ypart(vloc(__v6)))}
\fmfiequ{w5}{(xpart(vloc(__v7)),ypart(vloc(__v7)))}
\fmfiequ{w6}{(xpart(vloc(__v8)),ypart(vloc(__v8)))}
\fmfiequ{w7}{(xpart(vloc(__v9)),ypart(vloc(__v9)))}
\end{fmfchar*}}
\qquad\qquad
K_{3}^{(\ell)}=
\raisebox{\eqoff}{%
\begin{fmfchar*}(20,20)
\fmfleft{in}
\fmfright{out}
\fmf{phantom,tension=1}{in,vi}
\fmf{phantom,tension=1}{out,v7}
\fmfpoly{phantom}{v12,v11,v10,v9,v8,v7,v6,v5,v4,v3,v2,v1}
\fmffixed{(0.9w,0)}{v1,v7}
\fmffixed{(0.3w,0)}{vi,v0}
\fmffixed{(0,whatever)}{v0,v3}
\fmf{plain}{v3,v4}
\fmf{plain}{v4,v5}
\fmf{plain}{v5,v6}
\fmf{dashes}{v6,v7}
\fmf{dashes}{v7,v8}
\fmf{plain}{v8,v9}
\fmf{plain}{v9,v10}
\fmf{plain}{v10,v11}
\fmf{phantom}{v3,vi}
\fmf{phantom}{vi,v11}
\fmf{derplain}{v0,v3}
\fmf{plain}{v0,v4}
\fmf{plain}{v0,v5}
\fmf{plain}{v0,v6}
\fmf{dashes}{v0,v7}
\fmf{plain}{v0,v8}
\fmf{plain}{v0,v9}
\fmf{plain}{v0,v10}
\fmf{derplain}{v0,v11}
\fmffreeze
\fmfposition
\fmf{phantom}{vi,v0}
\fmfipair{w[]}
\fmfiequ{w1}{(xpart(vloc(__v3)),ypart(vloc(__v3)))}
\fmfiequ{w2}{(xpart(vloc(__v0)),ypart(vloc(__v0)))}
\fmfiequ{w3}{(xpart(vloc(__v11)),ypart(vloc(__v11)))}
\fmfiv{l=\footnotesize{$a$},l.a=200,l.d=4}{w1}
\fmfiv{l=\footnotesize{$b$},l.a=200,l.d=4}{w2}
\fmfiv{l=\footnotesize{$c$},l.a=200,l.d=4}{w3}
\end{fmfchar*}}
\end{align*}
\caption{Momentum integrals from diagrams with structure $\chi(1,\ell)$}
\label{int-open}
\end{figure}

When the length of the chain is reduced to $(\ell+1)$, the two impurities can interact directly. All the diagrams of the previous case are still allowed, with the same combinatorial coefficients, as the $\chi(1)$ block is forced to act onto one of the impurities at the boundaries, so the cancellation of a $Z$ line in the bulk of the chain has no consequences.
Moreover, the new compatible chiral structure $\chi(1,\ell)$ appears. So the leading finite-size correction to the $\ell$-loop matrix element for the length-$(\ell+1)$ chain is equal to the sum of the contributions of the Feynman supergraphs with structure $\chi(1,\ell)$, shown in Figure~\ref{chi-open}. Their structure is simple enough to allow to perform the D-algebra at generic loop order, and the results are shown in Figure~\ref{chi-open} as well. Note that the contribution of the diagram $\nwgraph{\ell}{2}{1,\ell}$, which is not symmetric under parity, must be doubled to take also its reflection into account.
The relevant momentum integrals are given in Figure~\ref{int-open}, where as usual the pair of arrows denotes the scalar product of the corresponding momenta in the numerator of the integral. Notice that the terms containing $K_{1}^{(\ell)}$ cancel, and the result depends only on $K_{3}^{(\ell)}$. 

According to the definition of the anomalous dimension, the contribution of a diagram to the mixing matrix is equal to the coefficient of the $1/\varepsilon$ pole of the result of D-algebra, multiplied by $(-2\ell)$. So, the leading $\ell$-loop finite-size correction reads
\begin{equation}
\delta\gamma(\tilde{O}_{\ell+1})=4\ell(g^2 N)^\ell \lim_{\varepsilon\to0}(\varepsilon\,K_{3}^{(\ell)}) \pnt
\end{equation}
The explicit value of $K_{3}^{(\ell)}$ as a function of $\ell$ can be determined by exploiting the fact that, thanks to the scalar product of momenta in the numerator, the ingoing and outgoing external momenta can be put at any pair of vertices without generating infrared divergences. In fact, were it not for the two momenta in the numerator, the choice of vertices $a$ and $c$ in Figure~\ref{int-open} would be forced. In the actual case, instead, points $a$ and $b$ can be chosen, greatly simplifying the computation. With this arrangement of the external momenta, in fact, all the integrations on the $\ell$ independent internal momenta can be performed directly. The result reads
\begin{equation}
K_{3}^{(\ell)}=-\frac{1}{(4\pi)^{2\ell}}G(1+(\ell-1)\varepsilon,1)\prod_{k=0}^{\ell-2}G^{(1)}(2+k\,\varepsilon,1)\frac{1}{(p^2)^{\ell\varepsilon}} \col
\end{equation}
where the functions $G$ and $G^{(1)}$ are defined in~\eqref{Gdef} in Appendix~\ref{app:triangles}.
Using the expression for $G^{(1)}$ in terms of $G$~\eqref{G1G} and the relations given in~\cite{Chetyrkin:1980pr}, it is now possible to write
\begin{equation}
G^{(1)}(2+k\,\varepsilon,1)=-\frac{(k+1)\varepsilon}{1-(k+2)\varepsilon}G(2+k\,\varepsilon,1)=\frac{(k+1)\varepsilon}{1+k\,\varepsilon}G(1+k\,\varepsilon,1) \col
\end{equation}
and since~\cite{Chetyrkin:1980pr}
\begin{equation}
G(1+k\,\varepsilon,1)=\frac{1}{(k+1)\varepsilon}+\mathcal{O}(1) \col
\end{equation}
the required integral is equal to
\begin{equation}
K_{3}^{(\ell)}=-\frac{1}{(4\pi)^{2\ell}}\frac{1}{\ell\,\varepsilon}\prod_{k=0}^{\ell-2}\frac{(k+1)\varepsilon}{1+k\,\varepsilon}\frac{1}{(k+1)\varepsilon}+\mathcal{O}(1)=-\frac{1}{(4\pi)^{2L}}\frac{1}{\ell\,\varepsilon}+\mathcal{O}(1) \pnt
\end{equation}
Thus, the $\ell$-loop finite-size correction for the $\tilde{\mathcal{O}}_{\ell+1}$ operator, written in terms of the 't~Hooft coupling, is
\begin{equation}
\delta\gamma(\tilde{O}_{\ell+1})=-4\lambda^{\ell} \col
\end{equation}
that is, its coefficient is the same whatever the perturbative order $\ell$. This result agrees with the one obtained in~\cite{Correa:2009mz} by means of the boundary thermodynamic Bethe ansatz.

\section{Conclusions}
The leading $\ell$-loop finite-size correction to the mixing matrix element of the length-$(\ell+1)$ operator describing an open chain in the $\sutwo$ sector of $\N=4$ SYM can be found for any value of $\ell$ in terms of Feynman diagrams. The result matches the existing calculation of the same quantity performed in the dual string theory by means of a version of the thermodynamic Bethe ansatz modified to take boundaries into account. The fact that the perturbative computation can be carried out completely at any order is another example of the power of superspace techniques.

\chapter{Conclusions}
\label{chapter:conclusions}
In this thesis, several perturbative computations concerning finite-size effects, both in $\N=4$ SYM and in its $\beta$-deformed version, have been presented. 
All of them have been made possible by the systematic use of $\N=1$ superspace techniques, which considerably simplified all the required steps.
First of all, the use of Feynman supergraphs allowed to recognize general cancellation results connected to supersymmetry, which revealed to be very powerful in reducing the number of relevant diagrams. Moreover, the total number of graphs requiring explicit analysis was greatly reduced with respect to typical calculations in the component-field formalism, since every supergraph encodes the information on several standard diagrams. As a further simplification, supergraphs contain only scalar and vector superfields, so that fermionic interactions never show up explicitly.

As a second technical remark, the possibility to discard finite contributions, together with the choice to compute anomalous dimensions staring from one-point functions instead of two-point ones, made the Gegenbauer polynomial $x$-space technique the ideal method for the analysis of multi-loop momentum integrals.

The four-loop anomalous dimension of the Konishi operator, presented in Chapter~\ref{chapter:fourloop}, was the first exact result to appear~\cite{us,uslong} about finite-size effects on the field theory side of the AdS/CFT correspondence. Its agreement with the later computation from the string side constitutes a non-trivial check of both the procedures and of the duality itself. This result ruled out all the previous conjectures on how to deal with wrapping interactions, based on the Hubbard model or on the analogy with the BFKL equation, and presents unexpected transcendentality properties.

Remaining in the framework of $\N=4$ SYM, the five-loop result of Chapter~\ref{chapter:fiveloop} is interesting because it provides another non-trivial check on the recently proposed Y-system. 
In order to perform the five-loop calculation along the general lines outlined in Chapter~\ref{chapter:fourloop}, the extension of the known five-loop dilatation operator to the case with a dressing phase, and the analysis of its behaviour under general similarity transformations, were required.

The found results at four and five loops in $\N=4$ SYM share the same transcendentality properties. In fact, in both cases wrapping corrections produce a term with a degree of transcendentality that is higher than the one which would be expected from the asymptotic regime. In the lower-transcendentality terms, finite-size effects mix with contributions from the dressing phase. Moreover, the rational part of the final result is modified too. This transcendentality pattern seems to be partially preserved when going beyond the critical order, according to the result of~\cite{Bajnok:2009vm}. It would be very interesting and important to be able to find the five-loop anomalous dimension of the Konishi operator, predicted in~\cite{Bajnok:2009vm} by means of the L\"uscher formalism, through a pure field-theoretical perturbative calculation like the ones of Chapters~\ref{chapter:fourloop} and~\ref{chapter:fiveloop}. This task, however, appears to be very difficult at the moment, because of the huge number and highly complicated structures of the supergraphs that should be considered. Hence, such a computation could become manageable only if new, more general and more powerful cancellation results, improving those presented in Appendix~\ref{app:non-maximal}, were found. In particular, one of the most challenging aspects of non-critical calculations is the appearance of three-vector vertices, which greatly increases the number of possible structures and makes the D-algebra procedure much more complicated.

When the $\beta$-deformed theory is considered, as in Chapter~\ref{chapter:wrapbetadef}, 
thanks to the reduced supersymmetry the new class of single-impurity operators can be studied. In that case, the reduced number and complexity of the relevant Feynman supergraphs allow to write down the anomalous dimension of a generic such operator at the critical order in terms of momentum integrals, that can be determined by means of an automated algorithm based on recurrence relations. The explicit results, computed up to eleven loops, show a very precise transcendentality pattern which is likely to be preserved at general loop order, as far as criticality is preserved. In particular, no rational part ever arises as a consequence of finite-size effects. For the special case of even length of the operators, and for a particular value of the deformation parameter $\beta$, the anomalous dimensions in the deformed theory can be calculated by means of the L\"uscher approach applied to the undeformed case, and the results agree with the perturbative analysis. 

In the deformed case too, it would be important to manage to push perturbative computations beyond the critical order. Indeed, the simplest example of such calculations would be found in this very theory, consisting in the four-loop anomalous dimension of the length-three single-impurity operator. Despite the simplifications with respect to the five-loop study of the Konishi operator in the undeformed case, namely the reduced number of loops and of relevant chiral structures, even this result is out of reach at the moment, requiring additional cancellation rules. Anyway, the $\beta$-deformed theory seems to be the ideal, simplified environment where the quest for more powerful supergraph identities should be undertaken.

The recent work on the possible inclusion of finite-size effects into a consistent integrable system led to the formulation of the Y-system, from which it should be possible to extract the whole spectrum of the theory. Therefore, it will be very important in the future to further test its validity by means of independent computations, especially of the kind of the above mentioned non-critical ones. 
In view of the great simplifications related to the possibility to study single-impurity states, the deformed theory may be the best environment for a deeper analysis of the new proposals. Thus, it would be interesting to extend the known Y-system to the $\beta$-deformed case, where an infinite series of results for the anomalous dimensions at the critical wrapping order is available, so that the validity of the technique could be checked in general on a whole class of operators.

Some kind of finite-size effects appear also on a new class of operators dual to open strings, and the leading correction to the anomalous dimensions can be computed at any order both on the gauge and on the string side of the correspondence.
Even though the origin of such effects is very different from the more standard wrapping case, it would be important to achieve a better understanding of them in order to gain new information on the spectrum.

In summary, in the last years a big progress has been achieved in the analysis of finite-size effects, on both sides of the AdS/CFT correspondence, and perturbative calculations on the gauge theory side played a fundamental role in the process, providing explicit results to check the new proposals. The recently found tools, which seem to be able to combine the power of integrability techniques with the correct description of finite-size corrections, may allow in the near future to compute the full spectra of $\N=4$ SYM and superstrings on $AdS_5\times S^5$, which would represent a very strong test of the correctness of the conjecture.

\end{mainmatter}

\begin{appendices}
\chapter{Cancellation results for diagrams of maximum range}
\label{app:non-maximal}

In this appendix the result of Section~\ref{sec:proof} is used to demonstrate useful cancellation properties for certain classes of planar diagrams involving vector interactions. 
Only the cases of diagrams with the maximum interaction range compatible with the loop number are studied, that comprise non-wrapping $\ell$-loop diagrams of range $(\ell+1)$ and $\ell$-loop wrapping ones for a length-$\ell$ operator.

\begin{figure}[t]
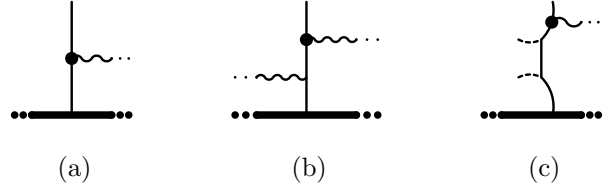

\capstart
\unitlength=0.75mm
\settoheight{\eqoff}{$\times$}%
\setlength{\eqoff}{0.5\eqoff}%
\addtolength{\eqoff}{-12.5\unitlength}%
\settoheight{\eqofftwo}{$\times$}%
\setlength{\eqofftwo}{0.5\eqofftwo}%
\addtolength{\eqofftwo}{-7.5\unitlength}%
\centering
\raisebox{\eqoff}{%
\subfigure[\hspace{-0.5cm}]{
\label{onevector}
\fmfframe(3,1)(1,4){%
\begin{fmfchar*}(20,20)
\fmftop{v1}
\fmfbottom{v6}
\fmfforce{(0w,h)}{v1}
\fmfforce{(0w,0)}{v6}
\fmffixed{(0.15w,0)}{v1,v2}
\fmffixed{(0.35w,0)}{v2,v3}
\fmffixed{(0.35w,0)}{v3,v4}
\fmffixed{(0.15w,0)}{v4,v5}
\fmffixed{(0.15w,0)}{v6,v7}
\fmffixed{(0.35w,0)}{v7,v8}
\fmffixed{(0.35w,0)}{v8,v9}
\fmffixed{(0.15w,0)}{v9,v10}
\fmf{plain}{v3,v8}
\fmf{phantom}{v2,v7}
\fmf{phantom}{v1,v6}
\fmf{phantom}{v4,v9}
\fmf{phantom}{v5,v10}
\fmf{plain,tension=0.5,right=0,width=1mm}{v7,v9}
\fmf{dots,tension=0.5,right=0,width=1mm}{v6,v7}
\fmf{dots,tension=0.5,right=0,width=1mm}{v9,v10}
\fmffreeze
\fmfposition
\fmfipath{p[]}
\fmfipair{w[]}
\fmfiset{p1}{vpath(__v3,__v8)}
\fmfiset{p2}{vpath(__v2,__v7)}
\fmfiset{p3}{vpath(__v4,__v9)}
\fmfiset{p4}{vpath(__v1,__v6)}
\fmfiset{p5}{vpath(__v5,__v10)}
\fmfiequ{w1}{point length(p1)/2 of p1}
\fmfiequ{w2}{point 2length(p1)/3 of p1}
\vvertex{w4}{w1}{p3}
\vvertex{w6}{w1}{p5}
\fmfi{wiggly}{w1..w4}
\fmfi{dots}{w4..w6}
\fmfiv{d.sh=circle,d.f=1,d.size=4}{w1}
\fmfposition
\end{fmfchar*}}}
\qquad
\subfigure[\hspace{-0.5cm}]{
\label{twovectors}
\fmfframe(3,1)(1,4){%
\begin{fmfchar*}(25,20)
\fmftop{v1}
\fmfbottom{v6}
\fmfforce{(0w,h)}{v1}
\fmfforce{(0w,0)}{v6}
\fmffixed{(0.15w,0)}{v1,v2}
\fmffixed{(0.35w,0)}{v2,v3}
\fmffixed{(0.35w,0)}{v3,v4}
\fmffixed{(0.15w,0)}{v4,v5}
\fmffixed{(0.15w,0)}{v6,v7}
\fmffixed{(0.35w,0)}{v7,v8}
\fmffixed{(0.35w,0)}{v8,v9}
\fmffixed{(0.15w,0)}{v9,v10}
\fmf{plain}{v3,v8}
\fmf{phantom}{v2,v7}
\fmf{phantom}{v1,v6}
\fmf{phantom}{v4,v9}
\fmf{phantom}{v5,v10}
\fmf{plain,tension=0.5,right=0,width=1mm}{v7,v9}
\fmf{dots,tension=0.5,right=0,width=1mm}{v6,v7}
\fmf{dots,tension=0.5,right=0,width=1mm}{v9,v10}
\fmffreeze
\fmfposition
\fmfipath{p[]}
\fmfipair{w[]}
\fmfiset{p1}{vpath(__v3,__v8)}
\fmfiset{p2}{vpath(__v2,__v7)}
\fmfiset{p3}{vpath(__v4,__v9)}
\fmfiset{p4}{vpath(__v1,__v6)}
\fmfiset{p5}{vpath(__v5,__v10)}
\fmfiequ{w1}{point length(p1)/3 of p1}
\fmfiequ{w2}{point 2length(p1)/3 of p1}
\vvertex{w3}{w2}{p2}
\vvertex{w4}{w1}{p3}
\vvertex{w5}{w2}{p4}
\vvertex{w6}{w1}{p5}
\fmfi{wiggly}{w3..w2}
\fmfi{wiggly}{w1..w4}
\fmfi{dots}{w5..w3}
\fmfi{dots}{w4..w6}
\fmfiv{d.sh=circle,d.f=1,d.size=4}{w1}
\fmfposition
\end{fmfchar*}}}
\qquad
\subfigure[\hspace{-0.5cm}]{
\fmfframe(3,1)(1,4){%
\begin{fmfchar*}(20,20)
\fmftop{v1}
\fmfbottom{v7}
\fmfforce{(0w,h)}{v1}
\fmfforce{(0w,0)}{v7}
\fmffixed{(0.15w,0)}{v1,v2}
\fmffixed{(0.25w,0)}{v2,v3}
\fmffixed{(0.2w,0)}{v3,v4}
\fmffixed{(0.25w,0)}{v4,v5}
\fmffixed{(0.15w,0)}{v5,v6}
\fmffixed{(0.15w,0)}{v7,v8}
\fmffixed{(0.25w,0)}{v8,v9}
\fmffixed{(0.2w,0)}{v9,v10}
\fmffixed{(0.25w,0)}{v10,v11}
\fmffixed{(0.15w,0)}{v11,v12}
\fmf{phantom}{v1,v7}
\fmf{phantom}{v2,v8}
\fmf{phantom}{v5,v11}
\fmf{phantom}{v6,v12}
\fmf{plain,tension=0.5,right=0,width=1mm}{v8,v11}
\fmf{dots,tension=0.5,right=0,width=1mm}{v7,v8}
\fmf{dots,tension=0.5,right=0,width=1mm}{v11,v12}
\fmf{phantom,tension=0.25,right=0.25}{v3,vc1}
\fmfset{dash_len}{1.5mm}
\fmf{plain,tension=0.25,left=0.25}{v4,vc1}
\fmf{phantom,tension=0.25,left=0.25}{v9,vc2}
\fmf{plain,tension=0.25,right=0.25}{v10,vc2}
\fmf{plain,tension=0.5}{vc1,vc2}
\fmffreeze
\fmffixed{(0.2w,0)}{vc3,vc1}
\fmffixed{(0.2w,0)}{vc4,vc2}
\fmf{dashes,tension=0.25,right=0.25}{vc3,vc1}
\fmf{dashes,tension=0.25,left=0.25}{vc4,vc2}
\fmfposition
\fmfipath{p[]}
\fmfipair{w[]}
\fmfiset{p1}{vpath(__vc1,__vc2)}
\fmfiset{p2}{vpath(__v5,__v11)}
\fmfiset{p3}{vpath(__v6,__v12)}
\fmfiset{p4}{vpath(__vc1,__v4)}
\svertex{w1}{p1}
\vvertex{w2}{w1}{p2}
\vvertex{w3}{w1}{p3}
\fmfiequ{w4}{point length(p4)/2 of p4}
\vvertex{w5}{w4}{p2}
\vvertex{w6}{w4}{p3}
\fmfi{wiggly}{w4..w5}
\fmfi{dots}{w5..w6}
\fmfiv{d.sh=circle,d.f=1,d.size=4}{w4}
\fmfposition
\end{fmfchar*}}}
}
\caption{Scalar line with one or two distinct single-vector vertices}
\label{startstruct}
\end{figure}

Consider a diagram that contains an outgoing scalar line where the farthest vertex from the insertion of the composite operator is a single-vector one. 
Because of the hypothesis on the range of the diagrams,
the only possible configurations are shown in Figure~\ref{startstruct}, where the relevant single-vector vertex has been marked with a circle. As far as the following considerations are concerned, these three cases behave exactly in the same way, so from now on only the second one will be explicitly taken into account. Note that the case of Figure~\ref{onevector} corresponds to non-maximal, $\ell$-loop diagrams with range $(\ell+1)$, where the requirement of maximum range forces the first or last line to interact with the rest of the diagram through a single vector interaction.

\begin{figure}[t]
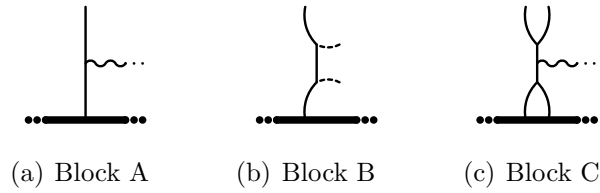

\capstart
\unitlength=0.75mm
\settoheight{\eqoff}{$\times$}%
\setlength{\eqoff}{0.5\eqoff}%
\addtolength{\eqoff}{-12.5\unitlength}%
\settoheight{\eqofftwo}{$\times$}%
\setlength{\eqofftwo}{0.5\eqofftwo}%
\addtolength{\eqofftwo}{-7.5\unitlength}%
\centering
\raisebox{\eqoff}{%
\subfigure[Block A]{
\label{block-A}
\fmfframe(3,1)(1,4){%
\begin{fmfchar*}(20,20)
\fmftop{v1}
\fmfbottom{v6}
\fmfforce{(0w,h)}{v1}
\fmfforce{(0w,0)}{v6}
\fmffixed{(0.15w,0)}{v1,v2}
\fmffixed{(0.35w,0)}{v2,v3}
\fmffixed{(0.35w,0)}{v3,v4}
\fmffixed{(0.15w,0)}{v4,v5}
\fmffixed{(0.15w,0)}{v6,v7}
\fmffixed{(0.35w,0)}{v7,v8}
\fmffixed{(0.35w,0)}{v8,v9}
\fmffixed{(0.15w,0)}{v9,v10}
\fmf{plain}{v3,v8}
\fmf{phantom}{v2,v7}
\fmf{phantom}{v1,v6}
\fmf{phantom}{v4,v9}
\fmf{phantom}{v5,v10}
\fmf{plain,tension=0.5,right=0,width=1mm}{v7,v9}
\fmf{dots,tension=0.5,right=0,width=1mm}{v6,v7}
\fmf{dots,tension=0.5,right=0,width=1mm}{v9,v10}
\fmffreeze
\fmfposition
\fmfipath{p[]}
\fmfipair{w[]}
\fmfiset{p1}{vpath(__v3,__v8)}
\fmfiset{p2}{vpath(__v2,__v7)}
\fmfiset{p3}{vpath(__v4,__v9)}
\fmfiset{p4}{vpath(__v1,__v6)}
\fmfiset{p5}{vpath(__v5,__v10)}
\fmfiequ{w1}{point length(p1)/2 of p1}
\fmfiequ{w2}{point 2length(p1)/3 of p1}
\vvertex{w4}{w1}{p3}
\vvertex{w6}{w1}{p5}
\fmfi{wiggly}{w1..w4}
\fmfi{dots}{w4..w6}
\fmfposition
\end{fmfchar*}}
}
\qquad
\subfigure[Block B]{
\label{block-B}
\fmfframe(3,1)(1,4){%
\begin{fmfchar*}(20,20)
\fmftop{v1}
\fmfbottom{v7}
\fmfforce{(0w,h)}{v1}
\fmfforce{(0w,0)}{v7}
\fmffixed{(0.15w,0)}{v1,v2}
\fmffixed{(0.25w,0)}{v2,v3}
\fmffixed{(0.2w,0)}{v3,v4}
\fmffixed{(0.25w,0)}{v4,v5}
\fmffixed{(0.15w,0)}{v5,v6}
\fmffixed{(0.15w,0)}{v7,v8}
\fmffixed{(0.25w,0)}{v8,v9}
\fmffixed{(0.2w,0)}{v9,v10}
\fmffixed{(0.25w,0)}{v10,v11}
\fmffixed{(0.15w,0)}{v11,v12}
\fmf{phantom}{v1,v7}
\fmf{phantom}{v2,v8}
\fmf{phantom}{v5,v11}
\fmf{phantom}{v6,v12}
\fmf{plain,tension=0.5,right=0,width=1mm}{v8,v11}
\fmf{dots,tension=0.5,right=0,width=1mm}{v7,v8}
\fmf{dots,tension=0.5,right=0,width=1mm}{v11,v12}
\fmf{plain,tension=0.25,right=0.25}{v3,vc1}
\fmfset{dash_len}{1.5mm}
\fmf{phantom,tension=0.25,left=0.25}{v4,vc1}
\fmf{plain,tension=0.25,left=0.25}{v9,vc2}
\fmf{phantom,tension=0.25,right=0.25}{v10,vc2}
\fmf{plain,tension=0.5}{vc1,vc2}
\fmffreeze
\fmffixed{(0.2w,0)}{vc1,vc3}
\fmffixed{(0.2w,0)}{vc2,vc4}
\fmf{dashes,tension=0.25,left=0.25}{vc3,vc1}
\fmf{dashes,tension=0.25,right=0.25}{vc4,vc2}
\fmfposition
\fmfipath{p[]}
\fmfipair{w[]}
\fmfiset{p1}{vpath(__vc1,__vc2)}
\fmfiset{p2}{vpath(__v5,__v11)}
\fmfiset{p3}{vpath(__v6,__v12)}
\svertex{w1}{p1}
\vvertex{w2}{w1}{p2}
\vvertex{w3}{w1}{p3}
\fmfposition
\end{fmfchar*}}}
\qquad
\subfigure[Block C]{
\label{block-C}
\fmfframe(3,1)(1,4){%
\begin{fmfchar*}(20,20)
\fmftop{v1}
\fmfbottom{v7}
\fmfforce{(0w,h)}{v1}
\fmfforce{(0w,0)}{v7}
\fmffixed{(0.15w,0)}{v1,v2}
\fmffixed{(0.25w,0)}{v2,v3}
\fmffixed{(0.2w,0)}{v3,v4}
\fmffixed{(0.25w,0)}{v4,v5}
\fmffixed{(0.15w,0)}{v5,v6}
\fmffixed{(0.15w,0)}{v7,v8}
\fmffixed{(0.25w,0)}{v8,v9}
\fmffixed{(0.2w,0)}{v9,v10}
\fmffixed{(0.25w,0)}{v10,v11}
\fmffixed{(0.15w,0)}{v11,v12}
\fmf{phantom}{v1,v7}
\fmf{phantom}{v2,v8}
\fmf{phantom}{v5,v11}
\fmf{phantom}{v6,v12}
\fmf{plain,tension=0.5,right=0,width=1mm}{v8,v11}
\fmf{dots,tension=0.5,right=0,width=1mm}{v7,v8}
\fmf{dots,tension=0.5,right=0,width=1mm}{v11,v12}
\fmfset{dash_len}{1.5mm}
\fmf{plain,tension=0.25,right=0.25}{v3,vc1}
\fmf{plain,tension=0.25,left=0.25}{v4,vc1}
\fmf{plain,tension=0.25,left=0.25}{v9,vc2}
\fmf{plain,tension=0.25,right=0.25}{v10,vc2}
\fmf{plain,tension=0.5}{vc1,vc2}
\fmffreeze
\fmfposition
\fmfipath{p[]}
\fmfipair{w[]}
\fmfiset{p1}{vpath(__vc1,__vc2)}
\fmfiset{p2}{vpath(__v5,__v11)}
\fmfiset{p3}{vpath(__v6,__v12)}
\svertex{w1}{p1}
\vvertex{w2}{w1}{p2}
\vvertex{w3}{w1}{p3}
\fmfi{wiggly}{w1..w2}
\fmfi{dots}{w2..w3}
\fmfposition
\end{fmfchar*}}
}
}
\caption{Building blocks for diagrams with vector interactions and maximum allowed range}
\end{figure}
\begin{figure}[t]
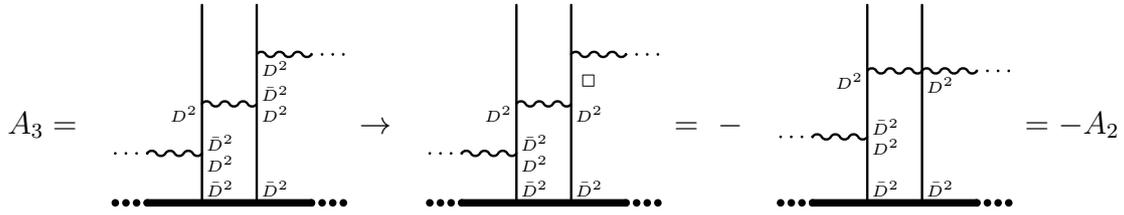

\capstart
\renewcommand*{\thesubfigure}{}
\unitlength=0.75mm
\settoheight{\eqoff}{$\times$}%
\setlength{\eqoff}{0.5\eqoff}%
\addtolength{\eqoff}{-12.5\unitlength}%
\settoheight{\eqofftwo}{$\times$}%
\setlength{\eqofftwo}{0.5\eqofftwo}%
\addtolength{\eqofftwo}{-7.5\unitlength}%
\centering
\raisebox{\eqoff}{%
\raisebox{-1.5\eqoff}{$A_3=$}
\subfigure[]{
\fmfframe(3,1)(1,4){%
\begin{fmfchar*}(40,35)
\fmftop{v1}
\fmfbottom{v7}
\fmfforce{(0w,h)}{v1}
\fmfforce{(0w,0)}{v7}
\fmffixed{(0.14w,0)}{v1,v2}
\fmffixed{(0.24w,0)}{v2,v3}
\fmffixed{(0.24w,0)}{v3,v4}
\fmffixed{(0.24w,0)}{v4,v5}
\fmffixed{(0.14w,0)}{v5,v6}
\fmffixed{(0.14w,0)}{v7,v8}
\fmffixed{(0.24w,0)}{v8,v9}
\fmffixed{(0.24w,0)}{v9,v10}
\fmffixed{(0.24w,0)}{v10,v11}
\fmffixed{(0.14w,0)}{v11,v12}
\fmf{phantom}{v1,v7}
\fmf{phantom}{v2,v8}
\fmf{plain}{v3,v9}
\fmf{plain}{v4,v10}
\fmf{phantom}{v5,v11}
\fmf{phantom}{v6,v12}
\fmf{plain,tension=0.5,right=0,width=1mm}{v8,v11}
\fmf{dots,tension=0.5,right=0,width=1mm}{v7,v8}
\fmf{dots,tension=0.5,right=0,width=1mm}{v11,v12}
\fmffreeze
\fmfposition
\fmfipath{p[]}
\fmfipair{w[]}
\fmfiset{p1}{vpath(__v1,__v7)}
\fmfiset{p2}{vpath(__v2,__v8)}
\fmfiset{p3}{vpath(__v3,__v9)}
\fmfiset{p4}{vpath(__v4,__v10)}
\fmfiset{p5}{vpath(__v5,__v11)}
\fmfiset{p6}{vpath(__v6,__v12)}
\fmfiequ{w1}{point length(p3)/4 of p3}
\fmfiequ{w2}{point length(p3)/2 of p3}
\fmfiequ{w3}{point 3length(p3)/4 of p3}
\vvertex{w4}{w3}{p1}
\vvertex{w5}{w3}{p2}
\vvertex{w6}{w2}{p4}
\vvertex{w7}{w1}{p4}
\vvertex{w8}{w1}{p5}
\vvertex{w9}{w1}{p6}
\fmfi{wiggly}{w5..w3}
\fmfi{wiggly}{w2..w6}
\fmfi{wiggly}{w7..w8}
\fmfi{dots}{w4..w5}
\fmfi{dots}{w8..w9}
\fmfiv{l=\tiny{$\bcovder^2$},l.a=55,l.d=3}{vloc(__v9)}
\fmfiv{l=\tiny{$\covder^2$},l.a=-30,l.d=2}{w3}
\fmfiv{l=\tiny{$\bcovder^2$},l.a=30,l.d=2}{w3}
\fmfiv{l=\tiny{$\covder^2$},l.a=-150,l.d=2}{w2}
\fmfiv{l=\tiny{$\bcovder^2$},l.a=55,l.d=3}{vloc(__v10)}
\fmfiv{l=\tiny{$\covder^2$},l.a=-30,l.d=2}{w6}
\fmfiv{l=\tiny{$\bcovder^2$},l.a=30,l.d=2}{w6}
\fmfiv{l=\tiny{$\covder^2$},l.a=-55,l.d=3}{w7}
\fmfposition
\end{fmfchar*}}}
\raisebox{-1.5\eqoff}{$\rightarrow$}
\subfigure[]{
\fmfframe(3,1)(1,4){%
\begin{fmfchar*}(40,35)
\fmftop{v1}
\fmfbottom{v7}
\fmfforce{(0w,h)}{v1}
\fmfforce{(0w,0)}{v7}
\fmffixed{(0.14w,0)}{v1,v2}
\fmffixed{(0.24w,0)}{v2,v3}
\fmffixed{(0.24w,0)}{v3,v4}
\fmffixed{(0.24w,0)}{v4,v5}
\fmffixed{(0.14w,0)}{v5,v6}
\fmffixed{(0.14w,0)}{v7,v8}
\fmffixed{(0.24w,0)}{v8,v9}
\fmffixed{(0.24w,0)}{v9,v10}
\fmffixed{(0.24w,0)}{v10,v11}
\fmffixed{(0.14w,0)}{v11,v12}
\fmf{phantom}{v1,v7}
\fmf{phantom}{v2,v8}
\fmf{plain}{v3,v9}
\fmf{plain}{v4,v10}
\fmf{phantom}{v5,v11}
\fmf{phantom}{v6,v12}
\fmf{plain,tension=0.5,right=0,width=1mm}{v8,v11}
\fmf{dots,tension=0.5,right=0,width=1mm}{v7,v8}
\fmf{dots,tension=0.5,right=0,width=1mm}{v11,v12}
\fmffreeze
\fmfposition
\fmfipath{p[]}
\fmfipair{w[]}
\fmfiset{p1}{vpath(__v1,__v7)}
\fmfiset{p2}{vpath(__v2,__v8)}
\fmfiset{p3}{vpath(__v3,__v9)}
\fmfiset{p4}{vpath(__v4,__v10)}
\fmfiset{p5}{vpath(__v5,__v11)}
\fmfiset{p6}{vpath(__v6,__v12)}
\fmfiequ{w1}{point length(p3)/4 of p3}
\fmfiequ{w2}{point length(p3)/2 of p3}
\fmfiequ{w3}{point 3length(p3)/4 of p3}
\vvertex{w4}{w3}{p1}
\vvertex{w5}{w3}{p2}
\vvertex{w6}{w2}{p4}
\vvertex{w7}{w1}{p4}
\vvertex{w8}{w1}{p5}
\vvertex{w9}{w1}{p6}
\fmfi{wiggly}{w5..w3}
\fmfi{wiggly}{w2..w6}
\fmfi{wiggly}{w7..w8}
\fmfi{dots}{w4..w5}
\fmfi{dots}{w8..w9}
\fmfiv{l=\tiny{$\bcovder^2$},l.a=55,l.d=3}{vloc(__v9)}
\fmfiv{l=\tiny{$\covder^2$},l.a=-30,l.d=2}{w3}
\fmfiv{l=\tiny{$\bcovder^2$},l.a=30,l.d=2}{w3}
\fmfiv{l=\tiny{$\covder^2$},l.a=-150,l.d=2}{w2}
\fmfiv{l=\tiny{$\bcovder^2$},l.a=55,l.d=3}{vloc(__v10)}
\fmfiv{l=\tiny{$\covder^2$},l.a=-30,l.d=2}{w6}
\fmfiv{l=\scriptsize{$\Box$},l.a=60,l.d=7}{w6}
\fmfposition
\end{fmfchar*}}}
\raisebox{-1.5\eqoff}{$=\ -$}
\subfigure[]{
\fmfframe(3,1)(1,4){%
\begin{fmfchar*}(40,35)
\fmftop{v1}
\fmfbottom{v7}
\fmfforce{(0w,h)}{v1}
\fmfforce{(0w,0)}{v7}
\fmffixed{(0.14w,0)}{v1,v2}
\fmffixed{(0.24w,0)}{v2,v3}
\fmffixed{(0.24w,0)}{v3,v4}
\fmffixed{(0.24w,0)}{v4,v5}
\fmffixed{(0.14w,0)}{v5,v6}
\fmffixed{(0.14w,0)}{v7,v8}
\fmffixed{(0.24w,0)}{v8,v9}
\fmffixed{(0.24w,0)}{v9,v10}
\fmffixed{(0.24w,0)}{v10,v11}
\fmffixed{(0.14w,0)}{v11,v12}
\fmf{phantom}{v1,v7}
\fmf{phantom}{v2,v8}
\fmf{plain}{v3,v9}
\fmf{plain}{v4,v10}
\fmf{phantom}{v5,v11}
\fmf{phantom}{v6,v12}
\fmf{plain,tension=0.5,right=0,width=1mm}{v8,v11}
\fmf{dots,tension=0.5,right=0,width=1mm}{v7,v8}
\fmf{dots,tension=0.5,right=0,width=1mm}{v11,v12}
\fmffreeze
\fmfposition
\fmfipath{p[]}
\fmfipair{w[]}
\fmfiset{p1}{vpath(__v1,__v7)}
\fmfiset{p2}{vpath(__v2,__v8)}
\fmfiset{p3}{vpath(__v3,__v9)}
\fmfiset{p4}{vpath(__v4,__v10)}
\fmfiset{p5}{vpath(__v5,__v11)}
\fmfiset{p6}{vpath(__v6,__v12)}
\fmfiequ{w1}{point length(p3)/3 of p3}
\fmfiequ{w2}{point 2length(p3)/3 of p3}
\vvertex{w3}{w2}{p1}
\vvertex{w4}{w2}{p2}
\vvertex{w5}{w1}{p4}
\vvertex{w6}{w1}{p5}
\vvertex{w7}{w1}{p6}
\vvertex{w8}{w1}{p4}
\fmfi{wiggly}{w1..w5}
\fmfi{wiggly}{w4..w2}
\fmfi{wiggly}{w8..w6}
\fmfi{dots}{w3..w4}
\fmfi{dots}{w6..w7}
\fmfiv{l=\tiny{$\bcovder^2$},l.a=55,l.d=3}{vloc(__v9)}
\fmfiv{l=\tiny{$\covder^2$},l.a=-30,l.d=2}{w2}
\fmfiv{l=\tiny{$\bcovder^2$},l.a=30,l.d=2}{w2}
\fmfiv{l=\tiny{$\covder^2$},l.a=-150,l.d=2}{w1}
\fmfiv{l=\tiny{$\bcovder^2$},l.a=55,l.d=3}{vloc(__v10)}
\fmfiv{l=\tiny{$\covder^2$},l.a=-55,l.d=3}{w8}
\fmfposition
\end{fmfchar*}}}
\raisebox{-1.5\eqoff}{$=-A_2$}
}
\caption{Partial D-algebra for the diagram $A_3$}
\label{A2A3algebra}
\end{figure}

In a diagram whose range is the largest compatible with the perturbative order, the propagator coming out from the single-vector vertex can be attached only to one out of three different structures, which are shown in Figures~\ref{block-A}, \ref{block-B} and \ref{block-C}. All the corresponding combinations are presented in Figures~\ref{diagrams-A}, ~\ref{diagrams-B} and \ref{diagrams-C}.
By analyzing the three possibilities separately, it is possible to show that the divergent parts of the corresponding classes of diagrams sum up to zero.
The proof relies on the observation that when the double covariant derivative $\covder^2$ at the single-vector vertex is integrated by parts, the result of Section~\ref{sec:proof} ensures that divergent contributions can appear in the end only if the whole $\covder^2$ is kept inside the graph, and thus is moved onto the vector propagator leaving the vertex. At the opposite end of the propagator, a second integration by parts is possible, which either brings the derivatives out of the diagram, or, after exploiting the standard D-algebra identities for covariant derivatives, modifies the structure of the underlying superspace integral. In the former case, the new structure is finite, whereas in the latter it turns out that the new integral is cancelled by one coming from a different original graph. As an example, the manipulations required in the case of the cancellation between the diagrams $A_2$ and $A_3$ of Figure~\ref{diagrams-A} are shown explicitly in Figure~\ref{A2A3algebra}. 
The three classes are now considered in detail.
\begin{itemize}
\item Class A (Fig.~\ref{diagrams-A}):\\
The diagram $A_1$ is finite, since a $\covder^2$ is forced to exit the graph. 
Performing part of the D-algebra for the diagram $A_3$, one immediately obtains the structure of the diagram $A_2$, with a different sign due to the $\Box=-p^2$ cancelling one propagator, as shown in Figure~\ref{A2A3algebra}. 
As the two diagrams have the same colour factor, their divergent parts sum up to zero.
\item Class B (Fig.~\ref{diagrams-B}):\\
A partial D-algebra computation shows that the diagrams $B_1$ and $B_2$ produce the same superspace integral, and
the opposite colour factors make the total divergent part vanish.
The diagram $B_3$ is finite, for the same reason as $A_1$.
\item Class C (Fig.~\ref{diagrams-C}):\\
For the diagrams $C_1$ and $C_2$, the situation is the same as for $B_1$ and $B_2$ and the divergent parts of the two diagrams cancel out.
For the diagrams $C_3$ and $C_4$, the cancellation occurs following the same pattern as for $A_2$ and $A_3$.
The diagram $C_5$ is finite, similarly to $A_1$ and $B_3$.
\end{itemize}

\begin{figure}[h]
\capstart
\unitlength=0.75mm
\settoheight{\eqoff}{$\times$}%
\setlength{\eqoff}{0.5\eqoff}%
\addtolength{\eqoff}{-12.5\unitlength}%
\settoheight{\eqofftwo}{$\times$}%
\setlength{\eqofftwo}{0.5\eqofftwo}%
\addtolength{\eqofftwo}{-7.5\unitlength}%
\centering
\raisebox{\eqoff}{%
\subfigure[$A_1$]{
\fmfframe(3,1)(1,4){%
\begin{fmfchar*}(30,20)
\fmftop{v1}
\fmfbottom{v7}
\fmfforce{(0w,h)}{v1}
\fmfforce{(0w,0)}{v7}
\fmffixed{(0.14w,0)}{v1,v2}
\fmffixed{(0.24w,0)}{v2,v3}
\fmffixed{(0.24w,0)}{v3,v4}
\fmffixed{(0.24w,0)}{v4,v5}
\fmffixed{(0.14w,0)}{v5,v6}
\fmffixed{(0.14w,0)}{v7,v8}
\fmffixed{(0.24w,0)}{v8,v9}
\fmffixed{(0.24w,0)}{v9,v10}
\fmffixed{(0.24w,0)}{v10,v11}
\fmffixed{(0.14w,0)}{v11,v12}
\fmf{phantom}{v1,v7}
\fmf{phantom}{v2,v8}
\fmf{plain}{v3,v9}
\fmf{plain}{v4,v10}
\fmf{phantom}{v5,v11}
\fmf{phantom}{v6,v12}
\fmf{plain,tension=0.5,right=0,width=1mm}{v8,v11}
\fmf{dots,tension=0.5,right=0,width=1mm}{v7,v8}
\fmf{dots,tension=0.5,right=0,width=1mm}{v11,v12}
\fmffreeze
\fmfposition
\fmfipath{p[]}
\fmfipair{w[]}
\fmfiset{p1}{vpath(__v1,__v7)}
\fmfiset{p2}{vpath(__v2,__v8)}
\fmfiset{p3}{vpath(__v3,__v9)}
\fmfiset{p4}{vpath(__v4,__v10)}
\fmfiset{p5}{vpath(__v5,__v11)}
\fmfiset{p6}{vpath(__v6,__v12)}
\fmfiequ{w1}{point length(p3)/3 of p3}
\fmfiequ{w2}{point 2length(p3)/3 of p3}
\vvertex{w3}{w2}{p1}
\vvertex{w4}{w2}{p2}
\vvertex{w5}{w1}{p4}
\vvertex{w6}{w2}{p5}
\vvertex{w7}{w2}{p6}
\vvertex{w8}{w2}{p4}
\fmfi{wiggly}{w1..w5}
\fmfi{wiggly}{w4..w2}
\fmfi{wiggly}{w8..w6}
\fmfi{dots}{w3..w4}
\fmfi{dots}{w6..w7}
\fmfposition
\end{fmfchar*}}
}
\qquad
\subfigure[$A_2$]{
\fmfframe(3,1)(1,4){%
\begin{fmfchar*}(30,20)
\fmftop{v1}
\fmfbottom{v7}
\fmfforce{(0w,h)}{v1}
\fmfforce{(0w,0)}{v7}
\fmffixed{(0.14w,0)}{v1,v2}
\fmffixed{(0.24w,0)}{v2,v3}
\fmffixed{(0.24w,0)}{v3,v4}
\fmffixed{(0.24w,0)}{v4,v5}
\fmffixed{(0.14w,0)}{v5,v6}
\fmffixed{(0.14w,0)}{v7,v8}
\fmffixed{(0.24w,0)}{v8,v9}
\fmffixed{(0.24w,0)}{v9,v10}
\fmffixed{(0.24w,0)}{v10,v11}
\fmffixed{(0.14w,0)}{v11,v12}
\fmf{phantom}{v1,v7}
\fmf{phantom}{v2,v8}
\fmf{plain}{v3,v9}
\fmf{plain}{v4,v10}
\fmf{phantom}{v5,v11}
\fmf{phantom}{v6,v12}
\fmf{plain,tension=0.5,right=0,width=1mm}{v8,v11}
\fmf{dots,tension=0.5,right=0,width=1mm}{v7,v8}
\fmf{dots,tension=0.5,right=0,width=1mm}{v11,v12}
\fmffreeze
\fmfposition
\fmfipath{p[]}
\fmfipair{w[]}
\fmfiset{p1}{vpath(__v1,__v7)}
\fmfiset{p2}{vpath(__v2,__v8)}
\fmfiset{p3}{vpath(__v3,__v9)}
\fmfiset{p4}{vpath(__v4,__v10)}
\fmfiset{p5}{vpath(__v5,__v11)}
\fmfiset{p6}{vpath(__v6,__v12)}
\fmfiequ{w1}{point length(p3)/3 of p3}
\fmfiequ{w2}{point 2length(p3)/3 of p3}
\vvertex{w3}{w2}{p1}
\vvertex{w4}{w2}{p2}
\vvertex{w5}{w1}{p4}
\vvertex{w6}{w1}{p5}
\vvertex{w7}{w1}{p6}
\vvertex{w8}{w1}{p4}
\fmfi{wiggly}{w1..w5}
\fmfi{wiggly}{w4..w2}
\fmfi{wiggly}{w8..w6}
\fmfi{dots}{w3..w4}
\fmfi{dots}{w6..w7}
\fmfposition
\end{fmfchar*}}
}
\qquad
\subfigure[$A_3$]{
\fmfframe(3,1)(1,4){%
\begin{fmfchar*}(30,20)
\fmftop{v1}
\fmfbottom{v7}
\fmfforce{(0w,h)}{v1}
\fmfforce{(0w,0)}{v7}
\fmffixed{(0.14w,0)}{v1,v2}
\fmffixed{(0.24w,0)}{v2,v3}
\fmffixed{(0.24w,0)}{v3,v4}
\fmffixed{(0.24w,0)}{v4,v5}
\fmffixed{(0.14w,0)}{v5,v6}
\fmffixed{(0.14w,0)}{v7,v8}
\fmffixed{(0.24w,0)}{v8,v9}
\fmffixed{(0.24w,0)}{v9,v10}
\fmffixed{(0.24w,0)}{v10,v11}
\fmffixed{(0.14w,0)}{v11,v12}
\fmf{phantom}{v1,v7}
\fmf{phantom}{v2,v8}
\fmf{plain}{v3,v9}
\fmf{plain}{v4,v10}
\fmf{phantom}{v5,v11}
\fmf{phantom}{v6,v12}
\fmf{plain,tension=0.5,right=0,width=1mm}{v8,v11}
\fmf{dots,tension=0.5,right=0,width=1mm}{v7,v8}
\fmf{dots,tension=0.5,right=0,width=1mm}{v11,v12}
\fmffreeze
\fmfposition
\fmfipath{p[]}
\fmfipair{w[]}
\fmfiset{p1}{vpath(__v1,__v7)}
\fmfiset{p2}{vpath(__v2,__v8)}
\fmfiset{p3}{vpath(__v3,__v9)}
\fmfiset{p4}{vpath(__v4,__v10)}
\fmfiset{p5}{vpath(__v5,__v11)}
\fmfiset{p6}{vpath(__v6,__v12)}
\fmfiequ{w1}{point length(p3)/4 of p3}
\fmfiequ{w2}{point length(p3)/2 of p3}
\fmfiequ{w3}{point 3length(p3)/4 of p3}
\vvertex{w4}{w3}{p1}
\vvertex{w5}{w3}{p2}
\vvertex{w6}{w2}{p4}
\vvertex{w7}{w1}{p4}
\vvertex{w8}{w1}{p5}
\vvertex{w9}{w1}{p6}
\fmfi{wiggly}{w5..w3}
\fmfi{wiggly}{w2..w6}
\fmfi{wiggly}{w7..w8}
\fmfi{dots}{w4..w5}
\fmfi{dots}{w8..w9}
\fmfposition
\end{fmfchar*}}}
}

\caption{Diagrams of class A}
\label{diagrams-A}
\end{figure}
\begin{figure}[h]
\capstart
\unitlength=0.75mm
\settoheight{\eqoff}{$\times$}%
\setlength{\eqoff}{0.5\eqoff}%
\addtolength{\eqoff}{-12.5\unitlength}%
\settoheight{\eqofftwo}{$\times$}%
\setlength{\eqofftwo}{0.5\eqofftwo}%
\addtolength{\eqofftwo}{-7.5\unitlength}%
\centering
\raisebox{\eqoff}{%
\subfigure[$B_1$]{
\fmfframe(3,1)(1,4){%
\begin{fmfchar*}(30,20)
\fmftop{v1}
\fmfbottom{v7}
\fmfforce{(0w,h)}{v1}
\fmfforce{(0w,0)}{v7}
\fmffixed{(0.15w,0)}{v1,v2}
\fmffixed{(0.25w,0)}{v2,v3}
\fmffixed{(0.25w,0)}{v3,v4}
\fmffixed{(0.2w,0)}{v4,v5}
\fmffixed{(0.15w,0)}{v5,v6}
\fmffixed{(0.15w,0)}{v7,v8}
\fmffixed{(0.25w,0)}{v8,v9}
\fmffixed{(0.25w,0)}{v9,v10}
\fmffixed{(0.2w,0)}{v10,v11}
\fmffixed{(0.15w,0)}{v11,v12}
\fmf{phantom}{v1,v7}
\fmf{phantom}{v2,v8}
\fmf{plain}{v3,v9}
\fmfset{dash_len}{1.5mm}
\fmf{plain,tension=0.25,right=0.25}{v4,vc1}
\fmf{phantom,tension=0.25,left=0.25}{v5,vc1}
\fmf{plain,tension=0.25,left=0.25}{v10,vc2}
\fmf{phantom,tension=0.25,right=0.25}{v11,vc2}
\fmf{plain,tension=0.5}{vc1,vc2}
\fmf{plain,tension=0.5,right=0,width=1mm}{v8,v11}
\fmf{dots,tension=0.5,right=0,width=1mm}{v7,v8}
\fmf{dots,tension=0.5,right=0,width=1mm}{v11,v12}
\fmffreeze
\fmfposition
\fmffixed{(0.1w,0)}{vc1,vc3}
\fmffixed{(0.1w,0)}{vc2,vc4}
\fmf{dashes,tension=0.25,left=0.25}{vc3,vc1}
\fmf{dashes,tension=0.25,right=0.25}{vc4,vc2}
\fmfipath{p[]}
\fmfipair{w[]}
\fmfiset{p1}{vpath(__v1,__v7)}
\fmfiset{p2}{vpath(__v2,__v8)}
\fmfiset{p3}{vpath(__v3,__v9)}
\fmfiset{p4}{vpath(__vc2,__v10)}
\fmfiequ{w1}{point length(p3)/3 of p3}
\fmfiequ{w2}{point 2length(p3)/3 of p3}
\vvertex{w3}{w2}{p1}
\vvertex{w4}{w2}{p2}
\svertex{w5}{p4}
\fmfi{wiggly}{w4..w2}
\fmfi{wiggly}{w1..w5}
\fmfi{dots}{w3..w4}
\fmfposition
\end{fmfchar*}}
}
\qquad
\subfigure[$B_2$]{
\fmfframe(3,1)(1,4){%
\begin{fmfchar*}(30,20)
\fmftop{v1}
\fmfbottom{v7}
\fmfforce{(0w,h)}{v1}
\fmfforce{(0w,0)}{v7}
\fmffixed{(0.15w,0)}{v1,v2}
\fmffixed{(0.25w,0)}{v2,v3}
\fmffixed{(0.25w,0)}{v3,v4}
\fmffixed{(0.2w,0)}{v4,v5}
\fmffixed{(0.15w,0)}{v5,v6}
\fmffixed{(0.15w,0)}{v7,v8}
\fmffixed{(0.25w,0)}{v8,v9}
\fmffixed{(0.25w,0)}{v9,v10}
\fmffixed{(0.2w,0)}{v10,v11}
\fmffixed{(0.15w,0)}{v11,v12}
\fmf{phantom}{v1,v7}
\fmf{phantom}{v2,v8}
\fmf{plain}{v3,v9}
\fmfset{dash_len}{1.5mm}
\fmf{plain,tension=0.25,right=0.25}{v4,vc1}
\fmf{phantom,tension=0.25,left=0.25}{v5,vc1}
\fmf{plain,tension=0.25,left=0.25}{v10,vc2}
\fmf{phantom,tension=0.25,right=0.25}{v11,vc2}
\fmf{plain,tension=0.5}{vc1,vc2}
\fmf{plain,tension=0.5,right=0,width=1mm}{v8,v11}
\fmf{dots,tension=0.5,right=0,width=1mm}{v7,v8}
\fmf{dots,tension=0.5,right=0,width=1mm}{v11,v12}
\fmffreeze
\fmfposition
\fmffixed{(0.1w,0)}{vc1,vc3}
\fmffixed{(0.1w,0)}{vc2,vc4}
\fmf{dashes,tension=0.25,left=0.25}{vc3,vc1}
\fmf{dashes,tension=0.25,right=0.25}{vc4,vc2}
\fmfipath{p[]}
\fmfipair{w[]}
\fmfiset{p1}{vpath(__v1,__v7)}
\fmfiset{p2}{vpath(__v2,__v8)}
\fmfiset{p3}{vpath(__v3,__v9)}
\fmfiset{p4}{vpath(__vc2,__vc1)}
\fmfiequ{w1}{point length(p3)/3 of p3}
\fmfiequ{w2}{point 2length(p3)/3 of p3}
\vvertex{w3}{w2}{p1}
\vvertex{w4}{w2}{p2}
\svertex{w5}{p4}
\fmfi{wiggly}{w4..w2}
\fmfi{wiggly}{w1..w5}
\fmfi{dots}{w3..w4}
\fmfposition
\end{fmfchar*}}
}
\qquad
\subfigure[$B_3$]{
\fmfframe(3,1)(1,4){%
\begin{fmfchar*}(30,20)
\fmftop{v1}
\fmfbottom{v7}
\fmfforce{(0w,h)}{v1}
\fmfforce{(0w,0)}{v7}
\fmffixed{(0.15w,0)}{v1,v2}
\fmffixed{(0.25w,0)}{v2,v3}
\fmffixed{(0.25w,0)}{v3,v4}
\fmffixed{(0.2w,0)}{v4,v5}
\fmffixed{(0.15w,0)}{v5,v6}
\fmffixed{(0.15w,0)}{v7,v8}
\fmffixed{(0.25w,0)}{v8,v9}
\fmffixed{(0.25w,0)}{v9,v10}
\fmffixed{(0.2w,0)}{v10,v11}
\fmffixed{(0.15w,0)}{v11,v12}
\fmf{phantom}{v1,v7}
\fmf{phantom}{v2,v8}
\fmf{plain}{v3,v9}
\fmfset{dash_len}{1.5mm}
\fmf{plain,tension=0.25,right=0.25}{v4,vc1}
\fmf{phantom,tension=0.25,left=0.25}{v5,vc1}
\fmf{plain,tension=0.25,left=0.25}{v10,vc2}
\fmf{phantom,tension=0.25,right=0.25}{v11,vc2}
\fmf{plain,tension=0.5}{vc1,vc2}
\fmf{plain,tension=0.5,right=0,width=1mm}{v8,v11}
\fmf{dots,tension=0.5,right=0,width=1mm}{v7,v8}
\fmf{dots,tension=0.5,right=0,width=1mm}{v11,v12}
\fmffreeze
\fmfposition
\fmffixed{(0.1w,0)}{vc1,vc3}
\fmffixed{(0.1w,0)}{vc2,vc4}
\fmf{dashes,tension=0.25,left=0.25}{vc3,vc1}
\fmf{dashes,tension=0.25,right=0.25}{vc4,vc2}
\fmfipath{p[]}
\fmfipair{w[]}
\fmfiset{p1}{vpath(__v1,__v7)}
\fmfiset{p2}{vpath(__v2,__v8)}
\fmfiset{p3}{vpath(__v3,__v9)}
\fmfiset{p4}{vpath(__v4,__vc1)}
\fmfiequ{w1}{point length(p3)/3 of p3}
\fmfiequ{w2}{point 2length(p3)/3 of p3}
\vvertex{w3}{w2}{p1}
\vvertex{w4}{w2}{p2}
\svertex{w5}{p4}
\fmfi{wiggly}{w4..w2}
\fmfi{wiggly}{w1..w5}
\fmfi{dots}{w3..w4}
\fmfposition
\end{fmfchar*}}}
}
\caption{Diagrams of class B}
\label{diagrams-B}
\end{figure}
\begin{figure}[h]
\capstart
\unitlength=0.75mm
\settoheight{\eqoff}{$\times$}%
\setlength{\eqoff}{0.5\eqoff}%
\addtolength{\eqoff}{-12.5\unitlength}%
\settoheight{\eqofftwo}{$\times$}%
\setlength{\eqofftwo}{0.5\eqofftwo}%
\addtolength{\eqofftwo}{-7.5\unitlength}%
\centering
\raisebox{\eqoff}{%
\subfigure[$C_1$]{
\fmfframe(3,1)(1,4){%
\begin{fmfchar*}(35,20)
\fmftop{v1}
\fmfbottom{v8}
\fmfforce{(0w,h)}{v1}
\fmfforce{(0w,0)}{v8}
\fmffixed{(0.1w,0)}{v1,v2}
\fmffixed{(0.2w,0)}{v2,v3}
\fmffixed{(0.2w,0)}{v3,v4}
\fmffixed{(0.2w,0)}{v4,v5}
\fmffixed{(0.2w,0)}{v5,v6}
\fmffixed{(0.1w,0)}{v6,v7}
\fmffixed{(0.1w,0)}{v8,v9}
\fmffixed{(0.2w,0)}{v9,v10}
\fmffixed{(0.2w,0)}{v10,v11}
\fmffixed{(0.2w,0)}{v11,v12}
\fmffixed{(0.2w,0)}{v12,v13}
\fmffixed{(0.1w,0)}{v13,v14}
\fmf{phantom}{v1,v8}
\fmf{phantom}{v2,v9}
\fmf{plain}{v3,v10}
\fmf{phantom}{v6,v13}
\fmf{phantom}{v7,v14}
\fmf{plain,tension=0.25,right=0.25}{v4,vc1}
\fmf{plain,tension=0.25,left=0.25}{v5,vc1}
\fmf{plain,tension=0.25,left=0.25}{v11,vc2}
\fmf{plain,tension=0.25,right=0.25}{v12,vc2}
\fmf{plain,tension=0.5}{vc1,vc2}
\fmf{plain,tension=0.5,right=0,width=1mm}{v9,v13}
\fmf{dots,tension=0.5,right=0,width=1mm}{v8,v9}
\fmf{dots,tension=0.5,right=0,width=1mm}{v13,v14}
\fmffreeze
\fmfposition
\fmfipath{p[]}
\fmfipair{w[]}
\fmfiset{p1}{vpath(__v1,__v8)}
\fmfiset{p2}{vpath(__v2,__v9)}
\fmfiset{p3}{vpath(__v3,__v10)}
\fmfiset{p4}{vpath(__v11,__vc2)}
\fmfiset{p5}{vpath(__v6,__v13)}
\fmfiset{p6}{vpath(__v7,__v14)}
\fmfiset{p7}{vpath(__vc1,__vc2)}
\fmfiequ{w1}{point length(p3)/3 of p3}
\fmfiequ{w2}{point 2length(p3)/3 of p3}
\vvertex{w3}{w2}{p1}
\vvertex{w4}{w2}{p2}
\svertex{w5}{p4}
\svertex{w6}{p7}
\vvertex{w7}{w6}{p5}
\vvertex{w8}{w6}{p6}
\fmfi{wiggly}{w4..w2}
\fmfi{wiggly}{w1..w5}
\fmfi{dots}{w3..w4}
\fmfi{wiggly}{w6..w7}
\fmfi{dots}{w7..w8}
\fmfposition
\end{fmfchar*}}
}
\qquad
\subfigure[$C_2$]{
\fmfframe(3,1)(1,4){%
\begin{fmfchar*}(35,20)
\fmftop{v1}
\fmfbottom{v8}
\fmfforce{(0w,h)}{v1}
\fmfforce{(0w,0)}{v8}
\fmffixed{(0.1w,0)}{v1,v2}
\fmffixed{(0.2w,0)}{v2,v3}
\fmffixed{(0.2w,0)}{v3,v4}
\fmffixed{(0.2w,0)}{v4,v5}
\fmffixed{(0.2w,0)}{v5,v6}
\fmffixed{(0.1w,0)}{v6,v7}
\fmffixed{(0.1w,0)}{v8,v9}
\fmffixed{(0.2w,0)}{v9,v10}
\fmffixed{(0.2w,0)}{v10,v11}
\fmffixed{(0.2w,0)}{v11,v12}
\fmffixed{(0.2w,0)}{v12,v13}
\fmffixed{(0.1w,0)}{v13,v14}
\fmf{phantom}{v1,v8}
\fmf{phantom}{v2,v9}
\fmf{plain}{v3,v10}
\fmf{phantom}{v6,v13}
\fmf{phantom}{v7,v14}
\fmf{plain,tension=0.25,right=0.25}{v4,vc1}
\fmf{plain,tension=0.25,left=0.25}{v5,vc1}
\fmf{plain,tension=0.25,left=0.25}{v11,vc2}
\fmf{plain,tension=0.25,right=0.25}{v12,vc2}
\fmf{plain,tension=0.5}{vc1,vc2}
\fmf{plain,tension=0.5,right=0,width=1mm}{v9,v13}
\fmf{dots,tension=0.5,right=0,width=1mm}{v8,v9}
\fmf{dots,tension=0.5,right=0,width=1mm}{v13,v14}
\fmffreeze
\fmfposition
\fmfipath{p[]}
\fmfipair{w[]}
\fmfiset{p1}{vpath(__v1,__v8)}
\fmfiset{p2}{vpath(__v2,__v9)}
\fmfiset{p3}{vpath(__v3,__v10)}
\fmfiset{p4}{vpath(__v11,__vc2)}
\fmfiset{p5}{vpath(__v6,__v13)}
\fmfiset{p6}{vpath(__v7,__v14)}
\fmfiset{p7}{vpath(__vc1,__vc2)}
\fmfiequ{w1}{point length(p3)/3 of p3}
\fmfiequ{w2}{point 2length(p3)/3 of p3}
\vvertex{w3}{w2}{p1}
\vvertex{w4}{w2}{p2}
\fmfiequ{w5}{point 2length(p3)/3 of p7}
\fmfiequ{w6}{point length(p3)/3 of p7}
\vvertex{w7}{w6}{p5}
\vvertex{w8}{w6}{p6}
\fmfi{wiggly}{w4..w2}
\fmfi{wiggly}{w1..w5}
\fmfi{dots}{w3..w4}
\fmfi{wiggly}{w6..w7}
\fmfi{dots}{w7..w8}
\fmfposition
\end{fmfchar*}}
}
\qquad
\subfigure[$C_3$]{
\fmfframe(3,1)(1,4){%
\begin{fmfchar*}(35,20)
\fmftop{v1}
\fmfbottom{v8}
\fmfforce{(0w,h)}{v1}
\fmfforce{(0w,0)}{v8}
\fmffixed{(0.1w,0)}{v1,v2}
\fmffixed{(0.2w,0)}{v2,v3}
\fmffixed{(0.2w,0)}{v3,v4}
\fmffixed{(0.2w,0)}{v4,v5}
\fmffixed{(0.2w,0)}{v5,v6}
\fmffixed{(0.1w,0)}{v6,v7}
\fmffixed{(0.1w,0)}{v8,v9}
\fmffixed{(0.2w,0)}{v9,v10}
\fmffixed{(0.2w,0)}{v10,v11}
\fmffixed{(0.2w,0)}{v11,v12}
\fmffixed{(0.2w,0)}{v12,v13}
\fmffixed{(0.1w,0)}{v13,v14}
\fmf{phantom}{v1,v8}
\fmf{phantom}{v2,v9}
\fmf{plain}{v3,v10}
\fmf{phantom}{v6,v13}
\fmf{phantom}{v7,v14}
\fmf{plain,tension=0.25,right=0.25}{v4,vc1}
\fmf{plain,tension=0.25,left=0.25}{v5,vc1}
\fmf{plain,tension=0.25,left=0.25}{v11,vc2}
\fmf{plain,tension=0.25,right=0.25}{v12,vc2}
\fmf{plain,tension=0.5}{vc1,vc2}
\fmf{plain,tension=0.5,right=0,width=1mm}{v9,v13}
\fmf{dots,tension=0.5,right=0,width=1mm}{v8,v9}
\fmf{dots,tension=0.5,right=0,width=1mm}{v13,v14}
\fmffreeze
\fmfposition
\fmfipath{p[]}
\fmfipair{w[]}
\fmfiset{p1}{vpath(__v1,__v8)}
\fmfiset{p2}{vpath(__v2,__v9)}
\fmfiset{p3}{vpath(__v3,__v10)}
\fmfiset{p4}{vpath(__v11,__vc2)}
\fmfiset{p5}{vpath(__v6,__v13)}
\fmfiset{p6}{vpath(__v7,__v14)}
\fmfiset{p7}{vpath(__vc1,__vc2)}
\fmfiequ{w1}{point length(p3)/3 of p3}
\fmfiequ{w2}{point 2length(p3)/3 of p3}
\vvertex{w3}{w2}{p1}
\vvertex{w4}{w2}{p2}
\fmfiequ{w5}{point length(p3)/2 of p7}
\fmfiequ{w6}{point length(p3)/2 of p7}
\vvertex{w7}{w6}{p5}
\vvertex{w8}{w6}{p6}
\fmfi{wiggly}{w4..w2}
\fmfi{wiggly}{w1..w5}
\fmfi{dots}{w3..w4}
\fmfi{wiggly}{w6..w7}
\fmfi{dots}{w7..w8}
\fmfposition
\end{fmfchar*}}}
}
\\
\raisebox{\eqoff}{%
\subfigure[$C_4$]{
\fmfframe(3,1)(1,4){%
\begin{fmfchar*}(35,20)
\fmftop{v1}
\fmfbottom{v8}
\fmfforce{(0w,h)}{v1}
\fmfforce{(0w,0)}{v8}
\fmffixed{(0.1w,0)}{v1,v2}
\fmffixed{(0.2w,0)}{v2,v3}
\fmffixed{(0.2w,0)}{v3,v4}
\fmffixed{(0.2w,0)}{v4,v5}
\fmffixed{(0.2w,0)}{v5,v6}
\fmffixed{(0.1w,0)}{v6,v7}
\fmffixed{(0.1w,0)}{v8,v9}
\fmffixed{(0.2w,0)}{v9,v10}
\fmffixed{(0.2w,0)}{v10,v11}
\fmffixed{(0.2w,0)}{v11,v12}
\fmffixed{(0.2w,0)}{v12,v13}
\fmffixed{(0.1w,0)}{v13,v14}
\fmf{phantom}{v1,v8}
\fmf{phantom}{v2,v9}
\fmf{plain}{v3,v10}
\fmf{phantom}{v6,v13}
\fmf{phantom}{v7,v14}
\fmf{plain,tension=0.25,right=0.25}{v4,vc1}
\fmf{plain,tension=0.25,left=0.25}{v5,vc1}
\fmf{plain,tension=0.25,left=0.25}{v11,vc2}
\fmf{plain,tension=0.25,right=0.25}{v12,vc2}
\fmf{plain,tension=0.5}{vc1,vc2}
\fmf{plain,tension=0.5,right=0,width=1mm}{v9,v13}
\fmf{dots,tension=0.5,right=0,width=1mm}{v8,v9}
\fmf{dots,tension=0.5,right=0,width=1mm}{v13,v14}
\fmffreeze
\fmfposition
\fmfipath{p[]}
\fmfipair{w[]}
\fmfiset{p1}{vpath(__v1,__v8)}
\fmfiset{p2}{vpath(__v2,__v9)}
\fmfiset{p3}{vpath(__v3,__v10)}
\fmfiset{p4}{vpath(__v11,__vc2)}
\fmfiset{p5}{vpath(__v6,__v13)}
\fmfiset{p6}{vpath(__v7,__v14)}
\fmfiset{p7}{vpath(__vc1,__vc2)}
\fmfiequ{w1}{point length(p3)/3 of p3}
\fmfiequ{w2}{point 2length(p3)/3 of p3}
\vvertex{w3}{w2}{p1}
\vvertex{w4}{w2}{p2}
\fmfiequ{w5}{point length(p3)/3 of p7}
\fmfiequ{w6}{point 2length(p3)/3 of p7}
\vvertex{w7}{w6}{p5}
\vvertex{w8}{w6}{p6}
\fmfi{wiggly}{w4..w2}
\fmfi{wiggly}{w1..w5}
\fmfi{dots}{w3..w4}
\fmfi{wiggly}{w6..w7}
\fmfi{dots}{w7..w8}
\fmfposition
\end{fmfchar*}}
}
\qquad
\subfigure[$C_5$]{
\fmfframe(3,1)(1,4){%
\begin{fmfchar*}(35,20)
\fmftop{v1}
\fmfbottom{v8}
\fmfforce{(0w,h)}{v1}
\fmfforce{(0w,0)}{v8}
\fmffixed{(0.1w,0)}{v1,v2}
\fmffixed{(0.2w,0)}{v2,v3}
\fmffixed{(0.2w,0)}{v3,v4}
\fmffixed{(0.2w,0)}{v4,v5}
\fmffixed{(0.2w,0)}{v5,v6}
\fmffixed{(0.1w,0)}{v6,v7}
\fmffixed{(0.1w,0)}{v8,v9}
\fmffixed{(0.2w,0)}{v9,v10}
\fmffixed{(0.2w,0)}{v10,v11}
\fmffixed{(0.2w,0)}{v11,v12}
\fmffixed{(0.2w,0)}{v12,v13}
\fmffixed{(0.1w,0)}{v13,v14}
\fmf{phantom}{v1,v8}
\fmf{phantom}{v2,v9}
\fmf{plain}{v3,v10}
\fmf{phantom}{v6,v13}
\fmf{phantom}{v7,v14}
\fmf{plain,tension=0.25,right=0.25}{v4,vc1}
\fmf{plain,tension=0.25,left=0.25}{v5,vc1}
\fmf{plain,tension=0.25,left=0.25}{v11,vc2}
\fmf{plain,tension=0.25,right=0.25}{v12,vc2}
\fmf{plain,tension=0.5}{vc1,vc2}
\fmf{plain,tension=0.5,right=0,width=1mm}{v9,v13}
\fmf{dots,tension=0.5,right=0,width=1mm}{v8,v9}
\fmf{dots,tension=0.5,right=0,width=1mm}{v13,v14}
\fmffreeze
\fmfposition
\fmfipath{p[]}
\fmfipair{w[]}
\fmfiset{p1}{vpath(__v1,__v8)}
\fmfiset{p2}{vpath(__v2,__v9)}
\fmfiset{p3}{vpath(__v3,__v10)}
\fmfiset{p4}{vpath(__v11,__vc2)}
\fmfiset{p5}{vpath(__v6,__v13)}
\fmfiset{p6}{vpath(__v7,__v14)}
\fmfiset{p7}{vpath(__vc1,__vc2)}
\fmfiset{p8}{vpath(__v4,__vc1)}
\fmfiequ{w1}{point length(p3)/3 of p3}
\fmfiequ{w2}{point 2length(p3)/3 of p3}
\vvertex{w3}{w2}{p1}
\vvertex{w4}{w2}{p2}
\svertex{w5}{p8}
\fmfiequ{w6}{point length(p3)/2 of p7}
\vvertex{w7}{w6}{p5}
\vvertex{w8}{w6}{p6}
\fmfi{wiggly}{w4..w2}
\fmfi{wiggly}{w1..w5}
\fmfi{dots}{w3..w4}
\fmfi{wiggly}{w6..w7}
\fmfi{dots}{w7..w8}
\fmfposition
\end{fmfchar*}}}
}
\caption{Diagrams of class C}
\label{diagrams-C}
\end{figure}

It is therefore possible to conclude that all the diagrams of maximum range containing one of the structures of Figure~\ref{startstruct} can be neglected for the computation of anomalous dimensions, where only divergent contributions are of interest. In particular, this is true for the $\ell$-loop, non-maximal diagrams with range $(\ell+1)$. Another useful consequence is that scalar lines free of three-scalar interactions can interact only through double-vector vertices.

\chapter{Four-loop range-five diagrams}
\label{app:fourR5}
From the explicit expression~\eqref{D4sub} for the subtracted four-loop dilatation operator, and from the matrix form of the chiral functions~\eqref{chiM}, it turns out that the only similarity coefficient required for the four-loop computation of Chapter~\ref{chapter:fourloop} is $\epsilon_{3a}$. In this appendix, its value in the chosen renormalization scheme will be determined by studying four-loop range-five diagrams. The explicit values of $\epsilon_{3b}$ and $\epsilon_{3c}$ will be found too, even though they are not strictly needed, and moreover several consistency checks on the procedure will be performed.

\section{Coefficients from supergraphs}
\label{sec:fourR5}
\begin{figure}[h]
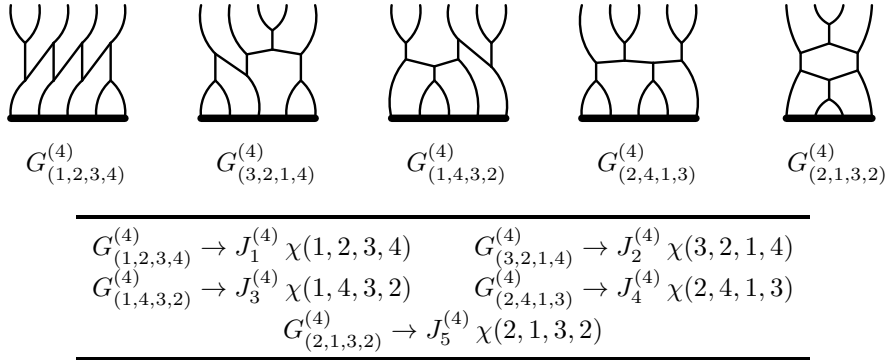

\capstart
\renewcommand*{\thesubfigure}{\ \ }
\footnotesize
\unitlength=0.75mm
\settoheight{\eqoff}{$\times$}%
\setlength{\eqoff}{0.5\eqoff}%
\addtolength{\eqoff}{-12.5\unitlength}%
\settoheight{\eqofftwo}{$\times$}%
\setlength{\eqofftwo}{0.5\eqofftwo}%
\addtolength{\eqofftwo}{-7.5\unitlength}%
\subfigure[$\nwgraph{4}{}{1,2,3,4}$]{
\raisebox{\eqoff}{%
\fmfframe(3,1)(1,4){%
\begin{fmfchar*}(20,20)
\fmftop{v1}
\fmfbottom{v5}
\fmfforce{(0w,h)}{v1}
\fmfforce{(0w,0)}{v5}
\fmffixed{(0.25w,0)}{v1,v2}
\fmffixed{(0.25w,0)}{v2,v3}
\fmffixed{(0.25w,0)}{v3,v4}
\fmffixed{(0.25w,0)}{v4,v9}
\fmffixed{(0.25w,0)}{v5,v6}
\fmffixed{(0.25w,0)}{v6,v7}
\fmffixed{(0.25w,0)}{v7,v8}
\fmffixed{(0.25w,0)}{v8,v10}
\fmffixed{(0,whatever)}{vc1,vc2}
\fmffixed{(0,whatever)}{vc3,vc4}
\fmffixed{(0,whatever)}{vc5,vc6}
\fmffixed{(0,whatever)}{vc7,vc8}
\fmf{plain,tension=0.25,right=0.25}{v1,vc1}
\fmf{plain,tension=0.25,left=0.25}{v2,vc1}
\fmf{plain,left=0.25}{v5,vc2}
\fmf{plain,tension=1,left=0.25}{v3,vc3}
\fmf{plain,tension=1,left=0.25}{v4,vc5}
\fmf{plain,tension=1,left=0.25}{v9,vc7}
\fmf{plain,left=0.25}{v7,vc6}
\fmf{plain,tension=0.25,left=0.25}{v8,vc8}
\fmf{plain,tension=0.25,right=0.25}{v10,vc8}
\fmf{plain,left=0.25}{v6,vc4}
\fmf{plain,tension=0.5}{vc1,vc2}
\fmf{plain,tension=0.5}{vc2,vc3}
\fmf{plain,tension=0.5}{vc3,vc4}
\fmf{plain,tension=0.5}{vc4,vc5}
\fmf{plain,tension=0.5}{vc5,vc6}
\fmf{plain,tension=0.5}{vc6,vc7}
\fmf{plain,tension=0.5}{vc7,vc8}
\fmf{plain,tension=0.5,right=0,width=1mm}{v5,v10}
\fmfposition
\end{fmfchar*}}}
}
\subfigspace
\subfigure[$\nwgraph{4}{}{3,2,1,4}$]{
\raisebox{\eqoff}{%
\fmfframe(3,1)(1,4){%
\begin{fmfchar*}(20,20)
\fmftop{v1}
\fmfbottom{v5}
\fmfforce{(0w,h)}{v1}
\fmfforce{(0w,0)}{v5}
\fmffixed{(0.25w,0)}{v1,v2}
\fmffixed{(0.25w,0)}{v2,v3}
\fmffixed{(0.25w,0)}{v3,v4}
\fmffixed{(0.25w,0)}{v4,v9}
\fmffixed{(0.25w,0)}{v5,v6}
\fmffixed{(0.25w,0)}{v6,v7}
\fmffixed{(0.25w,0)}{v7,v8}
\fmffixed{(0.25w,0)}{v8,v10}
\fmffixed{(0,whatever)}{vb1,vc2}
\fmffixed{(0,whatever)}{vc3,vb4}
\fmffixed{(0,whatever)}{vb7,vc8}
\fmf{plain,tension=0.25,left=0.25}{v5,vc2}
\fmf{plain,tension=0.25,right=0.25}{v6,vc2}
\fmf{phantom,tension=0.25,right=0.25}{v1,vb1}
\fmf{phantom,tension=0.25,left=0.25}{v2,vb1}
\fmf{phantom,tension=0.5}{vc2,vb1}
\fmf{plain,tension=0.25,left=0.25}{v8,vc8}
\fmf{plain,tension=0.25,right=0.25}{v10,vc8}
\fmf{phantom,tension=0.25,right=0.25}{v4,vb7}
\fmf{phantom,tension=0.25,left=0.25}{v9,vb7}
\fmf{phantom,tension=0.5}{vb7,vc8}
\fmf{plain,tension=0.25,right=0.25}{v3,vc3}
\fmf{plain,tension=0.25,left=0.25}{v4,vc3}
\fmf{phantom,tension=0.25,left=0.25}{v7,vb4}
\fmf{phantom,tension=0.25,right=0.25}{v8,vb4}
\fmf{phantom,tension=0.2}{vc3,vb4}
\fmffreeze
\fmffixed{(0,whatever)}{vc1,vc2}
\fmffixed{(0,whatever)}{vc3,vc4}
\fmffixed{(0,whatever)}{vc5,vc6}
\fmffixed{(0,whatever)}{vc7,vc8}
\fmffixed{(whatever,0)}{vc7,vc5}
\fmf{plain,tension=0.5,right=0.25}{v1,vc1}
\fmf{plain,tension=0.15,right=0.25}{v2,vc5}
\fmf{plain,tension=0.5,left=0.25}{v9,vc7}
\fmf{plain,tension=0.25}{vc3,vc4}
\fmf{plain,tension=0.5}{vc1,vc2}
\fmf{plain,tension=0.5}{vc6,vc1}
\fmf{plain,tension=0.5}{vc7,vc4}
\fmf{plain,tension=0.5}{vc4,vc5}
\fmf{plain,tension=0.5}{vc5,vc6}
\fmf{plain,tension=0.5,left=0.25}{vc6,v7}
\fmf{plain,tension=1}{vc7,vc8}
\fmf{plain,tension=0.5,right=0,width=1mm}{v5,v10}
\fmfposition
\fmfipath{p[]}
\end{fmfchar*}}}
}
\subfigspace
\subfigure[$\nwgraph{4}{}{1,4,3,2}$]{
\raisebox{\eqoff}{%
\fmfframe(3,1)(1,4){%
\begin{fmfchar*}(20,20)
\fmftop{v1}
\fmfbottom{v5}
\fmfforce{(0w,h)}{v1}
\fmfforce{(0w,0)}{v5}
\fmffixed{(0.25w,0)}{v1,v2}
\fmffixed{(0.25w,0)}{v2,v3}
\fmffixed{(0.25w,0)}{v3,v4}
\fmffixed{(0.25w,0)}{v4,v9}
\fmffixed{(0.25w,0)}{v5,v6}
\fmffixed{(0.25w,0)}{v6,v7}
\fmffixed{(0.25w,0)}{v7,v8}
\fmffixed{(0.25w,0)}{v8,v10}
\fmffixed{(0,whatever)}{vb1,vc1}
\fmffixed{(0,whatever)}{vb2,vc2}
\fmffixed{(0,whatever)}{vb7,vc7}
\fmf{plain,tension=0.25,right=0.25}{v1,vc1}
\fmf{plain,tension=0.25,left=0.25}{v2,vc1}
\fmf{phantom,tension=0.25,right=0.25}{v5,vb1}
\fmf{phantom,tension=0.25,left=0.25}{v6,vb1}
\fmf{phantom,tension=0.5}{vc1,vb1}
\fmf{plain,tension=0.25,left=0.25}{v6,vc2}
\fmf{plain,tension=0.25,right=0.25}{v7,vc2}
\fmf{phantom,tension=0.25,right=0.25}{v2,vb2}
\fmf{phantom,tension=0.25,left=0.25}{v3,vb2}
\fmf{phantom,tension=0.5}{vc2,vb2}
\fmf{plain,tension=0.25,right=0.25}{v4,vc7}
\fmf{plain,tension=0.25,left=0.25}{v9,vc7}
\fmf{phantom,tension=0.25,right=0.25}{v8,vb7}
\fmf{phantom,tension=0.25,left=0.25}{v10,vb7}
\fmf{phantom,tension=0.5}{vc7,vb7}
\fmffreeze
\fmffixed{(0,whatever)}{vc1,vc3}
\fmffixed{(0,whatever)}{vc2,vc4}
\fmffixed{(0,whatever)}{vc5,vc6}
\fmffixed{(0,whatever)}{vc7,vc8}
\fmffixed{(whatever,0)}{vc3,vc5}
\fmf{plain,tension=1}{vc1,vc3}
\fmf{plain,tension=0.25,left=0.25}{v5,vc3}
\fmf{plain,tension=0.5}{vc3,vc4}
\fmf{plain,tension=0.5}{vc2,vc4}
\fmf{plain,tension=0.5}{vc4,vc5}
\fmf{plain,tension=0.25,left=0.25}{vc5,v8}
\fmf{plain,tension=1}{vc5,vc6}
\fmf{plain,tension=1,right=0.25}{v3,vc6}
\fmf{plain,tension=0.5,right=0.25}{v10,vc8}
\fmf{plain,tension=0.5}{vc6,vc8}
\fmf{plain,tension=0.5}{vc7,vc8}
\fmf{plain,tension=0.5,right=0,width=1mm}{v5,v10}
\fmfposition
\fmfipath{p[]}
\end{fmfchar*}}}
}
\subfigspace
\subfigure[$\nwgraph{4}{}{2,4,1,3}$]{
\raisebox{\eqoff}{%
\fmfframe(3,1)(1,4){%
\begin{fmfchar*}(20,20)
\fmftop{v1}
\fmfbottom{v5}
\fmfforce{(0w,h)}{v1}
\fmfforce{(0w,0)}{v5}
\fmffixed{(0.25w,0)}{v1,v2}
\fmffixed{(0.25w,0)}{v2,v3}
\fmffixed{(0.25w,0)}{v3,v4}
\fmffixed{(0.25w,0)}{v4,v9}
\fmffixed{(0.25w,0)}{v5,v6}
\fmffixed{(0.25w,0)}{v6,v7}
\fmffixed{(0.25w,0)}{v7,v8}
\fmffixed{(0.25w,0)}{v8,v10}
\fmffixed{(0,whatever)}{vb2,vc2}
\fmffixed{(0,whatever)}{vb3,vc3}
\fmffixed{(0,whatever)}{vb5,vc5}
\fmffixed{(0,whatever)}{vb7,vc7}
\fmf{plain,tension=0.25,left=0.25}{v5,vc2}
\fmf{plain,tension=0.25,right=0.25}{v6,vc2}
\fmf{phantom,tension=0.25,right=0.25}{v1,vb2}
\fmf{phantom,tension=0.25,left=0.25}{v2,vb2}
\fmf{phantom,tension=0.5}{vc2,vb2}
\fmf{plain,tension=0.25,right=0.25}{v2,vc3}
\fmf{plain,tension=0.25,left=0.25}{v3,vc3}
\fmf{phantom,tension=0.25,right=0.25}{v6,vb3}
\fmf{phantom,tension=0.25,left=0.25}{v7,vb3}
\fmf{phantom,tension=0.5}{vc3,vb3}
\fmf{plain,tension=0.25,left=0.25}{v7,vc5}
\fmf{plain,tension=0.25,right=0.25}{v8,vc5}
\fmf{phantom,tension=0.25,right=0.25}{v3,vb5}
\fmf{phantom,tension=0.25,left=0.25}{v4,vb5}
\fmf{phantom,tension=0.5}{vc5,vb5}
\fmf{plain,tension=0.25,right=0.25}{v4,vc7}
\fmf{plain,tension=0.25,left=0.25}{v9,vc7}
\fmf{phantom,tension=0.25,right=0.25}{v8,vb7}
\fmf{phantom,tension=0.25,left=0.25}{v10,vb7}
\fmf{phantom,tension=0.5}{vc7,vb7}
\fmffreeze
\fmffixed{(0,whatever)}{vc1,vc2}
\fmffixed{(0,whatever)}{vc3,vc4}
\fmffixed{(0,whatever)}{vc5,vc6}
\fmffixed{(0,whatever)}{vc7,vc8}
\fmffixed{(whatever,0)}{vc1,vc6}
\fmffixed{(whatever,0)}{vc4,vc8}
\fmf{plain,tension=0.25,right=0.25}{v1,vc1}
\fmf{plain,tension=0.5}{vc1,vc2}
\fmf{plain,tension=0.5}{vc3,vc4}
\fmf{plain,tension=0.5}{vc1,vc4}
\fmf{plain,tension=0.5}{vc5,vc6}
\fmf{plain,tension=0.5}{vc4,vc6}
\fmf{plain,tension=0.25,right=0.25}{v10,vc8}
\fmf{plain,tension=0.5}{vc7,vc8}
\fmf{plain,tension=0.5}{vc6,vc8}
\fmf{plain,tension=0.5,right=0,width=1mm}{v5,v10}
\fmfposition
\fmfipath{p[]}
\end{fmfchar*}}}
}
\subfigspace
\subfigure[$\nwgraph{4}{}{2,1,3,2}$]{
\raisebox{\eqoff}{%
\fmfframe(3,1)(1,4){%
\begin{fmfchar*}(20,20)
\fmftop{v1}
\fmfbottom{v5}
\fmfforce{(0.125w,h)}{v1}
\fmfforce{(0.125w,0)}{v5}
\fmffixed{(0.25w,0)}{v1,v2}
\fmffixed{(0.25w,0)}{v2,v3}
\fmffixed{(0.25w,0)}{v3,v4}
\fmffixed{(0.25w,0)}{v5,v6}
\fmffixed{(0.25w,0)}{v6,v7}
\fmffixed{(0.25w,0)}{v7,v8}
\fmffixed{(0,whatever)}{vc1,vc5}
\fmffixed{(0,whatever)}{vc2,vc3}
\fmffixed{(0,whatever)}{vc3,vc6}
\fmffixed{(0,whatever)}{vc6,vc7}
\fmffixed{(0,whatever)}{vc4,vc8}
\fmffixed{(0.5w,0)}{vc1,vc4}
\fmffixed{(0.5w,0)}{vc5,vc8}
\fmf{plain,tension=1,right=0.125}{v1,vc1}
\fmf{plain,tension=0.25,right=0.25}{v2,vc2}
\fmf{plain,tension=0.25,left=0.25}{v3,vc2}
\fmf{plain,tension=1,left=0.125}{v4,vc4}
\fmf{plain,tension=1,left=0.125}{v5,vc5}
\fmf{plain,tension=0.25,left=0.25}{v6,vc6}
\fmf{plain,tension=0.25,right=0.25}{v7,vc6}
\fmf{plain,tension=1,right=0.125}{v8,vc8}
\fmf{plain,tension=0.5}{vc1,vc3}
\fmf{plain,tension=0.5}{vc2,vc3}
\fmf{plain,tension=0.5}{vc3,vc4}
\fmf{plain,tension=0.5}{vc5,vc7}
\fmf{plain,tension=0.5}{vc6,vc7}
\fmf{plain,tension=0.5}{vc7,vc8}
\fmf{plain,tension=2}{vc1,vc5}
\fmf{plain,tension=2}{vc4,vc8}
\fmf{phantom,tension=2}{vc5,vc4}
\fmffreeze
\fmfposition
\fmf{plain,tension=1,left=0,width=1mm}{v5,v8}
\fmffreeze
\end{fmfchar*}}}
}
\\[0.2cm]
\begin{tabular}{cm{12cm}}
\toprule
$\nwgraph{4}{}{1,2,3,4}\rightarrow \cf{4} \jint{4}{1}\,\chi(1,2,3,4) $ \qquad
$\nwgraph{4}{}{3,2,1,4}\rightarrow \cf{4} \jint{4}{2}\,\chi(3,2,1,4) $ \\
$\nwgraph{4}{}{1,4,3,2}\rightarrow \cf{4} \jint{4}{3}\,\chi(1,4,3,2) $ \qquad
$\nwgraph{4}{}{2,4,1,3}\rightarrow \cf{4} \jint{4}{4}\,\chi(2,4,1,3) $ \\
$\nwgraph{4}{}{2,1,3,2}\rightarrow \cf{4} \jint{4}{5}\,\chi(2,1,3,2) $ \\
\bottomrule
\end{tabular}
\normalsize
\caption{Diagrams with only chiral interactions}
\label{r5diagrams-chiral}
\end{figure}

\begin{figure}[h]
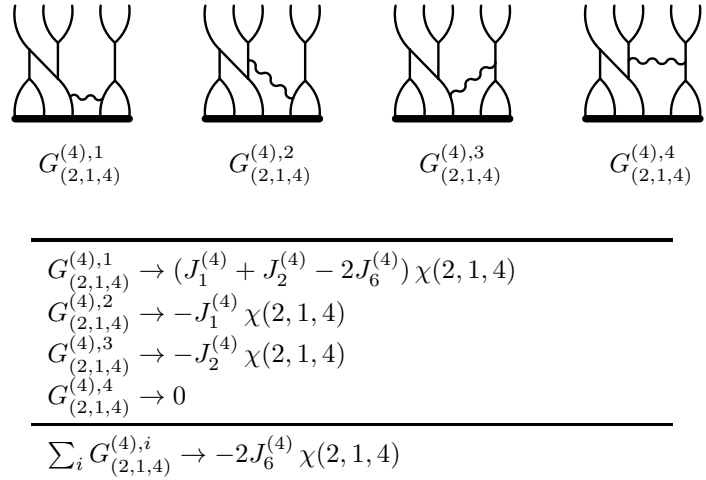

\capstart
\renewcommand*{\thesubfigure}{\ \ }
\footnotesize
\unitlength=0.75mm
\settoheight{\eqoff}{$\times$}%
\setlength{\eqoff}{0.5\eqoff}%
\addtolength{\eqoff}{-12.5\unitlength}%
\settoheight{\eqofftwo}{$\times$}%
\setlength{\eqofftwo}{0.5\eqofftwo}%
\addtolength{\eqofftwo}{-7.5\unitlength}%
\subfigure[$\nwgraph{4}{1}{2,1,4}$]{
\raisebox{\eqoff}{%
\fmfframe(3,1)(1,4){%
\begin{fmfchar*}(20,20)
\Wsevenplain
\fmfipair{w[]}
\svertex{w1}{p10}
\svertex{w2}{p6}
\fmfi{wiggly}{w1..w2}
\end{fmfchar*}}}
}
\subfigspace
\subfigure[$\nwgraph{4}{2}{2,1,4}$]{
\raisebox{\eqoff}{%
\fmfframe(3,1)(1,4){%
\begin{fmfchar*}(20,20)
\Wsevenplain
\fmfipair{w[]}
\svertex{w1}{p10}
\svertex{w2}{p5}
\fmfi{wiggly}{w1..w2}
\end{fmfchar*}}}
}
\subfigspace
\subfigure[$\nwgraph{4}{3}{2,1,4}$]{
\raisebox{\eqoff}{%
\fmfframe(3,1)(1,4){%
\begin{fmfchar*}(20,20)
\Wsevenplain
\fmfipair{w[]}
\svertex{w1}{p9}
\svertex{w2}{p6}
\fmfi{wiggly}{w1..w2}
\end{fmfchar*}}}
}
\subfigspace
\subfigure[$\nwgraph{4}{4}{2,1,4}$]{
\raisebox{\eqoff}{%
\fmfframe(3,1)(1,4){%
\begin{fmfchar*}(20,20)
\Wsevenplain
\fmfipair{w[]}
\svertex{w1}{p9}
\svertex{w2}{p5}
\fmfi{wiggly}{w1..w2}
\end{fmfchar*}}}
}
\\[0.5cm]
\begin{tabular}{m{8cm}}
\toprule
$\nwgraph{4}{1}{2,1,4}\rightarrow \cf{4} (\jint{4}{1}+\jint{4}{2}-2\jint{4}{6})\,\chi(2,1,4) $ \\
$\nwgraph{4}{2}{2,1,4}\rightarrow -\cf{4} \jint{4}{1}\,\chi(2,1,4) $ \\
$\nwgraph{4}{3}{2,1,4}\rightarrow -\cf{4} \jint{4}{2}\,\chi(2,1,4) $ \\
$\nwgraph{4}{4}{2,1,4}\rightarrow 0 $ \\
\midrule
$\sum_i \nwgraph{4}{i}{2,1,4}\rightarrow -2\cf{4} \jint{4}{6}\,\chi(2,1,4) $   \\
\bottomrule
\end{tabular}
\normalsize
\caption{Diagrams with structure $\chi(2,1,4)$}
\label{r5diagrams-214}
\end{figure}

The coefficient of each chiral structure in the four-loop asymptotic dilatation operator~\eqref{D4} can be computed directly from the Feynman supergraphs with that particular structure. More precisely, the coefficient $\mathcal{C}_4(\chi(\cdots))$ of the structure $\chi(\ldots)$ will be equal to the sum of the coefficients of the $1/\varepsilon$ poles of the diagrams with chiral structure $\chi(\ldots)$, multiplied by $(-8)$, according to the definition of the anomalous dimension~\eqref{anomalous}. 
Every structure must be completed to a planar, connected four-loop graph by adding the right number of vector propagators in all the possible ways.
Therefore, the number of diagrams grows with the number of vector interactions, and the coefficient of the five completely chiral structures are the easiest to calculate. Apart from this class, the next simplest computations involve the three disconnected structures with range five, which need a vector to produce connected diagrams. The independent structures of the first kind are shown in Figure~\ref{r5diagrams-chiral}, whereas those of the second kind are given in Figures~\ref{r5diagrams-214}, \ref{r5diagrams-143} and~\ref{r5diagrams-14}. When the underlying chiral structure is symmetric under parity, the symmetry coefficient $\scgraph{4}{i}{\ldots}$ of every diagram is given if different from 1.

For all the diagrams, the result of D-algebra is given in terms of standard four-loop momentum-space integrals $J_{i}^{(4)}$, whose values are listed in Table~\ref{table:integrals-r5}. The D-algebra always produces a $(g^2 N)^4$ colour factor, which can be combined with the $1/(4\pi)^8$ from the integrals to build the required power of the 't~Hooft coupling $\lambda^4$. Hence, in order not to clutter the results with such coefficients, they will not be shown explicitly.

A total of eight conditions on the coefficients of the dilatation operator, listed in Table~\ref{table:fourloop}, are found. Some of these relations allow to find the values of $\epsilon_{3a}$, $\epsilon_{3b}$ and $\epsilon_{3c}$ and to reproduce the known value for $\beta$. All the remaining constraints reduce to identities that are fulfilled, thus constituting non-trivial checks on the whole procedure. 
\begin{table}[h]
\capstart
\begin{tabular}{m{12cm}}
\toprule
$\mathcal{C}_4(\chi(1,2,3,4))=-10$ \\
$\mathcal{C}_4(\chi(3,2,1,4))+\mathcal{C}_4(\chi(1,4,3,2))=-2(8+\epsilon_{3a})=0\Rightarrow \epsilon_{3a}=-4$ \\
$\mathcal{C}_4(\chi(3,2,1,4))-\mathcal{C}_4(\chi(1,4,3,2))=4i\epsilon_{3b}=16/3\Rightarrow \epsilon_{3b}=-i4/3$ \\
$\mathcal{C}_4(\chi(2,4,1,3))=18+4\epsilon_{3a}=2$ \\
$\mathcal{C}_4(\chi(2,1,3,2))=-(12+2\beta+4\epsilon_{3a})=4-8\zeta(3)\Rightarrow \beta=4\zeta(3)$ \\
$\mathcal{C}_4(\chi(2,1,4))+\mathcal{C}_4(\chi(1,4,3))=-8$ \\
$\mathcal{C}_4(\chi(2,1,4))-\mathcal{C}_4(\chi(1,4,3))=8i\epsilon_{3b}+4i\epsilon_{3c}=16/3\Rightarrow \epsilon_{3c}=i4/3$ \\
$\mathcal{C}_4(\chi(1,4))=-4$ \\
\bottomrule
\end{tabular}
\caption{Conditions on the coefficients of the dilatation operator}
\label{table:fourloop}
\end{table}

In particular, the values of the three $\epsilon_{3x}$ coefficients are
\begin{equation}
\epsilon_{3a}=-4\col\qquad
\epsilon_{3b}=-i\frac{4}{3}\col\qquad
\epsilon_{3c}=i\frac{4}{3}\pnt
\end{equation}
A simple three-loop computation allows to determine also the coefficient $\epsilon_{2a}$, which enters the three-loop operator $\mathcal{D}_3$. In fact, from the expression for the three-loop dilatation operator given in~\eqref{Duptothree}, and taking~\eqref{invchistruc} into account, it is clear that the coefficient of the permutation operator $\{2,1,3\}$ will be equal to the one of $\chi(2,1,3)$ when the operator is written in the basis of chiral structures. This coefficient $\mathcal{C}_3(\chi(2,1,3))$ can be obtained from the contribution of the single diagram shown in Figure~\ref{diagram-threeloop}, multiplied by $(-6)$
\begin{equation}
\mathcal{C}_3(\chi(2,1,3))=4i\epsilon_{2a}=2\Rightarrow\epsilon_{2a}=-\frac{i}{2} \pnt
\end{equation}

\begin{figure}[h]
\capstart
\renewcommand*{\thesubfigure}{\ \ }
\footnotesize
\unitlength=0.75mm
\settoheight{\eqoff}{$\times$}%
\setlength{\eqoff}{0.5\eqoff}%
\addtolength{\eqoff}{-12.5\unitlength}%
\settoheight{\eqofftwo}{$\times$}%
\setlength{\eqofftwo}{0.5\eqofftwo}%
\addtolength{\eqofftwo}{-7.5\unitlength}%
\subfigure[$\nwgraph{4}{1}{1,4,3}$]{
\raisebox{\eqoff}{%
\fmfframe(3,1)(1,4){%
\begin{fmfchar*}(20,20)
\Weightplain
\fmfipair{w[]}
\svertex{w1}{p3}
\svertex{w2}{p10}
\fmfi{wiggly}{w1..w2}
\end{fmfchar*}}}
}
\subfigspace
\subfigure[$\nwgraph{4}{2}{1,4,3}$]{
\raisebox{\eqoff}{%
\fmfframe(3,1)(1,4){%
\begin{fmfchar*}(20,20)
\Weightplain
\fmfipair{w[]}
\svertex{w1}{p3}
\svertex{w2}{p9}
\fmfi{wiggly}{w1..w2}
\end{fmfchar*}}}
}
\subfigspace
\subfigure[$\nwgraph{4}{3}{1,4,3}$]{
\raisebox{\eqoff}{%
\fmfframe(3,1)(1,4){%
\begin{fmfchar*}(20,20)
\Weightplain
\fmfipair{w[]}
\svertex{w1}{p2}
\svertex{w2}{p10}
\fmfi{wiggly}{w1..w2}
\end{fmfchar*}}}
}
\subfigspace
\subfigure[$\nwgraph{4}{4}{1,4,3}$]{
\raisebox{\eqoff}{%
\fmfframe(3,1)(1,4){%
\begin{fmfchar*}(20,20)
\Weightplain
\fmfipair{w[]}
\svertex{w1}{p2}
\svertex{w2}{p9}
\fmfi{wiggly}{w1..w2}
\end{fmfchar*}}}
}
\\[0.5cm]
\begin{tabular}{m{8cm}}
\toprule
$\nwgraph{4}{1}{1,4,3}\rightarrow \cf{4} (\jint{4}{1}+\jint{4}{3}-2\jint{4}{7})\,\chi(1,4,3) $ \\
$\nwgraph{4}{2}{1,4,3}\rightarrow -\cf{4} \jint{4}{1}\,\chi(1,4,3) $ \\
$\nwgraph{4}{3}{1,4,3}\rightarrow -\cf{4} \jint{4}{3}\,\chi(1,4,3) $ \\
$\nwgraph{4}{4}{1,4,3}\rightarrow 0 $ \\
\midrule
$\sum_i \nwgraph{4}{i}{1,4,3}\rightarrow -2\cf{4} \jint{4}{7}\,\chi(1,4,3) $   \\
\bottomrule
\end{tabular}
\normalsize
\caption{Diagrams with structure $\chi(1,4,3)$}
\label{r5diagrams-143}
\end{figure}

\begin{figure}[h]
\capstart
\renewcommand*{\thesubfigure}{\ \ }
\footnotesize
\unitlength=0.75mm
\settoheight{\eqoff}{$\times$}%
\setlength{\eqoff}{0.5\eqoff}%
\addtolength{\eqoff}{-12.5\unitlength}%
\settoheight{\eqofftwo}{$\times$}%
\setlength{\eqofftwo}{0.5\eqofftwo}%
\addtolength{\eqofftwo}{-7.5\unitlength}%
\subfigure[$\nwgraph{4}{1}{1,4}$]{
\raisebox{\eqoff}{%
\fmfframe(3,1)(1,4){%
\begin{fmfchar*}(20,20)
\Wnineplain
\fmfipair{w[]}
\svertex{w1}{p4}
\svertex{w2}{p3}
\svertex{w3}{p7}
\svertex{w5}{p6}
\vvertex{w4}{w3}{p4}
\fmfi{wiggly}{w2..w4}
\fmfi{wiggly}{w3..w4}
\end{fmfchar*}}}
}
\subfigspace
\subfigure[$\nwgraph{4}{2}{1,4}$ $(\times 2)$]{
\raisebox{\eqoff}{%
\fmfframe(3,1)(1,4){%
\begin{fmfchar*}(20,20)
\Wnineplain
\fmfipair{w[]}
\svertex{w1}{p4}
\svertex{w2}{p3}
\svertex{w3}{p6}
\fmfi{wiggly}{w1..w2}
\fmfi{wiggly}{w3..w1}
\end{fmfchar*}}}
}
\subfigspace
\subfigure[$\nwgraph{4}{3}{1,4}$]{
\raisebox{\eqoff}{%
\fmfframe(3,1)(1,4){%
\begin{fmfchar*}(20,20)
\Wnineplain
\fmfipair{w[]}
\svertex{w1}{p4}
\svertex{w2}{p3}
\svertex{w3}{p2}
\svertex{w5}{p6}
\vvertex{w4}{w2}{p4}
\fmfi{wiggly}{w1..w3}
\fmfi{wiggly}{w5..w1}
\end{fmfchar*}}}
}
\\[0.5cm]
\begin{tabular}{m{8cm}}
\toprule
$\nwgraph{4}{1}{1,4}\rightarrow -2\cf{4} (\jint{4}{1}+\jint{4}{8})\,\chi(1,4) $ \\
$\nwgraph{4}{2}{1,4}\rightarrow \cf{4} \jint{4}{1}\,\chi(1,4) $ \\
$\nwgraph{4}{3}{1,4}\rightarrow 0 $ \\
\midrule
$\sum_i \scgraph{4}{i}{1,4} \nwgraph{4}{i}{1,4}\rightarrow -2\cf{4} \jint{4}{8}\,\chi(1,4) $   \\
\bottomrule
\end{tabular}
\normalsize
\caption{Diagrams with structure $\chi(1,4)$}
\label{r5diagrams-14}
\end{figure}

\begin{figure}[h]
\capstart
\renewcommand*{\thesubfigure}{}
\begin{equation*}
\unitlength=0.75mm
\settoheight{\eqoff}{$\times$}%
\setlength{\eqoff}{0.5\eqoff}%
\addtolength{\eqoff}{-12.5\unitlength}%
\settoheight{\eqofftwo}{$\times$}%
\setlength{\eqofftwo}{0.5\eqofftwo}%
\addtolength{\eqofftwo}{-7.5\unitlength}%
\raisebox{\eqoff}{%
\fmfframe(3,1)(1,4){%
\begin{fmfchar*}(20,20)
\Wsixplain
\fmfipair{wu[]}
\fmfipair{w[]}
\fmfipair{wd[]}
\svertex{w3}{p3}
\svertex{w6}{p6}
\end{fmfchar*}}
}
\rightarrow \cf{3} \jint{3}{1}\,\chi(2,1,3)
\end{equation*}
\caption{Three-loop diagram with structure $\chi(2,1,3)$}
\label{diagram-threeloop}
\end{figure}

\clearpage
\section{Integrals}
Table~\ref{table:integrals-r5} shows the divergent parts of the momentum integrals required for the computation of the previous section. The $1/(4\pi)^8$ factor in each diagram has been omitted.

\begin{table}[h]
\capstart
\settoheight{\eqoff}{$\times$}%
\setlength{\eqoff}{0.5\eqoff}%
\addtolength{\eqoff}{-7.5\unitlength}
\begin{equation*}
\begin{aligned}
\jint{4}{1}=
\raisebox{\eqoff}{%
\begin{fmfchar*}(20,15)
\fmfleft{in}
\fmfright{out}
\fmf{plain}{in,v1}
\fmf{plain,left=0.25}{v1,v2}
\fmf{plain,left=0}{v2,v4}
\fmf{plain,left=0.25}{v4,v3}
\fmf{plain,tension=0.5,right=0.25}{v1,v0,v1}
\fmf{plain,right=0.25}{v0,v3}
\fmf{plain}{v0,v2}
\fmf{plain}{v0,v4}
\fmf{plain}{v3,out}
\fmffixed{(0.9w,0)}{v1,v3}
\fmffixed{(0.4w,0)}{v2,v4}
\fmfpoly{phantom}{v4,v2,v0}
\fmffreeze
\end{fmfchar*}}
&=\intf{8}
-\frac{1}{24\varepsilon^4}+\frac{1}{4\varepsilon^3}
-\frac{19}{24\varepsilon^2}
+\frac{5}{4\varepsilon}
\\
\jint{4}{2}=
\raisebox{\eqoff}{%
\begin{fmfchar*}(20,15)
\fmfleft{in}
\fmfright{out}
\fmf{plain}{in,v1}
\fmf{plain,left=0}{v1,v2}
\fmf{plain,left=0}{v2,v4}
\fmf{plain,left=0}{v3,v4}
\fmf{plain,left=0}{v0,v2}
\fmf{plain,tension=0.35,right=0.25}{v1,v0,v1}
\fmf{plain,tension=0.25,right=0.25}{v3,v0,v3}
\fmf{plain}{v4,out}
\fmffixed{(0.9w,0)}{v1,v4}
\fmffixed{(whatever,0.5h)}{v0,v2}
\fmffreeze
\end{fmfchar*}}
&=\intf{8}
-\frac{1}{8\varepsilon^4}+\frac{1}{3\varepsilon^3}
-\frac{5}{24\varepsilon^2}
-\frac{1}{3\varepsilon}
\\
\jint{4}{3}=
\raisebox{\eqoff}{%
\begin{fmfchar*}(20,15)
\fmfleft{in}
\fmfright{out}
\fmf{plain}{in,v1}
\fmf{plain,left=0}{v1,v2}
\fmf{plain,tension=0.35,left=0}{v2,v3}
\fmf{plain,left=0}{v3,v4}
\fmf{plain,left=0}{v0,v1}
\fmf{plain,tension=0.35,left=0}{v0,v2}
\fmf{plain,left=0}{v0,v4}
\fmf{plain,tension=0.35,right=0.25}{v3,v0,v3}
\fmf{plain}{v4,out}
\fmffixed{(0.9w,0)}{v1,v4}
\fmffixed{(whatever,0.5h)}{v0,v3}
\fmffixed{(whatever,0.4h)}{v0,v2}
\fmffreeze
\end{fmfchar*}}
&=\intf{8}
-\frac{1}{8\varepsilon^4}+\frac{1}{2\varepsilon^3}
-\frac{7}{8\varepsilon^2}
+\frac{1}{3\varepsilon}
\\
\jint{4}{4}=
\raisebox{\eqoff}{%
\begin{fmfchar*}(20,15)
\fmfleft{in}
\fmfright{out}
\fmf{plain}{in,v1}
\fmf{plain,tension=0.5,left=0}{v1,v2}
\fmf{plain,tension=0,left=0}{v2,v4}
\fmf{plain,tension=0.5,left=0}{v1,v3}
\fmf{plain,left=0}{v0,v4}
\fmf{plain,tension=0.35,right=0.5}{v3,v0,v3}
\fmf{plain,tension=0.35,right=0.5}{v2,v0,v2}
\fmf{plain}{v4,out}
\fmffixed{(0.9w,0)}{v1,v4}
\fmffixed{(0,0.5h)}{v2,v3}
\fmffixed{(0,whatever)}{v0,v3}
\fmffreeze
\end{fmfchar*}}
&=\intf{8}
-\frac{5}{24\varepsilon^4}+\frac{5}{12\varepsilon^3}
+\frac{1}{24\varepsilon^2}
-\frac{1}{4\varepsilon}
\\
\jint{4}{5}=\raisebox{\eqoff}{%
\begin{fmfchar*}(20,15)
\fmfleft{in}
\fmfright{out}
\fmf{plain}{in,v1}
\fmf{plain,left=0.25}{v1,v2}
\fmf{plain,left=0.25}{v2,v3}
\fmf{plain,left=0.25}{v3,v4}
\fmf{plain,left=0.25}{v4,v1}
\fmf{plain,tension=0.5,right=0.25}{v1,v0,v1}
\fmf{phantom}{v0,v3}
\fmf{plain}{v2,v0}
\fmf{plain}{v0,v4}
\fmf{plain}{v3,out}
\fmffixed{(0.9w,0)}{v1,v3}
\fmffixed{(0,0.45w)}{v4,v2}
\fmffreeze
\end{fmfchar*}}
&=\intf{8}
-\frac{1}{12\varepsilon^4}+\frac{1}{3\varepsilon^3}
-\frac{5}{12\varepsilon^2}
-\frac{1}{\varepsilon}\Big(\frac{1}{2}-\zeta(3)\Big)
\\
\jint{4}{6}=
\settoheight{\eqoff}{$\times$}%
\setlength{\eqoff}{0.5\eqoff}%
\addtolength{\eqoff}{-7.5\unitlength}%
\raisebox{\eqoff}{%
\begin{fmfchar*}(20,15)
\fmfleft{in}
\fmfright{out}
\fmf{plain}{in,v1}
\fmf{phantom,tension=2,left=0.25}{v1,v2}
\fmf{plain,tension=2,left=0.25}{v2,v3}
\fmf{derplain,left=0.25}{v4,v1}
\fmf{plain,left=0.25}{v0,v4}
\fmf{plain,right=0}{v0,v1}
\fmf{plain,right=0.25}{v0,v5}
\fmf{plain,right=0.75}{v4,v5}
\fmf{phantom,right=0}{v3,v0}
\fmf{derplain,right=0.25}{v5,v3}
\fmf{plain}{v3,out}
\fmffixed{(0.9w,0)}{v1,v3}
\fmfpoly{phantom}{v2,v4,v5}
\fmffixed{(0.5w,0)}{v4,v5}
\fmf{plain,tension=0.25,right=0.25}{v2,v0}
\fmf{plain,tension=0.25,right=0.25}{v0,v2}
\fmffreeze
\fmfshift{(0,0.15w)}{in,out,v1,v2,v3,v4,v5,v0}
\end{fmfchar*}}
&=\intf{8}
\frac{1}{12\varepsilon^2}
-\frac{1}{12\varepsilon}
\\
\jint{4}{7}=
\raisebox{\eqoff}{%
\begin{fmfchar*}(20,15)
\fmfleft{in}
\fmfright{out}
\fmf{plain}{in,v1}
\fmf{plain,right=0.25}{v1,v4}
\fmf{plain,right=0.25}{v4,v0}
\fmf{plain,right=0.25}{v0,v1}
\fmf{plain,right=0.25}{v0,v5}
\fmf{derplain,left=0.25}{v6,v4}
\fmf{plain,left=0.25}{v5,v6}
\fmf{plain,left=0.25}{v0,v3}
\fmf{derplain,right=0.25}{v5,v3}
\fmf{plain}{v3,out}
\fmffixed{(0.9w,0)}{v1,v3}
\fmfpoly{phantom}{v0,v4,v6,v5}
\fmffixed{(0.45w,0)}{v4,v5}
\fmf{plain}{v6,v0}
\fmffreeze
\fmfshift{(0,0.15w)}{in,out,v1,v2,v3,v4,v5,v6,v0}
\end{fmfchar*}}
&=\intf{8}
\frac{1}{4\varepsilon^2}
-\frac{5}{12\varepsilon}
\\
\jint{4}{8}
=\raisebox{\eqoff}{%
\begin{fmfchar*}(20,15)
\fmfleft{in}
\fmfright{out}
\fmf{plain}{in,v1}
\fmf{plain,right=0.25}{v1,v4}
\fmf{plain,right=0.25}{v4,v0}
\fmf{derplain,right=0.25}{v0,v1}
\fmf{plain,right=0.25}{v0,v5}
\fmf{plain,right=0.25}{v4,v6}
\fmf{plain,left=0.25}{v5,v6}
\fmf{derplain,left=0.25}{v0,v3}
\fmf{plain,right=0.25}{v5,v3}
\fmf{plain}{v3,out}
\fmffixed{(0.9w,0)}{v1,v3}
\fmfpoly{phantom}{v0,v4,v6,v5}
\fmffixed{(0.45w,0)}{v4,v5}
\fmf{plain}{v6,v0}
\fmffreeze
\fmfshift{(0,0.15w)}{in,out,v1,v2,v3,v4,v5,v6,v0}
\end{fmfchar*}}
&=\intf{8}
-\frac{1}{4\varepsilon}
\\
\jint{3}{1}
=\raisebox{\eqoff}{%
\begin{fmfchar*}(20,15)
\fmfleft{in}
\fmfright{out}
\fmf{plain}{in,v1}
\fmf{plain,tension=0.5,left=0}{v1,v2}
\fmf{phantom,tension=0,left=0}{v2,v4}
\fmf{plain,tension=0.5,left=0}{v1,v3}
\fmf{phantom,left=0}{v0,v4}
\fmf{plain,tension=0.35,right=0.5}{v3,v0,v3}
\fmf{plain,tension=0.35,right=0.5}{v2,v0,v2}
\fmf{phantom}{v4,out}
\fmffixed{(0.9w,0)}{v1,v4}
\fmffixed{(0,0.7h)}{v2,v3}
\fmffixed{(0,whatever)}{v0,v3}
\fmffreeze
\end{fmfchar*}}
&=\intf{6}
\frac{1}{3\varepsilon^3}-\frac{1}{3\varepsilon^2}-\frac{1}{3\varepsilon}
\end{aligned}
\end{equation*}
\caption{Loop integrals from the diagrams of Section~\ref{sec:fourR5}. Arrows of the same type denote contracted momenta.}
\label{table:integrals-r5}
\end{table}

\chapter{Four-loop wrapping diagrams with vectors}
\label{app:fourloopwrap}
In this appendix the four-loop wrapping diagrams with vector interactions, which are needed for the computation of Chapter~\ref{chapter:fourloop}, are listed, together with the results of D-algebra. In Section~\ref{app:fourloopwrapintegrals}, the values of the required momentum integrals are shown. 

\section{Wrapping diagrams and D-algebra}
The first structure that requires one vector is $\chi(1,2,3)$ (Figure~\ref{diagrams-123}).
\begin{figure}[h]
\capstart
\renewcommand*{\thesubfigure}{\ \ }
\addtolength{\subfigcapskip}{5pt}
\footnotesize
\unitlength=0.75mm
\settoheight{\eqoff}{$\times$}%
\setlength{\eqoff}{0.5\eqoff}%
\addtolength{\eqoff}{-12.5\unitlength}%
\settoheight{\eqofftwo}{$\times$}%
\setlength{\eqofftwo}{0.5\eqofftwo}%
\addtolength{\eqofftwo}{-7.5\unitlength}%
\subfigure[$\wgraph{4}{1}{1,2,3}$]{
\fmfframe(3,1)(1,4){%
\begin{fmfchar*}(20,20)
\Wfourplain
\fmfipair{wu[]}
\fmfipair{w[]}
\fmfipair{wd[]}
\svertex{w6}{p6}
\vvertex{w7}{w6}{p3}
\wigglywrap{w7}{v5}{v8}{w6}
\end{fmfchar*}}}
\subfigspace
\subfigure[$\wgraph{4}{2}{1,2,3}$]{
\fmfframe(3,1)(1,4){%
\begin{fmfchar*}(20,20)
\Wfourplain
\fmfipair{wu[]}
\fmfipair{w[]}
\fmfipair{wd[]}
\svertex{w5}{p5}
\svertex{w3}{p3}
\wigglywrap{w3}{v5}{v8}{w5}
\end{fmfchar*}}}
\subfigspace
\subfigure[$\wgraph{4}{3}{1,2,3}$]{
\fmfframe(3,1)(1,4){%
\begin{fmfchar*}(20,20)
\Wfourplain
\fmfipair{wu[]}
\fmfipair{w[]}
\fmfipair{wd[]}
\svertex{w2}{p2}
\svertex{w6}{p6}
\wigglywrap{w2}{v5}{v8}{w6}
\end{fmfchar*}}}
\subfigspace
\subfigure[$\wgraph{4}{4}{1,2,3}$]{
\fmfframe(3,1)(1,4){%
\begin{fmfchar*}(20,20)
\Wfourplain
\fmfipair{wu[]}
\fmfipair{w[]}
\fmfipair{wd[]}
\svertex{w2}{p2}
\svertex{w5}{p5}
\wigglywrap{w2}{v5}{v8}{w5}
\end{fmfchar*}}}
\begin{tabular}{m{14cm}}
\toprule
$\wgraph{4}{1}{1,2,3}\rightarrow -\cf{4}(J_{10}^{(4)}+J_{12}^{(4)}+2J_{14}^{(4)})\chi(1,2,3) \rightarrow \cf{4}(J_{10}^{(4)}+J_{12}^{(4)}+2J_{14}^{(4)})M_4$ \\
$\wgraph{4}{2}{1,2,3}\rightarrow \cf{4}J_{10}^{(4)}\chi(1,2,3) \rightarrow -\cf{4}J_{10}^{(4)} M_4$ \\
$\wgraph{4}{3}{1,2,3}\rightarrow \cf{4}(J_{9}^{(4)}+J_{12}^{(4)}+2J_{16}^{(4)})\chi(1,2,3) \rightarrow -\cf{4}(J_{9}^{(4)}+J_{12}^{(4)}+2J_{16}^{(4)}) M_4$ \\
$\wgraph{4}{4}{1,2,3}\rightarrow -\cf{4}J_{9}^{(4)}\chi(1,2,3) \rightarrow \cf{4} J_{9}^{(4)} M_4$ \\
\midrule
$\sum_i \wgraph{4}{i}{1,2,3}\rightarrow -2\cf{4}(J_{14}^{(4)}-J_{16}^{(4)})\chi(1,2,3) \rightarrow 2\cf{4}(J_{14}^{(4)}-J_{16}^{(4)})M_4$   \\
\bottomrule
\end{tabular}
\normalsize
\caption{Wrapping diagrams with chiral structure $\chi(1,2,3)$}
\label{diagrams-123}
\end{figure}
\\
The reflected structure $\chi(3,2,1)$ gives the same contribution. The other two structures that need a single vector are $\chi(1,3,2)$ (Figure~\ref{diagrams-132}) and $\chi(3,2,1)$ (Figure~\ref{diagrams-213}), which are symmetric under parity. For such structures, the symmetry factor of each diagram is shown under the graph if different from 1.

\begin{figure}[!h]
\capstart
\renewcommand*{\thesubfigure}{}
\addtolength{\subfigcapskip}{5pt}
\footnotesize
\unitlength=0.75mm
\settoheight{\eqoff}{$\times$}%
\setlength{\eqoff}{0.5\eqoff}%
\addtolength{\eqoff}{-12.5\unitlength}%
\settoheight{\eqofftwo}{$\times$}%
\setlength{\eqofftwo}{0.5\eqofftwo}%
\addtolength{\eqofftwo}{-7.5\unitlength}%
\subfigure[$\wgraph{4}{1}{1,3,2}$]{
\fmfframe(3,1)(1,4){%
\begin{fmfchar*}(20,20)
\Wfiveplain
\fmfipair{wu[]}
\fmfipair{w[]}
\fmfipair{wd[]}
\svertex{w3}{p3}
\svertex{w6}{p6}
\wigglywrap{w3}{v5}{v8}{w6}
\end{fmfchar*}}}
\subfigspace
\subfigspace
\subfigure[$\wgraph{4}{2}{1,3,2}\!(\!\times2)$]{
\fmfframe(3,1)(1,4){%
\begin{fmfchar*}(20,20)
\Wfiveplain
\fmfipair{wu[]}
\fmfipair{w[]}
\fmfipair{wd[]}
\svertex{w2}{p2}
\svertex{w6}{p6}
\wigglywrap{w2}{v5}{v8}{w6}
\end{fmfchar*}}}
\subfigspace
\subfigspace
\subfigure[$\wgraph{4}{3}{1,3,2}$]{
\fmfframe(3,1)(1,4){%
\begin{fmfchar*}(20,20)
\Wfiveplain
\fmfipair{wu[]}
\fmfipair{w[]}
\fmfipair{wd[]}
\svertex{w2}{p2}
\svertex{w5}{p5}
\wigglywrap{w2}{v5}{v8}{w5}
\end{fmfchar*}}}
\\[0.2cm]
\begin{tabular}{m{13cm}}
\toprule
$\wgraph{4}{1}{1,3,2}\rightarrow -\cf{4}(2J_{9}^{(4)}+2J_{13}^{(4)})\chi(1,3,2) \rightarrow 4\cf{4}(J_{9}^{(4)}+J_{13}^{(4)})M_4$ \\
$\wgraph{4}{2}{1,3,2}\rightarrow \cf{4} J_{9}^{(4)}\chi(1,3,2) \rightarrow -2\cf{4} J_{9}^{(4)} M_4$ \\
$\wgraph{4}{3}{1,3,2}\rightarrow 0 $ \\
\midrule
$\sum_i \scgraph{4}{i}{1,3,2}\wgraph{4}{i}{1,3,2}\rightarrow -2\cf{4} J_{13}^{(4)} \chi(1,3,2) \rightarrow 4\cf{4} J_{13}^{(4)} M_4$   \\
\bottomrule
\end{tabular}
\normalsize
\caption{Wrapping diagrams with chiral structure $\chi(1,3,2)$}
\label{diagrams-132}
\end{figure}

\begin{figure}[!h]
\capstart
\renewcommand*{\thesubfigure}{}
\addtolength{\subfigcapskip}{5pt}
\footnotesize
\unitlength=0.75mm
\settoheight{\eqoff}{$\times$}%
\setlength{\eqoff}{0.5\eqoff}%
\addtolength{\eqoff}{-12.5\unitlength}%
\settoheight{\eqofftwo}{$\times$}%
\setlength{\eqofftwo}{0.5\eqofftwo}%
\addtolength{\eqofftwo}{-7.5\unitlength}%
\subfigure[$\wgraph{4}{1}{2,1,3}$]{
\fmfframe(3,1)(1,4){%
\begin{fmfchar*}(20,20)
\Wsixplain
\fmfipair{wu[]}
\fmfipair{w[]}
\fmfipair{wd[]}
\svertex{w3}{p3}
\svertex{w6}{p6}
\wigglywrap{w3}{v5}{v8}{w6}
\end{fmfchar*}}}
\subfigspace
\subfigspace
\subfigure[$\wgraph{4}{2}{2,1,3}\!(\!\times2)$]{
\fmfframe(3,1)(1,4){%
\begin{fmfchar*}(20,20)
\Wsixplain
\fmfipair{wu[]}
\fmfipair{w[]}
\fmfipair{wd[]}
\svertex{w2}{p2}
\svertex{w6}{p6}
\wigglywrap{w2}{v5}{v8}{w6}
\end{fmfchar*}}}
\subfigspace
\subfigspace
\subfigure[$\wgraph{4}{3}{2,1,3}$]{
\fmfframe(3,1)(1,4){%
\begin{fmfchar*}(20,20)
\Wsixplain
\fmfipair{wu[]}
\fmfipair{w[]}
\fmfipair{wd[]}
\svertex{w2}{p2}
\svertex{w5}{p5}
\wigglywrap{w2}{v5}{v8}{w5}
\end{fmfchar*}}}
\\[0.2cm]
\begin{tabular}{m{12cm}}
\toprule
$\wgraph{4}{1}{2,1,3}\rightarrow -2\cf{4}(J_{9}^{(4)}+J_{15}^{(4)})\chi(2,1,3) \rightarrow 4\cf{4}(J_{9}^{(4)}+J_{15}^{(4)})M_4$ \\
$\wgraph{4}{2}{2,1,3}\rightarrow \cf{4} J_{9}^{(4)} \chi(1,3,2) \rightarrow -2\cf{4} J_{9}^{(4)} M_4$ \\
$\wgraph{4}{3}{2,1,3}\rightarrow 0 $ \\
\midrule
$\sum_i \scgraph{4}{i}{2,1,3}\wgraph{4}{i}{2,1,3}\rightarrow -2\cf{4} J_{15}^{(4)} \chi(2,1,3) \rightarrow 4\cf{4} J_{15}^{(4)} M_4$   \\
\bottomrule
\end{tabular}
\normalsize
\caption{Wrapping diagrams with chiral structure $\chi(2,1,3)$}
\label{diagrams-213}
\end{figure}

\begin{figure}[!h]
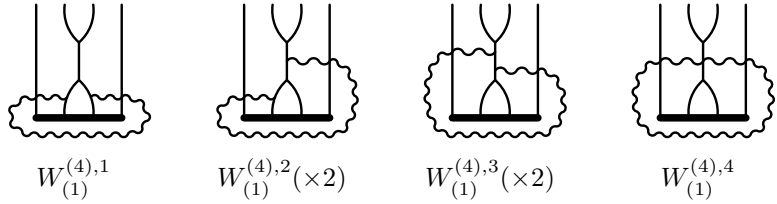

\capstart
\renewcommand*{\thesubfigure}{}
\addtolength{\subfigcapskip}{5pt}
\footnotesize
\unitlength=0.75mm
\settoheight{\eqoff}{$\times$}%
\setlength{\eqoff}{0.5\eqoff}%
\addtolength{\eqoff}{-12.5\unitlength}%
\settoheight{\eqofftwo}{$\times$}%
\setlength{\eqofftwo}{0.5\eqofftwo}%
\addtolength{\eqofftwo}{-7.5\unitlength}%
\subfigure[$\wgraph{4}{1}{2,1}$]{
\fmfframe(3,1)(1,4)
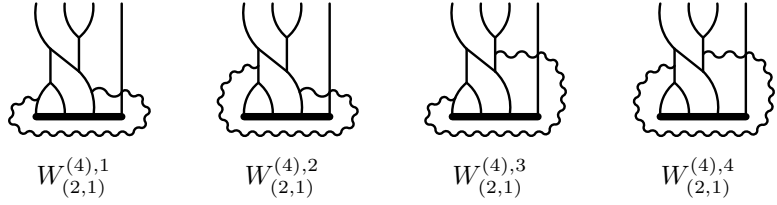
{%
\begin{fmfchar*}(20,20)
\Wthreeplain
\fmfipair{wu[]}
\fmfipair{w[]}
\fmfipair{wd[]}
\svertex{w3}{p3}
\svertex{w6}{p6}
\vvertex{w8}{w6}{p7}
\fmfi{wiggly}{w6..w8}
\wigglywrap{w3}{v5}{v8}{w8}
\end{fmfchar*}}}
\subfigspace
\subfigspace
\subfigure[$\wgraph{4}{2}{2,1}$]{
\fmfframe(3,1)(1,4){%
\begin{fmfchar*}(20,20)
\Wthreeplain
\fmfipair{wu[]}
\fmfipair{w[]}
\fmfipair{wd[]}
\svertex{w2}{p2}
\svertex{w6}{p6}
\vvertex{w8}{w6}{p7}
\fmfi{wiggly}{w6..w8}
\wigglywrap{w2}{v5}{v8}{w8}
\end{fmfchar*}}}
\subfigspace
\subfigspace
\subfigure[$\wgraph{4}{3}{2,1}$]{
\fmfframe(3,1)(1,4){%
\begin{fmfchar*}(20,20)
\Wthreeplain
\fmfipair{wu[]}
\fmfipair{w[]}
\fmfipair{wd[]}
\svertex{w3}{p3}
\svertex{w5}{p5}
\vvertex{w8}{w5}{p7}
\fmfi{wiggly}{w5..w8}
\wigglywrap{w3}{v5}{v8}{w8}
\end{fmfchar*}}}
\subfigspace
\subfigspace
\subfigure[$\wgraph{4}{4}{2,1}$]{
\fmfframe(3,1)(1,4){%
\begin{fmfchar*}(20,20)
\Wthreeplain
\fmfipair{wu[]}
\fmfipair{w[]}
\fmfipair{wd[]}
\svertex{w2}{p2}
\svertex{w5}{p5}
\vvertex{w8}{w5}{p7}
\fmfi{wiggly}{w5..w8}
\wigglywrap{w2}{v5}{v8}{w8}
\end{fmfchar*}}}
\\[0.2cm]
\begin{tabular}{m{12cm}}
\toprule
$\wgraph{4}{1}{2,1}\rightarrow -\cf{4}(J_{9}^{(4)}+J_{12}^{(4)}+2J_{16}^{(4)})\chi(2,1) \rightarrow -\cf{4}(J_{9}^{(4)}+J_{12}^{(4)}+2J_{16}^{(4)})M_4$ \\
$\wgraph{4}{2}{2,1}\rightarrow \cf{4} J_{9}^{(4)}\chi(2,1) \rightarrow \cf{4} J_{9}^{(4)} M_4$ \\
$\wgraph{4}{3}{2,1}\rightarrow \cf{4} (J_{10}^{(4)}+J_{12}^{(4)}+2J_{14}^{(4)})\chi(2,1) \rightarrow \cf{4} (J_{10}^{(4)}+J_{12}^{(4)}+2J_{14}^{(4)}) M_4$ \\
$\wgraph{4}{4}{2,1}\rightarrow -\cf{4} J_{10}^{(4)} \chi(2,1) \rightarrow -\cf{4} J_{10}^{(4)} M_4$ \\
\midrule
$\sum_i \wgraph{4}{i}{2,1}\rightarrow 2\cf{4} (J_{14}^{(4)}-J_{16}^{(4)}) \chi(2,1) \rightarrow 2\cf{4} (J_{14}^{(4)}-J_{16}^{(4)}) M_4$   \\
\bottomrule
\end{tabular}
\normalsize
\caption{Wrapping diagrams with chiral structure $\chi(2,1)$}
\label{diagrams-21}
\end{figure}

There are two structures requiring two vectors, $\chi(2,1)$ (Figure~\ref{diagrams-21}) and $\chi(1,3)$ (Figure~\ref{diagrams-13}), and only one that must be completed with three vectors, $\chi(1)$ (Figure~\ref{diagrams-1}).

\begin{figure}[!h]
\capstart
\renewcommand*{\thesubfigure}{}
\addtolength{\subfigcapskip}{5pt}
\footnotesize
\unitlength=0.75mm
\settoheight{\eqoff}{$\times$}%
\setlength{\eqoff}{0.5\eqoff}%
\addtolength{\eqoff}{-12.5\unitlength}%
\settoheight{\eqofftwo}{$\times$}%
\setlength{\eqofftwo}{0.5\eqofftwo}%
\addtolength{\eqofftwo}{-7.5\unitlength}%
\subfigure[$\wgraph{4}{1}{1}$]{
\fmfframe(3,1)(1,4){%
\begin{fmfchar*}(20,20)
\Woneplain
\fmfipair{wu[]}
\fmfipair{w[]}
\fmfipair{wd[]}
\svertex{w5}{p5}
\svertex{w6}{p6}
\svertex{w1}{p1}
\svertex{w7}{p7}
\vvertex{w8}{w5}{p1}
\vvertex{w9}{w5}{p7}
\fmfi{wiggly}{w8..w5}
\fmfi{wiggly}{w6..w9}
\wigglywrap{w8}{v5}{v8}{w9}
\end{fmfchar*}}}
\subfigspace
\subfigspace
\subfigure[$\wgraph{4}{2}{1}(\times 2)$]{
\fmfframe(3,1)(1,4){%
\begin{fmfchar*}(20,20)
\Woneplain
\fmfipair{wu[]}
\fmfipair{w[]}
\fmfipair{wd[]}
\svertex{w5}{p5}
\svertex{w4}{p4}
\svertex{w3}{p3}
\svertex{w7}{p7}
\vvertex{w8}{w5}{p1}
\vvertex{w9}{w3}{p7}
\fmfi{wiggly}{w8..w5}
\fmfi{wiggly}{w7..w4}
\wigglywrap{w8}{v5}{v8}{w7}
\end{fmfchar*}}}
\subfigspace
\subfigspace
\subfigure[$\wgraph{4}{3}{1}(\times 2)$]{
\fmfframe(3,1)(1,4){%
\begin{fmfchar*}(20,20)
\Woneplain
\fmfipair{wu[]}
\fmfipair{w[]}
\fmfipair{wd[]}
\svertex{w5}{p5}
\svertex{w3}{p3}
\svertex{w4}{p4}
\svertex{w6}{p6}
\svertex{w7}{p7}
\svertex{w6}{p6}
\fmfiequ{wu4}{point length(p4)/4 of p4}
\fmfiequ{wd4}{point 3length(p4)/4 of p4}
\vvertex{w8}{w5}{p1}
\vvertex{w9}{w6}{p7}
\vvertex{w10}{wu4}{p1}
\vvertex{w11}{wd4}{p7}
\fmfi{wiggly}{w10..wu4}
\fmfi{wiggly}{w11..wd4}
\wigglywrap{w10}{v5}{v8}{w11}
\end{fmfchar*}}}
\subfigspace
\subfigspace
\subfigure[$\wgraph{4}{4}{1}$]{
\fmfframe(3,1)(1,4){%
\begin{fmfchar*}(20,20)
\Woneplain
\fmfipair{wu[]}
\fmfipair{w[]}
\fmfipair{wd[]}
\svertex{w2}{p2}
\svertex{w1}{p1}
\svertex{w3}{p3}
\svertex{w4}{p4}
\svertex{w6}{p6}
\svertex{w7}{p7}
\svertex{w6}{p6}
\dvertex{wu4}{wd4}{p4}
\vvertex{w8}{w2}{p1}
\vvertex{w9}{w3}{p7}
\vvertex{w10}{wu4}{p1}
\vvertex{w11}{wd4}{p7}
\fmfi{wiggly}{w4..w7}
\fmfi{wiggly}{w1..w4}
\wigglywrap{w1}{v5}{v8}{w7}
\end{fmfchar*}}}
\\[0.2cm]
\begin{tabular}{m{12cm}}
\toprule
$\wgraph{4}{1}{1}\rightarrow 0 $ \\
$\wgraph{4}{2}{1}\rightarrow 2\cf{4} J_{1}^{(4)}\chi(1) \rightarrow -2\cf{4} J_{1}^{(4)} M_4$ \\
$\wgraph{4}{3}{1}\rightarrow -2\cf{4} J_{9}^{(4)}\chi(1) \rightarrow 2\cf{4} J_{9}^{(4)} M_4$ \\
$\wgraph{4}{4}{1}\rightarrow 0 $ \\
\midrule
$\sum_i \scgraph{4}{i}{1} \wgraph{4}{i}{1}\rightarrow 2\cf{4} (J_{1}^{(4)}-J_{9}^{(4)}) \chi(1) \rightarrow -2\cf{4} (J_{1}^{(4)}-J_{9}^{(4)}) M_4$   \\
\bottomrule
\end{tabular}
\normalsize
\caption{Wrapping diagrams with chiral structure $\chi(1)$}
\label{diagrams-1}
\end{figure}

\begin{figure}[p]
\capstart
\renewcommand*{\thesubfigure}{}
\addtolength{\subfigcapskip}{5pt}
\footnotesize
\unitlength=0.75mm
\settoheight{\eqoff}{$\times$}%
\setlength{\eqoff}{0.5\eqoff}%
\addtolength{\eqoff}{-12.5\unitlength}%
\settoheight{\eqofftwo}{$\times$}%
\setlength{\eqofftwo}{0.5\eqofftwo}%
\addtolength{\eqofftwo}{-7.5\unitlength}%
\subfigure[$\wgraph{4}{1}{1,3}$]{
\fmfframe(3,1)(1,4){%
\begin{fmfchar*}(20,20)
\Wtwoplain
\fmfipair{wu[]}
\fmfipair{w[]}
\fmfipair{wd[]}
\svertex{w5}{p5}
\svertex{w9}{p9}
\svertex{w4}{p4}
\svertex{w10}{p10}
\fmfi{wiggly}{w5..w9}
\wigglywrap{w4}{v5}{v8}{w10}
\end{fmfchar*}}}
\subfigspace
\subfigspace
\subfigure[$\wgraph{4}{2}{1,3}(\times 4)$]{
\fmfframe(3,1)(1,4){%
\begin{fmfchar*}(20,20)
\Wtwoplain
\fmfipair{wu[]}
\fmfipair{w[]}
\fmfipair{wd[]}
\svertex{w5}{p5}
\svertex{w9}{p9}
\svertex{w3}{p3}
\svertex{w10}{p10}
\fmfi{wiggly}{w5..w9}
\wigglywrap{w3}{v5}{v8}{w10}
\end{fmfchar*}}}
\subfigspace
\subfigspace
\subfigure[$\wgraph{4}{3}{1,3}(\times 4)$]{
\fmfframe(3,1)(1,4){%
\begin{fmfchar*}(20,20)
\Wtwoplain
\fmfipair{wu[]}
\fmfipair{w[]}
\fmfipair{wd[]}
\svertex{w5}{p5}
\svertex{w9}{p9}
\svertex{w3}{p3}
\svertex{w8}{p8}
\fmfi{wiggly}{w5..w9}
\wigglywrap{w3}{v5}{v8}{w8}
\end{fmfchar*}}}
\subfigspace
\subfigspace
\subfigure[$\wgraph{4}{4}{1,3}(\times 2)$]{
\fmfframe(3,1)(1,4){%
\begin{fmfchar*}(20,20)
\Wtwoplain
\fmfipair{wu[]}
\fmfipair{w[]}
\fmfipair{wd[]}
\svertex{w5}{p5}
\svertex{w8}{p8}
\svertex{w3}{p3}
\svertex{w10}{p10}
\fmfi{wiggly}{w5..w8}
\wigglywrap{w3}{v5}{v8}{w10}
\end{fmfchar*}}}
\subfigspace
\subfigspace
\subfigure[$\wgraph{4}{5}{1,3}(\times 4)$]{
\fmfframe(3,1)(1,4){%
\begin{fmfchar*}(20,20)
\Wtwoplain
\fmfipair{wu[]}
\fmfipair{w[]}
\fmfipair{wd[]}
\svertex{w5}{p5}
\svertex{w4}{p4}
\fmfiequ{wu8}{point length(p8)/4 of p8}
\fmfiequ{wd8}{point 3length(p8)/4 of p8}
\fmfi{wiggly}{w5..wu8}
\wigglywrap{w4}{v5}{v8}{wd8}
\end{fmfchar*}}}
\\
\vspace{-0.7cm}
\subfigspace
\subfigspace
\subfigure[$\wgraph{4}{6}{1,3}(\times 4)$]{
\fmfframe(3,1)(1,4){%
\begin{fmfchar*}(20,20)
\Wtwoplain
\fmfipair{wu[]}
\fmfipair{w[]}
\fmfipair{wd[]}
\svertex{w5}{p5}
\svertex{w3}{p3}
\fmfiequ{wu8}{point length(p8)/4 of p8}
\fmfiequ{wd8}{point 3length(p8)/4 of p8}
\fmfi{wiggly}{wu8..w5}
\wigglywrap{w3}{v5}{v8}{wd8}
\end{fmfchar*}}}
\subfigspace
\subfigspace
\subfigure[$\wgraph{4}{7}{1,3}(\times 2)$]{
\fmfframe(3,1)(1,4){%
\begin{fmfchar*}(20,20)
\Wtwoplain
\fmfipair{wu[]}
\fmfipair{w[]}
\fmfipair{wd[]}
\svertex{w5}{p5}
\svertex{w8}{p8}
\svertex{w4}{p4}
\fmfi{wiggly}{w5..w8}
\wigglywrap{w4}{v5}{v8}{w8}
\end{fmfchar*}}}
\subfigspace
\subfigspace
\subfigure[$\wgraph{4}{8}{1,3}(\times 4)$]{
\fmfframe(3,1)(1,4){%
\begin{fmfchar*}(20,20)
\Wtwoplain
\fmfipair{wu[]}
\fmfipair{w[]}
\fmfipair{wd[]}
\svertex{w5}{p5}
\svertex{w3}{p3}
\svertex{w8}{p8}
\fmfi{wiggly}{w5..w8}
\wigglywrap{w3}{v5}{v8}{w8}
\end{fmfchar*}}}
\subfigspace
\subfigspace
\subfigure[$\wgraph{4}{9}{1,3}(\times 4)$]{
\fmfframe(3,1)(1,4){%
\begin{fmfchar*}(20,20)
\Wtwoplain
\fmfipair{wu[]}
\fmfipair{w[]}
\fmfipair{wd[]}
\svertex{w5}{p5}
\svertex{w3}{p3}
\dvertex{wu8}{wd8}{p8}
\fmfi{wiggly}{w5..wd8}
\wigglywrap{w3}{v5}{v8}{wu8}
\end{fmfchar*}}}
\subfigspace
\subfigspace
\subfigure[$\wgraph{4}{10}{1,3}(\!\times 4)$]{
\fmfframe(3,1)(1,4){%
\begin{fmfchar*}(20,20)
\Wtwoplain
\fmfipair{wu[]}
\fmfipair{w[]}
\fmfipair{wd[]}
\fmfiequ{wu8}{point length(p8)/4 of p8}
\fmfiequ{wd8}{point 3length(p8)/4 of p8}
\vvertex{w9}{wu8}{p3}
\fmfi{wiggly}{w9..wu8}
\wigglywrap{w9}{v5}{v8}{wd8}
\end{fmfchar*}}}
\\
\vspace{0.1cm}
\subfigure[$\wgraph{4}{11}{1,3}(\!\times 2)$]{
\fmfframe(3,1)(1,4){%
\begin{fmfchar*}(20,20)
\Wtwoplain
\fmfipair{wu[]}
\fmfipair{w[]}
\fmfipair{wd[]}
\fmfiequ{wu8}{point length(p8)/4 of p8}
\fmfiequ{wd8}{point 3length(p8)/4 of p8}
\fmfiequ{wu3}{point length(p8)/4 of p3}
\fmfiequ{wd3}{point 3length(p8)/4 of p3}
\fmfi{wiggly}{wd3..wu8}
\wigglywrap{wu3}{v5}{v8}{wd8}
\end{fmfchar*}}}
\subfigspace
\subfigspace
\subfigure[$\wgraph{4}{12}{1,3}$]{
\fmfframe(3,1)(1,4){%
\begin{fmfchar*}(20,20)
\Wtwoplain
\fmfipair{wu[]}
\fmfipair{w[]}
\fmfipair{wd[]}
\svertex{w3}{p3}
\svertex{w8}{p8}
\fmfi{wiggly}{w3..w8}
\wigglywrap{w3}{v5}{v8}{w8}
\end{fmfchar*}}}
\subfigspace
\subfigspace
\subfigure[$\wgraph{4}{13}{1,3}(\!\times 2)$]{
\fmfframe(3,1)(1,4){%
\begin{fmfchar*}(20,20)
\Wtwoplain
\fmfipair{wu[]}
\fmfipair{w[]}
\fmfipair{wd[]}
\fmfiequ{wu8}{point length(p8)/4 of p8}
\fmfiequ{wd8}{point 3length(p8)/4 of p8}
\fmfiequ{wu3}{point length(p8)/4 of p3}
\fmfiequ{wd3}{point 3length(p8)/4 of p3}
\fmfi{wiggly}{wd3..wd8}
\wigglywrap{wu3}{v5}{v8}{wu8}
\end{fmfchar*}}}
\\[0.2cm]
\begin{tabular}{m{15cm}}
\toprule
$\wgraph{4}{1}{1,3}\rightarrow \cf{4}(J_{10}^{(4)}+J_{12}^{(4)}+4J_{14}^{(4)}+2J_{17}^{(4)})\chi(1,3) \rightarrow 2\cf{4}(J_{10}^{(4)}+J_{12}^{(4)}+4J_{14}^{(4)}+2J_{17}^{(4)})M_4$ \\
$\wgraph{4}{2}{1,3}\rightarrow -\frac{1}{2}\cf{4}(J_{10}^{(4)}+J_{12}^{(4)}+2J_{14}^{(4)})\chi(1,3) \rightarrow -\cf{4}(J_{10}^{(4)}+J_{12}^{(4)}+2J_{14}^{(4)})M_4$ \\
$\wgraph{4}{3}{1,3}\rightarrow 0 $ \\
$\wgraph{4}{4}{1,3}\rightarrow \frac{1}{2}\cf{4} (J_{11}^{(4)}+J_{12}^{(4)}+2J_{18}^{(4)}-4J_{19}^{(4)})\chi(1,3) \rightarrow \cf{4} (J_{11}^{(4)}+J_{12}^{(4)}+2J_{18}^{(4)}-4J_{19}^{(4)})M_4 $ \\
$\wgraph{4}{5}{1,3}\rightarrow \frac{1}{2}\cf{4} J_{10}^{(4)}\chi(1,3) \rightarrow \cf{4} J_{10}^{(4)} M_4 $ \\
$\wgraph{4}{6}{1,3}\rightarrow -\frac{1}{2}\cf{4} J_{11}^{(4)}\chi(1,3) \rightarrow -\cf{4} J_{11}^{(4)} M_4 $ \\
$\wgraph{4}{7}{1,3}\rightarrow -\frac{1}{2}\cf{4} J_{10}^{(4)}\chi(1,3) \rightarrow -\cf{4} J_{10}^{(4)} M_4 $ \\
$\wgraph{4}{8}{1,3}$, $\wgraph{4}{9}{1,3}$, $\wgraph{4}{10}{1,3}\rightarrow 0 $ \\
$\wgraph{4}{11}{1,3}\rightarrow \frac{1}{2}\cf{4} J_{11}^{(4)}\chi(1,3) \rightarrow \cf{4} J_{11}^{(4)} M_4 $ \\
$\wgraph{4}{12}{1,3}\rightarrow 0 $, $\wgraph{4}{13}{1,3}\rightarrow 0 $ \\
\midrule
$\sum_i \scgraph{4}{i}{1,3} \wgraph{4}{i}{1,3}\rightarrow 2\cf{4} (J_{17}^{(4)}+J_{18}^{(4)}-2J_{19}^{(4)}) \chi(1,3) \rightarrow 4\cf{4} (J_{17}^{(4)}+J_{18}^{(4)}-2J_{19}^{(4)}) M_4$   \\
\bottomrule
\end{tabular}
\normalsize
\caption{Wrapping diagrams with chiral structure $\chi(1,3)$}
\label{diagrams-13}
\end{figure}

\clearpage
\section{Integrals}
\label{app:fourloopwrapintegrals}
In this section, the values of the four-loop integrals required for the computation of Chapter~\ref{chapter:fourloop} are listed. One set, shown in Table~\ref{fourloopint-scalar}, contains integrals with trivial numerators, whereas the second one, presented in Table~\ref{fourloopint-der}, is made of integrals with scalar products of momenta in the numerators, denoted by pairs of arrows of the same type.
\begin{table}[h]
\capstart
\settoheight{\eqoff}{$\times$}%
\setlength{\eqoff}{0.5\eqoff}%
\addtolength{\eqoff}{-7.5\unitlength}
\begin{equation*}
\begin{aligned}
J_{9}^{(4)}=\raisebox{\eqoff}{%
\begin{fmfchar*}(20,15)
\fmfleft{in}
\fmfright{out}
\fmf{plain}{in,v1}
\fmf{plain,left=0.25}{v1,v2}
\fmf{plain,left=0.25}{v2,v3}
\fmf{plain,left=0.25}{v3,v4}
\fmf{plain,left=0.25}{v4,v1}
\fmf{plain,tension=0.5,right=0.5}{v2,v0,v2}
\fmf{phantom}{v0,v3}
\fmf{plain}{v1,v0}
\fmf{plain}{v0,v4}
\fmf{plain}{v3,out}
\fmffixed{(0.9w,0)}{v1,v3}
\fmffixed{(0,0.45w)}{v4,v2}
\fmffreeze
\end{fmfchar*}}
&=\intf{8}
-\frac{1}{24\varepsilon^4}+\frac{1}{4\varepsilon^3}
-\frac{19}{24\varepsilon^2}
+\frac{1}{\varepsilon}\Big(\frac{5}{4}-\zeta(3)\Big)
\\
J_{10}^{(4)}=\raisebox{\eqoff}{%
\begin{fmfchar*}(20,15)
\fmfleft{in}
\fmfright{out}
\fmf{plain}{in,v1}
\fmf{plain,left=0.25}{v1,v2}
\fmf{plain,left=0.25}{v2,v3}
\fmf{plain,left=0.25}{v3,v4}
\fmf{plain,left=0.25}{v4,v1}
\fmf{plain,tension=0.5,right=0.25}{v1,v0,v1}
\fmf{phantom}{v0,v3}
\fmf{plain}{v2,v0}
\fmf{plain}{v0,v4}
\fmf{plain}{v3,out}
\fmffixed{(0.9w,0)}{v1,v3}
\fmffixed{(0,0.45w)}{v4,v2}
\fmffreeze
\end{fmfchar*}}
&=\intf{8}
-\frac{1}{12\varepsilon^4}+\frac{1}{3\varepsilon^3}
-\frac{5}{12\varepsilon^2}
-\frac{1}{\varepsilon}\Big(\frac{1}{2}-\zeta(3)\Big)
\\
J_{11}^{(4)}=\raisebox{\eqoff}{%
\begin{fmfchar*}(20,15)
\fmfleft{in}
\fmfright{out}
\fmf{plain}{in,v1}
\fmf{plain,left=0.25}{v1,v2}
\fmf{plain,left=0.25}{v2,v3}
\fmf{plain,left=0.25}{v3,v4}
\fmf{plain,left=0.25}{v4,v1}
\fmf{plain,tension=0.5,right=0.5}{v2,v0,v2}
\fmf{plain,tension=0.5,right=0.5}{v0,v4,v0}
\fmf{plain}{v3,out}
\fmffixed{(0.9w,0)}{v1,v3}
\fmffixed{(0,0.45w)}{v4,v2}
\fmffreeze
\end{fmfchar*}}
&=\intf{8}
-\frac{1}{6\varepsilon^4}+\frac{1}{3\varepsilon^3}
+\frac{1}{3\varepsilon^2}
-\frac{1}{\varepsilon}(1-\zeta(3))
\\
J_{12}^{(4)}=\raisebox{\eqoff}{%
\begin{fmfchar*}(20,15)
\fmfleft{in}
\fmfright{out}
\fmf{plain}{in,v1}
\fmf{plain,left=0.25}{v1,v2}
\fmf{plain,left=0.25}{v2,v3}
\fmf{plain,left=0.25}{v3,v4}
\fmf{plain,left=0.25}{v4,v1}
\fmf{plain}{v2,v4}
\fmf{plain}{v3,v1}
\fmf{plain}{v3,out}
\fmffixed{(0.9w,0)}{v1,v3}
\fmffixed{(0,0.45w)}{v4,v2}
\fmffreeze
\end{fmfchar*}}
&=\intf{8}
\frac{1}{\varepsilon}5\zeta(5)
\end{aligned}
\end{equation*}
\caption{Four-loop integrals with trivial numerators coming from wrapping diagrams.}
\label{fourloopint-scalar}
\end{table}

\begin{table}[h]
\capstart
\settoheight{\eqoff}{$\times$}%
\setlength{\eqoff}{0.5\eqoff}%
\addtolength{\eqoff}{-7.5\unitlength}
\begin{equation*}
\begin{aligned}
J_{13}^{(4)}=\raisebox{\eqoff}{%
\begin{fmfchar*}(20,15)
\fmfleft{in}
\fmfright{out}
\fmf{plain}{in,v1}
\fmf{plain,left=0.25}{v1,v2}
\fmf{plain,left=0.25}{v2,v3}
\fmf{derplain,left=0.25}{v4,v1}
\fmf{plain,right=0.25}{v4,v0}
\fmf{plain,right=0.25}{v0,v5}
\fmf{plain,right=0.75}{v4,v5}
\fmf{derplain,right=0.25}{v5,v3}
\fmf{plain}{v3,out}
\fmffixed{(0.9w,0)}{v1,v3}
\fmfpoly{phantom}{v2,v4,v5}
\fmffixed{(0.5w,0)}{v4,v5}
\fmffixed{(0.5w,0)}{v4,v5}
\fmf{plain,tension=0.25,right=0.25}{v2,v0,v2}
\fmffreeze
\fmfshift{(0,0.1w)}{in,out,v1,v2,v3,v4,v5,v0}
\end{fmfchar*}}
&=\intf{8}
\frac{1}{12\varepsilon^2}
-\frac{7}{12\varepsilon}
\\
J_{14}^{(4)}=\raisebox{\eqoff}{%
\begin{fmfchar*}(20,15)
\fmfleft{in}
\fmfright{out}
\fmf{plain}{in,v1}
\fmf{plain,tension=2,left=0.25}{v1,v2}
\fmf{plain,tension=2,left=0.25}{v2,v3}
\fmf{derplain,left=0.25}{v4,v1}
\fmf{plain,right=0.25}{v4,v0}
\fmf{plain,right=0}{v0,v1}
\fmf{plain,right=0.25}{v0,v5}
\fmf{plain,right=0.75}{v4,v5}
\fmf{phantom,right=0}{v3,v0}
\fmf{derplain,right=0.25}{v5,v3}
\fmf{plain}{v3,out}
\fmffixed{(0.9w,0)}{v1,v3}
\fmfpoly{phantom}{v2,v4,v5}
\fmffixed{(0.5w,0)}{v4,v5}
\fmf{plain,tension=0.5}{v2,v0}
\fmffreeze
\fmfshift{(0,0.15w)}{in,out,v1,v2,v3,v4,v5,v0}
\end{fmfchar*}}
&=
\intf{8}
\frac{1}{\varepsilon}(-\zeta(3))
\\
J_{15}^{(4)}=\raisebox{\eqoff}{%
\begin{fmfchar*}(20,15)
\fmfleft{in}
\fmfright{out}
\fmf{plain}{in,v1}
\fmf{plain,tension=2,left=0.125}{v1,v2c}
\fmf{plain,tension=2,left=0.125}{v2c,v3}
\fmf{plain,tension=1}{v2c,v2}
\fmf{derplain,left=0.25}{v4,v1}
\fmf{plain,right=0.25}{v4,v0}
\fmf{plain,right=0}{v0,v1}
\fmf{plain,right=0.25}{v0,v5}
\fmf{plain,right=0.75}{v4,v5}
\fmf{plain,right=0}{v3,v0}
\fmf{derplain,right=0.25}{v5,v3}
\fmf{phantom}{v3,out}
\fmffixed{(0,0.05w)}{v2c,v2}
\fmffixed{(0.9w,0)}{v1,v3}
\fmfpoly{phantom}{v2c,v4,v5}
\fmffixed{(0.5w,0)}{v4,v5}
\fmffreeze
\fmfshift{(0,0.15w)}{in,out,v1,v2,v2c,v3,v4,v5,v0}
\end{fmfchar*}}
&=
\intf{8}
\frac{1}{4\varepsilon^2}
-\frac{11}{12\varepsilon}
\\
J_{16}^{(4)}=\raisebox{\eqoff}{%
\begin{fmfchar*}(20,15)
\fmfleft{in}
\fmfright{out}
\fmf{plain}{in,v1}
\fmf{derplain,tension=2,right=0.25}{v2,v1}
\fmf{derplain,tension=2,left=0.25}{v2,v3}
\fmf{plain,left=0.25}{v4,v1}
\fmf{plain,right=0.25}{v4,v0}
\fmf{plain,right=0}{v0,v1}
\fmf{plain,right=0.25}{v0,v5}
\fmf{plain,right=0.75}{v4,v5}
\fmf{phantom,right=0}{v3,v0}
\fmf{plain,right=0.25}{v5,v3}
\fmf{plain}{v3,out}
\fmffixed{(0.9w,0)}{v1,v3}
\fmfpoly{phantom}{v2,v4,v5}
\fmffixed{(0.5w,0)}{v4,v5}
\fmf{plain,tension=0.5}{v2,v0}
\fmffreeze
\fmfshift{(0,0.15w)}{in,out,v1,v2,v3,v4,v5,v0}
\end{fmfchar*}}
&=\intf{8}\frac{1}{\varepsilon}\Big(\frac{1}{2}\zeta(3)
-\frac{5}{2}\zeta(5)\Big)
\\
J_{17}^{(4)}=\raisebox{\eqoff}{%
\begin{fmfchar*}(20,15)
\fmfleft{in}
\fmfright{out}
\fmf{plain}{in,v1}
\fmf{derplain,tension=2,right=0.25}{v2,v1}
\fmf{derplain,tension=2,left=0.25}{v3,v4}
\fmf{derplainpt,tension=2,left=0.25}{v5,v1}
\fmf{derplainpt,tension=2,right=0.25}{v6,v4}
\fmf{plain}{v2,v0}
\fmf{plain}{v3,v0}
\fmf{plain}{v5,v0}
\fmf{plain}{v6,v0}
\fmf{plain}{v4,out}
\fmffixed{(0.9w,0)}{v1,v4}
\fmfpoly{phantom}{v3,v2,v5,v6}
\fmf{plain}{v2,v3}
\fmf{plain}{v5,v6}
\fmffixed{(0.4w,0)}{v5,v6}
\fmffreeze
\end{fmfchar*}}
&=
\intf{8}\frac{1}{\varepsilon}
\Big(-\frac{1}{2}-\frac{1}{2}\zeta(3)+\frac{5}{2}\zeta(5)\Big)
\\
J_{18}^{(4)}=\raisebox{\eqoff}{%
\begin{fmfchar*}(20,15)
\fmfleft{in}
\fmfright{out}
\fmf{plain}{in,v1}
\fmf{derplain,tension=2,right=0.25}{v2,v1}
\fmf{derplain,tension=2,left=0.25}{v3,v4}
\fmf{plain,tension=2,right=0.25}{v1,v5}
\fmf{plain,tension=2,right=0.25}{v6,v4}
\fmf{plain}{v2,v0}
\fmf{plain}{v3,v0}
\fmf{plain}{v5,v0}
\fmf{plain}{v0,v6}
\fmf{plain}{v4,out}
\fmffixed{(0.9w,0)}{v1,v4}
\fmfpoly{phantom}{v3,v2,v5,v6}
\fmf{derplainpt}{v2,v3}
\fmf{derplainpt}{v5,v6}
\fmffixed{(0.4w,0)}{v5,v6}
\fmffreeze
\end{fmfchar*}}
&=
\intf{8}\frac{1}{\varepsilon}
\Big(-\frac{1}{4}-\frac{3}{2}\zeta(3)
+\frac{5}{2}\zeta(5)\Big)
\\
J_{19}^{(4)}=\raisebox{\eqoff}{%
\begin{fmfchar*}(20,15)
\fmfleft{in}
\fmfright{out}
\fmf{plain}{in,v1}
\fmf{derplain,tension=2,right=0.25}{v2,v1}
\fmf{plain,tension=2,left=0.25}{v3,v4}
\fmf{plain,tension=2,right=0.25}{v1,v5}
\fmf{derplainpt,tension=2,right=0.25}{v6,v4}
\fmf{plain}{v2,v0}
\fmf{plain}{v3,v0}
\fmf{plain}{v5,v0}
\fmf{plain}{v0,v6}
\fmf{plain}{v4,out}
\fmffixed{(0.9w,0)}{v1,v4}
\fmfpoly{phantom}{v3,v2,v5,v6}
\fmf{derplain}{v3,v2}
\fmf{derplainpt}{v5,v6}
\fmffixed{(0.4w,0)}{v5,v6}
\fmffreeze
\end{fmfchar*}}
&=\intf{8}\frac{1}{\varepsilon}
\Big(-\frac{1}{8}-\frac{1}{4}\zeta(3)
+\frac{5}{4}\zeta(5)\Big)
\end{aligned}
\end{equation*}
\caption{Four-loop integrals with momenta in the numerator.}
\label{fourloopint-der}
\end{table}

\chapter{Range-six five-loop diagrams}
\label{app:fiveR6}
This appendix contains the explicit analysis of several classes of five-loop diagrams with range five and six. The results are used to compute the values of the similarity coefficients that are required in Chapter~\ref{chapter:fiveloop}.

\section{Coefficients from supergraphs}
\label{sec:fiveR6diag}
From~\eqref{D5subM} it follows that the five-loop computation of Chapter~\ref{chapter:fiveloop} requires the explicit values of the similarity coefficients $\epsilon_{4b}$ and $\epsilon_{4f}$ in the chosen renormalization scheme. This calculation is similar to the one performed in Appendix~\ref{app:fourR5} to determine the values of $\epsilon_{3a}$, $\epsilon_{3b}$ and $\epsilon_{3c}$ from four-loop graphs. In this case too, the analysis will not be limited to the minimum number of diagrams that are strictly required to find the needed coefficients. Instead, several simple classes of diagrams will be considered, in order to make some additional checks on the procedure. 

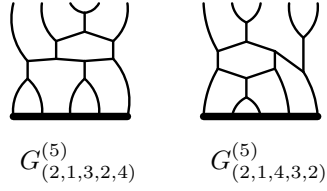
\begin{figure}[t]
\capstart
\renewcommand*{\thesubfigure}{\ \ }
\footnotesize
\centering
\unitlength=0.75mm
\settoheight{\eqoff}{$\times$}%
\setlength{\eqoff}{0.5\eqoff}%
\addtolength{\eqoff}{-12.5\unitlength}%
\settoheight{\eqofftwo}{$\times$}%
\setlength{\eqofftwo}{0.5\eqofftwo}%
\addtolength{\eqofftwo}{-7.5\unitlength}%
\subfigure[$\nwgraph{5}{}{1,2,3,4,5}$]{
\raisebox{\eqoff}{%
\fmfframe(3,1)(1,4){%
\begin{fmfchar*}(30,20)
\fmftop{v1}
\fmfbottom{v7}
\fmfforce{(0w,h)}{v1}
\fmfforce{(0w,0)}{v7}
\fmffixed{(0.2w,0)}{v1,v2}
\fmffixed{(0.2w,0)}{v2,v3}
\fmffixed{(0.2w,0)}{v3,v4}
\fmffixed{(0.2w,0)}{v4,v5}
\fmffixed{(0.2w,0)}{v5,v6}
\fmffixed{(0.2w,0)}{v7,v8}
\fmffixed{(0.2w,0)}{v8,v9}
\fmffixed{(0.2w,0)}{v9,v10}
\fmffixed{(0.2w,0)}{v10,v11}
\fmffixed{(0.2w,0)}{v11,v12}
\fmf{plain,tension=0.25,right=0.25}{v1,vc1}
\fmf{plain,tension=0.25,left=0.25}{v2,vc1}
\fmf{plain,left=0.25}{v7,vc2}
\fmf{plain,tension=1,left=0.25}{v3,vc3}
\fmf{plain,tension=1,left=0.25}{v4,vc5}
\fmf{plain,tension=1,left=0.25}{v5,vc7}
\fmf{plain,tension=1,left=0.25}{v6,vc9}
\fmf{plain,left=0.25}{v9,vc6}
\fmf{plain,tension=0.25,left=0.25}{v11,vc10}
\fmf{plain,tension=0.25,right=0.25}{v12,vc10}
\fmf{plain,left=0.25}{v10,vc8}
\fmf{plain,left=0.25}{v8,vc4}
\fmf{plain,tension=0.5}{vc1,vc2}
\fmf{plain,tension=0.5}{vc2,vc3}
\fmf{plain,tension=0.5}{vc3,vc4}
\fmf{plain,tension=0.5}{vc4,vc5}
\fmf{plain,tension=0.5}{vc5,vc6}
\fmf{plain,tension=0.5}{vc6,vc7}
\fmf{plain,tension=0.5}{vc7,vc8}
\fmf{plain,tension=0.5}{vc8,vc9}
\fmf{plain,tension=0.5}{vc9,vc10}
\fmffreeze
\fmfposition
\fmf{plain,tension=1,right=0,width=1mm}{v7,v12}
\fmffreeze
\end{fmfchar*}}}
}
\subfigspace
\subfigure[$\nwgraph{5}{}{2,1,4,3,5}$]{
\raisebox{\eqoff}{%
\fmfframe(3,1)(1,4){%
\begin{fmfchar*}(30,20)
\fmftop{v7}
\fmfbottom{v1}
\fmfforce{(0w,h)}{v7}
\fmfforce{(0w,0)}{v1}
\fmffixed{(0.2w,0)}{v1,v2}
\fmffixed{(0.2w,0)}{v2,v3}
\fmffixed{(0.2w,0)}{v3,v4}
\fmffixed{(0.2w,0)}{v4,v5}
\fmffixed{(0.2w,0)}{v5,v6}
\fmffixed{(0.2w,0)}{v7,v8}
\fmffixed{(0.2w,0)}{v8,v9}
\fmffixed{(0.2w,0)}{v9,v10}
\fmffixed{(0.2w,0)}{v10,v11}
\fmffixed{(0.2w,0)}{v11,v12}
\fmf{plain,tension=1,left=0.25}{v1,vc1}
\fmf{plain,tension=1,right=0.25}{v2,vc1}
\fmf{phantom,tension=1,left=0.25}{v8,va2}
\fmf{phantom,tension=1,right=0.25}{v7,va2}
\fmf{phantom,tension=2}{vc1,va2}
\fmf{phantom,tension=1,left=0.25}{v2,va3}
\fmf{phantom,tension=1,right=0.25}{v3,va3}
\fmf{plain,tension=1,left=0.25}{v9,vc4}
\fmf{plain,tension=1,right=0.25}{v8,vc4}
\fmf{phantom,tension=2}{va3,vc4}
\fmf{plain,tension=1,left=0.25}{v3,vc5}
\fmf{plain,tension=1,right=0.25}{v4,vc5}
\fmf{phantom,tension=1,left=0.25}{v10,va6}
\fmf{phantom,tension=1,right=0.25}{v9,va6}
\fmf{phantom,tension=2}{vc5,va6}
\fmf{phantom,tension=1,left=0.25}{v4,va7}
\fmf{phantom,tension=1,right=0.25}{v5,va7}
\fmf{plain,tension=1,left=0.25}{v11,vc8}
\fmf{plain,tension=1,right=0.25}{v10,vc8}
\fmf{phantom,tension=2}{va7,vc8}
\fmf{plain,tension=1,left=0.25}{v5,vc9}
\fmf{plain,tension=1,right=0.25}{v6,vc9}
\fmf{phantom,tension=1,left=0.25}{v12,va10}
\fmf{phantom,tension=1,right=0.25}{v11,va10}
\fmf{phantom,tension=2}{vc9,va10}
\fmffreeze
\fmf{plain,tension=1}{vc1,vc2}
\fmf{plain,tension=0.25}{vc5,vc6}
\fmf{plain,tension=1}{vc9,vc10}
\fmf{plain,tension=0.25}{vc3,vc4}
\fmf{plain,tension=0.25}{vc7,vc8}
\fmf{plain,tension=0.125}{vc2,vc3}
\fmf{plain,tension=0.125}{vc3,vc6}
\fmf{plain,tension=0.125}{vc6,vc7}
\fmf{plain,tension=0.125}{vc7,vc10}
\fmf{plain,tension=0.25,right=0.25}{v7,vc2}
\fmf{plain,tension=0.25,left=0.25}{v12,vc10}
\fmffreeze
\fmfposition
\fmf{plain,tension=1,right=0,width=1mm}{v1,v6}
\fmffreeze
\end{fmfchar*}}}
}
\subfigspace
\subfigure[$\nwgraph{5}{}{1,3,2,5,4}$]{
\raisebox{\eqoff}{%
\fmfframe(3,1)(1,4){%
\begin{fmfchar*}(30,20)
\fmftop{v1}
\fmfbottom{v7}
\fmfforce{(0w,h)}{v1}
\fmfforce{(0w,0)}{v7}
\fmffixed{(0.2w,0)}{v1,v2}
\fmffixed{(0.2w,0)}{v2,v3}
\fmffixed{(0.2w,0)}{v3,v4}
\fmffixed{(0.2w,0)}{v4,v5}
\fmffixed{(0.2w,0)}{v5,v6}
\fmffixed{(0.2w,0)}{v7,v8}
\fmffixed{(0.2w,0)}{v8,v9}
\fmffixed{(0.2w,0)}{v9,v10}
\fmffixed{(0.2w,0)}{v10,v11}
\fmffixed{(0.2w,0)}{v11,v12}
\fmf{plain,tension=1,right=0.25}{v1,vc1}
\fmf{plain,tension=1,left=0.25}{v2,vc1}
\fmf{phantom,tension=1,right=0.25}{v8,va2}
\fmf{phantom,tension=1,left=0.25}{v7,va2}
\fmf{phantom,tension=2}{vc1,va2}
\fmf{phantom,tension=1,right=0.25}{v2,va3}
\fmf{phantom,tension=1,left=0.25}{v3,va3}
\fmf{plain,tension=1,right=0.25}{v9,vc4}
\fmf{plain,tension=1,left=0.25}{v8,vc4}
\fmf{phantom,tension=2}{va3,vc4}
\fmf{plain,tension=1,right=0.25}{v3,vc5}
\fmf{plain,tension=1,left=0.25}{v4,vc5}
\fmf{phantom,tension=1,right=0.25}{v10,va6}
\fmf{phantom,tension=1,left=0.25}{v9,va6}
\fmf{phantom,tension=2}{vc5,va6}
\fmf{phantom,tension=1,right=0.25}{v4,va7}
\fmf{phantom,tension=1,left=0.25}{v5,va7}
\fmf{plain,tension=1,right=0.25}{v11,vc8}
\fmf{plain,tension=1,left=0.25}{v10,vc8}
\fmf{phantom,tension=2}{va7,vc8}
\fmf{plain,tension=1,right=0.25}{v5,vc9}
\fmf{plain,tension=1,left=0.25}{v6,vc9}
\fmf{phantom,tension=1,right=0.25}{v12,va10}
\fmf{phantom,tension=1,left=0.25}{v11,va10}
\fmf{phantom,tension=2}{vc9,va10}
\fmffreeze
\fmf{plain,tension=1}{vc1,vc2}
\fmf{plain,tension=0.25}{vc5,vc6}
\fmf{plain,tension=1}{vc9,vc10}
\fmf{plain,tension=0.25}{vc3,vc4}
\fmf{plain,tension=0.25}{vc7,vc8}
\fmf{plain,tension=0.125}{vc2,vc3}
\fmf{plain,tension=0.125}{vc3,vc6}
\fmf{plain,tension=0.125}{vc6,vc7}
\fmf{plain,tension=0.125}{vc7,vc10}
\fmf{plain,tension=0.25,left=0.25}{v7,vc2}
\fmf{plain,tension=0.25,right=0.25}{v12,vc10}
\fmffreeze
\fmfposition
\fmf{plain,tension=1,left=0,width=1mm}{v7,v12}
\fmffreeze
\end{fmfchar*}}}
}
\subfigspace
\subfigure[$\nwgraph{5}{}{1,2,5,4,3}$]{
\raisebox{\eqoff}{%
\fmfframe(3,1)(1,4){%
\begin{fmfchar*}(30,20)
\fmftop{v1}
\fmfbottom{v7}
\fmfforce{(0w,h)}{v1}
\fmfforce{(0w,0)}{v7}
\fmffixed{(0.2w,0)}{v1,v2}
\fmffixed{(0.2w,0)}{v2,v3}
\fmffixed{(0.2w,0)}{v3,v4}
\fmffixed{(0.2w,0)}{v4,v5}
\fmffixed{(0.2w,0)}{v5,v6}
\fmffixed{(0.2w,0)}{v7,v8}
\fmffixed{(0.2w,0)}{v8,v9}
\fmffixed{(0.2w,0)}{v9,v10}
\fmffixed{(0.2w,0)}{v10,v11}
\fmffixed{(0.2w,0)}{v11,v12}
\fmffixed{(0,whatever)}{vc1,vc2}
\fmffixed{(0,whatever)}{vc3,vc4}
\fmffixed{(0,whatever)}{vc5,vc6}
\fmffixed{(0,whatever)}{vc7,vc8}
\fmffixed{(0,whatever)}{vc9,vc10}
\fmf{plain,tension=0.5,right=0.25}{v1,vc1}
\fmf{plain,tension=0.5,left=0.25}{v2,vc1}
\fmf{plain,tension=1,left=0.25}{v7,vc2}
\fmf{plain,tension=1,left=0.25}{v3,vc3}
\fmf{plain,tension=1,left=0.25}{v8,vc4}
\fmf{plain,tension=0.25,left=0.25}{v9,vc6}
\fmf{plain,tension=0.25,right=0.25}{v10,vc6}
\fmf{plain,tension=1,right=0.25}{v4,vc7}
\fmf{plain,tension=1,right=0.25}{v11,vc8}
\fmf{plain,tension=1,right=0.25}{v12,vc10}
\fmf{plain,tension=0.5,right=0.25}{v5,vc9}
\fmf{plain,tension=0.5,left=0.25}{v6,vc9}
  \fmf{plain,tension=1}{vc1,vc2}
  \fmf{plain,tension=0.5}{vc2,vc3}
  \fmf{plain,tension=2}{vc3,vc4}
  \fmf{plain,tension=0.5}{vc4,vc5}
  \fmf{plain,tension=0.5}{vc5,vc6}
  \fmf{plain,tension=0.5}{vc5,vc8}
  \fmf{plain,tension=2}{vc7,vc8}
  \fmf{plain,tension=0.5}{vc7,vc10}
  \fmf{plain,tension=1}{vc9,vc10}
\fmffreeze
\fmfposition
\fmf{plain,tension=1,left=0,width=1mm}{v7,v12}
\fmffreeze
\end{fmfchar*}}}
}
\\
\subfigure[$\nwgraph{5}{}{1,2,4,3,5}$]{
\raisebox{\eqoff}{%
\fmfframe(3,1)(1,4){%
\begin{fmfchar*}(30,20)
\fmftop{v1}
\fmfbottom{v7}
\fmfforce{(0w,h)}{v1}
\fmfforce{(0w,0)}{v7}
\fmffixed{(0.2w,0)}{v1,v2}
\fmffixed{(0.2w,0)}{v2,v3}
\fmffixed{(0.2w,0)}{v3,v4}
\fmffixed{(0.2w,0)}{v4,v5}
\fmffixed{(0.2w,0)}{v5,v6}
\fmffixed{(0.2w,0)}{v7,v8}
\fmffixed{(0.2w,0)}{v8,v9}
\fmffixed{(0.2w,0)}{v9,v10}
\fmffixed{(0.2w,0)}{v10,v11}
\fmffixed{(0.2w,0)}{v11,v12}
\fmffixed{(0,whatever)}{vc1,vc2}
\fmffixed{(0,whatever)}{vc5,vc6}
\fmffixed{(0,whatever)}{vc7,vc8}
\fmffixed{(0,whatever)}{vc9,vc10}
\fmf{plain,tension=1,right=0.25}{v1,vc1}
\fmf{plain,tension=1,left=0.25}{v2,vc1}
\fmf{phantom,tension=0.25,right=0.25}{v8,vc2}
\fmf{phantom,tension=0.25,left=0.25}{v7,vc2}
\fmf{plain,tension=0.25,right=0.25}{v4,vc7}
\fmf{plain,tension=0.25,left=0.25}{v5,vc7}
\fmf{plain,tension=0.25,right=0.25}{v10,vc6}
\fmf{plain,tension=0.25,left=0.25}{v9,vc6}
\fmf{plain,tension=0.25,right=0.25}{v12,vc10}
\fmf{plain,tension=0.25,left=0.25}{v11,vc10}
\fmf{plain,tension=0.25,left=0.25}{v6,vc9}
\fmf{phantom,tension=0.25,right=0.25}{v3,vc5}
\fmf{phantom,tension=1}{vc1,vc2}
\fmf{plain,tension=1}{vc5,vc6}
\fmf{plain,tension=0.25}{vc7,vc8}
\fmf{plain,tension=0.125}{vc5,vc8}
\fmf{plain,tension=0.125}{vc8,vc9}
\fmf{plain,tension=1}{vc9,vc10}
\fmffreeze
\fmf{plain,tension=2}{vc1,vc2}
\fmf{plain,tension=0.5}{vc3,vc4}
\fmf{plain,tension=0.125}{vc2,vc3}
\fmf{plain,tension=0.25}{vc4,vc5}
\fmf{plain,tension=0.25,left=0.25}{v7,vc2}
\fmf{plain,tension=0.5,left=0.25}{v8,vc4}
\fmf{plain,tension=0.25,left=0.25}{v3,vc3}
\fmfposition
\fmf{plain,tension=1,left=0,width=1mm}{v7,v12}
\fmffreeze
\end{fmfchar*}}}
}
\subfigspace
\subfigure[$\nwgraph{5}{}{3,2,1,4,5}$]{
\raisebox{\eqoff}{%
\fmfframe(3,1)(1,4){%
\begin{fmfchar*}(30,20)
\fmftop{v1}
\fmfbottom{v7}
\fmfforce{(0w,h)}{v1}
\fmfforce{(0w,0)}{v7}
\fmffixed{(0.2w,0)}{v1,v2}
\fmffixed{(0.2w,0)}{v2,v3}
\fmffixed{(0.2w,0)}{v3,v4}
\fmffixed{(0.2w,0)}{v4,v5}
\fmffixed{(0.2w,0)}{v5,v6}
\fmffixed{(0.2w,0)}{v7,v8}
\fmffixed{(0.2w,0)}{v8,v9}
\fmffixed{(0.2w,0)}{v9,v10}
\fmffixed{(0.2w,0)}{v10,v11}
\fmffixed{(0.2w,0)}{v11,v12}
\fmf{plain,tension=0.5,right=0.25}{v2,vc3}
\fmf{phantom,tension=0.5,left=0.25}{v3,vc3}
\fmf{plain,right=0.25}{v1,vc1}
\fmf{plain,tension=0.5,left=0.25}{v7,vc2}
\fmf{plain,tension=0.5,right=0.25}{v8,vc2}
\fmf{plain,right=0.25}{v9,vc4}
\fmf{plain,tension=1}{vc1,vc2}
\fmf{plain,tension=0.5}{vc1,vc4}
\fmf{plain,tension=3}{vc3,vc4}
\fmf{phantom,tension=0.5,right=0.25}{v4,vc7}
\fmf{plain,tension=0.5,left=0.25}{v5,vc7}
\fmf{plain,left=0.25}{v6,vc9}
\fmf{plain,tension=0.5,left=0.25}{v11,vc10}
\fmf{plain,tension=0.5,right=0.25}{v12,vc10}
\fmf{plain,left=0.25}{v10,vc8}
\fmf{plain,tension=3}{vc7,vc8}
\fmf{plain,tension=0.5}{vc8,vc9}
\fmf{plain,tension=1}{vc9,vc10}
\fmf{plain,tension=0.5,right=0.25}{v3,vc5}
\fmf{plain,tension=0.5,left=0.25}{v4,vc5}
\fmf{phantom,tension=0.5,right=0.25}{v10,va2}
\fmf{phantom,tension=0.5,left=0.25}{v9,va2}
\fmf{phantom,tension=0.5}{vc5,va2}
\fmffreeze
\fmf{plain,tension=0.5}{vc5,vc6}
\fmf{plain,tension=0.5}{vc3,vc6}
\fmf{plain,tension=0.5}{vc6,vc7}
\fmffreeze
\fmfposition
\fmf{plain,tension=1,right=0,width=1mm}{v7,v12}
\fmffreeze
\end{fmfchar*}}}
}
\subfigspace
\subfigure[$\nwgraph{5}{}{1,3,2,4,5}$]{
\raisebox{\eqoff}{%
\fmfframe(3,1)(1,4){%
\begin{fmfchar*}(30,20)
\fmftop{v6}
\fmfbottom{v12}
\fmfforce{(0w,h)}{v6}
\fmfforce{(0w,0)}{v12}
\fmffixed{(0.2w,0)}{v6,v5}
\fmffixed{(0.2w,0)}{v5,v4}
\fmffixed{(0.2w,0)}{v4,v3}
\fmffixed{(0.2w,0)}{v3,v2}
\fmffixed{(0.2w,0)}{v2,v1}
\fmffixed{(0.2w,0)}{v12,v11}
\fmffixed{(0.2w,0)}{v11,v10}
\fmffixed{(0.2w,0)}{v10,v9}
\fmffixed{(0.2w,0)}{v9,v8}
\fmffixed{(0.2w,0)}{v8,v7}
\fmf{plain,tension=0.5,left=0.25}{v2,vc3}
\fmf{phantom,tension=0.5,right=0.25}{v3,vc3}
\fmf{plain,left=0.25}{v1,vc1}
\fmf{plain,tension=0.5,right=0.25}{v7,vc2}
\fmf{plain,tension=0.5,left=0.25}{v8,vc2}
\fmf{plain,left=0.25}{v9,vc4}
\fmf{plain,tension=1}{vc1,vc2}
\fmf{plain,tension=0.5}{vc1,vc4}
\fmf{plain,tension=3}{vc3,vc4}
\fmf{plain,tension=0.5,left=0.25}{v5,vc9}
\fmf{plain,tension=0.5,right=0.25}{v6,vc9}
\fmf{phantom,left=0.25}{v4,vc7}
\fmf{plain,tension=0.5,right=0.25}{v10,vc8}
\fmf{plain,tension=0.5,left=0.25}{v11,vc8}
\fmf{plain,left=0.25}{v12,vc10}
\fmf{plain,tension=1}{vc7,vc8}
\fmf{plain,tension=0.5}{vc7,vc10}
\fmf{plain,tension=1}{vc9,vc10}
\fmf{plain,tension=0.5,left=0.25}{v3,vc5}
\fmf{plain,tension=0.5,right=0.25}{v4,vc5}
\fmf{phantom,tension=0.5,left=0.25}{v10,va2}
\fmf{phantom,tension=0.5,right=0.25}{v9,va2}
\fmf{phantom,tension=0.5}{vc5,va2}
\fmffreeze
\fmf{plain,tension=0.5}{vc3,vc6}
\fmf{plain,tension=0.25}{vc5,vc6}
\fmf{plain,tension=0.5}{vc6,vc7}
\fmffreeze
\fmfposition
\fmf{plain,tension=1,right=0,width=1mm}{v7,v12}
\fmffreeze
\end{fmfchar*}}}
}
\subfigspace
\subfigure[$\nwgraph{5}{}{1,2,3,5,4}$]{
\raisebox{\eqoff}{%
\fmfframe(3,1)(1,4){%
\begin{fmfchar*}(30,20)
\fmftop{v1}
\fmfbottom{v7}
\fmfforce{(0w,h)}{v1}
\fmfforce{(0w,0)}{v7}
\fmffixed{(0.2w,0)}{v1,v2}
\fmffixed{(0.2w,0)}{v2,v3}
\fmffixed{(0.2w,0)}{v3,v4}
\fmffixed{(0.2w,0)}{v4,v5}
\fmffixed{(0.2w,0)}{v5,v6}
\fmffixed{(0.2w,0)}{v7,v8}
\fmffixed{(0.2w,0)}{v8,v9}
\fmffixed{(0.2w,0)}{v9,v10}
\fmffixed{(0.2w,0)}{v10,v11}
\fmffixed{(0.2w,0)}{v11,v12}
\fmffixed{(0,whatever)}{vc1,vc2}
\fmffixed{(0,whatever)}{vc3,vc4}
\fmffixed{(0,whatever)}{vc5,vc6}
\fmffixed{(0,whatever)}{va7,vc8}
\fmf{plain,tension=0.25,right=0.25}{v1,vc1}
\fmf{plain,tension=0.25,left=0.25}{v2,vc1}
\fmf{plain,left=0.25}{v7,vc2}
\fmf{plain,tension=1,left=0.25}{v3,vc3}
\fmf{plain,tension=1,left=0.25}{v4,vc5}
\fmf{phantom,tension=1,left=0.25}{v5,va7}
\fmf{plain,left=0.25}{v9,vc6}
\fmf{plain,tension=0.25,left=0.25}{v10,vc8}
\fmf{plain,tension=0.25,right=0.25}{v11,vc8}
\fmf{plain,left=0.25}{v8,vc4}
\fmf{plain,tension=0.5}{vc1,vc2}
\fmf{plain,tension=0.5}{vc2,vc3}
\fmf{plain,tension=0.5}{vc3,vc4}
\fmf{plain,tension=0.5}{vc4,vc5}
\fmf{plain,tension=1}{vc5,vc6}
\fmf{phantom,tension=0.5}{vc6,va7}
\fmf{phantom,tension=0.5}{va7,vc8}
\fmf{plain,tension=0.25,right=0.25}{v5,vc9}
\fmf{plain,tension=0.25,left=0.25}{v6,vc9}
\fmf{phantom,tension=0.25,right=0.25}{v12,va10}
\fmf{phantom,tension=0.25,left=0.25}{v11,va10}
\fmf{phantom,tension=0.5}{vc9,va10}
\fmffreeze
\fmf{plain,tension=0.5}{vc6,vc7}
\fmf{plain,tension=0.125}{vc7,vc8}
\fmf{plain,tension=0.5}{vc7,vc10}
\fmf{plain,tension=4}{vc9,vc10}
\fmf{plain,right=0.25}{v12,vc10}
\fmffreeze
\fmfposition
\fmf{plain,tension=1,right=0,width=1mm}{v7,v12}
\fmffreeze
\end{fmfchar*}}}
}
\\
\subfigure[$\nwgraph{5}{}{2,1,3,4,5}$]{
\raisebox{\eqoff}{%
\fmfframe(3,1)(1,4){%
\begin{fmfchar*}(30,20)
\fmftop{v1}
\fmfbottom{v7}
\fmfforce{(0w,h)}{v1}
\fmfforce{(0w,0)}{v7}
\fmffixed{(0.2w,0)}{v1,v2}
\fmffixed{(0.2w,0)}{v2,v3}
\fmffixed{(0.2w,0)}{v3,v4}
\fmffixed{(0.2w,0)}{v4,v5}
\fmffixed{(0.2w,0)}{v5,v6}
\fmffixed{(0.2w,0)}{v7,v8}
\fmffixed{(0.2w,0)}{v8,v9}
\fmffixed{(0.2w,0)}{v9,v10}
\fmffixed{(0.2w,0)}{v10,v11}
\fmffixed{(0.2w,0)}{v11,v12}
\fmf{phantom,tension=0.25,right=0.25}{v3,vc5}
\fmf{plain,tension=0.25,left=0.25}{v4,vc5}
\fmf{plain,left=0.25}{v9,vc6}
\fmf{plain,tension=1,left=0.25}{v5,vc7}
\fmf{plain,tension=1,left=0.25}{v6,vc9}
\fmf{plain,left=0.25}{v10,vc8}
\fmf{plain,tension=0.25,left=0.25}{v11,vc10}
\fmf{plain,tension=0.25,right=0.25}{v12,vc10}
\fmf{plain,tension=0.5}{vc5,vc6}
\fmf{plain,tension=0.5}{vc6,vc7}
\fmf{plain,tension=0.5}{vc7,vc8}
\fmf{plain,tension=0.5}{vc8,vc9}
\fmf{plain,tension=0.5}{vc9,vc10}
\fmf{plain,tension=0.25,right=0.25}{v8,vc2}
\fmf{plain,tension=0.25,left=0.25}{v7,vc2}
\fmf{phantom,tension=0.25,right=0.25}{v1,va1}
\fmf{phantom,tension=0.25,left=0.25}{v2,va1}
\fmf{phantom,tension=0.5}{va1,vc2}
\fmffreeze
\fmf{plain,tension=0.25,right=0.25}{v2,vc3}
\fmf{plain,tension=0.25,left=0.25}{v3,vc3}
\fmf{plain,tension=0.5}{vc1,vc4}
\fmf{plain,tension=0.5}{vc3,vc4}
\fmf{plain,tension=0.5}{vc4,vc5}
\fmf{plain,tension=0.5}{vc1,vc2}
\fmf{plain,tension=1,right=0.25}{v1,vc1}
\fmffreeze
\fmfposition
\fmf{plain,tension=1,right=0,width=1mm}{v7,v12}
\fmffreeze
\end{fmfchar*}}}
}
\subfigspace
\subfigure[$\nwgraph{5}{}{1,4,3,2,5}$]{
\raisebox{\eqoff}{%
\fmfframe(3,1)(1,4){%
\begin{fmfchar*}(30,20)
\fmftop{v1}
\fmfbottom{v7}
\fmfforce{(0w,h)}{v1}
\fmfforce{(0w,0)}{v7}
\fmffixed{(0.2w,0)}{v1,v2}
\fmffixed{(0.2w,0)}{v2,v3}
\fmffixed{(0.2w,0)}{v3,v4}
\fmffixed{(0.2w,0)}{v4,v5}
\fmffixed{(0.2w,0)}{v5,v6}
\fmffixed{(0.2w,0)}{v7,v8}
\fmffixed{(0.2w,0)}{v8,v9}
\fmffixed{(0.2w,0)}{v9,v10}
\fmffixed{(0.2w,0)}{v10,v11}
\fmffixed{(0.2w,0)}{v11,v12}
\fmffixed{(0,whatever)}{vc1,vc2}
\fmffixed{(0,whatever)}{vc3,vc4}
\fmffixed{(0,whatever)}{vc5,vc6}
\fmffixed{(0,whatever)}{vc7,vc8}
\fmffixed{(0,whatever)}{vc9,vc10}
\fmf{phantom,tension=1,right=0.25}{v1,vc1}
\fmf{phantom,tension=1,left=0.25}{v2,vc1}
\fmf{phantom,tension=1,right=0.25}{v3,vc5}
\fmf{phantom,tension=1,left=0.25}{v4,vc5}
\fmf{phantom,tension=1,left=0.25}{v7,vc2}
\fmf{phantom,tension=0.25,left=0.25}{v8,vc4}
\fmf{phantom,tension=0.25,right=0.25}{v9,vc4}
\fmf{phantom,tension=1,right=0.25}{v10,vc6}
\fmf{phantom,tension=1,right=0.25}{v5,vc9}
\fmf{phantom,tension=1,left=0.25}{v6,vc9}
\fmf{phantom,tension=1,left=0.25}{v11,vc10}
\fmf{phantom,tension=1,right=0.25}{v12,vc10}
  \fmf{phantom,tension=4}{vc1,vc2}
  \fmf{phantom,tension=4}{vc5,vc6}
  \fmf{phantom,tension=0.5}{vc2,vc3}
  \fmf{phantom,tension=0.5}{vc3,vc6}
  \fmf{phantom,tension=1}{vc3,vc4}
  \fmf{phantom,tension=1.5}{vc9,vc10}
\fmffreeze
\fmf{plain,tension=0.25,right=0.25}{v1,vc1}
\fmf{plain,tension=0.25,left=0.25}{v2,vc1}
\fmf{plain,tension=1,left=0.125}{v7,vc2}
\fmf{plain,tension=0.25,right=0.25}{v3,vc5}
\fmf{plain,tension=1,right=0.125}{v10,vc6}
\fmf{plain,tension=0.25,left=0.25}{v8,vc4}
\fmf{plain,tension=0.25,right=0.25}{v9,vc4}
\fmf{plain,tension=0.25,right=0.25}{v4,vc7}
\fmf{plain,tension=0.25,left=0.25}{v5,vc7}
\fmf{plain,tension=0.25,left=0.125}{v6,vc9}
\fmf{plain,tension=0.25,left=0.25}{v11,vc10}
\fmf{plain,tension=0.25,right=0.25}{v12,vc10}
  \fmf{plain,tension=0.5}{vc1,vc2}
  \fmf{plain,tension=0.5}{vc2,vc3}
  \fmf{plain,tension=0.5}{vc3,vc4}
  \fmf{plain,tension=0.5}{vc3,vc6}
  \fmf{plain,tension=0.5}{vc5,vc6}
  \fmf{plain,tension=0.5}{vc5,vc8}
  \fmf{plain,tension=0.5}{vc7,vc8}
  \fmf{plain,tension=0.5}{vc8,vc9}
  \fmf{plain,tension=0.5}{vc9,vc10}
\fmffreeze
\fmfposition
\fmf{plain,tension=1,left=0,width=1mm}{v7,v12}
\fmffreeze
\end{fmfchar*}}}
}
\\[0.5cm]
\begin{tabular}{m{12cm}}
\toprule
$\nwgraph{5}{}{1,2,3,4,5}\to \jint{5}{1}\,\chi(1,2,3,4,5)$  \qquad
$\nwgraph{5}{}{2,1,4,3,5}\to \jint{5}{2}\,\chi(2,1,4,3,5)$  \\
$\nwgraph{5}{}{1,3,2,5,4}\to \jint{5}{3}\,\chi(1,3,2,5,4)$  \qquad
$\nwgraph{5}{}{1,2,5,4,3}\to \jint{5}{4}\,\chi(1,2,5,4,3)$  \\
$\nwgraph{5}{}{1,2,4,3,5}\to \jint{5}{5}\,\chi(1,2,4,3,5)$  \qquad
$\nwgraph{5}{}{3,2,1,4,5}\to \jint{5}{6}\,\chi(3,2,1,4,5)$  \\
$\nwgraph{5}{}{1,3,2,4,5}\to \jint{5}{7}\,\chi(1,3,2,4,5)$  \qquad
$\nwgraph{5}{}{1,2,3,5,4}\to \jint{5}{8}\,\chi(1,2,3,5,4)$  \\
$\nwgraph{5}{}{2,1,3,4,5}\to \jint{5}{9}\,\chi(2,1,3,4,5)$  \qquad
$\nwgraph{5}{}{1,4,3,2,5}\to \jint{5}{10}\,\chi(1,4,3,2,5)$  \\
\bottomrule
\end{tabular}
\normalsize
\caption{Completely chiral range-six diagrams}
\label{r6-chiral}
\end{figure}

\begin{figure}[p]
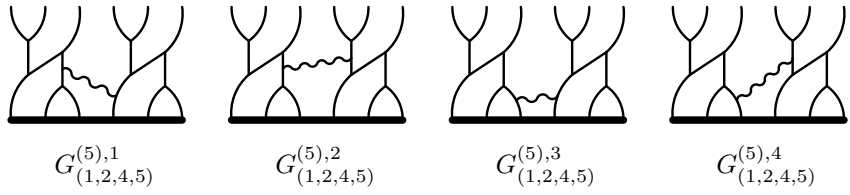

\capstart
\renewcommand*{\thesubfigure}{\ \ }
\footnotesize
\centering
\unitlength=0.75mm
\settoheight{\eqoff}{$\times$}%
\setlength{\eqoff}{0.5\eqoff}%
\addtolength{\eqoff}{-12.5\unitlength}%
\settoheight{\eqofftwo}{$\times$}%
\setlength{\eqofftwo}{0.5\eqofftwo}%
\addtolength{\eqofftwo}{-7.5\unitlength}%
\subfigure[$\nwgraph{5}{}{2,1,3,2,4}$]{
\raisebox{\eqoff}{%
\fmfframe(3,1)(1,4){%
\begin{fmfchar*}(20,20)
\fmftop{v1}
\fmfbottom{v5}
\fmfforce{(0w,h)}{v1}
\fmfforce{(0w,0)}{v5}
\fmffixed{(0.25w,0)}{v1,v2}
\fmffixed{(0.25w,0)}{v2,v3}
\fmffixed{(0.25w,0)}{v3,v4}
\fmffixed{(0.25w,0)}{v4,v9}
\fmffixed{(0.25w,0)}{v5,v6}
\fmffixed{(0.25w,0)}{v6,v7}
\fmffixed{(0.25w,0)}{v7,v8}
\fmffixed{(0.25w,0)}{v8,v10}
\fmffixed{(0,whatever)}{vb2,vc2}
\fmffixed{(0,whatever)}{vb3,vc3}
\fmffixed{(0,whatever)}{vb5,vc5}
\fmffixed{(0,whatever)}{vb7,vc7}
\fmf{plain,tension=0.25,left=0.25}{v5,vc2}
\fmf{plain,tension=0.25,right=0.25}{v6,vc2}
\fmf{phantom,tension=0.25,right=0.25}{v1,vb2}
\fmf{phantom,tension=0.25,left=0.25}{v2,vb2}
\fmf{phantom,tension=0.5}{vc2,vb2}
\fmf{plain,tension=0.25,right=0.25}{v2,vc3}
\fmf{phantom,tension=0.25,left=0.25}{v3,vc3}
\fmf{phantom,tension=0.25,right=0.25}{v6,vb3}
\fmf{phantom,tension=0.25,left=0.25}{v7,vb3}
\fmf{phantom,tension=0.5}{vc3,vb3}
\fmf{plain,tension=0.25,left=0.25}{v7,vc5}
\fmf{plain,tension=0.25,right=0.25}{v8,vc5}
\fmf{phantom,tension=0.25,right=0.25}{v3,vb5}
\fmf{phantom,tension=0.25,left=0.25}{v4,vb5}
\fmf{phantom,tension=0.5}{vc5,vb5}
\fmf{phantom,tension=0.25,right=0.25}{v4,vc7}
\fmf{plain,tension=0.25,left=0.25}{v9,vc7}
\fmf{phantom,tension=0.25,right=0.25}{v8,vb7}
\fmf{phantom,tension=0.25,left=0.25}{v10,vb7}
\fmf{phantom,tension=0.5}{vc7,vb7}
\fmffreeze
\fmf{plain,tension=0.25,right=0.25}{v3,vc9}
\fmf{plain,tension=0.25,left=0.25}{v4,vc9}
\fmf{plain,tension=0.25}{vc3,vc10}
\fmf{plain,tension=0.25}{vc7,vc10}
\fmf{plain,tension=0.25}{vc9,vc10}
\fmffixed{(0,whatever)}{vc1,vc2}
\fmffixed{(0,whatever)}{vc3,vc4}
\fmffixed{(0,whatever)}{vc5,vc6}
\fmffixed{(0,whatever)}{vc7,vc8}
\fmffixed{(whatever,0)}{vc1,vc6}
\fmffixed{(whatever,0)}{vc4,vc8}
\fmf{plain,tension=0.25,right=0.25}{v1,vc1}
\fmf{plain,tension=0.5}{vc1,vc2}
\fmf{plain,tension=0.5}{vc3,vc4}
\fmf{plain,tension=0.5}{vc1,vc4}
\fmf{plain,tension=0.5}{vc5,vc6}
\fmf{plain,tension=0.5}{vc4,vc6}
\fmf{plain,tension=0.25,right=0.25}{v10,vc8}
\fmf{plain,tension=0.5}{vc7,vc8}
\fmf{plain,tension=0.5}{vc6,vc8}
\fmf{plain,tension=0.5,right=0,width=1mm}{v5,v10}
\fmfposition
\fmfipath{p[]}
\end{fmfchar*}}}
}
\subfigspace
\subfigure[$\nwgraph{5}{}{2,1,4,3,2}$]{
\raisebox{\eqoff}{%
\fmfframe(3,1)(1,4){%
\begin{fmfchar*}(20,20)
\fmftop{v1}
\fmfbottom{v5}
\fmfforce{(0w,h)}{v1}
\fmfforce{(0w,0)}{v5}
\fmffixed{(0.25w,0)}{v1,v2}
\fmffixed{(0.25w,0)}{v2,v3}
\fmffixed{(0.25w,0)}{v3,v4}
\fmffixed{(0.25w,0)}{v4,v9}
\fmffixed{(0.25w,0)}{v5,v6}
\fmffixed{(0.25w,0)}{v6,v7}
\fmffixed{(0.25w,0)}{v7,v8}
\fmffixed{(0.25w,0)}{v8,v10}
\fmffixed{(0,whatever)}{vc1,vc5}
\fmffixed{(0,whatever)}{vc2,vc3}
\fmffixed{(0,whatever)}{vc3,vc6}
\fmffixed{(0,whatever)}{vc6,vc7}
\fmffixed{(0,whatever)}{vc4,vc8}
\fmffixed{(0,whatever)}{vc9,vc10}
\fmffixed{(0.5w,0)}{vc1,vc4}
\fmffixed{(0.5w,0)}{vc5,vc8}
\fmf{plain,tension=1,right=0.125}{v1,vc1}
\fmf{plain,tension=0.25,right=0.25}{v2,vc2}
\fmf{plain,tension=0.25,left=0.25}{v3,vc2}
\fmf{phantom,tension=1,left=0.125}{v4,vc4}
\fmf{plain,tension=1,left=0.125}{v5,vc5}
\fmf{plain,tension=0.25,left=0.25}{v6,vc6}
\fmf{plain,tension=0.25,right=0.25}{v7,vc6}
\fmf{plain,tension=1,right=0.125}{v8,vc8}
  \fmf{plain,tension=0.5}{vc1,vc3}
  \fmf{plain,tension=0.5}{vc2,vc3}
  \fmf{plain,tension=0.5}{vc3,vc4}
  \fmf{plain,tension=0.5}{vc5,vc7}
  \fmf{plain,tension=0.5}{vc6,vc7}
  \fmf{plain,tension=0.5}{vc7,vc8}
  \fmf{plain,tension=2}{vc1,vc5}
  \fmf{plain,tension=2}{vc4,vc8}
  \fmf{phantom,tension=2}{vc5,vc4}
\fmffreeze
\fmf{plain,tension=0.5,left=0.25}{v9,vc10}
\fmf{plain,tension=0.5,right=0.25}{v4,vc10}
\fmf{plain,tension=0.5}{vc4,vc9}
\fmf{plain,tension=1}{vc10,vc9}
\fmf{plain,tension=1,right=0.125}{v10,vc9}
\fmfposition
\fmf{plain,tension=1,left=0,width=1mm}{v5,v10}
\fmffreeze
\end{fmfchar*}}}
}
\\[0.5cm]
\begin{tabular}{m{7cm}}
\toprule
$\nwgraph{5}{}{2,1,3,2,4}\to \jint{5}{19}\,\chi(2,1,3,2,4)$  \\
$\nwgraph{5}{}{2,1,4,3,2}\to \jint{5}{18}\,\chi(2,1,4,3,2)$ \\ 
\bottomrule
\end{tabular}
\normalsize
\caption{Completely chiral range-five diagrams}
\label{r5-chiral}
\end{figure}

\begin{figure}[p]
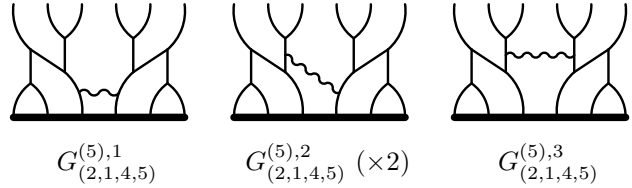

\capstart
\renewcommand*{\thesubfigure}{\ \ }
\footnotesize
\centering
\unitlength=0.75mm
\settoheight{\eqoff}{$\times$}%
\setlength{\eqoff}{0.5\eqoff}%
\addtolength{\eqoff}{-12.5\unitlength}%
\settoheight{\eqofftwo}{$\times$}%
\setlength{\eqofftwo}{0.5\eqofftwo}%
\addtolength{\eqofftwo}{-7.5\unitlength}%
\subfigure[$\nwgraph{5}{1}{2,1,4,5}$]{
\raisebox{\eqoff}{%
\fmfframe(3,1)(1,4){%
\begin{fmfchar*}(30,20)
\fmftop{v1}
\fmfbottom{v7}
\fmfforce{(0w,h)}{v1}
\fmfforce{(0w,0)}{v7}
\fmffixed{(0.2w,0)}{v1,v2}
\fmffixed{(0.2w,0)}{v2,v3}
\fmffixed{(0.2w,0)}{v3,v4}
\fmffixed{(0.2w,0)}{v4,v5}
\fmffixed{(0.2w,0)}{v5,v6}
\fmffixed{(0.2w,0)}{v7,v8}
\fmffixed{(0.2w,0)}{v8,v9}
\fmffixed{(0.2w,0)}{v9,v10}
\fmffixed{(0.2w,0)}{v10,v11}
\fmffixed{(0.2w,0)}{v11,v12}
\fmf{plain,tension=0.5,right=0.25}{v2,vc3}
\fmf{plain,tension=0.5,left=0.25}{v3,vc3}
\fmf{plain,right=0.25}{v1,vc1}
\fmf{plain,tension=0.5,left=0.25}{v7,vc2}
\fmf{plain,tension=0.5,right=0.25}{v8,vc2}
\fmf{plain,right=0.25}{v9,vc4}
\fmf{plain,tension=1}{vc1,vc2}
\fmf{plain,tension=0.5}{vc1,vc4}
\fmf{plain,tension=1}{vc3,vc4}
\fmf{plain,tension=0.5,right=0.25}{v4,vc7}
\fmf{plain,tension=0.5,left=0.25}{v5,vc7}
\fmf{plain,left=0.25}{v6,vc9}
\fmf{plain,tension=0.5,left=0.25}{v11,vc10}
\fmf{plain,tension=0.5,right=0.25}{v12,vc10}
\fmf{plain,left=0.25}{v10,vc8}
\fmf{plain,tension=1}{vc7,vc8}
\fmf{plain,tension=0.5}{vc8,vc9}
\fmf{plain,tension=1}{vc9,vc10}
\fmffreeze
\fmfposition
\fmf{plain,tension=1,right=0,width=1mm}{v7,v12}
\fmfipath{p[]}
\fmfiset{p1}{vpath(__v9,__vc4)}
\fmfiset{p2}{vpath(__v10,__vc8)}
\fmfipair{w[]}
\svertex{w1}{p1}
\svertex{w2}{p2}
\fmfi{wiggly}{w1..w2}
\fmffreeze
\end{fmfchar*}}}
}
\subfigure[$\nwgraph{5}{2}{2,1,4,5}$ $(\times 2)$]{
\raisebox{\eqoff}{%
\fmfframe(3,1)(1,4){%
\begin{fmfchar*}(30,20)
\fmftop{v1}
\fmfbottom{v7}
\fmfforce{(0w,h)}{v1}
\fmfforce{(0w,0)}{v7}
\fmffixed{(0.2w,0)}{v1,v2}
\fmffixed{(0.2w,0)}{v2,v3}
\fmffixed{(0.2w,0)}{v3,v4}
\fmffixed{(0.2w,0)}{v4,v5}
\fmffixed{(0.2w,0)}{v5,v6}
\fmffixed{(0.2w,0)}{v7,v8}
\fmffixed{(0.2w,0)}{v8,v9}
\fmffixed{(0.2w,0)}{v9,v10}
\fmffixed{(0.2w,0)}{v10,v11}
\fmffixed{(0.2w,0)}{v11,v12}
\fmf{plain,tension=0.5,right=0.25}{v2,vc3}
\fmf{plain,tension=0.5,left=0.25}{v3,vc3}
\fmf{plain,right=0.25}{v1,vc1}
\fmf{plain,tension=0.5,left=0.25}{v7,vc2}
\fmf{plain,tension=0.5,right=0.25}{v8,vc2}
\fmf{plain,right=0.25}{v9,vc4}
\fmf{plain,tension=1}{vc1,vc2}
\fmf{plain,tension=0.5}{vc1,vc4}
\fmf{plain,tension=1}{vc3,vc4}
\fmf{plain,tension=0.5,right=0.25}{v4,vc7}
\fmf{plain,tension=0.5,left=0.25}{v5,vc7}
\fmf{plain,left=0.25}{v6,vc9}
\fmf{plain,tension=0.5,left=0.25}{v11,vc10}
\fmf{plain,tension=0.5,right=0.25}{v12,vc10}
\fmf{plain,left=0.25}{v10,vc8}
\fmf{plain,tension=1}{vc7,vc8}
\fmf{plain,tension=0.5}{vc8,vc9}
\fmf{plain,tension=1}{vc9,vc10}
\fmffreeze
\fmfposition
\fmf{plain,tension=1,right=0,width=1mm}{v7,v12}
\fmfipath{p[]}
\fmfiset{p1}{vpath(__vc3,__vc4)}
\fmfiset{p2}{vpath(__v10,__vc8)}
\fmfipair{w[]}
\svertex{w1}{p1}
\svertex{w2}{p2}
\fmfi{wiggly}{w1..w2}
\fmffreeze
\end{fmfchar*}}}
}
\subfigure[$\nwgraph{5}{3}{2,1,4,5}$]{
\raisebox{\eqoff}{%
\fmfframe(3,1)(1,4){%
\begin{fmfchar*}(30,20)
\fmftop{v1}
\fmfbottom{v7}
\fmfforce{(0w,h)}{v1}
\fmfforce{(0w,0)}{v7}
\fmffixed{(0.2w,0)}{v1,v2}
\fmffixed{(0.2w,0)}{v2,v3}
\fmffixed{(0.2w,0)}{v3,v4}
\fmffixed{(0.2w,0)}{v4,v5}
\fmffixed{(0.2w,0)}{v5,v6}
\fmffixed{(0.2w,0)}{v7,v8}
\fmffixed{(0.2w,0)}{v8,v9}
\fmffixed{(0.2w,0)}{v9,v10}
\fmffixed{(0.2w,0)}{v10,v11}
\fmffixed{(0.2w,0)}{v11,v12}
\fmf{plain,tension=0.5,right=0.25}{v2,vc3}
\fmf{plain,tension=0.5,left=0.25}{v3,vc3}
\fmf{plain,right=0.25}{v1,vc1}
\fmf{plain,tension=0.5,left=0.25}{v7,vc2}
\fmf{plain,tension=0.5,right=0.25}{v8,vc2}
\fmf{plain,right=0.25}{v9,vc4}
\fmf{plain,tension=1}{vc1,vc2}
\fmf{plain,tension=0.5}{vc1,vc4}
\fmf{plain,tension=1}{vc3,vc4}
\fmf{plain,tension=0.5,right=0.25}{v4,vc7}
\fmf{plain,tension=0.5,left=0.25}{v5,vc7}
\fmf{plain,left=0.25}{v6,vc9}
\fmf{plain,tension=0.5,left=0.25}{v11,vc10}
\fmf{plain,tension=0.5,right=0.25}{v12,vc10}
\fmf{plain,left=0.25}{v10,vc8}
\fmf{plain,tension=1}{vc7,vc8}
\fmf{plain,tension=0.5}{vc8,vc9}
\fmf{plain,tension=1}{vc9,vc10}
\fmffreeze
\fmfposition
\fmf{plain,tension=1,right=0,width=1mm}{v7,v12}
\fmfipath{p[]}
\fmfiset{p1}{vpath(__vc3,__vc4)}
\fmfiset{p2}{vpath(__vc7,__vc8)}
\fmfipair{w[]}
\svertex{w1}{p1}
\svertex{w2}{p2}
\fmfi{wiggly}{w1..w2}
\fmffreeze
\end{fmfchar*}}}
}
\\[0.5cm]
\begin{tabular}{m{8cm}}
\toprule
$\nwgraph{5}{1}{2,1,4,5}\to -2(\jint{5}{9}+\jint{5}{15})\,\chi(2,1,4,5) $  \\
$\nwgraph{5}{2}{2,1,4,5}\to \jint{5}{9}\,\chi(2,1,4,5) $  \\
$\nwgraph{5}{3}{2,1,4,5}\to 0 $  \\
\midrule
$\sum_i \scgraph{5}{i}{2,1,4,5}\nwgraph{5}{i}{2,1,4,5} \to -2\jint{5}{15}\,\chi(2,1,4,5)  $ \\
\bottomrule
\end{tabular}
\normalsize
\caption{Range-six diagrams with structure $\chi(2,1,4,5)$}
\label{r6-2145}
\end{figure}

The first class that is analyzed here contains all the range-six diagrams with only scalar interactions, which are shown in Figure~\ref{r6-chiral}. The second class is made of the two range-five diagrams without vector interactions, listed in Figure~\ref{r5-chiral}. Then, the independent range-six structures requiring a single vector to become connected, presented in Figures~\ref{r6-1245}-\ref{r6-2145}, are studied.

The results from these diagrams put constraints on the coefficients of the asymptotic five-loop dilatation operator~\eqref{table:D5}. A total of 19 relations are found, some of which allow to find the values of a subset of the similarity coefficients, all the other ones serving as non-trivial consistency checks. The explicit values of the similarity coefficients are
\begin{equation}
\begin{aligned}
\epsilon_{4a} &=\frac{13}{64}i\col & \epsilon_{4b} &= -\frac{85}{192} \col & \epsilon_{4c} &= \frac{i}{96} \col \\
\epsilon_{4d} &= -\frac{5}{192}i \col & \epsilon_{4e} &= \frac{\pi^4-75}{960}i \col & \epsilon_{4f} &= \frac{35}{96} \col \\
& & \epsilon_{4g} &= -\frac{3}{32}i \pnt
\end{aligned}
\end{equation}
In particular, the values of $\epsilon_{4b}$ and $\epsilon_{4f}$ are fixed.

\begin{figure}[p]
\capstart
\footnotesize
\renewcommand*{\thesubfigure}{\ \ }
\centering
\unitlength=0.75mm
\settoheight{\eqoff}{$\times$}%
\setlength{\eqoff}{0.5\eqoff}%
\addtolength{\eqoff}{-12.5\unitlength}%
\settoheight{\eqofftwo}{$\times$}%
\setlength{\eqofftwo}{0.5\eqofftwo}%
\addtolength{\eqofftwo}{-7.5\unitlength}%
\subfigure[$\nwgraph{5}{1}{1,2,4,5}$]{
\raisebox{\eqoff}{%
\fmfframe(3,1)(1,4){%
\begin{fmfchar*}(30,20)
\fmftop{v1}
\fmfbottom{v7}
\fmfforce{(0w,h)}{v1}
\fmfforce{(0w,0)}{v7}
\fmffixed{(0.2w,0)}{v1,v2}
\fmffixed{(0.2w,0)}{v2,v3}
\fmffixed{(0.2w,0)}{v3,v4}
\fmffixed{(0.2w,0)}{v4,v5}
\fmffixed{(0.2w,0)}{v5,v6}
\fmffixed{(0.2w,0)}{v7,v8}
\fmffixed{(0.2w,0)}{v8,v9}
\fmffixed{(0.2w,0)}{v9,v10}
\fmffixed{(0.2w,0)}{v10,v11}
\fmffixed{(0.2w,0)}{v11,v12}
\fmf{plain,tension=0.5,right=0.25}{v1,vc1}
\fmf{plain,tension=0.5,left=0.25}{v2,vc1}
\fmf{plain,left=0.25}{v3,vc3}
\fmf{plain,tension=0.5,left=0.25}{v8,vc4}
\fmf{plain,tension=0.5,right=0.25}{v9,vc4}
\fmf{plain,left=0.25}{v7,vc2}
\fmf{plain,tension=1}{vc1,vc2}
\fmf{plain,tension=0.5}{vc2,vc3}
\fmf{plain,tension=1}{vc3,vc4}
\fmf{plain,tension=0.5,right=0.25}{v4,vc7}
\fmf{plain,tension=0.5,left=0.25}{v5,vc7}
\fmf{plain,left=0.25}{v6,vc9}
\fmf{plain,tension=0.5,left=0.25}{v11,vc10}
\fmf{plain,tension=0.5,right=0.25}{v12,vc10}
\fmf{plain,left=0.25}{v10,vc8}
\fmf{plain,tension=1}{vc7,vc8}
\fmf{plain,tension=0.5}{vc8,vc9}
\fmf{plain,tension=1}{vc9,vc10}
\fmffreeze
\fmfposition
\fmf{plain,tension=1,right=0,width=1mm}{v7,v12}
\fmfipath{p[]}
\fmfiset{p1}{vpath(__vc3,__vc4)}
\fmfiset{p2}{vpath(__v10,__vc8)}
\fmfipair{w[]}
\svertex{w1}{p1}
\svertex{w2}{p2}
\fmfi{wiggly}{w1..w2}
\fmffreeze
\end{fmfchar*}}}
}
\subfigure[$\nwgraph{5}{2}{1,2,4,5}$]{
\raisebox{\eqoff}{%
\fmfframe(3,1)(1,4){%
\begin{fmfchar*}(30,20)
\fmftop{v1}
\fmfbottom{v7}
\fmfforce{(0w,h)}{v1}
\fmfforce{(0w,0)}{v7}
\fmffixed{(0.2w,0)}{v1,v2}
\fmffixed{(0.2w,0)}{v2,v3}
\fmffixed{(0.2w,0)}{v3,v4}
\fmffixed{(0.2w,0)}{v4,v5}
\fmffixed{(0.2w,0)}{v5,v6}
\fmffixed{(0.2w,0)}{v7,v8}
\fmffixed{(0.2w,0)}{v8,v9}
\fmffixed{(0.2w,0)}{v9,v10}
\fmffixed{(0.2w,0)}{v10,v11}
\fmffixed{(0.2w,0)}{v11,v12}
\fmf{plain,tension=0.5,right=0.25}{v1,vc1}
\fmf{plain,tension=0.5,left=0.25}{v2,vc1}
\fmf{plain,left=0.25}{v3,vc3}
\fmf{plain,tension=0.5,left=0.25}{v8,vc4}
\fmf{plain,tension=0.5,right=0.25}{v9,vc4}
\fmf{plain,left=0.25}{v7,vc2}
\fmf{plain,tension=1}{vc1,vc2}
\fmf{plain,tension=0.5}{vc2,vc3}
\fmf{plain,tension=1}{vc3,vc4}
\fmf{plain,tension=0.5,right=0.25}{v4,vc7}
\fmf{plain,tension=0.5,left=0.25}{v5,vc7}
\fmf{plain,left=0.25}{v6,vc9}
\fmf{plain,tension=0.5,left=0.25}{v11,vc10}
\fmf{plain,tension=0.5,right=0.25}{v12,vc10}
\fmf{plain,left=0.25}{v10,vc8}
\fmf{plain,tension=1}{vc7,vc8}
\fmf{plain,tension=0.5}{vc8,vc9}
\fmf{plain,tension=1}{vc9,vc10}
\fmffreeze
\fmfposition
\fmf{plain,tension=1,right=0,width=1mm}{v7,v12}
\fmfipath{p[]}
\fmfiset{p1}{vpath(__vc3,__vc4)}
\fmfiset{p2}{vpath(__vc7,__vc8)}
\fmfipair{w[]}
\svertex{w1}{p1}
\svertex{w2}{p2}
\fmfi{wiggly}{w1..w2}
\fmffreeze
\end{fmfchar*}}}
}
\subfigure[$\nwgraph{5}{3}{1,2,4,5}$]{
\raisebox{\eqoff}{%
\fmfframe(3,1)(1,4){%
\begin{fmfchar*}(30,20)
\fmftop{v1}
\fmfbottom{v7}
\fmfforce{(0w,h)}{v1}
\fmfforce{(0w,0)}{v7}
\fmffixed{(0.2w,0)}{v1,v2}
\fmffixed{(0.2w,0)}{v2,v3}
\fmffixed{(0.2w,0)}{v3,v4}
\fmffixed{(0.2w,0)}{v4,v5}
\fmffixed{(0.2w,0)}{v5,v6}
\fmffixed{(0.2w,0)}{v7,v8}
\fmffixed{(0.2w,0)}{v8,v9}
\fmffixed{(0.2w,0)}{v9,v10}
\fmffixed{(0.2w,0)}{v10,v11}
\fmffixed{(0.2w,0)}{v11,v12}
\fmf{plain,tension=0.5,right=0.25}{v1,vc1}
\fmf{plain,tension=0.5,left=0.25}{v2,vc1}
\fmf{plain,left=0.25}{v3,vc3}
\fmf{plain,tension=0.5,left=0.25}{v8,vc4}
\fmf{plain,tension=0.5,right=0.25}{v9,vc4}
\fmf{plain,left=0.25}{v7,vc2}
\fmf{plain,tension=1}{vc1,vc2}
\fmf{plain,tension=0.5}{vc2,vc3}
\fmf{plain,tension=1}{vc3,vc4}
\fmf{plain,tension=0.5,right=0.25}{v4,vc7}
\fmf{plain,tension=0.5,left=0.25}{v5,vc7}
\fmf{plain,left=0.25}{v6,vc9}
\fmf{plain,tension=0.5,left=0.25}{v11,vc10}
\fmf{plain,tension=0.5,right=0.25}{v12,vc10}
\fmf{plain,left=0.25}{v10,vc8}
\fmf{plain,tension=1}{vc7,vc8}
\fmf{plain,tension=0.5}{vc8,vc9}
\fmf{plain,tension=1}{vc9,vc10}
\fmffreeze
\fmfposition
\fmf{plain,tension=1,right=0,width=1mm}{v7,v12}
\fmfipath{p[]}
\fmfiset{p1}{vpath(__v9,__vc4)}
\fmfiset{p2}{vpath(__v10,__vc8)}
\fmfipair{w[]}
\svertex{w1}{p1}
\svertex{w2}{p2}
\fmfi{wiggly}{w1..w2}
\fmffreeze
\end{fmfchar*}}}
}
\subfigure[$\nwgraph{5}{4}{1,2,4,5}$]{
\raisebox{\eqoff}{%
\fmfframe(3,1)(1,4){%
\begin{fmfchar*}(30,20)
\fmftop{v1}
\fmfbottom{v7}
\fmfforce{(0w,h)}{v1}
\fmfforce{(0w,0)}{v7}
\fmffixed{(0.2w,0)}{v1,v2}
\fmffixed{(0.2w,0)}{v2,v3}
\fmffixed{(0.2w,0)}{v3,v4}
\fmffixed{(0.2w,0)}{v4,v5}
\fmffixed{(0.2w,0)}{v5,v6}
\fmffixed{(0.2w,0)}{v7,v8}
\fmffixed{(0.2w,0)}{v8,v9}
\fmffixed{(0.2w,0)}{v9,v10}
\fmffixed{(0.2w,0)}{v10,v11}
\fmffixed{(0.2w,0)}{v11,v12}
\fmf{plain,tension=0.5,right=0.25}{v1,vc1}
\fmf{plain,tension=0.5,left=0.25}{v2,vc1}
\fmf{plain,left=0.25}{v3,vc3}
\fmf{plain,tension=0.5,left=0.25}{v8,vc4}
\fmf{plain,tension=0.5,right=0.25}{v9,vc4}
\fmf{plain,left=0.25}{v7,vc2}
\fmf{plain,tension=1}{vc1,vc2}
\fmf{plain,tension=0.5}{vc2,vc3}
\fmf{plain,tension=1}{vc3,vc4}
\fmf{plain,tension=0.5,right=0.25}{v4,vc7}
\fmf{plain,tension=0.5,left=0.25}{v5,vc7}
\fmf{plain,left=0.25}{v6,vc9}
\fmf{plain,tension=0.5,left=0.25}{v11,vc10}
\fmf{plain,tension=0.5,right=0.25}{v12,vc10}
\fmf{plain,left=0.25}{v10,vc8}
\fmf{plain,tension=1}{vc7,vc8}
\fmf{plain,tension=0.5}{vc8,vc9}
\fmf{plain,tension=1}{vc9,vc10}
\fmffreeze
\fmfposition
\fmf{plain,tension=1,right=0,width=1mm}{v7,v12}
\fmfipath{p[]}
\fmfiset{p1}{vpath(__v9,__vc4)}
\fmfiset{p2}{vpath(__vc7,__vc8)}
\fmfipair{w[]}
\svertex{w1}{p1}
\svertex{w2}{p2}
\fmfi{wiggly}{w1..w2}
\fmffreeze
\end{fmfchar*}}}
}
\\[0.5cm]
\begin{tabular}{m{8cm}}
\toprule
$\nwgraph{5}{1}{1,2,4,5}\to \jint{5}{10}\,\chi(1,2,4,5)$  \\
$\nwgraph{5}{2}{1,2,4,5}\to -\jint{5}{7}\,\chi(1,2,4,5)$  \\
$\nwgraph{5}{3}{1,2,4,5}\to -(\jint{5}{1}+\jint{5}{10}+2\jint{5}{13})\,\chi(1,2,4,5)$  \\
$\nwgraph{5}{4}{1,2,4,5}\to (\jint{5}{1}+\jint{5}{7})\,\chi(1,2,4,5)$  \\
\midrule
$\sum_i \nwgraph{5}{i}{1,2,4,5} \to -2\jint{5}{13}\,\chi(1,2,4,5)  $ \\
\bottomrule
\end{tabular}
\normalsize
\caption{Range-six diagrams with structure $\chi(1,2,4,5)$}
\label{r6-1245}
\end{figure}

\clearpage

\begin{figure}[p]
\capstart
\renewcommand*{\thesubfigure}{\ \ }
\footnotesize
\centering
\unitlength=0.75mm
\settoheight{\eqoff}{$\times$}%
\setlength{\eqoff}{0.5\eqoff}%
\addtolength{\eqoff}{-12.5\unitlength}%
\settoheight{\eqofftwo}{$\times$}%
\setlength{\eqofftwo}{0.5\eqofftwo}%
\addtolength{\eqofftwo}{-7.5\unitlength}%
\subfigure[$\nwgraph{5}{1}{1,5,4,3}$]{
\raisebox{\eqoff}{%
\fmfframe(3,1)(1,4){%
\begin{fmfchar*}(30,20)
\fmftop{v1}
\fmfbottom{v7}
\fmfforce{(0w,h)}{v1}
\fmfforce{(0w,0)}{v7}
\fmffixed{(0.2w,0)}{v1,v2}
\fmffixed{(0.2w,0)}{v2,v3}
\fmffixed{(0.2w,0)}{v3,v4}
\fmffixed{(0.2w,0)}{v4,v5}
\fmffixed{(0.2w,0)}{v5,v6}
\fmffixed{(0.2w,0)}{v7,v8}
\fmffixed{(0.2w,0)}{v8,v9}
\fmffixed{(0.2w,0)}{v9,v10}
\fmffixed{(0.2w,0)}{v10,v11}
\fmffixed{(0.2w,0)}{v11,v12}
\fmf{plain,tension=0.25,right=0.25}{v5,vc9}
\fmf{plain,tension=0.25,left=0.25}{v6,vc9}
\fmf{plain,right=0.25}{v11,vc8}
\fmf{plain,tension=1,right=0.25}{v3,vc5}
\fmf{plain,tension=1,right=0.25}{v4,vc7}
\fmf{plain,right=0.25}{v12,vc10}
\fmf{plain,tension=0.25,left=0.25}{v9,vc6}
\fmf{plain,tension=0.25,right=0.25}{v10,vc6}
\fmf{plain,tension=0.5}{vc5,vc6}
\fmf{plain,tension=0.5}{vc5,vc8}
\fmf{plain,tension=0.5}{vc7,vc8}
\fmf{plain,tension=0.5}{vc7,vc10}
\fmf{plain,tension=0.5}{vc9,vc10}
\fmf{plain,tension=0.25,right=0.25}{v1,vc1}
\fmf{plain,tension=0.25,left=0.25}{v2,vc1}
\fmf{plain,tension=0.25,left=0.25}{v7,vc2}
\fmf{plain,tension=0.25,right=0.25}{v8,vc2}
\fmf{plain,tension=0.5}{vc1,vc2}
\fmffreeze
\fmfposition
\fmf{plain,tension=1,right=0,width=1mm}{v7,v12}
\fmfipath{p[]}
\fmfiset{p1}{vpath(__vc1,__vc2)}
\fmfiset{p2}{vpath(__vc5,__vc6)}
\fmfipair{w[]}
\svertex{w1}{p1}
\svertex{w2}{p2}
\fmfi{wiggly}{w1..w2}
\fmffreeze
\end{fmfchar*}}}
}
\subfigure[$\nwgraph{5}{2}{1,5,4,3}$]{
\raisebox{\eqoff}{%
\fmfframe(3,1)(1,4){%
\begin{fmfchar*}(30,20)
\fmftop{v1}
\fmfbottom{v7}
\fmfforce{(0w,h)}{v1}
\fmfforce{(0w,0)}{v7}
\fmffixed{(0.2w,0)}{v1,v2}
\fmffixed{(0.2w,0)}{v2,v3}
\fmffixed{(0.2w,0)}{v3,v4}
\fmffixed{(0.2w,0)}{v4,v5}
\fmffixed{(0.2w,0)}{v5,v6}
\fmffixed{(0.2w,0)}{v7,v8}
\fmffixed{(0.2w,0)}{v8,v9}
\fmffixed{(0.2w,0)}{v9,v10}
\fmffixed{(0.2w,0)}{v10,v11}
\fmffixed{(0.2w,0)}{v11,v12}
\fmf{plain,tension=0.25,right=0.25}{v5,vc9}
\fmf{plain,tension=0.25,left=0.25}{v6,vc9}
\fmf{plain,right=0.25}{v11,vc8}
\fmf{plain,tension=1,right=0.25}{v3,vc5}
\fmf{plain,tension=1,right=0.25}{v4,vc7}
\fmf{plain,right=0.25}{v12,vc10}
\fmf{plain,tension=0.25,left=0.25}{v9,vc6}
\fmf{plain,tension=0.25,right=0.25}{v10,vc6}
\fmf{plain,tension=0.5}{vc5,vc6}
\fmf{plain,tension=0.5}{vc5,vc8}
\fmf{plain,tension=0.5}{vc7,vc8}
\fmf{plain,tension=0.5}{vc7,vc10}
\fmf{plain,tension=0.5}{vc9,vc10}
\fmf{plain,tension=0.25,right=0.25}{v1,vc1}
\fmf{plain,tension=0.25,left=0.25}{v2,vc1}
\fmf{plain,tension=0.25,left=0.25}{v7,vc2}
\fmf{plain,tension=0.25,right=0.25}{v8,vc2}
\fmf{plain,tension=0.5}{vc1,vc2}
\fmffreeze
\fmfposition
\fmf{plain,tension=1,right=0,width=1mm}{v7,v12}
\fmfipath{p[]}
\fmfiset{p1}{vpath(__vc1,__vc2)}
\fmfiset{p2}{vpath(__v9,__vc6)}
\fmfipair{w[]}
\svertex{w1}{p1}
\svertex{w2}{p2}
\fmfi{wiggly}{w1..w2}
\fmffreeze
\end{fmfchar*}}}
}
\subfigure[$\nwgraph{5}{3}{1,5,4,3}$]{
\raisebox{\eqoff}{%
\fmfframe(3,1)(1,4){%
\begin{fmfchar*}(30,20)
\fmftop{v1}
\fmfbottom{v7}
\fmfforce{(0w,h)}{v1}
\fmfforce{(0w,0)}{v7}
\fmffixed{(0.2w,0)}{v1,v2}
\fmffixed{(0.2w,0)}{v2,v3}
\fmffixed{(0.2w,0)}{v3,v4}
\fmffixed{(0.2w,0)}{v4,v5}
\fmffixed{(0.2w,0)}{v5,v6}
\fmffixed{(0.2w,0)}{v7,v8}
\fmffixed{(0.2w,0)}{v8,v9}
\fmffixed{(0.2w,0)}{v9,v10}
\fmffixed{(0.2w,0)}{v10,v11}
\fmffixed{(0.2w,0)}{v11,v12}
\fmf{plain,tension=0.25,right=0.25}{v5,vc9}
\fmf{plain,tension=0.25,left=0.25}{v6,vc9}
\fmf{plain,right=0.25}{v11,vc8}
\fmf{plain,tension=1,right=0.25}{v3,vc5}
\fmf{plain,tension=1,right=0.25}{v4,vc7}
\fmf{plain,right=0.25}{v12,vc10}
\fmf{plain,tension=0.25,left=0.25}{v9,vc6}
\fmf{plain,tension=0.25,right=0.25}{v10,vc6}
\fmf{plain,tension=0.5}{vc5,vc6}
\fmf{plain,tension=0.5}{vc5,vc8}
\fmf{plain,tension=0.5}{vc7,vc8}
\fmf{plain,tension=0.5}{vc7,vc10}
\fmf{plain,tension=0.5}{vc9,vc10}
\fmf{plain,tension=0.25,right=0.25}{v1,vc1}
\fmf{plain,tension=0.25,left=0.25}{v2,vc1}
\fmf{plain,tension=0.25,left=0.25}{v7,vc2}
\fmf{plain,tension=0.25,right=0.25}{v8,vc2}
\fmf{plain,tension=0.5}{vc1,vc2}
\fmffreeze
\fmfposition
\fmf{plain,tension=1,right=0,width=1mm}{v7,v12}
\fmfipath{p[]}
\fmfiset{p1}{vpath(__v8,__vc2)}
\fmfiset{p2}{vpath(__vc5,__vc6)}
\fmfipair{w[]}
\svertex{w1}{p1}
\svertex{w2}{p2}
\fmfi{wiggly}{w1..w2}
\fmffreeze
\end{fmfchar*}}}
}
\subfigure[$\nwgraph{5}{4}{1,5,4,3}$]{
\raisebox{\eqoff}{%
\fmfframe(3,1)(1,4){%
\begin{fmfchar*}(30,20)
\fmftop{v1}
\fmfbottom{v7}
\fmfforce{(0w,h)}{v1}
\fmfforce{(0w,0)}{v7}
\fmffixed{(0.2w,0)}{v1,v2}
\fmffixed{(0.2w,0)}{v2,v3}
\fmffixed{(0.2w,0)}{v3,v4}
\fmffixed{(0.2w,0)}{v4,v5}
\fmffixed{(0.2w,0)}{v5,v6}
\fmffixed{(0.2w,0)}{v7,v8}
\fmffixed{(0.2w,0)}{v8,v9}
\fmffixed{(0.2w,0)}{v9,v10}
\fmffixed{(0.2w,0)}{v10,v11}
\fmffixed{(0.2w,0)}{v11,v12}
\fmf{plain,tension=0.25,right=0.25}{v5,vc9}
\fmf{plain,tension=0.25,left=0.25}{v6,vc9}
\fmf{plain,right=0.25}{v11,vc8}
\fmf{plain,tension=1,right=0.25}{v3,vc5}
\fmf{plain,tension=1,right=0.25}{v4,vc7}
\fmf{plain,right=0.25}{v12,vc10}
\fmf{plain,tension=0.25,left=0.25}{v9,vc6}
\fmf{plain,tension=0.25,right=0.25}{v10,vc6}
\fmf{plain,tension=0.5}{vc5,vc6}
\fmf{plain,tension=0.5}{vc5,vc8}
\fmf{plain,tension=0.5}{vc7,vc8}
\fmf{plain,tension=0.5}{vc7,vc10}
\fmf{plain,tension=0.5}{vc9,vc10}
\fmf{plain,tension=0.25,right=0.25}{v1,vc1}
\fmf{plain,tension=0.25,left=0.25}{v2,vc1}
\fmf{plain,tension=0.25,left=0.25}{v7,vc2}
\fmf{plain,tension=0.25,right=0.25}{v8,vc2}
\fmf{plain,tension=0.5}{vc1,vc2}
\fmffreeze
\fmfposition
\fmf{plain,tension=1,right=0,width=1mm}{v7,v12}
\fmfipath{p[]}
\fmfiset{p1}{vpath(__v8,__vc2)}
\fmfiset{p2}{vpath(__v9,__vc6)}
\fmfipair{w[]}
\svertex{w1}{p1}
\svertex{w2}{p2}
\fmfi{wiggly}{w1..w2}
\fmffreeze
\end{fmfchar*}}}
}
\\[0.5cm]
\begin{tabular}{m{8cm}}
\toprule
$\nwgraph{5}{1}{1,5,4,3}\to 0 $  \\
$\nwgraph{5}{2}{1,5,4,3}\to \jint{5}{1}\,\chi(1,5,4,3)$  \\
$\nwgraph{5}{3}{1,5,4,3}\to \jint{5}{4}\,\chi(1,5,4,3)$  \\
$\nwgraph{5}{4}{1,5,4,3}\to -(\jint{5}{1}+\jint{5}{4}+2\jint{5}{11})\,\chi(1,5,4,3)$  \\
\midrule
$\sum_i \nwgraph{5}{i}{1,5,4,3} \to -2\jint{5}{11}\,\chi(1,5,4,3)  $ \\
\bottomrule
\end{tabular}
\normalsize
\caption{Range-six diagrams with structure $\chi(1,5,4,3)$}
\label{r6-1543}
\end{figure}

\begin{figure}[p]
\capstart
\renewcommand*{\thesubfigure}{\ \ }
\footnotesize
\centering
\unitlength=0.75mm
\settoheight{\eqoff}{$\times$}%
\setlength{\eqoff}{0.5\eqoff}%
\addtolength{\eqoff}{-12.5\unitlength}%
\settoheight{\eqofftwo}{$\times$}%
\setlength{\eqofftwo}{0.5\eqofftwo}%
\addtolength{\eqofftwo}{-7.5\unitlength}%
\subfigure[$\nwgraph{5}{1}{1,3,4,5}$]{
\raisebox{\eqoff}{%
\fmfframe(3,1)(1,4){%
\begin{fmfchar*}(30,20)
\fmftop{v1}
\fmfbottom{v7}
\fmfforce{(0w,h)}{v1}
\fmfforce{(0w,0)}{v7}
\fmffixed{(0.2w,0)}{v1,v2}
\fmffixed{(0.2w,0)}{v2,v3}
\fmffixed{(0.2w,0)}{v3,v4}
\fmffixed{(0.2w,0)}{v4,v5}
\fmffixed{(0.2w,0)}{v5,v6}
\fmffixed{(0.2w,0)}{v7,v8}
\fmffixed{(0.2w,0)}{v8,v9}
\fmffixed{(0.2w,0)}{v9,v10}
\fmffixed{(0.2w,0)}{v10,v11}
\fmffixed{(0.2w,0)}{v11,v12}
\fmf{plain,tension=0.25,right=0.25}{v3,vc5}
\fmf{plain,tension=0.25,left=0.25}{v4,vc5}
\fmf{plain,left=0.25}{v9,vc6}
\fmf{plain,tension=1,left=0.25}{v5,vc7}
\fmf{plain,tension=1,left=0.25}{v6,vc9}
\fmf{plain,left=0.25}{v10,vc8}
\fmf{plain,tension=0.25,left=0.25}{v11,vc10}
\fmf{plain,tension=0.25,right=0.25}{v12,vc10}
\fmf{plain,tension=0.5}{vc5,vc6}
\fmf{plain,tension=0.5}{vc6,vc7}
\fmf{plain,tension=0.5}{vc7,vc8}
\fmf{plain,tension=0.5}{vc8,vc9}
\fmf{plain,tension=0.5}{vc9,vc10}
\fmf{plain,tension=0.25,right=0.25}{v1,vc1}
\fmf{plain,tension=0.25,left=0.25}{v2,vc1}
\fmf{plain,tension=0.25,left=0.25}{v7,vc2}
\fmf{plain,tension=0.25,right=0.25}{v8,vc2}
\fmf{plain,tension=0.5}{vc1,vc2}
\fmffreeze
\fmfposition
\fmf{plain,tension=1,right=0,width=1mm}{v7,v12}
\fmfipath{p[]}
\fmfiset{p1}{vpath(__vc1,__vc2)}
\fmfiset{p2}{vpath(__v9,__vc6)}
\fmfipair{w[]}
\svertex{w1}{p1}
\svertex{w2}{p2}
\fmfi{wiggly}{w1..w2}
\fmffreeze
\end{fmfchar*}}}
}
\subfigure[$\nwgraph{5}{2}{1,3,4,5}$]{
\raisebox{\eqoff}{%
\fmfframe(3,1)(1,4){%
\begin{fmfchar*}(30,20)
\fmftop{v1}
\fmfbottom{v7}
\fmfforce{(0w,h)}{v1}
\fmfforce{(0w,0)}{v7}
\fmffixed{(0.2w,0)}{v1,v2}
\fmffixed{(0.2w,0)}{v2,v3}
\fmffixed{(0.2w,0)}{v3,v4}
\fmffixed{(0.2w,0)}{v4,v5}
\fmffixed{(0.2w,0)}{v5,v6}
\fmffixed{(0.2w,0)}{v7,v8}
\fmffixed{(0.2w,0)}{v8,v9}
\fmffixed{(0.2w,0)}{v9,v10}
\fmffixed{(0.2w,0)}{v10,v11}
\fmffixed{(0.2w,0)}{v11,v12}
\fmf{plain,tension=0.25,right=0.25}{v3,vc5}
\fmf{plain,tension=0.25,left=0.25}{v4,vc5}
\fmf{plain,left=0.25}{v9,vc6}
\fmf{plain,tension=1,left=0.25}{v5,vc7}
\fmf{plain,tension=1,left=0.25}{v6,vc9}
\fmf{plain,left=0.25}{v10,vc8}
\fmf{plain,tension=0.25,left=0.25}{v11,vc10}
\fmf{plain,tension=0.25,right=0.25}{v12,vc10}
\fmf{plain,tension=0.5}{vc5,vc6}
\fmf{plain,tension=0.5}{vc6,vc7}
\fmf{plain,tension=0.5}{vc7,vc8}
\fmf{plain,tension=0.5}{vc8,vc9}
\fmf{plain,tension=0.5}{vc9,vc10}
\fmf{plain,tension=0.25,right=0.25}{v1,vc1}
\fmf{plain,tension=0.25,left=0.25}{v2,vc1}
\fmf{plain,tension=0.25,left=0.25}{v7,vc2}
\fmf{plain,tension=0.25,right=0.25}{v8,vc2}
\fmf{plain,tension=0.5}{vc1,vc2}
\fmffreeze
\fmfposition
\fmf{plain,tension=1,right=0,width=1mm}{v7,v12}
\fmfipath{p[]}
\fmfiset{p1}{vpath(__vc1,__vc2)}
\fmfiset{p2}{vpath(__vc5,__vc6)}
\fmfipair{w[]}
\svertex{w1}{p1}
\svertex{w2}{p2}
\fmfi{wiggly}{w1..w2}
\fmffreeze
\end{fmfchar*}}}
}
\subfigure[$\nwgraph{5}{3}{1,3,4,5}$]{
\raisebox{\eqoff}{%
\fmfframe(3,1)(1,4){%
\begin{fmfchar*}(30,20)
\fmftop{v1}
\fmfbottom{v7}
\fmfforce{(0w,h)}{v1}
\fmfforce{(0w,0)}{v7}
\fmffixed{(0.2w,0)}{v1,v2}
\fmffixed{(0.2w,0)}{v2,v3}
\fmffixed{(0.2w,0)}{v3,v4}
\fmffixed{(0.2w,0)}{v4,v5}
\fmffixed{(0.2w,0)}{v5,v6}
\fmffixed{(0.2w,0)}{v7,v8}
\fmffixed{(0.2w,0)}{v8,v9}
\fmffixed{(0.2w,0)}{v9,v10}
\fmffixed{(0.2w,0)}{v10,v11}
\fmffixed{(0.2w,0)}{v11,v12}
\fmf{plain,tension=0.25,right=0.25}{v3,vc5}
\fmf{plain,tension=0.25,left=0.25}{v4,vc5}
\fmf{plain,left=0.25}{v9,vc6}
\fmf{plain,tension=1,left=0.25}{v5,vc7}
\fmf{plain,tension=1,left=0.25}{v6,vc9}
\fmf{plain,left=0.25}{v10,vc8}
\fmf{plain,tension=0.25,left=0.25}{v11,vc10}
\fmf{plain,tension=0.25,right=0.25}{v12,vc10}
\fmf{plain,tension=0.5}{vc5,vc6}
\fmf{plain,tension=0.5}{vc6,vc7}
\fmf{plain,tension=0.5}{vc7,vc8}
\fmf{plain,tension=0.5}{vc8,vc9}
\fmf{plain,tension=0.5}{vc9,vc10}
\fmf{plain,tension=0.25,right=0.25}{v1,vc1}
\fmf{plain,tension=0.25,left=0.25}{v2,vc1}
\fmf{plain,tension=0.25,left=0.25}{v7,vc2}
\fmf{plain,tension=0.25,right=0.25}{v8,vc2}
\fmf{plain,tension=0.5}{vc1,vc2}
\fmffreeze
\fmfposition
\fmf{plain,tension=1,right=0,width=1mm}{v7,v12}
\fmfipath{p[]}
\fmfiset{p1}{vpath(__v8,__vc2)}
\fmfiset{p2}{vpath(__vc5,__vc6)}
\fmfipair{w[]}
\svertex{w1}{p1}
\svertex{w2}{p2}
\fmfi{wiggly}{w1..w2}
\fmffreeze
\end{fmfchar*}}}
}
\subfigure[$\nwgraph{5}{4}{1,3,4,5}$]{
\raisebox{\eqoff}{%
\fmfframe(3,1)(1,4){%
\begin{fmfchar*}(30,20)
\fmftop{v1}
\fmfbottom{v7}
\fmfforce{(0w,h)}{v1}
\fmfforce{(0w,0)}{v7}
\fmffixed{(0.2w,0)}{v1,v2}
\fmffixed{(0.2w,0)}{v2,v3}
\fmffixed{(0.2w,0)}{v3,v4}
\fmffixed{(0.2w,0)}{v4,v5}
\fmffixed{(0.2w,0)}{v5,v6}
\fmffixed{(0.2w,0)}{v7,v8}
\fmffixed{(0.2w,0)}{v8,v9}
\fmffixed{(0.2w,0)}{v9,v10}
\fmffixed{(0.2w,0)}{v10,v11}
\fmffixed{(0.2w,0)}{v11,v12}
\fmf{plain,tension=0.25,right=0.25}{v3,vc5}
\fmf{plain,tension=0.25,left=0.25}{v4,vc5}
\fmf{plain,left=0.25}{v9,vc6}
\fmf{plain,tension=1,left=0.25}{v5,vc7}
\fmf{plain,tension=1,left=0.25}{v6,vc9}
\fmf{plain,left=0.25}{v10,vc8}
\fmf{plain,tension=0.25,left=0.25}{v11,vc10}
\fmf{plain,tension=0.25,right=0.25}{v12,vc10}
\fmf{plain,tension=0.5}{vc5,vc6}
\fmf{plain,tension=0.5}{vc6,vc7}
\fmf{plain,tension=0.5}{vc7,vc8}
\fmf{plain,tension=0.5}{vc8,vc9}
\fmf{plain,tension=0.5}{vc9,vc10}
\fmf{plain,tension=0.25,right=0.25}{v1,vc1}
\fmf{plain,tension=0.25,left=0.25}{v2,vc1}
\fmf{plain,tension=0.25,left=0.25}{v7,vc2}
\fmf{plain,tension=0.25,right=0.25}{v8,vc2}
\fmf{plain,tension=0.5}{vc1,vc2}
\fmffreeze
\fmfposition
\fmf{plain,tension=1,right=0,width=1mm}{v7,v12}
\fmfipath{p[]}
\fmfiset{p1}{vpath(__v8,__vc2)}
\fmfiset{p2}{vpath(__v9,__vc6)}
\fmfipair{w[]}
\svertex{w1}{p1}
\svertex{w2}{p2}
\fmfi{wiggly}{w1..w2}
\fmffreeze
\end{fmfchar*}}}
}
\\[0.5cm]
\begin{tabular}{m{8cm}}
\toprule
$\nwgraph{5}{1}{1,3,4,5}\to \jint{5}{3}\,\chi(1,3,4,5) $  \\
$\nwgraph{5}{2}{1,3,4,5}\to -\jint{5}{9}\,\chi(1,3,4,5) $  \\
$\nwgraph{5}{3}{1,3,4,5}\to (\jint{5}{1}+\jint{5}{9})\,\chi(1,3,4,5) $  \\
$\nwgraph{5}{4}{1,3,4,5}\to -(\jint{5}{1}+\jint{5}{3}+2\jint{5}{12})\,\chi(1,3,4,5) $  \\
\midrule
$\sum_i \nwgraph{5}{i}{1,3,4,5} \to -2\jint{5}{12}\,\chi(1,3,4,5)  $ \\
\bottomrule
\end{tabular}
\normalsize
\caption{Range-six diagrams with structure $\chi(1,3,4,5)$}
\label{r6-1345}
\end{figure}

\begin{figure}[p]
\capstart
\renewcommand*{\thesubfigure}{\ \ }
\footnotesize
\centering
\unitlength=0.75mm
\settoheight{\eqoff}{$\times$}%
\setlength{\eqoff}{0.5\eqoff}%
\addtolength{\eqoff}{-12.5\unitlength}%
\settoheight{\eqofftwo}{$\times$}%
\setlength{\eqofftwo}{0.5\eqofftwo}%
\addtolength{\eqofftwo}{-7.5\unitlength}%
\subfigure[$\nwgraph{5}{1}{1,4,3,5}$]{
\raisebox{\eqoff}{%
\fmfframe(3,1)(1,4){%
\begin{fmfchar*}(30,20)
\fmftop{v1}
\fmfbottom{v7}
\fmfforce{(0w,h)}{v1}
\fmfforce{(0w,0)}{v7}
\fmffixed{(0.2w,0)}{v1,v2}
\fmffixed{(0.2w,0)}{v2,v3}
\fmffixed{(0.2w,0)}{v3,v4}
\fmffixed{(0.2w,0)}{v4,v5}
\fmffixed{(0.2w,0)}{v5,v6}
\fmffixed{(0.2w,0)}{v7,v8}
\fmffixed{(0.2w,0)}{v8,v9}
\fmffixed{(0.2w,0)}{v9,v10}
\fmffixed{(0.2w,0)}{v10,v11}
\fmffixed{(0.2w,0)}{v11,v12}
\fmf{plain,tension=1,left=0.25}{v9,vc6}
\fmf{plain,tension=1,right=0.25}{v10,vc6}
\fmf{plain,tension=1,left=0.25}{v11,vc10}
\fmf{plain,tension=1,right=0.25}{v12,vc10}
\fmf{plain,tension=1,right=0.125}{v3,vc5}
\fmf{plain,tension=0.25,right=0.25}{v4,vc7}
\fmf{plain,tension=0.25,left=0.25}{v5,vc7}
\fmf{plain,tension=1,left=0.125}{v6,vc9}
\fmf{plain,tension=4}{vc5,vc6}
\fmf{plain,tension=4}{vc9,vc10}
\fmf{plain,tension=0.5}{vc5,vc8}
\fmf{plain,tension=0.5}{vc8,vc9}
\fmf{plain,tension=1}{vc7,vc8}
\fmf{plain,tension=1,right=0.25}{v1,vc1}
\fmf{plain,tension=1,left=0.25}{v2,vc1}
\fmf{plain,tension=1,left=0.25}{v7,vc2}
\fmf{plain,tension=1,right=0.25}{v8,vc2}
\fmf{plain,tension=1.5}{vc1,vc2}
\fmffreeze
\fmfposition
\fmf{plain,tension=1,right=0,width=1mm}{v7,v12}
\fmfipath{p[]}
\fmfiset{p1}{vpath(__v8,__vc2)}
\fmfiset{p2}{vpath(__v9,__vc6)}
\fmfipair{w[]}
\svertex{w1}{p1}
\svertex{w2}{p2}
\fmfi{wiggly}{w1..w2}
\fmffreeze
\end{fmfchar*}}}
}
\subfigure[$\nwgraph{5}{2}{1,4,3,5}$]{
\raisebox{\eqoff}{%
\fmfframe(3,1)(1,4){%
\begin{fmfchar*}(30,20)
\fmftop{v1}
\fmfbottom{v7}
\fmfforce{(0w,h)}{v1}
\fmfforce{(0w,0)}{v7}
\fmffixed{(0.2w,0)}{v1,v2}
\fmffixed{(0.2w,0)}{v2,v3}
\fmffixed{(0.2w,0)}{v3,v4}
\fmffixed{(0.2w,0)}{v4,v5}
\fmffixed{(0.2w,0)}{v5,v6}
\fmffixed{(0.2w,0)}{v7,v8}
\fmffixed{(0.2w,0)}{v8,v9}
\fmffixed{(0.2w,0)}{v9,v10}
\fmffixed{(0.2w,0)}{v10,v11}
\fmffixed{(0.2w,0)}{v11,v12}
\fmf{plain,tension=1,left=0.25}{v9,vc6}
\fmf{plain,tension=1,right=0.25}{v10,vc6}
\fmf{plain,tension=1,left=0.25}{v11,vc10}
\fmf{plain,tension=1,right=0.25}{v12,vc10}
\fmf{plain,tension=1,right=0.125}{v3,vc5}
\fmf{plain,tension=0.25,right=0.25}{v4,vc7}
\fmf{plain,tension=0.25,left=0.25}{v5,vc7}
\fmf{plain,tension=1,left=0.125}{v6,vc9}
\fmf{plain,tension=4}{vc5,vc6}
\fmf{plain,tension=4}{vc9,vc10}
\fmf{plain,tension=0.5}{vc5,vc8}
\fmf{plain,tension=0.5}{vc8,vc9}
\fmf{plain,tension=1}{vc7,vc8}
\fmf{plain,tension=1,right=0.25}{v1,vc1}
\fmf{plain,tension=1,left=0.25}{v2,vc1}
\fmf{plain,tension=1,left=0.25}{v7,vc2}
\fmf{plain,tension=1,right=0.25}{v8,vc2}
\fmf{plain,tension=1.5}{vc1,vc2}
\fmffreeze
\fmfposition
\fmf{plain,tension=1,right=0,width=1mm}{v7,v12}
\fmfipath{p[]}
\fmfiset{p1}{vpath(__v8,__vc2)}
\fmfiset{p2}{vpath(__vc5,__vc6)}
\fmfipair{w[]}
\svertex{w1}{p1}
\svertex{w2}{p2}
\fmfi{wiggly}{w1..w2}
\fmffreeze
\end{fmfchar*}}}
}
\subfigure[$\nwgraph{5}{3}{1,4,3,5}$]{
\raisebox{\eqoff}{%
\fmfframe(3,1)(1,4){%
\begin{fmfchar*}(30,20)
\fmftop{v1}
\fmfbottom{v7}
\fmfforce{(0w,h)}{v1}
\fmfforce{(0w,0)}{v7}
\fmffixed{(0.2w,0)}{v1,v2}
\fmffixed{(0.2w,0)}{v2,v3}
\fmffixed{(0.2w,0)}{v3,v4}
\fmffixed{(0.2w,0)}{v4,v5}
\fmffixed{(0.2w,0)}{v5,v6}
\fmffixed{(0.2w,0)}{v7,v8}
\fmffixed{(0.2w,0)}{v8,v9}
\fmffixed{(0.2w,0)}{v9,v10}
\fmffixed{(0.2w,0)}{v10,v11}
\fmffixed{(0.2w,0)}{v11,v12}
\fmf{plain,tension=1,left=0.25}{v9,vc6}
\fmf{plain,tension=1,right=0.25}{v10,vc6}
\fmf{plain,tension=1,left=0.25}{v11,vc10}
\fmf{plain,tension=1,right=0.25}{v12,vc10}
\fmf{plain,tension=1,right=0.125}{v3,vc5}
\fmf{plain,tension=0.25,right=0.25}{v4,vc7}
\fmf{plain,tension=0.25,left=0.25}{v5,vc7}
\fmf{plain,tension=1,left=0.125}{v6,vc9}
\fmf{plain,tension=4}{vc5,vc6}
\fmf{plain,tension=4}{vc9,vc10}
\fmf{plain,tension=0.5}{vc5,vc8}
\fmf{plain,tension=0.5}{vc8,vc9}
\fmf{plain,tension=1}{vc7,vc8}
\fmf{plain,tension=1,right=0.25}{v1,vc1}
\fmf{plain,tension=1,left=0.25}{v2,vc1}
\fmf{plain,tension=1,left=0.25}{v7,vc2}
\fmf{plain,tension=1,right=0.25}{v8,vc2}
\fmf{plain,tension=1.5}{vc1,vc2}
\fmffreeze
\fmfposition
\fmf{plain,tension=1,right=0,width=1mm}{v7,v12}
\fmfipath{p[]}
\fmfiset{p1}{vpath(__vc1,__vc2)}
\fmfiset{p2}{vpath(__v9,__vc6)}
\fmfipair{w[]}
\svertex{w1}{p1}
\svertex{w2}{p2}
\fmfi{wiggly}{w1..w2}
\fmffreeze
\end{fmfchar*}}}
}
\subfigure[$\nwgraph{5}{4}{1,4,3,5}$]{
\raisebox{\eqoff}{%
\fmfframe(3,1)(1,4){%
\begin{fmfchar*}(30,20)
\fmftop{v1}
\fmfbottom{v7}
\fmfforce{(0w,h)}{v1}
\fmfforce{(0w,0)}{v7}
\fmffixed{(0.2w,0)}{v1,v2}
\fmffixed{(0.2w,0)}{v2,v3}
\fmffixed{(0.2w,0)}{v3,v4}
\fmffixed{(0.2w,0)}{v4,v5}
\fmffixed{(0.2w,0)}{v5,v6}
\fmffixed{(0.2w,0)}{v7,v8}
\fmffixed{(0.2w,0)}{v8,v9}
\fmffixed{(0.2w,0)}{v9,v10}
\fmffixed{(0.2w,0)}{v10,v11}
\fmffixed{(0.2w,0)}{v11,v12}
\fmf{plain,tension=1,left=0.25}{v9,vc6}
\fmf{plain,tension=1,right=0.25}{v10,vc6}
\fmf{plain,tension=1,left=0.25}{v11,vc10}
\fmf{plain,tension=1,right=0.25}{v12,vc10}
\fmf{plain,tension=1,right=0.125}{v3,vc5}
\fmf{plain,tension=0.25,right=0.25}{v4,vc7}
\fmf{plain,tension=0.25,left=0.25}{v5,vc7}
\fmf{plain,tension=1,left=0.125}{v6,vc9}
\fmf{plain,tension=4}{vc5,vc6}
\fmf{plain,tension=4}{vc9,vc10}
\fmf{plain,tension=0.5}{vc5,vc8}
\fmf{plain,tension=0.5}{vc8,vc9}
\fmf{plain,tension=1}{vc7,vc8}
\fmf{plain,tension=1,right=0.25}{v1,vc1}
\fmf{plain,tension=1,left=0.25}{v2,vc1}
\fmf{plain,tension=1,left=0.25}{v7,vc2}
\fmf{plain,tension=1,right=0.25}{v8,vc2}
\fmf{plain,tension=1.5}{vc1,vc2}
\fmffreeze
\fmfposition
\fmf{plain,tension=1,right=0,width=1mm}{v7,v12}
\fmfipath{p[]}
\fmfiset{p1}{vpath(__vc1,__vc2)}
\fmfiset{p2}{vpath(__vc5,__vc6)}
\fmfipair{w[]}
\svertex{w1}{p1}
\svertex{w2}{p2}
\fmfi{wiggly}{w1..w2}
\fmffreeze
\end{fmfchar*}}}
}
\\[0.5cm]
\begin{tabular}{m{8cm}}
\toprule
$\nwgraph{5}{1}{1,4,3,5}\to -(\jint{5}{5}+\jint{5}{9}+2\jint{5}{16})\,\chi(1,4,3,5) $  \\
$\nwgraph{5}{2}{1,4,3,5}\to \jint{5}{5}\,\chi(1,4,3,5) $  \\
$\nwgraph{5}{3}{1,4,3,5}\to \jint{5}{9}\,\chi(1,4,3,5) $  \\
$\nwgraph{5}{4}{1,4,3,5}\to 0 $  \\
\midrule
$\sum_i \nwgraph{5}{i}{1,4,3,5} \to -2\jint{5}{16}\,\chi(1,4,3,5)  $ \\
\bottomrule
\end{tabular}
\normalsize
\caption{Range-six diagrams with structure $\chi(1,4,3,5)$}
\label{r6-1435}
\end{figure}

\begin{figure}[p]
\capstart
\renewcommand*{\thesubfigure}{\ \ }
\footnotesize
\centering
\unitlength=0.75mm
\settoheight{\eqoff}{$\times$}%
\setlength{\eqoff}{0.5\eqoff}%
\addtolength{\eqoff}{-12.5\unitlength}%
\settoheight{\eqofftwo}{$\times$}%
\setlength{\eqofftwo}{0.5\eqofftwo}%
\addtolength{\eqofftwo}{-7.5\unitlength}%
\subfigure[$\nwgraph{5}{1}{1,3,2,5}$]{
\raisebox{\eqoff}{%
\fmfframe(3,1)(1,4){%
\begin{fmfchar*}(30,20)
\fmftop{v1}
\fmfbottom{v7}
\fmfforce{(0w,h)}{v1}
\fmfforce{(0w,0)}{v7}
\fmffixed{(0.2w,0)}{v1,v2}
\fmffixed{(0.2w,0)}{v2,v3}
\fmffixed{(0.2w,0)}{v3,v4}
\fmffixed{(0.2w,0)}{v4,v5}
\fmffixed{(0.2w,0)}{v5,v6}
\fmffixed{(0.2w,0)}{v7,v8}
\fmffixed{(0.2w,0)}{v8,v9}
\fmffixed{(0.2w,0)}{v9,v10}
\fmffixed{(0.2w,0)}{v10,v11}
\fmffixed{(0.2w,0)}{v11,v12}
\fmf{plain,tension=1,right=0.25}{v3,vc5}
\fmf{plain,tension=1,left=0.25}{v4,vc5}
\fmf{plain,tension=1,right=0.25}{v5,vc9}
\fmf{plain,tension=1,left=0.25}{v6,vc9}
\fmf{plain,tension=1,left=0.125}{v9,vc6}
\fmf{plain,tension=0.25,left=0.25}{v10,vc8}
\fmf{plain,tension=0.25,right=0.25}{v11,vc8}
\fmf{plain,tension=1,right=0.125}{v12,vc10}
\fmf{plain,tension=4}{vc5,vc6}
\fmf{plain,tension=4}{vc9,vc10}
\fmf{plain,tension=0.5}{vc6,vc7}
\fmf{plain,tension=0.5}{vc7,vc10}
\fmf{plain,tension=1}{vc7,vc8}
\fmf{plain,tension=1,right=0.25}{v1,vc1}
\fmf{plain,tension=1,left=0.25}{v2,vc1}
\fmf{plain,tension=1,left=0.25}{v7,vc2}
\fmf{plain,tension=1,right=0.25}{v8,vc2}
\fmf{plain,tension=1.5}{vc1,vc2}
\fmffreeze
\fmfposition
\fmf{plain,tension=1,right=0,width=1mm}{v7,v12}
\fmfipath{p[]}
\fmfiset{p1}{vpath(__v8,__vc2)}
\fmfiset{p2}{vpath(__v9,__vc6)}
\fmfipair{w[]}
\svertex{w1}{p1}
\svertex{w2}{p2}
\fmfi{wiggly}{w1..w2}
\fmffreeze
\end{fmfchar*}}}
}
\subfigure[$\nwgraph{5}{2}{1,3,2,5}$]{
\raisebox{\eqoff}{%
\fmfframe(3,1)(1,4){%
\begin{fmfchar*}(30,20)
\fmftop{v1}
\fmfbottom{v7}
\fmfforce{(0w,h)}{v1}
\fmfforce{(0w,0)}{v7}
\fmffixed{(0.2w,0)}{v1,v2}
\fmffixed{(0.2w,0)}{v2,v3}
\fmffixed{(0.2w,0)}{v3,v4}
\fmffixed{(0.2w,0)}{v4,v5}
\fmffixed{(0.2w,0)}{v5,v6}
\fmffixed{(0.2w,0)}{v7,v8}
\fmffixed{(0.2w,0)}{v8,v9}
\fmffixed{(0.2w,0)}{v9,v10}
\fmffixed{(0.2w,0)}{v10,v11}
\fmffixed{(0.2w,0)}{v11,v12}
\fmf{plain,tension=1,right=0.25}{v3,vc5}
\fmf{plain,tension=1,left=0.25}{v4,vc5}
\fmf{plain,tension=1,right=0.25}{v5,vc9}
\fmf{plain,tension=1,left=0.25}{v6,vc9}
\fmf{plain,tension=1,left=0.125}{v9,vc6}
\fmf{plain,tension=0.25,left=0.25}{v10,vc8}
\fmf{plain,tension=0.25,right=0.25}{v11,vc8}
\fmf{plain,tension=1,right=0.125}{v12,vc10}
\fmf{plain,tension=4}{vc5,vc6}
\fmf{plain,tension=4}{vc9,vc10}
\fmf{plain,tension=0.5}{vc6,vc7}
\fmf{plain,tension=0.5}{vc7,vc10}
\fmf{plain,tension=1}{vc7,vc8}
\fmf{plain,tension=1,right=0.25}{v1,vc1}
\fmf{plain,tension=1,left=0.25}{v2,vc1}
\fmf{plain,tension=1,left=0.25}{v7,vc2}
\fmf{plain,tension=1,right=0.25}{v8,vc2}
\fmf{plain,tension=1.5}{vc1,vc2}
\fmffreeze
\fmfposition
\fmf{plain,tension=1,right=0,width=1mm}{v7,v12}
\fmfipath{p[]}
\fmfiset{p1}{vpath(__vc1,__vc2)}
\fmfiset{p2}{vpath(__v9,__vc6)}
\fmfipair{w[]}
\svertex{w1}{p1}
\svertex{w2}{p2}
\fmfi{wiggly}{w1..w2}
\fmffreeze
\end{fmfchar*}}}
}
\subfigure[$\nwgraph{5}{3}{1,3,2,5}$]{
\raisebox{\eqoff}{%
\fmfframe(3,1)(1,4){%
\begin{fmfchar*}(30,20)
\fmftop{v1}
\fmfbottom{v7}
\fmfforce{(0w,h)}{v1}
\fmfforce{(0w,0)}{v7}
\fmffixed{(0.2w,0)}{v1,v2}
\fmffixed{(0.2w,0)}{v2,v3}
\fmffixed{(0.2w,0)}{v3,v4}
\fmffixed{(0.2w,0)}{v4,v5}
\fmffixed{(0.2w,0)}{v5,v6}
\fmffixed{(0.2w,0)}{v7,v8}
\fmffixed{(0.2w,0)}{v8,v9}
\fmffixed{(0.2w,0)}{v9,v10}
\fmffixed{(0.2w,0)}{v10,v11}
\fmffixed{(0.2w,0)}{v11,v12}
\fmf{plain,tension=1,right=0.25}{v3,vc5}
\fmf{plain,tension=1,left=0.25}{v4,vc5}
\fmf{plain,tension=1,right=0.25}{v5,vc9}
\fmf{plain,tension=1,left=0.25}{v6,vc9}
\fmf{plain,tension=1,left=0.125}{v9,vc6}
\fmf{plain,tension=0.25,left=0.25}{v10,vc8}
\fmf{plain,tension=0.25,right=0.25}{v11,vc8}
\fmf{plain,tension=1,right=0.125}{v12,vc10}
\fmf{plain,tension=4}{vc5,vc6}
\fmf{plain,tension=4}{vc9,vc10}
\fmf{plain,tension=0.5}{vc6,vc7}
\fmf{plain,tension=0.5}{vc7,vc10}
\fmf{plain,tension=1}{vc7,vc8}
\fmf{plain,tension=1,right=0.25}{v1,vc1}
\fmf{plain,tension=1,left=0.25}{v2,vc1}
\fmf{plain,tension=1,left=0.25}{v7,vc2}
\fmf{plain,tension=1,right=0.25}{v8,vc2}
\fmf{plain,tension=1.5}{vc1,vc2}
\fmffreeze
\fmfposition
\fmf{plain,tension=1,right=0,width=1mm}{v7,v12}
\fmfipath{p[]}
\fmfiset{p1}{vpath(__v8,__vc2)}
\fmfiset{p2}{vpath(__vc5,__vc6)}
\fmfipair{w[]}
\svertex{w1}{p1}
\svertex{w2}{p2}
\fmfi{wiggly}{w1..w2}
\fmffreeze
\end{fmfchar*}}}
}
\subfigure[$\nwgraph{5}{4}{1,3,2,5}$]{
\raisebox{\eqoff}{%
\fmfframe(3,1)(1,4){%
\begin{fmfchar*}(30,20)
\fmftop{v1}
\fmfbottom{v7}
\fmfforce{(0w,h)}{v1}
\fmfforce{(0w,0)}{v7}
\fmffixed{(0.2w,0)}{v1,v2}
\fmffixed{(0.2w,0)}{v2,v3}
\fmffixed{(0.2w,0)}{v3,v4}
\fmffixed{(0.2w,0)}{v4,v5}
\fmffixed{(0.2w,0)}{v5,v6}
\fmffixed{(0.2w,0)}{v7,v8}
\fmffixed{(0.2w,0)}{v8,v9}
\fmffixed{(0.2w,0)}{v9,v10}
\fmffixed{(0.2w,0)}{v10,v11}
\fmffixed{(0.2w,0)}{v11,v12}
\fmf{plain,tension=1,right=0.25}{v3,vc5}
\fmf{plain,tension=1,left=0.25}{v4,vc5}
\fmf{plain,tension=1,right=0.25}{v5,vc9}
\fmf{plain,tension=1,left=0.25}{v6,vc9}
\fmf{plain,tension=1,left=0.125}{v9,vc6}
\fmf{plain,tension=0.25,left=0.25}{v10,vc8}
\fmf{plain,tension=0.25,right=0.25}{v11,vc8}
\fmf{plain,tension=1,right=0.125}{v12,vc10}
\fmf{plain,tension=4}{vc5,vc6}
\fmf{plain,tension=4}{vc9,vc10}
\fmf{plain,tension=0.5}{vc6,vc7}
\fmf{plain,tension=0.5}{vc7,vc10}
\fmf{plain,tension=1}{vc7,vc8}
\fmf{plain,tension=1,right=0.25}{v1,vc1}
\fmf{plain,tension=1,left=0.25}{v2,vc1}
\fmf{plain,tension=1,left=0.25}{v7,vc2}
\fmf{plain,tension=1,right=0.25}{v8,vc2}
\fmf{plain,tension=1.5}{vc1,vc2}
\fmffreeze
\fmfposition
\fmf{plain,tension=1,right=0,width=1mm}{v7,v12}
\fmfipath{p[]}
\fmfiset{p1}{vpath(__vc1,__vc2)}
\fmfiset{p2}{vpath(__vc5,__vc6)}
\fmfipair{w[]}
\svertex{w1}{p1}
\svertex{w2}{p2}
\fmfi{wiggly}{w1..w2}
\fmffreeze
\end{fmfchar*}}}
}
\\[0.5cm]
\begin{tabular}{m{8cm}}
\toprule
$\nwgraph{5}{1}{1,3,2,5}\to -(\jint{5}{7}+\jint{5}{8}+2\jint{5}{17})\,\chi(1,3,2,5) $  \\
$\nwgraph{5}{2}{1,3,2,5}\to \jint{5}{7}\,\chi(1,3,2,5) $  \\
$\nwgraph{5}{3}{1,3,2,5}\to \jint{5}{8}\,\chi(1,3,2,5) $  \\
$\nwgraph{5}{4}{1,3,2,5}\to 0 $  \\
\midrule
$\sum_i \nwgraph{5}{i}{1,3,2,5} \to -2\jint{5}{17}\,\chi(1,3,2,5)  $ \\
\bottomrule
\end{tabular}
\normalsize
\caption{Range-six diagrams with structure $\chi(1,3,2,5)$}
\label{r6-1325}
\end{figure}

\begin{figure}[p]
\capstart
\renewcommand*{\thesubfigure}{\ \ }
\footnotesize
\centering
\unitlength=0.75mm
\settoheight{\eqoff}{$\times$}%
\setlength{\eqoff}{0.5\eqoff}%
\addtolength{\eqoff}{-12.5\unitlength}%
\settoheight{\eqofftwo}{$\times$}%
\setlength{\eqofftwo}{0.5\eqofftwo}%
\addtolength{\eqofftwo}{-7.5\unitlength}%
\subfigure[$\nwgraph{5}{1}{1,2,5,4}$]{
\raisebox{\eqoff}{%
\fmfframe(3,1)(1,4){%
\begin{fmfchar*}(30,20)
\fmftop{v1}
\fmfbottom{v7}
\fmfforce{(0w,h)}{v1}
\fmfforce{(0w,0)}{v7}
\fmffixed{(0.2w,0)}{v1,v2}
\fmffixed{(0.2w,0)}{v2,v3}
\fmffixed{(0.2w,0)}{v3,v4}
\fmffixed{(0.2w,0)}{v4,v5}
\fmffixed{(0.2w,0)}{v5,v6}
\fmffixed{(0.2w,0)}{v7,v8}
\fmffixed{(0.2w,0)}{v8,v9}
\fmffixed{(0.2w,0)}{v9,v10}
\fmffixed{(0.2w,0)}{v10,v11}
\fmffixed{(0.2w,0)}{v11,v12}
\fmf{plain,tension=0.5,right=0.25}{v1,vc1}
\fmf{plain,tension=0.5,left=0.25}{v2,vc1}
\fmf{plain,left=0.25}{v3,vc3}
\fmf{plain,tension=0.5,left=0.25}{v8,vc4}
\fmf{plain,tension=0.5,right=0.25}{v9,vc4}
\fmf{plain,left=0.25}{v7,vc2}
\fmf{plain,tension=1}{vc1,vc2}
\fmf{plain,tension=0.5}{vc2,vc3}
\fmf{plain,tension=1}{vc3,vc4}
\fmf{plain,tension=0.5,right=0.25}{v5,vc9}
\fmf{plain,tension=0.5,left=0.25}{v6,vc9}
\fmf{plain,right=0.25}{v4,vc7}
\fmf{plain,tension=0.5,left=0.25}{v10,vc8}
\fmf{plain,tension=0.5,right=0.25}{v11,vc8}
\fmf{plain,right=0.25}{v12,vc10}
\fmf{plain,tension=1}{vc7,vc8}
\fmf{plain,tension=0.5}{vc7,vc10}
\fmf{plain,tension=1}{vc9,vc10}
\fmffreeze
\fmfposition
\fmf{plain,tension=1,right=0,width=1mm}{v7,v12}
\fmfipath{p[]}
\fmfiset{p1}{vpath(__v9,__vc4)}
\fmfiset{p2}{vpath(__v10,__vc8)}
\fmfipair{w[]}
\svertex{w1}{p1}
\svertex{w2}{p2}
\fmfi{wiggly}{w1..w2}
\fmffreeze
\end{fmfchar*}}}
}
\subfigure[$\nwgraph{5}{2}{1,2,5,4}$ $(\times 2)$]{
\raisebox{\eqoff}{%
\fmfframe(3,1)(1,4){%
\begin{fmfchar*}(30,20)
\fmftop{v1}
\fmfbottom{v7}
\fmfforce{(0w,h)}{v1}
\fmfforce{(0w,0)}{v7}
\fmffixed{(0.2w,0)}{v1,v2}
\fmffixed{(0.2w,0)}{v2,v3}
\fmffixed{(0.2w,0)}{v3,v4}
\fmffixed{(0.2w,0)}{v4,v5}
\fmffixed{(0.2w,0)}{v5,v6}
\fmffixed{(0.2w,0)}{v7,v8}
\fmffixed{(0.2w,0)}{v8,v9}
\fmffixed{(0.2w,0)}{v9,v10}
\fmffixed{(0.2w,0)}{v10,v11}
\fmffixed{(0.2w,0)}{v11,v12}
\fmf{plain,tension=0.5,right=0.25}{v1,vc1}
\fmf{plain,tension=0.5,left=0.25}{v2,vc1}
\fmf{plain,left=0.25}{v3,vc3}
\fmf{plain,tension=0.5,left=0.25}{v8,vc4}
\fmf{plain,tension=0.5,right=0.25}{v9,vc4}
\fmf{plain,left=0.25}{v7,vc2}
\fmf{plain,tension=1}{vc1,vc2}
\fmf{plain,tension=0.5}{vc2,vc3}
\fmf{plain,tension=1}{vc3,vc4}
\fmf{plain,tension=0.5,right=0.25}{v5,vc9}
\fmf{plain,tension=0.5,left=0.25}{v6,vc9}
\fmf{plain,right=0.25}{v4,vc7}
\fmf{plain,tension=0.5,left=0.25}{v10,vc8}
\fmf{plain,tension=0.5,right=0.25}{v11,vc8}
\fmf{plain,right=0.25}{v12,vc10}
\fmf{plain,tension=1}{vc7,vc8}
\fmf{plain,tension=0.5}{vc7,vc10}
\fmf{plain,tension=1}{vc9,vc10}
\fmffreeze
\fmfposition
\fmf{plain,tension=1,right=0,width=1mm}{v7,v12}
\fmfipath{p[]}
\fmfiset{p1}{vpath(__vc3,__vc4)}
\fmfiset{p2}{vpath(__v10,__vc8)}
\fmfipair{w[]}
\svertex{w1}{p1}
\svertex{w2}{p2}
\fmfi{wiggly}{w1..w2}
\fmffreeze
\end{fmfchar*}}}
}
\subfigure[$\nwgraph{5}{3}{1,2,5,4}$]{
\raisebox{\eqoff}{%
\fmfframe(3,1)(1,4){%
\begin{fmfchar*}(30,20)
\fmftop{v1}
\fmfbottom{v7}
\fmfforce{(0w,h)}{v1}
\fmfforce{(0w,0)}{v7}
\fmffixed{(0.2w,0)}{v1,v2}
\fmffixed{(0.2w,0)}{v2,v3}
\fmffixed{(0.2w,0)}{v3,v4}
\fmffixed{(0.2w,0)}{v4,v5}
\fmffixed{(0.2w,0)}{v5,v6}
\fmffixed{(0.2w,0)}{v7,v8}
\fmffixed{(0.2w,0)}{v8,v9}
\fmffixed{(0.2w,0)}{v9,v10}
\fmffixed{(0.2w,0)}{v10,v11}
\fmffixed{(0.2w,0)}{v11,v12}
\fmf{plain,tension=0.5,right=0.25}{v1,vc1}
\fmf{plain,tension=0.5,left=0.25}{v2,vc1}
\fmf{plain,left=0.25}{v3,vc3}
\fmf{plain,tension=0.5,left=0.25}{v8,vc4}
\fmf{plain,tension=0.5,right=0.25}{v9,vc4}
\fmf{plain,left=0.25}{v7,vc2}
\fmf{plain,tension=1}{vc1,vc2}
\fmf{plain,tension=0.5}{vc2,vc3}
\fmf{plain,tension=1}{vc3,vc4}
\fmf{plain,tension=0.5,right=0.25}{v5,vc9}
\fmf{plain,tension=0.5,left=0.25}{v6,vc9}
\fmf{plain,right=0.25}{v4,vc7}
\fmf{plain,tension=0.5,left=0.25}{v10,vc8}
\fmf{plain,tension=0.5,right=0.25}{v11,vc8}
\fmf{plain,right=0.25}{v12,vc10}
\fmf{plain,tension=1}{vc7,vc8}
\fmf{plain,tension=0.5}{vc7,vc10}
\fmf{plain,tension=1}{vc9,vc10}
\fmffreeze
\fmfposition
\fmf{plain,tension=1,right=0,width=1mm}{v7,v12}
\fmfipath{p[]}
\fmfiset{p1}{vpath(__vc3,__vc4)}
\fmfiset{p2}{vpath(__vc7,__vc8)}
\fmfipair{w[]}
\svertex{w1}{p1}
\svertex{w2}{p2}
\fmfi{wiggly}{w1..w2}
\fmffreeze
\end{fmfchar*}}}
}
\\[0.5cm]
\begin{tabular}{m{8cm}}
\toprule
$\nwgraph{5}{1}{1,2,5,4}\to -2(\jint{5}{8}+\jint{5}{14})\,\chi(1,2,5,4) $  \\
$\nwgraph{5}{2}{1,2,5,4}\to \jint{5}{8}\,\chi(1,2,5,4) $  \\
$\nwgraph{5}{3}{1,2,5,4}\to 0 $  \\
\midrule
$\sum_i \scgraph{5}{i}{1,2,5,4}\nwgraph{5}{i}{1,2,5,4} \to -2\jint{5}{14}\,\chi(1,2,5,4)  $ \\
\bottomrule
\end{tabular}
\normalsize
\caption{Range-six diagrams with structure $\chi(1,2,5,4)$}
\label{r6-1254}
\end{figure}

\clearpage
\section{Integrals}
The values of the integrals coming from the D-algebra applied to the supergraphs of the previous section are listed in Table~\ref{table:fiveloopintegrals}.

\settoheight{\eqoff}{$\times$}%
\setlength{\eqoff}{0.5\eqoff}%
\addtolength{\eqoff}{-7.5\unitlength}
\begin{table}[h]
\capstart
\begin{align*}
\jint{5}{1}=
\raisebox{\eqoff}{%
\begin{fmfchar*}(20,15)
\fmfleft{in}
\fmfright{out}
\fmf{plain}{in,v1}
\fmf{plain,left=0.25}{v1,v2}
\fmf{plain,left=0}{v2,v5}
\fmf{plain,left=0}{v5,v4}
\fmf{plain,left=0.25}{v4,v3}
\fmf{plain,tension=0.5,right=0.25}{v1,v0,v1}
\fmf{plain,right=0.25}{v0,v3}
\fmf{plain}{v0,v2}
\fmf{plain}{v0,v4}
\fmf{plain}{v0,v5}
\fmf{plain}{v3,out}
\fmffixed{(0.9w,0)}{v1,v3}
\fmffixed{(0.4w,0)}{v2,v4}
\fmffixed{(0.2w,0)}{v2,v5}
\fmffixed{(0.2w,0)}{v5,v4}
\fmfpoly{phantom}{v4,v2,v0}
\fmffreeze
\end{fmfchar*}}
&=
\frac{1}{120\varepsilon^5}-\frac{1}{12\varepsilon^4}+\frac{11}{24\varepsilon^3}-\frac{19}{12\varepsilon^2}+\frac{14}{5\varepsilon}
\\
\jint{5}{2}=\raisebox{\eqoff}{%
\begin{fmfchar*}(20,15)
\fmfleft{in}
\fmfright{out}
\fmf{plain}{in,v1}
\fmf{plain}{v5,out}
\fmf{plain}{v1,v2}
\fmf{plain}{v2,v3}
\fmf{plain}{v3,v4}
\fmf{plain}{v4,v5}
\fmffixed{(0,0.5h)}{v0,v3}
\fmffixed{(0.9w,0)}{v1,v5}
\fmffixed{(0.8w,0)}{v2,v4}
\fmffixed{(0,0.5h)}{v0,v3}
\fmf{plain,tension=0.25,right=0.25}{v2,v0,v2}
\fmf{plain,tension=0.25,right=0.25}{v4,v0,v4}
\fmf{plain,tension=0.25,right=0.25}{v3,v0,v3}
\fmffreeze
\end{fmfchar*}}
&=
\frac{2}{15\varepsilon^5}-\frac{4}{15\varepsilon^4}-\frac{1}{6\varepsilon^3}+\frac{3}{10\varepsilon^2}+\frac{4}{15\varepsilon}
\\
\jint{5}{3}=\raisebox{\eqoff}{%
\begin{fmfchar*}(20,15)
\fmfleft{in}
\fmfright{out}
\fmf{plain}{in,v1}
\fmf{plain}{v5,out}
\fmf{plain,tension=0.25}{v1,v2}
\fmf{plain}{v2,v3}
\fmf{plain}{v3,v4}
\fmf{plain,tension=0.25}{v4,v5}
\fmf{plain}{v1,v0}
\fmf{plain}{v0,v5}
\fmffixed{(0,0.5h)}{v0,v3}
\fmffixed{(0.9w,0)}{v1,v5}
\fmffixed{(0.6w,0)}{v2,v4}
\fmf{plain,tension=0.25,right=0.25}{v2,v0,v2}
\fmf{plain,tension=0.25,right=0.25}{v4,v0,v4}
\fmffreeze
\end{fmfchar*}}
&=
\frac{2}{15\varepsilon^5}-\frac{2}{5\varepsilon^4}+\frac{1}{5\varepsilon^3}+\frac{4}{15\varepsilon^2}-\frac{1}{15\varepsilon}
\\
\jint{5}{4}=\raisebox{\eqoff}{%
\begin{fmfchar*}(20,15)
\fmfleft{in}
\fmfright{out}
\fmf{plain}{in,v1}
\fmf{plain}{v5,out}
\fmf{plain}{v1,v2}
\fmf{plain}{v2,v3}
\fmf{plain}{v3,v4}
\fmf{plain}{v4,v5}
\fmf{plain}{v1,v0}
\fmf{plain}{v0,v5}
\fmffixed{(0,0.5h)}{v0,v3}
\fmffixed{(0.9w,0)}{v1,v5}
\fmffixed{(0.6w,0)}{v2,v4}
\fmf{plain,tension=0.25}{v0,v2}
\fmf{plain,tension=0.25}{v0,v4}
\fmf{plain,tension=0.5,right=0.25}{v3,v0,v3}
\fmffreeze
\end{fmfchar*}}
&=
\frac{1}{20\varepsilon^5}-\frac{3}{10\varepsilon^4}+\frac{17}{20\varepsilon^3}-\frac{14}{15\varepsilon^2}+\frac{2}{5\varepsilon}
\\
\jint{5}{5}=\raisebox{\eqoff}{%
\begin{fmfchar*}(20,15)
\fmfleft{in}
\fmfright{out}
\fmf{plain}{in,v1}
\fmf{plain}{v4,out}
\fmf{plain}{v1,v2}
\fmf{plain}{v2,v3}
\fmf{plain}{v3,v4}
\fmf{plain}{v4,v5}
\fmf{plain}{v0,v1}
\fmf{plain,tension=0.5}{v0,v2}
\fmffixed{(0.9w,0)}{v1,v4}
\fmffixed{(whatever,0)}{v1,in}
\fmffixed{(whatever,0)}{v4,out}
\fmf{plain,tension=0.5,right=0.25}{v3,v0,v3}
\fmf{plain,tension=0.5,right=0.25}{v5,v0,v5}
\fmffixed{(whatever,0.3h)}{v0,v5}
\fmffixed{(whatever,-0.3h)}{v0,v3}
\fmffixed{(0.5w,0)}{v1,v0}
\fmffixed{(0.25w,whatever)}{v1,v2}
\fmffreeze
\end{fmfchar*}}
&=
\frac{3}{40\varepsilon^5}-\frac{3}{10\varepsilon^4}+\frac{17}{40\varepsilon^3}+\frac{1}{30\varepsilon^2}-\frac{1}{2\varepsilon}
\\
\jint{5}{6}=\raisebox{\eqoff}{%
\begin{fmfchar*}(20,15)
\fmfleft{in}
\fmfright{out}
\fmf{plain}{in,v0}
\fmf{phantom,tension=0.5,left=0.25}{v1,v2}
\fmf{plain,tension=0.5,left=0.25}{v2,v3}
\fmf{plain,tension=0.5,left=0.25}{v3,v4}
\fmf{phantom,tension=0.5,left=0.25}{v4,v1}
\fmf{plain,tension=0.5,right=0.5}{v2,v0,v2}
\fmf{plain,tension=0.5,right=0.5}{v0,v4,v0}
\fmf{plain}{v3,out}
\fmffixed{(0.9w,0)}{v1,v3}
\fmffixed{(0,0.45w)}{v4,v2}
\fmffreeze
\fmfipath{px}
\fmfipath{py}
\fmfipair{w[]}
\fmfiset{px}{vpath(__v2,__v3)}
\fmfiset{py}{vpath(__v4,__v3)}
\fmfiequ{w1}{point length(px)/2 of px}
\fmfiequ{w2}{(xpart(vloc(__v0)),ypart(vloc(__v0)))}
\fmfiequ{w3}{point length(py)/2 of py}
\fmfi{plain}{w1..w2}
\fmfi{plain}{w3..w2}
\end{fmfchar*}}
&=
\frac{1}{20\varepsilon^5}-\frac{1}{5\varepsilon^4}+\frac{17}{60\varepsilon^3}-\frac{2}{15\varepsilon^2}+\frac{4}{15\varepsilon}
\\
\jint{5}{7}=\raisebox{\eqoff}{%
\begin{fmfchar*}(20,15)
\fmfleft{in}
\fmfright{out}
\fmf{plain}{in,v1}
\fmf{phantom,left=0.25}{v1,v2}
\fmf{plain,left=0.25}{v2,v3}
\fmf{plain,left=0.25}{v3,v4}
\fmf{plain,left=0.25}{v4,v1}
\fmf{plain,tension=0.5,right=0.5}{v2,v0,v2}
\fmf{plain,tension=0.5,right=0.5}{v0,v4,v0}
\fmf{plain}{v3,out}
\fmffixed{(0.9w,0)}{v1,v3}
\fmffixed{(0,0.45w)}{v4,v2}
\fmffreeze
\fmf{plain}{v1,v0}
\fmfipath{px}
\fmfipair{w[]}
\fmfiset{px}{vpath(__v2,__v3)}
\fmfiequ{w1}{point length(px)/2 of px}
\fmfiequ{w2}{(xpart(vloc(__v0)),ypart(vloc(__v0)))}
\fmfi{plain,left=0.25}{w1..w2}
\end{fmfchar*}}
&=
\frac{3}{40\varepsilon^5}-\frac{17}{60\varepsilon^4}+\frac{43}{120\varepsilon^3}+\frac{7}{60\varepsilon^2}-\frac{8}{15\varepsilon}
\\
\jint{5}{8}=
\raisebox{\eqoff}{%
\begin{fmfchar*}(20,15)
\fmfleft{in}
\fmfright{out}
\fmf{plain}{in,v1}
\fmf{plain,left=0}{v1,v2}
\fmf{plain,tension=0.35,left=0}{v2,v3}
\fmf{plain,left=0}{v3,v4}
\fmf{plain,left=0}{v0,v1}
\fmf{plain,tension=0.35,left=0}{v0,v2}
\fmf{plain,left=0}{v0,v4}
\fmf{plain,tension=0.35,right=0.25}{v3,v0,v3}
\fmf{plain}{v4,out}
\fmffixed{(0.9w,0)}{v1,v4}
\fmffixed{(whatever,0.5h)}{v0,v3}
\fmffixed{(whatever,0.4h)}{v0,v2}
\fmffreeze
\fmfipath{px}
\fmfipair{w[]}
\fmfiset{px}{vpath(__v2,__v3)}
\fmfiequ{w1}{point length(px)/2 of px}
\fmfiequ{w2}{(xpart(vloc(__v0)),ypart(vloc(__v0)))}
\fmfi{plain}{w1..w2}
\end{fmfchar*}}
&=
\frac{1}{30\varepsilon^5}-\frac{7}{30\varepsilon^4}+\frac{5}{6\varepsilon^3}-\frac{3}{2\varepsilon^2}+\frac{11}{30\varepsilon}
\end{align*}
\caption{Five-loop momentum integrals for the diagrams of Section~\ref{sec:fiveR6diag}}
\label{table:fiveloopintegrals}
\end{table}

\addtocounter{table}{-1}
\begin{table}
\capstart
\begin{align*}
\jint{5}{9}=
\raisebox{\eqoff}{%
\begin{fmfchar*}(20,15)
\fmfleft{in}
\fmfright{out}
\fmf{plain}{in,v1}
\fmf{plain,left=0}{v1,v2}
\fmf{plain,left=0}{v2,v4}
\fmf{plain,left=0}{v3,v4}
\fmf{plain,left=0}{v0,v2}
\fmf{plain,tension=0.35,right=0.25}{v1,v0,v1}
\fmf{plain,tension=0.25,right=0.25}{v3,v0,v3}
\fmf{plain}{v4,out}
\fmffixed{(0.9w,0)}{v1,v4}
\fmffixed{(whatever,0.5h)}{v0,v2}
\fmffreeze
\fmfipath{px}
\fmfipair{w[]}
\fmfiset{px}{vpath(__v2,__v4)}
\fmfiequ{w1}{point length(px)/3 of px}
\fmfiequ{w2}{(xpart(vloc(__v0)),ypart(vloc(__v0)))}
\fmfi{plain}{w1..w2}
\end{fmfchar*}}
&=
\frac{1}{30\varepsilon^5}-\frac{11}{60\varepsilon^4}+\frac{7}{15\varepsilon^3}-\frac{19}{60\varepsilon^2}-\frac{14}{15\varepsilon}
\\
\jint{5}{10}=\raisebox{\eqoff}{%
\begin{fmfchar*}(20,15)
\fmfleft{in}
\fmfright{out}
\fmf{plain}{in,v1}
\fmf{plain,left=0.25}{v1,v2}
\fmf{plain,left=0.25}{v2,v3}
\fmf{plain,left=0.25}{v3,v4}
\fmf{phantom,left=0.25}{v4,v1}
\fmf{plain,tension=0.5,right=0.5}{v2,v0,v2}
\fmf{plain,tension=0.5,right=0.5}{v0,v4,v0}
\fmf{plain}{v3,out}
\fmffixed{(0.9w,0)}{v1,v3}
\fmffixed{(0,0.45w)}{v4,v2}
\fmffreeze
\fmf{plain}{v1,v0}
\fmfipath{px}
\fmfipair{w[]}
\fmfiset{px}{vpath(__v2,__v3)}
\fmfiequ{w1}{point length(px)/2 of px}
\fmfiequ{w2}{(xpart(vloc(__v0)),ypart(vloc(__v0)))}
\fmfi{plain,left=0.25}{w1..w2}
\end{fmfchar*}}
&=
\frac{11}{120\varepsilon^5}-\frac{1}{3\varepsilon^4}+\frac{3}{8\varepsilon^3}+\frac{1}{6\varepsilon^2}-\frac{1}{30\varepsilon}
\\
\jint{5}{11}=\raisebox{\eqoff}{%
\begin{fmfchar*}(20,15)
\fmfleft{in}
\fmfright{out}
\fmf{plain}{in,v1}
\fmf{plain}{v6,out}
\fmffixed{(0.9w,0)}{v1,v6}
\fmffixed{(whatever,0)}{v1,in}
\fmffixed{(whatever,0)}{out,v6}
\fmf{derplain}{v2,v1}
\fmf{plain}{v2,v3}
\fmf{derplain}{v3,v4}
\fmf{plain}{v4,v5}
\fmf{plain}{v5,v6}
\fmf{plain}{v0,v1}
\fmf{plain,tension=0.5}{v0,v2}
\fmf{plain}{v0,v3}
\fmf{plain}{v0,v4}
\fmf{plain,tension=0.5}{v0,v5}
\fmf{plain}{v0,v6}
\fmffixed{(whatever,0.5h)}{v0,v3}
\fmffixed{(0.25w,0)}{v3,v4}
\fmffixed{(0.65w,0)}{v2,v5}
\fmffixed{(whatever,0.3h)}{v0,v2}
\fmffreeze
\end{fmfchar*}}
&=
-\frac{1}{10\varepsilon^3}+\frac{13}{30\varepsilon^2}-\frac{23}{30\varepsilon}
\\
\jint{5}{12}=\raisebox{\eqoff}{%
\begin{fmfchar*}(20,15)
\fmfleft{in}
\fmfright{out}
\fmf{plain}{in,v1}
\fmf{plain}{v4,out}
\fmffixed{(0.9w,0)}{v1,v4}
\fmffixed{(whatever,0)}{v1,in}
\fmffixed{(whatever,0)}{out,v4}
\fmf{derplain}{v2,v1}
\fmf{plain}{v2,v3}
\fmf{derplain}{v3,v4}
\fmf{plain}{v4,v5}
\fmf{plain,tension=0.5}{v5,v6}
\fmf{plain}{v0,v1}
\fmf{plain,tension=0.5}{v0,v2}
\fmf{plain}{v0,v3}
\fmf{plain,tension=0.5}{v0,v5}
\fmf{plain,tension=0.25,right=0.25}{v6,v0,v6}
\fmffixed{(whatever,0.5h)}{v0,v3}
\fmffixed{(whatever,0.3h)}{v6,v0}
\fmffreeze
\end{fmfchar*}}
&=
-\frac{1}{60\varepsilon^3}+\frac{1}{12\varepsilon^2}-\frac{2}{5\varepsilon}
\\
\jint{5}{13}=
\settoheight{\eqoff}{$\times$}%
\setlength{\eqoff}{0.5\eqoff}%
\addtolength{\eqoff}{-7.5\unitlength}%
\raisebox{\eqoff}{%
\begin{fmfchar*}(20,15)
  \fmfleft{in}
  \fmfright{out}
  \fmf{plain}{in,v1}
  \fmf{phantom,tension=2,left=0.25}{v1,v2}
  \fmf{plain,tension=2,left=0.25}{v2,v3}
  \fmf{plain,left=0.25}{v4,v1}
  \fmf{plain,left=0.25}{v0,v4}
  \fmf{plain,right=0}{v0,v1}
  \fmf{plain,right=0.25}{v0,v5}
  \fmf{derplain,left=0.25}{v6,v4}
  \fmf{plain,right=0.25}{v6,v5}
  \fmf{phantom,right=0}{v3,v0}
  \fmf{derplain,right=0.25}{v5,v3}
  \fmf{plain}{v3,out}
\fmffixed{(0.9w,0)}{v1,v3}
\fmfpoly{phantom}{v2,v4,v5}
\fmffixed{(0.5w,0)}{v4,v5}
\fmf{plain}{v0,v6}
\fmf{plain,tension=0.25,right=0.25}{v2,v0}
\fmf{plain,tension=0.25,right=0.25}{v0,v2}
\fmffreeze
\fmffreeze
\fmfshift{(0,0.15w)}{in,out,v1,v2,v3,v4,v5,v0}
\end{fmfchar*}}
&=
-\frac{1}{20\varepsilon^3}+\frac{3}{20\varepsilon^2}-\frac{13}{30\varepsilon}
\\
\jint{5}{14}=\raisebox{\eqoff}{%
\begin{fmfchar*}(20,15)
\fmfleft{in}
\fmfright{out}
\fmf{plain}{in,v1}
\fmf{plain}{v6,out}
\fmffixed{(0.9w,0)}{v1,v6}
\fmffixed{(whatever,0)}{v1,in}
\fmffixed{(whatever,0)}{out,v6}
\fmf{plain}{v2,v1}
\fmf{derplain}{v3,v2}
\fmf{plain}{v3,v4}
\fmf{derplain}{v4,v5}
\fmf{plain}{v5,v6}
\fmf{plain}{v0,v1}
\fmf{plain,tension=0.5}{v0,v2}
\fmf{plain}{v0,v3}
\fmf{plain}{v0,v4}
\fmf{plain,tension=0.5}{v0,v5}
\fmf{plain}{v0,v6}
\fmffixed{(whatever,0.5h)}{v0,v3}
\fmffixed{(0.25w,0)}{v3,v4}
\fmffixed{(0.65w,0)}{v2,v5}
\fmffixed{(whatever,0.3h)}{v0,v2}
\fmffreeze
\end{fmfchar*}}
&=
-\frac{1}{5\varepsilon^3}+\frac{2}{3\varepsilon^2}-\frac{4}{15\varepsilon}
\\
\jint{5}{15}=\raisebox{\eqoff}{%
\begin{fmfchar*}(20,15)
\fmfleft{in}
\fmfright{out}
\fmf{plain}{in,v1}
\fmf{plain}{v4,out}
\fmffixed{(0.9w,0)}{v1,v4}
\fmffixed{(whatever,0)}{v1,in}
\fmffixed{(whatever,0)}{out,v4}
\fmf{derplain}{v2,v1}
\fmf{plain}{v2,v3}
\fmf{derplain}{v3,v4}
\fmf{plain}{v4,v5}
\fmf{plain}{v6,v1}
\fmf{plain}{v0,v2}
\fmf{plain}{v0,v3}
\fmf{plain,tension=0.25,right=0.25}{v5,v0,v5}
\fmf{plain,tension=0.25,right=0.25}{v6,v0,v6}
\fmffixed{(whatever,0.5h)}{v0,v2}
\fmffixed{(whatever,0)}{v2,v3}
\fmffreeze
\end{fmfchar*}}
&=
-\frac{1}{30\varepsilon^3}+\frac{1}{30\varepsilon^2}+\frac{1}{3\varepsilon}
\\
\jint{5}{16}=\raisebox{\eqoff}{%
\begin{fmfchar*}(20,15)
\fmfleft{in}
\fmfright{out}
\fmf{plain}{in,v1}
\fmf{plain}{v5,out}
\fmffixed{(0.9w,0)}{v1,v5}
\fmffixed{(whatever,0)}{v1,in}
\fmffixed{(whatever,0)}{out,v5}
\fmf{derplain}{v2,v1}
\fmf{plain}{v2,v3}
\fmf{derplain,tension=0.5}{v3,v4}
\fmf{plain}{v4,v5}
\fmf{plain}{v5,v6}
\fmf{plain}{v0,v1}
\fmf{plain}{v0,v2}
\fmf{plain}{v0,v3}
\fmf{plain}{v0,v4}
\fmf{plain,tension=0.25,right=0.25}{v6,v0,v6}
\fmffixed{(whatever,0.5h)}{v0,v3}
\fmffixed{(whatever,0.4h)}{v0,v2}
\fmffixed{(whatever,0.4h)}{v0,v4}
\fmffreeze
\end{fmfchar*}}
&=
-\frac{3}{20\varepsilon^3}+\frac{11}{60\varepsilon^2}+\frac{3}{10\varepsilon}
\\
\jint{5}{17}=\raisebox{\eqoff}{%
\begin{fmfchar*}(20,15)
\fmfleft{in}
\fmfright{out}
\fmf{plain}{in,v1}
\fmf{plain}{v4,out}
\fmffixed{(0.9w,0)}{v1,v4}
\fmffixed{(whatever,0)}{v1,in}
\fmffixed{(whatever,0)}{out,v4}
\fmf{derplain}{v2,v1}
\fmf{plain}{v2,v3}
\fmf{derplain}{v3,v4}
\fmf{plain}{v5,v6}
\fmf{plain}{v6,v1}
\fmf{plain}{v0,v2}
\fmf{plain}{v0,v3}
\fmf{plain}{v0,v4}
\fmf{plain}{v0,v5}
\fmf{plain,tension=0.25,right=0.25}{v6,v0,v6}
\fmffixed{(whatever,0.5h)}{v0,v3}
\fmffixed{(whatever,0.4h)}{v0,v2}
\fmffixed{(whatever,0.2h)}{v5,v0}
\fmffreeze
\end{fmfchar*}}
&=
-\frac{1}{15\varepsilon^3}+\frac{2}{15\varepsilon^2}+\frac{1}{15\varepsilon}
\\
\jint{5}{18}=\raisebox{\eqoff}{%
\begin{fmfchar*}(20,15)
\fmfleft{in}
\fmfright{out}
\fmf{plain}{in,v1}
\fmf{plain}{v3,out}
\fmffixed{(0.9w,0)}{v1,v3}
\fmffixed{(whatever,0)}{v1,in}
\fmffixed{(whatever,0)}{out,v3}
\fmf{plain}{v1,v2}
\fmf{plain}{v2,v3}
\fmf{plain}{v3,v4}
\fmf{plain}{v4,v5}
\fmf{plain}{v5,v1}
\fmf{plain}{v5,v2}
\fmf{plain}{v0,v2}
\fmf{plain}{v0,v4}
\fmf{plain,tension=0.25,right=0.25}{v5,v0,v5}
\fmffixed{(whatever,0.5h)}{v2,v4}
\fmffixed{(0.3w,whatever)}{v5,v0}
\fmffreeze
\end{fmfchar*}}
&=
\frac{1}{24\varepsilon^5}-\frac{1}{4\varepsilon^4}+\frac{5}{8\varepsilon^3}-\frac{1}{\varepsilon^2}\Big(\frac{1}{12}+\frac{4\zeta(3)}{5}\Big)-\frac{1}{\varepsilon}\Big(\frac{5}{6}-\frac{4\zeta(3)}{5}-\frac{\pi^4}{300}\Big)
\\
\jint{5}{19}=\raisebox{\eqoff}{%
\begin{fmfchar*}(20,15)
\fmfleft{in}
\fmfright{out}
\fmf{plain}{in,v1}
\fmf{plain}{v3,out}
\fmffixed{(0.9w,0)}{v1,v3}
\fmffixed{(whatever,0)}{in,v1}
\fmffixed{(whatever,0)}{out,v3}
\fmf{plain}{v1,v2}
\fmf{plain}{v2,v3}
\fmf{plain}{v1,v4}
\fmf{plain}{v4,v5}
\fmf{plain}{v4,v0}
\fmf{plain}{v2,v0}
\fmf{plain,tension=0.25,right=0.25}{v5,v0,v5}
\fmf{plain,tension=0.25,right=0.25}{v5,v3,v5}
\fmffixed{(whatever,0.5h)}{v2,v4}
\fmffixed{(0.28w,whatever)}{v0,v5}
\fmffreeze
\end{fmfchar*}}
&=
\frac{1}{24\varepsilon^5}-\frac{1}{6\varepsilon^4}+\frac{1}{8\varepsilon^3}+\frac{1}{\varepsilon^2}\Big(\frac{1}{3}-\frac{\zeta(3)}{5}\Big)-\frac{1}{\varepsilon}\Big(\frac{1}{3}-\frac{4\zeta(5)}{5}+\frac{\pi^4}{300}\Big)
\end{align*}
\caption{Five-loop momentum integrals for the diagrams of Section~\ref{sec:fiveR6diag} \textit{(continued)}}
\end{table}

\chapter{Five-loop wrapping diagrams}
\label{app:fivewrap}
In this appendix, the results of the computation of all the five-loop wrapping diagrams that are needed for the calculation of Chapter~\ref{chapter:fiveloop} are presented.
As usual, the symmetry factors $\scgraph{5}{i}{\cdots}$ of diagrams built starting from parity-symmetric chiral structures will be given when they are different from 1. Moreover, the colour factors $(g^2 N)^5$ from the D-algebra and the $1/(4\pi)^{10}$ ones from the momentum integrals will not be shown explicitly. The results of D-algebra are written in terms of the momentum-space integrals listed in Section~\ref{section:integrals5wrap}.

\section{The supergraphs}

\begin{figure}[t]
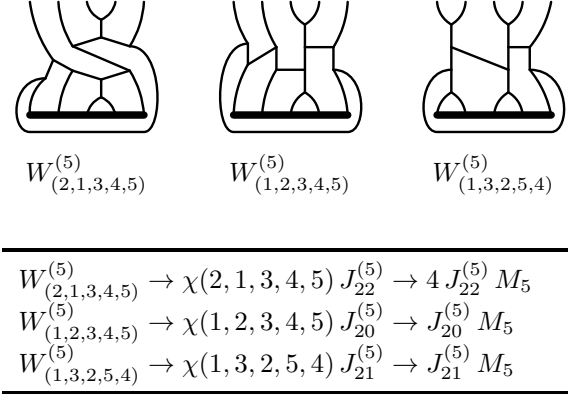

\capstart
\footnotesize
\addtolength{\subfigcapskip}{5pt}
\renewcommand*{\thesubfigure}{}
\renewcommand{\subfigspace}{\phantom{***}}
\centering
\unitlength=0.75mm
\settoheight{\eqoff}{$\times$}%
\setlength{\eqoff}{0.5\eqoff}%
\addtolength{\eqoff}{-12.5\unitlength}%
\settoheight{\eqofftwo}{$\times$}%
\setlength{\eqofftwo}{0.5\eqofftwo}%
\addtolength{\eqofftwo}{-7.5\unitlength}%
\subfigure[$\wgraph{5}{}{2,1,3,4,5}$]{
\raisebox{\eqoff}{%
\fmfframe(3,1)(1,4){%
\begin{fmfchar*}(20,20)
\fmftop{v1}
\fmfbottom{v6}
\fmfforce{(0,h)}{v1}
\fmfforce{(0,0)}{v6}
\fmffixed{(0.25w,0)}{v1,v2}
\fmffixed{(0.25w,0)}{v2,v3}
\fmffixed{(0.25w,0)}{v3,v4}
\fmffixed{(0.25w,0)}{v4,v5}
\fmffixed{(0.25w,0)}{v6,v7}
\fmffixed{(0.25w,0)}{v7,v8}
\fmffixed{(0.25w,0)}{v8,v9}
\fmffixed{(0.25w,0)}{v9,v10}
\fmffixed{(0,whatever)}{vc1,vc5}
\fmffixed{(0,whatever)}{vc2,vc3}
\fmffixed{(0,whatever)}{vc3,vc6}
\fmffixed{(0,whatever)}{vc6,vc7}
\fmffixed{(0,whatever)}{vc4,vc8}
\fmffixed{(0.5w,0)}{vc1,vc4}
\fmffixed{(0.5w,0)}{vc5,vc8}
\fmf{plain,tension=1,right=0.125}{v2,vc1}
\fmf{plain,tension=0.25,right=0.25}{v3,vc2}
\fmf{plain,tension=0.25,left=0.25}{v4,vc2}
\fmf{plain,tension=1,left=0.125}{v5,vc4}
\fmf{plain,tension=1,left=0.125}{v7,vc5}
\fmf{plain,tension=0.25,left=0.25}{v8,vc6}
\fmf{plain,tension=0.25,right=0.25}{v9,vc6}
\fmf{plain,tension=1,right=0.125}{v10,vc8}
\fmf{plain,tension=0.5}{vc1,vc3}
\fmf{plain,tension=0.5}{vc2,vc3}
\fmf{plain,tension=0.5}{vc3,vc4}
\fmf{plain,tension=0.5}{vc5,vc7}
\fmf{plain,tension=0.5}{vc6,vc7}
\fmf{plain,tension=0.5}{vc7,vc8}
\fmf{plain,tension=2}{vc1,vc8}
\fmf{phantom,tension=2}{vc5,vc4}
\fmffreeze
\fmfposition
\fmf{plain,tension=2}{vc5,vc9}
\fmf{plain,tension=2}{vc9,vc10}
\fmf{plain,tension=1,right=0.125}{v1,vc9}
\fmf{plain,tension=1,left=0.25}{v6,vc10}
\fmffreeze
\plainwrap{vc10}{v6}{v10}{vc4}
\fmf{plain,tension=1,left=0,width=1mm}{v6,v10}
\fmffreeze
\end{fmfchar*}}}
}
\subfigspace
\subfigure[$\wgraph{5}{}{1,2,3,4,5}$]{
\raisebox{\eqoff}{%
\fmfframe(3,1)(1,4){%
\begin{fmfchar*}(20,20)
\fmftop{v1}
\fmfbottom{v6}
\fmfforce{(0,h)}{v1}
\fmfforce{(0,0)}{v6}
\fmffixed{(0.25w,0)}{v1,v2}
\fmffixed{(0.25w,0)}{v2,v3}
\fmffixed{(0.25w,0)}{v3,v4}
\fmffixed{(0.25w,0)}{v4,v5}
\fmffixed{(0.25w,0)}{v6,v7}
\fmffixed{(0.25w,0)}{v7,v8}
\fmffixed{(0.25w,0)}{v8,v9}
\fmffixed{(0.25w,0)}{v9,v10}
\fmffixed{(whatever,0)}{vc1,vc3}
\fmffixed{(whatever,0)}{vc5,vc7}
\fmffixed{(whatever,0)}{vc3,vc4}
\fmffixed{(whatever,0)}{vc7,vc8}
\fmf{plain,tension=1,right=0.125}{v2,vc1}
\fmf{plain,tension=0.5,right=0.25}{v3,vc2}
\fmf{plain,tension=0.5,left=0.25}{v4,vc2}
\fmf{plain,tension=1,left=0.125}{v5,vc4}
\fmf{plain,tension=1,left=0.125}{v7,vc5}
\fmf{plain,tension=0.5,left=0.25}{v8,vc6}
\fmf{plain,tension=0.5,right=0.25}{v9,vc6}
\fmf{plain,tension=1,right=0.125}{v10,vc8}
\fmf{plain}{vc1,vc5}
\fmf{plain}{vc4,vc8}
\fmf{plain}{vc2,vc3}
\fmf{plain}{vc6,vc7}
\fmf{plain,tension=3}{vc3,vc7}
\fmf{plain,tension=0.5}{vc3,vc4}
\fmf{plain,tension=0.5}{vc5,vc7}
\fmf{phantom,tension=0.5}{vc7,vc8}
\fmf{phantom,tension=0.5}{vc1,vc3}
\fmf{plain,tension=0.5,right=0,width=1mm}{v6,v10}
\fmffreeze
\fmf{plain,tension=0.5,right=0.125}{v1,vc9}
\fmf{phantom,tension=0.5,left=0.125}{v2,vc9}
\fmf{plain,tension=1,left=0.125}{v6,vc10}
\fmf{plain,tension=0.5}{vc1,vc10}
\fmf{plain,tension=2}{vc9,vc10}
\fmffreeze
\fmfposition
\plainwrap{vc9}{v6}{v10}{vc8}
\fmffreeze
\end{fmfchar*}}}
}
\subfigspace
\subfigure[$\wgraph{5}{}{1,3,2,5,4}$]{
\raisebox{\eqoff}{%
\fmfframe(3,1)(1,4){%
\begin{fmfchar*}(20,20)
\fmftop{v1}
\fmfbottom{v6}
\fmfforce{(0,h)}{v1}
\fmfforce{(0,0)}{v6}
\fmffixed{(0.25w,0)}{v1,v2}
\fmffixed{(0.25w,0)}{v2,v3}
\fmffixed{(0.25w,0)}{v3,v4}
\fmffixed{(0.25w,0)}{v4,v5}
\fmffixed{(0.25w,0)}{v6,v7}
\fmffixed{(0.25w,0)}{v7,v8}
\fmffixed{(0.25w,0)}{v8,v9}
\fmffixed{(0.25w,0)}{v9,v10}
\fmffixed{(0,0.9w)}{v6,vh1}
\fmf{plain,tension=0.5,right=0.25}{v1,vc1}
\fmf{plain,tension=0.5,left=0.25}{v2,vc1}
\fmf{plain,tension=0.5,right=0.25}{v3,vc2}
\fmf{plain,tension=0.5,left=0.25}{v4,vc2}
\fmf{plain}{vc1,vc3}
\fmf{plain}{vc3,vc7}
\fmf{plain}{vc7,vc5}
\fmf{plain}{vc2,vc8}
\fmf{plain}{vc8,vc4}
\fmf{plain}{vc4,vc6}
\fmf{plain,tension=0}{vc3,vc4}
\fmf{plain,tension=0.5,left=0.25}{v6,vc5}
\fmf{plain,tension=0.5,right=0.25}{v7,vc5}
\fmf{plain,tension=0.5,left=0.25}{v8,vc6}
\fmf{plain,tension=0.5,right=0.25}{v9,vc6}
\fmf{plain,tension=0.5,right=0,width=1mm}{v6,v10}
\fmffreeze
\fmfposition
\fmf{plain,tension=0.5,left=0.125}{v5,vc9}
\fmf{plain,tension=0.5,right=0.125}{v10,vc10}
\fmf{plain,tension=0.5}{vc8,vc9}
\fmf{phantom,tension=0.5}{vc4,vc10}
\fmf{plain}{vc9,vc10}
\fmffreeze
\plainwrap{vc7}{v6}{v10}{vc10}
\end{fmfchar*}}}
}
\\[0.5cm]
\begin{tabular}{m{7cm}}
\toprule
$\wgraph{5}{}{2,1,3,4,5}\rightarrow \chi(2,1,3,4,5)\,\jint{5}{22} \rightarrow 4\,\jint{5}{22}\,M_5$  \\
$\wgraph{5}{}{1,2,3,4,5}\rightarrow \chi(1,2,3,4,5)\,\jint{5}{20} \rightarrow \jint{5}{20}\,M_5$ \\
$\wgraph{5}{}{1,3,2,5,4}\rightarrow \chi(1,3,2,5,4)\,\jint{5}{21} \rightarrow \jint{5}{21}\,M_5$ \\
\bottomrule
\end{tabular}
\normalsize

\caption{Wrapping diagrams with completely chiral structure}
\label{wrap5-chiral}
\end{figure}

The wrapping diagrams containing only scalar interactions are shown in Figure~\ref{wrap5-chiral}. Since none of them is symmetric under parity, the corresponding reflections have to be considered too, and the total contribution of this class of graphs becomes
\begin{equation}
\begin{aligned}
\chi_{\mathrm{chiral}}&\to \jint{5}{22}\,(\chi(2,1,3,4,5)+\chi(4,5,3,2,1))+\jint{5}{20}\,(\chi(1,2,3,4,5)+\chi(5,4,3,2,1))\\
&\phantom{\to}+\jint{5}{21}\,(\chi(1,3,2,5,4)+\chi(5,3,4,1,2)) \\
&\to 2(\jint{5}{20}+\jint{5}{21}+4\jint{5}{22})\,M_5 \pnt
\end{aligned}
\end{equation}

Among the structures of the minimal independent set described in Chapter~\ref{chapter:fiveloop}, which require at least one vector line, the non-symmetric ones are $\chi(2,4,1,3)$, $\chi(3,2,1,4)$, $\chi(1,2,3,4)$, $\chi(1,4,3,2)$, $\chi(1,2,3)$, $\chi(2,1,4)$ and $\chi(2,1)$. The corresponding reflections under parity are
\begin{align}
&\chi(2,4,1,3)\to\chi(1,3,2,4) \col &&\chi(3,2,1,4)\to\chi(2,1,3,4) \col \notag \\
&\chi(1,2,3,4)\to\chi(4,3,2,1) \col &&\chi(1,4,3,2)\to\chi(1,2,4,3) \col \\
&\chi(1,2,3)\to\chi(3,2,1) \col &&\chi(2,1,4)\to\chi(1,3,4) \col \notag \\
&\chi(2,1)\to\chi(1,2) \pnt \notag
\end{align}
For these structures, one must remember to add the corresponding reflected structure with the same coefficient when assembling the total wrapping contribution. When the structures are restricted to the length-five subsector, this is equivalent to doubling the matrix contribution from every non-symmetric structure of the independent set. The diagrams associated to the structures of the minimal set are listed in Figures~\ref{wrap5-2413}-\ref{wrap5-14}.

Remember that, as discussed in Chapter~\ref{chapter:fiveloop}, on the length-five subsector the structures $\chi(1,2,4)$ and $\chi(1,4,3)$ are the same as $\chi(2,1,4)$ and its reflection $\chi(1,3,4)$. Similarly, $\chi(1,3)$ is the same structure as $\chi(1,4)$. 

\begin{figure}[p]
\capstart
\addtolength{\subfigcapskip}{5pt}
\footnotesize
\renewcommand*{\thesubfigure}{}
\renewcommand{\subfigspace}{\phantom{****}}
\centering
\unitlength=0.75mm
\settoheight{\eqoff}{$\times$}%
\setlength{\eqoff}{0.5\eqoff}%
\addtolength{\eqoff}{-12.5\unitlength}%
\settoheight{\eqofftwo}{$\times$}%
\setlength{\eqofftwo}{0.5\eqofftwo}%
\addtolength{\eqofftwo}{-7.5\unitlength}%
\subfigure[$\wgraph{5}{1}{2,4,1,3}$]{
\raisebox{\eqoff}{%
\fmfframe(3,1)(1,4){%
\begin{fmfchar*}(16,20)
\Wdqut
\fmfipair{wa[]}
\fmfipair{wb[]}
\fmfipair{wc[]}
\fmfipair{wd[]}
\fmfiequ{wa0}{point 1*length(p0)/2 of p0}
\fmfiv{d.shape=circle,d.size=2}{wa0}
\fmfiequ{wa3}{point 1*length(p3)/2 of p3}
\fmfiv{d.shape=circle,d.size=2}{wa3}
\fmfforce{(-0w,-0h)}{va0}
\fmfforce{(w,-0h)}{vb0}
\wigglywrap{wa0}{va0}{vb0}{wa3}
\end{fmfchar*}}}
}
\subfigspace
\subfigure[$\wgraph{5}{2}{2,4,1,3}$]{
\raisebox{\eqoff}{%
\fmfframe(3,1)(1,4){%
\begin{fmfchar*}(16,20)
\Wdqut
\fmfipair{wa[]}
\fmfipair{wb[]}
\fmfipair{wc[]}
\fmfipair{wd[]}
\fmfiequ{wa0}{point 1*length(p0)/2 of p0}
\fmfiv{d.shape=circle,d.size=2}{wa0}
\fmfiequ{wa4}{point 1*length(p4)/2 of p4}
\fmfiv{d.shape=circle,d.size=2}{wa4}
\fmfforce{(-0w,-0h)}{va0}
\fmfforce{(w,-0h)}{vb0}
\wigglywrap{wa0}{va0}{vb0}{wa4}
\end{fmfchar*}}}
}
\subfigspace
\subfigure[$\wgraph{5}{3}{2,4,1,3}$]{
\raisebox{\eqoff}{%
\fmfframe(3,1)(1,4){%
\begin{fmfchar*}(16,20)
\Wdqut
\fmfipair{wa[]}
\fmfipair{wb[]}
\fmfipair{wc[]}
\fmfipair{wd[]}
\fmfiequ{wa1}{point 1*length(p1)/2 of p1}
\fmfiv{d.shape=circle,d.size=2}{wa1}
\fmfiequ{wa3}{point 1*length(p3)/2 of p3}
\fmfiv{d.shape=circle,d.size=2}{wa3}
\fmfforce{(-0w,-0h)}{va0}
\fmfforce{(w,-0h)}{vb0}
\wigglywrap{wa1}{va0}{vb0}{wa3}
\end{fmfchar*}}}
}
\subfigspace
\subfigure[$\wgraph{5}{4}{2,4,1,3}$]{
\raisebox{\eqoff}{%
\fmfframe(3,1)(1,4){%
\begin{fmfchar*}(16,20)
\Wdqut
\fmfipair{wa[]}
\fmfipair{wb[]}
\fmfipair{wc[]}
\fmfipair{wd[]}
\fmfiequ{wa1}{point 1*length(p1)/2 of p1}
\fmfiv{d.shape=circle,d.size=2}{wa1}
\fmfiequ{wa4}{point 1*length(p4)/2 of p4}
\fmfiv{d.shape=circle,d.size=2}{wa4}
\fmfforce{(-0w,-0h)}{va0}
\fmfforce{(w,-0h)}{vb0}
\wigglywrap{wa1}{va0}{vb0}{wa4}
\end{fmfchar*}}}
}
\\[0.5cm]
\begin{tabular}{m{12cm}}
\toprule
$\wgraph{5}{1}{2,4,1,3}\rightarrow -(\jint{5}{20}+\jint{5}{21}+2\jint{5}{24})\,\chi(2,4,1,3) \rightarrow -(\jint{5}{20}+\jint{5}{21}+2\jint{5}{24})\,M_5$  \\
$\wgraph{5}{2}{2,4,1,3}\rightarrow \jint{5}{20}\,\chi(2,4,1,3) \rightarrow \jint{5}{20}\,M_5$ \\
$\wgraph{5}{3}{2,4,1,3}\rightarrow \jint{5}{21}\,\chi(2,4,1,3) \rightarrow \jint{5}{21}\,M_5$ \\
$\wgraph{5}{4}{2,4,1,3}\rightarrow0$  \\
\midrule
$\sum_{i}\wgraph{5}{i}{2,4,1,3}\rightarrow -2\,\jint{5}{24}\,\chi(2,4,1,3) \rightarrow -2\,\jint{5}{24}\,M_5$   \\
\bottomrule
\end{tabular}
\normalsize
\caption{Wrapping diagrams with structure $\chi(2,4,1,3)$}
\label{wrap5-2413}
\end{figure}



\begin{figure}[p]
\capstart
\addtolength{\subfigcapskip}{5pt}
\footnotesize
\renewcommand*{\thesubfigure}{}
\renewcommand{\subfigspace}{\phantom{****}}
\centering
\unitlength=0.75mm
\settoheight{\eqoff}{$\times$}%
\setlength{\eqoff}{0.5\eqoff}%
\addtolength{\eqoff}{-12.5\unitlength}%
\settoheight{\eqofftwo}{$\times$}%
\setlength{\eqofftwo}{0.5\eqofftwo}%
\addtolength{\eqofftwo}{-7.5\unitlength}%
\subfigure[$\wgraph{5}{1}{3,2,1,4}$]{
\raisebox{\eqoff}{%
\fmfframe(3,1)(1,4){%
\begin{fmfchar*}(16,20)
\Wtduq
\fmfipair{wa[]}
\fmfipair{wb[]}
\fmfipair{wc[]}
\fmfipair{wd[]}
\fmfiequ{wa0}{point 1*length(p0)/2 of p0}
\fmfiv{d.shape=circle,d.size=2}{wa0}
\fmfiequ{wa3}{point 1*length(p3)/2 of p3}
\fmfiv{d.shape=circle,d.size=2}{wa3}
\fmfforce{(-0w,-0h)}{va0}
\fmfforce{(w,-0h)}{vb0}
\wigglywrap{wa0}{va0}{vb0}{wa3}
\end{fmfchar*}}}
}
\subfigspace
\subfigure[$\wgraph{5}{2}{3,2,1,4}$]{
\raisebox{\eqoff}{%
\fmfframe(3,1)(1,4){%
\begin{fmfchar*}(16,20)
\Wtduq
\fmfipair{wa[]}
\fmfipair{wb[]}
\fmfipair{wc[]}
\fmfipair{wd[]}
\fmfiequ{wa0}{point 1*length(p0)/2 of p0}
\fmfiv{d.shape=circle,d.size=2}{wa0}
\fmfiequ{wa4}{point 1*length(p4)/2 of p4}
\fmfiv{d.shape=circle,d.size=2}{wa4}
\fmfforce{(-0w,-0h)}{va0}
\fmfforce{(w,-0h)}{vb0}
\wigglywrap{wa0}{va0}{vb0}{wa4}
\end{fmfchar*}}}
}
\subfigspace
\subfigure[$\wgraph{5}{3}{3,2,1,4}$]{
\raisebox{\eqoff}{%
\fmfframe(3,1)(1,4){%
\begin{fmfchar*}(16,20)
\Wtduq
\fmfipair{wa[]}
\fmfipair{wb[]}
\fmfipair{wc[]}
\fmfipair{wd[]}
\fmfiequ{wa1}{point 1*length(p1)/2 of p1}
\fmfiv{d.shape=circle,d.size=2}{wa1}
\fmfiequ{wa3}{point 1*length(p3)/2 of p3}
\fmfiv{d.shape=circle,d.size=2}{wa3}
\fmfforce{(-0w,-0h)}{va0}
\fmfforce{(w,-0h)}{vb0}
\wigglywrap{wa1}{va0}{vb0}{wa3}
\end{fmfchar*}}}
}
\subfigspace
\subfigure[$\wgraph{5}{4}{3,2,1,4}$]{
\raisebox{\eqoff}{%
\fmfframe(3,1)(1,4){%
\begin{fmfchar*}(16,20)
\Wtduq
\fmfipair{wa[]}
\fmfipair{wb[]}
\fmfipair{wc[]}
\fmfipair{wd[]}
\fmfiequ{wa1}{point 1*length(p1)/2 of p1}
\fmfiv{d.shape=circle,d.size=2}{wa1}
\fmfiequ{wa4}{point 1*length(p4)/2 of p4}
\fmfiv{d.shape=circle,d.size=2}{wa4}
\fmfforce{(-0w,-0h)}{va0}
\fmfforce{(w,-0h)}{vb0}
\wigglywrap{wa1}{va0}{vb0}{wa4}
\end{fmfchar*}}}
}
\\[0.5cm]
\begin{tabular}{m{12cm}}
\toprule
$\wgraph{5}{1}{3,2,1,4}\rightarrow -(\jint{5}{20}+\jint{5}{22}+2\jint{5}{25})\,\chi(3,2,1,4) \rightarrow 2(\jint{5}{20}+\jint{5}{22}+2\jint{5}{25})\,M_5$  \\
$\wgraph{5}{2}{3,2,1,4}\rightarrow \jint{5}{20}\,\chi(3,2,1,4) \rightarrow -2\jint{5}{20}\,M_5$ \\
$\wgraph{5}{3}{3,2,1,4}\rightarrow \jint{5}{22}\,\chi(3,2,1,4) \rightarrow -2\jint{5}{22}\,M_5$ \\
$\wgraph{5}{4}{3,2,1,4}\rightarrow0$  \\
\midrule
$\sum_{i}\wgraph{5}{i}{3,2,1,4}\rightarrow -2\jint{5}{25}\chi(3,2,1,4) \rightarrow 4\,\jint{5}{25}\,M_5$   \\
\bottomrule
\end{tabular}
\normalsize

\caption{Wrapping diagrams with structure $\chi(3,2,1,4)$}
\label{wrap5-3214}
\end{figure}



\begin{figure}[p]
\capstart
\addtolength{\subfigcapskip}{5pt}
\footnotesize
\renewcommand*{\thesubfigure}{}
\renewcommand{\subfigspace}{\phantom{****}}
\centering
\unitlength=0.75mm
\settoheight{\eqoff}{$\times$}%
\setlength{\eqoff}{0.5\eqoff}%
\addtolength{\eqoff}{-12.5\unitlength}%
\settoheight{\eqofftwo}{$\times$}%
\setlength{\eqofftwo}{0.5\eqofftwo}%
\addtolength{\eqofftwo}{-7.5\unitlength}%
\subfigure[$\wgraph{5}{1}{1,2,3,4}$]{
\raisebox{\eqoff}{%
\fmfframe(3,1)(1,4){%
\begin{fmfchar*}(16,20)
\Wudtq
\fmfipair{wa[]}
\fmfipair{wb[]}
\fmfipair{wc[]}
\fmfipair{wd[]}
\fmfiequ{wa0}{point 1*length(p0)/2 of p0}
\fmfiv{d.shape=circle,d.size=2}{wa0}
\fmfiequ{wa3}{point 1*length(p3)/2 of p3}
\fmfiv{d.shape=circle,d.size=2}{wa3}
\fmfforce{(-0w,-0h)}{va0}
\fmfforce{(w,-0h)}{vb0}
\wigglywrap{wa0}{va0}{vb0}{wa3}
\end{fmfchar*}}}
}
\subfigspace
\subfigure[$\wgraph{5}{2}{1,2,3,4}$]{
\raisebox{\eqoff}{%
\fmfframe(3,1)(1,4){%
\begin{fmfchar*}(16,20)
\Wudtq
\fmfipair{wa[]}
\fmfipair{wb[]}
\fmfipair{wc[]}
\fmfipair{wd[]}
\fmfiequ{wa0}{point 1*length(p0)/2 of p0}
\fmfiv{d.shape=circle,d.size=2}{wa0}
\fmfiequ{wa4}{point 1*length(p4)/2 of p4}
\fmfiv{d.shape=circle,d.size=2}{wa4}
\fmfforce{(-0w,-0h)}{va0}
\fmfforce{(w,-0h)}{vb0}
\wigglywrap{wa0}{va0}{vb0}{wa4}
\end{fmfchar*}}}
}
\subfigspace
\subfigure[$\wgraph{5}{3}{1,2,3,4}$]{
\raisebox{\eqoff}{%
\fmfframe(3,1)(1,4){%
\begin{fmfchar*}(16,20)
\Wudtq
\fmfipair{wa[]}
\fmfipair{wb[]}
\fmfipair{wc[]}
\fmfipair{wd[]}
\fmfiequ{wa1}{point 1*length(p1)/2 of p1}
\fmfiv{d.shape=circle,d.size=2}{wa1}
\fmfiequ{wa3}{point 1*length(p3)/2 of p3}
\fmfiv{d.shape=circle,d.size=2}{wa3}
\fmfforce{(-0w,-0h)}{va0}
\fmfforce{(w,-0h)}{vb0}
\wigglywrap{wa1}{va0}{vb0}{wa3}
\end{fmfchar*}}}
}
\subfigspace
\subfigure[$\wgraph{5}{4}{1,2,3,4}$]{
\raisebox{\eqoff}{%
\fmfframe(3,1)(1,4){%
\begin{fmfchar*}(16,20)
\Wudtq
\fmfipair{wa[]}
\fmfipair{wb[]}
\fmfipair{wc[]}
\fmfipair{wd[]}
\fmfiequ{wa1}{point 1*length(p1)/2 of p1}
\fmfiv{d.shape=circle,d.size=2}{wa1}
\fmfiequ{wa4}{point 1*length(p4)/2 of p4}
\fmfiv{d.shape=circle,d.size=2}{wa4}
\fmfforce{(-0w,-0h)}{va0}
\fmfforce{(w,-0h)}{vb0}
\wigglywrap{wa1}{va0}{vb0}{wa4}
\end{fmfchar*}}}
}
\\[0.5cm]
\begin{tabular}{m{12cm}}
\toprule
$\wgraph{5}{1}{1,2,3,4}\rightarrow -(\jint{5}{22}+\jint{5}{23}+2\jint{5}{26})\,\chi(1,2,3,4) \rightarrow 2(\jint{5}{22}+\jint{5}{23}+2\jint{5}{26})\,M_5$  \\
$\wgraph{5}{2}{1,2,3,4}\rightarrow \jint{5}{22}\,\chi(1,2,3,4) \rightarrow -2\jint{5}{22}\,M_5$ \\
$\wgraph{5}{3}{1,2,3,4}\rightarrow \jint{5}{1}\,\chi(1,2,3,4) \rightarrow -2\jint{5}{1}\,M_5$ \\
$\wgraph{5}{4}{1,2,3,4}\rightarrow -\jint{5}{20}\,\chi(1,2,3,4) \rightarrow 2\jint{5}{20}\,M_5$  \\
\midrule
$\sum_{i}\wgraph{5}{i}{1,2,3,4}\rightarrow (\jint{5}{1}-\jint{5}{20}-\jint{5}{23}-2\jint{5}{26})\,\chi(1,2,3,4)$  \\
$\phantom{\chi(1,2,3,4)}\rightarrow -2(\jint{5}{1}-\jint{5}{20}-\jint{5}{23}-2\jint{5}{26})\,M_5$  \\
\bottomrule
\end{tabular}
\normalsize

\caption{Wrapping diagrams with structure $\chi(1,2,3,4)$}
\label{wrap5-1234}
\end{figure}



\begin{figure}[p]
\capstart
\addtolength{\subfigcapskip}{5pt}
\footnotesize
\renewcommand*{\thesubfigure}{}
\renewcommand{\subfigspace}{\phantom{****}}
\centering
\unitlength=0.75mm
\settoheight{\eqoff}{$\times$}%
\setlength{\eqoff}{0.5\eqoff}%
\addtolength{\eqoff}{-12.5\unitlength}%
\settoheight{\eqofftwo}{$\times$}%
\setlength{\eqofftwo}{0.5\eqofftwo}%
\addtolength{\eqofftwo}{-7.5\unitlength}%
\subfigure[$\wgraph{5}{1}{1,4,3,2}$]{
\raisebox{\eqoff}{%
\fmfframe(3,1)(1,4){%
\begin{fmfchar*}(16,20)
\Wuqtd
\fmfipair{wa[]}
\fmfipair{wb[]}
\fmfipair{wc[]}
\fmfipair{wd[]}
\fmfiequ{wa0}{point 1*length(p0)/2 of p0}
\fmfiv{d.shape=circle,d.size=2}{wa0}
\fmfiequ{wa3}{point 1*length(p3)/2 of p3}
\fmfiv{d.shape=circle,d.size=2}{wa3}
\fmfforce{(-0w,-0h)}{va0}
\fmfforce{(w,-0h)}{vb0}
\wigglywrap{wa0}{va0}{vb0}{wa3}
\end{fmfchar*}}}
}
\subfigspace
\subfigure[$\wgraph{5}{2}{1,4,3,2}$]{
\raisebox{\eqoff}{%
\fmfframe(3,1)(1,4){%
\begin{fmfchar*}(16,20)
\Wuqtd
\fmfipair{wa[]}
\fmfipair{wb[]}
\fmfipair{wc[]}
\fmfipair{wd[]}
\fmfiequ{wa0}{point 1*length(p0)/2 of p0}
\fmfiv{d.shape=circle,d.size=2}{wa0}
\fmfiequ{wa4}{point 1*length(p4)/2 of p4}
\fmfiv{d.shape=circle,d.size=2}{wa4}
\fmfforce{(-0w,-0h)}{va0}
\fmfforce{(w,-0h)}{vb0}
\wigglywrap{wa0}{va0}{vb0}{wa4}
\end{fmfchar*}}}
}
\subfigspace
\subfigure[$\wgraph{5}{3}{1,4,3,2}$]{
\raisebox{\eqoff}{%
\fmfframe(3,1)(1,4){%
\begin{fmfchar*}(16,20)
\Wuqtd
\fmfipair{wa[]}
\fmfipair{wb[]}
\fmfipair{wc[]}
\fmfipair{wd[]}
\fmfiequ{wa1}{point 1*length(p1)/2 of p1}
\fmfiv{d.shape=circle,d.size=2}{wa1}
\fmfiequ{wa3}{point 1*length(p3)/2 of p3}
\fmfiv{d.shape=circle,d.size=2}{wa3}
\fmfforce{(-0w,-0h)}{va0}
\fmfforce{(w,-0h)}{vb0}
\wigglywrap{wa1}{va0}{vb0}{wa3}
\end{fmfchar*}}}
}
\subfigspace
\subfigure[$\wgraph{5}{4}{1,4,3,2}$]{
\raisebox{\eqoff}{%
\fmfframe(3,1)(1,4){%
\begin{fmfchar*}(16,20)
\Wuqtd
\fmfipair{wa[]}
\fmfipair{wb[]}
\fmfipair{wc[]}
\fmfipair{wd[]}
\fmfiequ{wa1}{point 1*length(p1)/2 of p1}
\fmfiv{d.shape=circle,d.size=2}{wa1}
\fmfiequ{wa4}{point 1*length(p4)/2 of p4}
\fmfiv{d.shape=circle,d.size=2}{wa4}
\fmfforce{(-0w,-0h)}{va0}
\fmfforce{(w,-0h)}{vb0}
\wigglywrap{wa1}{va0}{vb0}{wa4}
\end{fmfchar*}}}
}
\\[0.5cm]
\begin{tabular}{m{12cm}}
\toprule
$\wgraph{5}{1}{1,4,3,2}\rightarrow -(\jint{5}{20}+\jint{5}{22}+2\jint{5}{27})\,\chi(1,4,3,2) \rightarrow 2(\jint{5}{20}+\jint{5}{22}+2\jint{5}{27})\,M_5$  \\
$\wgraph{5}{2}{1,4,3,2}\rightarrow \jint{5}{20}\,\chi(1,4,3,2) \rightarrow -2\jint{5}{20}\,M_5$ \\
$\wgraph{5}{3}{1,4,3,2}\rightarrow \jint{5}{22}\,\chi(1,4,3,2) \rightarrow -2\jint{5}{22}\,M_5$ \\
$\wgraph{5}{4}{1,4,3,2}\rightarrow 0$  \\
\midrule
$\sum_{i}\wgraph{5}{i}{1,4,3,2}\rightarrow -2\jint{5}{27}\,\chi(1,4,3,2) \rightarrow 4\jint{5}{27}\,M_5$   \\
\bottomrule
\end{tabular}
\normalsize

\caption{Wrapping diagrams with structure $\chi(1,4,3,2)$}
\label{wrap5-1432}
\end{figure}


\newpage


\begin{figure}[p]
\capstart
\addtolength{\subfigcapskip}{5pt}
\footnotesize
\renewcommand*{\thesubfigure}{}
\renewcommand{\subfigspace}{\phantom{****}}
\centering
\unitlength=0.75mm
\settoheight{\eqoff}{$\times$}%
\setlength{\eqoff}{0.5\eqoff}%
\addtolength{\eqoff}{-12.5\unitlength}%
\settoheight{\eqofftwo}{$\times$}%
\setlength{\eqofftwo}{0.5\eqofftwo}%
\addtolength{\eqofftwo}{-7.5\unitlength}%
\subfigure[\protect\phantom{aa}$\wgraph{5}{1}{1,3,2}\protect\phantom{aa}$]{
\raisebox{\eqoff}{%
\fmfframe(3,1)(1,4){%
\begin{fmfchar*}(16,20)
\Wutd
\fmfipair{wa[]}
\fmfipair{wb[]}
\fmfipair{wc[]}
\fmfipair{wd[]}
\fmfiequ{wa0}{point 1*length(p0)/2 of p0}
\fmfiv{d.shape=circle,d.size=2}{wa0}
\fmfiequ{wa3}{point 1*length(p3)/2 of p3}
\fmfiv{d.shape=circle,d.size=2}{wa3}
\fmfiequ{wa6}{point 1*length(p6)/2 of p6}
\fmfiv{d.shape=circle,d.size=2}{wa6}
\fmfforce{(-0w,-0h)}{va0}
\fmfforce{(1w,-0h)}{vb0}
\wigglywrap{wa0}{va0}{vb0}{wa6}
\fmfi{wiggly}{wa3..wa6}
\end{fmfchar*}}}
}
\subfigspace
\subfigure[$\wgraph{5}{2}{1,3,2}(\times2)$]{
\raisebox{\eqoff}{%
\fmfframe(3,1)(1,4){%
\begin{fmfchar*}(16,20)
\Wutd
\fmfipair{wa[]}
\fmfipair{wb[]}
\fmfipair{wc[]}
\fmfipair{wd[]}
\fmfiequ{wa0}{point 1*length(p0)/2 of p0}
\fmfiv{d.shape=circle,d.size=2}{wa0}
\fmfiequ{wa4}{point 1*length(p4)/2 of p4}
\fmfiv{d.shape=circle,d.size=2}{wa4}
\fmfiequ{wa6}{point 1*length(p6)/2 of p6}
\fmfiv{d.shape=circle,d.size=2}{wa6}
\fmfforce{(-0w,-0h)}{va0}
\fmfforce{(1w,-0h)}{vb0}
\wigglywrap{wa0}{va0}{vb0}{wa6}
\fmfi{wiggly}{wa4..wa6}
\end{fmfchar*}}}
}
\subfigspace
\subfigure[\protect\phantom{aa}$\wgraph{5}{3}{1,3,2}$\protect\phantom{aa}]{
\raisebox{\eqoff}{%
\fmfframe(3,1)(1,4){%
\begin{fmfchar*}(16,20)
\Wutd
\fmfipair{wa[]}
\fmfipair{wb[]}
\fmfipair{wc[]}
\fmfipair{wd[]}
\fmfiequ{wa1}{point 1*length(p1)/2 of p1}
\fmfiv{d.shape=circle,d.size=2}{wa1}
\fmfiequ{wa4}{point 1*length(p4)/2 of p4}
\fmfiv{d.shape=circle,d.size=2}{wa4}
\fmfiequ{wa6}{point 1*length(p6)/2 of p6}
\fmfiv{d.shape=circle,d.size=2}{wa6}
\fmfforce{(-0w,-0h)}{va0}
\fmfforce{(1w,-0h)}{vb0}
\wigglywrap{wa1}{va0}{vb0}{wa6}
\fmfi{wiggly}{wa4..wa6}
\end{fmfchar*}}}
}
\\[0.2cm]
\begin{tabular}{m{12cm}}
\toprule
$\wgraph{5}{1}{1,3,2}\rightarrow -2(\jint{5}{20}+\jint{5}{29})\,\chi(1,3,2) \rightarrow 4(\jint{5}{20}+\jint{5}{29})\,M_5$  \\
$\wgraph{5}{2}{1,3,2}\rightarrow \jint{5}{20}\,\chi(1,3,2) \rightarrow -2\jint{5}{20}\,M_5$ \\
$\wgraph{5}{3}{1,3,2}\rightarrow 0$  \\
\midrule
$\sum_{i}\scgraph{5}{i}{1,3,2}\wgraph{5}{i}{1,3,2}\rightarrow -2\jint{5}{29}\chi(1,3,2) \rightarrow 4\jint{5}{29}\,M_5$   \\
\bottomrule
\end{tabular}
\normalsize

\caption{Wrapping diagrams with structure $\chi(1,3,2)$}
\label{wrap5-132}
\end{figure}



\begin{figure}[p]
\capstart
\addtolength{\subfigcapskip}{5pt}
\footnotesize
\renewcommand*{\thesubfigure}{}
\renewcommand{\subfigspace}{\phantom{****}}
\centering
\unitlength=0.75mm
\settoheight{\eqoff}{$\times$}%
\setlength{\eqoff}{0.5\eqoff}%
\addtolength{\eqoff}{-12.5\unitlength}%
\settoheight{\eqofftwo}{$\times$}%
\setlength{\eqofftwo}{0.5\eqofftwo}%
\addtolength{\eqofftwo}{-7.5\unitlength}%
\subfigure[\protect\phantom{aa}$\wgraph{5}{1}{2,1,3}$\protect\phantom{aa}]{
\raisebox{\eqoff}{%
\fmfframe(3,1)(1,4){%
\begin{fmfchar*}(16,20)
\Wdut
\fmfipair{wa[]}
\fmfipair{wb[]}
\fmfipair{wc[]}
\fmfipair{wd[]}
\fmfiequ{wa0}{point 1*length(p0)/2 of p0}
\fmfiv{d.shape=circle,d.size=2}{wa0}
\fmfiequ{wa3}{point 1*length(p3)/2 of p3}
\fmfiv{d.shape=circle,d.size=2}{wa3}
\fmfiequ{wa6}{point 1*length(p6)/2 of p6}
\fmfiv{d.shape=circle,d.size=2}{wa6}
\fmfforce{(-0w,-0h)}{va0}
\fmfforce{(1w,-0h)}{vb0}
\wigglywrap{wa0}{va0}{vb0}{wa6}
\fmfi{wiggly}{wa3..wa6}
\end{fmfchar*}}}
}
\subfigspace
\subfigure[$\wgraph{5}{2}{2,1,3}(\times2)$]{
\raisebox{\eqoff}{%
\fmfframe(3,1)(1,4){%
\begin{fmfchar*}(16,20)
\Wdut
\fmfipair{wa[]}
\fmfipair{wb[]}
\fmfipair{wc[]}
\fmfipair{wd[]}
\fmfiequ{wa0}{point 1*length(p0)/2 of p0}
\fmfiv{d.shape=circle,d.size=2}{wa0}
\fmfiequ{wa4}{point 1*length(p4)/2 of p4}
\fmfiv{d.shape=circle,d.size=2}{wa4}
\fmfiequ{wa6}{point 1*length(p6)/2 of p6}
\fmfiv{d.shape=circle,d.size=2}{wa6}
\fmfforce{(-0w,-0h)}{va0}
\fmfforce{(1w,-0h)}{vb0}
\wigglywrap{wa0}{va0}{vb0}{wa6}
\fmfi{wiggly}{wa4..wa6}
\end{fmfchar*}}}
}
\subfigspace
\subfigure[\protect\phantom{aa}$\wgraph{5}{3}{2,1,3}$\protect\phantom{aa}]{
\raisebox{\eqoff}{%
\fmfframe(3,1)(1,4){%
\begin{fmfchar*}(16,20)
\Wdut
\fmfipair{wa[]}
\fmfipair{wb[]}
\fmfipair{wc[]}
\fmfipair{wd[]}
\fmfiequ{wa1}{point 1*length(p1)/2 of p1}
\fmfiv{d.shape=circle,d.size=2}{wa1}
\fmfiequ{wa4}{point 1*length(p4)/2 of p4}
\fmfiv{d.shape=circle,d.size=2}{wa4}
\fmfiequ{wa6}{point 1*length(p6)/2 of p6}
\fmfiv{d.shape=circle,d.size=2}{wa6}
\fmfforce{(-0w,-0h)}{va0}
\fmfforce{(1w,-0h)}{vb0}
\wigglywrap{wa1}{va0}{vb0}{wa6}
\fmfi{wiggly}{wa4..wa6}
\end{fmfchar*}}}
}
\\[0.2cm]
\begin{tabular}{m{12cm}}
\toprule
$\wgraph{5}{1}{2,1,3}\rightarrow -2(\jint{5}{20}+\jint{5}{28})\,\chi(2,1,3) \rightarrow 4(\jint{5}{20}+\jint{5}{28})\,M_5$  \\
$\wgraph{5}{2}{2,1,3}\rightarrow \jint{5}{20}\,\chi(2,1,3) \rightarrow -2\jint{5}{20}\,M_5$ \\
$\wgraph{5}{3}{2,1,3}\rightarrow 0$  \\
\midrule
$\sum_{i}\scgraph{5}{i}{2,1,3}\wgraph{5}{i}{2,1,3}\rightarrow -2\jint{5}{28}\chi(2,1,3) \rightarrow 4\jint{5}{28}\,M_5$   \\
\bottomrule
\end{tabular}
\normalsize

\caption{Wrapping diagrams with structure $\chi(2,1,3)$}
\label{wrap5-213}
\end{figure}



\begin{figure}[p]
\capstart
\addtolength{\subfigcapskip}{5pt}
\footnotesize
\renewcommand*{\thesubfigure}{}
\renewcommand{\subfigspace}{\phantom{****}}
\centering
\unitlength=0.75mm
\settoheight{\eqoff}{$\times$}%
\setlength{\eqoff}{0.5\eqoff}%
\addtolength{\eqoff}{-12.5\unitlength}%
\settoheight{\eqofftwo}{$\times$}%
\setlength{\eqofftwo}{0.5\eqofftwo}%
\addtolength{\eqofftwo}{-7.5\unitlength}%
\subfigure[$\wgraph{5}{1}{1,2,3}$]{
\raisebox{\eqoff}{%
\fmfframe(3,1)(1,4){%
\begin{fmfchar*}(16,20)
\Wudt
\fmfipair{wa[]}
\fmfipair{wb[]}
\fmfipair{wc[]}
\fmfipair{wd[]}
\fmfiequ{wa0}{point 1*length(p0)/2 of p0}
\fmfiv{d.shape=circle,d.size=2}{wa0}
\fmfiequ{wa3}{point 1*length(p3)/2 of p3}
\fmfiv{d.shape=circle,d.size=2}{wa3}
\fmfiequ{wa6}{point 1*length(p6)/2 of p6}
\fmfiv{d.shape=circle,d.size=2}{wa6}
\fmfforce{(-0w,-0h)}{va0}
\fmfforce{(1w,-0h)}{vb0}
\wigglywrap{wa0}{va0}{vb0}{wa6}
\fmfi{wiggly}{wa3..wa6}
\end{fmfchar*}}}
}
\subfigspace
\subfigure[$\wgraph{5}{2}{1,2,3}$]{
\raisebox{\eqoff}{%
\fmfframe(3,1)(1,4){%
\begin{fmfchar*}(16,20)
\Wudt
\fmfipair{wa[]}
\fmfipair{wb[]}
\fmfipair{wc[]}
\fmfipair{wd[]}
\fmfiequ{wa0}{point 1*length(p0)/2 of p0}
\fmfiv{d.shape=circle,d.size=2}{wa0}
\fmfiequ{wa4}{point 1*length(p4)/2 of p4}
\fmfiv{d.shape=circle,d.size=2}{wa4}
\fmfiequ{wa6}{point 1*length(p6)/2 of p6}
\fmfiv{d.shape=circle,d.size=2}{wa6}
\fmfforce{(-0w,-0h)}{va0}
\fmfforce{(1w,-0h)}{vb0}
\wigglywrap{wa0}{va0}{vb0}{wa6}
\fmfi{wiggly}{wa4..wa6}
\end{fmfchar*}}}
}
\subfigspace
\subfigure[$\wgraph{5}{3}{1,2,3}$]{
\raisebox{\eqoff}{%
\fmfframe(3,1)(1,4){%
\begin{fmfchar*}(16,20)
\Wudt
\fmfipair{wa[]}
\fmfipair{wb[]}
\fmfipair{wc[]}
\fmfipair{wd[]}
\fmfiequ{wa1}{point 1*length(p1)/2 of p1}
\fmfiv{d.shape=circle,d.size=2}{wa1}
\fmfiequ{wa3}{point 1*length(p3)/2 of p3}
\fmfiv{d.shape=circle,d.size=2}{wa3}
\fmfiequ{wa6}{point 1*length(p6)/2 of p6}
\fmfiv{d.shape=circle,d.size=2}{wa6}
\fmfforce{(-0w,-0h)}{va0}
\fmfforce{(1w,-0h)}{vb0}
\wigglywrap{wa1}{va0}{vb0}{wa6}
\fmfi{wiggly}{wa3..wa6}
\end{fmfchar*}}}
}
\subfigspace
\subfigure[$\wgraph{5}{4}{1,2,3}$]{
\raisebox{\eqoff}{%
\fmfframe(3,1)(1,4){%
\begin{fmfchar*}(16,20)
\Wudt
\fmfipair{wa[]}
\fmfipair{wb[]}
\fmfipair{wc[]}
\fmfipair{wd[]}
\fmfiequ{wa1}{point 1*length(p1)/2 of p1}
\fmfiv{d.shape=circle,d.size=2}{wa1}
\fmfiequ{wa4}{point 1*length(p4)/2 of p4}
\fmfiv{d.shape=circle,d.size=2}{wa4}
\fmfiequ{wa6}{point 1*length(p6)/2 of p6}
\fmfiv{d.shape=circle,d.size=2}{wa6}
\fmfforce{(-0w,-0h)}{va0}
\fmfforce{(1w,-0h)}{vb0}
\wigglywrap{wa1}{va0}{vb0}{wa6}
\fmfi{wiggly}{wa4..wa6}
\end{fmfchar*}}}
}
\\[0.2cm]
\begin{tabular}{m{12cm}}
\toprule
$\wgraph{5}{1}{1,2,3}\rightarrow -(\jint{5}{22}+\jint{5}{23}+2\jint{5}{26})\,\chi(1,2,3) \rightarrow -(\jint{5}{22}+\jint{5}{23}+2\jint{5}{26})\,M_5$  \\
$\wgraph{5}{2}{1,2,3}\rightarrow \jint{5}{22}\,\chi(1,2,3) \rightarrow \jint{5}{22}\,M_5$ \\
$\wgraph{5}{3}{1,2,3}\rightarrow (\jint{5}{22}+\jint{5}{23}+2\jint{5}{26})\,\chi(1,2,3) \rightarrow (\jint{5}{22}+\jint{5}{23}+2\jint{5}{26})\,M_5$ \\
$\wgraph{5}{4}{1,2,3}\rightarrow -\jint{5}{22}\,\chi(1,2,3) \rightarrow -\jint{5}{22}\,M_5$  \\
\midrule
$\sum_{i}\wgraph{5}{i}{1,2,3}\rightarrow 0$   \\
\bottomrule
\end{tabular}
\normalsize

\caption{Wrapping diagrams with structure $\chi(1,2,3)$}
\label{wrap5-123}
\end{figure}



\begin{figure}[p]
\capstart
\addtolength{\subfigcapskip}{5pt}
\footnotesize
\renewcommand*{\thesubfigure}{}
\renewcommand{\subfigspace}{\phantom{****}}
\centering
\unitlength=0.75mm
\settoheight{\eqoff}{$\times$}%
\setlength{\eqoff}{0.5\eqoff}%
\addtolength{\eqoff}{-12.5\unitlength}%
\settoheight{\eqofftwo}{$\times$}%
\setlength{\eqofftwo}{0.5\eqofftwo}%
\addtolength{\eqofftwo}{-7.5\unitlength}%
\subfigure[$\wgraph{5}{1}{2,1}$]{
\raisebox{\eqoff}{%
\fmfframe(3,1)(1,4){%
\begin{fmfchar*}(16,20)
\Wdu
\fmfipair{wa[]}
\fmfipair{wb[]}
\fmfipair{wc[]}
\fmfipair{wd[]}
\fmfiequ{wa0}{point 1*length(p0)/2 of p0}
\fmfiv{d.shape=circle,d.size=2}{wa0}
\fmfiequ{wa3}{point 1*length(p3)/2 of p3}
\fmfiv{d.shape=circle,d.size=2}{wa3}
\fmfiequ{wa6}{point 1*length(p6)/2 of p6}
\fmfiv{d.shape=circle,d.size=2}{wa6}
\fmfiequ{wa7}{point 1*length(p7)/2 of p7}
\fmfiv{d.shape=circle,d.size=2}{wa7}
\fmfforce{(-0w,-0h)}{va0}
\fmfforce{(1w,-0h)}{vb0}
\wigglywrap{wa0}{va0}{vb0}{wa7}
\fmfi{wiggly}{wa3..wa6}
\fmfi{wiggly}{wa6..wa7}
\end{fmfchar*}}}
}
\subfigspace
\subfigure[$\wgraph{5}{2}{2,1}$]{
\raisebox{\eqoff}{%
\fmfframe(3,1)(1,4){%
\begin{fmfchar*}(16,20)
\Wdu
\fmfipair{wa[]}
\fmfipair{wb[]}
\fmfipair{wc[]}
\fmfipair{wd[]}
\fmfiequ{wa0}{point 1*length(p0)/2 of p0}
\fmfiv{d.shape=circle,d.size=2}{wa0}
\fmfiequ{wa4}{point 1*length(p4)/2 of p4}
\fmfiv{d.shape=circle,d.size=2}{wa4}
\fmfiequ{wa6}{point 1*length(p6)/2 of p6}
\fmfiv{d.shape=circle,d.size=2}{wa6}
\fmfiequ{wa7}{point 1*length(p7)/2 of p7}
\fmfiv{d.shape=circle,d.size=2}{wa7}
\fmfforce{(-0w,-0h)}{va0}
\fmfforce{(1w,-0h)}{vb0}
\wigglywrap{wa0}{va0}{vb0}{wa7}
\fmfi{wiggly}{wa4..wa6}
\fmfi{wiggly}{wa6..wa7}
\end{fmfchar*}}}
}
\subfigspace
\subfigure[$\wgraph{5}{3}{2,1}$]{
\raisebox{\eqoff}{%
\fmfframe(3,1)(1,4){%
\begin{fmfchar*}(16,20)
\Wdu
\fmfipair{wa[]}
\fmfipair{wb[]}
\fmfipair{wc[]}
\fmfipair{wd[]}
\fmfiequ{wa1}{point 1*length(p1)/2 of p1}
\fmfiv{d.shape=circle,d.size=2}{wa1}
\fmfiequ{wa3}{point 1*length(p3)/2 of p3}
\fmfiv{d.shape=circle,d.size=2}{wa3}
\fmfiequ{wa6}{point 1*length(p6)/2 of p6}
\fmfiv{d.shape=circle,d.size=2}{wa6}
\fmfiequ{wa7}{point 1*length(p7)/2 of p7}
\fmfiv{d.shape=circle,d.size=2}{wa7}
\fmfforce{(-0w,-0h)}{va0}
\fmfforce{(1w,-0h)}{vb0}
\wigglywrap{wa1}{va0}{vb0}{wa7}
\fmfi{wiggly}{wa3..wa6}
\fmfi{wiggly}{wa6..wa7}
\end{fmfchar*}}}
}
\subfigspace
\subfigure[$\wgraph{5}{4}{2,1}$]{
\raisebox{\eqoff}{%
\fmfframe(3,1)(1,4){%
\begin{fmfchar*}(16,20)
\Wdu
\fmfipair{wa[]}
\fmfipair{wb[]}
\fmfipair{wc[]}
\fmfipair{wd[]}
\fmfiequ{wa1}{point 1*length(p1)/2 of p1}
\fmfiv{d.shape=circle,d.size=2}{wa1}
\fmfiequ{wa4}{point 1*length(p4)/2 of p4}
\fmfiv{d.shape=circle,d.size=2}{wa4}
\fmfiequ{wa6}{point 1*length(p6)/2 of p6}
\fmfiv{d.shape=circle,d.size=2}{wa6}
\fmfiequ{wa7}{point 1*length(p7)/2 of p7}
\fmfiv{d.shape=circle,d.size=2}{wa7}
\fmfforce{(-0w,-0h)}{va0}
\fmfforce{(1w,-0h)}{vb0}
\wigglywrap{wa1}{va0}{vb0}{wa7}
\fmfi{wiggly}{wa4..wa6}
\fmfi{wiggly}{wa6..wa7}
\end{fmfchar*}}}
}
\\[0.5cm]
\begin{tabular}{m{12cm}}
\toprule
$\wgraph{5}{1}{2,1}\rightarrow -\jint{5}{1}\,\chi(2,1) \rightarrow -\jint{5}{1}\,M_5$  \\
$\wgraph{5}{2}{2,1}\rightarrow (\jint{5}{22}+\jint{5}{23}+2\jint{5}{26})\,\chi(2,1) \rightarrow (\jint{5}{22}+\jint{5}{23}+2\jint{5}{26})\,M_5$ \\
$\wgraph{5}{3}{2,1}\rightarrow \jint{5}{20}\,\chi(2,1) \rightarrow \jint{5}{20}\,M_5$ \\
$\wgraph{5}{4}{2,1}\rightarrow -\jint{5}{22}\,\chi(2,1) \rightarrow -\jint{5}{22}\,M_5$  \\
\midrule
$\sum_{i}\wgraph{5}{i}{2,1}\rightarrow -(\jint{5}{1}-\jint{5}{20}-\jint{5}{23}-2\jint{5}{26})\,\chi(2,1)$  \\
$\phantom{\chi(2,1)}\rightarrow -(\jint{5}{1}-\jint{5}{20}-\jint{5}{23}-2\jint{5}{26})\,M_5$  \\
\bottomrule
\end{tabular}
\normalsize

\caption{Wrapping diagrams with structure $\chi(2,1)$}
\label{wrap5-21}
\end{figure}



\begin{figure}[p]
\capstart
\addtolength{\subfigcapskip}{5pt}
\footnotesize
\renewcommand*{\thesubfigure}{}
\renewcommand{\subfigspace}{\phantom{****}}
\centering
\unitlength=0.75mm
\settoheight{\eqoff}{$\times$}%
\setlength{\eqoff}{0.5\eqoff}%
\addtolength{\eqoff}{-12.5\unitlength}%
\settoheight{\eqofftwo}{$\times$}%
\setlength{\eqofftwo}{0.5\eqofftwo}%
\addtolength{\eqofftwo}{-7.5\unitlength}%
\subfigure[$\wgraph{5}{1}{1}(\times2)$]{
\raisebox{\eqoff}{%
\fmfframe(3,1)(1,4){%
\begin{fmfchar*}(16,20)
\Wu
\fmfipair{wa[]}
\fmfipair{wb[]}
\fmfipair{wc[]}
\fmfipair{wd[]}
\fmfiequ{wa0}{point 1*length(p0)/2 of p0}
\fmfiv{d.shape=circle,d.size=2}{wa0}
\fmfiequ{wa1}{point 1*length(p1)/2 of p1}
\fmfiv{d.shape=circle,d.size=2}{wa1}
\fmfiequ{wa2}{point 1*length(p2)/2 of p2}
\fmfiv{d.shape=circle,d.size=2}{wa2}
\fmfiequ{wa6}{point 1*length(p6)/2 of p6}
\fmfiv{d.shape=circle,d.size=2}{wa6}
\fmfiequ{wa7}{point 1*length(p7)/2 of p7}
\fmfiv{d.shape=circle,d.size=2}{wa7}
\fmfi{wiggly}{wa0..wa1}
\fmfforce{(-0w,-0h)}{va0}
\fmfforce{(1w,-0h)}{vb0}
\wigglywrap{wa0}{va0}{vb0}{wa7}
\fmfi{wiggly}{wa2..wa6}
\fmfi{wiggly}{wa6..wa7}
\end{fmfchar*}}}
}
\subfigspace
\subfigure[$\wgraph{5}{2}{1}(\times2)$]{
\raisebox{\eqoff}{%
\fmfframe(3,1)(1,4){%
\begin{fmfchar*}(16,20)
\Wu
\fmfipair{wa[]}
\fmfipair{wb[]}
\fmfipair{wc[]}
\fmfipair{wd[]}
\fmfiequ{wa0}{point 1*length(p0)/2 of p0}
\fmfiv{d.shape=circle,d.size=2}{wa0}
\fmfiequ{wa2}{point 1*length(p2)/3 of p2}
\fmfiv{d.shape=circle,d.size=2}{wa2}
\fmfiequ{wb2}{point 2*length(p2)/3 of p2}
\fmfiv{d.shape=circle,d.size=2}{wb2}
\fmfiequ{wa6}{point 1*length(p6)/2 of p6}
\fmfiv{d.shape=circle,d.size=2}{wa6}
\fmfiequ{wa7}{point 1*length(p7)/2 of p7}
\fmfiv{d.shape=circle,d.size=2}{wa7}
\fmfi{wiggly}{wa0..wa2}
\fmfforce{(-0w,-0h)}{va0}
\fmfforce{(1w,-0h)}{vb0}
\wigglywrap{wa0}{va0}{vb0}{wa7}
\fmfi{wiggly}{wb2..wa6}
\fmfi{wiggly}{wa6..wa7}
\end{fmfchar*}}}
}
\\[0.5cm]
\begin{tabular}{m{12cm}}
\toprule
$\wgraph{5}{1}{1}\rightarrow \jint{5}{1}\,\chi(1) \rightarrow-2\jint{5}{1}\,M_5$  \\
$\wgraph{5}{2}{1}\rightarrow -\jint{5}{20}\,\chi(1) \rightarrow 2\jint{5}{20}\,M_5$  \\
\midrule
$\sum_{i}\scgraph{5}{i}{1}\wgraph{5}{i}{1}\rightarrow 2(\jint{5}{1}-\jint{5}{20})\,\chi(1) \rightarrow -4(\jint{5}{1}-\jint{5}{20})\,M_5$   \\ 
\bottomrule
\end{tabular}
\normalsize

\caption{Wrapping diagrams with structure $\chi(1)$}
\label{wrap5-1}
\end{figure}



\begin{figure}[p]
\capstart
\footnotesize
\renewcommand*{\thesubfigure}{}
\renewcommand*{\subfigspace}{\phantom{-}}
\vspace{-1cm}
\centering
\unitlength=0.75mm
\settoheight{\eqoff}{$\times$}%
\setlength{\eqoff}{0.5\eqoff}%
\addtolength{\eqoff}{-12.5\unitlength}%
\settoheight{\eqofftwo}{$\times$}%
\setlength{\eqofftwo}{0.5\eqofftwo}%
\addtolength{\eqofftwo}{-7.5\unitlength}%
\subfigure[$\wgraph{5}{1}{2,1,4}$]{
\raisebox{\eqoff}{%
\fmfframe(3,1)(1,4){%
\begin{fmfchar*}(16,20)
\Wduq
\fmfipair{wa[]}
\fmfipair{wb[]}
\fmfipair{wc[]}
\fmfipair{wd[]}
\fmfiequ{wa0}{point 1*length(p0)/2 of p0}
\fmfiv{d.shape=circle,d.size=2}{wa0}
\fmfiequ{wa3}{point 1*length(p3)/2 of p3}
\fmfiv{d.shape=circle,d.size=2}{wa3}
\fmfiequ{wa6}{point 1*length(p6)/2 of p6}
\fmfiv{d.shape=circle,d.size=2}{wa6}
\fmfiequ{wa7}{point 1*length(p7)/2 of p7}
\fmfiv{d.shape=circle,d.size=2}{wa7}
\fmfforce{(-0w,-0h)}{va0}
\fmfforce{(1w,-0h)}{vb0}
\wigglywrap{wa0}{va0}{vb0}{wa7}
\fmfi{wiggly}{wa3..wa6}
\end{fmfchar*}}}
}
\subfigspace
\subfigure[$\wgraph{5}{2}{2,1,4}$]{
\raisebox{\eqoff}{%
\fmfframe(3,1)(1,4){%
\begin{fmfchar*}(16,20)
\Wduq
\fmfipair{wa[]}
\fmfipair{wb[]}
\fmfipair{wc[]}
\fmfipair{wd[]}
\fmfiequ{wa0}{point 1*length(p0)/2 of p0}
\fmfiv{d.shape=circle,d.size=2}{wa0}
\fmfiequ{wa3}{point 1*length(p3)/2 of p3}
\fmfiv{d.shape=circle,d.size=2}{wa3}
\fmfiequ{wa7}{point 1*length(p7)/2 of p7}
\fmfiv{d.shape=circle,d.size=2}{wa7}
\fmfforce{(-0w,-0h)}{va0}
\fmfforce{(1w,-0h)}{vb0}
\wigglywrap{wa0}{va0}{vb0}{wa7}
\fmfi{wiggly}{wa3..wa7}
\end{fmfchar*}}}
}
\subfigspace
\subfigure[$\wgraph{5}{3}{2,1,4}$]{
\raisebox{\eqoff}{%
\fmfframe(3,1)(1,4){%
\begin{fmfchar*}(16,20)
\Wduq
\fmfipair{wa[]}
\fmfipair{wb[]}
\fmfipair{wc[]}
\fmfipair{wd[]}
\fmfiequ{wa0}{point 1*length(p0)/2 of p0}
\fmfiv{d.shape=circle,d.size=2}{wa0}
\fmfiequ{wa3}{point 1*length(p3)/2 of p3}
\fmfiv{d.shape=circle,d.size=2}{wa3}
\fmfiequ{wa7}{point 1*length(p7)/3 of p7}
\fmfiv{d.shape=circle,d.size=2}{wa7}
\fmfiequ{wb7}{point 2*length(p7)/3 of p7}
\fmfiv{d.shape=circle,d.size=2}{wb7}
\fmfforce{(-0w,-0h)}{va0}
\fmfforce{(1w,-0h)}{vb0}
\wigglywrap{wa0}{va0}{vb0}{wa7}
\fmfi{wiggly}{wa3..wb7}
\end{fmfchar*}}}
}
\subfigspace
\subfigure[$\wgraph{5}{4}{2,1,4}$]{
\raisebox{\eqoff}{%
\fmfframe(3,1)(1,4){%
\begin{fmfchar*}(16,20)
\Wduq
\fmfipair{wa[]}
\fmfipair{wb[]}
\fmfipair{wc[]}
\fmfipair{wd[]}
\fmfiequ{wa0}{point 1*length(p0)/2 of p0}
\fmfiv{d.shape=circle,d.size=2}{wa0}
\fmfiequ{wa4}{point 1*length(p4)/2 of p4}
\fmfiv{d.shape=circle,d.size=2}{wa4}
\fmfiequ{wa6}{point 1*length(p6)/2 of p6}
\fmfiv{d.shape=circle,d.size=2}{wa6}
\fmfiequ{wa7}{point 1*length(p7)/2 of p7}
\fmfiv{d.shape=circle,d.size=2}{wa7}
\fmfforce{(-0w,-0h)}{va0}
\fmfforce{(1w,-0h)}{vb0}
\wigglywrap{wa0}{va0}{vb0}{wa7}
\fmfi{wiggly}{wa4..wa6}
\end{fmfchar*}}}
}
\subfigspace
\subfigure[$\wgraph{5}{5}{2,1,4}$]{
\raisebox{\eqoff}{%
\fmfframe(3,1)(1,4){%
\begin{fmfchar*}(16,20)
\Wduq
\fmfipair{wa[]}
\fmfipair{wb[]}
\fmfipair{wc[]}
\fmfipair{wd[]}
\fmfiequ{wa0}{point 1*length(p0)/2 of p0}
\fmfiv{d.shape=circle,d.size=2}{wa0}
\fmfiequ{wa4}{point 1*length(p4)/2 of p4}
\fmfiv{d.shape=circle,d.size=2}{wa4}
\fmfiequ{wa7}{point 1*length(p7)/2 of p7}
\fmfiv{d.shape=circle,d.size=2}{wa7}
\fmfforce{(-0w,-0h)}{va0}
\fmfforce{(1w,-0h)}{vb0}
\wigglywrap{wa0}{va0}{vb0}{wa7}
\fmfi{wiggly}{wa4..wa7}
\end{fmfchar*}}}
}
\subfigspace
\subfigure[$\wgraph{5}{6}{2,1,4}$]{
\raisebox{\eqoff}{%
\fmfframe(3,1)(1,4){%
\begin{fmfchar*}(16,20)
\Wduq
\fmfipair{wa[]}
\fmfipair{wb[]}
\fmfipair{wc[]}
\fmfipair{wd[]}
\fmfiequ{wa0}{point 1*length(p0)/2 of p0}
\fmfiv{d.shape=circle,d.size=2}{wa0}
\fmfiequ{wa4}{point 1*length(p4)/2 of p4}
\fmfiv{d.shape=circle,d.size=2}{wa4}
\fmfiequ{wa7}{point 1*length(p7)/3 of p7}
\fmfiv{d.shape=circle,d.size=2}{wa7}
\fmfiequ{wb7}{point 2*length(p7)/3 of p7}
\fmfiv{d.shape=circle,d.size=2}{wb7}
\fmfforce{(-0w,-0h)}{va0}
\fmfforce{(1w,-0h)}{vb0}
\wigglywrap{wa0}{va0}{vb0}{wa7}
\fmfi{wiggly}{wa4..wb7}
\end{fmfchar*}}}
}
\subfigspace
\subfigure[$\wgraph{5}{7}{2,1,4}$]{
\raisebox{\eqoff}{%
\fmfframe(3,1)(1,4){%
\begin{fmfchar*}(16,20)
\Wduq
\fmfipair{wa[]}
\fmfipair{wb[]}
\fmfipair{wc[]}
\fmfipair{wd[]}
\fmfiequ{wa0}{point 1*length(p0)/2 of p0}
\fmfiv{d.shape=circle,d.size=2}{wa0}
\fmfiequ{wa3}{point 1*length(p3)/2 of p3}
\fmfiv{d.shape=circle,d.size=2}{wa3}
\fmfiequ{wa7}{point 1*length(p7)/3 of p7}
\fmfiv{d.shape=circle,d.size=2}{wa7}
\fmfiequ{wb7}{point 2*length(p7)/3 of p7}
\fmfiv{d.shape=circle,d.size=2}{wb7}
\fmfforce{(-0w,-0h)}{va0}
\fmfforce{(1w,-0h)}{vb0}
\wigglywrap{wa0}{va0}{vb0}{wb7}
\fmfi{wiggly}{wa3..wa7}
\end{fmfchar*}}}
}
\subfigspace
\subfigure[$\wgraph{5}{8}{2,1,4}$]{
\raisebox{\eqoff}{%
\fmfframe(3,1)(1,4){%
\begin{fmfchar*}(16,20)
\Wduq
\fmfipair{wa[]}
\fmfipair{wb[]}
\fmfipair{wc[]}
\fmfipair{wd[]}
\fmfiequ{wa0}{point 1*length(p0)/2 of p0}
\fmfiv{d.shape=circle,d.size=2}{wa0}
\fmfiequ{wa4}{point 1*length(p4)/2 of p4}
\fmfiv{d.shape=circle,d.size=2}{wa4}
\fmfiequ{wa7}{point 1*length(p7)/3 of p7}
\fmfiv{d.shape=circle,d.size=2}{wa7}
\fmfiequ{wb7}{point 2*length(p7)/3 of p7}
\fmfiv{d.shape=circle,d.size=2}{wb7}
\fmfforce{(-0w,-0h)}{va0}
\fmfforce{(1w,-0h)}{vb0}
\wigglywrap{wa0}{va0}{vb0}{wb7}
\fmfi{wiggly}{wa4..wa7}
\end{fmfchar*}}}
}
\\
\subfigure[$\wgraph{5}{9}{2,1,4}$]{
\raisebox{\eqoff}{%
\fmfframe(3,1)(1,4){%
\begin{fmfchar*}(16,20)
\Wduq
\fmfipair{wa[]}
\fmfipair{wb[]}
\fmfipair{wc[]}
\fmfipair{wd[]}
\fmfiequ{wa0}{point 1*length(p0)/2 of p0}
\fmfiv{d.shape=circle,d.size=2}{wa0}
\fmfiequ{wa3}{point 1*length(p3)/2 of p3}
\fmfiv{d.shape=circle,d.size=2}{wa3}
\fmfiequ{wa6}{point 1*length(p6)/2 of p6}
\fmfiv{d.shape=circle,d.size=2}{wa6}
\fmfiequ{wa9}{point 1*length(p9)/2 of p9}
\fmfiv{d.shape=circle,d.size=2}{wa9}
\fmfforce{(-0w,-0h)}{va0}
\fmfforce{(1w,-0h)}{vb0}
\wigglywrap{wa0}{va0}{vb0}{wa9}
\fmfi{wiggly}{wa3..wa6}
\end{fmfchar*}}}
}
\subfigspace
\subfigure[$\wgraph{5}{10}{2,1,4}$]{
\raisebox{\eqoff}{%
\fmfframe(3,1)(1,4){%
\begin{fmfchar*}(16,20)
\Wduq
\fmfipair{wa[]}
\fmfipair{wb[]}
\fmfipair{wc[]}
\fmfipair{wd[]}
\fmfiequ{wa0}{point 1*length(p0)/2 of p0}
\fmfiv{d.shape=circle,d.size=2}{wa0}
\fmfiequ{wa3}{point 1*length(p3)/2 of p3}
\fmfiv{d.shape=circle,d.size=2}{wa3}
\fmfiequ{wa7}{point 1*length(p7)/2 of p7}
\fmfiv{d.shape=circle,d.size=2}{wa7}
\fmfiequ{wa9}{point 1*length(p9)/2 of p9}
\fmfiv{d.shape=circle,d.size=2}{wa9}
\fmfforce{(-0w,-0h)}{va0}
\fmfforce{(1w,-0h)}{vb0}
\wigglywrap{wa0}{va0}{vb0}{wa9}
\fmfi{wiggly}{wa3..wa7}
\end{fmfchar*}}}
}
\subfigspace
\subfigure[$\wgraph{5}{11}{2,1,4}$]{
\raisebox{\eqoff}{%
\fmfframe(3,1)(1,4){%
\begin{fmfchar*}(16,20)
\Wduq
\fmfipair{wa[]}
\fmfipair{wb[]}
\fmfipair{wc[]}
\fmfipair{wd[]}
\fmfiequ{wa0}{point 1*length(p0)/2 of p0}
\fmfiv{d.shape=circle,d.size=2}{wa0}
\fmfiequ{wa4}{point 1*length(p4)/2 of p4}
\fmfiv{d.shape=circle,d.size=2}{wa4}
\fmfiequ{wa6}{point 1*length(p6)/2 of p6}
\fmfiv{d.shape=circle,d.size=2}{wa6}
\fmfiequ{wa9}{point 1*length(p9)/2 of p9}
\fmfiv{d.shape=circle,d.size=2}{wa9}
\fmfforce{(-0w,-0h)}{va0}
\fmfforce{(1w,-0h)}{vb0}
\wigglywrap{wa0}{va0}{vb0}{wa9}
\fmfi{wiggly}{wa4..wa6}
\end{fmfchar*}}}
}
\subfigspace
\subfigure[$\wgraph{5}{12}{2,1,4}$]{
\raisebox{\eqoff}{%
\fmfframe(3,1)(1,4){%
\begin{fmfchar*}(16,20)
\Wduq
\fmfipair{wa[]}
\fmfipair{wb[]}
\fmfipair{wc[]}
\fmfipair{wd[]}
\fmfiequ{wa0}{point 1*length(p0)/2 of p0}
\fmfiv{d.shape=circle,d.size=2}{wa0}
\fmfiequ{wa4}{point 1*length(p4)/2 of p4}
\fmfiv{d.shape=circle,d.size=2}{wa4}
\fmfiequ{wa7}{point 1*length(p7)/2 of p7}
\fmfiv{d.shape=circle,d.size=2}{wa7}
\fmfiequ{wa9}{point 1*length(p9)/2 of p9}
\fmfiv{d.shape=circle,d.size=2}{wa9}
\fmfforce{(-0w,-0h)}{va0}
\fmfforce{(1w,-0h)}{vb0}
\wigglywrap{wa0}{va0}{vb0}{wa9}
\fmfi{wiggly}{wa4..wa7}
\end{fmfchar*}}}
}
\subfigspace
\subfigure[$\wgraph{5}{13}{2,1,4}$]{
\raisebox{\eqoff}{%
\fmfframe(3,1)(1,4){%
\begin{fmfchar*}(16,20)
\Wduq
\fmfipair{wa[]}
\fmfipair{wb[]}
\fmfipair{wc[]}
\fmfipair{wd[]}
\fmfiequ{wa1}{point 1*length(p1)/2 of p1}
\fmfiv{d.shape=circle,d.size=2}{wa1}
\fmfiequ{wa3}{point 1*length(p3)/2 of p3}
\fmfiv{d.shape=circle,d.size=2}{wa3}
\fmfiequ{wa6}{point 1*length(p6)/2 of p6}
\fmfiv{d.shape=circle,d.size=2}{wa6}
\fmfiequ{wa7}{point 1*length(p7)/2 of p7}
\fmfiv{d.shape=circle,d.size=2}{wa7}
\fmfforce{(-0w,-0h)}{va0}
\fmfforce{(1w,-0h)}{vb0}
\wigglywrap{wa1}{va0}{vb0}{wa7}
\fmfi{wiggly}{wa3..wa6}
\end{fmfchar*}}}
}
\subfigspace
\subfigure[$\wgraph{5}{14}{2,1,4}$]{
\raisebox{\eqoff}{%
\fmfframe(3,1)(1,4){%
\begin{fmfchar*}(16,20)
\Wduq
\fmfipair{wa[]}
\fmfipair{wb[]}
\fmfipair{wc[]}
\fmfipair{wd[]}
\fmfiequ{wa1}{point 1*length(p1)/2 of p1}
\fmfiv{d.shape=circle,d.size=2}{wa1}
\fmfiequ{wa3}{point 1*length(p3)/2 of p3}
\fmfiv{d.shape=circle,d.size=2}{wa3}
\fmfiequ{wa7}{point 1*length(p7)/2 of p7}
\fmfiv{d.shape=circle,d.size=2}{wa7}
\fmfforce{(-0w,-0h)}{va0}
\fmfforce{(1w,-0h)}{vb0}
\wigglywrap{wa1}{va0}{vb0}{wa7}
\fmfi{wiggly}{wa3..wa7}
\end{fmfchar*}}}
}
\subfigspace
\subfigure[$\wgraph{5}{15}{2,1,4}$]{
\raisebox{\eqoff}{%
\fmfframe(3,1)(1,4){%
\begin{fmfchar*}(16,20)
\Wduq
\fmfipair{wa[]}
\fmfipair{wb[]}
\fmfipair{wc[]}
\fmfipair{wd[]}
\fmfiequ{wa1}{point 1*length(p1)/2 of p1}
\fmfiv{d.shape=circle,d.size=2}{wa1}
\fmfiequ{wa3}{point 1*length(p3)/2 of p3}
\fmfiv{d.shape=circle,d.size=2}{wa3}
\fmfiequ{wa7}{point 1*length(p7)/3 of p7}
\fmfiv{d.shape=circle,d.size=2}{wa7}
\fmfiequ{wb7}{point 2*length(p7)/3 of p7}
\fmfiv{d.shape=circle,d.size=2}{wb7}
\fmfforce{(-0w,-0h)}{va0}
\fmfforce{(1w,-0h)}{vb0}
\wigglywrap{wa1}{va0}{vb0}{wa7}
\fmfi{wiggly}{wa3..wb7}
\end{fmfchar*}}}
}
\subfigspace
\subfigure[$\wgraph{5}{16}{2,1,4}$]{
\raisebox{\eqoff}{%
\fmfframe(3,1)(1,4){%
\begin{fmfchar*}(16,20)
\Wduq
\fmfipair{wa[]}
\fmfipair{wb[]}
\fmfipair{wc[]}
\fmfipair{wd[]}
\fmfiequ{wa1}{point 1*length(p1)/2 of p1}
\fmfiv{d.shape=circle,d.size=2}{wa1}
\fmfiequ{wa4}{point 1*length(p4)/2 of p4}
\fmfiv{d.shape=circle,d.size=2}{wa4}
\fmfiequ{wa6}{point 1*length(p6)/2 of p6}
\fmfiv{d.shape=circle,d.size=2}{wa6}
\fmfiequ{wa7}{point 1*length(p7)/2 of p7}
\fmfiv{d.shape=circle,d.size=2}{wa7}
\fmfforce{(-0w,-0h)}{va0}
\fmfforce{(1w,-0h)}{vb0}
\wigglywrap{wa1}{va0}{vb0}{wa7}
\fmfi{wiggly}{wa4..wa6}
\end{fmfchar*}}}
}
\\
\subfigure[$\wgraph{5}{17}{2,1,4}$]{
\raisebox{\eqoff}{%
\fmfframe(3,1)(1,4){%
\begin{fmfchar*}(16,20)
\Wduq
\fmfipair{wa[]}
\fmfipair{wb[]}
\fmfipair{wc[]}
\fmfipair{wd[]}
\fmfiequ{wa1}{point 1*length(p1)/2 of p1}
\fmfiv{d.shape=circle,d.size=2}{wa1}
\fmfiequ{wa4}{point 1*length(p4)/2 of p4}
\fmfiv{d.shape=circle,d.size=2}{wa4}
\fmfiequ{wa7}{point 1*length(p7)/2 of p7}
\fmfiv{d.shape=circle,d.size=2}{wa7}
\fmfforce{(-0w,-0h)}{va0}
\fmfforce{(1w,-0h)}{vb0}
\wigglywrap{wa1}{va0}{vb0}{wa7}
\fmfi{wiggly}{wa4..wa7}
\end{fmfchar*}}}
}
\subfigspace
\subfigure[$\wgraph{5}{18}{2,1,4}$]{
\raisebox{\eqoff}{%
\fmfframe(3,1)(1,4){%
\begin{fmfchar*}(16,20)
\Wduq
\fmfipair{wa[]}
\fmfipair{wb[]}
\fmfipair{wc[]}
\fmfipair{wd[]}
\fmfiequ{wa1}{point 1*length(p1)/2 of p1}
\fmfiv{d.shape=circle,d.size=2}{wa1}
\fmfiequ{wa4}{point 1*length(p4)/2 of p4}
\fmfiv{d.shape=circle,d.size=2}{wa4}
\fmfiequ{wa7}{point 1*length(p7)/3 of p7}
\fmfiv{d.shape=circle,d.size=2}{wa7}
\fmfiequ{wb7}{point 2*length(p7)/3 of p7}
\fmfiv{d.shape=circle,d.size=2}{wb7}
\fmfforce{(-0w,-0h)}{va0}
\fmfforce{(1w,-0h)}{vb0}
\wigglywrap{wa1}{va0}{vb0}{wa7}
\fmfi{wiggly}{wa4..wb7}
\end{fmfchar*}}}
}
\subfigspace
\subfigure[$\wgraph{5}{19}{2,1,4}$]{
\raisebox{\eqoff}{%
\fmfframe(3,1)(1,4){%
\begin{fmfchar*}(16,20)
\Wduq
\fmfipair{wa[]}
\fmfipair{wb[]}
\fmfipair{wc[]}
\fmfipair{wd[]}
\fmfiequ{wa1}{point 1*length(p1)/2 of p1}
\fmfiv{d.shape=circle,d.size=2}{wa1}
\fmfiequ{wa3}{point 1*length(p3)/2 of p3}
\fmfiv{d.shape=circle,d.size=2}{wa3}
\fmfiequ{wa7}{point 1*length(p7)/3 of p7}
\fmfiv{d.shape=circle,d.size=2}{wa7}
\fmfiequ{wb7}{point 2*length(p7)/3 of p7}
\fmfiv{d.shape=circle,d.size=2}{wb7}
\fmfforce{(-0w,-0h)}{va0}
\fmfforce{(1w,-0h)}{vb0}
\wigglywrap{wa1}{va0}{vb0}{wb7}
\fmfi{wiggly}{wa3..wa7}
\end{fmfchar*}}}
}
\subfigspace
\subfigure[$\wgraph{5}{20}{2,1,4}$]{
\raisebox{\eqoff}{%
\fmfframe(3,1)(1,4){%
\begin{fmfchar*}(16,20)
\Wduq
\fmfipair{wa[]}
\fmfipair{wb[]}
\fmfipair{wc[]}
\fmfipair{wd[]}
\fmfiequ{wa1}{point 1*length(p1)/2 of p1}
\fmfiv{d.shape=circle,d.size=2}{wa1}
\fmfiequ{wa4}{point 1*length(p4)/2 of p4}
\fmfiv{d.shape=circle,d.size=2}{wa4}
\fmfiequ{wa7}{point 1*length(p7)/3 of p7}
\fmfiv{d.shape=circle,d.size=2}{wa7}
\fmfiequ{wb7}{point 2*length(p7)/3 of p7}
\fmfiv{d.shape=circle,d.size=2}{wb7}
\fmfforce{(-0w,-0h)}{va0}
\fmfforce{(1w,-0h)}{vb0}
\wigglywrap{wa1}{va0}{vb0}{wb7}
\fmfi{wiggly}{wa4..wa7}
\end{fmfchar*}}}
}
\subfigspace
\subfigure[$\wgraph{5}{21}{2,1,4}$]{
\raisebox{\eqoff}{%
\fmfframe(3,1)(1,4){%
\begin{fmfchar*}(16,20)
\Wduq
\fmfipair{wa[]}
\fmfipair{wb[]}
\fmfipair{wc[]}
\fmfipair{wd[]}
\fmfiequ{wa1}{point 1*length(p1)/2 of p1}
\fmfiv{d.shape=circle,d.size=2}{wa1}
\fmfiequ{wa3}{point 1*length(p3)/2 of p3}
\fmfiv{d.shape=circle,d.size=2}{wa3}
\fmfiequ{wa6}{point 1*length(p6)/2 of p6}
\fmfiv{d.shape=circle,d.size=2}{wa6}
\fmfiequ{wa9}{point 1*length(p9)/2 of p9}
\fmfiv{d.shape=circle,d.size=2}{wa9}
\fmfforce{(-0w,-0h)}{va0}
\fmfforce{(1w,-0h)}{vb0}
\wigglywrap{wa1}{va0}{vb0}{wa9}
\fmfi{wiggly}{wa3..wa6}
\end{fmfchar*}}}
}
\subfigspace
\subfigure[$\wgraph{5}{22}{2,1,4}$]{
\raisebox{\eqoff}{%
\fmfframe(3,1)(1,4){%
\begin{fmfchar*}(16,20)
\Wduq
\fmfipair{wa[]}
\fmfipair{wb[]}
\fmfipair{wc[]}
\fmfipair{wd[]}
\fmfiequ{wa1}{point 1*length(p1)/2 of p1}
\fmfiv{d.shape=circle,d.size=2}{wa1}
\fmfiequ{wa3}{point 1*length(p3)/2 of p3}
\fmfiv{d.shape=circle,d.size=2}{wa3}
\fmfiequ{wa7}{point 1*length(p7)/2 of p7}
\fmfiv{d.shape=circle,d.size=2}{wa7}
\fmfiequ{wa9}{point 1*length(p9)/2 of p9}
\fmfiv{d.shape=circle,d.size=2}{wa9}
\fmfforce{(-0w,-0h)}{va0}
\fmfforce{(1w,-0h)}{vb0}
\wigglywrap{wa1}{va0}{vb0}{wa9}
\fmfi{wiggly}{wa3..wa7}
\end{fmfchar*}}}
}
\subfigspace
\subfigure[$\wgraph{5}{23}{2,1,4}$]{
\raisebox{\eqoff}{%
\fmfframe(3,1)(1,4){%
\begin{fmfchar*}(16,20)
\Wduq
\fmfipair{wa[]}
\fmfipair{wb[]}
\fmfipair{wc[]}
\fmfipair{wd[]}
\fmfiequ{wa1}{point 1*length(p1)/2 of p1}
\fmfiv{d.shape=circle,d.size=2}{wa1}
\fmfiequ{wa4}{point 1*length(p4)/2 of p4}
\fmfiv{d.shape=circle,d.size=2}{wa4}
\fmfiequ{wa6}{point 1*length(p6)/2 of p6}
\fmfiv{d.shape=circle,d.size=2}{wa6}
\fmfiequ{wa9}{point 1*length(p9)/2 of p9}
\fmfiv{d.shape=circle,d.size=2}{wa9}
\fmfforce{(-0w,-0h)}{va0}
\fmfforce{(1w,-0h)}{vb0}
\wigglywrap{wa1}{va0}{vb0}{wa9}
\fmfi{wiggly}{wa4..wa6}
\end{fmfchar*}}}
}
\subfigspace
\subfigure[$\wgraph{5}{24}{2,1,4}$]{
\raisebox{\eqoff}{%
\fmfframe(3,1)(1,4){%
\begin{fmfchar*}(16,20)
\Wduq
\fmfipair{wa[]}
\fmfipair{wb[]}
\fmfipair{wc[]}
\fmfipair{wd[]}
\fmfiequ{wa1}{point 1*length(p1)/2 of p1}
\fmfiv{d.shape=circle,d.size=2}{wa1}
\fmfiequ{wa4}{point 1*length(p4)/2 of p4}
\fmfiv{d.shape=circle,d.size=2}{wa4}
\fmfiequ{wa7}{point 1*length(p7)/2 of p7}
\fmfiv{d.shape=circle,d.size=2}{wa7}
\fmfiequ{wa9}{point 1*length(p9)/2 of p9}
\fmfiv{d.shape=circle,d.size=2}{wa9}
\fmfforce{(-0w,-0h)}{va0}
\fmfforce{(1w,-0h)}{vb0}
\wigglywrap{wa1}{va0}{vb0}{wa9}
\fmfi{wiggly}{wa4..wa7}
\end{fmfchar*}}}
}
\\[0.5cm]
\begin{tabular}{m{14cm}}
\toprule
$\wgraph{5}{1}{2,1,4}\rightarrow -(\jint{5}{22}+\jint{5}{23}+2\jint{5}{26})\,\chi(2,1,4) \rightarrow -(\jint{5}{22}+\jint{5}{23}+2\jint{5}{26})\,M_5$  \\
$\wgraph{5}{2}{2,1,4}\rightarrow -\jint{5}{22}\,\chi(2,1,4) \rightarrow -\jint{5}{22}\,M_5$ \\
$\wgraph{5}{3}{2,1,4}\rightarrow \jint{5}{22}\,\chi(2,1,4) \rightarrow \jint{5}{22}\,M_5$ \\
$\wgraph{5}{4}{2,1,4}\rightarrow (\jint{5}{21}+\jint{5}{23}+2(\jint{5}{32}-\jint{5}{33}-\jint{5}{34}+i\,\epsilon_{\mu\nu\rho\sigma}\jintind{5}{35}{\mu\sigma\rho\nu}))\,\chi(2,1,4)$ \\
$\phantom{\wgraph{5}{4}{2,1,4}}\rightarrow (\jint{5}{21}+\jint{5}{23}+2(\jint{5}{32}-\jint{5}{33}-\jint{5}{34}+i\,\epsilon_{\mu\nu\rho\sigma}\jintind{5}{35}{\mu\sigma\rho\nu}))\,M_5$ \\
$\wgraph{5}{5}{2,1,4}\rightarrow \jint{5}{20}\,\chi(2,1,4) \rightarrow \jint{5}{20}\,M_5$ \\
$\wgraph{5}{6}{2,1,4}\rightarrow -\jint{5}{20}\,\chi(2,1,4) \rightarrow -\jint{5}{20}\,M_5$ \\
$\wgraph{5}{7}{2,1,4}\rightarrow \jint{5}{22}\,\chi(2,1,4) \rightarrow \jint{5}{22}\,M_5$ \\
$\wgraph{5}{8}{2,1,4}\rightarrow -\jint{5}{21}\,\chi(2,1,4) \rightarrow -\jint{5}{21}\,M_5$ \\
$\wgraph{5}{9}{2,1,4}\rightarrow (\jint{5}{20}+2\jint{5}{22}+\jint{5}{23}+2\jint{5}{25}+4\jint{5}{26}+2\jint{5}{27}+4\jint{5}{30})\,\chi(2,1,4)$ \\
$\phantom{\wgraph{5}{9}{2,1,4}}\rightarrow (\jint{5}{20}+2\jint{5}{22}+\jint{5}{23}+2\jint{5}{25}+4\jint{5}{26}+2\jint{5}{27}+4\jint{5}{30})\,M_5$ \\
$\wgraph{5}{10}{2,1,4}\rightarrow -(\jint{5}{20}+\jint{5}{22}+2\jint{5}{25})\,\chi(2,1,4) \rightarrow -(\jint{5}{20}+\jint{5}{22}+2\jint{5}{25})\,M_5$ \\
$\wgraph{5}{11}{2,1,4}\rightarrow -(\jint{5}{22}+\jint{5}{23}+2\jint{5}{26})\,\chi(2,1,4) \rightarrow -(\jint{5}{22}+\jint{5}{23}+2\jint{5}{26})\,M_5$ \\
$\wgraph{5}{16}{2,1,4}\rightarrow -\jint{5}{21}\,\chi(2,1,4) \rightarrow -\jint{5}{21}\,M_5$ \\
$\wgraph{5}{20}{2,1,4}\rightarrow \jint{5}{21}\,\chi(2,1,4) \rightarrow \jint{5}{21}\,M_5$ \\
$\wgraph{5}{21}{2,1,4}\rightarrow -(\jint{5}{20}+\jint{5}{22}+2\jint{5}{27})\,\chi(2,1,4) \rightarrow -(\jint{5}{20}+\jint{5}{22}+2\jint{5}{27})\,M_5$ \\
$\wgraph{5}{22}{2,1,4}\rightarrow \jint{5}{20}\,\chi(2,1,4) \rightarrow \jint{5}{20}\,M_5$ \\
$\wgraph{5}{23}{2,1,4}\rightarrow \jint{5}{22}\,\chi(2,1,4) \rightarrow \jint{5}{22}\,M_5$ \\
$\wgraph{5}{12}{2,1,4}=\wgraph{5}{13}{2,1,4}=\wgraph{5}{14}{2,1,4}=\wgraph{5}{15}{2,1,4}=\wgraph{5}{17}{2,1,4}=\wgraph{5}{18}{2,1,4}=\wgraph{5}{19}{2,1,4}=\wgraph{5}{24}{2,1,4}\rightarrow 0$ \\
\midrule
$\sum_{i}\wgraph{5}{i}{2,1,4}\rightarrow 2(2\jint{5}{30}+\jint{5}{32}-\jint{5}{33}-\jint{5}{34}+i\,\epsilon_{\mu\nu\rho\sigma}\jintind{5}{35}{\mu\sigma\rho\nu})\,\chi(2,1,4) $  \\
$\phantom{\sum_{i}\wgraph{5}{i}{2,1,4}}\rightarrow 2(2\jint{5}{30}+\jint{5}{32}-\jint{5}{33}-\jint{5}{34}+i\,\epsilon_{\mu\nu\rho\sigma}\jintind{5}{35}{\mu\sigma\rho\nu})\,M_5$  \\
\bottomrule
\end{tabular}
\normalsize
\caption{Wrapping diagrams with structure $\chi(2,1,4)$}
\label{wrap5-214}
\end{figure}



\begin{figure}[p]
\capstart
\footnotesize
\renewcommand*{\thesubfigure}{}
\centering
\unitlength=0.75mm
\settoheight{\eqoff}{$\times$}%
\setlength{\eqoff}{0.5\eqoff}%
\addtolength{\eqoff}{-12.5\unitlength}%
\settoheight{\eqofftwo}{$\times$}%
\setlength{\eqofftwo}{0.5\eqofftwo}%
\addtolength{\eqofftwo}{-7.5\unitlength}%
\subfigure[$\wgraph{5}{1}{1,4}(\times 2)$]{
\raisebox{\eqoff}{%
\fmfframe(3,1)(1,4){%
\begin{fmfchar*}(16,20)
\Wuq
\fmfipair{wa[]}
\fmfipair{wb[]}
\fmfipair{wc[]}
\fmfipair{wd[]}
\fmfiequ{wa0}{point 1*length(p0)/2 of p0}
\fmfiv{d.shape=circle,d.size=2}{wa0}
\fmfiequ{wa1}{point 1*length(p1)/2 of p1}
\fmfiv{d.shape=circle,d.size=2}{wa1}
\fmfiequ{wa5}{point 1*length(p5)/2 of p5}
\fmfiv{d.shape=circle,d.size=2}{wa5}
\fmfiequ{wa6}{point 1*length(p6)/2 of p6}
\fmfiv{d.shape=circle,d.size=2}{wa6}
\fmfiequ{wa7}{point 1*length(p7)/2 of p7}
\fmfiv{d.shape=circle,d.size=2}{wa7}
\fmfforce{(-0w,-0h)}{va0}
\fmfforce{(1w,-0h)}{vb0}
\wigglywrap{wa0}{va0}{vb0}{wa7}
\fmfi{wiggly}{wa1..wa5}
\fmfi{wiggly}{wa5..wa6}
\end{fmfchar*}}}
}
\subfigspace
\subfigure[$\wgraph{5}{2}{1,4}(\times2)$]{
\raisebox{\eqoff}{%
\fmfframe(3,1)(1,4){%
\begin{fmfchar*}(16,20)
\Wuq
\fmfipair{wa[]}
\fmfipair{wb[]}
\fmfipair{wc[]}
\fmfipair{wd[]}
\fmfiequ{wa0}{point 1*length(p0)/2 of p0}
\fmfiv{d.shape=circle,d.size=2}{wa0}
\fmfiequ{wa1}{point 1*length(p1)/2 of p1}
\fmfiv{d.shape=circle,d.size=2}{wa1}
\fmfiequ{wa5}{point 1*length(p5)/2 of p5}
\fmfiv{d.shape=circle,d.size=2}{wa5}
\fmfiequ{wa7}{point 1*length(p7)/2 of p7}
\fmfiv{d.shape=circle,d.size=2}{wa7}
\fmfforce{(-0w,-0h)}{va0}
\fmfforce{(1w,-0h)}{vb0}
\wigglywrap{wa0}{va0}{vb0}{wa7}
\fmfi{wiggly}{wa1..wa5}
\fmfi{wiggly}{wa5..wa7}
\end{fmfchar*}}}
}
\subfigspace
\subfigure[$\wgraph{5}{3}{1,4}(\times2)$]{
\raisebox{\eqoff}{%
\fmfframe(3,1)(1,4){%
\begin{fmfchar*}(16,20)
\Wuq
\fmfipair{wa[]}
\fmfipair{wb[]}
\fmfipair{wc[]}
\fmfipair{wd[]}
\fmfiequ{wa0}{point 1*length(p0)/2 of p0}
\fmfiv{d.shape=circle,d.size=2}{wa0}
\fmfiequ{wa1}{point 1*length(p1)/2 of p1}
\fmfiv{d.shape=circle,d.size=2}{wa1}
\fmfiequ{wa5}{point 1*length(p5)/2 of p5}
\fmfiv{d.shape=circle,d.size=2}{wa5}
\fmfiequ{wa7}{point 1*length(p7)/3 of p7}
\fmfiv{d.shape=circle,d.size=2}{wa7}
\fmfiequ{wb7}{point 2*length(p7)/3 of p7}
\fmfiv{d.shape=circle,d.size=2}{wb7}
\fmfforce{(-0w,-0h)}{va0}
\fmfforce{(1w,-0h)}{vb0}
\wigglywrap{wa0}{va0}{vb0}{wa7}
\fmfi{wiggly}{wa1..wa5}
\fmfi{wiggly}{wa5..wb7}
\end{fmfchar*}}}
}
\subfigspace
\subfigure[$\wgraph{5}{4}{1,4}(\times2)$]{
\raisebox{\eqoff}{%
\fmfframe(3,1)(1,4){%
\begin{fmfchar*}(16,20)
\Wuq
\fmfipair{wa[]}
\fmfipair{wb[]}
\fmfipair{wc[]}
\fmfipair{wd[]}
\fmfiequ{wa0}{point 1*length(p0)/2 of p0}
\fmfiv{d.shape=circle,d.size=2}{wa0}
\fmfiequ{wa3}{point 1*length(p3)/2 of p3}
\fmfiv{d.shape=circle,d.size=2}{wa3}
\fmfiequ{wa5}{point 1*length(p5)/2 of p5}
\fmfiv{d.shape=circle,d.size=2}{wa5}
\fmfiequ{wa6}{point 1*length(p6)/2 of p6}
\fmfiv{d.shape=circle,d.size=2}{wa6}
\fmfiequ{wa7}{point 1*length(p7)/2 of p7}
\fmfiv{d.shape=circle,d.size=2}{wa7}
\fmfforce{(-0w,-0h)}{va0}
\fmfforce{(1w,-0h)}{vb0}
\wigglywrap{wa0}{va0}{vb0}{wa7}
\fmfi{wiggly}{wa3..wa5}
\fmfi{wiggly}{wa5..wa6}
\end{fmfchar*}}}
}
\subfigspace
\subfigure[$\wgraph{5}{5}{1,4}(\times2)$]{
\raisebox{\eqoff}{%
\fmfframe(3,1)(1,4){%
\begin{fmfchar*}(16,20)
\Wuq
\fmfipair{wa[]}
\fmfipair{wb[]}
\fmfipair{wc[]}
\fmfipair{wd[]}
\fmfiequ{wa0}{point 1*length(p0)/2 of p0}
\fmfiv{d.shape=circle,d.size=2}{wa0}
\fmfiequ{wa3}{point 1*length(p3)/2 of p3}
\fmfiv{d.shape=circle,d.size=2}{wa3}
\fmfiequ{wa5}{point 1*length(p5)/2 of p5}
\fmfiv{d.shape=circle,d.size=2}{wa5}
\fmfiequ{wa7}{point 1*length(p7)/2 of p7}
\fmfiv{d.shape=circle,d.size=2}{wa7}
\fmfforce{(-0w,-0h)}{va0}
\fmfforce{(1w,-0h)}{vb0}
\wigglywrap{wa0}{va0}{vb0}{wa7}
\fmfi{wiggly}{wa3..wa5}
\fmfi{wiggly}{wa5..wa7}
\end{fmfchar*}}}
}
\subfigspace
\subfigure[$\wgraph{5}{6}{1,4}(\times2)$]{
\raisebox{\eqoff}{%
\fmfframe(3,1)(1,4){%
\begin{fmfchar*}(16,20)
\Wuq
\fmfipair{wa[]}
\fmfipair{wb[]}
\fmfipair{wc[]}
\fmfipair{wd[]}
\fmfiequ{wa0}{point 1*length(p0)/2 of p0}
\fmfiv{d.shape=circle,d.size=2}{wa0}
\fmfiequ{wa3}{point 1*length(p3)/2 of p3}
\fmfiv{d.shape=circle,d.size=2}{wa3}
\fmfiequ{wa5}{point 1*length(p5)/2 of p5}
\fmfiv{d.shape=circle,d.size=2}{wa5}
\fmfiequ{wa7}{point 1*length(p7)/3 of p7}
\fmfiv{d.shape=circle,d.size=2}{wa7}
\fmfiequ{wb7}{point 2*length(p7)/3 of p7}
\fmfiv{d.shape=circle,d.size=2}{wb7}
\fmfforce{(-0w,-0h)}{va0}
\fmfforce{(1w,-0h)}{vb0}
\wigglywrap{wa0}{va0}{vb0}{wa7}
\fmfi{wiggly}{wa3..wa5}
\fmfi{wiggly}{wa5..wb7}
\end{fmfchar*}}}
}
\subfigspace
\subfigure[$\wgraph{5}{7}{1,4}(\times2)$]{
\raisebox{\eqoff}{%
\fmfframe(3,1)(1,4){%
\begin{fmfchar*}(16,20)
\Wuq
\fmfipair{wa[]}
\fmfipair{wb[]}
\fmfipair{wc[]}
\fmfipair{wd[]}
\fmfiequ{wa0}{point 1*length(p0)/2 of p0}
\fmfiv{d.shape=circle,d.size=2}{wa0}
\fmfiequ{wa1}{point 1*length(p1)/2 of p1}
\fmfiv{d.shape=circle,d.size=2}{wa1}
\fmfiequ{wa5}{point 1*length(p5)/2 of p5}
\fmfiv{d.shape=circle,d.size=2}{wa5}
\fmfiequ{wa7}{point 1*length(p7)/3 of p7}
\fmfiv{d.shape=circle,d.size=2}{wa7}
\fmfiequ{wb7}{point 2*length(p7)/3 of p7}
\fmfiv{d.shape=circle,d.size=2}{wb7}
\fmfforce{(-0w,-0h)}{va0}
\fmfforce{(1w,-0h)}{vb0}
\wigglywrap{wa0}{va0}{vb0}{wb7}
\fmfi{wiggly}{wa1..wa5}
\fmfi{wiggly}{wa5..wa7}
\end{fmfchar*}}}
}
\\
\subfigure[$\wgraph{5}{8}{1,4}(\times2)$]{
\raisebox{\eqoff}{%
\fmfframe(3,1)(1,4){%
\begin{fmfchar*}(16,20)
\Wuq
\fmfipair{wa[]}
\fmfipair{wb[]}
\fmfipair{wc[]}
\fmfipair{wd[]}
\fmfiequ{wa0}{point 1*length(p0)/2 of p0}
\fmfiv{d.shape=circle,d.size=2}{wa0}
\fmfiequ{wa3}{point 1*length(p3)/2 of p3}
\fmfiv{d.shape=circle,d.size=2}{wa3}
\fmfiequ{wa5}{point 1*length(p5)/2 of p5}
\fmfiv{d.shape=circle,d.size=2}{wa5}
\fmfiequ{wa7}{point 1*length(p7)/3 of p7}
\fmfiv{d.shape=circle,d.size=2}{wa7}
\fmfiequ{wb7}{point 2*length(p7)/3 of p7}
\fmfiv{d.shape=circle,d.size=2}{wb7}
\fmfforce{(-0w,-0h)}{va0}
\fmfforce{(1w,-0h)}{vb0}
\wigglywrap{wa0}{va0}{vb0}{wb7}
\fmfi{wiggly}{wa3..wa5}
\fmfi{wiggly}{wa5..wa7}
\end{fmfchar*}}}
}
\subfigspace
\subfigure[$\wgraph{5}{9}{1,4}(\times2)$]{
\raisebox{\eqoff}{%
\fmfframe(3,1)(1,4){%
\begin{fmfchar*}(16,20)
\Wuq
\fmfipair{wa[]}
\fmfipair{wb[]}
\fmfipair{wc[]}
\fmfipair{wd[]}
\fmfiequ{wa0}{point 1*length(p0)/2 of p0}
\fmfiv{d.shape=circle,d.size=2}{wa0}
\fmfiequ{wa1}{point 1*length(p1)/2 of p1}
\fmfiv{d.shape=circle,d.size=2}{wa1}
\fmfiequ{wa5}{point 1*length(p5)/2 of p5}
\fmfiv{d.shape=circle,d.size=2}{wa5}
\fmfiequ{wa6}{point 1*length(p6)/2 of p6}
\fmfiv{d.shape=circle,d.size=2}{wa6}
\fmfiequ{wa9}{point 1*length(p9)/2 of p9}
\fmfiv{d.shape=circle,d.size=2}{wa9}
\fmfforce{(-0w,-0h)}{va0}
\fmfforce{(1w,-0h)}{vb0}
\wigglywrap{wa0}{va0}{vb0}{wa9}
\fmfi{wiggly}{wa1..wa5}
\fmfi{wiggly}{wa5..wa6}
\end{fmfchar*}}}
}
\subfigspace
\subfigure[\protect\phantom{aa}$\wgraph{5}{10}{1,4}$\protect\phantom{aa}]{
\raisebox{\eqoff}{%
\fmfframe(3,1)(1,4){%
\begin{fmfchar*}(16,20)
\Wuq
\fmfipair{wa[]}
\fmfipair{wb[]}
\fmfipair{wc[]}
\fmfipair{wd[]}
\fmfiequ{wa0}{point 1*length(p0)/2 of p0}
\fmfiv{d.shape=circle,d.size=2}{wa0}
\fmfiequ{wa1}{point 1*length(p1)/2 of p1}
\fmfiv{d.shape=circle,d.size=2}{wa1}
\fmfiequ{wa5}{point 1*length(p5)/2 of p5}
\fmfiv{d.shape=circle,d.size=2}{wa5}
\fmfiequ{wa7}{point 1*length(p7)/2 of p7}
\fmfiv{d.shape=circle,d.size=2}{wa7}
\fmfiequ{wa9}{point 1*length(p9)/2 of p9}
\fmfiv{d.shape=circle,d.size=2}{wa9}
\fmfforce{(-0w,-0h)}{va0}
\fmfforce{(1w,-0h)}{vb0}
\wigglywrap{wa0}{va0}{vb0}{wa9}
\fmfi{wiggly}{wa1..wa5}
\fmfi{wiggly}{wa5..wa7}
\end{fmfchar*}}}
}
\subfigspace
\subfigure[\protect\phantom{aa}$\wgraph{5}{11}{1,4}$\protect\phantom{aa}]{
\raisebox{\eqoff}{%
\fmfframe(3,1)(1,4){%
\begin{fmfchar*}(16,20)
\Wuq
\fmfipair{wa[]}
\fmfipair{wb[]}
\fmfipair{wc[]}
\fmfipair{wd[]}
\fmfiequ{wa0}{point 1*length(p0)/2 of p0}
\fmfiv{d.shape=circle,d.size=2}{wa0}
\fmfiequ{wa3}{point 1*length(p3)/2 of p3}
\fmfiv{d.shape=circle,d.size=2}{wa3}
\fmfiequ{wa5}{point 1*length(p5)/2 of p5}
\fmfiv{d.shape=circle,d.size=2}{wa5}
\fmfiequ{wa6}{point 1*length(p6)/2 of p6}
\fmfiv{d.shape=circle,d.size=2}{wa6}
\fmfiequ{wa9}{point 1*length(p9)/2 of p9}
\fmfiv{d.shape=circle,d.size=2}{wa9}
\fmfforce{(-0w,-0h)}{va0}
\fmfforce{(1w,-0h)}{vb0}
\wigglywrap{wa0}{va0}{vb0}{wa9}
\fmfi{wiggly}{wa3..wa5}
\fmfi{wiggly}{wa5..wa6}
\end{fmfchar*}}}
}
\subfigspace
\subfigure[$\wgraph{5}{12}{1,4}(\times2)$]{
\raisebox{\eqoff}{%
\fmfframe(3,1)(1,4){%
\begin{fmfchar*}(16,20)
\Wuq
\fmfipair{wa[]}
\fmfipair{wb[]}
\fmfipair{wc[]}
\fmfipair{wd[]}
\fmfiequ{wa1}{point 1*length(p1)/2 of p1}
\fmfiv{d.shape=circle,d.size=2}{wa1}
\fmfiequ{wa5}{point 1*length(p5)/2 of p5}
\fmfiv{d.shape=circle,d.size=2}{wa5}
\fmfiequ{wa6}{point 1*length(p6)/2 of p6}
\fmfiv{d.shape=circle,d.size=2}{wa6}
\fmfiequ{wa7}{point 1*length(p7)/2 of p7}
\fmfiv{d.shape=circle,d.size=2}{wa7}
\fmfi{wiggly}{wa1..wa5}
\fmfforce{(-0w,-0h)}{va0}
\fmfforce{(1w,-0h)}{vb0}
\wigglywrap{wa1}{va0}{vb0}{wa7}
\fmfi{wiggly}{wa5..wa6}
\end{fmfchar*}}}
}
\subfigspace
\subfigure[$\wgraph{5}{13}{1,4}(\times2)$]{
\raisebox{\eqoff}{%
\fmfframe(3,1)(1,4){%
\begin{fmfchar*}(16,20)
\Wuq
\fmfipair{wa[]}
\fmfipair{wb[]}
\fmfipair{wc[]}
\fmfipair{wd[]}
\fmfiequ{wa1}{point 1*length(p1)/2 of p1}
\fmfiv{d.shape=circle,d.size=2}{wa1}
\fmfiequ{wa5}{point 1*length(p5)/2 of p5}
\fmfiv{d.shape=circle,d.size=2}{wa5}
\fmfiequ{wa7}{point 1*length(p7)/3 of p7}
\fmfiv{d.shape=circle,d.size=2}{wa7}
\fmfiequ{wb7}{point 2*length(p7)/3 of p7}
\fmfiv{d.shape=circle,d.size=2}{wb7}
\fmfi{wiggly}{wa1..wa5}
\fmfforce{(-0w,-0h)}{va0}
\fmfforce{(1w,-0h)}{vb0}
\wigglywrap{wa1}{va0}{vb0}{wa7}
\fmfi{wiggly}{wa5..wb7}
\end{fmfchar*}}}
}
\subfigspace
\subfigure[$\wgraph{5}{14}{1,4}(\times2)$]{
\raisebox{\eqoff}{%
\fmfframe(3,1)(1,4){%
\begin{fmfchar*}(16,20)
\Wuq
\fmfipair{wa[]}
\fmfipair{wb[]}
\fmfipair{wc[]}
\fmfipair{wd[]}
\fmfiequ{wa1}{point 1*length(p1)/2 of p1}
\fmfiv{d.shape=circle,d.size=2}{wa1}
\fmfiequ{wa5}{point 1*length(p5)/2 of p5}
\fmfiv{d.shape=circle,d.size=2}{wa5}
\fmfiequ{wa7}{point 1*length(p7)/3 of p7}
\fmfiv{d.shape=circle,d.size=2}{wa7}
\fmfiequ{wb7}{point 2*length(p7)/3 of p7}
\fmfiv{d.shape=circle,d.size=2}{wb7}
\fmfi{wiggly}{wa1..wa5}
\fmfforce{(-0w,-0h)}{va0}
\fmfforce{(1w,-0h)}{vb0}
\wigglywrap{wa1}{va0}{vb0}{wb7}
\fmfi{wiggly}{wa5..wa7}
\end{fmfchar*}}}
}
\\
\subfigure[$\wgraph{5}{15}{1,4}(\times2)$]{
\raisebox{\eqoff}{%
\fmfframe(3,1)(1,4){%
\begin{fmfchar*}(16,20)
\Wuq
\fmfipair{wa[]}
\fmfipair{wb[]}
\fmfipair{wc[]}
\fmfipair{wd[]}
\fmfiequ{wa1}{point 1*length(p1)/3 of p1}
\fmfiv{d.shape=circle,d.size=2}{wa1}
\fmfiequ{wb1}{point 2*length(p1)/3 of p1}
\fmfiv{d.shape=circle,d.size=2}{wb1}
\fmfiequ{wa5}{point 1*length(p5)/2 of p5}
\fmfiv{d.shape=circle,d.size=2}{wa5}
\fmfiequ{wa6}{point 1*length(p6)/2 of p6}
\fmfiv{d.shape=circle,d.size=2}{wa6}
\fmfiequ{wa7}{point 1*length(p7)/2 of p7}
\fmfiv{d.shape=circle,d.size=2}{wa7}
\fmfi{wiggly}{wa1..wa5}
\fmfforce{(-0w,-0h)}{va0}
\fmfforce{(1w,-0h)}{vb0}
\wigglywrap{wb1}{va0}{vb0}{wa7}
\fmfi{wiggly}{wa5..wa6}
\end{fmfchar*}}}
}
\subfigspace
\subfigure[$\wgraph{5}{16}{1,4}(\times2)$]{
\raisebox{\eqoff}{%
\fmfframe(3,1)(1,4){%
\begin{fmfchar*}(16,20)
\Wuq
\fmfipair{wa[]}
\fmfipair{wb[]}
\fmfipair{wc[]}
\fmfipair{wd[]}
\fmfiequ{wa1}{point 1*length(p1)/3 of p1}
\fmfiv{d.shape=circle,d.size=2}{wa1}
\fmfiequ{wb1}{point 2*length(p1)/3 of p1}
\fmfiv{d.shape=circle,d.size=2}{wb1}
\fmfiequ{wa5}{point 1*length(p5)/2 of p5}
\fmfiv{d.shape=circle,d.size=2}{wa5}
\fmfiequ{wa7}{point 1*length(p7)/3 of p7}
\fmfiv{d.shape=circle,d.size=2}{wa7}
\fmfiequ{wb7}{point 2*length(p7)/3 of p7}
\fmfiv{d.shape=circle,d.size=2}{wb7}
\fmfi{wiggly}{wa1..wa5}
\fmfforce{(-0w,-0h)}{va0}
\fmfforce{(1w,-0h)}{vb0}
\wigglywrap{wb1}{va0}{vb0}{wa7}
\fmfi{wiggly}{wa5..wb7}
\end{fmfchar*}}}
}
\subfigspace
\subfigure[\protect\phantom{aa}$\wgraph{5}{17}{1,4}$\protect\phantom{aa}]{
\raisebox{\eqoff}{%
\fmfframe(3,1)(1,4){%
\begin{fmfchar*}(16,20)
\Wuq
\fmfipair{wa[]}
\fmfipair{wb[]}
\fmfipair{wc[]}
\fmfipair{wd[]}
\fmfiequ{wa1}{point 1*length(p1)/3 of p1}
\fmfiv{d.shape=circle,d.size=2}{wa1}
\fmfiequ{wb1}{point 2*length(p1)/3 of p1}
\fmfiv{d.shape=circle,d.size=2}{wb1}
\fmfiequ{wa5}{point 1*length(p5)/2 of p5}
\fmfiv{d.shape=circle,d.size=2}{wa5}
\fmfiequ{wa7}{point 1*length(p7)/3 of p7}
\fmfiv{d.shape=circle,d.size=2}{wa7}
\fmfiequ{wb7}{point 2*length(p7)/3 of p7}
\fmfiv{d.shape=circle,d.size=2}{wb7}
\fmfi{wiggly}{wa1..wa5}
\fmfforce{(-0w,-0h)}{va0}
\fmfforce{(1w,-0h)}{vb0}
\wigglywrap{wb1}{va0}{vb0}{wb7}
\fmfi{wiggly}{wa5..wa7}
\end{fmfchar*}}}
}
\subfigspace
\subfigure[$\wgraph{5}{18}{1,4}(\times2)$]{
\raisebox{\eqoff}{%
\fmfframe(3,1)(1,4){%
\begin{fmfchar*}(16,20)
\Wuq
\fmfipair{wa[]}
\fmfipair{wb[]}
\fmfipair{wc[]}
\fmfipair{wd[]}
\fmfiequ{wa1}{point 1*length(p1)/3 of p1}
\fmfiv{d.shape=circle,d.size=2}{wa1}
\fmfiequ{wb1}{point 2*length(p1)/3 of p1}
\fmfiv{d.shape=circle,d.size=2}{wb1}
\fmfiequ{wa5}{point 1*length(p5)/2 of p5}
\fmfiv{d.shape=circle,d.size=2}{wa5}
\fmfiequ{wa6}{point 1*length(p6)/2 of p6}
\fmfiv{d.shape=circle,d.size=2}{wa6}
\fmfiequ{wa7}{point 1*length(p7)/2 of p7}
\fmfiv{d.shape=circle,d.size=2}{wa7}
\fmfi{wiggly}{wb1..wa5}
\fmfforce{(-0w,-0h)}{va0}
\fmfforce{(1w,-0h)}{vb0}
\wigglywrap{wa1}{va0}{vb0}{wa7}
\fmfi{wiggly}{wa5..wa6}
\end{fmfchar*}}}
}
\subfigspace
\subfigure[\protect\phantom{aa}$\wgraph{5}{19}{1,4}$\protect\phantom{aa}]{
\raisebox{\eqoff}{%
\fmfframe(3,1)(1,4){%
\begin{fmfchar*}(16,20)
\Wuq
\fmfipair{wa[]}
\fmfipair{wb[]}
\fmfipair{wc[]}
\fmfipair{wd[]}
\fmfiequ{wa1}{point 1*length(p1)/3 of p1}
\fmfiv{d.shape=circle,d.size=2}{wa1}
\fmfiequ{wb1}{point 2*length(p1)/3 of p1}
\fmfiv{d.shape=circle,d.size=2}{wb1}
\fmfiequ{wa5}{point 1*length(p5)/2 of p5}
\fmfiv{d.shape=circle,d.size=2}{wa5}
\fmfiequ{wa7}{point 1*length(p7)/3 of p7}
\fmfiv{d.shape=circle,d.size=2}{wa7}
\fmfiequ{wb7}{point 2*length(p7)/3 of p7}
\fmfiv{d.shape=circle,d.size=2}{wb7}
\fmfi{wiggly}{wb1..wa5}
\fmfforce{(-0w,-0h)}{va0}
\fmfforce{(1w,-0h)}{vb0}
\wigglywrap{wa1}{va0}{vb0}{wa7}
\fmfi{wiggly}{wa5..wb7}
\end{fmfchar*}}}
}
\subfigspace
\subfigure[\protect\phantom{aa}$\wgraph{5}{20}{1,4}$\protect\phantom{aa}]{
\raisebox{\eqoff}{%
\fmfframe(3,1)(1,4){%
\begin{fmfchar*}(16,20)
\Wuq
\fmfipair{wa[]}
\fmfipair{wb[]}
\fmfipair{wc[]}
\fmfipair{wd[]}
\fmfiequ{wa1}{point 1*length(p1)/2 of p1}
\fmfiv{d.shape=circle,d.size=2}{wa1}
\fmfiequ{wa3}{point 1*length(p3)/2 of p3}
\fmfiv{d.shape=circle,d.size=2}{wa3}
\fmfiequ{wa5}{point 1*length(p5)/2 of p5}
\fmfiv{d.shape=circle,d.size=2}{wa5}
\fmfiequ{wa6}{point 1*length(p6)/2 of p6}
\fmfiv{d.shape=circle,d.size=2}{wa6}
\fmfiequ{wa7}{point 1*length(p7)/2 of p7}
\fmfiv{d.shape=circle,d.size=2}{wa7}
\fmfforce{(-0w,-0h)}{va0}
\fmfforce{(1w,-0h)}{vb0}
\wigglywrap{wa1}{va0}{vb0}{wa7}
\fmfi{wiggly}{wa3..wa5}
\fmfi{wiggly}{wa5..wa6}
\end{fmfchar*}}}
}
\\[0.5cm]
\begin{tabular}{m{13cm}}
\toprule
$\wgraph{5}{1}{1,4}\rightarrow (\jint{5}{21}+\jint{5}{23}+2(\jint{5}{32}-\jint{5}{33}-\jint{5}{34}+i\,\epsilon_{\mu\nu\rho\sigma}\jintind{5}{35}{\mu\nu\rho\sigma}))\,\chi(1,4)$  \\
$\phantom{\wgraph{5}{1}{1,4}}\rightarrow (\jint{5}{21}+\jint{5}{23}+2(\jint{5}{32}-\jint{5}{33}-\jint{5}{34}+i\,\epsilon_{\mu\nu\rho\sigma}\jintind{5}{35}{\mu\nu\rho\sigma}))\,M_5$  \\
$\wgraph{5}{2}{1,4}\rightarrow 0$ \\
$\wgraph{5}{3}{1,4}\rightarrow -\jint{5}{21}\,\chi(1,4) \rightarrow -\jint{5}{21}\,M_5$ \\
$\wgraph{5}{4}{1,4}\rightarrow -(\jint{5}{22}+\jint{5}{23}+2\jint{5}{26})\,\chi(1,4) \rightarrow -(\jint{5}{22}+\jint{5}{23}+2\jint{5}{26})\,M_5$ \\
$\wgraph{5}{5}{1,4}\rightarrow -\jint{5}{22}\,\chi(1,4) \rightarrow -\jint{5}{22}\,M_5$ \\
$\wgraph{5}{6}{1,4}\rightarrow \jint{5}{22}\,\chi(1,4) \rightarrow \jint{5}{22}\,M_5$ \\
$\wgraph{5}{7}{1,4}\rightarrow 0$ \\
$\wgraph{5}{8}{1,4}\rightarrow \jint{5}{22}\,\chi(1,4) \rightarrow \jint{5}{22}\,M_5$ \\
$\wgraph{5}{9}{1,4}\rightarrow -(\jint{5}{22}+\jint{5}{23}+2\jint{5}{26})\,\chi(1,4) \rightarrow -(\jint{5}{22}+\jint{5}{23}+2\jint{5}{26})\,M_5$ \\
$\wgraph{5}{10}{1,4}\rightarrow 0$ \\
$\wgraph{5}{11}{1,4}\rightarrow 2(\jint{5}{22}+\jint{5}{23}+4\jint{5}{26}+2\jint{5}{31})\,\chi(1,4) \rightarrow 2(\jint{5}{22}+\jint{5}{23}+4\jint{5}{26}+2\jint{5}{31})\,M_5$ \\
$\wgraph{5}{12}{1,4}=\wgraph{5}{13}{1,4}=\wgraph{5}{14}{1,4}\rightarrow 0$ \\
$\wgraph{5}{15}{1,4}\rightarrow -\jint{5}{21}\,\chi(1,4) \rightarrow -\jint{5}{21}\,M_5$ \\
$\wgraph{5}{16}{1,4}\rightarrow \jint{5}{21}\,\chi(1,4) \rightarrow \jint{5}{21}\,M_5$ \\
$\wgraph{5}{17}{1,4}=\wgraph{5}{18}{1,4}=\wgraph{5}{19}{1,4}=\wgraph{5}{20}{1,4}\rightarrow 0$  \\
\midrule
$\sum_{i}\wgraph{5}{i}{1,4}\rightarrow 4(\jint{5}{31}+\jint{5}{32}-\jint{5}{33}-\jint{5}{34}+i\,\epsilon_{\mu\nu\rho\sigma}\jintind{5}{35}{\mu\nu\rho\sigma})\,\chi(1,4)$  \\
$\phantom{\sum_{i}\wgraph{5}{i}{1,4}}\rightarrow 4(\jint{5}{31}+\jint{5}{32}-\jint{5}{33}-\jint{5}{34}+i\,\epsilon_{\mu\nu\rho\sigma}\jintind{5}{35}{\mu\nu\rho\sigma})\,M_5$  \\
\bottomrule
\end{tabular}
\normalsize

\caption{Wrapping diagrams with structure $\chi(1,4)$} 
\label{wrap5-14}
\end{figure}

\clearpage

\section{Integrals}
\label{section:integrals5wrap}
In this section the divergent parts of the required five-loop momentum integrals are listed. 
All the computations have been performed using the GPXT, and
the factor $1/(4\pi)^{10}$ in each integral has been omitted.
Note that the explicit value of integral $\jintind{5}{35}{\mu\nu\rho\sigma}$ is not needed since this integral appears only for the structures $\chi(2,1,4)$ and $\chi(1,4)$, and the two contributions exactly cancel each other thanks to the antisymmetry of $\epsilon_{\mu\nu\rho\sigma}$.
\begin{table}[h!]
\capstart
\begin{align*}
\jint{5}{20}=\raisebox{\eqoff}{%
\begin{fmfchar*}(20,15)
\fmfleft{in}
\fmfright{out}
\fmf{plain}{in,v1}
\fmf{plain,left=0.25}{v1,v2}
\fmf{plain,left=0.25}{v2,v3}
\fmffixed{(0.9w,0)}{v1,v3}
\fmffixed{(0,0.45w)}{v6,v2}
\fmffixed{(0.15w,0)}{v4,v6}
\fmffixed{(0.45w,0)}{v1,v0}
\fmf{plain,left=0.25}{v4,v1}
\fmf{plain,tension=0.5,right=0.5}{v2,v0,v2}
\fmf{plain}{v4,v5}
\fmf{plain,left=0.25}{v3,v5}
\fmffixed{(0.3w,0)}{v4,v5}
\fmf{phantom}{v0,v3}
\fmf{plain}{v1,v0}
\fmf{plain}{v0,v4}
\fmf{plain}{v0,v5}
\fmf{plain}{v3,out}
\fmffreeze
\end{fmfchar*}}
&=
\frac{1}{120\varepsilon^5}-\frac{1}{12\varepsilon^4}+\frac{11}{24\varepsilon^3}-\frac{19}{12\varepsilon^2}+\frac{1}{\varepsilon}\Big(\frac{14}{5}-4\zeta(5)\Big)
\\
\jint{5}{21}=\raisebox{\eqoff}{%
\begin{fmfchar*}(20,15)
\fmfleft{in}
\fmfright{out}
\fmf{plain}{in,v1}
\fmf{plain,left=0.25}{v1,v2}
\fmf{plain,left=0.25}{v2,v3}
\fmf{plain,left=0.25}{v3,v4}
\fmf{plain,left=0.25}{v4,v1}
\fmf{plain,tension=0.5,right=0.5}{v2,v0,v2}
\fmf{plain,tension=0.5,right=0.5}{v0,v4,v0}
\fmf{plain}{v3,out}
\fmffixed{(0.9w,0)}{v1,v3}
\fmffixed{(0,0.45w)}{v4,v2}
\fmffreeze
\fmfipath{px}
\fmfipair{w[]}
\fmfiset{px}{vpath(__v2,__v3)}
\fmfiequ{w1}{point length(px)/2 of px}
\fmfiequ{w2}{(xpart(vloc(__v0)),ypart(vloc(__v0)))}
\fmfi{plain,left=0.25}{w1..w2}
\end{fmfchar*}}
&=
\frac{1}{15\varepsilon^5}-\frac{1}{4\varepsilon^4}+\frac{1}{5\varepsilon^3}+\frac{29}{60\varepsilon^2}-\frac{1}{\varepsilon}\Big(\frac{19}{30}-\frac{4\zeta(3)}{5}\Big)
\\
\jint{5}{22}=\raisebox{\eqoff}{%
\begin{fmfchar*}(20,15)
\fmfleft{in}
\fmfright{out}
\fmf{plain}{in,v1}
\fmf{plain,left=0.25}{v1,v2}
\fmf{plain,left=0.25}{v2,v3}
\fmf{plain,left=0.25}{v3,v4}
\fmf{plain,left=0.25}{v4,v1}
\fmf{plain,tension=0.5,right=0.25}{v1,v0,v1}
\fmf{phantom}{v0,v3}
\fmf{plain}{v2,v0}
\fmf{plain}{v0,v4}
\fmf{plain}{v3,out}
\fmffixed{(0.9w,0)}{v1,v3}
\fmffixed{(0,0.45w)}{v4,v2}
\fmffreeze
\fmfipath{px}
\fmfipair{w[]}
\fmfiset{px}{vpath(__v3,__v4)}
\fmfiequ{w1}{point length(px)/2 of px}
\fmfiequ{w2}{(xpart(vloc(__v0)),ypart(vloc(__v0)))}
\fmfi{plain}{w1..w2}
\end{fmfchar*}}
&=
\frac{1}{40\varepsilon^5}-\frac{1}{6\varepsilon^4}+\frac{61}{120\varepsilon^3}-\frac{17}{30\varepsilon^2}-\frac{1}{\varepsilon}\Big(\frac{1}{6}-\frac{4\zeta(3)}{5}\Big)
\\
\jint{5}{23}=\raisebox{\eqoff}{%
\begin{fmfchar*}(20,15)
\fmfleft{in}
\fmfright{out}
\fmfpoly{plain}{v5,v4,v3,v2,v1}
\fmf{plain}{in,v1}
\fmf{plain}{v3,out}
\fmffixed{(0.8w,0)}{v1,v3}
\fmf{plain}{v0,v1}
\fmf{plain}{v0,v2}
\fmf{plain}{v0,v3}
\fmf{plain}{v0,v4}
\fmf{plain}{v0,v5}
\fmffreeze
\end{fmfchar*}}
&=
\frac{14}{\varepsilon}\zeta(7)
\\
\jint{5}{24}=\raisebox{\eqoff}{%
\begin{fmfchar*}(20,15)
\fmfleft{in}
\fmfright{out}
\fmf{plain}{in,v1}
\fmf{plain}{v5,out}
\fmffixed{(0.9w,0)}{v1,v5}
\fmffixed{(whatever,0)}{v1,in}
\fmffixed{(whatever,0)}{out,v5}
\fmf{derplain}{v2,v1}
\fmf{plain,tension=0.5}{v2,v3}
\fmf{derplain,tension=0.5}{v3,v4}
\fmf{plain}{v4,v5}
\fmf{plain}{v5,v6}
\fmf{plain}{v6,v1}
\fmf{plain}{v0,v2}
\fmf{plain}{v0,v3}
\fmf{plain}{v0,v4}
\fmf{plain,tension=0.25,right=0.25}{v6,v0,v6}
\fmffixed{(whatever,0.5h)}{v6,v4}
\fmffixed{(whatever,0)}{v4,v2}
\fmffixed{(whatever,0.3h)}{v0,v3}
\fmffixed{(whatever,0.3h)}{v6,v0}
\fmffreeze
\end{fmfchar*}}
&=
-\frac{1}{20\varepsilon^3}+\frac{3}{20\varepsilon^2}+\frac{1}{\varepsilon}\Big(\frac{1}{6}+\frac{6\zeta(3)}{5}-2\zeta(5)\Big)
\\
\jint{5}{25}=\raisebox{\eqoff}{%
\begin{fmfchar*}(20,15)
\fmfleft{in}
\fmfright{out}
\fmf{plain}{in,v1}
\fmf{plain}{v4,out}
\fmffixed{(0.8w,0)}{v1,v4}
\fmffixed{(whatever,0)}{v1,in}
\fmffixed{(whatever,0)}{out,v4}
\fmfpoly{phantom}{v1,v2,v3,v4,v5,v6}
\fmf{plain}{v2,v1}
\fmf{derplain}{v2,v3}
\fmf{plain}{v3,v4}
\fmf{plain}{v5,v4}
\fmf{plain}{v5,v6}
\fmf{derplain}{v1,v6}
\fmf{plain}{v0,v1}
\fmf{plain}{v0,v2}
\fmf{plain}{v0,v3}
\fmf{plain}{v0,v5}
\fmf{plain}{v0,v6}
\fmffixed{(0.4w,0)}{v1,v0}
\fmffreeze
\end{fmfchar*}}
&=
-\frac{1}{10\varepsilon^3}+\frac{13}{30\varepsilon^2}-\frac{1}{\varepsilon}\Big(\frac{1}{6}-\frac{6\zeta(3)}{5}+2\zeta(5)\Big)
\\
\jint{5}{26}=\raisebox{\eqoff}{%
\begin{fmfchar*}(20,15)
\fmfleft{in}
\fmfright{out}
\fmf{plain}{in,v1}
\fmf{plain}{v4,out}
\fmffixed{(0.8w,0)}{v1,v4}
\fmffixed{(whatever,0)}{v1,in}
\fmffixed{(whatever,0)}{out,v4}
\fmfpoly{phantom}{v1,v2,v3,v4,v5,v6}
\fmf{plain}{v2,v1}
\fmf{plain}{v2,v3}
\fmf{plain}{v3,v4}
\fmf{derplain}{v5,v4}
\fmf{plain}{v5,v6}
\fmf{derplain}{v6,v1}
\fmf{plain}{v0,v1}
\fmf{plain}{v0,v2}
\fmf{plain}{v0,v3}
\fmf{plain}{v0,v5}
\fmf{plain}{v0,v6}
\fmffixed{(0.4w,0)}{v1,v0}
\fmffreeze
\end{fmfchar*}}
&=
-\frac{2\zeta(5)}{\varepsilon}
\\
\jint{5}{27}=\raisebox{\eqoff}{%
\begin{fmfchar*}(20,15)
\fmfleft{in}
\fmfright{out}
\fmf{plain}{in,v1}
\fmf{plain}{v4,out}
\fmffixed{(0.8w,0)}{v1,v4}
\fmffixed{(whatever,0)}{v1,in}
\fmffixed{(whatever,0)}{out,v4}
\fmf{plain}{v1,v2}
\fmf{plain,tension=0.5}{v2,v3}
\fmf{plain}{v3,v4}
\fmf{plain}{v4,v5}
\fmf{plain}{v5,v6}
\fmf{plain}{v6,v1}
\fmf{plain,tension=0.4}{v0,v3}
\fmf{plain}{v0,v5}
\fmf{plain}{v0,v6}
\fmf{plain,tension=0.25,right=0.25}{v2,v0,v2}
\fmffixed{(whatever,0)}{v5,v6}
\fmffixed{(whatever,0.6h)}{v5,v2}
\fmffixed{(0,whatever)}{v0,v2}
\fmffreeze
\end{fmfchar*}}
&=
-\frac{1}{60\varepsilon^3}+\frac{1}{12\varepsilon^2}+\frac{1}{\varepsilon}\Big(\frac{1}{5}+\frac{6\zeta(3)}{5}-2\zeta(5)\Big)
\end{align*}
\caption{Momentum integrals for five-loop wrapping diagrams}
\label{table:fiveloopwrapintegrals}
\end{table}

\addtocounter{table}{-1}
\begin{table}
\capstart
\begin{align*}
\jint{5}{28}=\raisebox{\eqoff}{%
\begin{fmfchar*}(20,15)
\fmfleft{in}
\fmfright{out}
\fmf{plain}{in,v1}
\fmf{plain}{v4,out}
\fmffixed{(0.8w,0)}{v1,v4}
\fmffixed{(whatever,0)}{v1,in}
\fmffixed{(whatever,0)}{out,v4}
\fmfpoly{phantom}{v1,v2,v3,v4,v5,v6}
\fmf{plain}{v2,v1}
\fmf{derplain}{v2,v3}
\fmf{plain}{v3,v4}
\fmf{plain}{v5,v4}
\fmf{derplain}{v6,v5}
\fmf{plain}{v6,v1}
\fmf{plain}{v0,v1}
\fmf{plain}{v0,v2}
\fmf{plain}{v0,v3}
\fmf{plain}{v0,v5}
\fmf{plain}{v0,v6}
\fmffixed{(0.4w,0)}{v1,v0}
\fmffreeze
\end{fmfchar*}}
&=
\frac{1}{5\varepsilon^2}-\frac{1}{\varepsilon}\Big(\frac{19}{20}+\frac{4\zeta(3)}{5}\Big)
\\
\jint{5}{29}=\raisebox{\eqoff}{%
\begin{fmfchar*}(20,15)
\fmfleft{in}
\fmfright{out}
\fmf{plain}{in,v1}
\fmf{plain}{v3,out}
\fmffixed{(0.8w,0)}{v1,v3}
\fmffixed{(whatever,0)}{v1,in}
\fmffixed{(whatever,0)}{out,v3}
\fmf{plain}{v1,v2}
\fmf{plain}{v2,v3}
\fmf{derplain}{v4,v3}
\fmf{plain}{v4,v5}
\fmf{plain}{v5,v6}
\fmf{derplain}{v6,v1}
\fmf{plain}{v0,v4}
\fmf{plain}{v0,v5}
\fmf{plain}{v0,v6}
\fmf{plain,tension=0.25,right=0.25}{v2,v0,v2}
\fmffixed{(0,0.4h)}{v0,v2}
\fmffixed{(0,0.3h)}{v5,v0}
\fmffixed{(0.4w,0)}{v6,v4}
\fmffreeze
\end{fmfchar*}}
&=
\frac{1}{20\varepsilon^2}-\frac{1}{\varepsilon}\Big(\frac{13}{20}+\frac{4\zeta(3)}{5}\Big)
\\
\jint{5}{30}=\raisebox{\eqoff}{%
\begin{fmfchar*}(20,15)
\fmfleft{in}
\fmfright{out}
\fmf{plain}{in,v1}
\fmf{plain}{v4,out}
\fmffixed{(0.8w,0)}{v1,v4}
\fmffixed{(whatever,0)}{v1,in}
\fmffixed{(whatever,0)}{out,v4}
\fmfpoly{phantom}{v1,v2,v3,v4,v5,v6}
\fmffreeze
\fmf{derplainpt}{v2,v1}
\fmf{plain}{v2,v7}
\fmf{derplainpt}{v7,v3}
\fmf{plain}{v3,v4}
\fmf{derplain}{v5,v4}
\fmf{plain}{v5,v6}
\fmf{derplain}{v6,v1}
\fmf{plain}{v0,v2}
\fmf{plain}{v0,v7}
\fmf{plain}{v0,v3}
\fmf{plain}{v0,v5}
\fmf{plain}{v0,v6}
\fmffixed{(0.4w,0)}{v1,v0}
\fmffixed{(whatever,0)}{v2,v7}
\fmffreeze
\end{fmfchar*}}
&=
\frac{1}{\varepsilon}\Big(\frac{9}{10}+\frac{11\zeta(3)}{5}-2\zeta(5)\Big)
\\
\jint{5}{31}=\raisebox{\eqoff}{%
\begin{fmfchar*}(20,15)
\fmfleft{in}
\fmfright{out}
\fmf{plain}{in,v1}
\fmf{plain}{v4,out}
\fmffixed{(0.8w,0)}{v1,v4}
\fmffixed{(whatever,0)}{v1,in}
\fmffixed{(whatever,0)}{out,v4}
\fmfpoly{phantom}{v1,v2,v3,v4,v5,v6}
\fmffreeze
\fmf{derplainpt}{v2,v1}
\fmf{plain}{v2,v7}
\fmf{plain}{v7,v3}
\fmf{derplainpt}{v3,v4}
\fmf{derplain}{v5,v4}
\fmf{plain}{v5,v6}
\fmf{derplain}{v6,v1}
\fmf{plain}{v0,v2}
\fmf{plain}{v0,v7}
\fmf{plain}{v0,v3}
\fmf{plain}{v0,v5}
\fmf{plain}{v0,v6}
\fmffixed{(0.4w,0)}{v1,v0}
\fmffixed{(whatever,0)}{v2,v7}
\fmffreeze
\end{fmfchar*}}
&=
-\frac{1}{\varepsilon}\Big(
\frac{9}{10}+2\zeta(3)-7\zeta(7)\Big)
\\
\jint{5}{32}=\raisebox{\eqoff}{%
\begin{fmfchar*}(20,15)
\fmfleft{in}
\fmfright{out}
\fmf{plain}{in,v1}
\fmf{plain}{v4,out}
\fmffixed{(0.8w,0)}{v1,v4}
\fmffixed{(whatever,0)}{v1,in}
\fmffixed{(whatever,0)}{out,v4}
\fmfpoly{phantom}{v1,v2,v3,v4,v5,v6}
\fmffreeze
\fmf{derplain}{v1,v2}
\fmf{derplainpt}{v7,v2}
\fmf{plain}{v7,v3}
\fmf{plain}{v3,v4}
\fmf{derplain}{v4,v5}
\fmf{derplainpt}{v6,v5}
\fmf{plain}{v6,v1}
\fmf{plain}{v0,v2}
\fmf{plain}{v0,v7}
\fmf{plain}{v0,v3}
\fmf{plain}{v0,v5}
\fmf{plain}{v0,v6}
\fmffixed{(0.4w,0)}{v1,v0}
\fmffixed{(whatever,0)}{v2,v7}
\fmffreeze
\end{fmfchar*}}
&=
-\frac{1}{\varepsilon}\Big(
\frac{1}{5}+\frac{2\zeta(3)}{5}+2\zeta(5)-\frac{7\zeta(7)}{2}\Big)
\\
\jint{5}{33}=\raisebox{\eqoff}{%
\begin{fmfchar*}(20,15)
\fmfleft{in}
\fmfright{out}
\fmf{plain}{in,v1}
\fmf{plain}{v4,out}
\fmffixed{(0.8w,0)}{v1,v4}
\fmffixed{(whatever,0)}{v1,in}
\fmffixed{(whatever,0)}{out,v4}
\fmfpoly{phantom}{v1,v2,v3,v4,v5,v6}
\fmffreeze
\fmf{derplainpt}{v1,v2}
\fmf{derplain}{v7,v2}
\fmf{plain}{v7,v3}
\fmf{plain}{v3,v4}
\fmf{derplain}{v4,v5}
\fmf{derplainpt}{v6,v5}
\fmf{plain}{v6,v1}
\fmf{plain}{v0,v2}
\fmf{plain}{v0,v7}
\fmf{plain}{v0,v3}
\fmf{plain}{v0,v5}
\fmf{plain}{v0,v6}
\fmffixed{(0.4w,0)}{v1,v0}
\fmffixed{(whatever,0)}{v2,v7}
\fmffreeze
\end{fmfchar*}}
&=
-\frac{1}{\varepsilon}\Big(
\frac{3}{10}+\frac{3\zeta(3)}{5}-2\zeta(5)\Big)
\\
\jint{5}{34}=\raisebox{\eqoff}{%
\begin{fmfchar*}(20,15)
\fmfleft{in}
\fmfright{out}
\fmf{plain}{in,v1}
\fmf{plain}{v4,out}
\fmffixed{(0.8w,0)}{v1,v4}
\fmffixed{(whatever,0)}{v1,in}
\fmffixed{(whatever,0)}{out,v4}
\fmfpoly{phantom}{v1,v2,v3,v4,v5,v6}
\fmffreeze
\fmf{derplain}{v1,v2}
\fmf{derplain}{v7,v2}
\fmf{plain}{v7,v3}
\fmf{plain}{v3,v4}
\fmf{derplainpt}{v4,v5}
\fmf{derplainpt}{v6,v5}
\fmf{plain}{v6,v1}
\fmf{plain}{v0,v2}
\fmf{plain}{v0,v7}
\fmf{plain}{v0,v3}
\fmf{plain}{v0,v5}
\fmf{plain}{v0,v6}
\fmffixed{(0.4w,0)}{v1,v0}
\fmffixed{(whatever,0)}{v2,v7}
\fmffreeze
\end{fmfchar*}}
&=
\frac{1}{\varepsilon}\Big(
\frac{1}{10}+\frac{\zeta(3)}{5}-2\zeta(5)+\frac{7\zeta(7)}{2}\Big)
\\
& \\
\jintind{5}{35}{\mu\nu\rho\sigma}=\raisebox{\eqoff}{%
\begin{fmfchar*}(20,15)
\fmfleft{in}
\fmfright{out}
\fmf{plain}{in,v1}
\fmf{plain}{v4,out}
\fmffixed{(0.8w,0)}{v1,v4}
\fmffixed{(whatever,0)}{v1,in}
\fmffixed{(whatever,0)}{out,v4}
\fmfpoly{phantom}{v1,v2,v3,v4,v5,v6}
\fmffreeze
\fmf{derplain}{v1,v2}
\fmf{derplain}{v7,v2}
\fmf{plain}{v7,v3}
\fmf{plain}{v3,v4}
\fmf{derplain}{v4,v5}
\fmf{derplain}{v6,v5}
\fmf{plain}{v6,v1}
\fmf{plain}{v0,v2}
\fmf{plain}{v0,v7}
\fmf{plain}{v0,v3}
\fmf{plain}{v0,v5}
\fmf{plain}{v0,v6}
\fmffixed{(0.4w,0)}{v1,v0}
\fmffixed{(whatever,0)}{v2,v7}
\fmffreeze
\fmfipath{p[]}
\fmfiset{p1}{vpath(__v1,__v2)}
\fmfiset{p2}{vpath(__v7,__v2)}
\fmfiset{p3}{vpath(__v4,__v5)}
\fmfiset{p4}{vpath(__v5,__v6)}
\fmfipair{w[]}
\fmfiequ{w1}{point length(p1)/2 of p1}
\fmfiequ{w2}{point length(p2)/2 of p2}
\fmfiequ{w3}{point length(p3)/2 of p3}
\fmfiequ{w4}{point length(p4)/2 of p4}
\fmfiv{l=\footnotesize{$\partial_\mu$},l.a=200,l.d=4}{w1}
\fmfiv{l=\footnotesize{$\partial_\nu$},l.a=-90,l.d=5}{w2}
\fmfiv{l=\footnotesize{$\partial_\rho$},l.a=45,l.d=3}{w3}
\fmfiv{l=\footnotesize{$\partial_\sigma$},l.a=90,l.d=4}{w4}
\end{fmfchar*}}
\end{align*}
\caption{Momentum integrals for five-loop wrapping diagrams \textit{(continued)}}
\end{table}

\chapter{The Gegenbauer Polynomial \texorpdfstring{$x$}{x}-space Technique}
\label{app:gegenbauer}
In this appendix, the main features of the Gegenbauer Polynomial $x$-space Technique (GPXT)~\cite{Chetyrkin:1980pr} for the computation of integrals will be reviewed. First of all, the details of the steps that must be performed for a typical integral will be described. Then, an explicit example will be presented. The GPXT has been used for all the integrals required in this thesis.

\section{Description of the technique}
The power of the GPXT for the analysis of integrals related to Feynman diagrams derives from the fact that in $x$-space (that is, coordinate space) a propagator always depends on the difference of only two vectors. In momentum space ($p$-space), on the contrary, diagrams exist such that it is impossible to choose the integration momenta so that every propagator depends only on two variables. 

The first step in the calculation of momentum integrals by means of the GPXT is their transformation to $x$-space, which is realized by means of the Fourier transform.
All the computations are performed using dimensional regularization in a spacetime with $\stdim=4-2\varepsilon$ dimensions. It is useful to introduce the quantity $\lambda=\stdim/2-1=1-\varepsilon$.

\subsection{Fourier transform}
A generic $L$-loop momentum integral $I$ with $P$ propagators, written in momentum space and with a single non-vanishing external momentum, 
\begin{equation}
I=\frac{1}{(2\pi)^{Ld}}\int\frac{\mathrm{d}^d k_1\cdots\mathrm{d}^d k_L}{\Pi_1\cdots\Pi_P} \col
\end{equation}
can be transformed to coordinate space through the Fourier transform
\begin{equation}
\label{gegenbauer-fourier}
\frac{1}{k^2}=\frac{\Gamma(\lambda)}{\pi^{\lambda+1}}\int\frac{e^{2i k\cdot x}}{x^{2\lambda}}\mathrm{d}^\stdim x \col
\end{equation}
and the result reads
\begin{equation}
I=\frac{\Gamma(\lambda)^P}{(2^{2L}\pi^P)^{\lambda+1}}\int\frac{e^{2ip\cdot(x_\mathrm{out}-x_\mathrm{in})}}{X_1\cdots X_P}\mathrm{d}^\stdim x_1\cdots\mathrm{d}^\stdim x_{P-L} \col
\end{equation}
where $p$ is the external momentum, $x_\mathrm{in}$ and $x_\mathrm{out}$ are the vertices where the total external momentum enters and leaves the graph, and the $X_i$ are the scalar propagators as functions of the $x$-space coordinates. 

Momenta in the numerator of the integral correspond to derivatives in $x$-space, and can be simplified according to
\begin{equation}
\label{gegenbauer-fourierderiv}
\frac{k^\mu}{k^2}=\frac{\Gamma(\lambda)}{\pi^{\lambda+1}}\frac{1}{2i}\int\frac{\partial_\mu e^{2i k\cdot x}}{x^{2\lambda}}\mathrm{d}^\stdim x =
-i\lambda\frac{\Gamma(\lambda)}{\pi^{\lambda+1}}\int e^{2i k\cdot x}\frac{x^\mu}{x^{2(\lambda+1)}}\mathrm{d}^\stdim x \col 
\end{equation}
\begin{equation}
\frac{k^\mu k^\nu}{k^2}=-\frac{\Gamma(\lambda)}{\pi^{\lambda+1}}\lambda(\lambda+1)\int e^{2i k\cdot x}\frac{x^{(\mu}x^{\nu)}}{x^{2(\lambda+2)}}\mathrm{d}^\stdim x \col
\end{equation}
where $x_1^{(\mu}x_2^{\nu)}$ is the traceless product, defined in terms of $x_1$, $x_2$ and of the spacetime metric $g^{\mu\nu}$ as
\begin{equation}
x_1^{(\mu}x_2^{\nu)}=x_1^{\mu}x_2^{\nu}-\frac{x_1\cdot x_2}{\stdim}g^{\mu\nu} \pnt
\end{equation}
For a given $d$-dimensional vector $x$, it is useful to denote its angular part, that is the unit vector with the same direction, as $\hat{x}$, and the square of the radial part as $r=x^2$. With these definitions, the integration measure $\mathrm{d}^d x$ can be written as
\begin{equation}
\mathrm{d}^\stdim x=\frac{1}{2}\Omega_{\stdim-1}r^\lambda\mathrm{d}r\,\mathrm{d}\hat{x} \col
\end{equation}
where  
\begin{equation}
\Omega_{\stdim-1}=\frac{2\pi^\frac{\stdim}{2}}{\Gamma(\frac{\stdim}{2})}
\end{equation}
is the volume of the $(\stdim-1)$-dimensional unit sphere.

\subsection{Properties of the Gegenbauer polynomials}
After the transformation of the integral from momentum space to $x$-space, the fundamental step in the GPXT is the expansion of the scalar propagators in series of Gegenbauer polynomials $C_n^\lambda$. Such expansion allows to separate the dependence of the propagators on radial and angular variables
\begin{equation}
\label{gegenbauer-prop}
\frac{1}{(x_1-x_2)^{2\lambda}}=\frac{1}{\mathrm{max}(r_1,r_2)^\lambda}\sum_{n=0}^\infty C_n^\lambda(\hat{x}_1\cdot\hat{x}_2)\left[\frac{\mathrm{min}(r_1,r_2)}{\mathrm{max}(r_1,r_2)}\right]^\frac{n}{2} \pnt
\end{equation}
Thanks to translation symmetry, it is possible to put one of the points of the diagram at the origin $x=0$ of the space. All the propagators starting from this node, which is named the \emph{root vertex}, will be of the simplified form $1/x_i^{2\lambda}=1/r_i^\lambda$, and therefore there will be no need to expand them using~\eqref{gegenbauer-prop}. For this reason, a wise choice of the root vertex can simplify the whole calculation, by reducing the number of infinite summations coming from~\eqref{gegenbauer-prop}. The vertex of the graph with the highest number of outgoing lines is typically the best choice for the root vertex. If derivatives are present in some numerators, they should be moved onto lines connected to the root vertex by means of repeated integration by parts, thus reducing the original problem to the computation of a set of simpler integrals.

The Gegenbauer polynomials are orthogonal, and the orthogonality identity can be written as
\begin{equation}
\int_{-1}^1(1-z^2)^{\lambda-\frac{1}{2}}C_m^\lambda(z)C_n^\lambda(z)\mathrm{d} z=\frac{\pi 2^{1-2\lambda}\Gamma(n+2\lambda)}{n!(n+\lambda)\Gamma(\lambda)^2}\delta_{mn} \col
\end{equation}
or, in terms of angular variables, in the simplified form
\begin{equation}
\label{gegenbauer-orthogonality}
\int C_m^\lambda(\hat{x}_1\cdot\hat{x})C_n^\lambda(\hat{x}\cdot\hat{x}_2)\mathrm{d}\hat{x}=\frac{\lambda}{\lambda+n}C_n^\lambda(\hat{x}_1\cdot\hat{x}_2)\delta_{mn} \pnt
\end{equation}
As $\hat{x}\cdot\hat{x}=1$, the values of the polynomials for $z=1$ are often needed
\begin{equation}
C_n^\lambda(1)=\frac{\Gamma(n+2\lambda)}{n!\Gamma(2\lambda)} \pnt
\end{equation}
The scalar product $k_1\cdot k_2$ of two different momenta can be written in terms of $x_1$ and $x_2$ using~\eqref{gegenbauer-fourierderiv}, and simplified according to
\begin{equation}
\label{gegenbauer-xprod}
x_1^{(\mu_1,\ldots\mu_n)}x_2^{(\mu_1,\ldots\mu_n)}=\frac{n!\Gamma(\lambda)}{2^n\Gamma(\lambda+n)}C_n^\lambda(\hat{x}_1\cdot\hat{x}_2)(r_1 r_2)^\frac{n}{2} \pnt
\end{equation}
The exponential function too can be expanded in a series of Gegenbauer polynomials
\begin{equation}
\label{gegenbauer-exp}
e^{2i p\cdot x}=\Gamma(\lambda)\sum_{n=0}^\infty i^n(n+\lambda)C_n^\lambda(\hat{p}\cdot\hat{x})(p^2 r)^\frac{n}{2}j_{n+\lambda}(p^2r) \col
\end{equation}
where the functions $j_{n+\lambda}(z)$ are defined in terms of the standard Bessel ones $J_n$ as
\begin{equation}
j_{n+\lambda}(z)=\frac{1}{\sqrt{z}}J_n(2\sqrt{z}) \pnt
\end{equation}

Once all the propagators, scalar products and exponential terms have been expanded into series of Gegenbauer polynomials, the dependency of the whole integral on the angular integration variables is completely separated from the one on radial variables.

\subsection{Computation of the \texorpdfstring{$x$}{x}-space integral}
To proceed with the computation, the integration with respect to each variable must be split into its angular and radial parts. Then it is possible to perform all the angular integrations, by repeatedly using~\eqref{gegenbauer-orthogonality}. Every Kronecker delta appearing as a consequence of the application of~\eqref{gegenbauer-orthogonality} can be used to eliminate one of the infinite summations produced by the expansions~\eqref{gegenbauer-prop} and~\eqref{gegenbauer-exp}. If some of the $\hat{x}_i$ appear in the arguments of more than two polynomials, in order to be able to apply~\eqref{gegenbauer-orthogonality} one must first decompose products of Gegenbauer polynomials in Clebsch-Gordan series according to
\begin{equation}
\label{gegenbauer-clebsch}
C_m^\lambda(z)C_n^\lambda(z)=\sum_{\substack{j=|m-n| \\ \frac{j+m+n}{2}\in\mathds{N}}}^{m+n}D_\lambda(m,n,j)C_j^\lambda(z) \col
\end{equation}
where
\begin{equation}
D_\lambda(m,n,j)=
\frac{j!(j+\lambda)\Gamma(\frac{m+n+j}{2}\!+\!2\lambda)}
{\Gamma(\lambda)^2\Gamma(\frac{m+n+j}{2}\!+\!\lambda+1)\Gamma(j\!+\!2\lambda)}
\frac{\Gamma(\frac{-m+n+j}{2}\!+\!\lambda)\Gamma(\frac{m-n+j}{2}\!+\!\lambda)
\Gamma(\frac{m+n-j}{2}\!+\!\lambda)}
{\Gamma(\frac{-m+n+j}{2}\!+\!1)\Gamma(\frac{m-n+j}{2}\!+\!1)
\Gamma(\frac{m+n-j}{2}\!+\!1)}\pnt
\end{equation}
Since $\lambda>0$, the coefficients $D_\lambda(m,n,j)$ never become singular for $\varepsilon\to0$.

As far as the radial integrations are concerned, because of the presence of the $\mathrm{max}$ and $\mathrm{min}$ functions in the integrand, the integration domain must be split into $(P-L)!$ subdomains according to all the possible permutations of the radial variables. The radial integral is thus reduced to the sum of $(P-L)!$ integrals of standard functions free of $\mathrm{max}$ and $\mathrm{min}$ functions. It is therefore important to recognize all the possible symmetries of the integrand in order to reduce the number of independent domains that must be actually considered explicitly. In any case, the huge number of required domains makes the whole method feasible only in a computer implementation, where the task of radial integrations is completely automated.

The outcome of the angular and radial integrations is a function of $\lambda$ and of the indices of the infinite summations that survived after all the previous steps. Once these summations have been performed, the final result is found.
If only the divergent part of the result is required, it is in general easier to expand the partial result in powers of $1/\varepsilon$ before the final summations.

\subsection{Infrared regularization}
If the final goal of the computation of the integral is only the determination of its divergent part, it is possible to achieve a great simplification by neglecting the exponential function, which does not alter the ultraviolet behaviour of the integral~\cite{Chetyrkin:1980pr}. As this is equivalent to putting the external momentum to zero, which would introduce infrared divergences, a regulator is required. The simplest possibility is to introduce an infrared cutoff $R$ on the radial integrations. After all the radial integrations have been performed, the result can be expanded into a Laurent series, where the coefficients of the divergent terms are independent of $R$, once subdivergences have been subtracted. The $R$-dependent finite parts must be discarded.

The cancellation of the external momentum simplifies the integration because it prevents the appearance of the Bessel functions and reduces by one the number of infinite summations. Thanks to the absence of the Bessel functions, the integrand in each subdomain is a simple monomial in the $r_i$ variables, and the automatic processing of the integrals can be made much more efficient.
Obviously, if subdivergences are present they must be dealt with by means of the same cutoff, so that they can be subtracted consistently in the end. 

For integrals with more than one derivative on the same line, an additional regularization is required to obtain the correct extension of the traceless products to arbitrary dimensions. In~\cite{uslong}, an explicit example of this more complicated situation is analyzed in detail.

\section{Example}
\begin{figure}[t]
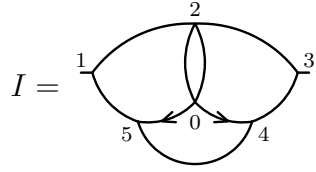

\capstart
\label{I1twoder}
\settoheight{\eqoff}{$\times$}
\setlength{\eqoff}{-2\eqoff}
\addtolength{\eqoff}{-7.5\unitlength}
$I=$
\raisebox{\eqoff}{
\begin{fmfchar*}(30,25)
\fmfleft{in}
\fmfright{out}
\fmf{plain}{in,v1}
\fmf{plain,left=0.25}{v1,v2}
\fmf{plain,left=0.25}{v2,v3}
\fmf{plain,right=0.25}{v1,v4}
\fmf{derplain,left=0.25}{v0,v4}
\fmf{derplain,right=0.25}{v0,v5}
\fmf{plain,right=0.75}{v4,v5}
\fmf{plain,right=0.25}{v5,v3}
\fmf{plain}{v3,out}
\fmffixed{(0.9w,0)}{v1,v3}
\fmfpoly{phantom}{v2,v4,v5}
\fmffixed{(0.5w,0)}{v4,v5}
\fmffixed{(0.5w,0)}{v4,v5}
\fmf{plain,tension=0.25,right=0.25}{v2,v0,v2}
\fmffreeze
\fmfshift{(0,0.1w)}{in,out,v1,v2,v3,v4,v5,v0}
\fmffreeze
\fmfiv{l=\scriptsize{1},l.a=135,l.d=3}{vloc(__v1)}
\fmfiv{l=\scriptsize{2},l.a=90,l.d=3}{vloc(__v2)}
\fmfiv{l=\scriptsize{3},l.a=45,l.d=3}{vloc(__v3)}
\fmfiv{l=\scriptsize{4},l.a=-45,l.d=3}{vloc(__v5)}
\fmfiv{l=\scriptsize{5},l.a=-135,l.d=3}{vloc(__v4)}
\fmfiv{l=\scriptsize{0},l.a=-90,l.d=5}{vloc(__v0)}
\end{fmfchar*}}
\caption{Example four-loop integral with non-trivial numerator}
\label{exampleint}
\end{figure}

As an explicit example, the GPXT will now be applied to find the pole part of the four-loop integral with non-trivial numerator of Figure~\ref{exampleint}~\cite{uslong}. This integral is required for the calculation of $J_{13}^{(4)}$, being one of the terms that are produced when the momenta in $J_{13}^{(4)}$ are moved by means of integrations by parts.

The pair of arrows of the same type represents as usual the contraction of the momenta flowing along the corresponding lines. 
In this case, the best choice for the root vertex is the node labeled by 0 in~\eqref{I1twoder}.
Using~\eqref{gegenbauer-fourier} and~\eqref{gegenbauer-fourierderiv}, the integral $I$ in $x$-space reads
\begin{equation}
I=-\lambda^2\frac{\Gamma(\lambda)^9}{(2^8\pi^9)^{1+\lambda}}
\int\frac{\de^\stdim x_1\de^\stdim x_2\de^\stdim x_3\de^\stdim x_4\de^\stdim x_5\,
(x_4\cdot x_5)e^{2ip\cdot(x_3-x_1)}}
{x_2^{4\lambda}(x_4^2x_5^2)^{1+\lambda}
(\Delta_{1,2}^2\Delta_{2,3}^2\Delta_{3,4}^2\Delta_{4,5}^2\Delta_{5,1}^2)^\lambda}
\col
\end{equation}
where $\Delta_{i,j}^2$ has been defined as
\begin{equation}
\Delta_{i,j}^2=\frac{1}{(x_i-x_j)^{2\lambda}} \pnt
\end{equation}
It is now possible to split the five integrations into their angular and radial parts. Note that the simple propagators $1/r_i^\lambda$ with trivial numerator are cancelled by the corresponding $r_i^\lambda$ factors from the integration measure. 

The scalar product $(x_4\cdot x_5)$ can be simplified thanks to~\eqref{gegenbauer-xprod} with $n=1$, and the integral becomes
\begin{equation}
I=-\frac{\lambda}{2}\frac{\Gamma(\lambda)^9}{(2^8\pi^9)^{1+\lambda}}\left(\frac{\Omega_{\stdim-1}}{2}\right)^5 \sum_{j,k,l,n,q=0}^\infty \mathcal{R}_\lambda(j,k,l,n,q)\mathcal{A}_\lambda(j,k,l,n,q) \col
\end{equation}
where $\mathcal{R}_\lambda(j,k,l,n,q)$ and $\mathcal{A}_\lambda(j,k,l,n,q)$ are the radial and angular parts respectively. 
With the introduction of the notation ${m}_{i,j}=\mathrm{min}(r_i,r_j)$, ${M}_{i,j}=\mathrm{max}(r_i,r_j)$, the explicit expression for the radial part is
\begin{multline}
\mathcal{R}_\lambda(j,k,l,n,q)=\int_0^R\!\!\!\di r_1\!\int_0^R\!\!\!\di r_2\!\int_0^R\!\!\!\di r_3\!\int_0^R\!\!\!\di r_4\!\int_0^R\!\!\!\di r_5\frac{(r_1r_3)^\lambda}{r_2^\lambda(r_4r_5)^{1/2}(M_{1,2}M_{2,3}M_{3,4}M_{4,5}M_{5,1})^\lambda} \\
\times
\left(\frac{m_{1,2}}{M_{1,2}}\right)^\frac{j}{2}
\left(\frac{m_{2,3}}{M_{2,3}}\right)^\frac{k}{2}
\left(\frac{m_{3,4}}{M_{3,4}}\right)^\frac{l}{2}
\left(\frac{m_{4,5}}{M_{4,5}}\right)^\frac{n}{2}
\left(\frac{m_{5,1}}{M_{5,1}}\right)^\frac{q}{2}
\pnt
\end{multline}
As for the angular part, according to~\eqref{gegenbauer-xprod} the scalar product $(x_4\cdot x_5)$ produces a factor $C_1^\lambda(\hat{x}_4\cdot\hat{x}_5)$ in addition to the $C_n^\lambda(\hat{x}_4\cdot\hat{x}_5)$ coming from the expansion of the propagator. Since both $\hat{x}_4$ and $\hat{x}_5$ also enter other polynomials, the product of these two Gegenbauer polynomials must be reduced using~\eqref{gegenbauer-clebsch} and $\mathcal{A}_\lambda(j,k,l,n,q)$ reads
\begin{multline}
\mathcal{A}_\lambda(j,k,l,n,q)
=\sum_{\substack{s=|n-1| \\ s\neq n}}^{n+1}D_\lambda(1,n,s)
\int\de\hat x_1\dots
\de\hat x_5\,\\
C_j^\lambda(\hat x_1\cdot\hat x_2)C_k^\lambda(\hat x_2\cdot\hat x_3)
C_l^\lambda(\hat x_3\cdot\hat x_4)
C_s^\lambda(\hat x_4\cdot\hat x_5)C_q^\lambda(\hat x_5\cdot\hat x_1) \pnt
\end{multline}
This expression can be simplified by means of the repeated application of~\eqref{gegenbauer-orthogonality}
\begin{equation}
\mathcal{A}_\lambda(j,k,l,n,q)=\delta_{jk}\delta_{jl}\delta_{jq}
(\delta_{n,j+1}+\delta_{n,j-1})
\frac{\lambda^4}{(j+\lambda)^4}
D_\lambda(1,n,j)C_j^\lambda(1)
\pnt
\end{equation}
The Kronecker deltas in $\mathcal{A}_\lambda(j,k,l,n,q)$ reduce the original five series to a single infinite summation on the index $j$ and a finite sum on $n$, which for every value of $j$ can assume only the values $j\pm 1$. Note that in general it is easier to perform the radial integrations by exploiting the appearance of such deltas, thus reducing the number of indices on which the radial integrand depends.

The result of the radial integrations is a complicated rational function of $\lambda$, $j$ and $n$, proportional to $R^{4\varepsilon}$. Defining
\begin{equation}
\mathcal{K}_\lambda(j,n)=-\frac{\lambda}{2}\frac{\Gamma(\lambda)^9}{(2^8\pi^9)^{1+\lambda}}\left(\frac{\Omega_{\stdim-1}}{2}\right)^5\frac{\lambda^4}{(j+\lambda)^4} D_\lambda(1,n,j)C_j^\lambda(1)\mathcal{R}(j,j,j,n,j) \col
\end{equation}
the integral can be written as
\begin{equation}
I=\mathcal{K}_\lambda(0,1)+\sum_{j=1}^\infty[\mathcal{K}_\lambda(j,j+1)+\mathcal{K}_\lambda(j,j-1)] \pnt
\end{equation}
To perform the summation explicitly, it is necessary to expand each term in a Laurent series for $\varepsilon\to0$, keeping only the pole part.
The poles $1/\varepsilon^k$ with $k>1$ come only from the term $\mathcal{K}(0,1)$, whereas all the other terms contribute only to the coefficient of the first-order pole $1/\varepsilon$. 
At this point the series can be summed and the final result for the divergent part of $I$ is found, as a sum of poles from $1/\varepsilon^4$ to $1/\varepsilon$. Owing to the presence of subdivergences, the coefficients in this expansion are rather complicated expressions involving powers of $\mathrm{log}R$. The final step to make is the subtraction of subdivergences, which removes all the dependence on $R$, producing the final result
\begin{equation}
I\to\frac{1}{(4\pi)^8}\left[\frac{1}{6\varepsilon^4}-\frac{1}{6\varepsilon^3}-\frac{3}{4\varepsilon^2}+\frac{1}{\varepsilon}\left(\frac{11}{12}-\zeta(3)\right)\right] \pnt
\end{equation}

\chapter{Generalized triangle rules}
\label{app:triangles}
This appendix is about the generalization of the standard triangle rule~\cite{Broadhurst:1985vq,Chetyrkin:1981qh} for integration by parts to the case of integrals with non-trivial numerators. The generalized rules are first derived following the method of~\cite{Broadhurst:1985vq}. Then, an example of their application to the analysis of the integrals required in Chapter~\ref{chapter:wrapbetadef} is presented. Finally, the results for the set of integrals with up to seven loops that are needed to compute all the integrals of Chapter~\ref{chapter:wrapbetadef} by means of recurrence relations, are given.

\section{Derivation of the rules}
In order to study triangle identities, it is useful to work with integrals in momentum space, rather than in $x$-space. The fundamental relations in this case are~\cite{Chetyrkin:1980pr}
\vspace{0.5cm}
\begin{equation}
\label{Gdef}
\begin{aligned}
\settoheight{\eqoff}{$\times$}%
\setlength{\eqoff}{0.5\eqoff}%
\addtolength{\eqoff}{-7.5\unitlength}%
\settoheight{\eqofftwo}{$\times$}%
\setlength{\eqofftwo}{0.5\eqofftwo}%
\addtolength{\eqofftwo}{-3\unitlength}%
\scriptsize
\raisebox{\eqofftwo}{%
\begin{fmfchar*}(15,6)
  \fmfleft{in}
  \fmfright{out}
  \fmf{plain}{in,v1}
  \fmf{plain}{v2,out}
  \fmf{plain,label=$\beta$,right=0.4}{v1,v2}
  \fmf{plain,label=$\alpha$,label.side=left,left=0.4}{v1,v2}
\fmffixed{(0.9w,0)}{v1,v2}
\fmffreeze
\end{fmfchar*}}
\normalsize
=\frac{1}{(4\pi)^\stdim}\int\frac{\mathrm{d}^\stdim l}{l^{2\alpha}(l+p)^{2\beta}}=G(\alpha,\beta)\frac{1}{(p^2)^{\alpha+\beta-\stdim/2}}=
G(\alpha,\beta)\ 
\scriptsize
\raisebox{\eqofftwo}{%
\begin{fmfchar*}(15,6)
  \fmfleft{in}
  \fmfright{out}
  \fmf{plain}{in,v1}
  \fmf{plain}{v2,out}
  \fmf{plain,label=$g(\alpha,,\beta)$,label.side=left}{v1,v2}
\fmffixed{(0.9w,0)}{v1,v2}
\fmffreeze
\end{fmfchar*}}
\scriptsize
\col 
\end{aligned}
\end{equation}
\begin{equation}
\begin{aligned}
\settoheight{\eqoff}{$\times$}%
\setlength{\eqoff}{0.5\eqoff}%
\addtolength{\eqoff}{-7.5\unitlength}%
\settoheight{\eqofftwo}{$\times$}%
\setlength{\eqofftwo}{0.5\eqofftwo}%
\addtolength{\eqofftwo}{-3\unitlength}%
\scriptsize
\raisebox{\eqofftwo}{%
\begin{fmfchar*}(15,6)
  \fmfleft{in}
  \fmfright{out}
  \fmf{plain}{in,v1}
  \fmf{plain}{v2,out}
  \fmf{plain,label=$\beta$,right=0.4}{v1,v2}
  \fmf{derplain,label=$\alpha$,label.side=left,left=0.4}{v1,v2}
\fmffixed{(0.9w,0)}{v1,v2}
\fmffreeze
\end{fmfchar*}}
\normalsize
=\frac{1}{(4\pi)^\stdim}\int\frac{l^\mu\mathrm{d}^\stdim l}{l^{2\alpha}(l+p)^{2\beta}}=G^{(1)}(\alpha,\beta)\frac{p^\mu}{(p^2)^{\alpha+\beta-\stdim/2}}=
G^{(1)}(\alpha,\beta)\ 
\scriptsize
\raisebox{\eqofftwo}{%
\begin{fmfchar*}(15,6)
  \fmfleft{in}
  \fmfright{out}
  \fmf{plain}{in,v1}
  \fmf{plain}{v2,out}
  \fmf{derplain,label=$g(\alpha,,\beta)$,label.side=left}{v1,v2}
\fmffixed{(0.9w,0)}{v1,v2}
\fmffreeze
\end{fmfchar*}}
\scriptsize
\col
\end{aligned}
\end{equation}
\begin{equation}
\begin{aligned}
\settoheight{\eqoff}{$\times$}%
\setlength{\eqoff}{0.5\eqoff}%
\addtolength{\eqoff}{-7.5\unitlength}%
\settoheight{\eqofftwo}{$\times$}%
\setlength{\eqofftwo}{0.5\eqofftwo}%
\addtolength{\eqofftwo}{-3\unitlength}%
\scriptsize
\raisebox{\eqofftwo}{%
\begin{fmfchar*}(15,6)
  \fmfleft{in}
  \fmfright{out}
  \fmf{plain}{in,v1}
  \fmf{plain}{v2,out}
  \fmf{plain,label=$\alpha$,label.side=left}{v1,v3}
  \fmf{plain,label=$\beta$,label.side=left,label.dist=4}{v3,v2}
\fmffixed{(0.9w,0)}{v1,v2}
\fmffreeze
\fmfiv{d.sh=circle,d.f=1,d.size=2}{vloc(__v3)}
\end{fmfchar*}}
\normalsize
=\frac{1}{(p^2)^\alpha}\frac{1}{(p^2)^\beta}=
\scriptsize
\raisebox{\eqofftwo}{%
\begin{fmfchar*}(15,6)
  \fmfleft{in}
  \fmfright{out}
  \fmf{plain}{in,v1}
  \fmf{plain}{v2,out}
  \fmf{plain,label=$\alpha+\beta$,label.side=left}{v1,v2}
\fmffixed{(0.9w,0)}{v1,v2}
\fmffreeze
\end{fmfchar*}}
\scriptsize
\col
\end{aligned}
\end{equation}
where $g(\alpha,\beta)=\alpha+\beta-\stdim/2$ and
\begin{equation}
G(\alpha,\beta)=\frac{\Gamma(\alpha+\beta-\stdim/2)\Gamma(\stdim/2-\alpha)\Gamma(\stdim/2-\beta)}{(4\pi)^{\stdim/2}\Gamma(\alpha)\Gamma(\beta)\Gamma(\stdim-\alpha-\beta)} \col
\end{equation}
\vspace{-1cm}
\begin{equation}
\label{G1G}
G^{(1)}(\alpha,\beta)=\frac{2-\alpha-\varepsilon}{4-2\varepsilon-\alpha-\beta}G(\alpha,\beta) \pnt
\end{equation}

The standard triangle rule was derived in~\cite{Broadhurst:1985vq,Chetyrkin:1981qh} using the technique of integration by parts. This rule allows to simplify the computation of integrals by exploiting the possible existence of particular triangular subgraphs, and reads
\vspace{0.5cm}
\newcommand{\figwidth}{18}
\newcommand{\figheight}{13}
\begin{equation}
\label{scalar-triangle}
\scriptsize
\settoheight{\eqoff}{$\times$}%
\setlength{\eqoff}{0.5\eqoff}%
\addtolength{\eqoff}{-6.5\unitlength}%
\raisebox{\eqoff}{%
\fmfframe(0,0)(0,0){%
\begin{fmfchar*}(\figwidth,\figheight)
\fmfleft{in}
\fmfright{out}
\fmf{plain,tension=1}{in,v1}
\fmfforce{(0,0.5h)}{in}
\fmfforce{(1.1w,0.5h)}{out}
\fmfforce{(0.5w,0)}{v3}
\fmfforce{(0.5w,h)}{v2}
\fmfforce{(0.03w,0.5h)}{v1}
\fmfforce{(w,0.5h)}{v4}
\fmfforce{(1.5w,0.5h)}{v5}
\fmfforce{(1.7w,0.5h)}{v6}
\fmffixed{(0.5w,0)}{v2,v7}
\fmffixed{(0.5w,0)}{v3,v8}
\fmffreeze
\fmf{plain,label=$\alpha$,l.dist=2,l.side=left}{v1,v2}
\fmf{plain,label=$\beta$,l.dist=2,l.side=right}{v1,v3}
\fmf{plain}{v2,v3}
\fmf{plain}{v2,v7}
\fmf{plain}{v3,v8}
\end{fmfchar*}}}\ \ 
\normalsize
= \Delta(\alpha,\beta)\ \ 
\raisebox{\eqoff}{%
\fmfframe(0,0)(0,0){%
\scriptsize
\begin{fmfchar*}(\figwidth,\figheight)
\fmfleft{in}
\fmfright{out}
\fmf{plain,label=$h(\alpha,,\beta)$,label.side=left,label.dist=2,tension=1}{in,v1}
\fmfforce{(0,0.5h)}{in}
\fmfforce{(1.1w,0.5h)}{out}
\fmfforce{(w,0)}{v3}
\fmfforce{(w,h)}{v2}
\fmfforce{(0.5w,0.5h)}{v1}
\fmfforce{(w,0.5h)}{v4}
\fmfforce{(1.5w,0.5h)}{v5}
\fmfforce{(1.7w,0.5h)}{v6}
\fmffixed{(0.5w,0)}{v2,v7}
\fmffixed{(0.5w,0)}{v3,v8}
\fmffreeze
\fmf{plain}{v1,v2}
\fmf{plain}{v1,v3}
\end{fmfchar*}}}\ \ 
\normalsize
+C(\alpha,\beta)\ \ 
\raisebox{\eqoff}{%
\fmfframe(0,0)(0,0){%
\scriptsize
\begin{fmfchar*}(\figwidth,\figheight)
\fmfleft{in}
\fmfright{out}
\fmf{plain,tension=1}{in,v1}
\fmfforce{(0,0.5h)}{in}
\fmfforce{(1.1w,0.5h)}{out}
\fmfforce{(0.5w,0)}{v3}
\fmfforce{(0.5w,h)}{v2}
\fmfforce{(0.03w,0.5h)}{v1}
\fmfforce{(w,0.5h)}{v4}
\fmfforce{(1.5w,0.5h)}{v5}
\fmfforce{(1.7w,0.5h)}{v6}
\fmffixed{(0.5w,0)}{v2,v7}
\fmffixed{(0.5w,0)}{v3,v8}
\fmffreeze
\fmf{plain,label=$\alpha+1$,l.dist=2,l.side=left}{v1,v2}
\fmf{plain,label=$\beta$,l.dist=2,l.side=right}{v1,v3}
\fmf{plain}{v2,v3}
\fmf{plain}{v3,v8}
\end{fmfchar*}}}\ \ 
\normalsize
+C(\beta,\alpha)\ \ 
\raisebox{\eqoff}{%
\fmfframe(0,0)(0,0){%
\scriptsize
\begin{fmfchar*}(\figwidth,\figheight)
\fmfleft{in}
\fmfright{out}
\fmf{plain,tension=1}{in,v1}
\fmfforce{(0,0.5h)}{in}
\fmfforce{(1.1w,0.5h)}{out}
\fmfforce{(0.5w,0)}{v3}
\fmfforce{(0.5w,h)}{v2}
\fmfforce{(0.03w,0.5h)}{v1}
\fmfforce{(w,0.5h)}{v4}
\fmfforce{(1.5w,0.5h)}{v5}
\fmfforce{(1.7w,0.5h)}{v6}
\fmffixed{(0.5w,0)}{v2,v7}
\fmffixed{(0.5w,0)}{v3,v8}
\fmffreeze
\fmf{plain,label=$\alpha$,l.dist=2,l.side=left}{v1,v2}
\fmf{plain,label=$\beta+1$,l.dist=2,l.side=right}{v1,v3}
\fmf{plain}{v2,v3}
\fmf{plain}{v2,v7}
\end{fmfchar*}}}
\col
\vspace{0.5cm}
\end{equation}
where $h(\alpha,\beta)=\alpha+\beta+1-d/2$ and the functions $\Delta(\alpha,\beta)$ and $C(\alpha,\beta)$ are defined as
\begin{equation}
\begin{aligned}
\Delta(\alpha,\beta)&=-\frac{\alpha G(\alpha+1,\beta)+\beta G(\alpha,\beta+1)}{\alpha+\beta+2-\stdim}\col\qquad
C(\alpha,\beta)=\frac{\alpha}{\alpha+\beta+2-\stdim}\pnt\\
\end{aligned}
\end{equation}

Here, the same method is extended to the case where some momenta appear in the numerator of the integral.
All the rules can be obtained from the following integration-by-parts identity, which is valid for $\alpha+\beta+1-\stdim/2-\mathrm{dim}(f)/2>0$
\begin{equation}
\begin{aligned}
0 &=\!\int\mathrm{d}^\stdim l\frac{\partial}{\partial l^{\mu}}\frac{f(l)l^{\mu}}{(l+p_1)^{2\alpha}(l+p_2)^{2\beta}l^2}\\
&=\!\int\mathrm{d}^\stdim l\Big(\partial_{\mu}f(l)l^{\mu}+\stdim\, f(l)\\
&\quad-2f(l)l^{\mu}\Big(\alpha\frac{(l+p_1)_{\mu}}{(l+p_1)^{2}}+\beta\frac{(l+p_2)_{\mu}}{(l+p_2)^2}+\frac{l^{\mu}}{l^2}\Big)\Big)\frac{1}{(l+p_1)^{2\alpha}(l+p_2)^{2\beta}l^2}\\
&=\!\int\mathrm{d}^\stdim l\Big[\partial_{\mu}f(l)l^{\mu}\\
&\quad-f(l)\Big(\alpha\frac{l^2+(l+p_1)^2-p_1^2}{(l+p1)^2}+\beta\frac{l^2+(l+p_2)^2-p_2^2}{(l+p_2)^2}-\stdim+2\Big)\Big]\frac{1}{(l+p_1)^{2\alpha}(l+p_2)^{2\beta}l^2}\\
&=\!\int\mathrm{d}^\stdim l\Big[\partial_{\mu}f(l)l^{\mu}\!-\!\Big(\alpha\frac{l^2-p_1^2}{(l+p_1)^2}+\beta\frac{l^2-p_2^2}{(l+p_2)^2}+\alpha+\beta+2-\stdim\Big)\Big]\frac{f(l)}{(l+p_1)^{2\alpha}(l+p_2)^{2\beta}l^2}
\pnt
\end{aligned}
\end{equation}
For $f(l)=1$, the standard rule~\eqref{scalar-triangle} is recovered. Taking $f(l)=l^\mu$, one finds the first generalized rule
\begin{equation}
\label{der1-triangle}
\begin{aligned}
\scriptsize
\settoheight{\eqoff}{$\times$}%
\setlength{\eqoff}{0.5\eqoff}%
\addtolength{\eqoff}{-6.5\unitlength}%
\settoheight{\eqofftwo}{$\times$}%
\setlength{\eqofftwo}{0.5\eqofftwo}%
\addtolength{\eqofftwo}{-7.5\unitlength}%
\raisebox{\eqoff}{%
\begin{fmfchar*}(\figwidth,\figheight)
\fmfleft{in}
\fmfright{out}
\fmf{plain,tension=1}{in,v1}
\fmfforce{(0,0.5h)}{in}
\fmfforce{(1.1w,0.5h)}{out}
\fmfforce{(0.5w,0)}{v3}
\fmfforce{(0.5w,h)}{v2}
\fmfforce{(0.03w,0.5h)}{v1}
\fmfforce{(w,0.5h)}{v4}
\fmfforce{(1.5w,0.5h)}{v5}
\fmfforce{(1.7w,0.5h)}{v6}
\fmffixed{(0.5w,0)}{v2,v7}
\fmffixed{(0.5w,0)}{v3,v8}
\fmffreeze
\fmf{plain,label=$\alpha$,l.dist=2,l.side=left}{v1,v2}
\fmf{plain,label=$\beta$,l.dist=2,l.side=right}{v1,v3}
\fmf{derplain}{v2,v3}
\fmf{plain}{v2,v7}
\fmf{plain}{v3,v8}
\end{fmfchar*}}\ \  
&= -\frac{1}{2}[\Delta(\alpha,\beta)+\tilde{\Delta}(\alpha,\beta)]\Big(\ \ 
\settoheight{\eqoff}{$\times$}%
\setlength{\eqoff}{0.5\eqoff}%
\addtolength{\eqoff}{-6.5\unitlength}%
\raisebox{\eqoff}{%
\scriptsize
\begin{fmfchar*}(\figwidth,\figheight)
\fmfleft{in}
\fmfright{out}
\fmf{plain,label=$h(\alpha,,\beta)$,label.side=left,label.dist=2,tension=1}{in,v1}
\fmfforce{(0,0.5h)}{in}
\fmfforce{(1.1w,0.5h)}{out}
\fmfforce{(w,0)}{v3}
\fmfforce{(w,h)}{v2}
\fmfforce{(0.5w,0.5h)}{v1}
\fmfforce{(w,0.5h)}{v4}
\fmfforce{(1.5w,0.5h)}{v5}
\fmfforce{(1.7w,0.5h)}{v6}
\fmffixed{(0.5w,0)}{v2,v7}
\fmffixed{(0.5w,0)}{v3,v8}
\fmffreeze
\fmf{plain}{v1,v2}
\fmf{derplain}{v3,v1}
\end{fmfchar*}}\ \ 
\normalsize
+\ \ 
\raisebox{\eqoff}{%
\scriptsize
\begin{fmfchar*}(\figwidth,\figheight)
\fmfleft{in}
\fmfright{out}
\fmf{plain,label=$h(\alpha,,\beta)$,label.side=left,label.dist=2,tension=1}{in,v1}
\fmfforce{(0,0.5h)}{in}
\fmfforce{(1.1w,0.5h)}{out}
\fmfforce{(w,0)}{v3}
\fmfforce{(w,h)}{v2}
\fmfforce{(0.5w,0.5h)}{v1}
\fmfforce{(w,0.5h)}{v4}
\fmfforce{(1.5w,0.5h)}{v5}
\fmfforce{(1.7w,0.5h)}{v6}
\fmffixed{(0.5w,0)}{v2,v7}
\fmffixed{(0.5w,0)}{v3,v8}
\fmffreeze
\fmf{derplain}{v1,v2}
\fmf{plain}{v3,v1}
\end{fmfchar*}}\ \ 
\Big)
\\
\\
&\qquad+\frac{1}{2}\Sigma(\alpha,\beta)\ \ 
\settoheight{\eqoff}{$\times$}%
\setlength{\eqoff}{0.5\eqoff}%
\addtolength{\eqoff}{-6.5\unitlength}%
\raisebox{\eqoff}{%
\scriptsize
\begin{fmfchar*}(\figwidth,\figheight)
\fmfleft{in}
\fmfright{out}
\fmf{derplain,label=$h(\alpha,,\beta)$,label.side=left,label.dist=3,tension=1}{in,v1}
\fmfforce{(0,0.5h)}{in}
\fmfforce{(1.1w,0.5h)}{out}
\fmfforce{(w,0)}{v3}
\fmfforce{(w,h)}{v2}
\fmfforce{(0.5w,0.5h)}{v1}
\fmfforce{(w,0.5h)}{v4}
\fmfforce{(1.5w,0.5h)}{v5}
\fmfforce{(1.7w,0.5h)}{v6}
\fmffixed{(0.5w,0)}{v2,v7}
\fmffixed{(0.5w,0)}{v3,v8}
\fmffreeze
\fmf{plain}{v1,v2}
\fmf{plain}{v3,v1}
\end{fmfchar*}}
\normalsize
\quad+[C(\alpha,\beta)+\tilde{C}(\alpha,\beta)]\ \ 
\settoheight{\eqoff}{$\times$}%
\setlength{\eqoff}{0.5\eqoff}%
\addtolength{\eqoff}{-6.5\unitlength}%
\raisebox{\eqoff}{%
\scriptsize
\begin{fmfchar*}(\figwidth,\figheight)
\fmfleft{in}
\fmfright{out}
\fmf{plain,tension=1}{in,v1}
\fmfforce{(0,0.5h)}{in}
\fmfforce{(1.1w,0.5h)}{out}
\fmfforce{(0.5w,0)}{v3}
\fmfforce{(0.5w,h)}{v2}
\fmfforce{(0.03w,0.5h)}{v1}
\fmfforce{(w,0.5h)}{v4}
\fmfforce{(1.5w,0.5h)}{v5}
\fmfforce{(1.7w,0.5h)}{v6}
\fmffixed{(0.5w,0)}{v2,v7}
\fmffixed{(0.5w,0)}{v3,v8}
\fmffreeze
\fmf{plain,label=$\alpha+1$,l.dist=2,l.side=left}{v1,v2}
\fmf{plain,label=$\beta$,l.dist=2,l.side=right}{v1,v3}
\fmf{derplain}{v2,v3}
\fmf{plain}{v3,v8}
\end{fmfchar*}}\ \ 
\normalsize
\\
\\
&\qquad+[C(\beta,\alpha)+\tilde{C}(\beta,\alpha)]\ \ 
\raisebox{\eqoff}{%
\scriptsize
\begin{fmfchar*}(\figwidth,\figheight)
\fmfleft{in}
\fmfright{out}
\fmf{plain,tension=1}{in,v1}
\fmfforce{(0,0.5h)}{in}
\fmfforce{(1.1w,0.5h)}{out}
\fmfforce{(0.5w,0)}{v3}
\fmfforce{(0.5w,h)}{v2}
\fmfforce{(0.03w,0.5h)}{v1}
\fmfforce{(w,0.5h)}{v4}
\fmfforce{(1.5w,0.5h)}{v5}
\fmfforce{(1.7w,0.5h)}{v6}
\fmffixed{(0.5w,0)}{v2,v7}
\fmffixed{(0.5w,0)}{v3,v8}
\fmffreeze
\fmf{plain,label=$\alpha$,l.dist=2,l.side=left}{v1,v2}
\fmf{plain,label=$\beta+1$,l.dist=2,l.side=right}{v1,v3}
\fmf{derplain}{v2,v3}
\fmf{plain}{v2,v7}
\end{fmfchar*}}
\col
\end{aligned}
\vspace{0.5cm}
\end{equation}
where 
\begin{equation}
\label{rules-def}
\begin{aligned}
\Dt(\alpha,\beta)&=\frac{\D(\alpha,\beta)}{\alpha+\beta+1-\stdim} \col \\
\Sigma(\alpha,\beta)&=\frac{\alpha G(\alpha+1,\beta-1)-(\alpha-\beta)G(\alpha,\beta)-\beta G(\alpha-1,\beta+1)}{\alpha+\beta+1-\stdim} \col \\
\tilde{C}(\alpha,\beta)&=\frac{C(\alpha,\beta)}{\alpha+\beta+1-\stdim}
\pnt
\end{aligned}
\end{equation}

A second, useful rule can be found through the combination of~\eqref{scalar-triangle} and~\eqref{der1-triangle}
\begin{equation}
\label{der2-triangle}
\begin{aligned}
\scriptsize
\settoheight{\eqoff}{$\times$}%
\setlength{\eqoff}{0.5\eqoff}%
\addtolength{\eqoff}{-6.5\unitlength}%
\settoheight{\eqofftwo}{$\times$}%
\setlength{\eqofftwo}{0.5\eqofftwo}%
\addtolength{\eqofftwo}{-7.5\unitlength}%
\raisebox{\eqoff}{%
\begin{fmfchar*}(\figwidth,\figheight)
\fmfleft{in}
\fmfright{out}
\fmf{plain,tension=1}{in,v1}
\fmfforce{(0,0.5h)}{in}
\fmfforce{(1.1w,0.5h)}{out}
\fmfforce{(0.5w,0)}{v3}
\fmfforce{(0.5w,h)}{v2}
\fmfforce{(0.03w,0.5h)}{v1}
\fmfforce{(w,0.5h)}{v4}
\fmfforce{(1.5w,0.5h)}{v5}
\fmfforce{(1.7w,0.5h)}{v6}
\fmffixed{(0.5w,0)}{v2,v7}
\fmffixed{(0.5w,0)}{v3,v8}
\fmffreeze
\fmf{derplain,label=$\alpha$,l.dist=2,l.side=left}{v1,v2}
\fmf{plain,label=$\beta$,l.dist=2,l.side=right}{v1,v3}
\fmf{plain}{v2,v3}
\fmf{plain}{v2,v7}
\fmf{plain}{v3,v8}
\end{fmfchar*}}\ \  
&= \frac{1}{2}\big(\Delta(\alpha,\beta)+\Sigma(\alpha,\beta)\big)\ \ 
\settoheight{\eqoff}{$\times$}%
\setlength{\eqoff}{0.5\eqoff}%
\addtolength{\eqoff}{-6.5\unitlength}%
\raisebox{\eqoff}{%
\scriptsize
\begin{fmfchar*}(\figwidth,\figheight)
\fmfleft{in}
\fmfright{out}
\fmf{derplain,label=$h(\alpha,,\beta)$,label.side=left,label.dist=2,tension=1}{in,v1}
\fmfforce{(0,0.5h)}{in}
\fmfforce{(1.1w,0.5h)}{out}
\fmfforce{(w,0)}{v3}
\fmfforce{(w,h)}{v2}
\fmfforce{(0.5w,0.5h)}{v1}
\fmfforce{(w,0.5h)}{v4}
\fmfforce{(1.5w,0.5h)}{v5}
\fmfforce{(1.7w,0.5h)}{v6}
\fmffixed{(0.5w,0)}{v2,v7}
\fmffixed{(0.5w,0)}{v3,v8}
\fmffreeze
\fmf{plain}{v1,v2}
\fmf{plain}{v3,v1}
\end{fmfchar*}}
\\
\normalsize
&\qquad- \ \frac{1}{2}\tilde{\Delta}(\alpha,\beta)\Big(\ \ 
\settoheight{\eqoff}{$\times$}%
\setlength{\eqoff}{0.5\eqoff}%
\addtolength{\eqoff}{-6.5\unitlength}%
\raisebox{\eqoff}{%
\scriptsize
\begin{fmfchar*}(\figwidth,\figheight)
\fmfleft{in}
\fmfright{out}
\fmf{plain,label=$h(\alpha,,\beta)$,label.side=left,label.dist=2,tension=1}{in,v1}
\fmfforce{(0,0.5h)}{in}
\fmfforce{(1.1w,0.5h)}{out}
\fmfforce{(w,0)}{v3}
\fmfforce{(w,h)}{v2}
\fmfforce{(0.5w,0.5h)}{v1}
\fmfforce{(w,0.5h)}{v4}
\fmfforce{(1.5w,0.5h)}{v5}
\fmfforce{(1.7w,0.5h)}{v6}
\fmffixed{(0.5w,0)}{v2,v7}
\fmffixed{(0.5w,0)}{v3,v8}
\fmffreeze
\fmf{derplain}{v1,v2}
\fmf{plain}{v3,v1}
\end{fmfchar*}}\ \ 
\normalsize
+ \ \  
\settoheight{\eqoff}{$\times$}%
\setlength{\eqoff}{0.5\eqoff}%
\addtolength{\eqoff}{-6.5\unitlength}%
\raisebox{\eqoff}{%
\scriptsize
\begin{fmfchar*}(\figwidth,\figheight)
\fmfleft{in}
\fmfright{out}
\fmf{plain,label=$h(\alpha,,\beta)$,label.side=left,label.dist=2,tension=1}{in,v1}
\fmfforce{(0,0.5h)}{in}
\fmfforce{(1.1w,0.5h)}{out}
\fmfforce{(w,0)}{v3}
\fmfforce{(w,h)}{v2}
\fmfforce{(0.5w,0.5h)}{v1}
\fmfforce{(w,0.5h)}{v4}
\fmfforce{(1.5w,0.5h)}{v5}
\fmfforce{(1.7w,0.5h)}{v6}
\fmffixed{(0.5w,0)}{v2,v7}
\fmffixed{(0.5w,0)}{v3,v8}
\fmffreeze
\fmf{plain}{v1,v2}
\fmf{derplain}{v3,v1}
\end{fmfchar*}}
\ \ \Big)
\\
\\
\normalsize
&\qquad+\tilde{C}(\alpha,\beta)\ \ 
\settoheight{\eqoff}{$\times$}%
\setlength{\eqoff}{0.5\eqoff}%
\addtolength{\eqoff}{-6.5\unitlength}%
\raisebox{\eqoff}{%
\scriptsize
\begin{fmfchar*}(\figwidth,\figheight)
\fmfleft{in}
\fmfright{out}
\fmf{plain,tension=1}{in,v1}
\fmfforce{(0,0.5h)}{in}
\fmfforce{(1.1w,0.5h)}{out}
\fmfforce{(0.5w,0)}{v3}
\fmfforce{(0.5w,h)}{v2}
\fmfforce{(0.03w,0.5h)}{v1}
\fmfforce{(w,0.5h)}{v4}
\fmfforce{(1.5w,0.5h)}{v5}
\fmfforce{(1.7w,0.5h)}{v6}
\fmffixed{(0.5w,0)}{v2,v7}
\fmffixed{(0.5w,0)}{v3,v8}
\fmffreeze
\fmf{plain,label=$\alpha+1$,l.dist=2,l.side=left}{v1,v2}
\fmf{plain,label=$\beta$,l.dist=2,l.side=right}{v1,v3}
\fmf{derplain}{v2,v3}
\fmf{plain}{v3,v8}
\end{fmfchar*}}\ \ 
\normalsize
+\tilde{C}(\beta,\alpha)\ \ 
\raisebox{\eqoff}{%
\scriptsize
\begin{fmfchar*}(\figwidth,\figheight)
\fmfleft{in}
\fmfright{out}
\fmf{plain,tension=1}{in,v1}
\fmfforce{(0,0.5h)}{in}
\fmfforce{(1.1w,0.5h)}{out}
\fmfforce{(0.5w,0)}{v3}
\fmfforce{(0.5w,h)}{v2}
\fmfforce{(0.03w,0.5h)}{v1}
\fmfforce{(w,0.5h)}{v4}
\fmfforce{(1.5w,0.5h)}{v5}
\fmfforce{(1.7w,0.5h)}{v6}
\fmffixed{(0.5w,0)}{v2,v7}
\fmffixed{(0.5w,0)}{v3,v8}
\fmffreeze
\fmf{plain,label=$\alpha$,l.dist=2,l.side=left}{v1,v2}
\fmf{plain,label=$\beta+1$,l.dist=2,l.side=right}{v1,v3}
\fmf{derplain}{v2,v3}
\fmf{plain}{v2,v7}
\end{fmfchar*}}
\\
\\
\normalsize
&\qquad+C(\alpha,\beta)\ \ 
\settoheight{\eqoff}{$\times$}%
\setlength{\eqoff}{0.5\eqoff}%
\addtolength{\eqoff}{-6.5\unitlength}%
\raisebox{\eqoff}{%
\scriptsize
\begin{fmfchar*}(\figwidth,\figheight)
\fmfleft{in}
\fmfright{out}
\fmf{plain,tension=1}{in,v1}
\fmfforce{(0,0.5h)}{in}
\fmfforce{(1.1w,0.5h)}{out}
\fmfforce{(0.5w,0)}{v3}
\fmfforce{(0.5w,h)}{v2}
\fmfforce{(0.03w,0.5h)}{v1}
\fmfforce{(w,0.5h)}{v4}
\fmfforce{(1.5w,0.5h)}{v5}
\fmfforce{(1.7w,0.5h)}{v6}
\fmffixed{(0.5w,0)}{v2,v7}
\fmffixed{(0.5w,0)}{v3,v8}
\fmffreeze
\fmf{derplain,label=$\alpha+1$,l.dist=2,l.side=left}{v1,v2}
\fmf{plain,label=$\beta$,l.dist=2,l.side=right}{v1,v3}
\fmf{plain}{v2,v3}
\fmf{plain}{v3,v8}
\end{fmfchar*}}\ \ 
\normalsize
+C(\beta,\alpha)\ \ 
\raisebox{\eqoff}{%
\scriptsize
\begin{fmfchar*}(\figwidth,\figheight)
\fmfleft{in}
\fmfright{out}
\fmf{plain,tension=1}{in,v1}
\fmfforce{(0,0.5h)}{in}
\fmfforce{(1.1w,0.5h)}{out}
\fmfforce{(0.5w,0)}{v3}
\fmfforce{(0.5w,h)}{v2}
\fmfforce{(0.03w,0.5h)}{v1}
\fmfforce{(w,0.5h)}{v4}
\fmfforce{(1.5w,0.5h)}{v5}
\fmfforce{(1.7w,0.5h)}{v6}
\fmffixed{(0.5w,0)}{v2,v7}
\fmffixed{(0.5w,0)}{v3,v8}
\fmffreeze
\fmf{derplain,label=$\alpha$,l.dist=2,l.side=left}{v1,v2}
\fmf{plain,label=$\beta+1$,l.dist=2,l.side=right}{v1,v3}
\fmf{plain}{v2,v3}
\fmf{plain}{v2,v7}
\end{fmfchar*}}
\pnt
\end{aligned}
\vspace{0.5cm}
\end{equation}
In all the triangle rules, the cancellation of a line correspond to the shrinking of the corresponding propagator to a point, so that in the end the two vertices originally joined by the cancelled line coincide.

\section{Application to single-impurity integrals}
The generalized triangle rules derived in the previous section can be used to find the values of the integrals $\sint{j}{L}$ of Figure~\ref{integrals}, which are needed to study single-impurity operators in the deformed theory. Note that these integrals have the same topology as the ones considered in~\cite{Broadhurst:1985vq} but with the additional presence of the scalar product of a pair of momenta in the numerator. 
Thanks to this special topology, the triangle identities are particularly effective, because their application always results in the cancellation of a loop.
Moreover, the computation is simplified by the fact that the integrals are free of subdivergences.

Consider the more general integral $\tsint{j}{L}(\alpha_1,\ldots,\alpha_L)$, with the same topology and numerator as $\sint{j}{L}$ but with propagator weights $\alpha_1,\ldots\alpha_L$ on the radial lines, so that
\begin{equation}
\sint{j}{L}=\tsint{j}{L}(1,\ldots,1) \pnt
\end{equation}
Take the propagator weights in the form 
\begin{equation}
\label{weights}
\alpha_i=1+(a_i-1)\varepsilon \col
\end{equation}
with integer values of the $a_i$. 
When the standard triangle rule~\eqref{scalar-triangle} is applied to any of the triangles of $\tsint{j}{L}(\alpha_1,\ldots,\alpha_L)$ with vertices $(0,n-1,n)$, with $j+3\leq n\leq L-1$, three terms are generated: 
\begin{equation}
\begin{aligned}
{}&\tsint{j}{L}(\alpha_1,\ldots,\alpha_L)=\Delta(\alpha_{n-1},\alpha_{n})\tsint{j}{L-1}(\alpha_1,\ldots,\alpha_{n-2},h(a_{n-1},\alpha_{n}),\alpha_{n+1},\ldots,\alpha_{L})\\
&\qquad\quad+C(\alpha_{n-1},\alpha_{n})G(\alpha_{n-2},\alpha_{n-1}+1)\tsint{j}{L-1}(\alpha_1,\ldots,g(\alpha_{n-2},\alpha_{n-1}+1),\alpha_{n},\ldots,\alpha_L)\\
&\qquad\quad+C(\alpha_{n},\alpha_{n-1})G(\alpha_{n}+1,\alpha_{n+1})\tsint{j}{L-1}(\alpha_1,\ldots,\alpha_{n-1},g(\alpha_{n}+1,\alpha_{n+1},\ldots,\alpha_L))
\pnt
\end{aligned}
\end{equation}
In all of them the chosen triangle has been cancelled, possibly after the application of~\eqref{Gdef} which introduces a $G(\alpha,\beta)$ factor, and therefore they all reduce to $(L-1)$-loop integrals of the form $\tsint{j}{L-1}(\beta_1,\ldots,\beta_{L-1})$, all with the same structure as the original $L$-loop one and where the new propagator weights $\beta_i$ are still of the form~\eqref{weights}.
A similar situation occurs when the rule is applied to one of the triangles with $3\leq n\leq j-2$, with the single difference that in the new integrals the arrow is now on line $j-1$. 
Hence, by recursively applying the triangle rule~\eqref{scalar-triangle} on both sides of the radial line with the arrow, the original $L$-loop integral can be reduced to a combination of lower-loop ones. This procedure stops when the standard rule cannot be applied any longer without modifying the topology of the integrals or introducing a $G^{(1)}(\alpha,\beta)$ function. It is easy to see that in the most general case, where $3<j<L-2$, this happens for seven-loop integrals, whereas in the particular cases $j\leq3$ and $j\geq L-2$ the final number of loops is even lower. Thus, any integral $\sint{j}{L}$ can be written in terms of the same \emph{finite} set of master integrals with at most seven loops and arbitrary propagator weights. These master integrals can be fully determined by means of the generalized triangle identities, and the results shown in Table~\ref{table:masterint} are obtained. 

Once the master integrals have been found as functions of the propagator weights, the method is particularly suitable for a computer implementation. In principle, the computation can be performed for any vales of $L$ and $j$. The actual time and memory requirements may however be huge, in view of the two main drawbacks of the method: the first one is the fact that every application of the triangle rule produces three terms, so that for integrals with a high number of loops the total number of contributions grows exponentially. The second one is the fact that this procedure gives the \emph{exact} value of the integral, requiring a series expansion of the final result to extract the divergent part.

\begin{table}
\capstart
\footnotesize
\begin{align*}
&F(\alpha,\beta)=
\scriptsize
\settoheight{\eqoff}{$\times$}%
\setlength{\eqoff}{0.5\eqoff}%
\addtolength{\eqoff}{-7\unitlength}%
\raisebox{\eqoff}{%
\fmfframe(0,0)(0,0){%
\begin{fmfchar*}(18,14)
\fmfleft{in}
\fmfright{out}
\fmf{plain,tension=1}{in,v1}
\fmf{plain,tension=1}{v4,out}
\fmfforce{(0,0.5h)}{in}
\fmfforce{(1.1w,0.5h)}{out}
\fmfforce{(0.5w,0)}{v3}
\fmfforce{(0.5w,h)}{v2}
\fmfforce{(0.03w,0.5h)}{v1}
\fmfforce{(0.97w,0.5h)}{v4}
\fmfforce{(1.5w,0.5h)}{v5}
\fmfforce{(1.7w,0.5h)}{v6}
\fmffreeze
\fmf{plain,label=$\alpha$,l.side=left,l.dist=3}{v1,v2}
\fmf{plain,label=$\beta$,l.dist=3}{v1,v3}
\fmf{plain}{v3,v4}
\fmf{plain}{v2,v4}
\fmf{plain}{v2,v3}
\end{fmfchar*}}}\ \ \ 
\footnotesize
=G(1,1)[\D(\a,\b)+C(\a,\b)G(\a+1,\b+\e)+C(\b,\a)G(\a+\e,\b+1)]
\\
&\rint{1}{2}(\alpha,\beta)=
\scriptsize
\settoheight{\eqoff}{$\times$}%
\setlength{\eqoff}{0.5\eqoff}%
\addtolength{\eqoff}{-7\unitlength}%
\raisebox{\eqoff}{%
\fmfframe(0,0)(0,0){%
\begin{fmfchar*}(18,14)
\fmfleft{in}
\fmfright{out}
\fmf{plain,tension=1}{in,v1}
\fmf{derplain,tension=1}{v4,out}
\fmfforce{(0,0.5h)}{in}
\fmfforce{(1.1w,0.5h)}{out}
\fmfforce{(0.5w,0)}{v3}
\fmfforce{(0.5w,h)}{v2}
\fmfforce{(0.03w,0.5h)}{v1}
\fmfforce{(0.97w,0.5h)}{v4}
\fmfforce{(1.5w,0.5h)}{v5}
\fmfforce{(1.7w,0.5h)}{v6}
\fmffreeze
\fmf{plain,label=$\alpha$,l.side=left,l.dist=3}{v1,v2}
\fmf{plain,label=$\beta$,l.dist=3}{v1,v3}
\fmf{plain}{v3,v4}
\fmf{plain}{v2,v4}
\fmf{derplain}{v2,v3}
\end{fmfchar*}}}\ \ \ 
\footnotesize
\begin{aligned}
&\phantom{\gu}\\
&=\frac{1}{2}\S(\a,\b)G(1,1)-[C(\a,\b)+\tilde{C}(\a,\b)]G^{(1)}(1,1)G^{(1)}(\b+\e,\a+1) \\
&\qquad+[C(\b,\a)+\tilde{C}(\b,\a)]G^{(1)}(1,1)G^{(1)}(\a+\e,\b+1)
\end{aligned}
\\[-0.7cm]
&\rint{2}{2}(\alpha,\beta)=
\scriptsize
\settoheight{\eqoff}{$\times$}%
\setlength{\eqoff}{0.5\eqoff}%
\addtolength{\eqoff}{-7\unitlength}%
\raisebox{\eqoff}{%
\fmfframe(0,0)(0,0){%
\begin{fmfchar*}(18,14)
\fmfleft{in}
\fmfright{out}
\fmf{plain,tension=1}{in,v1}
\fmf{derplain,tension=1}{v4,out}
\fmfforce{(0,0.5h)}{in}
\fmfforce{(1.1w,0.5h)}{out}
\fmfforce{(0.5w,0)}{v3}
\fmfforce{(0.5w,h)}{v2}
\fmfforce{(0.03w,0.5h)}{v1}
\fmfforce{(0.97w,0.5h)}{v4}
\fmfforce{(1.5w,0.5h)}{v5}
\fmfforce{(1.7w,0.5h)}{v6}
\fmffreeze
\fmf{derplain,label=$\alpha$,l.side=left,l.dist=3}{v1,v2}
\fmf{plain,label=$\beta$,l.dist=3}{v1,v3}
\fmf{plain}{v3,v4}
\fmf{plain}{v2,v4}
\fmf{plain}{v2,v3}
\end{fmfchar*}}}\ \ \ 
\footnotesize
\begin{aligned}
&\phantom{\gu}\\
&\phantom{\gu}\\
&=\frac{1}{2}[\D(\a,\b)+\S(\a,\b)]G(1,1)+C(\a,\b)G(1,1)\gu(\a+1,\b+\e) \\
&\quad+C(\b,\a)G(1,1)\gu(\a+\e,\b+1)-\Ct(\a,\b)\gu(1,1)\gu(\b+\e,\a+1) \\
&\quad+\Ct(\b,\a)\gu(1,1)\gu(\a+\e,\b+1)
\end{aligned}
\\[-2cm]
&\rint{3}{2}(\alpha,\beta)=
\scriptsize
\settoheight{\eqoff}{$\times$}%
\setlength{\eqoff}{0.5\eqoff}%
\addtolength{\eqoff}{-7\unitlength}%
\raisebox{\eqoff}{%
\fmfframe(0,0)(0,0){%
\begin{fmfchar*}(18,14)
\fmfleft{in}
\fmfright{out}
\fmf{plain,tension=1}{in,v1}
\fmf{plain,tension=1}{v4,out}
\fmfforce{(0,0.5h)}{in}
\fmfforce{(w,0.5h)}{out}
\fmfforce{(0.5w,0)}{v3}
\fmfforce{(0.5w,h)}{v2}
\fmfforce{(0.03w,0.5h)}{v1}
\fmfforce{(0.97w,0.5h)}{v4}
\fmfforce{(1.5w,0.5h)}{v5}
\fmfforce{(1.7w,0.5h)}{v6}
\fmffreeze
\fmf{plain,label=$\alpha$,l.side=left,l.dist=3}{v1,v2}
\fmf{plain,label=$\beta$,l.dist=3}{v1,v3}
\fmf{plain}{v3,v4}
\fmf{derplain}{v2,v4}
\fmf{derplain}{v3,v2}
\end{fmfchar*}}}\ \ \ 
\footnotesize
\begin{aligned}
&\phantom{\frac{1}{2}}\\
&\phantom{\frac{1}{2}}\\
&\phantom{\gu}\\
&=-\frac{1}{4}[\D(\a,\b)+\Dt(\a,\b)]G(1,1)-\frac{1}{2}\S(\a,\b)\gu(1,1) \\
&\quad-\frac{1}{2}[C(\b,\a)+\Ct(\b,\a)]G(1,1)G(\a-1+\e,\b+1)\\
&\quad+[C(\a,\b)+\Ct(\a,\b)][\gu(1,1)\gu(\b+\e,\a+1) \\
&\quad-\frac{1}{2}G(1,1)G(\a+1,\b-1+\e)]\\
\end{aligned}
\\
&\rint{4}{2}(\alpha,\beta)=
\scriptsize
\settoheight{\eqoff}{$\times$}%
\setlength{\eqoff}{0.5\eqoff}%
\addtolength{\eqoff}{-7\unitlength}%
\raisebox{\eqoff}{%
\fmfframe(0,0)(0,0){%
\begin{fmfchar*}(18,14)
\fmfleft{in}
\fmfright{out}
\fmf{plain,tension=1}{in,v1}
\fmf{plain,tension=1}{v4,out}
\fmfforce{(0,0.5h)}{in}
\fmfforce{(w,0.5h)}{out}
\fmfforce{(0.5w,0)}{v3}
\fmfforce{(0.5w,h)}{v2}
\fmfforce{(0.03w,0.5h)}{v1}
\fmfforce{(0.97w,0.5h)}{v4}
\fmfforce{(1.5w,0.5h)}{v5}
\fmfforce{(1.7w,0.5h)}{v6}
\fmffreeze
\fmf{derplain,label=$\alpha$,l.side=right,l.dist=3}{v2,v1}
\fmf{plain,label=$\beta$,l.dist=3}{v1,v3}
\fmf{plain}{v3,v4}
\fmf{derplain}{v2,v4}
\fmf{plain}{v3,v2}
\end{fmfchar*}}}\ \ \
\footnotesize
=\rint{3}{2}(\a,\b)-G(1,1)G(\a,\b+\e)
\\[0.5cm]
&\rint{1}{3}(\alpha,\beta,\g)=
\scriptsize
\settoheight{\eqoff}{$\times$}%
\setlength{\eqoff}{0.5\eqoff}%
\addtolength{\eqoff}{-7\unitlength}%
\raisebox{\eqoff}{%
\fmfframe(0,0)(0,0){%
\begin{fmfchar*}(18,14)
\fmfleft{in}
\fmfright{out}
\fmf{plain,tension=1}{in,v1}
\fmf{plain,tension=1}{v2,out}
\fmfforce{(0,0.5h)}{in}
\fmfforce{(0.5w,1.05h)}{out}
\fmfforce{(0.5w,0)}{v3}
\fmfforce{(0.5w,h)}{v2}
\fmfforce{(0.03w,0.5h)}{v1}
\fmfforce{(0.97w,0.5h)}{v4}
\fmfforce{(0.5w,0.5h)}{v5}
\fmffreeze
\fmf{plain,label=$\alpha$,l.side=right,l.dist=2}{v2,v1}
\fmf{plain}{v1,v5}
\fmf{plain,label=$\g$,l.dist=2}{v1,v3}
\fmf{phantom}{v3,v4}
\fmf{phantom}{v2,v4}
\fmf{plain}{v3,v2}
\fmf{plain,right=0.5}{v3,v2}
\fmfipath{p[]}
\fmfiset{p1}{vpath(__v1,__v5)}
\fmfipair{w[]}
\fmfiequ{w1}{point 5length(p1)/7 of p1}
\fmfiv{l=$\beta$,l.a=90,l.d=1}{w1}
\end{fmfchar*}}}\ \ \
\footnotesize
\begin{aligned}
&\phantom{A}\\
&\phantom{A}\\
&=\D(\b,\g)G(1,1)G(\a,f(\b,\g)+\e)\\
&\quad+C(\b,\g)G(1,1)G(\a,\b+1)G(g(\a,\b+1),\g+\e)\\
&\quad+C(\g,\b)G(1,1)G(\a,\g+1)G(\b+\e,g(\a,\g+1))
\end{aligned}
\\[0.0cm]
&\rint{2}{3}(\alpha,\beta,\g)=
\scriptsize
\settoheight{\eqoff}{$\times$}%
\setlength{\eqoff}{0.5\eqoff}%
\addtolength{\eqoff}{-7\unitlength}%
\raisebox{\eqoff}{%
\fmfframe(0,0)(0,0){%
\begin{fmfchar*}(18,14)
\fmfleft{in}
\fmfright{out}
\fmf{plain,tension=1}{in,v1}
\fmf{plain,tension=1}{v4,out}
\fmfforce{(0,0.5h)}{in}
\fmfforce{(w,0.5h)}{out}
\fmfforce{(0.5w,0)}{v3}
\fmfforce{(0.5w,h)}{v2}
\fmfforce{(0.03w,0.5h)}{v1}
\fmfforce{(0.97w,0.5h)}{v4}
\fmfforce{(0.5w,0.5h)}{v5}
\fmffreeze
\fmf{plain,label=$\alpha$,l.side=right,l.dist=2}{v2,v1}
\fmf{plain}{v1,v5}
\fmf{plain,label=$\g$,l.dist=2}{v1,v3}
\fmf{plain}{v3,v4}
\fmf{plain}{v2,v4}
\fmf{plain}{v3,v2}
\fmfipath{p[]}
\fmfiset{p1}{vpath(__v1,__v5)}
\fmfipair{w[]}
\fmfiequ{w1}{point 5length(p1)/7 of p1}
\fmfiv{l=$\beta$,l.a=90,l.d=1}{w1}
\end{fmfchar*}}}\ \ \
\footnotesize
\begin{aligned}
&\qquad\phantom{\gu}\\
&=\D(\b,\g)F(\a,f(\b,\g))+C(\b,\g)G(\a,\b+1)F(g(\a,\b+1),\g) \\
&\quad+C(\g,\b)\rint{1}{3}(\g+1,\b,\a)\\
\end{aligned}
\end{align*}
\normalsize
\caption{Master integrals required for the computation of $\sint{j}{L}$}
\label{table:masterint}
\end{table}

\addtocounter{table}{-1}
\begin{table}
\footnotesize
\begin{align*}
\\[-6cm]
&\rint{3}{3}(\alpha,\beta,\g)=
\scriptsize
\settoheight{\eqoff}{$\times$}%
\setlength{\eqoff}{0.5\eqoff}%
\addtolength{\eqoff}{-7\unitlength}%
\raisebox{\eqoff}{%
\fmfframe(0,0)(0,0){%
\begin{fmfchar*}(18,14)
\fmfleft{in}
\fmfright{out}
\fmf{plain,tension=1}{in,v1}
\fmf{plain,tension=1}{v3,out}
\fmfforce{(0,0.5h)}{in}
\fmfforce{(0.5w,-0.05h)}{out}
\fmfforce{(0.5w,0)}{v3}
\fmfforce{(0.5w,h)}{v2}
\fmfforce{(0.03w,0.5h)}{v1}
\fmfforce{(0.97w,0.5h)}{v4}
\fmfforce{(0.5w,0.5h)}{v5}
\fmffreeze
\fmf{plain,label=$\alpha$,l.side=right,l.dist=2}{v2,v1}
\fmf{plain}{v1,v5}
\fmf{plain,label=$\g$,l.dist=2}{v1,v3}
\fmf{phantom}{v3,v4}
\fmf{phantom}{v2,v4}
\fmf{plain}{v3,v2}
\fmf{derplain,left=0.5}{v2,v3}
\fmf{derplain}{v5,v2}
\fmfipath{p[]}
\fmfiset{p1}{vpath(__v1,__v5)}
\fmfipair{w[]}
\fmfiequ{w1}{point 5length(p1)/7 of p1}
\fmfiv{l=$\beta$,l.a=90,l.d=1}{w1}
\end{fmfchar*}}}\ \ \
\footnotesize
\begin{aligned}
&\phantom{A}\\
&\phantom{\gu}\\
&\phantom{\frac{1}{2}}\\
&\phantom{\frac{1}{2}}\\
&\phantom{\frac{1}{2}}\\
&\phantom{\frac{1}{2}}\\
&=-\frac{1}{4}[\D(\a,\b)+\Dt(\a,\b)]G(1,1)G(f(\a,\b)-1+\e,\g)\\
&\quad-\frac{1}{2}\S(\a,\b)\gu(1,1)G(f(\a,\b)-1+\e,\g)\\
&\quad+[C(\a,\b)+\Ct(\a,\b)][-\frac{1}{2}G(1,1)G(\a+1,\g)G(g(\a+1,\g),\b-1+\e)\\
&\quad+\gu(1,1)G(\a+1,\g)\gu(\b+\e,g(\a+1,\g))\\
&\quad+\frac{1}{2}\gu(1,1)\gu(\g,\a+1)(G(g(\g,\a+1),\b-1+\e)\\
&\quad+G(g(\g,\a+1)-1,\b+\e)-G(g(\g,\a+1),\b+\e))]\\
&\quad-\frac{1}{2}G(1,1)[C(\b,\a)+\Ct(\b,\a)]G(\b+1,\g)G(\a-1+\e,g(\b+1,\g))
\end{aligned}
\\[-0cm]
&\rint{4}{3}(\alpha,\beta,\g)=
\scriptsize
\settoheight{\eqoff}{$\times$}%
\setlength{\eqoff}{0.5\eqoff}%
\addtolength{\eqoff}{-7\unitlength}%
\raisebox{\eqoff}{%
\fmfframe(0,0)(0,0){%
\begin{fmfchar*}(18,14)
\fmfleft{in}
\fmfright{out}
\fmf{plain,tension=1}{in,v1}
\fmf{plain,tension=1}{v4,out}
\fmfforce{(0,0.5h)}{in}
\fmfforce{(w,0.5h)}{out}
\fmfforce{(0.5w,0)}{v3}
\fmfforce{(0.5w,h)}{v2}
\fmfforce{(0.03w,0.5h)}{v1}
\fmfforce{(0.97w,0.5h)}{v4}
\fmfforce{(0.5w,0.5h)}{v5}
\fmffreeze
\fmf{plain,label=$\alpha$,l.side=right,l.dist=2}{v2,v1}
\fmf{plain}{v1,v5}
\fmf{plain,label=$\g$,l.dist=2}{v1,v3}
\fmf{plain}{v3,v4}
\fmf{derplain}{v2,v4}
\fmf{plain}{v3,v2}
\fmf{derplain}{v5,v2}
\fmfipath{p[]}
\fmfiset{p1}{vpath(__v1,__v5)}
\fmfipair{w[]}
\fmfiequ{w1}{point 5length(p1)/7 of p1}
\fmfiv{l=$\beta$,l.a=90,l.d=1}{w1}
\end{fmfchar*}}}\ \ \
\footnotesize
\begin{aligned}
&\phantom{\gu}\\
&=\D(\b,\g)\rint{3}{2}(\a,f(\b,\g))+C(\b,\g)[\gu(\a,\b+1)\rint{4}{2}(g(\a,\b+1),\g)\\
&\quad+G(1,1)G(\a,\b+1)G(g(\a,\b+1),\g+\e)]
+C(\g,\b)\rint{3}{3}(\a,\b,\g+1)
\end{aligned}
\\
&\rint{5}{3}(\alpha,\beta,\g)=
\scriptsize
\settoheight{\eqoff}{$\times$}%
\setlength{\eqoff}{0.5\eqoff}%
\addtolength{\eqoff}{-7\unitlength}%
\raisebox{\eqoff}{%
\fmfframe(0,0)(0,0){%
\begin{fmfchar*}(18,14)
\fmfleft{in}
\fmfright{out}
\fmf{plain,tension=1}{in,v1}
\fmf{derplain,tension=1}{v2,out}
\fmfforce{(0,0.5h)}{in}
\fmfforce{(0.5w,1.1h)}{out}
\fmfforce{(0.5w,0)}{v3}
\fmfforce{(0.5w,h)}{v2}
\fmfforce{(0.03w,0.5h)}{v1}
\fmfforce{(0.97w,0.5h)}{v4}
\fmfforce{(0.5w,0.5h)}{v5}
\fmffreeze
\fmf{plain,label=$\alpha$,l.side=right,l.dist=2}{v2,v1}
\fmf{plain}{v1,v5}
\fmf{plain,label=$\g$,l.dist=2}{v1,v3}
\fmf{phantom}{v3,v4}
\fmf{phantom}{v2,v4}
\fmf{plain}{v3,v2}
\fmf{derplain}{v5,v3}
\fmf{plain,right=0.5}{v3,v2}
\fmfipath{p[]}
\fmfiset{p1}{vpath(__v1,__v5)}
\fmfipair{w[]}
\fmfiequ{w1}{point 5length(p1)/7 of p1}
\fmfiv{l=$\beta$,l.a=90,l.d=2}{w1}
\end{fmfchar*}}}\ \ \ 
\footnotesize
\begin{aligned}
&\phantom{\gu}\\
&\phantom{\gu}\\
&=\frac{1}{2}\S(\b,\g)G(1,1)\gu(f(\b,\g)+\e,\a)\\
&\quad-[C(\b,\g)+\Ct(\b,\g)]\gu(1,1)G(\a,\b+1)\gu(\g+\e,g(\a,\b+1))\\
&\quad+[C(\g,\b)+\Ct(\g,\b)]\gu(1,1)G(\a,\g+1)\gu(\b+\e,g(\a,\g+1))
\end{aligned}
\\[-1cm]
&\rint{6}{3}(\alpha,\beta,\g)=
\scriptsize
\settoheight{\eqoff}{$\times$}%
\setlength{\eqoff}{0.5\eqoff}%
\addtolength{\eqoff}{-7\unitlength}%
\raisebox{\eqoff}{%
\fmfframe(0,0)(0,0){%
\begin{fmfchar*}(18,14)
\fmfleft{in}
\fmfright{out}
\fmf{plain,tension=1}{in,v1}
\fmf{derplain,tension=1}{v4,out}
\fmfforce{(0,0.5h)}{in}
\fmfforce{(1.1w,0.5h)}{out}
\fmfforce{(0.5w,0)}{v3}
\fmfforce{(0.5w,h)}{v2}
\fmfforce{(0.03w,0.5h)}{v1}
\fmfforce{(0.97w,0.5h)}{v4}
\fmfforce{(0.5w,0.5h)}{v5}
\fmffreeze
\fmf{plain,label=$\alpha$,l.side=right,l.dist=2}{v2,v1}
\fmf{plain}{v1,v5}
\fmf{plain,label=$\g$,l.dist=2}{v1,v3}
\fmf{plain}{v3,v4}
\fmf{plain}{v2,v4}
\fmf{plain}{v3,v2}
\fmf{derplain}{v5,v3}
\fmfipath{p[]}
\fmfiset{p1}{vpath(__v1,__v5)}
\fmfipair{w[]}
\fmfiequ{w1}{point 5length(p1)/7 of p1}
\fmfiv{l=$\beta$,l.a=90,l.d=1}{w1}
\end{fmfchar*}}}\ \ \
\footnotesize
\begin{aligned}
&\phantom{A}\\
&\phantom{\frac{1}{2}}\\
&\phantom{\gu}\\
&=\D(\a,\b)\rint{1}{2}(f(\a,\b),\g)+C(\a,\b)\rint{5}{3}(\a+1,\b,\g)\\
&\quad+C(\b,\a)[-\gu(\g,\b+1)\rint{2}{2}(g(\g,\b+1),\a)\\
&\quad+G(\g,\b+1)\frac{1}{2}(F(\a,g(\g,\b+1))+G(1,1)G(g(\g,\b+1),\a+\e)\\
&\quad-G(1,1)G(g(\g,\b+1)+\e,\a))]
\end{aligned}
\\
&\rint{7}{3}(\alpha,\beta,\g)=
\scriptsize
\settoheight{\eqoff}{$\times$}%
\setlength{\eqoff}{0.5\eqoff}%
\addtolength{\eqoff}{-7\unitlength}%
\raisebox{\eqoff}{%
\fmfframe(0,0)(0,0){%
\begin{fmfchar*}(18,14)
\fmfleft{in}
\fmfright{out}
\fmf{plain,tension=1}{in,v1}
\fmf{plain,tension=1}{v4,out}
\fmfforce{(0,0.5h)}{in}
\fmfforce{(w,0.5h)}{out}
\fmfforce{(0.5w,0)}{v3}
\fmfforce{(0.5w,h)}{v2}
\fmfforce{(0.03w,0.5h)}{v1}
\fmfforce{(0.97w,0.5h)}{v4}
\fmfforce{(0.5w,0.5h)}{v5}
\fmffreeze
\fmf{plain,label=$\alpha$,l.side=right,l.dist=2}{v2,v1}
\fmf{derplain}{v1,v5}
\fmf{plain,label=$\g$,l.dist=2}{v1,v3}
\fmf{plain}{v3,v4}
\fmf{derplain}{v2,v4}
\fmf{plain}{v3,v2}
\fmf{plain}{v5,v3}
\fmfipath{p[]}
\fmfiset{p1}{vpath(__v1,__v5)}
\fmfipair{w[]}
\fmfiequ{w1}{point 5length(p1)/7 of p1}
\fmfiv{l=$\beta$,l.a=90,l.d=2}{w1}
\end{fmfchar*}}}\ \ \
\footnotesize
=\rint{4}{3}(\a,\b,\g)-\rint{4}{3}(\g,\b,\a)+\rint{6}{3}(\a,\b,\g)\\
\\
&\rint{8}{3}(\alpha,\beta,\g)=
\scriptsize
\settoheight{\eqoff}{$\times$}%
\setlength{\eqoff}{0.5\eqoff}%
\addtolength{\eqoff}{-7\unitlength}%
\raisebox{\eqoff}{%
\fmfframe(0,0)(0,0){%
\begin{fmfchar*}(18,14)
\fmfleft{in}
\fmfright{out}
\fmf{plain,tension=1}{in,v1}
\fmf{plain,tension=1}{v4,out}
\fmfforce{(0,0.5h)}{in}
\fmfforce{(w,0.5h)}{out}
\fmfforce{(0.5w,0)}{v3}
\fmfforce{(0.5w,h)}{v2}
\fmfforce{(0.03w,0.5h)}{v1}
\fmfforce{(0.97w,0.5h)}{v4}
\fmfforce{(0.5w,0.5h)}{v5}
\fmffreeze
\fmf{plain,label=$\alpha$,l.side=right,l.dist=2}{v2,v1}
\fmf{plain}{v1,v5}
\fmf{plain,label=$\g$,l.dist=1}{v1,v3}
\fmf{plain}{v3,v4}
\fmf{derplain}{v2,v4}
\fmf{plain}{v3,v2}
\fmf{derplain}{v5,v3}
\fmfipath{p[]}
\fmfiset{p1}{vpath(__v1,__v5)}
\fmfipair{w[]}
\fmfiequ{w1}{point 5length(p1)/7 of p1}
\fmfiv{l=$\beta$,l.a=90,l.d=1}{w1}
\end{fmfchar*}}}\ \ \
\footnotesize
=-\rint{4}{3}(\g,\b,\a)+\rint{6}{3}(\a,\b,\g)\\
\\
&\rint{9}{3}(\alpha,\beta,\g)=
\scriptsize
\settoheight{\eqoff}{$\times$}%
\setlength{\eqoff}{0.5\eqoff}%
\addtolength{\eqoff}{-7\unitlength}%
\raisebox{\eqoff}{%
\fmfframe(0,0)(0,0){%
\begin{fmfchar*}(18,14)
\fmfleft{in}
\fmfright{out}
\fmf{plain,tension=1}{in,v1}
\fmf{plain,tension=1}{v4,out}
\fmfforce{(0,0.5h)}{in}
\fmfforce{(w,0.5h)}{out}
\fmfforce{(0.5w,0)}{v3}
\fmfforce{(0.5w,h)}{v2}
\fmfforce{(0.03w,0.5h)}{v1}
\fmfforce{(0.97w,0.5h)}{v4}
\fmfforce{(0.5w,0.5h)}{v5}
\fmffreeze
\fmf{derplain,label=$\alpha$,l.side=left,l.dist=2}{v1,v2}
\fmf{plain}{v1,v5}
\fmf{plain,label=$\g$,l.dist=2}{v1,v3}
\fmf{plain}{v3,v4}
\fmf{derplain}{v2,v4}
\fmf{plain}{v3,v2}
\fmf{plain}{v5,v3}
\fmfipath{p[]}
\fmfiset{p1}{vpath(__v1,__v5)}
\fmfipair{w[]}
\fmfiequ{w1}{point 5length(p1)/7 of p1}
\fmfiv{l=$\beta$,l.a=90,l.d=1}{w1}
\end{fmfchar*}}}\ \ \
\footnotesize
=-\rint{4}{3}(\a,\b,\g)+\rint{1}{3}(\a,\b,\g)\\
\\
&\rint{10}{3}(\alpha,\beta,\g)=
\scriptsize
\settoheight{\eqoff}{$\times$}%
\setlength{\eqoff}{0.5\eqoff}%
\addtolength{\eqoff}{-7\unitlength}%
\raisebox{\eqoff}{%
\fmfframe(0,0)(0,0){%
\begin{fmfchar*}(18,14)
\fmfleft{in}
\fmfright{out}
\fmf{plain,tension=1}{in,v1}
\fmf{derplain,tension=1}{v4,out}
\fmfforce{(0,0.5h)}{in}
\fmfforce{(1.1w,0.5h)}{out}
\fmfforce{(0.5w,0)}{v3}
\fmfforce{(0.5w,h)}{v2}
\fmfforce{(0.03w,0.5h)}{v1}
\fmfforce{(0.97w,0.5h)}{v4}
\fmfforce{(0.5w,0.5h)}{v5}
\fmffreeze
\fmf{derplain,label=$\alpha$,l.side=left,l.dist=2}{v1,v2}
\fmf{plain}{v1,v5}
\fmf{plain,label=$\g$,l.dist=2}{v1,v3}
\fmf{plain}{v3,v4}
\fmf{plain}{v2,v4}
\fmf{plain}{v3,v2}
\fmf{plain}{v5,v3}
\fmfipath{p[]}
\fmfiset{p1}{vpath(__v1,__v5)}
\fmfipair{w[]}
\fmfiequ{w1}{point 5length(p1)/7 of p1}
\fmfiv{l=$\beta$,l.a=90,l.d=1}{w1}
\end{fmfchar*}}}\ \ \
\footnotesize
=-\rint{6}{3}(\g,\b,\a)+\frac{1}{2}[\rint{2}{3}(\a,\b,\g)+\rint{1}{3}(\a,\b,\g)-\rint{1}{3}(\g,\b,\a)]
\end{align*}
\normalsize
\caption{Master integrals required for the computation of $\sint{j}{L}$ \textit{(continued)}}
\end{table}

\addtocounter{table}{-1}
\begin{table}
\footnotesize
\begin{align*}
\\[-4cm]
&\rint{1}{4}=
\scriptsize
\settoheight{\eqoff}{$\times$}%
\setlength{\eqoff}{0.5\eqoff}%
\addtolength{\eqoff}{-12\unitlength}%
\raisebox{\eqoff}{%
\fmfframe(0,0)(0,0){%
\begin{fmfchar*}(32,24)
\fmfleft{in}
\fmfright{out}
\fmf{plain,tension=1}{in,v1}
\fmf{plain,tension=1}{v4,out}
\fmfforce{(0,0.5h)}{in}
\fmfforce{(w,0.5h)}{out}
\fmfforce{(0.5w,0)}{v3}
\fmfforce{(0.5w,h)}{v2}
\fmfforce{(0.03w,0.5h)}{v1}
\fmfforce{(0.97w,0.5h)}{v4}
\fmfforce{(0.5w,0.66h)}{v5}
\fmfforce{(0.5w,0.33h)}{v6}
\fmffreeze
\fmf{plain}{v1,v2}
\fmf{derplain}{v1,v5}
\fmf{plain}{v1,v6}
\fmf{plain}{v1,v3}
\fmf{plain}{v3,v4}
\fmf{derplain}{v2,v4}
\fmf{plain}{v3,v2}
\fmfipath{p[]}
\fmfiset{p1}{vpath(__v1,__v2)}
\fmfiset{p2}{vpath(__v1,__v5)}
\fmfiset{p3}{vpath(__v1,__v6)}
\fmfiset{p4}{vpath(__v1,__v3)}
\fmfipair{w[]}
\fmfiequ{w1}{point 3length(p1)/4 of p1}
\fmfiequ{w2}{point 3length(p2)/4 of p2}
\fmfiequ{w3}{point 3length(p3)/4 of p3}
\fmfiequ{w4}{point 3length(p4)/4 of p4}
\fmfiv{l=$\alpha_1$,l.a=90,l.d=4}{w1}
\fmfiv{l=$\alpha_2$,l.a=90,l.d=3}{w2}
\fmfiv{l=$\alpha_3$,l.a=-90,l.d=3}{w3}
\fmfiv{l=$\alpha_4$,l.a=-90,l.d=4}{w4}
\end{fmfchar*}}}\ \ \
\footnotesize
\begin{aligned}
&\phantom{\rint{9}{3}}\\
&\phantom{\rint{9}{3}}\\
&\phantom{\rint{9}{3}}\\
&\phantom{\rint{9}{3}}\\
&\phantom{\frac{1}{2}}\\
&=\frac{1}{2}[\D(\a_2,\a_3)+\S(\a_2,\a_3)]\rint{7}{3}(\a_1,f(\a_2,\a_3),\a_4)\\
&\quad+C(\a_2,\a_3)\gu(\a_2+1,\a_1)\rint{9}{3}(g(\a_2+1,\a_1),\a_3,\a_4)\\
&\quad+C(\a_3,\a_2)G(\a_3+1,\a_4)\rint{7}{3}(\a_1,\a_2,g(\a_3+1,\a_4))\\
&\quad-\frac{1}{2}\Dt(\a_2,\a_3)[\rint{4}{3}(\a_1,f(\a_2,\a_3),\a_4)-\rint{8}{3}(\a_1,f(\a_2,\a_3),\a_4)]\\
&\quad-\Ct(\a_2,\a_3)G(\a_2+1,\a_1)\rint{4}{3}(g(\a_2+1,\a_1),\a_3,\a_4)\\
&\quad+\Ct(\a_3,\a_2)G(\a_3+1,\a_4)\rint{8}{3}(\a_1,\a_2,g(\a_3+1,\a_4))
\end{aligned}
\\
&\rint{2}{4}=
\scriptsize
\settoheight{\eqoff}{$\times$}%
\setlength{\eqoff}{0.5\eqoff}%
\addtolength{\eqoff}{-12\unitlength}%
\raisebox{\eqoff}{%
\fmfframe(0,0)(0,0){%
\begin{fmfchar*}(32,24)
\fmfleft{in}
\fmfright{out}
\fmf{plain,tension=1}{in,v1}
\fmf{plain,tension=1}{v4,out}
\fmfforce{(0,0.5h)}{in}
\fmfforce{(w,0.5h)}{out}
\fmfforce{(0.5w,0)}{v3}
\fmfforce{(0.5w,h)}{v2}
\fmfforce{(0.03w,0.5h)}{v1}
\fmfforce{(0.97w,0.5h)}{v4}
\fmfforce{(0.5w,0.66h)}{v5}
\fmfforce{(0.5w,0.33h)}{v6}
\fmffreeze
\fmf{plain}{v1,v2}
\fmf{plain}{v1,v5}
\fmf{plain}{v1,v6}
\fmf{plain}{v1,v3}
\fmf{plain}{v3,v4}
\fmf{derplain}{v2,v4}
\fmf{plain}{v3,v2}
\fmf{derplain}{v5,v2}
\fmfipath{p[]}
\fmfiset{p1}{vpath(__v1,__v2)}
\fmfiset{p2}{vpath(__v1,__v5)}
\fmfiset{p3}{vpath(__v1,__v6)}
\fmfiset{p4}{vpath(__v1,__v3)}
\fmfipair{w[]}
\fmfiequ{w1}{point 3length(p1)/4 of p1}
\fmfiequ{w2}{point 3length(p2)/4 of p2}
\fmfiequ{w3}{point 3length(p3)/4 of p3}
\fmfiequ{w4}{point 3length(p4)/4 of p4}
\fmfiv{l=$\alpha_1$,l.a=90,l.d=4}{w1}
\fmfiv{l=$\alpha_2$,l.a=90,l.d=3}{w2}
\fmfiv{l=$\alpha_3$,l.a=-90,l.d=3}{w3}
\fmfiv{l=$\alpha_4$,l.a=-90,l.d=4}{w4}
\end{fmfchar*}}}\ \ \
\footnotesize
\begin{aligned}
&\phantom{\rint{1}{3}}\\
&\phantom{\rint{4}{3}}\\
&\phantom{\rint{9}{3}}\\
&=\D(\a_2,\a_3)\rint{4}{3}(\a_1,f(\a_2,\a_3),\a_4)\\
&\quad+C(\a_3,\a_2)G(\a_3+1,\a_4)\rint{4}{3}(\a_1,\a_2,g(\a_3+1,\a_4))\\
&\quad+C(\a_2,\a_3)[-\gu(\a_1,\a_2+1)\rint{9}{3}(g(\a_2+1,\a_1),\a_3,\a_4)\\
&\quad+G(\a_1,\a_2+1)\rint{1}{3}(g(\a_1,\a_2+1),\a_3,\a_4)]
\end{aligned}
\\[-2.5cm]
&\rint{3}{4}=
\scriptsize
\settoheight{\eqoff}{$\times$}%
\setlength{\eqoff}{0.5\eqoff}%
\addtolength{\eqoff}{-12\unitlength}%
\raisebox{\eqoff}{%
\fmfframe(0,0)(0,0){%
\begin{fmfchar*}(32,24)
\fmfleft{in}
\fmfright{out}
\fmf{plain,tension=1}{in,v1}
\fmf{derplain,tension=1}{v4,out}
\fmfforce{(0,0.5h)}{in}
\fmfforce{(1.1w,0.5h)}{out}
\fmfforce{(0.5w,0)}{v3}
\fmfforce{(0.5w,h)}{v2}
\fmfforce{(0.03w,0.5h)}{v1}
\fmfforce{(0.97w,0.5h)}{v4}
\fmfforce{(0.5w,0.66h)}{v5}
\fmfforce{(0.5w,0.33h)}{v6}
\fmffreeze
\fmf{plain}{v1,v2}
\fmf{plain}{v1,v5}
\fmf{derplain}{v6,v1}
\fmf{plain}{v1,v3}
\fmf{plain}{v3,v4}
\fmf{plain}{v2,v4}
\fmf{plain}{v3,v2}
\fmf{plain}{v5,v2}
\fmfipath{p[]}
\fmfiset{p1}{vpath(__v1,__v2)}
\fmfiset{p2}{vpath(__v1,__v5)}
\fmfiset{p3}{vpath(__v6,__v1)}
\fmfiset{p4}{vpath(__v1,__v3)}
\fmfipair{w[]}
\fmfiequ{w1}{point 3length(p1)/4 of p1}
\fmfiequ{w2}{point 3length(p2)/4 of p2}
\fmfiequ{w3}{point length(p3)/4 of p3}
\fmfiequ{w4}{point 3length(p4)/4 of p4}
\fmfiv{l=$\alpha_1$,l.a=90,l.d=4}{w1}
\fmfiv{l=$\alpha_2$,l.a=90,l.d=3}{w2}
\fmfiv{l=$\alpha_3$,l.a=-90,l.d=3}{w3}
\fmfiv{l=$\alpha_4$,l.a=-90,l.d=4}{w4}
\end{fmfchar*}}}\ \ \
\footnotesize
\begin{aligned}
&\phantom{\rint{2}{3}}\\
&\phantom{\rint{2}{3}}\\
&\phantom{\rint{2}{3}}\\
&\phantom{\rint{2}{3}}\\
&\phantom{\rint{2}{3}}\\
&\phantom{\rint{2}{3}}\\
&\phantom{\frac{1}{2}}\\
&=-\frac{1}{2}[\D(\a_3,\a_2)\!+\!\S(\a_3,\a_2)][\rint{6}{3}(\a_4,f(\a_3,\a_2),\a_1)\\
&\quad+\rint{6}{3}(\a_1,f(\a_3,\a_2),\a_4)]\\
&\quad-C(\a_3,\a_2)\gu(\a_3+1,\a_4)\rint{10}{3}(g(\a_3+1,\a_4),\a_2,\a_1)\\
&\quad-C(\a_2,\a_3)G(\a_1,\a_2+1)[\rint{6}{3}(\a_4,\a_3,g(\a_1,\a_2+1))\\
&\quad+\rint{6}{3}(g(\a_1,\a_2+1),\a_3,\a_4)]\\
&\quad+\frac{1}{2}\Dt(\a_3,\a_2)[\rint{6}{3}(\a_1,f(\a_3,\a_2),\a_4)-\rint{6}{3}(\a_4,f(\a_3,\a_2),\a_1)]\\
&\quad+\Ct(\a_3,\a_2)G(\a_3+1,\a_4)\rint{6}{3}(\a_1,\a_2,g(\a_3+1,\a_4))\\
&\quad-\Ct(\a_2,\a_3)G(\a_1,\a_2+1)\rint{6}{3}(\a_4,\a_3,g(\a_1,\a_2+1))
\end{aligned}
\\[-1cm]
&\rint{4}{4}=
\scriptsize
\settoheight{\eqoff}{$\times$}%
\setlength{\eqoff}{0.5\eqoff}%
\addtolength{\eqoff}{-12\unitlength}%
\raisebox{\eqoff}{%
\fmfframe(0,0)(0,0){%
\begin{fmfchar*}(32,24)
\fmfleft{in}
\fmfright{out}
\fmf{plain,tension=1}{in,v1}
\fmf{derplain,tension=1}{v4,out}
\fmfforce{(0,0.5h)}{in}
\fmfforce{(1.1w,0.5h)}{out}
\fmfforce{(0.5w,0)}{v3}
\fmfforce{(0.5w,h)}{v2}
\fmfforce{(0.03w,0.5h)}{v1}
\fmfforce{(0.97w,0.5h)}{v4}
\fmfforce{(0.5w,0.66h)}{v5}
\fmfforce{(0.5w,0.33h)}{v6}
\fmffreeze
\fmf{plain}{v1,v2}
\fmf{plain}{v1,v5}
\fmf{plain}{v1,v6}
\fmf{plain}{v1,v3}
\fmf{plain}{v3,v4}
\fmf{plain}{v2,v4}
\fmf{plain}{v3,v2}
\fmf{derplain}{v6,v3}
\fmfipath{p[]}
\fmfiset{p1}{vpath(__v1,__v2)}
\fmfiset{p2}{vpath(__v1,__v5)}
\fmfiset{p3}{vpath(__v1,__v6)}
\fmfiset{p4}{vpath(__v1,__v3)}
\fmfipair{w[]}
\fmfiequ{w1}{point 3length(p1)/4 of p1}
\fmfiequ{w2}{point 3length(p2)/4 of p2}
\fmfiequ{w3}{point 3length(p3)/4 of p3}
\fmfiequ{w4}{point 3length(p4)/4 of p4}
\fmfiv{l=$\alpha_1$,l.a=90,l.d=4}{w1}
\fmfiv{l=$\alpha_2$,l.a=90,l.d=3}{w2}
\fmfiv{l=$\alpha_3$,l.a=-90,l.d=3}{w3}
\fmfiv{l=$\alpha_4$,l.a=-90,l.d=4}{w4}
\end{fmfchar*}}}\ \ \ 
\footnotesize
\begin{aligned}
&\phantom{\rint{2}{3}}\\
&\phantom{\rint{2}{3}}\\
&\phantom{\rint{2}{3}}\\
&\phantom{\frac{1}{2}}\\
&=\D(\a_2,\a_3)\rint{6}{3}(\a_1,f(\a_2,\a_3),\a_4)\\
&\quad+C(\a_2,\a_3)G(\a_1,\a_2+1)\rint{6}{3}(g(\a_1,\a_2+1),\a_3,\a_4)\\
&\quad+C(\a_3,\a_2)[-\gu(\a_4,\a_3+1)\rint{10}{3}(g(\a_4,\a_3+1),\a_2,\a_1)\\
&\quad+\frac{1}{2}G(\a_4,\a_3+1)(\rint{2}{3}(\a_1,\a_2,g(\a_4,\a_3+1))\\
&\quad+\rint{1}{3}(g(\a_4,\a_3+1),\a_2,\a_1)-\rint{1}{3}(\a_1,\a_2,g(\a_4,\a_3+1)))]
\end{aligned}
\end{align*}
\normalsize
\caption{Master integrals required for the computation of $\sint{j}{L}$ \textit{(continued)}}
\end{table}

\addtocounter{table}{-1}
\begin{table}
\vspace{-1cm}
\footnotesize
\begin{align*}
&\rint{5}{4}=
\scriptsize
\settoheight{\eqoff}{$\times$}%
\setlength{\eqoff}{0.5\eqoff}%
\addtolength{\eqoff}{-12\unitlength}%
\raisebox{\eqoff}{%
\fmfframe(0,0)(0,0){%
\begin{fmfchar*}(32,24)
\fmfleft{in}
\fmfright{out}
\fmf{plain,tension=1}{in,v1}
\fmf{derplain,tension=1}{v4,out}
\fmfforce{(0,0.5h)}{in}
\fmfforce{(1.1w,0.5h)}{out}
\fmfforce{(0.5w,0)}{v3}
\fmfforce{(0.5w,h)}{v2}
\fmfforce{(0.03w,0.5h)}{v1}
\fmfforce{(0.97w,0.5h)}{v4}
\fmfforce{(0.5w,0.66h)}{v5}
\fmfforce{(0.5w,0.33h)}{v6}
\fmffreeze
\fmf{plain}{v1,v2}
\fmf{plain}{v1,v5}
\fmf{plain}{v1,v6}
\fmf{plain}{v1,v3}
\fmf{plain}{v3,v4}
\fmf{plain}{v2,v4}
\fmf{plain}{v3,v2}
\fmf{derplain}{v5,v6}
\fmfipath{p[]}
\fmfiset{p1}{vpath(__v1,__v2)}
\fmfiset{p2}{vpath(__v1,__v5)}
\fmfiset{p3}{vpath(__v1,__v6)}
\fmfiset{p4}{vpath(__v1,__v3)}
\fmfipair{w[]}
\fmfiequ{w1}{point 3length(p1)/4 of p1}
\fmfiequ{w2}{point 3length(p2)/4 of p2}
\fmfiequ{w3}{point 3length(p3)/4 of p3}
\fmfiequ{w4}{point 3length(p4)/4 of p4}
\fmfiv{l=$\alpha_1$,l.a=90,l.d=4}{w1}
\fmfiv{l=$\alpha_2$,l.a=90,l.d=3}{w2}
\fmfiv{l=$\alpha_3$,l.a=-90,l.d=3}{w3}
\fmfiv{l=$\alpha_4$,l.a=-90,l.d=4}{w4}
\end{fmfchar*}}}\ \ \
\footnotesize
=\rint{3}{4}(\a_1,\a_2,\a_3,\a_4)+\rint{4}{4}(\a_1,\a_2,\a_3,\a_4)\\
\\
&\rint{6}{4}=
\scriptsize
\settoheight{\eqoff}{$\times$}%
\setlength{\eqoff}{0.5\eqoff}%
\addtolength{\eqoff}{-12\unitlength}%
\raisebox{\eqoff}{%
\fmfframe(0,0)(0,0){%
\begin{fmfchar*}(32,24)
\fmfleft{in}
\fmfright{out}
\fmf{plain,tension=1}{in,v1}
\fmf{plain,tension=1}{v4,out}
\fmfforce{(0,0.5h)}{in}
\fmfforce{(w,0.5h)}{out}
\fmfforce{(0.5w,0)}{v3}
\fmfforce{(0.5w,h)}{v2}
\fmfforce{(0.03w,0.5h)}{v1}
\fmfforce{(0.97w,0.5h)}{v4}
\fmfforce{(0.5w,0.66h)}{v5}
\fmfforce{(0.5w,0.33h)}{v6}
\fmffreeze
\fmf{derplain}{v1,v2}
\fmf{plain}{v1,v5}
\fmf{plain}{v1,v6}
\fmf{plain}{v1,v3}
\fmf{plain}{v3,v4}
\fmf{derplain}{v2,v4}
\fmf{plain}{v3,v2}
\fmf{plain}{v5,v6}
\fmfipath{p[]}
\fmfiset{p1}{vpath(__v1,__v2)}
\fmfiset{p2}{vpath(__v1,__v5)}
\fmfiset{p3}{vpath(__v1,__v6)}
\fmfiset{p4}{vpath(__v1,__v3)}
\fmfipair{w[]}
\fmfiequ{w1}{point 3length(p1)/4 of p1}
\fmfiequ{w2}{point 3length(p2)/4 of p2}
\fmfiequ{w3}{point 3length(p3)/4 of p3}
\fmfiequ{w4}{point 3length(p4)/4 of p4}
\fmfiv{l=$\alpha_1$,l.a=90,l.d=4}{w1}
\fmfiv{l=$\alpha_2$,l.a=90,l.d=3}{w2}
\fmfiv{l=$\alpha_3$,l.a=-90,l.d=3}{w3}
\fmfiv{l=$\alpha_4$,l.a=-90,l.d=4}{w4}
\end{fmfchar*}}}\ \ \
\footnotesize
\begin{aligned}
&\phantom{\rint{2}{3}}\\
&\phantom{\rint{2}{3}}\\
&=\D(\a_2,\a_3)\rint{9}{3}(\a_1,f(\a_2,\a_3),\a_4)\\
&\quad+C(\a_2,\a_3)\gu(\a_1,\a_2+1)\rint{9}{3}(g(\a_1,\a_2+1),\a_3,\a_4)\\
&\quad+C(\a_3,\a_2)G(\a_3+1,\a_4)\rint{9}{3}(\a_1,\a_2,g(\a_3+1,\a_4))
\end{aligned}
\\
\\
&\rint{1}{5}=
\scriptsize
\settoheight{\eqoff}{$\times$}%
\setlength{\eqoff}{0.5\eqoff}%
\addtolength{\eqoff}{-12\unitlength}%
\raisebox{\eqoff}{%
\fmfframe(0,0)(0,0){%
\begin{fmfchar*}(32,24)
\fmfleft{in}
\fmfright{out}
\fmf{plain,tension=1}{in,v1}
\fmf{plain,tension=1}{v4,out}
\fmfforce{(0,0.5h)}{in}
\fmfforce{(w,0.5h)}{out}
\fmfforce{(0.5w,0)}{v3}
\fmfforce{(0.5w,h)}{v2}
\fmfforce{(0.03w,0.5h)}{v1}
\fmfforce{(0.97w,0.5h)}{v4}
\fmfforce{(0.5w,0.75h)}{v5}
\fmfforce{(0.5w,0.5h)}{v6}
\fmfforce{(0.5w,0.25h)}{v7}
\fmffreeze
\fmf{plain}{v1,v2}
\fmf{derplain}{v1,v5}
\fmf{plain}{v1,v6}
\fmf{plain}{v1,v7}
\fmf{plain}{v1,v3}
\fmf{plain}{v3,v4}
\fmf{derplain}{v2,v4}
\fmf{plain}{v3,v2}
\fmfipath{p[]}
\fmfiset{p1}{vpath(__v1,__v2)}
\fmfiset{p2}{vpath(__v1,__v5)}
\fmfiset{p3}{vpath(__v1,__v6)}
\fmfiset{p4}{vpath(__v1,__v7)}
\fmfiset{p5}{vpath(__v1,__v3)}
\fmfipair{w[]}
\fmfiequ{w1}{point 3length(p1)/4 of p1}
\fmfiequ{w2}{point 3length(p2)/4 of p2}
\fmfiequ{w3}{point 3length(p3)/4 of p3}
\fmfiequ{w4}{point 3length(p4)/4 of p4}
\fmfiequ{w5}{point 3length(p5)/4 of p5}
\fmfiv{l=$\alpha_1$,l.a=90,l.d=4}{w1}
\fmfiv{l=$\alpha_2$,l.a=90,l.d=3}{w2}
\fmfiv{l=$\alpha_3$,l.a=90,l.d=2}{w3}
\fmfiv{l=$\alpha_4$,l.a=-90,l.d=3}{w4}
\fmfiv{l=$\alpha_5$,l.a=-90,l.d=4}{w5}
\end{fmfchar*}}}\ \ \
\footnotesize
\begin{aligned}
&\phantom{\rint{1}{4}}\\
&\phantom{\rint{1}{4}}\\
&=\D(\a_3,\a_4)\rint{1}{4}(\a_1,\a_2,f(\a_3,\a_4),\a_5)\\
&\quad+C(\a_3,\a_4)\gu(\a_2,\a_3+1)\rint{1}{4}(\a_1,g(\a_2,\a_3+1),\a_4,\a_5)\\
&\quad+C(\a_4,\a_3)G(\a_4+1,\a_5)\rint{1}{4}(\a_1,\a_2,\a_3,g(\a_4+1,\a_5))
\end{aligned}
\\
\\[-1cm]
&\rint{2}{5}=
\scriptsize
\settoheight{\eqoff}{$\times$}%
\setlength{\eqoff}{0.5\eqoff}%
\addtolength{\eqoff}{-12\unitlength}%
\raisebox{\eqoff}{%
\fmfframe(0,0)(0,0){%
\begin{fmfchar*}(32,24)
\fmfleft{in}
\fmfright{out}
\fmf{plain,tension=1}{in,v1}
\fmf{plain,tension=1}{v4,out}
\fmfforce{(0,0.5h)}{in}
\fmfforce{(w,0.5h)}{out}
\fmfforce{(0.5w,0)}{v3}
\fmfforce{(0.5w,h)}{v2}
\fmfforce{(0.03w,0.5h)}{v1}
\fmfforce{(0.97w,0.5h)}{v4}
\fmfforce{(0.5w,0.75h)}{v5}
\fmfforce{(0.5w,0.5h)}{v6}
\fmfforce{(0.5w,0.25h)}{v7}
\fmffreeze
\fmf{plain}{v1,v2}
\fmf{plain}{v1,v5}
\fmf{plain}{v1,v6}
\fmf{plain}{v1,v7}
\fmf{plain}{v1,v3}
\fmf{plain}{v3,v4}
\fmf{derplain}{v2,v4}
\fmf{plain}{v2,v3}
\fmf{derplain}{v6,v5}
\fmfipath{p[]}
\fmfiset{p1}{vpath(__v1,__v2)}
\fmfiset{p2}{vpath(__v1,__v5)}
\fmfiset{p3}{vpath(__v1,__v6)}
\fmfiset{p4}{vpath(__v1,__v7)}
\fmfiset{p5}{vpath(__v1,__v3)}
\fmfipair{w[]}
\fmfiequ{w1}{point 3length(p1)/4 of p1}
\fmfiequ{w2}{point 3length(p2)/4 of p2}
\fmfiequ{w3}{point 3length(p3)/4 of p3}
\fmfiequ{w4}{point 3length(p4)/4 of p4}
\fmfiequ{w5}{point 3length(p5)/4 of p5}
\fmfiv{l=$\alpha_1$,l.a=90,l.d=4}{w1}
\fmfiv{l=$\alpha_2$,l.a=90,l.d=3}{w2}
\fmfiv{l=$\alpha_3$,l.a=90,l.d=2}{w3}
\fmfiv{l=$\alpha_4$,l.a=-90,l.d=3}{w4}
\fmfiv{l=$\alpha_5$,l.a=-90,l.d=4}{w5}
\end{fmfchar*}}}\ \ \
\footnotesize
\begin{aligned}
&\phantom{\rint{1}{4}}\\
&\phantom{\rint{1}{4}}\\
&\phantom{\rint{1}{4}}\\
&\phantom{\rint{1}{4}}\\
&=\D(\a_3,\a_4)[\rint{2}{4}(\a_1,\a_2,f(\a_3,\a_4),\a_5)-\rint{1}{4}(\a_1,\a_2,f(\a_3,\a_4),\a_5)]\\
&\quad+C(\a_3,\a_4)[-\gu(\a_2,\a_3+1)\rint{1}{4}(\a_1,g(\a_2,\a_3+1),\a_4,\a_5)\\
&\quad+G(\a_2,\a_3+1)\rint{2}{4}(\a_1,g(\a_2,\a_3+1),\a_4,\a_5)]\\
&\quad+C(\a_4,\a_3)G(\a_4+1,\a_5)[\rint{2}{4}(\a_1,\a_2,\a_3,g(\a_4+1,\a_5))\\
&\quad-\rint{1}{4}(\a_1,\a_2,\a_3,g(\a_4+1,\a_5))]
\end{aligned}
\\
\\[-1cm]
&\rint{3}{5}=
\scriptsize
\settoheight{\eqoff}{$\times$}%
\setlength{\eqoff}{0.5\eqoff}%
\addtolength{\eqoff}{-12\unitlength}%
\raisebox{\eqoff}{%
\fmfframe(0,0)(0,0){%
\begin{fmfchar*}(32,24)
\fmfleft{in}
\fmfright{out}
\fmf{plain,tension=1}{in,v1}
\fmf{derplain,tension=1}{v4,out}
\fmfforce{(0,0.5h)}{in}
\fmfforce{(1.1w,0.5h)}{out}
\fmfforce{(0.5w,0)}{v3}
\fmfforce{(0.5w,h)}{v2}
\fmfforce{(0.03w,0.5h)}{v1}
\fmfforce{(0.97w,0.5h)}{v4}
\fmfforce{(0.5w,0.75h)}{v5}
\fmfforce{(0.5w,0.5h)}{v6}
\fmfforce{(0.5w,0.25h)}{v7}
\fmffreeze
\fmf{plain}{v1,v2}
\fmf{plain}{v1,v5}
\fmf{plain}{v1,v6}
\fmf{plain}{v1,v7}
\fmf{plain}{v1,v3}
\fmf{plain}{v3,v4}
\fmf{plain}{v2,v4}
\fmf{plain}{v2,v3}
\fmf{derplain}{v6,v7}
\fmfipath{p[]}
\fmfiset{p1}{vpath(__v1,__v2)}
\fmfiset{p2}{vpath(__v1,__v5)}
\fmfiset{p3}{vpath(__v1,__v6)}
\fmfiset{p4}{vpath(__v1,__v7)}
\fmfiset{p5}{vpath(__v1,__v3)}
\fmfipair{w[]}
\fmfiequ{w1}{point 3length(p1)/4 of p1}
\fmfiequ{w2}{point 3length(p2)/4 of p2}
\fmfiequ{w3}{point 3length(p3)/4 of p3}
\fmfiequ{w4}{point 3length(p4)/4 of p4}
\fmfiequ{w5}{point 3length(p5)/4 of p5}
\fmfiv{l=$\alpha_1$,l.a=90,l.d=4}{w1}
\fmfiv{l=$\alpha_2$,l.a=90,l.d=3}{w2}
\fmfiv{l=$\alpha_3$,l.a=90,l.d=2}{w3}
\fmfiv{l=$\alpha_4$,l.a=-90,l.d=3}{w4}
\fmfiv{l=$\alpha_5$,l.a=-90,l.d=4}{w5}
\end{fmfchar*}}}\ \ \
\footnotesize
\begin{aligned}
&\phantom{\rint{5}{4}}\\
&\phantom{\rint{5}{4}}\\
&\phantom{\rint{5}{4}}\\
&=\D(\a_2,\a_3)\rint{5}{4}(\a_1,f(\a_2,\a_3),\a_4,\a_5)\\
&\quad+C(\a_2,\a_3)G(\a_1,\a_2+1)\rint{5}{4}(g(\a_1,\a_2+1),\a_3,\a_4,\a_5)\\
&\quad+C(\a_3,\a_2)[\gu(\a_4,\a_3+1)\rint{3}{4}(\a_1,\a_2,g(\a_4,\a_3+1),\a_5)\\
&\quad+G(\a_4,\a_3+1)\rint{4}{4}(\a_1,\a_2,g(\a_4,\a_3+1),\a_5)]
\end{aligned}
\\
&\rint{4}{5}=
\scriptsize
\settoheight{\eqoff}{$\times$}%
\setlength{\eqoff}{0.5\eqoff}%
\addtolength{\eqoff}{-12\unitlength}%
\raisebox{\eqoff}{%
\fmfframe(0,0)(0,0){%
\begin{fmfchar*}(32,24)
\fmfleft{in}
\fmfright{out}
\fmf{plain,tension=1}{in,v1}
\fmf{plain,tension=1}{v4,out}
\fmfforce{(0,0.5h)}{in}
\fmfforce{(w,0.5h)}{out}
\fmfforce{(0.5w,0)}{v3}
\fmfforce{(0.5w,h)}{v2}
\fmfforce{(0.03w,0.5h)}{v1}
\fmfforce{(0.97w,0.5h)}{v4}
\fmfforce{(0.5w,0.75h)}{v5}
\fmfforce{(0.5w,0.5h)}{v6}
\fmfforce{(0.5w,0.25h)}{v7}
\fmffreeze
\fmf{plain}{v1,v2}
\fmf{plain}{v1,v5}
\fmf{derplain}{v1,v6}
\fmf{plain}{v1,v7}
\fmf{plain}{v1,v3}
\fmf{plain}{v3,v4}
\fmf{derplain}{v2,v4}
\fmf{plain}{v3,v2}
\fmfipath{p[]}
\fmfiset{p1}{vpath(__v1,__v2)}
\fmfiset{p2}{vpath(__v1,__v5)}
\fmfiset{p3}{vpath(__v1,__v6)}
\fmfiset{p4}{vpath(__v1,__v7)}
\fmfiset{p5}{vpath(__v1,__v3)}
\fmfipair{w[]}
\fmfiequ{w1}{point 3length(p1)/4 of p1}
\fmfiequ{w2}{point 3length(p2)/4 of p2}
\fmfiequ{w3}{point 3length(p3)/4 of p3}
\fmfiequ{w4}{point 3length(p4)/4 of p4}
\fmfiequ{w5}{point 3length(p5)/4 of p5}
\fmfiv{l=$\alpha_1$,l.a=90,l.d=4}{w1}
\fmfiv{l=$\alpha_2$,l.a=90,l.d=3}{w2}
\fmfiv{l=$\alpha_3$,l.a=90,l.d=2}{w3}
\fmfiv{l=$\alpha_4$,l.a=-90,l.d=3}{w4}
\fmfiv{l=$\alpha_5$,l.a=-90,l.d=4}{w5}
\end{fmfchar*}}}\ \ \
\footnotesize
\begin{aligned}
&\phantom{\rint{1}{4}}\\
&=\rint{2}{5}(\a_1,\a_2,\a_3,\a_4,\a_5)-\rint{2}{5}(\a_5,\a_4,\a_3,\a_2,\a_1)\\
&\quad+\rint{3}{5}(\a_1,\a_2,\a_3,\a_4,\a_5)\\
\end{aligned}
\end{align*}
\normalsize
\caption{Master integrals required for the computation of $\sint{j}{L}$ \textit{(continued)}}
\end{table}

\addtocounter{table}{-1}
\begin{table}
\footnotesize
\begin{align*}
&\rint{1}{6}=
\scriptsize
\settoheight{\eqoff}{$\times$}%
\setlength{\eqoff}{0.5\eqoff}%
\addtolength{\eqoff}{-14\unitlength}%
\raisebox{\eqoff}{%
\fmfframe(0,0)(0,0){%
\begin{fmfchar*}(38,29)
\fmfleft{in}
\fmfright{out}
\fmf{plain,tension=1}{in,v1}
\fmf{plain,tension=1}{v4,out}
\fmfforce{(0,0.5h)}{in}
\fmfforce{(w,0.5h)}{out}
\fmfforce{(0.5w,0)}{v3}
\fmfforce{(0.5w,h)}{v2}
\fmfforce{(0.03w,0.5h)}{v1}
\fmfforce{(0.97w,0.5h)}{v4}
\fmfforce{(0.5w,0.8h)}{v5}
\fmfforce{(0.5w,0.6h)}{v6}
\fmfforce{(0.5w,0.4h)}{v7}
\fmfforce{(0.5w,0.2h)}{v8}
\fmffreeze
\fmf{plain}{v1,v2}
\fmf{plain}{v1,v5}
\fmf{derplain}{v1,v6}
\fmf{plain}{v1,v7}
\fmf{plain}{v1,v8}
\fmf{plain}{v1,v3}
\fmf{plain}{v3,v4}
\fmf{derplain}{v2,v4}
\fmf{plain}{v2,v3}
\fmfipath{p[]}
\fmfiset{p1}{vpath(__v1,__v2)}
\fmfiset{p2}{vpath(__v1,__v5)}
\fmfiset{p3}{vpath(__v1,__v6)}
\fmfiset{p4}{vpath(__v1,__v7)}
\fmfiset{p5}{vpath(__v1,__v8)}
\fmfiset{p6}{vpath(__v1,__v3)}
\fmfipair{w[]}
\fmfiequ{w1}{point 3length(p1)/4 of p1}
\fmfiequ{w2}{point 3length(p2)/4 of p2}
\fmfiequ{w3}{point 3length(p3)/4 of p3}
\fmfiequ{w4}{point 3length(p4)/4 of p4}
\fmfiequ{w5}{point 3length(p5)/4 of p5}
\fmfiequ{w6}{point 3length(p5)/4 of p6}
\fmfiv{l=$\alpha_1$,l.a=90,l.d=4}{w1}
\fmfiv{l=$\alpha_2$,l.a=90,l.d=3}{w2}
\fmfiv{l=$\alpha_3$,l.a=90,l.d=3}{w3}
\fmfiv{l=$\alpha_4$,l.a=-90,l.d=3}{w4}
\fmfiv{l=$\alpha_5$,l.a=-90,l.d=3}{w5}
\fmfiv{l=$\alpha_6$,l.a=-90,l.d=4}{w6}
\end{fmfchar*}}}\ \ \
\footnotesize
\begin{aligned}
&\phantom{A}\\
&\phantom{\gu}\\
&=\D(\a_4,\a_5)\rint{4}{5}(\a_1,\a_2,\a_3,f(\a_4,\a_5),\a_6)\\
&\quad+C(\a_4,\a_5)\gu(\a_3,\a_4+1)\rint{4}{5}(\a_1,\a_2,g(\a_3,\a_4+1),\a_5,\a_6)\\
&\quad+C(\a_5,\a_4)G(\a_5+1,\a_6)\rint{4}{5}(\a_1,\a_2,\a_3,\a_4,g(\a_5+1,\a_6))
\end{aligned}
\\
\\[1cm]
&\rint{1}{7}=
\scriptsize
\settoheight{\eqoff}{$\times$}%
\setlength{\eqoff}{0.5\eqoff}%
\addtolength{\eqoff}{-18\unitlength}%
\raisebox{\eqoff}{%
\fmfframe(0,0)(0,0){%
\begin{fmfchar*}(38,36)
\fmfleft{in}
\fmfright{out}
\fmf{plain,tension=1}{in,v1}
\fmf{plain,tension=1}{v4,out}
\fmfforce{(0,0.5h)}{in}
\fmfforce{(w,0.5h)}{out}
\fmfforce{(0.5w,0)}{v3}
\fmfforce{(0.5w,h)}{v2}
\fmfforce{(0.03w,0.5h)}{v1}
\fmfforce{(0.97w,0.5h)}{v4}
\fmfforce{(0.5w,0.83h)}{v5}
\fmfforce{(0.5w,0.66h)}{v6}
\fmfforce{(0.5w,0.5h)}{v7}
\fmfforce{(0.5w,0.34h)}{v8}
\fmfforce{(0.5w,0.17h)}{v9}
\fmffreeze
\fmf{plain}{v1,v2}
\fmf{plain}{v1,v5}
\fmf{plain}{v1,v6}
\fmf{derplain}{v1,v7}
\fmf{plain}{v1,v8}
\fmf{plain}{v1,v9}
\fmf{plain}{v1,v3}
\fmf{plain}{v3,v4}
\fmf{derplain}{v2,v4}
\fmf{plain}{v2,v3}
\fmfipath{p[]}
\fmfiset{p1}{vpath(__v1,__v2)}
\fmfiset{p2}{vpath(__v1,__v5)}
\fmfiset{p3}{vpath(__v1,__v6)}
\fmfiset{p4}{vpath(__v1,__v7)}
\fmfiset{p5}{vpath(__v1,__v8)}
\fmfiset{p6}{vpath(__v1,__v9)}
\fmfiset{p7}{vpath(__v1,__v3)}
\fmfipair{w[]}
\fmfiequ{w1}{point 3length(p1)/4 of p1}
\fmfiequ{w2}{point 3length(p2)/4 of p2}
\fmfiequ{w3}{point 3length(p3)/4 of p3}
\fmfiequ{w4}{point 3length(p4)/4 of p4}
\fmfiequ{w5}{point 3length(p5)/4 of p5}
\fmfiequ{w6}{point 3length(p5)/4 of p6}
\fmfiequ{w7}{point 3length(p5)/4 of p7}
\fmfiv{l=$\alpha_1$,l.a=90,l.d=4}{w1}
\fmfiv{l=$\alpha_2$,l.a=90,l.d=3}{w2}
\fmfiv{l=$\alpha_3$,l.a=90,l.d=3}{w3}
\fmfiv{l=$\alpha_4$,l.a=90,l.d=2}{w4}
\fmfiv{l=$\alpha_5$,l.a=-90,l.d=3}{w5}
\fmfiv{l=$\alpha_6$,l.a=-90,l.d=3}{w6}
\fmfiv{l=$\alpha_7$,l.a=-90,l.d=4}{w7}
\end{fmfchar*}}}\ \ \
\footnotesize
\begin{aligned}
&\phantom{A}\\
&\phantom{\gu}\\
&=\D(\a_2,\a_3)\rint{1}{6}(\a_1,f(\a_2,\a_3),\a_4,\a_5,\a_6,\a_7)\\
&\quad+C(\a_2,\a_3)G(\a_1,\a_2+1)\rint{1}{6}(g(\a_1,\a_2+1),\a_3,\a_4,\a_5,\a_6,\a_7)\\
&\quad+C(\a_3,\a_2)\gu(\a_4,\a_3+1)\rint{1}{6}(\a_1,\a_2,g(\a_4,\a_3+1),\a_5,\a_6,\a_7)
\end{aligned}
\end{align*}
\normalsize
\caption{Master integrals required for the computation of $\sint{j}{L}$ \textit{(continued)}}
\end{table}

\end{appendices}

\cleardoubleemptypage
\thispagestyle{empty}

\begin{backmatter}
\bibliographystyle{JHEP}
\bibliography{references}
\end{backmatter}

\end{fmffile}

\end{document}